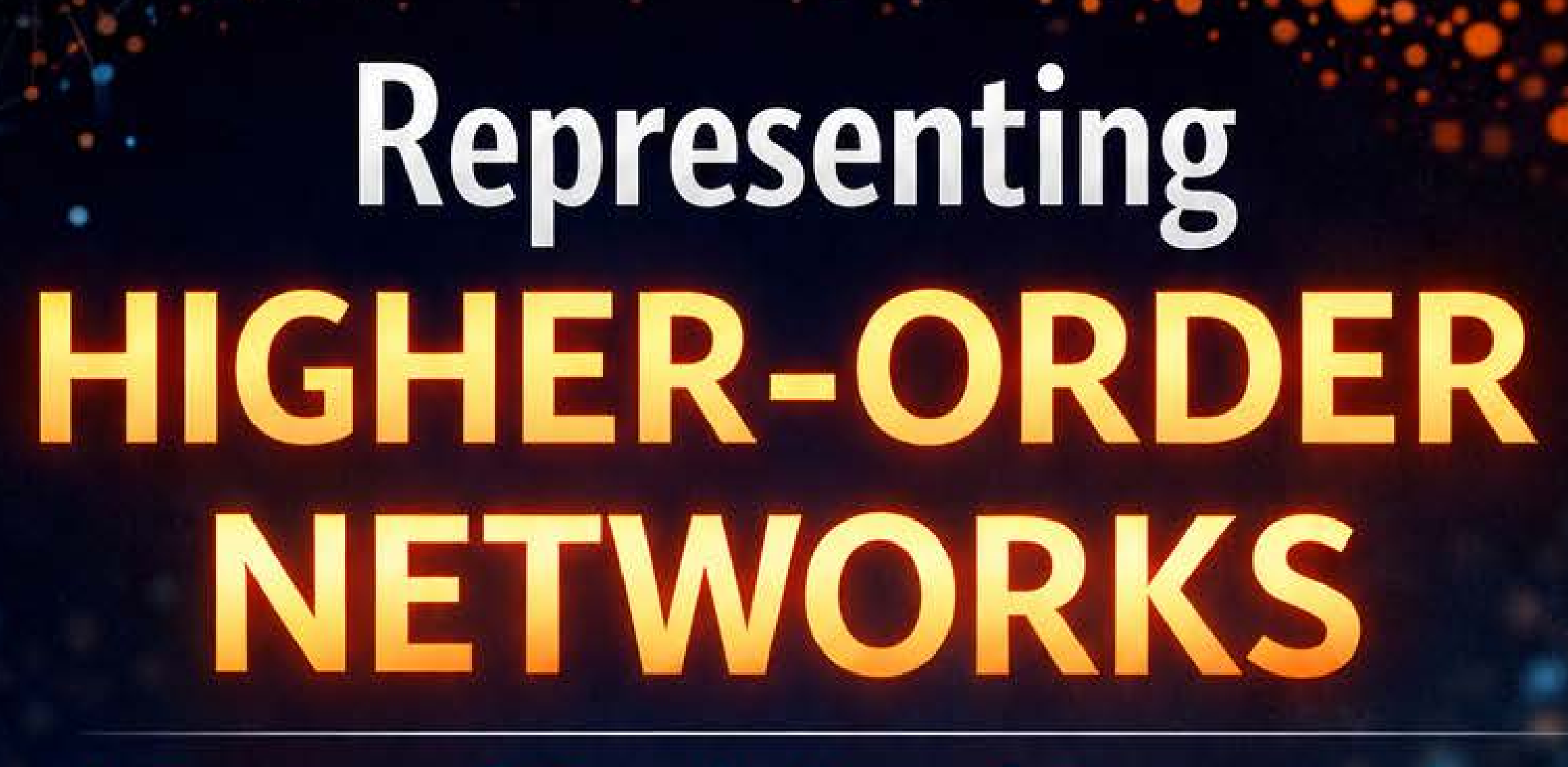

## A Survey of Graph-Based Frameworks

### Fourth Edition

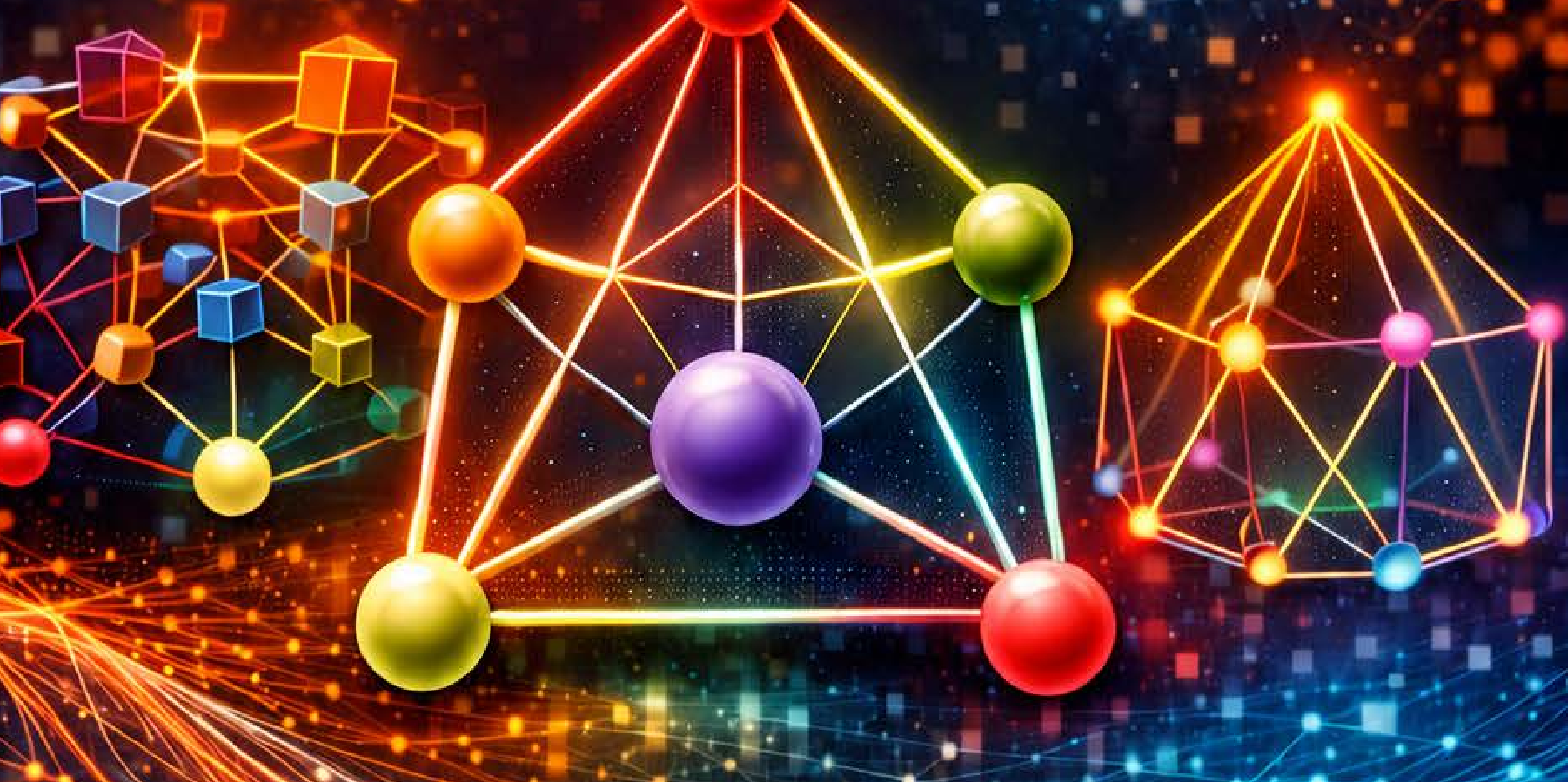

Takaaki Fujita
Florentin Smarandache

**Takaaki Fujita, Florentin Smarandache**

# Representing Higher-Order Networks: A Survey of Graph-Based Frameworks

(Fourth Edition)

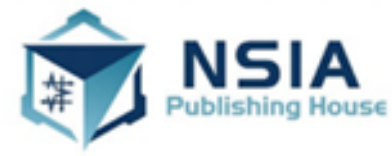

# Representing Higher-Order Networks: A Survey of Graph-Based Frameworks
# (Edition 4.0)

**Takaaki Fujita [1] * and Florentin Smarandache[2]**
[1] Independent Researcher, Tokyo, Japan.
Email: Takaaki.fujita060@gmail.com
[2] University of New Mexico, Gallup Campus, NM 87301, USA.
Email: fsmarandache@gmail.com

## Abstract

Many real-world phenomena are naturally modeled by graphs and networks. However, classical graph models are often limited to pairwise interactions and may not adequately capture the richer structures that arise in practice. Higher-order graph formalisms extend this framework by incorporating multiway, hierarchical, temporal, multilayer, recursive, and tensor-based interactions, thereby providing more expressive representations of complex systems.

This book presents a comprehensive overview of mathematical notions that can be used to model higher-order networks. It surveys foundational concepts, established extensions, and newly introduced formalisms, with an emphasis on their structural principles, relationships, and modeling roles. The aim is to provide a unified perspective that helps readers compare diverse higher-order network models and identify appropriate tools for theoretical study and practical applications.

This book is Edition 4.0. It mainly includes the addition of several concepts, as well as corrections and improvements of typographical errors and explanations. Compared with Edition 1.0, it contains substantial additions. Since Edition 3.0, typographical errors have been corrected, and some originally intended additions that had been omitted have now been included.

# 1 Introduction

## 1.1 Higher-Order Graphs

It is well known that many real-world phenomena can be modeled using graphs and networks [1, 2]. However, many such systems exhibit structures that go beyond pairwise interactions: they may involve multiway relations, hierarchical organization, nested or recursive dependencies, temporal evolution, or multilayer coupling. Classical graph models are often insufficient to represent these features in a mathematically faithful way.

To address this limitation, a wide range of higher-order formalisms has been developed, including hypergraphs [3, 4, 5], superhypergraphs [6], metagraph-based models [7, 8], simplicial and cell-complex-based frameworks, multilayer and temporal networks, and more recent category-theoretic or semantic approaches. For instance, superhypergraphs extend higher-order network models by allowing the *vertex domain itself* to be hierarchical, thereby enabling set-valued and iterated structures to be encoded directly in the object domain [6]. As a result, higher-order graph theory has grown into a broad and heterogeneous landscape, with many concepts arising from different mathematical viewpoints and modeling goals (cf. [9, 10, 11, 12, 13, 14]).

The concept of Higher-Order Networks (or Higher-Order Structures) has been applied in fields such as the following. Of course, the range of applications is not limited to these alone:

- **Transportation and logistics networks (cf. [15, 16, 17, 18, 19]):** Higher-order graphs can represent multi-stop delivery plans, hub coordination, shared routes, and group-wise flow constraints more naturally than ordinary pairwise graphs.
- **Social network analysis (cf. [20, 21, 22, 23, 24, 25]):** They model group conversations, team interactions, overlapping communities, and hierarchical memberships, going beyond simple person-to-person links.
- **Knowledge representation and semantic networks (cf. [26, 27, 28, 29]):** They are useful for encoding multi-entity relations, typed facts, contextual associations, and hierarchical semantic structures in knowledge systems.
- **Higher-Order Networks in Physics (cf. [30, 31, 32]):** They are useful for modeling multi-particle interactions, collective dynamics, higher-body correlations, lattice structures, and tensor-network representations in complex physical systems.
- **Molecular and chemical structure analysis (cf. [33, 34, 35, 36]):** They can describe multi-atom interactions, reaction mechanisms, molecular complexes, and higher-order structural dependencies in chemical systems.
- **Neuroscience and brain networks (cf.[37, 38, 39]):** They support the modeling of collective neural interactions, multi-region synchronization, layered brain connectivity, and time-dependent functional organization.
- **Machine learning and graph neural networks (cf. [40, 41, 42, 13, 43]):** Higher-order graphs are applied to learning tasks where relations involve groups, hierarchies, or nested structures, such as HyperGraph Neural Networks and related models.
- **Recommendation systems (cf.[44, 45, 46, 47, 48]):** They can simultaneously capture users, items, contexts, time, and attribute interactions, providing richer relational representations than ordinary bipartite graphs.
- **Supply chains and organizational systems (cf.[49, 50, 51]):** They model multi-party dependencies among suppliers, resources, departments, and processes, including hierarchical and cross-level coordination structures.
- **Communication and information networks (cf.[52, 53, 54, 55]):** They are suitable for representing multicast communication, layered protocols, group transmission, and dynamically changing higher-order connectivity patterns.

- **Decision-making and operations research (cf.[56, 57, 58, 59]):** They can express interacting criteria, grouped alternatives, hierarchical evaluation structures, and uncertainty-aware relational dependencies in complex decision problems.

- **VLSI[60, 61, 62, 63, 64]:** Higher-order networks represent multi-terminal nets, module interactions, routing constraints, circuit hierarchy, and collective dependencies, supporting VLSI placement, partitioning, optimization, and design analysis at multiple scales.

Beyond the field of graph theory, a substantial body of research on Higher-Order Networks has been conducted beyond the studies mentioned above [10, 65, 66, 67, 68, 69, 70, 71, 72, 73, 74]. Such frameworks have been applied to modeling increasingly complex real-world systems and phenomena.

## 1.2 Our Contributions

A wide variety of mathematical frameworks for representing higher-order networks has already been developed. However, these frameworks are often dispersed across different mathematical traditions, terminologies, and application areas, which makes systematic comparison difficult. For this reason, we consider it valuable to compile a survey-style book that brings these concepts together within a single coherent reference.

Accordingly, this book provides a broad and structured overview of mathematical notions that can be used to model higher-order networks. Its purpose is to offer a unified point of entry to these formalisms, to clarify their foundational ideas, and to highlight both their common features and their essential differences. In this way, the book is intended to support further theoretical development as well as applications in areas such as AI and related disciplines.

It is important to note, however, that the concepts collected here are not "higher-order" in one single uniform sense. Some frameworks generalize graphs by increasing the arity of interactions, as in hypergraph-type models. Others introduce hierarchy, nesting, or recursion, as in superhypergraph-type constructions. Still others encode higher-orderness through layers, temporal indexing, or multi-aspect organization, as in multilayer and temporal networks. Finally, some approaches arise from different mathematical semantics altogether, including operadic, monoidal, relational, tensor-based, closure-based, and coalgebraic viewpoints.

To make this diversity easier to understand and compare, the concepts in this book are organized into four broad families, in accordance with the practical classification adopted in the summary tables:

1. combinatorial, set-theoretic, and order-theoretic structures,

2. geometric, topological, and complex-based structures,

3. factorization-, constraint-, layered-, temporal-, and tensor-based structures, and

4. semantic, compositional, knowledge-based, and logical structures.

As a reference, Table 1.1 presents a practical four-family organization of the higher-order network concepts discussed in this book. This organization is not intended as a strict classification according to the intrinsic mathematical nature of the underlying objects. Rather, each concept is grouped according to its principal structural or modeling viewpoint. Since several frameworks combine ideas from more than one mathematical tradition, some concepts could reasonably be placed in more than one family.

Table 1.1: A practical four-family organization of the principal higher-order network concepts discussed in this book.

| **Family** | **Principal organizing viewpoint** | **Principal concepts discussed in this book** |
|---|---|---|
| **I. Combinatorial, set-theoretic, and order-theoretic** | This family emphasizes higher-order organization arising from combinatorial incidence, set systems, containment, recursion, iteration, hierarchical membership, multiplicity, order relations, hyperoperations, matroidal dependence, submodular splitting, and related algebraic or set-theoretic constructions. | Hyperstructure and Superhyperstructure; HyperGraph and SuperHyperGraph; MultiGraph and Iterated MultiGraph; $h$-model; Chain-Free Subsets; Power Set Graph; Johnson Graph; Kneser Graph; Meta-Graph and Iterated Meta-Graph; Meta-HyperGraph and Meta-SuperHyperGraph; Nested HyperGraph and Nested SuperHyperGraph; Multi-HyperGraph and Multi-SuperHyperGraph; Line Graph and Iterated Line Graph; Iterated Total Graph; Hierarchical SuperHyperGraph; Recursive HyperGraph and Recursive SuperHyperGraph; Tree-Vertex Graph; MultiMeta-Graph; Transfinite SuperHyperGraph; Graded SuperHyperGraph; HyperMatroid; SuperHyperMatroid; Kneser SuperHyperGraphs; Iterated Multi-Edge Graph; Iterated Multi-Recursive Graph; Multi-Axis SuperHyperGraph; MetaStructure and Iterated MetaStructure; MultiStructure and Iterative MultiStructure; TreeStructure; HyperCube, Meta-HyperCube, and $n$-SuperHyperCube; Ubergraph; Hb-Graph; Chygraph; Chy-SuperHyperGraph; $H_v$-Structure and $SH_v$-Structure; Submodular HyperGraph and Submodular $n$-SuperHyperGraph; Depth-$N$ Incidence SuperHyperGraph; Structured Quotient-Based Graph; Granular Graph; DAG-Based Graph; Constructor-Selectable Graph; Edge-Iterated SuperHyperGraph; Edge-Iterated HyperCube; Recursive MetaGraph and Recursive Iterated MetaGraph; Multiunion-Graph; Iterated SuperGraph. |
| **II. Geometric, topological, and complex-based** | This family emphasizes higher-order organization arising from geometric, topological, and complex-based objects, including simplices, cells, cubes, polyhedra, paths, ranked complexes, local-to-global structures, topological realizations, and multilevel refinement procedures. | Abstract Simplicial Complex; Simplicial Set; Cell Complex; CW Complex; Polyhedral Complex; Dowker Complex; Cubical Complex; Path Complex; Cellular Sheaf; Meta Simplicial Complex; Simplicial SuperHypercomplex; Combinatorial Complex; Depth-$r$ Iterated Subdivisions of Polyhedral Complexes; Topological SuperHyperGraph. |

*Table 1.1 (continued)*

| Family | Principal organizing viewpoint | Principal concepts discussed in this book |
|---|---|---|
| **III. Factorization, constraint, layered, temporal, and tensor-based** | This family emphasizes higher-order organization arising from factorization, coding and constraint relations, layer and time indexing, Cartesian-product constructions, multiscale organization, tensorial representations, contraction patterns, and quantum-state constructions. | Factor Graph; Tanner Graph; Tanner HyperGraph; Tanner $n$-SuperHyperGraph; Multilayer Network; Temporal Network; Temporal $n$-SuperHyperGraph; MultiDimensional Graph; Iterated MultiDimensional Graph; Tensor Network Graph; MultiTensor and Iterated MultiTensor Network; Tensor HyperNetwork and Tensor SuperHyperNetwork; Tensor Train; Tree Tensor Network (TTN); Projected Entangled Pair State (PEPS); Projected Entangled Simplex State (PESS); Adjacency-Tensor Network (ATN); HyperTensor Network Graph; SuperHyperTensor Network Graph; Quantum $n$-SuperHyperGraph; $n$-Filtrated Graph; Multiscale Graph; $m$-Multidynamic Graph; Multiinfinite Graph and Multiinfinite $n$-SuperHyperGraph; Multiuniverse-Graph. |
| **IV. Semantic, compositional, knowledge-based, and logical** | This family emphasizes meaning, typing, knowledge representation, interfaces, composition, transformation, logical implication, closure, uncertainty, parameterization, linguistic and similarity information, quantitative enrichment, decision processes, and learning-oriented forms of higher-order organization. | Open HyperGraph and Open SuperHyperGraph; Heterogeneous Graph, HyperGraph, and SuperHyperGraph; Knowledge Graph, HyperGraph, and SuperHyperGraph; Petri Net; Port Graph; Port HyperGraph and Port SuperHyperGraph; Combinatorial Map; Cognitive HyperGraphs and Cognitive SuperHyperGraphs; Multimodal Graph, HyperGraph, and SuperHyperGraph; Operadic Interaction Graph (OIG); Symmetric Monoidal Wiring Graph (SMWG); Relational-Arity Graph (RAG); Closure-Implication Graph (CIG); Coalgebraic Nested-Neighborhood Graph (CNNG); Curried Graph; Sheaf HyperGraph and Sheaf SuperHyperGraph; Fibered HyperGraph and Fibered SuperHyperGraph; Galois HyperGraph and Galois SuperHyperGraph; Rewrite HyperGraph and Rewrite SuperHyperGraph; Uncertain SuperHyperGraph; Functorial SuperHyperGraph; Motif HyperGraphs and Motif SuperHyperGraphs; Molecular SuperHyperGraphs; Soft $n$-SuperHyperGraph; Rough $n$-SuperHyperGraph; Decision $n$-SuperHyperTree; Weighted $n$-SuperHyperGraph; Similarity SuperHyperGraph; Linguistic Graph, Linguistic HyperGraph, and Linguistic SuperHyperGraph; $n$-SuperHyperGraph Neural Network; Random Forest, Random HyperForest, and Random SuperHyperForest; Compositional Graph; $n$-Iterated Labeling Graph; Structure-Vertexed Graph. |

This book is Edition 4.0. It mainly includes the addition of several new concepts, along with corrections of typographical errors and improvements in the explanations. Compared with Edition 1.0, it contains substantial additions. Since Edition 3.0, typographical errors have been corrected, and some originally intended additions that had previously been omitted have now been included. In Edition 4.0, several new concepts are introduced. A new chapter has been added for this purpose, which may serve as a useful reference.

## 1.3 Master Cross-Comparison of Higher-Order Network Concepts

To complement the family-wise organization adopted throughout this book, Table 1.2 provides a compact cross-comparison of the principal higher-order network concepts discussed in the main chapters.

Because these concepts originate from different mathematical traditions, the entries in the table should be interpreted as schematic summaries of their *principal structural characteristics*, rather than as exhaustive formal classifications. In particular, the final column lists a typical structure-preserving map appropriate to each framework; such a map need not constitute a uniquely standardized notion of morphism in the corresponding literature.

**Notation used in the table.**

- **VD** (*principal vertex or object domain*):

$$B = \text{base elements}, \qquad S = \text{subsets}, \qquad I = \text{iterated subsets},$$

$$T = \text{typed or structured objects}, \qquad N = \text{network or state nodes}.$$

  The symbol $T$ includes structured carriers such as graphs, tensors, cells, ports, algebraic structures, complexes, and similar mathematical objects. A notation such as $B/I$ indicates that different members of the grouped family use different domain types.

- **Ar** (*interaction arity or incidence form*):

$$2 = \text{pairwise}, \qquad \text{var} = \text{variable finite arity},$$

$$\text{fix} = \text{fixed finite arity}, \qquad \text{bip} = \text{bipartite incidence},$$

$$\text{loc} = \text{local attachment or contraction pattern}, \qquad \text{rec} = \text{recursive or nested interaction}.$$

- **H** (*principal locus of hierarchy or organization*):

$$- = \text{no explicit hierarchy}, \qquad V = \text{vertex or object side}, \qquad E = \text{edge or relation side},$$

$$C = \text{context, interface, layer, scale, time, or auxiliary structure}.$$

  Combined symbols such as $V/E$ or $V/E/C$ indicate that several structural components carry the relevant organization.

- **Feat.** (*additional structural feature*):

$$\text{Tm} = \text{temporal structure}, \qquad \text{Ly} = \text{layer, scale, grading, or filtration},$$

$$\text{Lg} = \text{logical, semantic, linguistic, typing, or inference structure},$$

$$\text{Cp} = \text{composition, interface, gluing, or contraction}, \qquad \text{Ts} = \text{tensorial structure},$$

$$\text{Wt} = \text{weights, cut costs, or quantitative valuations}, \qquad \text{Un} = \text{uncertainty structure},$$

$$\text{Qm} = \text{quantum-state or entanglement structure}, \qquad \text{ML} = \text{learning or message-passing structure},$$

  with $-$ indicating that none of these additional features is essential to the basic definition.

Table 1.2: Master cross-comparison of the principal higher-order network concepts treated in this book.

| Concept | VD | Ar | H | Feat. | Typical structure-preserving map |
|---|---|---|---|---|---|
| HyperGraph and SuperHyperGraph | $B/I$ | var | $E/V$ | – | Incidence-preserving HyperGraph or SuperHyperGraph morphism |
| MultiGraph and Iterated MultiGraph | $B/T$ | 2 | $V$ | – | Multiplicity-preserving MultiGraph homomorphism or level-preserving iterative map |
| $h$-model | $T$ | var | $C$ | Lg | Semantic-interpretation-preserving map |

*Table 1.2 (continued)*

| Concept | VD | Ar | H | Feat. | Typical structure-preserving map |
|---|---|---|---|---|---|
| Chain-Free Subsets | $S$ | var | $V$ | – | Order-preserving map compatible with the prescribed chain constraints |
| Power Set Graph | $S$ | 2 | $V$ | – | Graph homomorphism compatible with the subset construction |
| Johnson Graph | $S$ | 2 | $V$ | – | Graph homomorphism compatible with fixed-cardinality subset adjacency |
| Kneser Graph | $S$ | 2 | $V$ | – | Graph homomorphism compatible with the disjointness relation |
| Meta-Graph and Iterated Meta-Graph | $T$ | 2 | $V$ | – | Relation-preserving Meta-Graph morphism |
| Meta-HyperGraph and Meta-SuperHyperGraph | $T$ | var | $V/E$ | – | Incidence-preserving meta-level morphism |
| Nested HyperGraph and Nested SuperHyperGraph | $B/I$ | rec | $E/V$ | Cp | Nesting- and incidence-preserving morphism |
| Multi-HyperGraph and Multi-SuperHyperGraph | $B/I$ | var | $E/V$ | – | Multiplicity- and incidence-preserving morphism |
| Line Graph and Iterated Line Graph | $T$ | 2 | $V$ | – | Incidence-induced graph homomorphism |
| Iterated Total Graph | $T$ | 2 | $V/E$ | – | Total-graph-compatible homomorphism |
| Hierarchical SuperHyperGraph | $I$ | var | $V$ | Ly | Level- and incidence-preserving morphism |
| Recursive HyperGraph and Recursive SuperHyperGraph | $B/I$ | rec | $E/V$ | Cp | Recursion- and incidence-preserving morphism |
| Tree-Vertex Graph | $T$ | 2 | $V$ | Ly | Tree-structure-preserving graph homomorphism |
| MultiMeta-Graph | $T$ | 2 | $V$ | – | Meta-Graph morphism preserving family-valued vertices |
| Transfinite SuperHyperGraph | $I$ | var | $V$ | Ly | Ordinal- and level-preserving incidence morphism |
| Multi-Axis SuperHyperGraph | $I$ | var | $V$ | Ly | Multi-index- and incidence-preserving morphism |
| Iterated Multi-Edge Graph | $T$ | rec | $E$ | – | Edge-level multiplicity-preserving map |
| Iterated Multi-Recursive Graph | $T$ | rec | $V/E$ | – | Recursive multiplicity-preserving graph morphism |
| HyperMatroid | $B$ | var | $E$ | – | Matroid strong map or explicitly circuit-preserving map |
| SuperHyperMatroid | $I$ | var | $V/E$ | Ly | Level- and supercircuit-preserving map |
| Kneser SuperHyperGraphs | $I$ | 2 | $V$ | – | Flattened-support- and disjointness-preserving morphism |
| Graded SuperHyperGraph | $I$ | var | $V$ | Ly | Grade- and incidence-preserving morphism |
| Hyperstructures and Superhyperstructures | $B/I$ | var | $V$ | – | Hyperoperation-preserving homomorphism |
| MetaStructure and Iterated MetaStructure | $T$ | var | $V$ | Cp | Structure- and meta-operation-preserving morphism |
| MultiStructure and Iterative MultiStructure | $T$ | var | $V$ | – | Multiplicity- and multi-operation-preserving homomorphism |
| TreeStructure | $T$ | var | $V$ | Ly | Rooted-order- and operation-preserving homomorphism |
| HyperCube | $T$ | 2 | – | – | Cubical map or cube-graph homomorphism |
| Meta-HyperCube | $T$ | var | $V$ | Cp | Meta-operation-preserving map between HyperCube structures |
| $n$-SuperHyperCube | $I$ | 2 | $V$ | Ly | Level- and adjacency-preserving SuperHyperCube map |
| Ubergraph | $B/T$ | rec | $E$ | Cp | Membership-, nesting-, and incidence-preserving morphism |

*Table 1.2 (continued)*

| Concept | VD | Ar | H | Feat. | Typical structure-preserving map |
|---|---|---|---|---|---|
| Hb-Graph | $B$ | var | $E$ | – | Vertex-multiplicity- and incidence-preserving Hb-Graph morphism |
| Chygraph | $T$ | rec | $V/C$ | Cp | Atom-, complex-, and internal-incidence-preserving morphism |
| Chy-SuperHyperGraph | $T/I$ | rec | $V/E/C$ | Ly, Cp | Level-, support-, complex-, and incidence-preserving morphism |
| $H_v$-Structure and $SH_v$-Structure | $B/I$ | 2 | $V$ | – | Weak-hyperoperation-preserving homomorphism |
| Submodular HyperGraph and Submodular $n$-SuperHyperGraph | $B/I$ | var | $E/V$ | Wt | Incidence- and splitting-function-preserving morphism |
| Abstract Simplicial Complex | $B$ | var | $E$ | – | Simplicial map |
| Simplicial Set | $T$ | var | $C$ | – | Simplicial-set morphism commuting with face and degeneracy maps |
| Cell Complex | $T$ | var | $C$ | – | Cellular map |
| CW Complex | $T$ | var | $C$ | – | Cellular map |
| Polyhedral Complex | $T$ | var | $C$ | – | Polyhedral-complex map |
| Dowker Complex | $B$ | var | $C$ | – | Relation-induced simplicial map |
| Cubical Complex | $T$ | var | $C$ | – | Cubical map |
| Path Complex | $B$ | var | $E/C$ | – | Path-complex-preserving map |
| Cellular Sheaf | $T$ | var | $C$ | Cp | Cellular-sheaf morphism |
| Meta Simplicial Complex | $T$ | var | $V/C$ | – | Simplicial meta-morphism |
| Simplicial SuperHypercomplex | $I$ | var | $V/E$ | Ly | Simplicial-, level-, and incidence-preserving morphism |
| Combinatorial Complex | $B$ | var | $C$ | Ly | Cell-, inclusion-, and rank-compatible map |
| Depth-$r$ Iterated Subdivisions of Polyhedral Complexes | $T$ | var | $C$ | Ly, Cp | Subdivision-compatible cellular or polyhedral map |
| Topological SuperHyperGraph | $I$ | var | $C/V$ | – | Topology- and incidence-preserving morphism |
| Factor Graph | $N$ | bip | $C$ | Cp | Bipartition- and factor-incidence-preserving graph morphism |
| Tanner Graph | $N$ | bip | $C$ | Lg | Bipartition- and parity-incidence-preserving Tanner-graph morphism |
| Tanner HyperGraph | $B$ | var | $E$ | Lg | Incidence-preserving HyperGraph morphism |
| Tanner $n$-SuperHyperGraph | $I$ | var | $V/E$ | Lg | Level- and incidence-preserving morphism |
| Multilayer Network | $N$ | 2 | $C$ | Ly | Layer-preserving network morphism |
| Temporal Network | $N$ | 2 | $C$ | Tm | Time-respecting network morphism |
| MultiDimensional Graph (Cartesian-Product Graph) | $N$ | 2 | $C$ | Ly | Product-structure-preserving graph homomorphism |
| Iterated MultiDimensional Graph | $N/T$ | 2 | $C$ | Ly, Cp | Product-decomposition- and level-preserving graph homomorphism |
| Tensor Network Graph | $T$ | loc | $C$ | Ts, Cp | Index- and contraction-compatible tensor-network morphism |
| MultiTensor and Iterated MultiTensor Network | $T$ | loc | $V$ | Ts, Cp | Multiplicity-, level-, and contraction-compatible tensor-network map |
| Tensor Hypernetwork and Tensor Superhypernetwork | $T/I$ | var | $E/V$ | Ts, Cp | Incidence- and contraction-compatible morphism |
| Tensor Train | $T$ | loc | $C$ | Ts, Cp | Bond-index-compatible tensor-network map |
| Tree Tensor Network (TTN) | $T$ | loc | $V$ | Ts, Cp | Tree- and contraction-compatible tensor-network map |
| Projected Entangled Pair State (PEPS) | $T$ | loc | $C$ | Ts, Cp | Local-index- and contraction-compatible tensor-network map |

*Table 1.2 (continued)*

| Concept | VD | Ar | H | Feat. | Typical structure-preserving map |
|---|---|---|---|---|---|
| Projected Entangled Simplex State (PESS) | $T$ | loc | $E/C$ | Ts, Cp | Simplex- and contraction-compatible tensor-network map |
| Adjacency-Tensor Network (ATN) | $N$ | var | $C$ | Ts | Vertex-relabeling map compatible with the adjacency-tensor family |
| Temporal $n$-SuperHyperGraph | $I$ | var | $C/V$ | Tm | Time-, level-, and incidence-preserving morphism |
| Quantum $n$-SuperHyperGraph | $I$ | var | $V$ | Qm | SuperHyperGraph isomorphism together with the induced qubit relabeling |
| HyperTensor and SuperHyperTensor Network Graphs | $T$ | loc | $V/C$ | Ts, Cp | Contraction-compatible map preserving the lifted coefficient structure |
| Heterogeneous Graph, HyperGraph, and SuperHyperGraph | $T/I$ | 2/var | $C/V$ | Lg | Type- and incidence-preserving relational morphism |
| Knowledge Graph, HyperGraph, and SuperHyperGraph | $T/I$ | 2/var | $C/V$ | Lg | Typed relation- and incidence-preserving morphism |
| Petri Net | $T$ | bip | $C$ | Lg, Cp | Place-, transition-, and incidence-preserving Petri-net morphism |
| Port Graph | $T$ | 2 | $C$ | Cp | Port-preserving graph morphism |
| Port HyperGraph and Port SuperHyperGraph | $T/I$ | var | $C/V$ | Cp | Port- and incidence-preserving morphism |
| Open HyperGraph and Open SuperHyperGraph | $T/I$ | var | $C/V$ | Cp | Boundary- and interface-preserving open-structure morphism |
| Combinatorial Map | $B$ | 2 | $E$ | Lg | Color- and incidence-preserving combinatorial-map morphism |
| Cognitive HyperGraphs and Cognitive SuperHyperGraphs | $T/I$ | var | $C/V$ | Lg | Semantics- and incidence-preserving cognitive-network morphism |
| Multimodal Graph, HyperGraph, and SuperHyperGraph | $N/T/I$ | 2/var | $C/V$ | Lg | Modality- and incidence-preserving morphism |
| Operadic Interaction Graph (OIG) | $T$ | var | $C$ | Cp | Operad-compatible morphism |
| Symmetric Monoidal Wiring Graph (SMWG) | $T$ | var | $C$ | Cp | Symmetric-monoidal-structure-preserving morphism |
| Relational-Arity Graph (RAG) | $T$ | var | $C$ | Lg | Arity- and relation-preserving morphism |
| Closure-Implication Graph (CIG) | $T$ | 2/var | $C$ | Lg | Closure- and implication-preserving morphism |
| Coalgebraic Nested-Neighborhood Graph (CNNG) | $T$ | rec | $C$ | Lg | Coalgebra morphism or nested-neighborhood-preserving map |
| Curried Graph | $T$ | fix/var | $C$ | Cp | Currying- and evaluation-compatible morphism |
| Sheaf HyperGraph and Sheaf SuperHyperGraph | $T/I$ | var | $C/V$ | Lg | Sheaf- and incidence-preserving morphism |
| Fibered HyperGraph and Fibered SuperHyperGraph | $T/I$ | var | $C/V$ | Cp | Fiber- and incidence-preserving morphism |
| Galois HyperGraph and Galois SuperHyperGraph | $T/I$ | var | $C/V$ | Lg | Galois-connection- and incidence-preserving morphism |
| Rewrite HyperGraph and Rewrite SuperHyperGraph | $T/I$ | var | $C/V$ | Cp | Rewrite-compatible incidence morphism |
| Uncertain SuperHyperGraph | $I$ | var | $V/E$ | Un | Uncertainty- and incidence-preserving morphism |
| Functorial SuperHyperGraph | $I$ | var | $C/V$ | Cp | Functor-compatible incidence morphism |
| Motif HyperGraphs and Motif SuperHyperGraphs | $B/I$ | var | $E/V$ | – | Motif- and incidence-preserving morphism |

*Table 1.2 (continued)*

| **Concept** | **VD** | **Ar** | **H** | **Feat.** | **Typical structure-preserving map** |
|---|---|---|---|---|---|
| Molecular SuperHyperGraphs | $I$ | var | $V$ | Lg | Molecular-label- and incidence-preserving morphism |
| Soft $n$-SuperHyperGraph | $I$ | var | $C/V$ | Un | Parameter- and incidence-preserving soft morphism |
| Rough $n$-SuperHyperGraph | $I$ | var | $C/V$ | Un | Approximation- and incidence-preserving rough morphism |
| Decision $n$-SuperHyperTree | $I$ | var | $V/E$ | Lg, Wt | Root-, branching-, type-, probability-, utility-, and incidence-preserving morphism |
| Weighted $n$-SuperHyperGraph | $I$ | var | $V$ | Wt | Weight- and incidence-preserving morphism |
| Linguistic Graph, HyperGraph, and SuperHyperGraph | $T/I$ | 2/var | $C/V$ | Lg | Linguistic-label- and incidence-preserving morphism |
| $n$-SuperHyperGraph Neural Network | $I$ | var | $V/E$ | ML | Incidence-preserving relabeling compatible with feature propagation |
| Similarity SuperHyperGraph | $I$ | var | $V$ | Wt | Support-, similarity-, threshold-, and incidence-compatible map |
| $n$-Filtrated Graph | $N$ | 2 | $C$ | Ly | Filtration-preserving graph morphism |
| Depth-$N$ Incidence SuperHyperGraph | $I$ | rec | $V/E$ | Ly | Depth- and incidence-preserving morphism |
| Structured Quotient-Based Graph | $T$ | 2 | $C$ | – | Quotient-compatible graph morphism |
| Granular Graph | $S$ | 2 | $C$ | Ly | Granulation-preserving graph morphism |
| Multiscale Graph | $N$ | 2 | $C$ | Ly | Scale- and transition-preserving graph morphism |
| DAG-Based Graph | $N/T$ | 2 | $C$ | Ly | DAG- and transition-compatible graph morphism |
| Compositional Graph | $T$ | var | $C$ | Cp | Composition- and interface-preserving graph morphism |
| $n$-Iterated Labeling Graph | $T$ | 2 | $C$ | Ly, Lg | Level- and label-preserving graph morphism |
| $m$-Multidynamic Graph | $N$ | 2/var | $C$ | Tm, Ly | Multi-time-respecting graph morphism |
| Structure-Vertexed Graph | $T$ | 2 | $V$ | Lg | Relation-preserving map between structure-valued vertices |
| Constructor-Selectable Graph | $T$ | 2 | $V$ | Cp | Constructor-compatible graph morphism |
| Multiinfinite Graph and Multiinfinite $n$-SuperHyperGraph | $N/I$ | 2/var | $C/V$ | Ly | Axis-, transition-, level-, and incidence-preserving morphism |
| Edge-Iterated SuperHyperGraph | $I$ | rec | $V/E$ | Ly | Level-, flattening-, and incidence-preserving morphism on both hierarchy sides |
| Edge-Iterated HyperCube | $I/T$ | rec | $E/V$ | Ly | Level-, flattening-, and cube-incidence-preserving map |
| Recursive MetaGraph and Recursive Iterated MetaGraph | $T$ | rec | $V/E$ | Cp | Recursion-preserving MetaGraph morphism |
| Multiuniverse-Graph | $T$ | 2 | $C$ | Ly | Universe- and adjacency-preserving graph morphism |
| Multiunion-Graph | $S/T$ | 2 | $V$ | – | Support- and adjacency-preserving graph morphism |
| Iterated SuperGraph | $T$ | 2 | $V$ | Ly | Supergraph-containment- and level-preserving digraph morphism |
| Random Forest, Random HyperForest, and Random SuperHyperForest | $T/I$ | 2/var | $V/E$ | ML | Treewise structure-preserving maps compatible with the ensemble aggregation rule |
| Complete Higher-Graphic Structure (CHGS) | $T$ | var | $V/E/C$ | Tm, Ly, Lg, Cp, Wt | CHGS morphism, strong embedding, or CHGS isomorphism |

# 2 Combinatorial, Set-Theoretic, and Order-Theoretic Family

This chapter surveys higher-order structures whose principal organizing mechanisms arise from combinatorial incidence, set systems, containment, multiplicity, recursion, hierarchy, order relations, hyperoperations, and related algebraic or set-theoretic constructions. These frameworks generalize ordinary graph structures in different directions, including multiway incidence, iterated powerset hierarchies, nested edges, multiset-valued interactions, structure-valued vertices, and weak hyperalgebraic operations.

For reference, the principal combinatorial, set-theoretic, and order-theoretic higher-order structures discussed in this book are summarized in Table 2.1.

Table 2.1: Principal combinatorial, set-theoretic, and order-theoretic higher-order structures discussed in this book.

| **Concept** | **Concise description** |
|---|---|
| HyperGraph and SuperHyperGraph | HyperGraphs represent multiway relations through hyperedges containing arbitrary nonempty subsets of vertices, whereas $n$-SuperHyperGraphs extend the vertex domain to hierarchical objects drawn from iterated powerset levels. |
| MultiGraph and Iterated MultiGraph | MultiGraphs permit repeated edges through multiplicities, while Iterated MultiGraphs extend multiplicity-based organization recursively through successive multiset constructions. |
| $h$-model | A semantic hypergraph-based framework in which propositional atoms are interpreted by structured hypergraphs over a prescribed common domain. |
| Chain-Free Subsets | Families of subsets of a partially ordered set that avoid chains of a prescribed length, providing order-theoretic representations of incomparability and hierarchical constraints. |
| Power Set Graph | A graph whose vertices are selected subsets of an underlying set and whose adjacency is determined by a prescribed set-theoretic relation, such as proper inclusion. |
| Johnson Graph | A graph whose vertices are fixed-cardinality subsets of a finite set, with two vertices adjacent when their subsets differ by exactly one element. |
| Kneser Graph | A graph whose vertices are fixed-cardinality subsets of a finite set, with two vertices adjacent precisely when the corresponding subsets are disjoint. |
| Meta-Graph and Iterated Meta-Graph | Higher-level graph constructions in which graphs themselves serve as vertices, with the Meta-Graph construction iterated through successive levels. |
| Meta-HyperGraph and Meta-SuperHyperGraph | Higher-order incidence structures in which HyperGraphs or SuperHyperGraphs themselves serve as higher-level vertex objects. |
| Nested HyperGraph and Nested SuperHyperGraph | Recursive incidence structures in which hyperedges or superhyperedges may contain lower-level edge objects, subject to an appropriate well-founded nesting condition. |

*Table 2.1 (continued)*

| **Concept** | **Concise description** |
|---|---|
| Multi-Hypergraph and Multi-SuperHypergraph | HyperGraph and SuperHyperGraph variants in which hyperedges or superhyperedges may occur with nontrivial multiplicities. |
| Line Graph and Iterated Line Graph | Graph transformations in which edges become vertices and adjacency is induced by incidence, with the line-graph construction applied repeatedly in the iterated case. |
| Iterated Total Graph | Repeated total-graph constructions in which both vertices and edges at one level become vertices at the next, with adjacency determined by the original adjacency and incidence relations. |
| Hierarchical SuperHyperGraph | A SuperHyperGraph containing supervertices from multiple iterated powerset levels, thereby allowing incidence relations within and across different hierarchical levels. |
| Recursive HyperGraph and Recursive SuperHyperGraph | HyperGraph-type structures whose incidence objects may recursively contain lower-level hyperedges, supervertices, or superhyperedges. |
| Tree-Vertex Graph | A graph whose vertex objects are organized by an underlying rooted-tree hierarchy, allowing ordinary graph relations to coexist with explicit parent–child organization. |
| MultiMeta-Graph | A higher-level graph whose vertices are finite nonempty families of graphs, with adjacency determined by prescribed relations between such families. |
| Transfinite SuperHyperGraph | A SuperHyperGraph whose hierarchical supervertices may occur at ordinally indexed powerset levels, thereby extending finite powerset hierarchies to transfinite ranks. |
| Multi-Axis SuperHyperGraph | A SuperHyperGraph whose hierarchical supervertices are indexed along several independent iterated powerset axes, producing a multi-indexed higher-order organization. |
| Iterated Multi-Edge Graph | A graph-like structure in which edge objects are represented by iterated finite multisets of endpoint pairs, thereby encoding multiplicity at several levels. |
| Iterated Multi-Recursive Graph | A recursive graph-like structure combining iterated multiset-valued vertex objects with iterated multiset-valued edge objects. |
| HyperMatroid | A matroid represented through its circuit HyperGraph, whose hyperedges are the minimal dependent subsets satisfying the matroid circuit axioms. |
| SuperHyperMatroid | A matroid-type dependence structure whose ground objects are hierarchical supervertices and whose supercircuits satisfy corresponding minimality and circuit-elimination conditions. |
| Kneser SuperHypergraphs | SuperHyperGraph extensions of Kneser-type constructions in which disjointness relations are defined through the flattened supports of hierarchical supervertices. |
| Graded SuperHyperGraph | A SuperHyperGraph whose supervertices carry explicit grades or levels, allowing incidence relations to distinguish or connect different positions in the powerset hierarchy. |
| Hyperstructure and Superhyperstructure | Algebraic structures based on set-valued hyperoperations, together with hierarchical extensions whose carriers or operations are lifted through iterated powerset levels. |
| MetaStructure and Iterated MetaStructure | Frameworks in which complete mathematical structures are treated as higher-level objects and uniform meta-operations act on them, with the construction repeatedly lifted to successive meta-levels. |

*Table 2.1 (continued)*

| Concept | Concise description |
|---|---|
| MultiStructure and Iterative MultiStructure | Algebraic structures whose operations return finite multisets, together with iterative versions whose operations act successively on higher multiset levels. |
| TreeStructure | A finite rooted partially ordered set whose principal ideals are chains, optionally equipped with finitary operations acting on its tree-ordered carrier. |
| HyperCube, Meta-HyperCube, and $n$-SuperHyperCube | The HyperCube provides a canonical multidimensional cube structure; Meta-HyperCubes treat HyperCubes as structure-valued objects, while $n$-SuperHyperCubes replace ordinary coordinates by hierarchical coordinates drawn from an iterated powerset level. |
| Ubergraph | A finite recursively nested hypergraph-like structure in which an edge may contain fundamental vertices or previously constructed edges, subject to a finite-depth or well-founded membership condition. |
| Hb-Graph | A HyperGraph generalization in which every hb-edge is a multiset-valued edge represented by a multiplicity function on the vertex set, allowing a vertex to occur with nonbinary multiplicity within one interaction. |
| Chygraph | A recursively structured system of atoms and complexes in which every complex is assigned an internal HyperGraph whose vertices may themselves be atoms or other complex labels. |
| Chy-SuperHyperGraph | A hierarchical extension of a Chygraph in which each complex is assigned an internal $n$-SuperHyperGraph, with the SuperHyperGraph level allowed to vary from one complex to another. |
| $H_v$-Structure and $SH_v$-Structure | Weak hyperalgebraic structures in which selected equalities are replaced by nonempty-intersection conditions; $SH_v$-Structures extend this principle to carriers formed from iterated powerset levels. |
| Submodular Hypergraph and Submodular $n$-SuperHyperGraph | Higher-order cut structures assigning a symmetric submodular splitting function to each hyperedge or superhyperedge incidence set, allowing different partitions of the same interaction to have different costs. |

## 2.1 HyperGraph and SuperHyperGraph

Superhypergraphs extend higher–order network models by allowing the *vertex domain itself* to be hierarchical[6]. Concretely, one starts from a base set and iterates the powerset operation; vertices (often called *supervertices*) may then be set-valued objects living at a prescribed level of this iteration, while (super)edges encode incidence among these higher-level vertices [6]. Related hierarchical constructions have been explored in applications [75, 76, 77, 78].

In addition, several extensions of hypergraphs are known, including fuzzy hypergraphs[79, 80, 81, 82, 83], intuitionistic fuzzy hypergraphs [84, 85, 86, 87], neutrosophic hypergraphs[88, 89, 90], soft hypergraphs[91, 92], and plithogenic hypergraphs [93]. In particular, Fuzzy HyperGraphs have been extensively studied in a wide variety of research papers[94, 95]. Likewise, extensions of superhypergraphs such as fuzzy superhypergraphs [96, 97, 58], neutrosophic superhypergraphs [98, 99, 100], and plithogenic superhypergraphs [101, 102, 103] have been studied. Additionally, as oriented graph concepts, the following are known: Directed HyperGraph[104, 105], Bidirected HyperGraph[106], Directed SuperHyperGraph [107, 108], Oriented Hypergraph[109, 110, 111, 112], Oriented SuperHypergraph, and Bidirected SuperHyperGraph [113, 106]. A wide range of studies has explored various applications of HyperGraphs and SuperHyperGraphs. For example, concepts and applications such as hypergraph partitioning [60, 114, 115, 116], hypergraph clustering [117, 118], and hypergraph matching[119, 120] have been extensively studied in the field of HyperGraphs. For a broader overview, we refer the reader to the survey monograph [121].

**Definition 2.1.1** (Iterated powerset). [122] For $k \in \mathbb{N}_0$, define iterated powersets recursively by

$$\mathcal{P}^{0}(X) := X, \qquad \mathcal{P}^{k+1}(X) := \mathcal{P}\big(\mathcal{P}^{k}(X)\big).$$

For the nonempty variant, set

$$\big(\mathcal{P}^{*}\big)^{0}(X) := X, \qquad \big(\mathcal{P}^{*}\big)^{k+1}(X) := \mathcal{P}^{*}\big((\mathcal{P}^{*})^{k}(X)\big),$$

where $\mathcal{P}^{*}(Y) := \mathcal{P}(Y) \setminus \{\varnothing\}$.

**Definition 2.1.2** (Hypergraph). [3, 123] A *hypergraph* is a pair $H = (V(H), E(H))$ such that $V(H) \neq \varnothing$ and

$$E(H) \subseteq \mathcal{P}^{*}(V(H)).$$

Throughout this book, both $V(H)$ and $E(H)$ are assumed finite.

**Example 2.1.3** (Team-based bug triage as a HyperGraph). Consider a software-development team consisting of five engineers,

$$V(H) = \{\mathsf{Alice}, \mathsf{Bob}, \mathsf{Chen}, \mathsf{Dina}, \mathsf{Eli}\}.$$

Suppose that several software bugs must be examined by different groups of engineers. Since a bug may require simultaneous participation by more than two engineers, an ordinary graph is not always sufficient to represent the resulting collaboration structure. We therefore model each bug-triage group as a hyperedge.

Assume that the following four bug reports are under consideration:

$$\begin{aligned}
e_1 &= \{\mathsf{Alice}, \mathsf{Bob}, \mathsf{Chen}\}, && \text{authentication failure,}\\
e_2 &= \{\mathsf{Bob}, \mathsf{Dina}\}, && \text{database synchronization error,}\\
e_3 &= \{\mathsf{Chen}, \mathsf{Dina}, \mathsf{Eli}\}, && \text{API performance degradation,}\\
e_4 &= \{\mathsf{Alice}, \mathsf{Eli}\}, && \text{release-deployment failure.}
\end{aligned}$$

Hence,

$$E(H) = \{e_1, e_2, e_3, e_4\} \subseteq \mathcal{P}^{*}(V(H)),$$

and therefore

$$H = (V(H), E(H))$$

is a finite HyperGraph.

For example, the hyperedge $e_1$ represents a triage meeting involving Alice, Bob, and Chen simultaneously, whereas $e_3$ represents another three-person interaction involving Chen, Dina, and Eli. Thus, the model retains multi-person collaborations as single relational objects rather than decomposing them into several independent pairwise edges.

Figure 2.1 gives a graphical representation of this HyperGraph.

The example also illustrates overlapping responsibilities. In particular, Bob belongs to both $e_1$ and $e_2$, Chen belongs to both $e_1$ and $e_3$, Dina belongs to both $e_2$ and $e_3$, and Alice belongs to both $e_1$ and $e_4$. Such overlaps are naturally represented in the HyperGraph framework.

**Definition 2.1.4** ($n$-SuperHyperGraph). [6, 124] Fix a finite base set $V_0$ and an integer $n \in \mathbb{N}_0$. An *$n$-SuperHyperGraph over $V_0$* is a triple

$$\mathrm{SHG}^{(n)} = (V, E, \partial),$$

where

- $V \subseteq \mathcal{P}^{n}(V_0)$ is a finite set of *$n$-supervertices*;
- $E$ is a finite set of *(super)edge identifiers*;
- $\partial : E \to \mathcal{P}^{*}(V)$ is an *incidence map* such that $\partial(e) \subseteq V$ is a nonempty finite set for every $e \in E$.

The set $\partial(e)$ is called the *incidence set* (or *superincidence set*) of $e$.

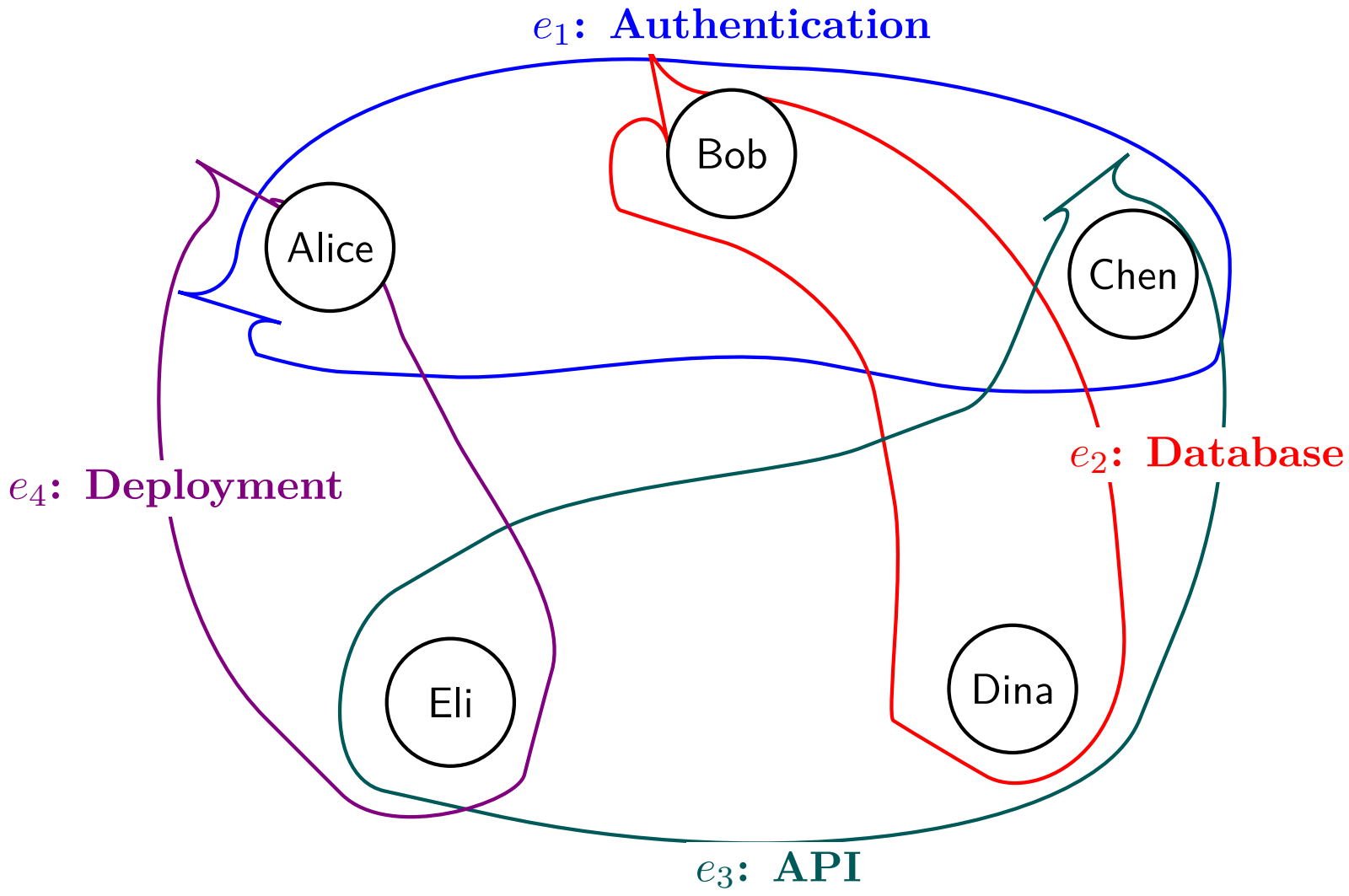

Each closed region represents one bug-triage hyperedge.
An engineer may participate in several hyperedges.

Figure 2.1: A team-based bug-triage HyperGraph. Each hyperedge represents the group of engineers jointly responsible for investigating one bug report.

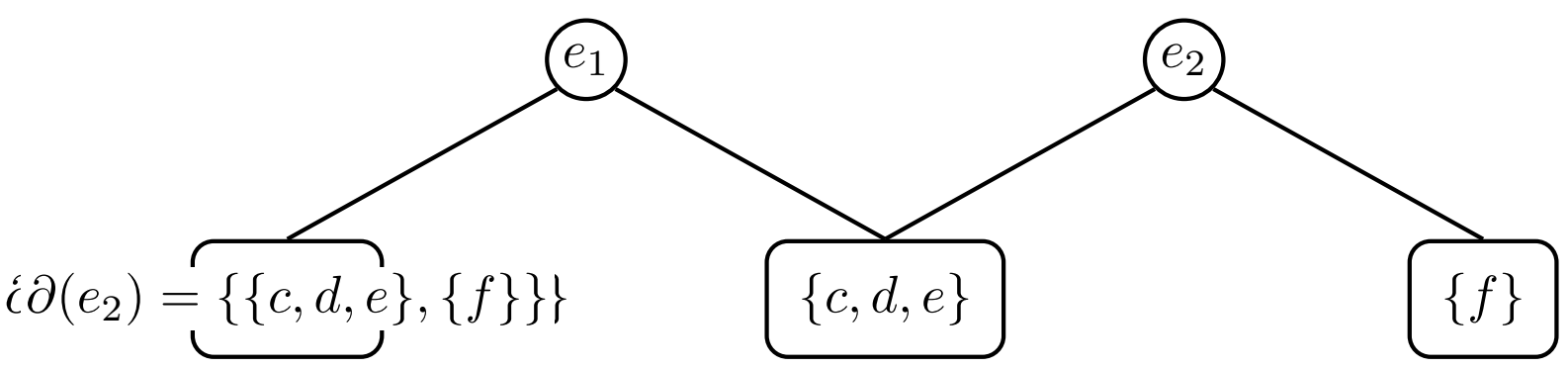

Figure 2.2: A two-level organization chart modeled as a 1-SuperHyperGraph $\mathrm{SHG}^{(1)} = (V, E, \partial)$. Teams are 1-supervertices, and $e_1, e_2$ are superedges with incidence given by $\partial$.

**Example 2.1.5** (Two-level organization chart as a 1-SuperHyperGraph)**.** Let the base set list individual employees:

$$V_0 = \{a, b, c, d, e, f\}.$$

Consider *teams* as 1-supervertices (subsets of $V_0$). Define

$$V = \{\{a, b\},\ \{c, d, e\},\ \{f\}\} \subseteq \mathcal{P}^1(V_0) = \mathcal{P}(V_0).$$

Now define a set of superedge identifiers $E = \{e_1, e_2\}$ and the incidence map

$$\partial : E \to \mathcal{P}^*(V)$$

by

$$\partial(e_1) = \{\{a, b\}, \{c, d, e\}\}, \qquad \partial(e_2) = \{\{c, d, e\}, \{f\}\}.$$

Then $\mathrm{SHG}^{(1)} = (V, E, \partial)$ is a 1-SuperHyperGraph over $V_0$ in the sense of Definition 2.1.4. Here $e_1$ encodes a collaboration between the two teams $\{a, b\}$ and $\{c, d, e\}$, while $e_2$ encodes a coordination link between the team $\{c, d, e\}$ and the individual team $\{f\}$. A reference illustration for this example is provided in Fig. 2.2.

**Example 2.1.6** (A 2-SuperHyperGraph with nested supervertices)**.** Let the base set be

$$V_0 = \{a, b, c\}.$$

For $n = 2$, we have $\mathcal{P}^2(V_0) = \mathcal{P}(\mathcal{P}(V_0))$, so a 2-supervertex is a *set of subsets* of $V_0$. Define the 2-supervertex set

$$V = \Big\{ v_1 = \{\{a\}, \{a, b\}\}, \quad v_2 = \{\{b\}, \{b, c\}\}, \quad v_3 = \{\{c\}\} \Big\} \ \subseteq\ \mathcal{P}^2(V_0).$$

Let the superedge identifier set be

$$E = \{e_1, e_2\},$$

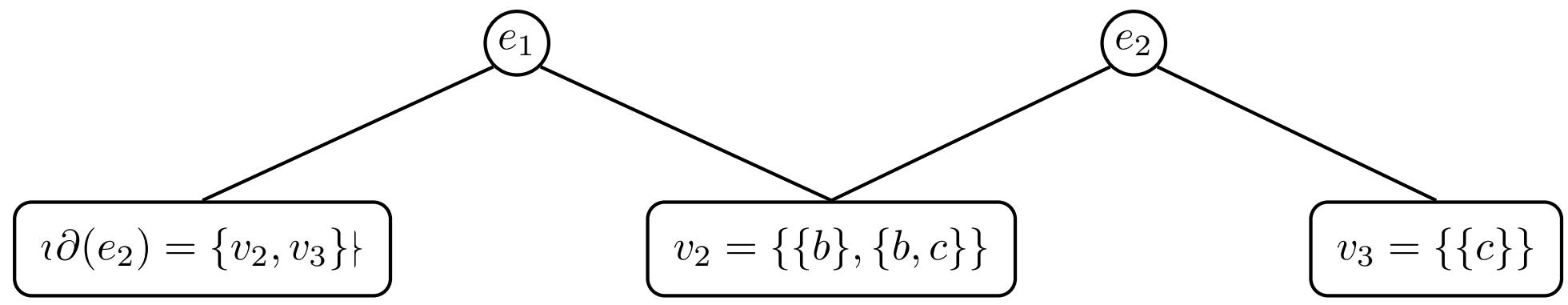

Figure 2.3: A 2-SuperHyperGraph $\mathrm{SHG}^{(2)} = (V, E, \partial)$ over $V_0 = \{a, b, c\}$. Each 2-supervertex is a set of subsets of $V_0$, and $e_1, e_2$ are superedges with incidence given by $\partial$.

and define the incidence map $\partial : E \to \mathcal{P}^*(V)$ by

$$\partial(e_1) = \{v_1, v_2\}, \qquad \partial(e_2) = \{v_2, v_3\}.$$

Then $\mathrm{SHG}^{(2)} = (V, E, \partial)$ is a 2-SuperHyperGraph over $V_0$ in the sense of Definition 2.1.4. Here $e_1$ links the two nested supervertices $v_1$ and $v_2$, while $e_2$ links $v_2$ and $v_3$.

A $(m, n)$-SuperHyperGraph is a mathematical structure in which each vertex corresponds to an $(m, n)$-superhyperfunction defined on a base set, while the hyperedges group such functions together to represent higher-order relationships and contextual connections. An $(h, k)$-ary $(m, n)$-SuperHyperGraph further generalizes this idea by taking vertices as $(h, k)$-ary $(m, n)$-superhyperfunctions.

**Notation 2.1.7.** For a nonempty base set $S$ define

$$\mathcal{P}_0(S) := S, \qquad \mathcal{P}_{m+1}(S) := \mathcal{P}\big(\mathcal{P}_m(S)\big) \quad (m \in \mathbb{N}_0),$$

so $\mathcal{P}_1(S) = \mathcal{P}(S)$, $\mathcal{P}_2(S) = \mathcal{P}(\mathcal{P}(S))$, etc. We also use the Cartesian power $X^h := \underbrace{X \times \cdots \times X}_{h \text{ copies}}$ for $h \in \mathbb{N}$.

**Definition 2.1.8** (*$(m, n)$-superhyperfunction*)**.** [125, 126] Let $m, n \in \mathbb{N}$ and $S \neq \varnothing$. An *$(m, n)$-superhyperfunction on $S$* is a map

$$f : \; \mathcal{P}_m(S) \;\longrightarrow\; \mathcal{P}_n(S).$$

Equivalently, $f \in \mathrm{Hom}\big(\mathcal{P}_m(S),\, \mathcal{P}_n(S)\big)$ as functions of sets.

**Definition 2.1.9** (*$(m, n)$-SuperHyperGraph*)**.** Fix $m, n \in \mathbb{N}$ and a nonempty base set $S$. Let

$$\mathfrak{F}_{m,n}(S) \;:=\; \Big\{\, f : \mathcal{P}_m(S) \to \mathcal{P}_n(S) \,\Big\}.$$

An *$(m, n)$-SuperHyperGraph* is a pair

$$\mathsf{SHG}^{(m,n)} \;:=\; (V,\; \mathcal{E}),$$

where $V \subseteq \mathfrak{F}_{m,n}(S)$ is a nonempty set of vertices (each vertex is a concrete $(m, n)$-superhyperfunction) and

$$\varnothing \neq \mathcal{E} \;\subseteq\; \mathcal{P}(V) \setminus \{\varnothing\}$$

is a nonempty family of nonempty *hyperedges*. Each hyperedge $E \in \mathcal{E}$ groups a finite, nonempty set of superhyperfunctions to encode higher-order relations/constraints among them.

**Example 2.1.10** (A concrete $(2, 1)$-SuperHyperGraph on $S = \{a, b\}$)**.** Let the base set be

$$S = \{a, b\}.$$

Then

$$\mathcal{P}_1(S) = \mathcal{P}(S) = \{\emptyset, \{a\}, \{b\}, \{a, b\}\}, \qquad \mathcal{P}_2(S) = \mathcal{P}(\mathcal{P}(S)).$$

We construct a small family of $(2, 1)$-superhyperfunctions, i.e. maps

$$f : \mathcal{P}_2(S) \to \mathcal{P}_1(S) = \mathcal{P}(S).$$

Define three functions $f_1, f_2, f_3 \in \mathfrak{F}_{2,1}(S)$ as follows. For any $X \in \mathcal{P}_2(S) = \mathcal{P}(\mathcal{P}(S))$ (so $X$ is a set of subsets of $S$), set

$$f_1(X) := \bigcup_{A \in X} A,$$

Table 2.2: A concise comparison of graphs, hypergraphs, $n$-superhypergraphs, and $(m, n)$-superhypergraphs.

| Structure | Vertex domain | Edge / incidence object | What it captures (one line) |
|---|---|---|---|
| Graph $G = (V, E)$ | $V$ is a (finite) set of *atomic* vertices | $E \subseteq \mathcal{P}_2^*(V) = \{\{u, v\} \subseteq V : u \neq v\}$ (pairwise edges) | Pairwise interactions only (binary relations). |
| Hypergraph $H = (V, E)$ | $V$ is a (finite) set of atomic vertices | $E \subseteq \mathcal{P}^*(V)$ (hyperedges are nonempty subsets of $V$) | Multiway interactions among arbitrary-size groups. |
| $n$-SuperHyperGraph $\mathrm{SHG}^{(n)} = (V, E, \partial)$ | $V \subseteq \mathcal{P}^n(V_0)$ (vertices are $n$-*level* set-valued objects over a base set $V_0$) | $E$ is a set of edge identifiers and $\partial : E \to \mathcal{P}^*(V)$ (incidence among supervertices) | Higher-order interactions *and* hierarchical/nested vertex semantics via iterated powersets. |
| $(m, n)$-SuperHyperGraph $\mathsf{SHG}^{(m,n)} = (V, \mathcal{E})$ | $V \subseteq \mathfrak{F}_{m,n}(S) = \{f : \mathcal{P}_m(S) \to \mathcal{P}_n(S)\}$ (vertices are *superhyperfunctions*) | $\varnothing \neq \mathcal{E} \subseteq \mathcal{P}(V) \setminus \{\varnothing\}$ (hyperedges group such functions) | Higher-order relations among *operators* between hierarchical domains/codomains (contextual constraints on maps). |

(the union of all subsets in $X$),

$$f_2(X) \coloneqq \bigcap_{A \in X} A,$$

(with the convention $\bigcap \emptyset = S$), and

$$f_3(X) \coloneqq \begin{cases} \{a\}, & \text{if } \{a\} \in X, \\ \emptyset, & \text{otherwise.} \end{cases}$$

Let the vertex set be the nonempty set of concrete $(2, 1)$-superhyperfunctions

$$V = \{f_1, f_2, f_3\} \subseteq \mathfrak{F}_{2,1}(S).$$

Define a nonempty family of hyperedges by

$$\mathcal{E} = \{E_1, E_2\} \subseteq \mathcal{P}(V) \setminus \{\emptyset\}, \qquad E_1 = \{f_1, f_2\}, \quad E_2 = \{f_2, f_3\}.$$

Then

$$\mathsf{SHG}^{(2,1)} = (V, \mathcal{E})$$

is a concrete $(2, 1)$-SuperHyperGraph in the sense of Definition 2.1.9. Here $E_1$ groups the "aggregation" superhyperfunctions $f_1$ (union) and $f_2$ (intersection), while $E_2$ groups $f_2$ with the indicator-type superhyperfunction $f_3$, representing a different contextual relationship among superhyperfunctions.

For reference, a comparison of graphs, hypergraphs, n-superhypergraphs, and $(m, n)$-superhypergraphs is provided in Table 2.2.

Note that the above graph is finite, but one may also consider infinite graphs. An Infinite SuperHyperGraph is a superhypergraph with infinitely many base elements, supervertices, or superhyperedges, while preserving the hierarchical powerset-based higher-order incidence structure. However, the graphs discussed in this book are basically finite.

**Definition 2.1.11** (Infinite $n$-SuperHyperGraph)**.** Let $V_0 \neq \varnothing$ be a (possibly infinite) base set, and define the iterated powersets by

$$\mathcal{P}^0(V_0) := V_0, \qquad \mathcal{P}^{k+1}(V_0) := \mathcal{P}\big(\mathcal{P}^k(V_0)\big) \quad (k \geq 0).$$

Fix $n \in \mathbb{N}_0$. An *Infinite $n$-SuperHyperGraph* on $V_0$ is a pair

$$\mathrm{SHG}_{\infty}^{(n)} = (V, E),$$

such that

$$V \subseteq \mathcal{P}^n(V_0)$$

and

$$E \subseteq \mathcal{P}(V) \setminus \{\varnothing\}.$$

The elements of $V$ are called *$n$-supervertices*, and the elements of $E$ are called *$n$-superhyperedges.*

If at least one of $V_0$, $V$, or $E$ is infinite, then

$$\mathrm{SHG}_{\infty}^{(n)}$$

is called a *genuinely infinite $n$-SuperHyperGraph.*

## 2.2 MultiGraph and Iterated MultiGraph

A multigraph is a graph allowing parallel edges and loops; formally edges are a multiset with multiplicities between vertices possibly [127, 128, 129]. As extensions, concepts such as fuzzy multigraphs [130, 131, 132], bipartite multigraphs[133, 134], complete multigraphs[135, 136], intuitionistic fuzzy multigraphs[137, 132], neutrosophic multigraphs [127, 138], soft multigraphs[139], and directed multigraphs [140] are known. An iterated multigraph uses iterated multisets as vertex objects, so vertices themselves can be multisets nested to depth n recursively.

**Definition 2.2.1** (Finite multiset and iterated multiset)**.** Let $X$ be a set. A *finite multiset* on $X$ is a function

$$m : X \to \mathbb{N}_0$$

whose *support*

$$\operatorname{supp}(m) := \{x \in X \mid m(x) > 0\}$$

is finite. We write $\mathsf{M}(X)$ for the set of all finite multisets on $X$.

For $n \geq 0$, define the *n-fold iterated multiset sets* recursively by

$$\mathsf{M}^0(X) := X, \qquad \mathsf{M}^{n+1}(X) := \mathsf{M}\big(\mathsf{M}^n(X)\big).$$

An element of $\mathsf{M}^n(X)$ is called an *n-fold iterated multiset over X*.

**Definition 2.2.2** (MultiGraph (undirected multigraph))**.** Let $V$ be a finite set. Denote by

$$\binom{V}{2}^{m} := \big\{\{\{u, v\}\} \;\big|\; u, v \in V\big\}$$

the set of *unordered pairs with repetition* (i.e. 2-element multisets), so that $\{\{v, v\}\}$ represents a loop at $v$.

A *MultiGraph* (undirected multigraph) on $V$ is a pair

$$G = (V, \mu),$$

where $\mu : \binom{V}{2}^m \to \mathbb{N}_0$ is an *edge-multiplicity function*. For $e \in \binom{V}{2}^m$, the value $\mu(e)$ is the number of parallel edges of type $e$. Equivalently, one may specify a finite multiset $E \in \mathsf{M}\Big(\binom{V}{2}^m\Big)$ and write $G = (V, E)$, where $\mu$ is the multiplicity function associated to $E$.

**Remark 2.2.3** (Directed variant)**.** *[141] A* directed *multigraph can be defined similarly by a multiplicity map $\mu : V \times V \to \mathbb{N}_0$, where $\mu(u, v)$ counts the number of directed edges from u to v.*

**Example 2.2.4** (Public transport routes as a MultiGraph)**.** Let

$$V = \{\mathsf{A}, \mathsf{B}, \mathsf{C}\}$$

be three stations. Suppose there are *two* distinct bus lines between A and B, *three* distinct train services between B and C, and one circular shuttle at A (a loop). Define the multiplicity map

$$\mu : \binom{V}{2}^m \to \mathbb{N}_0$$

by

$$\mu(\{\{\mathsf{A}, \mathsf{B}\}\}) = 2, \qquad \mu(\{\{\mathsf{B}, \mathsf{C}\}\}) = 3, \qquad \mu(\{\{\mathsf{A}, \mathsf{A}\}\}) = 1,$$

and $\mu(e) = 0$ for all other $e \in \binom{V}{2}^m$. Then $G = (V, \mu)$ is an undirected multigraph in the sense of Definition 2.2.2. For reference, an overview diagram is provided in Fig. 2.4.

**Definition 2.2.5** (Iterated MultiGraph of order $n$)**.** Let $X$ be a nonempty base set and let $n \geq 0$. Set $\mathsf{M}^n(X)$ as in Definition 2.2.1. An *Iterated MultiGraph of order n over X* is an undirected multigraph whose vertex objects are $n$-fold iterated multisets over $X$; concretely, it is a pair

$$G^{(n)} = (V^{(n)}, \mu^{(n)})$$

such that:

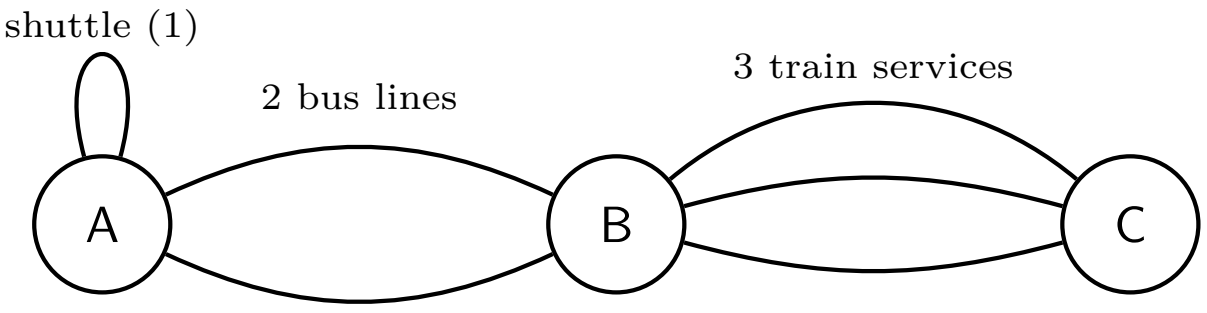

Figure 2.4: Public transport routes modeled as an undirected multigraph $G = (V, \mu)$ in Example 2.2.4: two parallel edges between A and B, three parallel edges between B and C, and a loop at A.

1. $V^{(n)} \in \mathsf{M}^n(X)$ is an $n$-fold iterated multiset, interpreted as a *vertex multiset*; let

$$\underline{V}^{(n)} := \mathrm{supp}(V^{(n)}) \subseteq \mathsf{M}^n(X)$$

   be the underlying *set* of distinct vertices.

2. $\mu^{(n)} : \binom{\underline{V}^{(n)}}{2}^m \to \mathbb{N}_0$ is an edge-multiplicity function on unordered pairs (with repetition) of distinct vertices.

Equivalently, one may specify an edge multiset $E^{(n)} \in \mathsf{M}\left(\binom{\underline{V}^{(n)}}{2}^m\right)$ and write $G^{(n)} = (V^{(n)}, E^{(n)})$.

**Remark 2.2.6** (Order 0 recovers ordinary multigraphs)**.** *When $n = 0$, we have $\mathsf{M}^0(X) = X$, so an Iterated MultiGraph of order 0 is just a multigraph whose vertices lie in the base set $X$ (up to the choice of vertex multiset versus vertex set).*

**Example 2.2.7** (Iterated MultiGraph of order 1 (multiset-vertices))**.** Let the base set be

$$X = \{a, b, c\}.$$

Consider the following 1-fold iterated multiset (a multiset of elements of $X$) as the vertex multiset:

$$V^{(1)} = \{\!\{\ \{\!\{a\}\!\},\ \{\!\{a\}\!\},\ \{\!\{b\}\!\},\ \{\!\{c\}\!\}\ \}\!\} \in \mathsf{M}^1(X) = \mathsf{M}(X).$$

Hence the underlying set of distinct vertices is

$$\underline{V}^{(1)} = \mathrm{supp}(V^{(1)}) = \{\{\!\{a\}\!\},\ \{\!\{b\}\!\},\ \{\!\{c\}\!\}\} \subseteq \mathsf{M}(X),$$

where we view each vertex as a 1-multiset (e.g. $\{\!\{a\}\!\}$).

Define an edge-multiplicity function

$$\mu^{(1)} : \binom{\underline{V}^{(1)}}{2}^m \to \mathbb{N}_0$$

by

$$\mu^{(1)}(\{\!\{\{\!\{a\}\!\}, \{\!\{b\}\!\}\}\!\}) = 2, \qquad \mu^{(1)}(\{\!\{\{\!\{a\}\!\}, \{\!\{c\}\!\}\}\!\}) = 1,$$

and $\mu^{(1)}(e) = 0$ for all other $e$. Then

$$G^{(1)} = \left(V^{(1)}, \mu^{(1)}\right)$$

is an Iterated MultiGraph of order 1 over $X$ in the sense of Definition 2.2.5. Intuitively, the vertices are *multiset-objects* built from $X$, and edges connect these multiset-vertices with possible parallel multiplicities.

A concise comparison of Graph, MultiGraph, and Iterated MultiGraph is given in Table 2.3.

## 2.3 h-model

An *h-model* is a structure $\langle S, \mathcal{H}, I\rangle$ consisting of a base set $S$, a finite collection $\mathcal{H}$ of hypergraphs on $S$, and an interpretation map $I$ assigning each propositional atom a hypergraph in $\mathcal{H}$ [142].

**Definition 2.3.1** ($h$-model)**.** [142] Let $S$ be a nonempty set and write

$$\mathcal{P}^*(S) := \mathcal{P}(S) \setminus \{\emptyset\}.$$

A *(simple) hypergraph on $S$* is a finite family $H \subseteq \mathcal{P}^*(S)$.

Let At be a set of propositional atoms. An *h-model* is a triple

$$\mathfrak{M} = \langle S, \mathcal{H}, I\rangle$$

such that

Table 2.3: A concise comparison of Graph, MultiGraph, and Iterated MultiGraph.

| **Concept** | **Vertex domain** | **Edge structure** | **Main feature** |
|---|---|---|---|
| Graph | Ordinary vertices | Ordinary edges between vertex pairs | Represents simple pairwise adjacency without edge multiplicity. |
| MultiGraph | Ordinary vertices | Edges with multiplicity between the same vertex pair (and possibly loops) | Allows repeated pairwise connections while keeping the vertex set classical. |
| Iterated MultiGraph | Iterated multiset-based vertices | Ordinary multiset edges on those higher-level vertex objects | Extends multigraph structure by introducing recursive multiset organization on the vertex side. |

1. $\mathcal{H} \subseteq \mathcal{P}(\mathcal{P}^*(S))$ is a finite collection of simple hypergraphs on $S$;
2. $I : \mathrm{At} \to \mathcal{H}$ assigns to each atom $p \in \mathrm{At}$ a hypergraph $I(p) \in \mathcal{H}$.

**Example 2.3.2** ($h$-model for small-group access policies)**.** Let the base set of users be

$$S = \{\mathsf{Alice}, \mathsf{Bob}, \mathsf{Carol}\}.$$

Define two simple hypergraphs on $S$:

$$H_{\mathsf{pair}} = \big\{\{\mathsf{Alice}, \mathsf{Bob}\}, \{\mathsf{Alice}, \mathsf{Carol}\}, \{\mathsf{Bob}, \mathsf{Carol}\}\big\},$$

$$H_{\mathsf{team}} = \big\{\{\mathsf{Alice}, \mathsf{Bob}, \mathsf{Carol}\}, \{\mathsf{Alice}, \mathsf{Bob}\}\big\}.$$

Let

$$\mathcal{H} = \{H_{\mathsf{pair}}, H_{\mathsf{team}}\} \subseteq \mathcal{P}(\mathcal{P}^*(S)).$$

Take a set of propositional atoms

$$\mathrm{At} = \{p, q\},$$

and define the interpretation map $I : \mathrm{At} \to \mathcal{H}$ by

$$I(p) = H_{\mathsf{pair}}, \qquad I(q) = H_{\mathsf{team}}.$$

Then $\mathfrak{M} = \langle S, \mathcal{H}, I\rangle$ is an $h$-model in the sense of Definition 2.3.1. Intuitively, the atom $p$ is interpreted as the hypergraph of all two-person approvals, while $q$ is interpreted as a hypergraph of larger team approvals. An overview diagram of this example is provided in Fig. 2.5.

An *sh-model* may be regarded as a SuperHyperGraph-based analogue of an $h$-model: instead of assigning a hypergraph to each propositional atom, it assigns a finite $n$-SuperHyperGraph over a fixed base set.

**Definition 2.3.3** ((Recall) Finite $n$-SuperHyperGraph over a base set)**.** Let $S$ be a nonempty set and let $n \in \mathbb{N}_0$. Write

$$\mathcal{P}^0(S) := S, \qquad \mathcal{P}^{k+1}(S) := \mathcal{P}(\mathcal{P}^k(S)) \quad (k \geq 0),$$

and

$$\mathcal{P}^*(X) := \mathcal{P}(X) \setminus \{\varnothing\}.$$

A *finite $n$-SuperHyperGraph over $S$* is a triple

$$G = (V, E, \partial),$$

such that

1. $V \subseteq \mathcal{P}^n(S)$ is a finite set;
2. $E$ is a finite set;

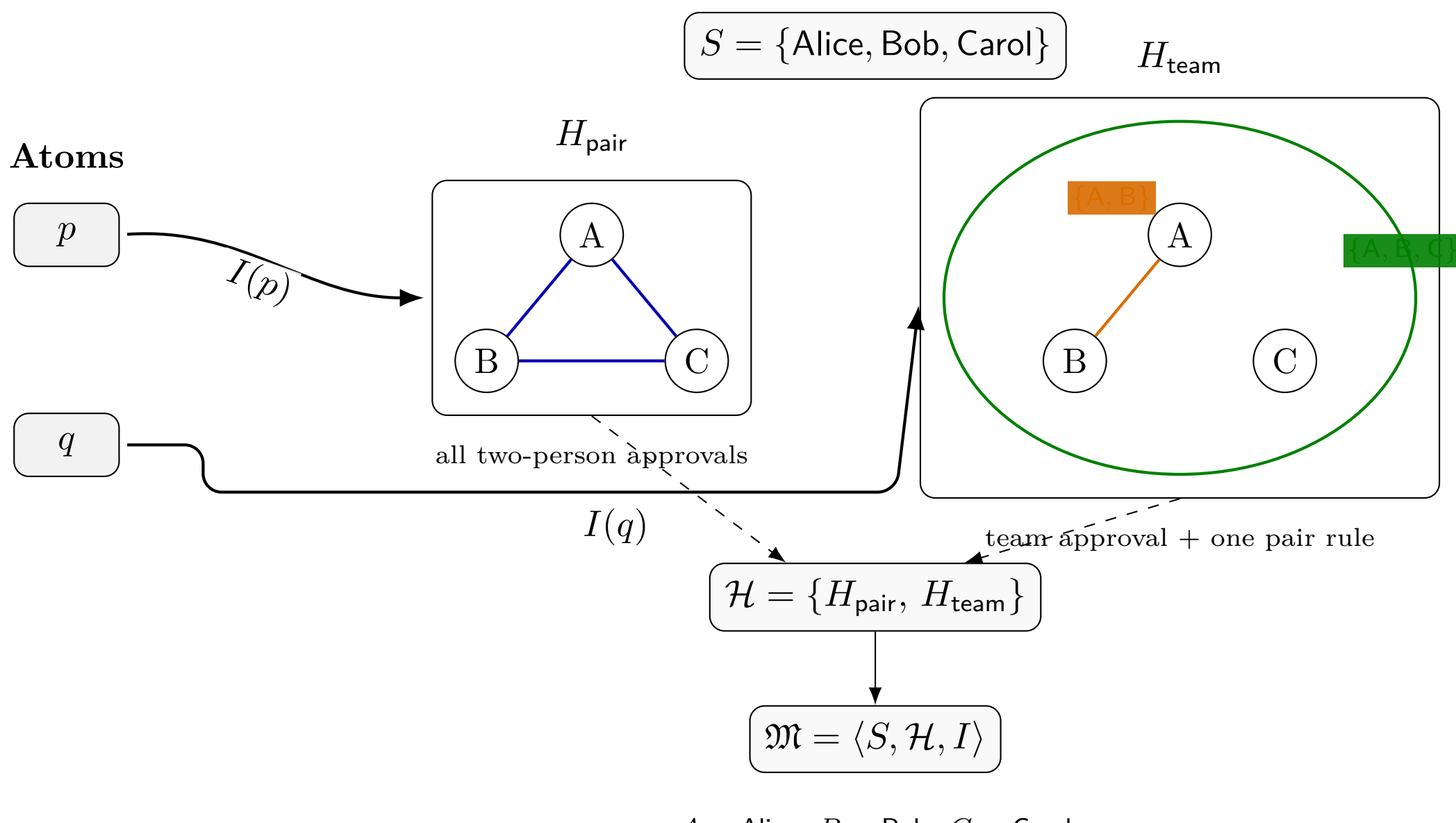

Figure 2.5: An illustration of the $h$-model in Example 2.3.2.

3.
$$\partial : E \longrightarrow \mathcal{P}^{*}(V)$$
is a map.

We denote by
$$\mathbf{SHG}_n(S)$$
the class of all finite $n$-SuperHyperGraphs over $S$.

**Definition 2.3.4** ($sh$-model)**.** Let $S$ be a nonempty set, let $n \in \mathbb{N}_0$, and let At be a set of propositional atoms. An *sh-model of level* $n$ is a quadruple
$$\mathfrak{M}^{(n)} = \langle S, \mathbf{SH}, I, n \rangle$$
such that

1.
$$\mathbf{SH} \subseteq \mathbf{SHG}_n(S)$$
is a finite collection of finite $n$-SuperHyperGraphs over $S$;

2.
$$I : \mathrm{At} \longrightarrow \mathbf{SH}$$
is an interpretation map assigning to each atom $p \in \mathrm{At}$ a member $I(p) \in \mathbf{SH}$.

**Remark 2.3.5.** *If $n = 0$, then $\mathcal{P}^0(S) = S$, and each finite 0-SuperHyperGraph over $S$ is simply a finite hypergraph written in incidence-map form. Hence an sh-model of level 0 reduces to a hypergraph-based model of the same general type as an h-model.*

**Theorem 2.3.6** (Well-definedness of $sh$-models)**.** *Let $S$ be a nonempty set, let $n \in \mathbb{N}_0$, and let* At *be a set of propositional atoms. Suppose that*
$$\mathbf{SH} \subseteq \mathbf{SHG}_n(S)$$
*is a finite family of finite n-SuperHyperGraphs over $S$, and let*
$$I : \mathrm{At} \to \mathbf{SH}$$
*be any map. Then*
$$\mathfrak{M}^{(n)} = \langle S, \mathbf{SH}, I, n \rangle$$
*is a well-defined sh-model in the sense of Definition 2.3.4.*

*Proof.* We verify the two clauses of Definition 2.3.4.

Since $n \in \mathbb{N}_0$ and $S \neq \varnothing$, the iterated powerset $\mathcal{P}^n(S)$ is well-defined by recursion. Hence the notion of a finite $n$-SuperHyperGraph over $S$ is meaningful by Definition 2.3.3. Therefore $\mathbf{SHG}_n(S)$ is a well-defined class of objects, and the assumption

$$\mathbf{SH} \subseteq \mathbf{SHG}_n(S)$$

means precisely that every member of $\mathbf{SH}$ is a finite $n$-SuperHyperGraph over $S$. By hypothesis, $\mathbf{SH}$ is finite, so condition (1) of Definition 2.3.4 holds.

Next, the map

$$I : \mathrm{At} \to \mathbf{SH}$$

is assumed to be a function. Hence for each propositional atom $p \in \mathrm{At}$, there exists a unique value $I(p) \in \mathbf{SH}$. Thus condition (2) of Definition 2.3.4 also holds.

Therefore all constituents of

$$\mathfrak{M}^{(n)} = \langle S, \mathbf{SH}, I, n \rangle$$

are well-defined and satisfy the required properties. Consequently, $\mathfrak{M}^{(n)}$ is a well-defined *sh*-model. □

## 2.4 Chain-Free Subsets

A *k-chain-free subset* of a poset is a subset that contains no strictly increasing chain of length $k$; equivalently, it avoids $k$ mutually comparable elements [143].

**Definition 2.4.1** ($k$-chain)**.** [143] Let $P = (X, \leq)$ be a finite poset and let $k \in \mathbb{N}$. A *chain of length k* (a *k-chain*) is a $k$-tuple $(x_1, \dots, x_k)$ of elements of $X$ such that

$$x_1 < x_2 < \cdots < x_k.$$

**Definition 2.4.2** ($k$-chain-free subset and $\mathrm{Forb}_k(P)$)**.** [143] Let $P = (X, \leq)$ be a finite poset and let $k \geq 2$. A subset $A \subseteq X$ is *k-chain-free* if it contains no $k$-chain; i.e., there do not exist distinct $a_1, \dots, a_k \in A$ with

$$a_1 < a_2 < \cdots < a_k.$$

We denote by

$$\mathrm{Forb}_k(P) := \{\, A \subseteq X \mid A \text{ is } k\text{-chain-free and maximal w.r.t. inclusion} \,\}$$

the family of inclusion-maximal $k$-chain-free subsets.

**Example 2.4.3** (2-chain-free subsets in a Boolean poset)**.** Let $P = (X, \subseteq)$ be the Boolean poset on

$$X = \mathcal{P}(\{1, 2, 3\})$$

ordered by inclusion. Take $k = 2$. Then a subset $A \subseteq X$ is 2-*chain-free* precisely when it contains no comparable pair (i.e., no $A_1, A_2 \in A$ with $A_1 \subset A_2$). For instance, the *middle layer*

$$A = \big\{\{1, 2\}, \{1, 3\}, \{2, 3\}\big\}$$

is 2-chain-free, because any two distinct 2-subsets of $\{1, 2, 3\}$ are incomparable. Moreover, $A$ is maximal with respect to inclusion among 2-chain-free subsets of $X$, hence

$$A \in \mathrm{Forb}_2(P).$$

An overview diagram of this example is provided in Fig. 2.6.

The iterated notion of a $k$-chain-free subset can also be defined as follows.

**Definition 2.4.4** (Iterated $k$-chain-free subset (depth $r$))**.** Let $P = (X, \leq)$ be a finite poset and let $k \geq 2$. Define recursively a sequence of posets

$$P_k^{(r)} = \big(U_k^{(r)},\ \preceq^{(r)}\big) \qquad (r \in \mathbb{N}_0)$$

as follows:

1. **(Depth 0)** Set $U_k^{(0)} := X$ and $\preceq^{(0)} := \leq$.

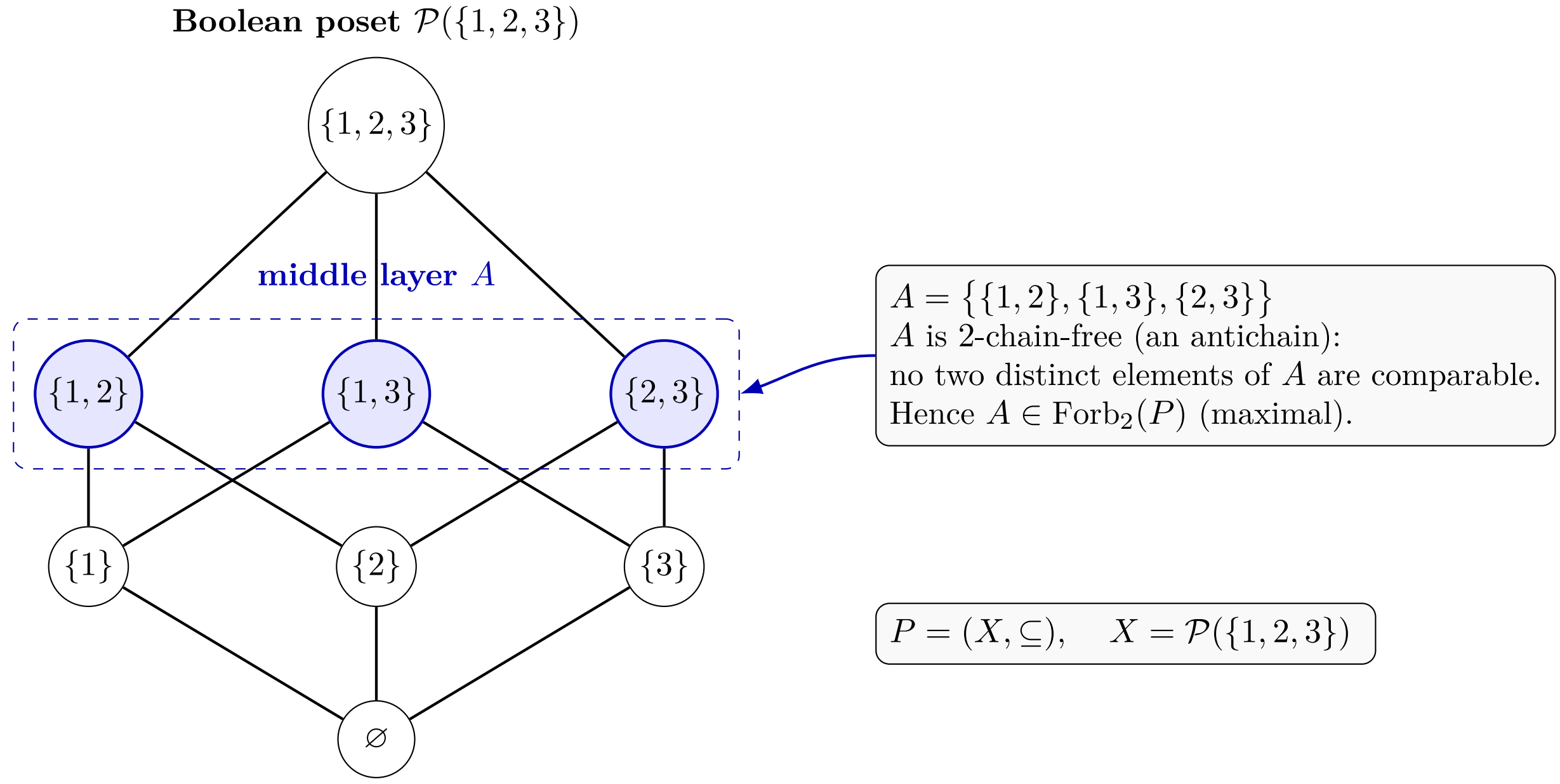

Figure 2.6: Hasse diagram of the Boolean poset on $\mathcal{P}(\{1,2,3\})$, highlighting the 2-chain-free middle layer $A$.

2. **(Depth $r+1$)** Having defined $P_k^{(r)} = (U_k^{(r)}, \preceq^{(r)})$, let

$$U_k^{(r+1)} := \Big\{ A \subseteq U_k^{(r)} \;\Big|\; A \text{ is } k\text{-chain-free in the poset } (U_k^{(r)}, \preceq^{(r)}) \Big\},$$

and equip $U_k^{(r+1)}$ with the inclusion order

$$A \preceq^{(r+1)} B \quad :\Longleftrightarrow \quad A \subseteq B \qquad (A, B \in U_k^{(r+1)}).$$

An element of $U_k^{(r)}$ is called an *iterated $k$-chain-free subset of depth $r$* (over $P$). Equivalently, it is a "$k$-chain-free set of $k$-chain-free sets of $\cdots$ of $k$-chain-free sets" ($r$ iterations), where the first iteration uses $\leq$ on $X$, and higher iterations use inclusion.

**Remark 2.4.5** (Maximal (iterated) $k$-chain-free families)**.** *If one prefers inclusion-*maximal *objects at each depth, define*

$$\mathrm{Forb}_k^{(r)}(P) := \Big\{ A \in U_k^{(r)} \;\Big|\; A \textit{ is maximal in } (U_k^{(r)}, \subseteq) \Big\} \qquad (r \geq 1),$$

*so* $\mathrm{Forb}_k^{(1)}(P)$ *coincides with the family of maximal $k$-chain-free subsets of $P$.*

**Example 2.4.6** (An iterated 2-chain-free subset of depth 2)**.** Let $P = (X, \subseteq)$ again be the Boolean poset on $X = \mathcal{P}(\{1,2,3\})$, and let $k = 2$. As in Example 2.4.3, elements of $U_2^{(1)}$ are 2-chain-free families of subsets of $\{1,2,3\}$. Consider the following three maximal 2-chain-free families (each is an antichain in $X$):

$$A_1 = \{\{1\},\{2\},\{3\}\}, \qquad A_2 = \{\{1,2\},\{1,3\},\{2,3\}\}, \qquad A_3 = \{\emptyset, \{1,2,3\}\}.$$

Each $A_i$ is 2-chain-free in $(X, \subseteq)$, so $A_1, A_2, A_3 \in U_2^{(1)}$. Now form a depth-2 object (a set of 2-chain-free sets) by

$$B = \{A_1, A_2\} \subseteq U_2^{(1)}.$$

Since $U_2^{(1)}$ is ordered by inclusion of families, and $A_1$ and $A_2$ are incomparable (neither is a subset of the other), the set $B$ contains no 2-chain in the poset $(U_2^{(1)}, \subseteq)$. Hence $B$ is 2-chain-free in $(U_2^{(1)}, \subseteq)$, and therefore

$$B \in U_2^{(2)}.$$

Thus $B$ is an iterated 2-chain-free subset of depth 2 in the sense of Definition 2.4.4.

**Theorem 2.4.7** (Well-definedness of iterated $k$-chain-free subsets)**.** *Let $P = (X, \leq)$ be a finite poset and let $k \geq 2$. Define $P_k^{(r)} = (U_k^{(r)}, \preceq^{(r)})$ recursively as in Definition 2.4.4. Then for every $r \in \mathbb{N}_0$:*

(i) *$U_k^{(r)}$ is a well-defined finite set, and in particular $U_k^{(r+1)} \subseteq \mathcal{P}\big(U_k^{(r)}\big)$.*

(ii) *$\preceq^{(r)}$ is a well-defined partial order on $U_k^{(r)}$; hence $P_k^{(r)}$ is a finite poset.*

*Consequently, the phrase "iterated $k$-chain-free subset of depth $r$" is well-defined.*

*Proof.* We argue by induction on $r$.

*Base case $r = 0$.* By definition, $U_k^{(0)} = X$ and $\preceq^{(0)}=\leq$. Since $P = (X, \leq)$ is assumed to be a finite poset, $U_k^{(0)}$ is a finite set and $\preceq^{(0)}$ is a partial order on it. Thus $P_k^{(0)}$ is well-defined as a finite poset.

*Inductive step.* Assume that for some $r \in \mathbb{N}_0$ the structure $P_k^{(r)} = (U_k^{(r)}, \preceq^{(r)})$ is a well-defined finite poset. We must show that $P_k^{(r+1)} = (U_k^{(r+1)}, \preceq^{(r+1)})$ is a well-defined finite poset.

*(i) $U_k^{(r+1)}$ is well-defined and finite.* Because $(U_k^{(r)}, \preceq^{(r)})$ is a poset by the induction hypothesis, the notion "$A$ is $k$-chain-free in $(U_k^{(r)}, \preceq^{(r)})$" is meaningful; explicitly, $A \subseteq U_k^{(r)}$ is $k$-chain-free if there do not exist elements $x_1, \ldots, x_k \in A$ such that

$$x_1 \prec^{(r)} x_2 \prec^{(r)} \cdots \prec^{(r)} x_k.$$

Hence the specification

$$U_k^{(r+1)} \;=\; \{\, A \subseteq U_k^{(r)} \mid A \text{ is } k\text{-chain-free in } (U_k^{(r)}, \preceq^{(r)}) \,\}$$

defines a bona fide subset of the power set $\mathcal{P}(U_k^{(r)})$. Moreover, since $U_k^{(r)}$ is finite, $\mathcal{P}(U_k^{(r)})$ is finite, and therefore $U_k^{(r+1)}$ is also finite.

*(ii) $\preceq^{(r+1)}$ is a partial order.* By definition, for $A, B \in U_k^{(r+1)}$ we set

$$A \preceq^{(r+1)} B \quad\Longleftrightarrow\quad A \subseteq B.$$

Since inclusion $\subseteq$ is reflexive, antisymmetric, and transitive on *all* subsets of $U_k^{(r)}$, its restriction to the subcollection $U_k^{(r+1)} \subseteq \mathcal{P}(U_k^{(r)})$ is again reflexive, antisymmetric, and transitive. Thus $\preceq^{(r+1)}$ is a well-defined partial order on $U_k^{(r+1)}$.

Combining (i) and (ii), $P_k^{(r+1)}$ is a well-defined finite poset. This completes the induction and proves the theorem. □

## 2.5 Power Set Graph

A *Power Set Graph* has vertices given by the nonempty proper subsets of a set, with edges between comparable subsets by inclusion [144]. An *Iterated Power Set Graph* repeatedly applies the nontrivial powerset construction; at each depth, vertices are nested subsets, adjacent when comparable by inclusion.

**Definition 2.5.1** (Power set graph)**.** [144] Let $A$ be a finite set with $|A| \geq 2$. The *power set graph* of $A$, denoted $\Gamma(\mathcal{P}(A))$, is the simple undirected graph

$$\Gamma(\mathcal{P}(A)) = (V, E),$$

where

$$V = \mathcal{P}(A) \setminus \{\emptyset, A\}, \qquad E = \big\{\{X, Y\} \subseteq V \;\big|\; X \subset Y \text{ or } Y \subset X\big\}.$$

Equivalently, vertices are the nonempty proper subsets of $A$, and two vertices are adjacent iff one subset is contained in the other.

**Example 2.5.2** (Power set graph on a 3-element set)**.** Let

$$A = \{1, 2, 3\}.$$

Then the vertex set of the power set graph $\Gamma(\mathcal{P}(A))$ is

$$V = \mathcal{P}(A) \setminus \{\emptyset, A\} = \big\{\{1\}, \{2\}, \{3\}, \{1, 2\}, \{1, 3\}, \{2, 3\}\big\}.$$

Two vertices are adjacent iff one is a subset of the other. For instance,

$$\{1\} \sim \{1, 2\}, \qquad \{1\} \sim \{1, 3\}, \qquad \{2\} \sim \{1, 2\},$$

while $\{1, 2\} \not\sim \{1, 3\}$ because neither contains the other. Hence $\Gamma(\mathcal{P}(A))$ is the comparability graph of the inclusion poset on nonempty proper subsets of $A$, as in Definition 2.5.1.

**Definition 2.5.3** (Iterated Power Set Graph). Let $A$ be a finite set with $|A| \geq 2$, and write

$$\mathcal{P}^{*}(S) \coloneqq \mathcal{P}(S) \setminus \{\emptyset, S\}$$

for the family of *nonempty proper* subsets of a finite set $S$.

Define recursively the *iterated nontrivial powerset universes* $(U_r(A))_{r\geq 0}$ by

$$U_0(A) \coloneqq A, \qquad U_{r+1}(A) \coloneqq \mathcal{P}^{*}\big(U_r(A)\big) \quad (r \geq 0).$$

For each $r \geq 1$, the *Iterated Power Set Graph of depth* $r$ on $A$ is the simple undirected graph

$$\Gamma_r(A) = (V_r, E_r),$$

where

$$V_r \coloneqq U_r(A), \qquad E_r \coloneqq \Big\{\{X, Y\} \subseteq V_r \;\Big|\; X \subset Y \text{ or } Y \subset X\Big\}.$$

Thus, vertices at depth $r$ are nonempty proper subsets of $U_{r-1}(A)$, and two vertices are adjacent if and only if they are comparable by inclusion.

In particular, $\Gamma_1(A)$ is the usual power set graph on $A$. (Optionally, one may set $\Gamma_0(A)$ to be the edgeless graph on vertex set $A$.)

**Example 2.5.4** (Iterated power set graph of depth 2). Let

$$A = \{1, 2, 3\}.$$

At depth 1,

$$U_1(A) = \mathcal{P}^{*}(A) = \big\{\{1\}, \{2\}, \{3\}, \{1, 2\}, \{1, 3\}, \{2, 3\}\big\},$$

so $\Gamma_1(A)$ is the power set graph from Example 2.5.2. At depth 2, the vertex universe is

$$V_2 = U_2(A) = \mathcal{P}^{*}(U_1(A)),$$

whose elements are the nonempty proper *subsets of* $U_1(A)$. For example, the following are vertices of $\Gamma_2(A)$:

$$X = \big\{\{1\}, \{1, 2\}\big\} \in U_2(A), \qquad Y = \big\{\{1\}, \{1, 2\}, \{2\}\big\} \in U_2(A),$$

and they are adjacent in $\Gamma_2(A)$ because $X \subset Y$. By contrast,

$$Z = \big\{\{3\}, \{2, 3\}\big\} \in U_2(A)$$

is not adjacent to $Y$ because neither $Z \subset Y$ nor $Y \subset Z$ holds. Thus $\Gamma_2(A)$ is a graph whose vertices are *sets of nontrivial subsets of* $A$, adjacent when comparable by inclusion, exactly as in Definition 2.5.3. An overview diagram of this example is provided in Fig. 2.7.

## 2.6 Johnson Graph

A *Johnson graph* has vertices given by the $w$-subsets of $[n]$, with edges joining pairs that differ by one element [145, 146, 147]. Related notions of the Johnson graph are also known, such as the generalized Johnson graph[148, 149].

**Definition 2.6.1** (Johnson graph). [150, 147] Let $n, w \in \mathbb{N}$ with $0 < w < n$, and write $[n] \coloneqq \{1, 2, \ldots, n\}$. The *Johnson graph* $J(n, w)$ is the simple graph

$$J(n, w) = (V, E),$$

where

$$V = \{\, X \subseteq [n] \mid |X| = w \,\}, \qquad E = \big\{\{X, Y\} \subseteq V \;\big|\; |X \cap Y| = w - 1\big\}.$$

Thus two $w$-subsets are adjacent exactly when they differ by one element.

**Example 2.6.2** (The Johnson graph $J(4, 2)$). Let $n = 4$ and $w = 2$. Then the vertex set of the Johnson graph $J(4, 2)$ consists of all 2-subsets of $[4] = \{1, 2, 3, 4\}$:

$$V = \big\{\{1, 2\}, \{1, 3\}, \{1, 4\}, \{2, 3\}, \{2, 4\}, \{3, 4\}\big\}.$$

Two vertices are adjacent if they differ by exactly one element (equivalently, their intersection has size 1). For instance,

$$\{1, 2\} \sim \{1, 3\}, \qquad \{1, 2\} \sim \{2, 4\}, \qquad \{2, 3\} \sim \{3, 4\},$$

since

$$|\{1, 2\} \cap \{1, 3\}| = 1, \quad |\{1, 2\} \cap \{2, 4\}| = 1, \quad |\{2, 3\} \cap \{3, 4\}| = 1.$$

By contrast, $\{1, 2\} \not\sim \{3, 4\}$ because $\{1, 2\} \cap \{3, 4\} = \emptyset$. Hence $J(4, 2)$ is the graph on 2-subsets of $[4]$ with edges joining pairs that differ in exactly one element, as in Definition 2.6.1.

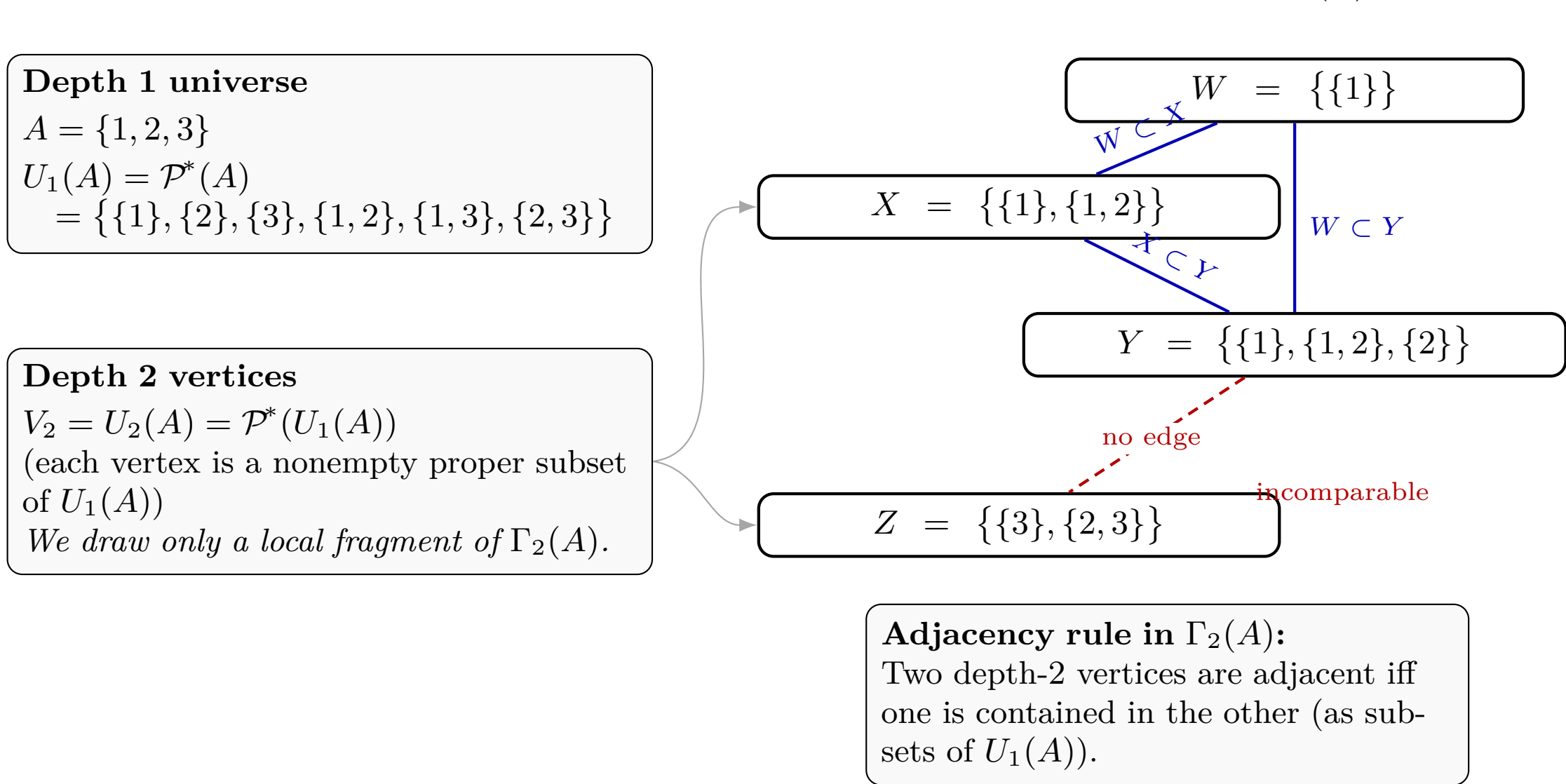

Figure 2.7: Illustration of an iterated power set graph of depth 2 for $A = \{1, 2, 3\}$. The figure shows a local induced fragment of $\Gamma_2(A)$ with example vertices $X, Y, Z, W$.

## 2.7 Kneser Graph

A *Kneser graph* has vertices given by the $k$-subsets of $[n]$, with edges joining pairs that are disjoint [151, 152, 153, 154]. Related concepts are also known, such as Kneser hypergraphs [155, 156], bipartite kneser graphs[157, 158], stable kneser graphs[159, 160], and generalized Kneser graphs [161, 162].

**Definition 2.7.1** (Kneser graph)**.** [151, 152] Let $n, k \in \mathbb{N}$ with $n \geq 2k$, and write $[n] \coloneqq \{1, 2, \ldots, n\}$. The *Kneser graph* $\mathrm{KG}_{n,k}$ is the simple graph

$$\mathrm{KG}_{n,k} = (V, E),$$

where

$$V = \{\, X \subseteq [n] \mid |X| = k \,\}, \qquad E = \big\{ \{X, Y\} \subseteq V \;\big|\; X \cap Y = \emptyset \big\}.$$

Thus two $k$-subsets are adjacent exactly when they are disjoint.

**Example 2.7.2** (The Kneser graph $\mathrm{KG}_{5,2}$ (the Petersen graph))**.** Let $n = 5$ and $k = 2$. Then the vertex set of the Kneser graph $\mathrm{KG}_{5,2}$ consists of all 2-subsets of $[5] = \{1, 2, 3, 4, 5\}$:

$$V = \big\{\{1, 2\}, \{1, 3\}, \{1, 4\}, \{1, 5\}, \{2, 3\}, \{2, 4\}, \{2, 5\}, \{3, 4\}, \{3, 5\}, \{4, 5\}\big\}.$$

Two vertices are adjacent if and only if the corresponding 2-subsets are disjoint. For example,

$$\{1, 2\} \sim \{3, 4\}, \qquad \{1, 5\} \sim \{2, 3\}, \qquad \{2, 4\} \sim \{1, 3\},$$

since

$$\{1, 2\} \cap \{3, 4\} = \emptyset, \quad \{1, 5\} \cap \{2, 3\} = \emptyset, \quad \{2, 4\} \cap \{1, 3\} = \emptyset.$$

By contrast, $\{1, 2\} \not\sim \{2, 5\}$ because $\{1, 2\} \cap \{2, 5\} = \{2\} \neq \emptyset$. Hence $\mathrm{KG}_{5,2}$ is the graph on 2-subsets of $[5]$ with edges joining disjoint pairs, as in Definition 2.7.1. (It is well known that $\mathrm{KG}_{5,2}$ is isomorphic to the Petersen graph.)

## 2.8 Meta-Graph and Iterated Meta-Graph

A meta-graph has graphs as vertices; edges represent labeled relations between those graphs, satisfying relation-defined incidence constraints (cf.[163, 164, 165, 166, 167]). A meta-graph is also called a graph of graphs [168, 169, 170]. An iterated meta-graph repeats the construction: vertices are meta-graphs of lower depth, with lifted relations linking them recursively [163]. A similar concept known as the composite graph has also been studied [171, 172, 173, 174]. Concepts such as graph clustering can likewise be regarded as operations for constructing a metagraph [175, 176, 177, 178, 179].

**Definition 2.8.1** (Meta-Graph (Metagraph; graph of graphs))**.** [163] Fix a nonempty universe $\mathcal{G}$ of finite graphs (undirected and loopless by default), and fix a nonempty family $\mathcal{R}$ of binary relations on $\mathcal{G}$, i.e.

$$\mathcal{R} \subseteq \mathcal{P}(\mathcal{G} \times \mathcal{G}).$$

A *Meta-Graph* (or *metagraph*) over $(\mathcal{G}, \mathcal{R})$ is a directed, $\mathcal{R}$-labeled multigraph

$$M = (V, E, s, t, \lambda)$$

such that

$$V \subseteq \mathcal{G}, \qquad s, t : E \to V, \qquad \lambda : E \to \mathcal{R},$$

and the following *incidence constraint* holds:

$$\forall e \in E : \quad \big(s(e), t(e)\big) \in \lambda(e).$$

Elements of $V$ are called *meta-vertices* (each meta-vertex is itself a graph in $\mathcal{G}$), and each $e \in E$ is a *meta-edge* labeled by the relation $\lambda(e)$. If $\mathcal{R} = \{R\}$ is a singleton, the label map may be omitted; if each $R \in \mathcal{R}$ is symmetric, one may view $M$ as an undirected labeled multigraph.

**Example 2.8.2** (A Meta-Graph of module-dependency graphs linked by a compatibility relation)**.** Let $\mathcal{G}$ be a universe consisting of three finite (simple, undirected) graphs that represent dependency structures of three software modules:

$$G_A = (V_A, E_A), \qquad G_B = (V_B, E_B), \qquad G_C = (V_C, E_C) \ \in \ \mathcal{G}.$$

For concreteness, take

$$V_A = \{a_1, a_2, a_3\}, \ E_A = \big\{\{a_1, a_2\}, \{a_2, a_3\}\big\},$$
$$V_B = \{b_1, b_2, b_3\}, \ E_B = \big\{\{b_1, b_2\}, \{b_1, b_3\}\big\},$$
$$V_C = \{c_1, c_2\}, \ E_C = \big\{\{c_1, c_2\}\big\}.$$

Define two binary relations on $\mathcal{G}$:

$$R_{\text{api}} := \big\{(G_i, G_j) \in \mathcal{G} \times \mathcal{G} \ \big| \ G_i \text{ and } G_j \text{ have mutually compatible APIs}\big\},$$
$$R_{\text{data}} := \big\{(G_i, G_j) \in \mathcal{G} \times \mathcal{G} \ \big| \ G_i \text{ can safely consume data produced by } G_j\big\}.$$

Let $\mathcal{R} = \{R_{\text{api}}, R_{\text{data}}\} \subseteq \mathcal{P}(\mathcal{G} \times \mathcal{G})$.

Now define a directed $\mathcal{R}$-labeled multigraph

$$M = (V, E, s, t, \lambda)$$

by taking

$$V = \{G_A, G_B, G_C\} \subseteq \mathcal{G}, \qquad E = \{e_1, e_2\},$$

and setting

$$s(e_1) = G_A, \quad t(e_1) = G_B, \quad \lambda(e_1) = R_{\text{api}},$$
$$s(e_2) = G_B, \quad t(e_2) = G_C, \quad \lambda(e_2) = R_{\text{data}}.$$

Assume that $(G_A, G_B) \in R_{\text{api}}$ and $(G_B, G_C) \in R_{\text{data}}$ hold, i.e. the API-compatibility and data-consumption conditions are satisfied for these pairs. Then the incidence constraint

$$\forall e \in E : \ (s(e), t(e)) \in \lambda(e)$$

holds, and hence $M$ is a Meta-Graph over $(\mathcal{G}, \mathcal{R})$ in the sense of Definition 2.8.1. Intuitively, $M$ is a "graph of graphs" that records how entire dependency graphs relate to each other under different semantic relations. An overview diagram of this example is provided in Fig. 2.8.

**Definition 2.8.3** (Iterated Meta-Graph (depth $t$))**.** [163] Fix $(\mathcal{G}, \mathcal{R})$ as in Definition 2.8.1. Define recursively, for $t \in \mathbb{N}_0$, a universe of *level-$t$ objects* $\mathcal{G}^{(t)}$ and a family of *level-$t$ relations* $\mathcal{R}^{(t)}$ as follows:

$$\mathcal{G}^{(0)} := \mathcal{G}, \qquad \mathcal{R}^{(0)} := \mathcal{R}.$$

Assume $\mathcal{G}^{(t)}$ and $\mathcal{R}^{(t)}$ are defined. Let $\mathcal{G}^{(t+1)}$ be the class of all finite metagraphs over $(\mathcal{G}^{(t)}, \mathcal{R}^{(t)})$ (i.e. all tuples $M = (V(M), E(M), s_M, t_M, \lambda_M)$ satisfying Definition 2.8.1 with $\mathcal{G}, \mathcal{R}$ replaced by $\mathcal{G}^{(t)}, \mathcal{R}^{(t)}$).

$\mathcal{G}$ (universe of module-dependency graphs)

$G_A$ $G_B$ $G_C$

$a_2$ $a_1$ $a_3$ $\lambda(e_1) = R_{\text{api}}$ $b_1$ $b_2$ $b_3$ $\lambda(e_2) = R_{\text{data}}$ $c_1$ $c_2$

$s(e_1) = G_A,\ t(e_1) = G_B$ $s(e_2) = G_B,\ t(e_2) = G_C$

$M = (V, E, s, t, \lambda), \quad V = \{G_A, G_B, G_C\}, \quad E = \{e_1, e_2\}$

Blue: API compatibility Orange: safe data-consumption relation

Figure 2.8: An illustration of the Meta-Graph in Example 2.8.2: each meta-vertex is itself a dependency graph, and meta-edges encode semantic relations between them.

For each relation $R \in \mathcal{R}^{(t)}$, define its *lift* $R^{\uparrow} \subseteq \mathcal{G}^{(t+1)} \times \mathcal{G}^{(t+1)}$ by

$$(M_1, M_2) \in R^{\uparrow} \quad :\Longleftrightarrow \quad \exists\, x \in V(M_1),\ \exists\, y \in V(M_2) \text{ such that } (x, y) \in R.$$

Set

$$\mathcal{R}^{(t+1)} := \{\, R^{\uparrow} \mid R \in \mathcal{R}^{(t)} \,\}.$$

An *Iterated Meta-Graph of depth* $t$ is then a metagraph

$$M^{(t)} = (V^{(t)}, E^{(t)}, s^{(t)}, t^{(t)}, \lambda^{(t)})$$

over $\left(\mathcal{G}^{(t)}, \mathcal{R}^{(t)}\right)$, i.e.

$$V^{(t)} \subseteq \mathcal{G}^{(t)}, \qquad \lambda^{(t)} : E^{(t)} \to \mathcal{R}^{(t)}, \qquad \forall e \in E^{(t)} :\ \left(s^{(t)}(e), t^{(t)}(e)\right) \in \lambda^{(t)}(e).$$

In particular, depth 0 iterated meta-graphs are ordinary metagraphs, and depth $t \geq 1$ iterated meta-graphs have vertices that are themselves metagraphs, recursively, up to $t$ levels.

**Example 2.8.4** (A depth-1 Iterated Meta-Graph (a metagraph of metagraphs))**.** Continue with the universe $(\mathcal{G}, \mathcal{R})$ from Example 2.8.2. A level-1 object is a finite metagraph over $(\mathcal{G}, \mathcal{R})$.

Define two metagraphs $M_1, M_2 \in \mathcal{G}^{(1)}$ as follows. Let $M_1$ be the metagraph from Example 2.8.2, i.e.

$$M_1 = (\{G_A, G_B, G_C\},\ \{e_1, e_2\},\ s_1, t_1, \lambda_1),$$

with $\lambda_1(e_1) = R_{\text{api}}$ and $\lambda_1(e_2) = R_{\text{data}}$.

Let $M_2$ be another metagraph on the same vertex set

$$V(M_2) = \{G_A, G_B, G_C\}$$

with a single meta-edge $f$ defined by

$$s_2(f) = G_A, \quad t_2(f) = G_C, \quad \lambda_2(f) = R_{\text{api}},$$

assuming $(G_A, G_C) \in R_{\text{api}}$ (module $A$ is API-compatible with module $C$).

**Lifted relations.** By Definition 2.8.3, each relation $R \in \mathcal{R}^{(0)} = \mathcal{R}$ induces a lifted relation $R^{\uparrow} \in \mathcal{R}^{(1)}$ on $\mathcal{G}^{(1)}$:

$$(M_i, M_j) \in R^{\uparrow} \iff \exists\, x \in V(M_i), \exists\, y \in V(M_j) \text{ with } (x, y) \in R.$$

In particular, since $M_1$ contains the pair $(G_A, G_B) \in R_{\text{api}}$ and $M_2$ contains the pair $(G_A, G_C) \in R_{\text{api}}$, we have

$$(M_1, M_2) \in R^{\uparrow}_{\text{api}}.$$

Table 2.4: A concise comparison of Graph, Meta-Graph, and Iterated Meta-Graph.

| **Concept** | **Vertex domain** | **Edge meaning** | **Main feature** |
|---|---|---|---|
| Graph | Ordinary vertices | Ordinary adjacency between vertices | The classical pairwise network model. |
| Meta-Graph | Graphs | Labeled relations between graphs | A graph of graphs, where each vertex is itself a graph-object. |
| Iterated Meta-Graph | Meta-graphs of lower depth | Lifted relations between meta-graphs | A recursive graph-of-graphs-of-graphs construction across multiple meta-levels. |

Now define a level-1 metagraph (an iterated metagraph of depth 1)

$$M^{(1)} = (V^{(1)}, E^{(1)}, s^{(1)}, t^{(1)}, \lambda^{(1)})$$

by

$$V^{(1)} = \{M_1, M_2\} \subseteq \mathcal{G}^{(1)}, \qquad E^{(1)} = \{g\},$$
$$s^{(1)}(g) = M_1, \quad t^{(1)}(g) = M_2, \quad \lambda^{(1)}(g) = R^{\uparrow}_{\mathrm{api}}.$$

Because $(M_1, M_2) \in R^{\uparrow}_{\mathrm{api}}$, the incidence constraint holds:

$$\big(s^{(1)}(g), t^{(1)}(g)\big) \in \lambda^{(1)}(g).$$

Therefore $M^{(1)}$ is an Iterated Meta-Graph of depth 1 in the sense of Definition 2.8.3. Intuitively, it is a "metagraph of metagraphs" that relates two meta-level system descriptions whenever they contain underlying module pairs connected by $R_{\mathrm{api}}$. An overview diagram of this example is provided in Fig. 2.9.

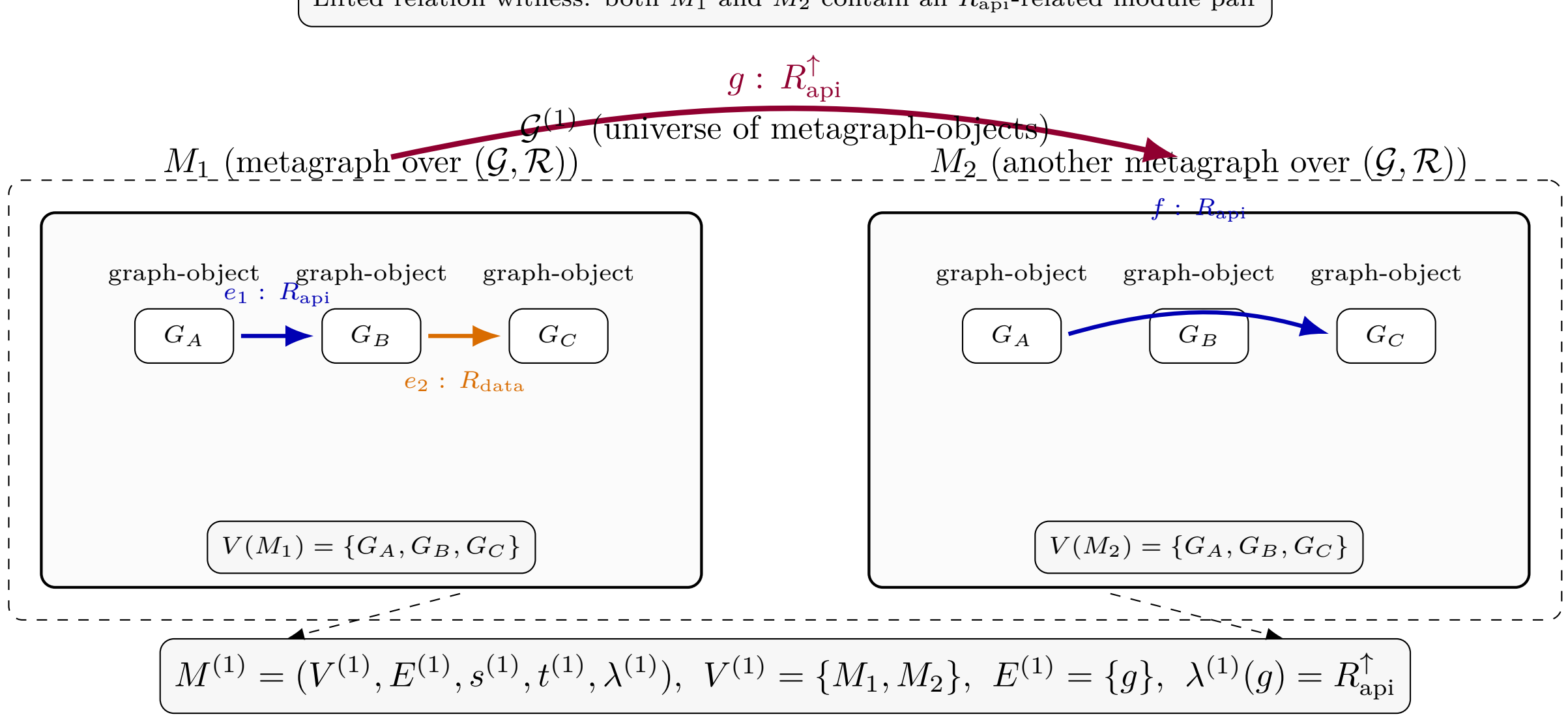

Figure 2.9: A depth-1 Iterated Meta-Graph: the vertices $M_1, M_2$ are themselves metagraphs, and the top-level edge $g$ is labeled by the lifted relation $R^{\uparrow}_{\mathrm{api}}$.

A concise comparison of Graph, Meta-Graph, and Iterated Meta-Graph is given in Table 2.4.

## 2.9 Meta-HyperGraph and Meta-SuperHyperGraph

A meta-hypergraph is a hypergraph whose vertices are objects; each hyperedge relates finite vertex sets via labeled relations (cf. [8, 180]). It can also be described as a *hypergraph of hypergraphs*. A meta-superhypergraph has vertices that are superhypergraphs; hyperedges relate finite collections of them, constrained by relation labels [8]. It can also be described as a *superhypergraph of superhypergraphs*.

**Definition 2.9.1** (MetaHyperGraph over $(U, \mathcal{R})$)**.** [8] Let $U$ be a nonempty universe of objects and let

$$\mathcal{R} \subseteq \mathcal{P}\big(\mathcal{P}_{\mathrm{fin}}(U) \times \mathcal{P}_{\mathrm{fin}}(U)\big)$$

be a nonempty family of *admissible set–relations*. A *MetaHyperGraph* over $(U, \mathcal{R})$ is a labelled directed hypergraph

$$M = (V, E, T, Hd, \lambda)$$

such that

$$V \subseteq U, \qquad T, Hd : E \to \mathcal{P}_{\mathrm{fin}}(V), \qquad \lambda : E \to \mathcal{R},$$

and the *incidence constraint* holds:

$$\forall e \in E : \ (T(e), Hd(e)) \in \lambda(e).$$

The elements of $V$ are called *meta-vertices*. If $U$ is chosen to be a universe of (finite) hypergraphs, then $M$ may be read as a *hypergraph of hypergraphs*.

**Example 2.9.2** (A MetaHyperGraph of hypergraphs linked by an overlap relation)**.** Let the universe $U$ be a collection of finite hypergraphs describing co-purchase groups in three cities:

$$U = \{H_T, H_O, H_N\},$$

where each $H_\bullet = (V_\bullet, E_\bullet)$ is a finite hypergraph on a product set $V_\bullet$. Define an admissible set–relation $R_{\mathrm{ov}} \in \mathcal{P}\big(\mathcal{P}_{\mathrm{fin}}(U) \times \mathcal{P}_{\mathrm{fin}}(U)\big)$ by

$$(A, B) \in R_{\mathrm{ov}} \quad :\Longleftrightarrow \quad \exists\, H \in A,\ \exists\, H' \in B \text{ such that } \frac{|V(H) \cap V(H')|}{|V(H) \cup V(H')|} \ \geq\ \theta,$$

for a fixed threshold $\theta \in (0, 1)$, where $V(H)$ denotes the vertex set of a hypergraph $H$. (Thus $R_{\mathrm{ov}}$ certifies that the two finite families contain at least one pair of hypergraphs whose vertex sets have sufficiently large Jaccard overlap.) Let

$$\mathcal{R} = \{R_{\mathrm{ov}}\}.$$

Now form a labelled directed hypergraph

$$M = (V, E, T, Hd, \lambda)$$

as follows:

$$V = \{H_T, H_O, H_N\} \subseteq U, \qquad E = \{e_1, e_2\}.$$

Define tail and head maps $T, Hd : E \to \mathcal{P}_{\mathrm{fin}}(V)$ by

$$T(e_1) = \{H_T, H_O\}, \quad Hd(e_1) = \{H_N\}, \qquad T(e_2) = \{H_O\}, \quad Hd(e_2) = \{H_T, H_N\},$$

and define the label map $\lambda : E \to \mathcal{R}$ by

$$\lambda(e_1) = R_{\mathrm{ov}}, \qquad \lambda(e_2) = R_{\mathrm{ov}}.$$

Assume that the chosen threshold $\theta$ is such that

$$\big(T(e_i), Hd(e_i)\big) \in R_{\mathrm{ov}} \qquad (i = 1, 2),$$

i.e., each tail–head pair satisfies the overlap criterion. Then the incidence constraint

$$\forall e \in E : \ (T(e), Hd(e)) \in \lambda(e)$$

holds, and therefore $M$ is a MetaHyperGraph over $(U, \mathcal{R})$ in the sense of Definition 2.9.1. Intuitively, $M$ is a "hypergraph of hypergraphs" whose meta-hyperedges assert that certain *families* of city-level hypergraphs overlap strongly in their product catalogs. An overview diagram of this example is provided in Fig. 2.10.

**Definition 2.9.3** (MetaSuperHyperGraph over $(\mathsf{S}_n, \mathcal{R})$)**.** [8] Fix $n \in \mathbb{N}_0$ and let $\mathsf{S}_n$ denote the class of all finite directed $n$-SuperHyperGraphs (over arbitrary base sets). Let

$$\mathcal{R} \subseteq \mathcal{P}\big(\mathcal{P}_{\mathrm{fin}}(\mathsf{S}_n) \times \mathcal{P}_{\mathrm{fin}}(\mathsf{S}_n)\big)$$

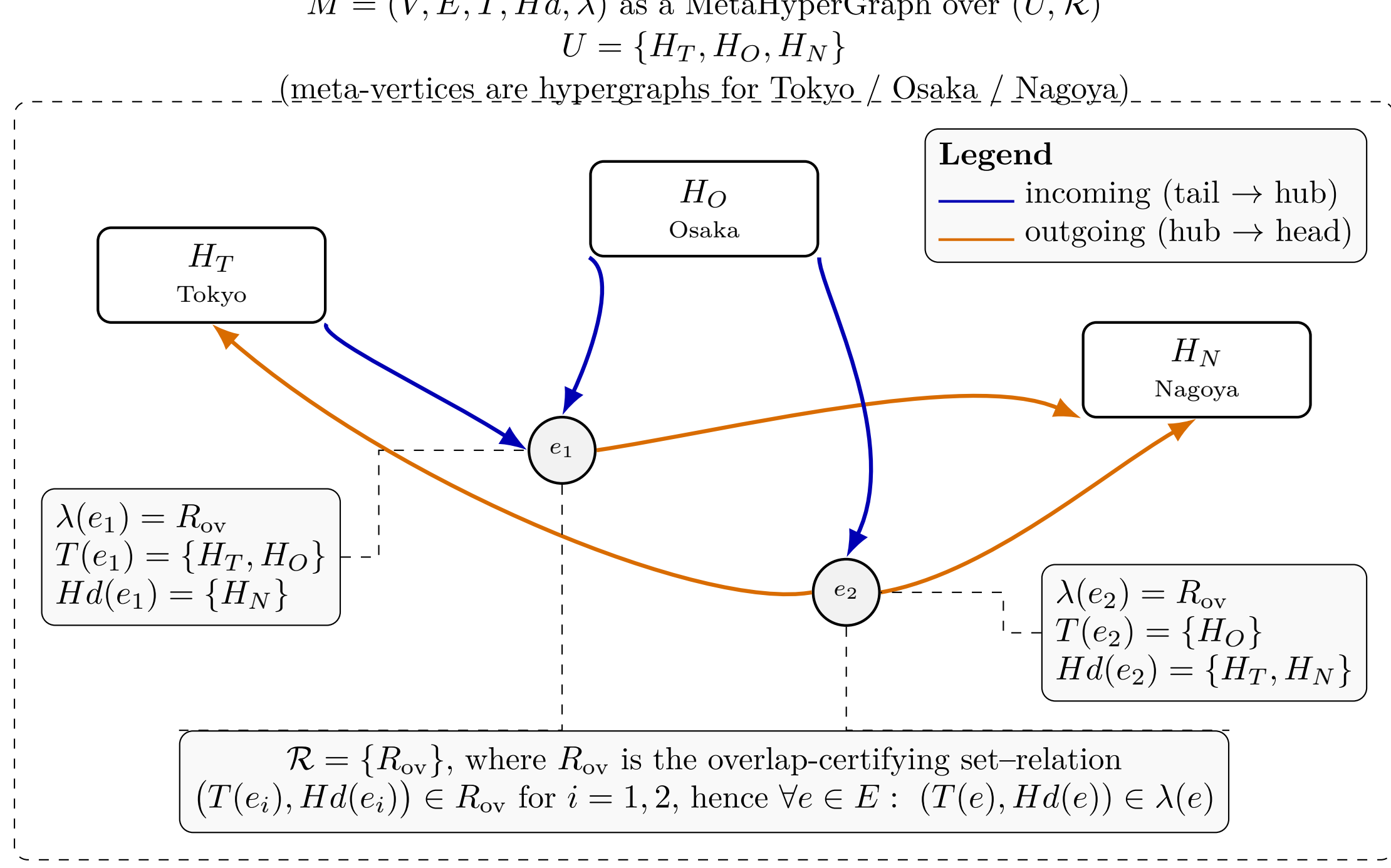

Figure 2.10: An illustration of the MetaHyperGraph in Example 2.9.2.

be a nonempty family of admissible set–relations on $\mathsf{S}_n$. A *MetaSuperHyperGraph* (abbrev. *MSHG*) over $(\mathsf{S}_n, \mathcal{R})$ is a labelled directed hypergraph

$$M = (V, E, T, Hd, \lambda)$$

such that

$$V \subseteq \mathsf{S}_n, \qquad T, Hd : E \to \mathcal{P}_{\mathrm{fin}}(V), \qquad \lambda : E \to \mathcal{R},$$

and the incidence constraint holds:

$$\forall e \in E :\ (T(e), Hd(e)) \in \lambda(e).$$

Thus vertices of $M$ are (finite) $n$-SuperHyperGraphs, and each meta-hyperedge relates finite families of such vertices, certified by its label $\lambda(e) \in \mathcal{R}$.

**Example 2.9.4** (A MetaSuperHyperGraph relating 1-SuperHyperGraphs by a refinement relation)**.** Fix $n = 1$. Let $V_0 = \{1, 2, 3, 4\}$ be a base set of atomic items. Consider three finite directed 1-SuperHyperGraphs (here written with undirected incidence for simplicity)

$$G_1 = (V_1^{(1)}, E_1^{(1)}, \partial_1), \qquad G_2 = (V_2^{(1)}, E_2^{(1)}, \partial_2), \qquad G_3 = (V_3^{(1)}, E_3^{(1)}, \partial_3),$$

over (possibly) different supervertex sets, where each $V_i^{(1)} \subseteq \mathcal{P}(V_0)$ is a finite set of 1-supervertices.

Define an admissible set–relation $R_{\mathrm{ref}} \in \mathcal{P}\big(\mathcal{P}_{\mathrm{fin}}(\mathsf{S}_1) \times \mathcal{P}_{\mathrm{fin}}(\mathsf{S}_1)\big)$ by

$$(A, B) \in R_{\mathrm{ref}} \quad :\Longleftrightarrow \quad \forall G \in A\ \exists G' \in B \text{ such that } G' \text{ is a supervertex-refinement of } G,$$

where "$G'$ is a supervertex-refinement of $G$" means: for every supervertex $v' \in V(G')$ there exists $v \in V(G)$ with $v' \subseteq v$ (as subsets of $V_0$), and for every superedge $e' \in E(G')$ there exists $e \in E(G)$ such that

$$\bigcup \partial_{G'}(e') \ \subseteq\ \bigcup \partial_G(e) \quad \text{(as subsets of } V_0\text{)}.$$

Let

$$\mathcal{R} = \{R_{\mathrm{ref}}\}.$$

Now define a labelled directed hypergraph

$$M = (V, E, T, Hd, \lambda)$$

by taking

$$V = \{G_1, G_2, G_3\} \subseteq \mathsf{S}_1, \qquad E = \{e\}.$$

Set

$$T(e) = \{G_1\}, \qquad Hd(e) = \{G_2, G_3\}, \qquad \lambda(e) = R_{\mathrm{ref}}.$$

Assume that $G_2$ or $G_3$ (or both) refines $G_1$ in the above sense, so that

$$\big(T(e), Hd(e)\big) \in R_{\mathrm{ref}}.$$

Then the incidence constraint holds, hence $M$ is a MetaSuperHyperGraph over $(\mathsf{S}_1, \mathcal{R})$ in the sense of Definition 2.9.3. Intuitively, the single meta-hyperedge asserts that the family $\{G_2, G_3\}$ contains refinements of $G_1$.

## 2.10 Nested HyperGraph and Nested SuperHyperGraph

A nested hypergraph allows hyperedges to contain other hyperedges as elements, with ranks enforcing well-founded nesting without cycles [181]. A nested superhypergraph uses supervertices from iterated powersets; superhyperedges may contain other superhyperedges, ranked to avoid cycles[181].

**Definition 2.10.1** (Nested HyperGraph)**.** [181] Let $V$ be a finite nonempty set (the vertex set). A *nested hypergraph* on $V$ is a triple

$$H = (V, E, \rho),$$

where $E$ is a finite set (the set of hyperedges) and

$$\rho : V \uplus E \to \mathbb{N}$$

is a *rank function* such that:

1. $\rho(v) = 0$ for all $v \in V$;

2. for every $e \in E$, the hyperedge $e$ is a nonempty finite set satisfying

$$e \subseteq V \uplus E,$$

and for every $x \in e$ one has

$$\rho(x) < \rho(e).$$

A hyperedge $e \in E$ is called a *nesting hyperedge* if $e \cap E \neq \emptyset$ (i.e., $e$ contains at least one hyperedge as a member). We say that $H$ is *(nontrivially) nested* if it has at least one nesting hyperedge.

**Remark 2.10.2** (Well-foundedness)**.** *The strict inequality $\rho(x) < \rho(e)$ for $x \in e$ excludes membership cycles among hyperedges, so the nesting relation is well-founded.*

**Example 2.10.3** (A nested hypergraph with a hyperedge containing another hyperedge)**.** Let the vertex set be

$$V = \{a, b, c\}.$$

Introduce two hyperedges $E = \{e_1, e_2\}$ and define them as elements of $V \uplus E$ by

$$e_1 = \{a, b\} \subseteq V, \qquad e_2 = \{e_1, c\} \subseteq V \uplus E.$$

Define a rank function $\rho : V \uplus E \to \mathbb{N}$ by

$$\rho(a) = \rho(b) = \rho(c) = 0, \qquad \rho(e_1) = 1, \qquad \rho(e_2) = 2.$$

Then for every $x \in e_1 = \{a, b\}$ we have $\rho(x) = 0 < 1 = \rho(e_1)$, and for every $x \in e_2 = \{e_1, c\}$ we have

$$\rho(e_1) = 1 < 2 = \rho(e_2), \qquad \rho(c) = 0 < 2 = \rho(e_2).$$

Hence $H = (V, E, \rho)$ is a nested hypergraph in the sense of Definition 2.10.1. Moreover, $e_2$ is a nesting hyperedge because $e_2 \cap E = \{e_1\} \neq \emptyset$. An overview diagram of this example is provided in Fig. 2.11.

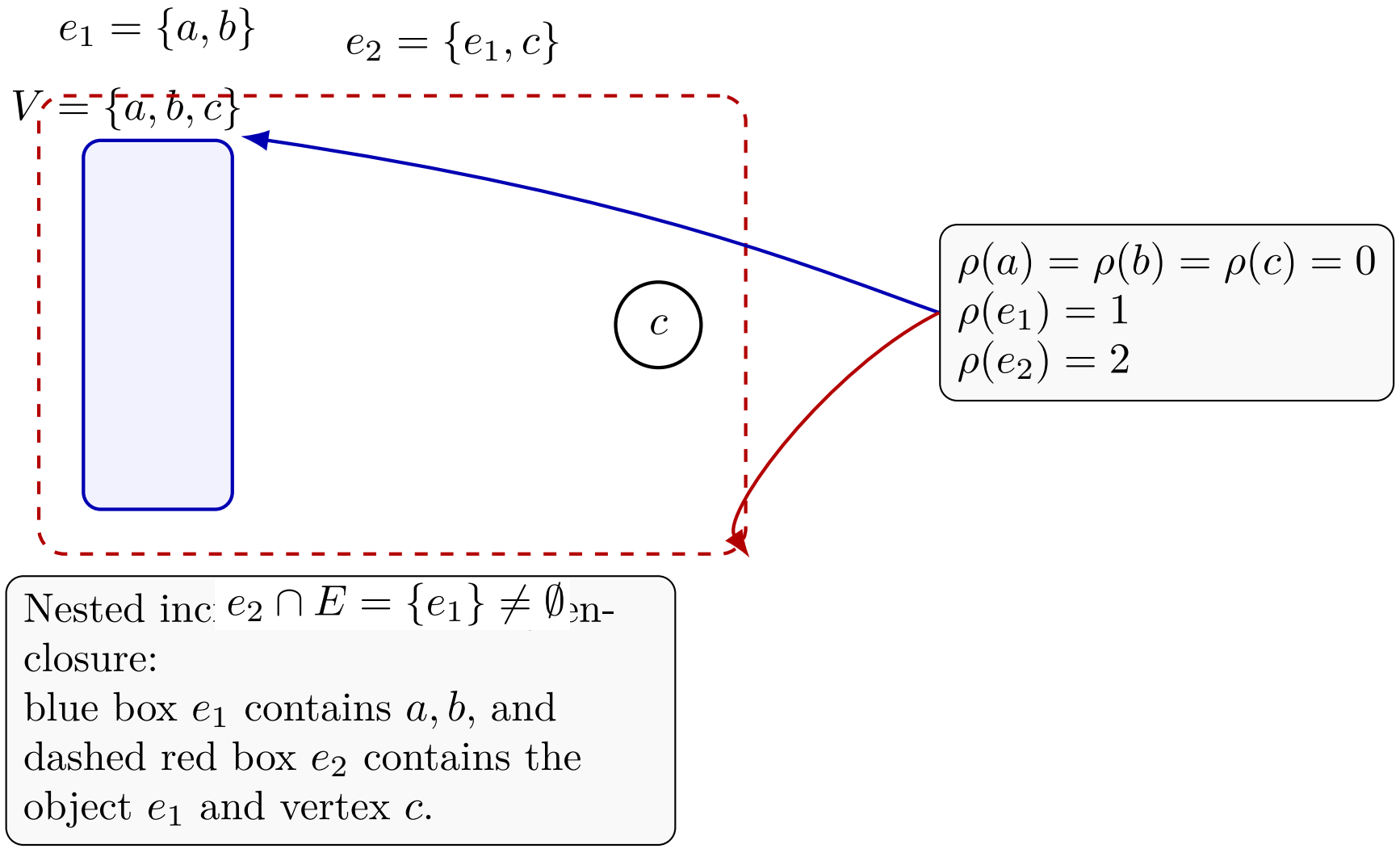

Figure 2.11: A nested hypergraph where the hyperedge $e_2$ contains another hyperedge $e_1$.

**Definition 2.10.4** (Nested level-$n$ SuperHyperGraph)**.** Fix a finite nonempty base set $V_0$ and an integer $n \geq 0$. Define iterated powersets recursively by

$$\mathcal{P}^0(V_0) := V_0, \qquad \mathcal{P}^{k+1}(V_0) := \mathcal{P}\big(\mathcal{P}^k(V_0)\big) \quad (k \geq 0).$$

Let

$$V_n \subseteq \mathcal{P}^n(V_0)$$

be a finite set; its elements are called *$n$-supervertices*. A *nested level-$n$ SuperHyperGraph* is a triple

$$H^{(n)} = (V_n, E_n, \rho),$$

where $E_n$ is a finite set (of superhyperedges) and

$$\rho : V_n \uplus E_n \to \mathbb{N}$$

is a rank function such that:

1. $\rho(v) = 0$ for all $v \in V_n$;
2. for every $e \in E_n$, the superhyperedge $e$ is a nonempty finite set satisfying

$$e \subseteq V_n \uplus E_n,$$

and for every $x \in e$ one has

$$\rho(x) < \rho(e).$$

A superhyperedge $e \in E_n$ is called *simple* if $e \subseteq V_n$, and it is called *nesting* if $e \cap E_n \neq \emptyset$ (i.e., $e$ contains at least one superhyperedge as a member).

**Example 2.10.5** (A nested level-1 SuperHyperGraph)**.** Let the base set be

$$V_0 = \{1, 2, 3\},$$

and take $n = 1$. Then $\mathcal{P}^1(V_0) = \mathcal{P}(V_0)$. Define the 1-supervertex set

$$V_1 = \big\{\{1\}, \{2\}, \{1, 2\}, \{3\}\big\} \subseteq \mathcal{P}(V_0).$$

Introduce two superhyperedges $E_1 = \{E_a, E_b\}$ and define them as elements of $V_1 \uplus E_1$ by

$$E_a = \big\{\{1, 2\}, \{3\}\big\} \subseteq V_1, \qquad E_b = \big\{E_a, \{1\}\big\} \subseteq V_1 \uplus E_1.$$

Define a rank function $\rho : V_1 \uplus E_1 \to \mathbb{N}$ by

$$\rho(v) = 0 \;\; (v \in V_1), \qquad \rho(E_a) = 1, \qquad \rho(E_b) = 2.$$

Then $H^{(1)} = (V_1, E_1, \rho)$ is a nested level-1 SuperHyperGraph in the sense of Definition 2.10.4: indeed, each endpoint of $E_a$ has rank $0 < 1$, and each endpoint of $E_b$ has rank $< 2$. Here $E_a$ is *simple* (since $E_a \subseteq V_1$), while $E_b$ is *nesting* because it contains $E_a$.

## 2.11 Multi-Hypergraph and Multi-Superhypergraph

A *Multi-Hypergraph* is a hypergraph that allows repeated hyperedges; edges form a multiset of nonempty vertex subsets with multiplicity counts [182, 183]. It is known that a multi-hypergraph generalizes both hypergraphs and multigraphs. A *Multi-Superhypergraph* is a superhypergraph that allows repeated superedges; vertices are nested-set objects, and superedges carry multiplicities.

**Definition 2.11.1** (Multi-Hypergraph)**.** [182, 183] Let $V$ be a finite nonempty set. Write

$$\mathcal{P}^{*}(V) := \mathcal{P}(V) \setminus \{\emptyset\}.$$

A *multi-hypergraph* is a triple

$$H = (V, E, \mu),$$

where

1. $E$ is a (multi)set of hyperedges, each hyperedge being a nonempty subset of $V$ (i.e. $e \in \mathcal{P}^{*}(V)$ for all $e \in E$), and
2. $\mu : E \to \mathbb{N}_{>0}$ is a multiplicity function, where $\mu(e)$ is the number of times the hyperedge $e$ occurs.

Equivalently, one may regard $E$ itself as a multiset in which each hyperedge $e$ appears with multiplicity $\mu(e)$.

**Example 2.11.2** (Repeated group interactions as a multi-hypergraph)**.** Let

$$V = \{\mathsf{A}, \mathsf{B}, \mathsf{C}, \mathsf{D}\}$$

be four participants. Suppose the same three-person meeting $\{\mathsf{A}, \mathsf{B}, \mathsf{C}\}$ occurs three times during a week, while the pair meeting $\{\mathsf{B}, \mathsf{D}\}$ occurs twice. Define the (multi)set of hyperedges

$$E = \big\{\, e_1 = \{\mathsf{A}, \mathsf{B}, \mathsf{C}\},\ e_2 = \{\mathsf{B}, \mathsf{D}\} \,\big\} \subseteq \mathcal{P}^{*}(V)$$

together with the multiplicity function $\mu : E \to \mathbb{N}_{>0}$ given by

$$\mu(e_1) = 3, \qquad \mu(e_2) = 2.$$

Then $H = (V, E, \mu)$ is a multi-hypergraph in the sense of Definition 2.11.1. An overview diagram of this example is provided in Fig. 2.12.

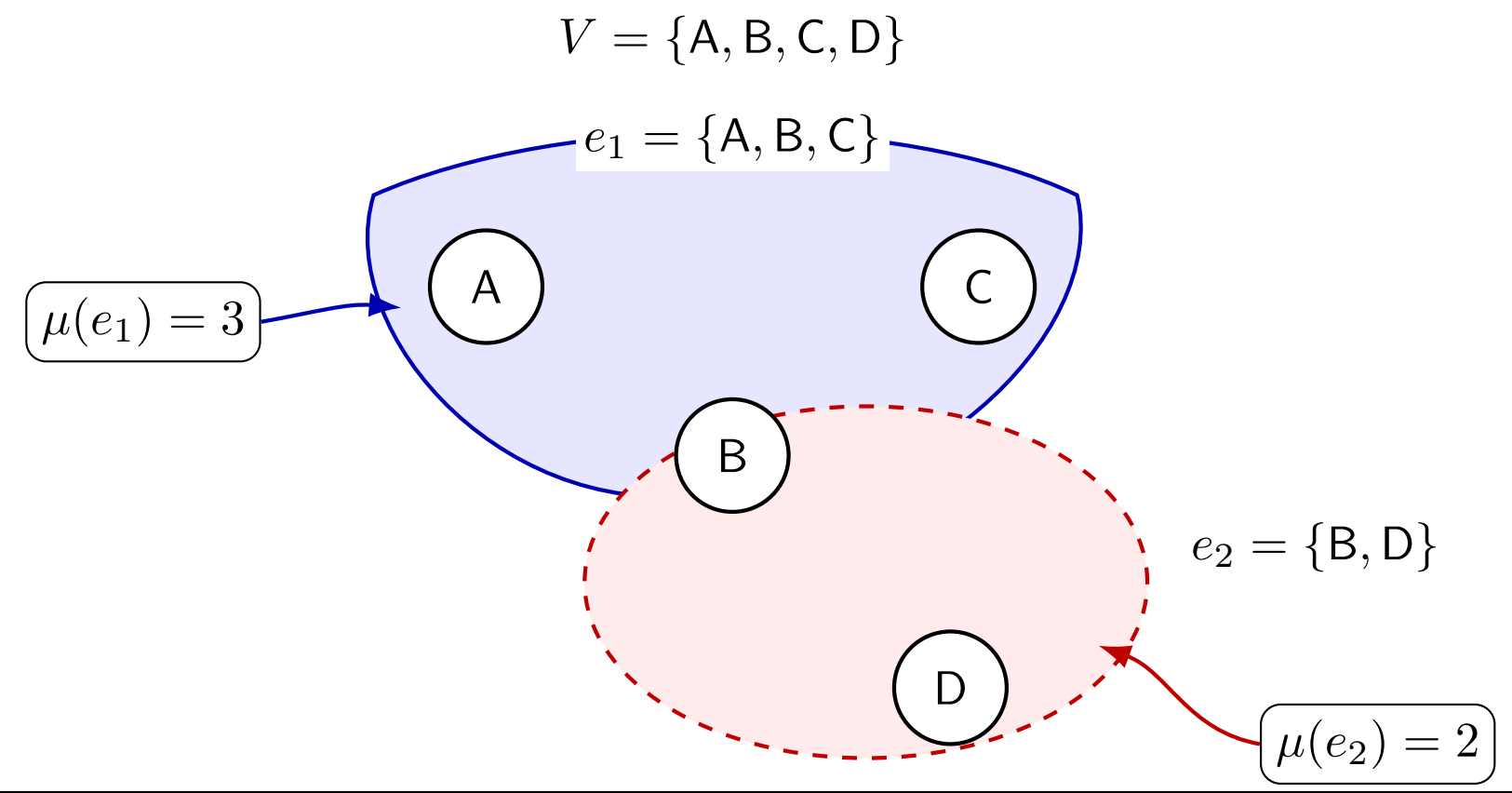

Figure 2.12: A multi-hypergraph for repeated group interactions (Example 2.11.2).

**Definition 2.11.3** (Multi $n$-SuperHyperGraph)**.** Let $V_0$ be a finite nonempty set and let $n \in \mathbb{N}_0$. Define the iterated powersets by

$$\mathcal{P}^{0}(V_0) := V_0, \qquad \mathcal{P}^{k+1}(V_0) := \mathcal{P}\big(\mathcal{P}^{k}(V_0)\big) \quad (k \geq 0).$$

A *Multi $n$-SuperHyperGraph* is a triple

$$\mathrm{MSHG}^{(n)} = (V, \mathcal{E}, \mu),$$

such that

1. $V \subseteq \mathcal{P}^n(V_0)$ (the set of $n$-supervertices);
2. $\mathcal{E} \subseteq \mathcal{P}(V) \setminus \{\varnothing\}$ (the set of $n$-superedges);
3. $\mu : \mathcal{E} \to \mathbb{N}_{>0}$ is a multiplicity function.

For each $e \in \mathcal{E}$, the value $\mu(e)$ indicates that the superedge $e$ occurs $\mu(e)$ times.

Elements of $V$ are called *$n$-supervertices*, and elements of $\mathcal{E}$ are called *$n$-superedges*.

**Example 2.11.4** (A Multi 1-SuperHyperGraph with repeated superedges)**.** Let the base set be

$$V_0 = \{1, 2, 3, 4\},$$

and take $n = 1$, so $\mathcal{P}^1(V_0) = \mathcal{P}(V_0)$.

Define the 1-supervertex set

$$V = \big\{\{1, 2\}, \{3\}, \{4\}\big\} \subseteq \mathcal{P}(V_0).$$

Define two 1-superedges (that is, nonempty subsets of $V$)

$$e_1 = \big\{\{1, 2\}, \{3\}\big\}, \qquad e_2 = \big\{\{1, 2\}, \{4\}\big\}.$$

Then indeed $e_1, e_2 \in \mathcal{P}(V)$, and hence

$$\mathcal{E} = \{e_1, e_2\} \subseteq \mathcal{P}(V).$$

Assign multiplicities by

$$\mu(e_1) = 2, \qquad \mu(e_2) = 4.$$

Therefore,

$$\mathrm{MSHG}^{(1)} = (V, \mathcal{E}, \mu)$$

is a Multi 1-SuperHyperGraph in the sense of Definition 2.11.3, where $e_1$ occurs twice and $e_2$ occurs four times.

## 2.12 Line Graph and Iterated Line Graph

The *line graph* construction converts edges into vertices: each edge of a graph becomes a vertex in the new graph, and two such vertices are adjacent if and only if the corresponding original edges share a common endpoint [184, 185, 186, 187]. Related notions are also known, including *Fuzzy Line Graphs* [186, 188], Line Digraph[189, 190, 191], Signed Line Graph[192, 193], Line HyperGraph[194], and *Neutrosophic Line Graphs* [195, 196]. By iterating this operation, one obtains the *iterated line graphs*: starting from $L^0(G) = G$, define $L^{k+1}(G) = L(L^k(G))$, producing a sequence that reflects increasingly higher-order edge-incidence patterns [197, 198, 199, 200, 201, 202]. Recently, this line-graph iteration paradigm has been further extended to *iterated line hypergraphs* and *iterated line superhypergraphs*, and active developments have been reported in these directions [203].

**Definition 2.12.1** (Line graph)**.** [185] Let $G = (V, E)$ be a finite simple undirected graph. The *line graph* of $G$ is the graph

$$L(G) \;:=\; (V_L, E_L)$$

defined by

$$V_L \;:=\; E, \qquad E_L \;:=\; \big\{\{e, f\} \in \tbinom{E}{2} \;\big|\; e \cap f \neq \emptyset\big\}.$$

Equivalently, vertices of $L(G)$ correspond to edges of $G$, and two vertices are adjacent precisely when the corresponding edges of $G$ share a common endpoint.

**Example 2.12.2** (Line graph of a path on four vertices)**.** Let $G = (V, E)$ be the path graph $P_4$ with

$$V = \{1, 2, 3, 4\}, \qquad E = \big\{\{1, 2\}, \{2, 3\}, \{3, 4\}\big\}.$$

Denote the three edges of $G$ by

$$e_{12} = \{1, 2\}, \quad e_{23} = \{2, 3\}, \quad e_{34} = \{3, 4\}.$$

Then the line graph $L(G) = (V_L, E_L)$ has vertex set

$$V_L = E = \{e_{12}, e_{23}, e_{34}\},$$

and two vertices are adjacent exactly when the corresponding edges in $G$ share an endpoint:

$$E_L = \big\{\{e_{12}, e_{23}\}, \{e_{23}, e_{34}\}\big\}.$$

Hence $L(P_4)$ is again a path on three vertices, i.e. $L(P_4) \cong P_3$. This illustrates Definition 2.12.1.

Table 2.5: Concise overview of a graph, its line graph, and iterated line graphs.

| **Object** | **Vertices** | **Adjacency / meaning** |
|---|---|---|
| Graph $G = (V, E)$ | Base objects: vertices $V$. | Edges encode *vertex adjacency*. For a simple undirected graph, $E \subseteq \binom{V}{2}$ and $\{u, v\} \in E$ means $u$ and $v$ are adjacent. |
| Line graph $L(G)$ | Edges of $G$ become vertices: $V(L(G)) := E(G)$. | Two vertices $e, f \in E(G)$ are adjacent in $L(G)$ iff the corresponding edges in $G$ share an endpoint (i.e., $e \cap f \neq \emptyset$). Thus $L(G)$ encodes *edge-incidence adjacency* of $G$. |
| Iterated line graph $L^k(G)$ | Vertices are edges of the previous iterate: $V(L^k(G)) := E(L^{k-1}(G))$ for $k \geq 1$. | Defined recursively by $L^0(G) = G$ and $L^{k+1}(G) = L(L^k(G))$. This yields a hierarchy capturing progressively higher-order patterns in how edges touch one another across iterations. |

**Remark 2.12.3** (Directed variant (line digraph))**.** *If $G$ is directed, one often considers an* oriented *line graph (also called a* line digraph*), whose vertices are the directed edges (arcs) and in which $(u \to v)$ is adjacent to $(v \to w)$, optionally with a non-backtracking constraint $w \neq u$. This oriented construction is closely related to non-backtracking operators used in graph learning.*

**Definition 2.12.4** (Iterated line graph)**.** [197, 198] Define the iterated line-graph operator $L^k$ recursively by

$$L^0(G) \;:=\; G, \qquad L^{k+1}(G) \;:=\; L\big(L^k(G)\big) \quad (k \in \mathbb{N}).$$

Thus $L^1(G) = L(G)$, $L^2(G) = L(L(G))$, and so forth. The sequence $\{L^k(G)\}_{k\geq 0}$ is called the *line-graph iteration* (or *line-graph hierarchy*) of $G$.

**Example 2.12.5** (Iterated line graph of a cycle $C_4$)**.** Let $G = C_4$ be the 4-cycle with vertex set $\{1, 2, 3, 4\}$ and edge set

$$E = \big\{\{1, 2\}, \{2, 3\}, \{3, 4\}, \{4, 1\}\big\}.$$

Each edge in $C_4$ meets exactly two other edges, so the line graph $L(C_4)$ is again a 4-cycle:

$$L(C_4) \cong C_4.$$

Consequently, the line-graph iteration stabilizes:

$$L^k(C_4) \cong C_4 \qquad \text{for all } k \geq 1.$$

This provides a concrete example of Definition 2.12.4.

For reference, an overview of a graph, its line graph, and iterated line graphs is presented in Table 2.5 (cf.[194]).

## 2.13 Iterated Total Graph

A total graph has vertices corresponding to both the vertices and the edges of the original graph, with adjacency determined by the adjacency and incidence relations of the original graph (cf. [204, 205]). It is also well known that total graphs are closely related to other classical graph transformations, particularly line graphs and middle graphs [206, 207, 208].

**Definition 2.13.1** (Total graph)**.** [209, 210] Let $G = (V(G), E(G))$ be a loopless simple undirected graph. The *total graph* $T(G)$ is the (simple) graph with

$$V\big(T(G)\big) = V(G) \,\dot{\cup}\, E(G),$$

where two distinct vertices $x, y \in V(G) \cup E(G)$ are adjacent in $T(G)$ iff one of the following holds:

1. (vertex–vertex) $x, y \in V(G)$ and $xy \in E(G)$;

2. (edge–edge) $x, y \in E(G)$ and $x \cap y \neq \emptyset$ in $G$;
3. (vertex–edge) $\{x, y\} = \{v, e\}$ with $v \in V(G)$, $e \in E(G)$, and $v \in e$.

**Example 2.13.2** (Total graph of the path $P_3$)**.** Let $G = P_3$ be the path on three vertices

$$V(G) = \{1, 2, 3\}, \qquad E(G) = \{\{1, 2\}, \{2, 3\}\}.$$

Write $e_{12} = \{1, 2\}$ and $e_{23} = \{2, 3\}$. Then the total graph $T(G)$ has vertex set

$$V\bigl(T(G)\bigr) = \{1, 2, 3\} \,\dot{\cup}\, \{e_{12}, e_{23}\}.$$

Adjacency in $T(G)$ arises from the three rules:

- *vertex–vertex:* $1 \sim 2$ and $2 \sim 3$ (since $\{1, 2\}, \{2, 3\} \in E(G)$);
- *edge–edge:* $e_{12} \sim e_{23}$ (since $e_{12} \cap e_{23} = \{2\} \neq \emptyset$ in $G$);
- *vertex–edge:* $1 \sim e_{12}$, $2 \sim e_{12}$, $2 \sim e_{23}$, $3 \sim e_{23}$ (because each vertex is adjacent to its incident edges).

Thus $T(P_3)$ is a simple graph on five vertices encoding both adjacency and incidence information of $P_3$, as in the definition of the total graph.

An iterated total graph repeatedly applies the total graph operation to a graph, incorporating vertices, edges, and all incidence relationships at each stage [205, 204].

**Definition 2.13.3** (Iterated total graphs)**.** Define $T^0(G) := G$, and for each integer $k \geq 1$ set

$$T^k(G) \;:=\; T\bigl(T^{k-1}(G)\bigr).$$

(Thus $T^1(G) = T(G)$, $T^2(G) = T(T(G))$, etc.) This notation is also used in the literature.

**Example 2.13.4** (Iterated total graphs starting from $K_2$)**.** Let $G = K_2$ be the complete graph on two vertices:

$$V(G) = \{1, 2\}, \qquad E(G) = \{\{1, 2\}\}.$$

Let $e = \{1, 2\}$ denote the unique edge. Then the total graph $T(G)$ has vertex set

$$V\bigl(T(G)\bigr) = \{1, 2, e\}$$

and edges

$$\{1, 2\},\; \{1, e\},\; \{2, e\},$$

so $T(K_2) \cong K_3$.

Applying the total operation again yields

$$T^2(K_2) = T\bigl(T(K_2)\bigr) = T(K_3).$$

Since $K_3$ has three vertices and three edges, $T(K_3)$ has 6 vertices and encodes all vertex–vertex, edge–edge, and vertex–edge incidences of $K_3$. Hence $(T^k(K_2))_{k\geq 0}$ provides a concrete iterated total-graph hierarchy

$$K_2 \;=\; T^0(K_2) \;\longmapsto\; T^1(K_2) \cong K_3 \;\longmapsto\; T^2(K_2) = T(K_3) \;\longmapsto\; \cdots,$$

illustrating the definition of iterated total graphs.

## 2.14 Hierarchical SuperHyperGraph

A hierarchical superhypergraph is a superhypergraph whose vertices live across multiple powerset levels, with edges allowed to join mixed-level supervertices, while maintaining downward-closure coherence (cf.[211, 212, 213]).

**Definition 2.14.1** (Support map and edge-support)**.** Define the support map $\mathrm{supp} : V \to \mathcal{P}(V_0) \setminus \{\emptyset\}$ recursively by

$$\mathrm{supp}(v) \coloneqq \{v\} \quad (v \in V_0), \qquad \mathrm{supp}(X) \coloneqq \bigcup_{y \in X} \mathrm{supp}(y) \quad (X \in V \setminus V_0),$$

and for each $e \in E$ define its *base support* (a hyperedge on $V_0$) by

$$\sigma(e) \;\coloneqq\; \bigcup_{x \in e} \mathrm{supp}(x) \;\in\; \mathcal{P}(V_0) \setminus \{\emptyset\}.$$

**Definition 2.14.2** (Hierarchical SuperHyperGraph of height $r$)**.** (cf.[211, 213]) Let $V_0$ be a finite, nonempty base set. For $k \geq 0$ define iterated powersets

$$\mathcal{P}^0(V_0) := V_0, \qquad \mathcal{P}^{k+1}(V_0) := \mathcal{P}\big(\mathcal{P}^k(V_0)\big),$$

and fix an integer $r \geq 0$. Set the *hierarchical universe*

$$\mathcal{U}_r(V_0) \ := \ \bigcup_{k=0}^{r} \Big(\mathcal{P}^k(V_0) \setminus \{\emptyset\}\Big).$$

For $x \in \mathcal{U}_r(V_0)$, define its *level*

$$\ell(x) \ := \ \min\{\, k \in \{0, 1, \dots, r\} : x \in \mathcal{P}^k(V_0) \,\}.$$

A *hierarchical superhypergraph of height* $r$ on $V_0$ is a pair

$$\mathbb{H}^{\langle r \rangle} = (V, E)$$

such that

(H1) (*Hierarchical vertex set*) $V$ is a finite nonempty set with

$$V \subseteq \mathcal{U}_r(V_0).$$

Elements of $V$ are called *hierarchical supervertices.*

(H2) (*Cross-level edges*) $E$ is a finite family of nonempty subsets of $V$:

$$E \subseteq \mathcal{P}(V) \setminus \{\emptyset\}.$$

Elements of $E$ are called *hierarchical superhyperedges.* In particular, a superhyperedge may contain supervertices of *different* levels.

(H3) (*Coherence / downward closure*) If $X \in V$ and $\ell(X) \geq 1$, then

$$X \subseteq V.$$

Equivalently, whenever a higher-level supervertex is present, all its immediate constituents are also present as supervertices.

For each $k \in \{0, \dots, r\}$ we define the *k-th layer* by

$$V_k \ := \ \{\, x \in V : \ \ell(x) = k \,\}, \qquad \text{so that} \qquad V = \dot{\bigcup}_{k=0}^{r} V_k.$$

**Example 2.14.3** (A hierarchical SuperHyperGraph of height 2 with cross-level edges)**.** Let the base set be

$$V_0 = \{a, b, c, d\}.$$

Consider the hierarchical universe

$$\mathcal{U}_2(V_0) = \big(V_0\big) \ \cup \ \big(\mathcal{P}(V_0) \setminus \{\emptyset\}\big) \ \cup \ \big(\mathcal{P}(\mathcal{P}(V_0)) \setminus \{\emptyset\}\big).$$

Define hierarchical supervertices (across levels $0, 1, 2$) by

$$X = \{a, b\} \in \mathcal{P}(V_0), \qquad Y = \{c, d\} \in \mathcal{P}(V_0), \qquad U = \{X\} = \{\{a, b\}\} \in \mathcal{P}(\mathcal{P}(V_0)).$$

Set

$$V = \{a, b, c, d, \ X, \ Y, \ U\} \subseteq \mathcal{U}_2(V_0).$$

This $V$ satisfies the coherence condition (H3): since $X, Y \in V$ (level 1), their elements $a, b, c, d$ belong to $V$; and since $U \in V$ (level 2), its element $X$ belongs to $V$.

Next, define a family of hierarchical superhyperedges

$$E = \{e_1, e_2\}, \qquad e_1 = \{X, Y\}, \qquad e_2 = \{U, c\}.$$

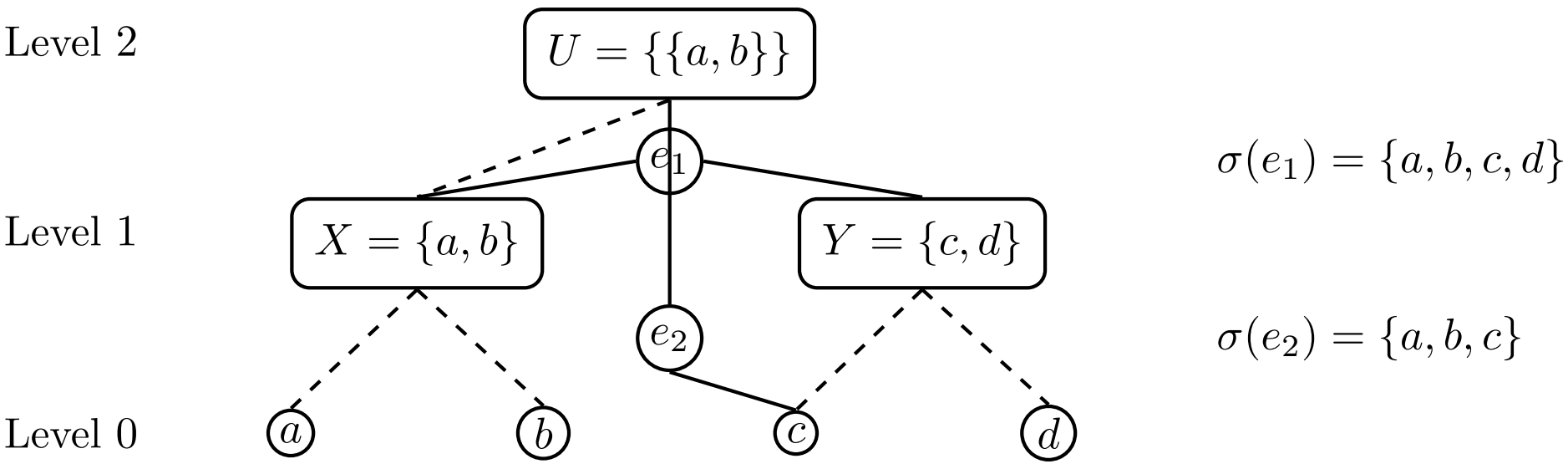

Figure 2.13: A hierarchical superhypergraph of height 2: vertices may live on different levels and edges may cross levels. Dashed lines indicate constituent relations (coherence), while $e_1, e_2$ are superhyperedges.

Thus $e_1$ links two level-1 supervertices (two teams), while $e_2$ links a level-2 supervertex (a nested unit) to a level-0 vertex (an individual), illustrating a cross-level edge as allowed by (H2). Therefore

$$\mathbb{H}^{\langle 2\rangle} = (V, E)$$

is a hierarchical superhypergraph of height 2 in the sense of Definition 2.14.2.

**Support map and edge-support.** Using Definition 2.14.1, the support map Supp satisfies

$$\mathrm{Supp}(a) = \{a\},\ \mathrm{Supp}(b) = \{b\},$$

$$\mathrm{Supp}(c) = \{c\},\ \mathrm{Supp}(d) = \{d\},$$

$$\mathrm{Supp}(X) = \mathrm{Supp}(\{a, b\}) = \{a, b\},$$

$$\mathrm{Supp}(Y) = \{c, d\},$$

$$\mathrm{Supp}(U) = \mathrm{Supp}(\{X\}) = \mathrm{Supp}(X) = \{a, b\}.$$

Hence the base supports of the two edges are

$$\sigma(e_1) = \mathrm{Supp}(X) \cup \mathrm{Supp}(Y) = \{a, b, c, d\},$$

$$\sigma(e_2) = \mathrm{Supp}(U) \cup \mathrm{Supp}(c) = \{a, b, c\}.$$

An overview diagram of this example is provided in Fig. 2.13.

## 2.15 Recursive HyperGraph and Recursive SuperHyperGraph

A *recursive hypergraph* is a hypergraph-like object in which a hyperedge may contain ordinary vertices and also lower-level hyperedges as elements, thereby permitting nested incidence up to a prescribed recursion depth [214, 215, 216, 217].

**Definition 2.15.1** (Depth-$k$ powerset universe)**.** [214, 217] Let $S$ be a nonempty set and let $k \in \mathbb{N} \cup \{0\}$. Define a hierarchy of sets $(S_i)_{i\geq 0}$ by

$$S_0 := S, \qquad S_i := \mathcal{P}\Big(\bigcup_{j=0}^{i-1} S_j\Big) \quad (i \geq 1).$$

Set $U_{S,k} := \bigcup_{i=0}^{k} S_i$. The *depth-$k$ powerset universe* over $S$ is

$$2_{S,k} \ := \ \mathcal{P}(U_{S,k}).$$

**Definition 2.15.2** ($k$-recursive hypergraph)**.** [214, 217] Let $V$ be a finite vertex set and let $k \in \mathbb{N} \cup \{0\}$. A *$k$-recursive hypergraph* is a pair

$$H = (V, E)$$

such that

$$E \subseteq 2_{V,k} \setminus \{\emptyset\},$$

where $2_{V,k}$ is the depth-$k$ powerset universe from Definition 2.15.1 applied to $S = V$.

In particular, for $k = 0$ one has $2_{V,0} = \mathcal{P}(V)$ and thus $E \subseteq \mathcal{P}(V) \setminus \{\emptyset\}$, i.e., $H$ reduces to an ordinary hypergraph.

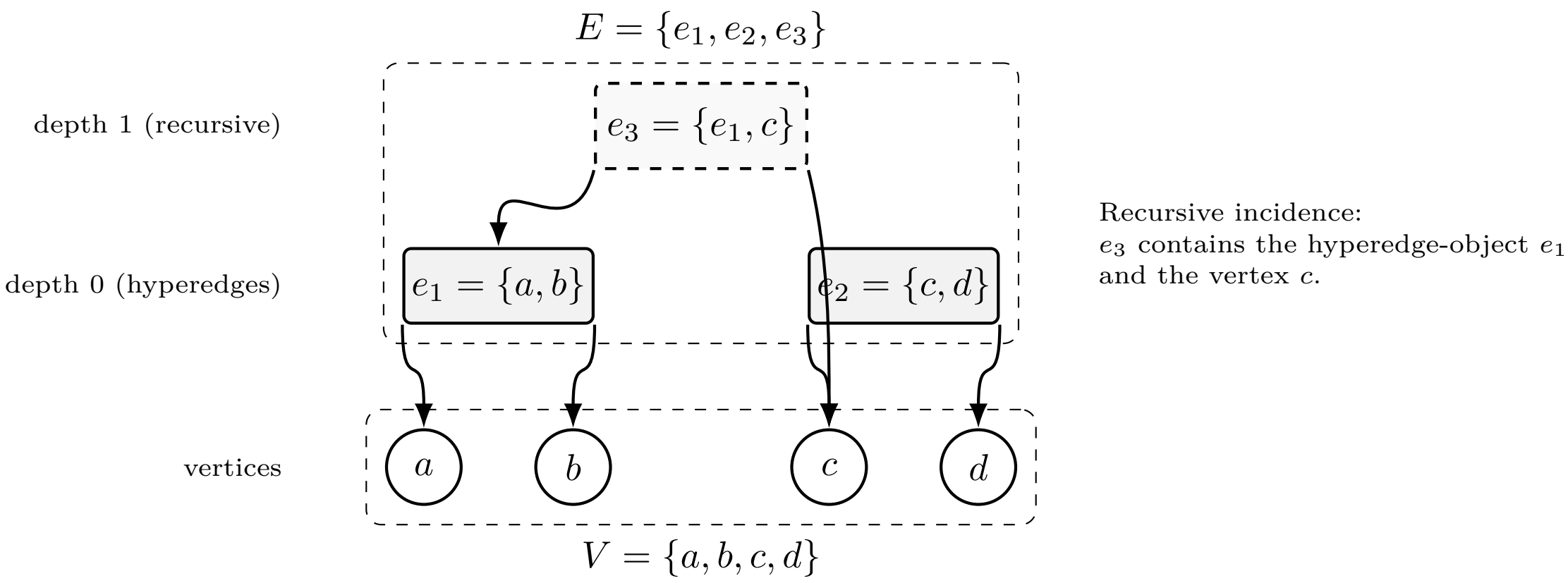

Figure 2.14: A schematic illustration of the 1-recursive hypergraph in Example 2.15.3. The recursive hyperedge $e_3$ contains a lower-level hyperedge $e_1$ and a vertex $c$.

**Example 2.15.3** (A 1-recursive hypergraph (edge-of-edges for task bundles))**.** Let

$$V = \{a, b, c, d\}$$

be four tasks. Consider two ordinary hyperedges (task bundles)

$$e_1 = \{a, b\}, \qquad e_2 = \{c, d\}.$$

A 1-*recursive* hyperedge may contain vertices and also lower-level hyperedges as elements. Define

$$e_3 = \{e_1, c\},$$

which represents a higher-level bundle "do the bundle $\{a, b\}$ together with task $c$." Let

$$E = \{e_1, e_2, e_3\}.$$

Then $H = (V, E)$ is a 1-recursive hypergraph in the sense of Definition 2.15.2: indeed, $e_1, e_2 \in 2_{V,0} = \mathcal{P}(V)$ and $e_3 \in 2_{V,1}$ because it mixes a depth-0 hyperedge $e_1$ with a vertex $c$. An overview diagram of this example is provided in Fig. 2.14.

An $(n, k)$-recursive SuperHyperGraph combines hierarchical supervertices (via iterated powersets) with recursive superhyperedges of bounded depth $k$, allowing edges to contain supervertices and nested lower-level edges as elements [218, 212, 219, 220, 221]. Related concepts, such as fuzzy recursive SuperHyperGraphs, have also been investigated [222, 223].

**Definition 2.15.4** ($(n, k)$-recursive SuperHyperGraph)**.** [224, 225] Fix a finite nonempty base set $V_0$ and let $n, k \in \mathbb{N} \cup \{0\}$. An $(n, k)$-*recursive SuperHyperGraph* is a pair

$$\mathrm{RSHG}^{(n,k)} = (V, E)$$

satisfying:

(i) *(Hierarchical supervertex set).* $V \subseteq \mathcal{P}^n(V_0)$.

(ii) *(Recursive superhyperedge family).* $E \subseteq 2_{V,k} \setminus \{\emptyset\}$, where $2_{V,k}$ is the depth-$k$ powerset universe constructed from $S = V$ as in Definition 2.15.1.

**Example 2.15.5** (A $(1, 1)$-recursive SuperHyperGraph (teams with a recursive superedge))**.** Let the base set be

$$V_0 = \{1, 2, 3, 4\},$$

and take $n = 1$, so $\mathcal{P}^1(V_0) = \mathcal{P}(V_0)$. Define the 1-supervertex set (teams)

$$V = \big\{\{1, 2\}, \{3\}, \{4\}\big\} \subseteq \mathcal{P}(V_0).$$

Consider two ordinary superhyperedges (depth 0 edges in $\mathcal{P}(V)$):

$$e_1 = \big\{\{1, 2\}, \{3\}\big\}, \qquad e_2 = \big\{\{1, 2\}, \{4\}\big\}.$$

Now form a recursive superhyperedge of depth 1 by letting

$$e_3 = \{e_1, \{4\}\} \in 2_{V,1},$$

which can be read as a higher-level interaction "activate the collaboration $e_1$ and also involve the team $\{4\}$." Let

$$E = \{e_1, e_2, e_3\} \subseteq 2_{V,1} \setminus \{\emptyset\}.$$

Then $\mathrm{RSHG}^{(1,1)} = (V, E)$ is a $(1,1)$-recursive SuperHyperGraph in the sense of Definition 2.15.4.

## 2.16 Tree-Vertex Graph

A tree-vertex graph labels tree nodes by nested vertex-sets; edges connect nodes within levels, representing relations across abstraction depths [226].

**Definition 2.16.1** (Depth-$n$ Tree-Vertex Graph with level edges)**.** [226] Let $V_0$ be a finite, nonempty set of *base vertices* and let $n \in \mathbb{N}$. A *depth-$n$ Tree-Vertex Graph (TVG) on* $V_0$ is a quadruple

$$\mathrm{TVG}^{(n)} \;=\; \bigl(V_0,\; T,\; \eta,\; \{E^{(k)}\}_{k=0}^{n}\bigr),$$

where:

(i) $T = (\mathcal{N}, \mathcal{F}, r)$ is a rooted tree whose leaves have depth 0 and whose root has depth $n$. For each $k \in \{0, \ldots, n\}$ define the level set

$$\mathcal{N}_k := \{u \in \mathcal{N} \mid \mathrm{depth}(u) = k\}.$$

(ii) $\eta$ is a *nested labeling* (level-typed label map)

$$\eta : \mathcal{N} \longrightarrow \bigcup_{k=0}^{n} \mathrm{PS}^k(V_0) \setminus \{\emptyset\}$$

satisfying:

(a) (*Level-typing*) For every $u \in \mathcal{N}_k$, one has

$$\eta(u) \in \mathrm{PS}^k(V_0) \setminus \{\emptyset\}.$$

(b) (*Leaf grounding*) The restriction $\eta|_{\mathcal{N}_0} : \mathcal{N}_0 \to V_0$ is a bijection (so each leaf represents exactly one base vertex).

(c) (*Recursive nesting along the tree*) For every internal node $u \in \mathcal{N}_k$ with $k \geq 1$,

$$\eta(u) \;=\; \{\eta(v) \mid v \in \mathrm{Ch}(u)\},$$

so the label of $u$ is literally the set of its children's labels (hence lives in the next iterated powerset).

(iii) The *support* (flattened base-vertex set) of a tree-vertex $u \in \mathcal{N}_k$ is defined by

$$\lambda(u) := \mathrm{Flat}_k(\eta(u)) \subseteq V_0$$

.

(iv) For each level $k \in \{0, \ldots, n\}$, $E^{(k)}$ is a set of *level-$k$ graph edges*:

$$E^{(k)} \subseteq \bigl\{\{u, v\} \subseteq \mathcal{N}_k \mid u \neq v\bigr\}.$$

An edge $\{u, v\} \in E^{(k)}$ encodes a binary relation *within the same abstraction depth $k$* between the hierarchical units represented by $u$ and $v$.

**Example 2.16.2** (A depth-2 Tree-Vertex Graph for a small organization)**.** Let the base vertices be four employees

$$V_0 = \{a, b, c, d\}.$$

We construct a depth-2 Tree-Vertex Graph $\mathrm{TVG}^{(2)}$ in the sense of Definition 2.16.1.

**(1) The rooted tree.** Let $T = (\mathcal{N}, \mathcal{F}, r)$ be the rooted tree with node set

$$\mathcal{N} = \{r, u_1, u_2, a_0, b_0, c_0, d_0\},$$

root $r$ at depth 2, internal nodes $u_1, u_2$ at depth 1, and leaves

$$\mathcal{N}_0 = \{a_0, b_0, c_0, d_0\}, \quad \mathcal{N}_1 = \{u_1, u_2\}, \quad \mathcal{N}_2 = \{r\}.$$

Define the parent–child relations by

$$\mathrm{Ch}(r) = \{u_1, u_2\}, \qquad \mathrm{Ch}(u_1) = \{a_0, b_0\}, \qquad \mathrm{Ch}(u_2) = \{c_0, d_0\}.$$

Thus the tree groups employees into two teams $\{a, b\}$ and $\{c, d\}$, and then into one department.

**(2) Nested labeling.** Define $\eta : \mathcal{N} \to \bigcup_{k=0}^{2} \mathrm{PS}^k(V_0) \setminus \{\emptyset\}$ by

$$\eta(a_0) = a, \ \eta(b_0) = b, \ \eta(c_0) = c, \ \eta(d_0) = d,$$

$$\eta(u_1) = \{\eta(a_0), \eta(b_0)\} = \{a, b\}, \qquad \eta(u_2) = \{\eta(c_0), \eta(d_0)\} = \{c, d\},$$
$$\eta(r) = \{\eta(u_1), \eta(u_2)\} = \big\{\{a, b\}, \{c, d\}\big\}.$$

By construction, $\eta|_{\mathcal{N}_0}$ is a bijection onto $V_0$, and each internal label is exactly the set of its children's labels.

**(3) Supports.** Let $\lambda(u) = \mathrm{Flat}_k(\eta(u)) \subseteq V_0$ be the flattened support. Then

$$\lambda(a_0) = \{a\}, \ \lambda(b_0) = \{b\}, \ \lambda(c_0) = \{c\}, \ \lambda(d_0) = \{d\},$$

$$\lambda(u_1) = \{a, b\}, \qquad \lambda(u_2) = \{c, d\}, \qquad \lambda(r) = \{a, b, c, d\}.$$

**(4) Level edges.** Define level-$k$ edge sets by

$$E^{(0)} = \big\{\{a_0, b_0\}, \{c_0, d_0\}\big\}, \qquad E^{(1)} = \big\{\{u_1, u_2\}\big\}, \qquad E^{(2)} = \emptyset.$$

Here $E^{(0)}$ encodes pairwise communication links inside each team at the employee level, while $E^{(1)}$ encodes a coordination link between the two teams.

Therefore,

$$\mathrm{TVG}^{(2)} = \big(V_0, \ T, \ \eta, \ \{E^{(k)}\}_{k=0}^{2}\big)$$

is a depth-2 Tree-Vertex Graph on $V_0$. An overview diagram of this example is provided in Fig. 2.15.

## 2.17 MultiMeta-Graph

Unlike an ordinary metagraph, where each vertex is a single graph, a MultiMetaGraph allows each vertex to represent a finite nonempty family of graphs. Thus, one higher-level vertex may encode a module, portfolio, layer, cluster, or any other grouped unit consisting of several graphs.

**Notation 2.17.1.** For a set $X$, let $\mathcal{P}_{\mathrm{fin}}(X)$ denote the family of all finite subsets of $X$, and let

$$\mathcal{P}^*_{\mathrm{fin}}(X) := \mathcal{P}_{\mathrm{fin}}(X) \setminus \{\emptyset\}.$$

**Definition 2.17.2** (MultiMetaGraph)**.** Fix a nonempty universe $\mathfrak{G}$ of finite graphs (undirected, loopless by default) and a nonempty family

$$\mathcal{R} \subseteq \mathcal{P}\Big(\mathcal{P}^*_{\mathrm{fin}}(\mathfrak{G}) \times \mathcal{P}^*_{\mathrm{fin}}(\mathfrak{G})\Big)$$

of admissible binary relations on finite nonempty families of graphs. A *MultiMetaGraph* over $(\mathfrak{G}, \mathcal{R})$ is a directed labelled multigraph

$$\mathcal{M} = (V, E, s, t, \lambda)$$

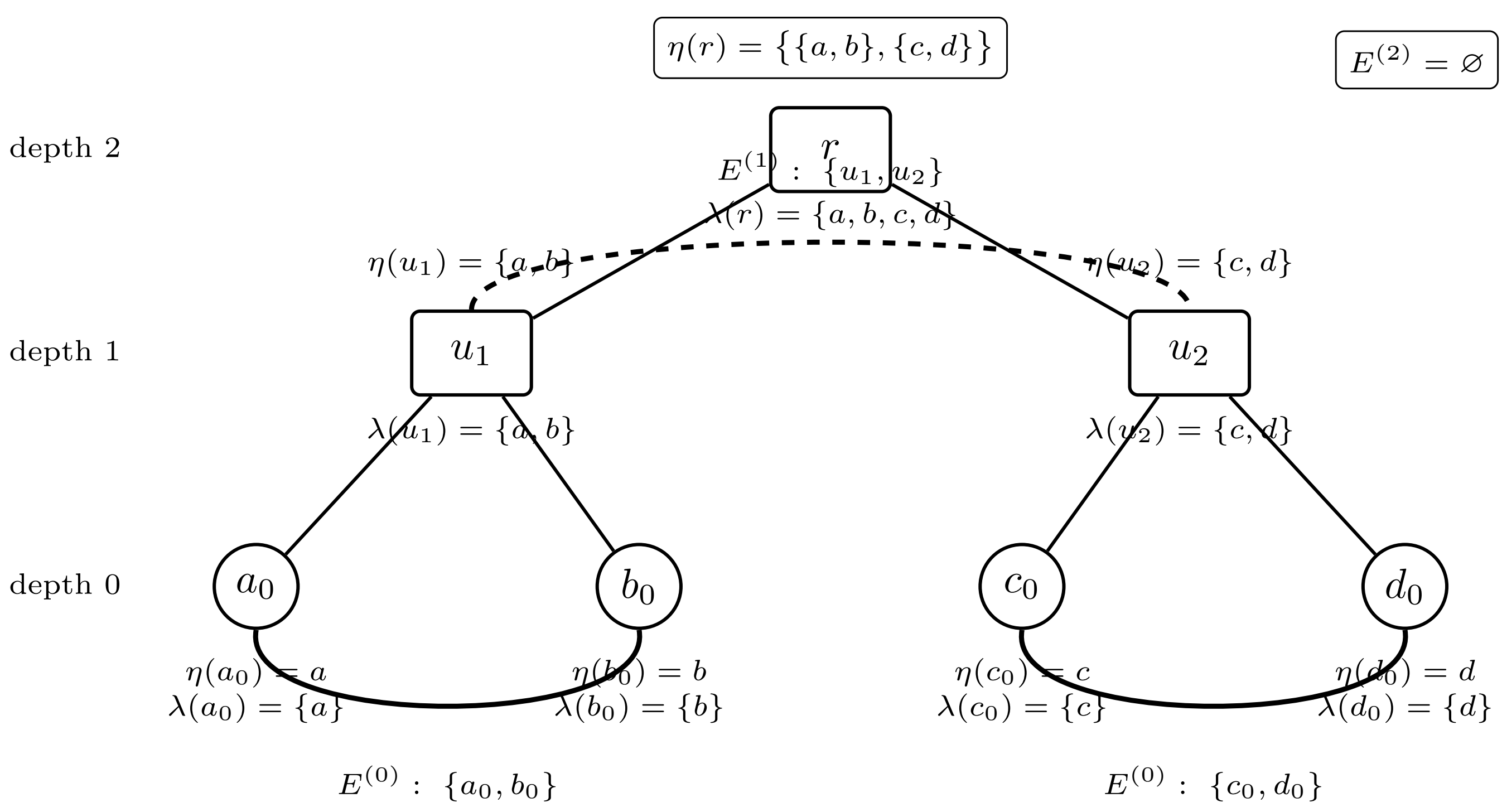

Figure 2.15: A depth-2 Tree-Vertex Graph for a small organization (Example 2.16.2). Leaves represent employees, internal nodes represent teams, and the root represents the department.

such that

$$V \subseteq \mathcal{P}^*_{\mathrm{fin}}(\mathfrak{G}), \qquad s,t : E \to V, \qquad \lambda : E \to \mathcal{R},$$

and

$$\forall e \in E : \quad \big(s(e), t(e)\big) \in \lambda(e).$$

Each element $v \in V$ is called a *multi-meta-vertex*; it is itself a finite nonempty family of graphs. If

$$v = \{G_1, \dots, G_m\},$$

then we say that $v$ *contains* the graphs $G_1, \dots, G_m$. For $e \in E$ with $\lambda(e) = R$, we write

$$s(e) \xrightarrow{R} t(e)$$

and call $e$ a *multi-meta-edge*.

**Remark 2.17.3.** *A MultiMetaGraph is still a graph-like object at the top level, but its vertices are no longer single graphs; they are finite graph-families. The same base graph may belong to several distinct multi-meta-vertices unless one explicitly imposes disjointness. If every relation in $\mathcal{R}$ is symmetric, then the MultiMetaGraph may be regarded as an undirected labelled multigraph.*

**Example 2.17.4** (A concrete MultiMetaGraph of graph-families)**.** Let $\mathfrak{G}$ be a universe of finite simple graphs containing the following four graphs:

$$G_1 = P_2, \qquad G_2 = P_3, \qquad G_3 = K_3, \qquad G_4 = P_4,$$

where $P_n$ denotes the path graph on $n$ vertices and $K_3$ denotes the triangle graph.

Define two admissible binary relations on finite nonempty families of graphs:

$$R_{\mathrm{com}} := \Big\{ (A,B) \in \mathcal{P}^*_{\mathrm{fin}}(\mathfrak{G}) \times \mathcal{P}^*_{\mathrm{fin}}(\mathfrak{G}) \;\Big|\; A \cap B \neq \emptyset \Big\},$$

and

$$R_{\mathrm{inc}} := \Big\{(A,B) \in \mathcal{P}^*_{\mathrm{fin}}(\mathfrak{G}) \times \mathcal{P}^*_{\mathrm{fin}}(\mathfrak{G}) \;\Big|\; \exists\, H \in A,\ \exists\, K \in B \text{ such that } |V(K)| = |V(H)| + 1\Big\}.$$

Let

$$\mathcal{R} = \{R_{\mathrm{com}}, R_{\mathrm{inc}}\}.$$

Now define three multi-meta-vertices by

$$M_1 = \{G_1, G_2\}, \qquad M_2 = \{G_2, G_3\}, \qquad M_3 = \{G_4\}.$$

Then

$$V = \{M_1, M_2, M_3\} \subseteq \mathcal{P}^*_{\mathrm{fin}}(\mathfrak{G}).$$

Next, define the multi-meta-edge set

$$E = \{e_1, e_2\},$$

together with source, target, and label maps by

$$s(e_1) = M_1, \qquad t(e_1) = M_2, \qquad \lambda(e_1) = R_{\mathrm{com}},$$

and

$$s(e_2) = M_2, \qquad t(e_2) = M_3, \qquad \lambda(e_2) = R_{\mathrm{inc}}.$$

We verify the incidence condition. Since

$$M_1 \cap M_2 = \{G_2\} \neq \emptyset,$$

we have

$$(M_1, M_2) \in R_{\mathrm{com}}.$$

Also, $G_3 \in M_2$ has 3 vertices and $G_4 \in M_3$ has 4 vertices, so

$$|V(G_4)| = |V(G_3)| + 1,$$

hence

$$(M_2, M_3) \in R_{\mathrm{inc}}.$$

Therefore,

$$\forall e \in E: \quad (s(e), t(e)) \in \lambda(e).$$

Consequently,

$$\mathcal{M} = (V, E, s, t, \lambda)$$

is a MultiMetaGraph over $(\mathfrak{G}, \mathcal{R})$.

The multi-meta-vertex $M_1$ groups the two graphs $P_2$ and $P_3$, while $M_2$ groups $P_3$ and $K_3$. Thus the edge

$$M_1 \xrightarrow{R_{\mathrm{com}}} M_2$$

records that the two graph-families share a common graph object. The edge

$$M_2 \xrightarrow{R_{\mathrm{inc}}} M_3$$

records that a graph in $M_3$ has one more vertex than some graph in $M_2$. An illustration is given in Fig. 2.16.

## 2.18 Transfinite SuperHyperGraph

A Transfinite Graph extends ordinary infinite graphs through ordinal-ranked nodes and paths, enabling connections across successive finite and transfinite ranks [227, 228, 229]. A transfinite superhypergraph models hierarchical set-based vertices with downward closure, connecting objects across infinitely many ordinally organized transfinite levels coherently.

$$\mathcal{P}^*(X) := \mathcal{P}(X) \setminus \{\emptyset\}, \qquad [V]^2 := \big\{\{x,y\} \subseteq V : x \neq y\big\}.$$

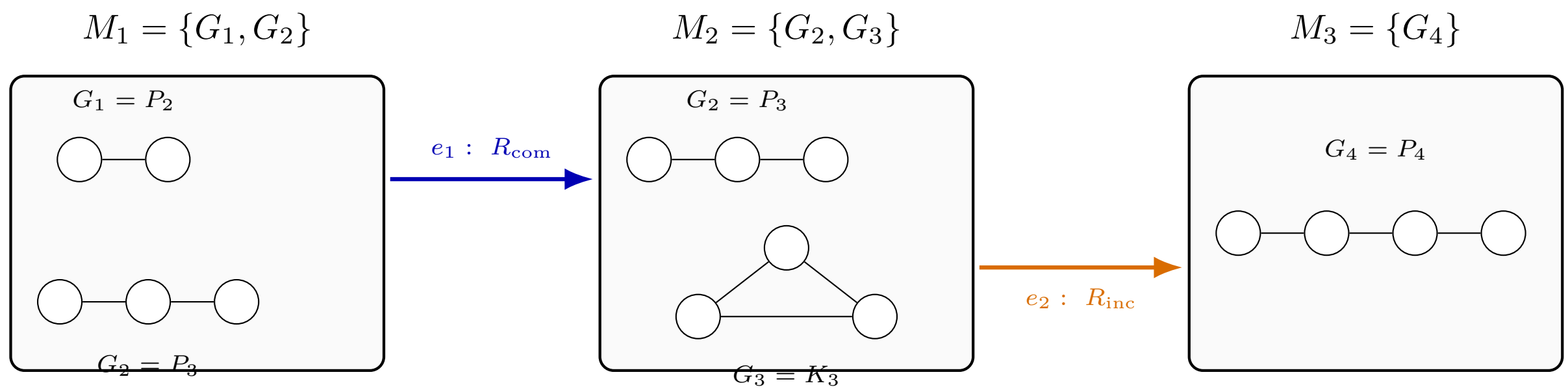

$R_{\mathrm{com}} : (A, B) \in R_{\mathrm{com}} \iff A \cap B \neq \emptyset$ $R_{\mathrm{inc}} : (A, B) \in R_{\mathrm{inc}} \iff \exists H \in A, \exists K \in B, |V(K)| = |V(H)| + 1$

Here $M_1 \cap M_2 = \{G_2\}$, and $G_3 \in M_2$, $G_4 \in M_3$ satisfy $4 = 3 + 1$.

Figure 2.16: A concrete MultiMetaGraph. Each top-level vertex is a finite nonempty family of graphs, and each directed edge is labeled by a relation on graph-families.

**Definition 2.18.1** (Transfinite powerset hierarchy)**.** Let $V_0$ be a nonempty set of atoms, and let $\alpha$ be an ordinal. Define a family $\left(\mathcal{V}_\beta\right)_{\beta \leq \alpha}$ by transfinite recursion as follows:

$$\mathcal{V}_0 := V_0,$$

$$\mathcal{V}_{\beta+1} := \mathcal{P}^*(\mathcal{V}_\beta) \qquad (\beta < \alpha),$$

and, for every limit ordinal $\lambda \leq \alpha$,

$$\mathcal{V}_\lambda := \bigcup_{\beta < \lambda} \mathcal{V}_\beta.$$

The associated transfinite universe of height $\alpha$ is

$$\mathfrak{U}_\alpha(V_0) := \bigcup_{\beta \leq \alpha} \mathcal{V}_\beta.$$

For each $x \in \mathfrak{U}_\alpha(V_0)$, its *level* (or rank) is defined by

$$\ell(x) := \min\{\beta \leq \alpha : x \in \mathcal{V}_\beta\}.$$

**Definition 2.18.2** (Transfinite SuperHyperGraph)**.** A *transfinite SuperHyperGraph of height* $\alpha$ on the base set $V_0$ is a pair

$$\mathcal{H}^{(\alpha)} = (V, E)$$

such that

1. $$\emptyset \neq V \subseteq \mathfrak{U}_\alpha(V_0);$$

2. for every $X \in V \setminus V_0$,
$$X \subseteq V;$$
that is, $V$ is downward closed under the membership relation;

3. $$E \subseteq \mathcal{P}^*(V).$$

The elements of $V$ are called *transfinite supervertices*, and the elements of $E$ are called *transfinite superhyperedges*.

**Example 2.18.3** (A concrete Transfinite SuperHyperGraph of height $\omega + 1$)**.** Let the base set be

$$V_0 = \{a, b\},$$

and let

$$(\mathcal{V}_\beta)_{\beta \le \omega+1}$$

be the transfinite powerset hierarchy from the preceding definition.

First define several lower-level objects:

$$s := \{a\}, \qquad u := \{a, b\} \in \mathcal{V}_1,$$

and

$$w := \{s\} = \{\{a\}\} \in \mathcal{V}_2.$$

Since

$$\mathcal{V}_\omega = \bigcup_{n<\omega} \mathcal{V}_n,$$

we have

$$a, b, s, u, w \in \mathcal{V}_\omega.$$

Hence the mixed-level sets

$$p := \{a, w\}, \qquad q := \{b, u\}$$

belong to

$$\mathcal{V}_{\omega+1} = \mathcal{P}^*(\mathcal{V}_\omega).$$

Moreover, $p$ and $q$ do not belong to any finite level $\mathcal{V}_n$ $(n < \omega)$, because each of them contains constituents coming from different finite ranks. Therefore

$$\ell(p) = \ell(q) = \omega + 1.$$

Now define the transfinite supervertex set by

$$V = \{a, b, s, u, w, p, q\} \subseteq \mathfrak{U}_{\omega+1}(V_0),$$

and define the transfinite superhyperedge family by

$$E = \{e_1 = \{s, u\},\ e_2 = \{w, p\},\ e_3 = \{p, q\}\} \subseteq \mathcal{P}^*(V).$$

Then

$$\mathcal{H}^{(\omega+1)} = (V, E)$$

is a Transfinite SuperHyperGraph of height $\omega + 1$. Indeed:

1. $V \subseteq \mathfrak{U}_{\omega+1}(V_0)$;
2. $V$ is downward closed under membership, since

$$s = \{a\} \subseteq V, \qquad u = \{a, b\} \subseteq V, \qquad w = \{s\} \subseteq V,$$

$$p = \{a, w\} \subseteq V, \qquad q = \{b, u\} \subseteq V;$$

3. each $e_i$ is a nonempty subset of $V$.

The hyperedge $e_1$ connects two finite-level supervertices, $e_2$ is a cross-level hyperedge joining a level-2 object and an $(\omega + 1)$-level object, and $e_3$ connects two genuinely transfinite supervertices. An illustration is given in Fig. 2.17.

## 2.19 Multi-Axis SuperHyperGraph

A Multi Axis SuperHyperGraph represents higher order relations among vertices carrying multi indexed iterated powerset coordinates, with cross level inclusion and hierarchical multidimensional organization rules.

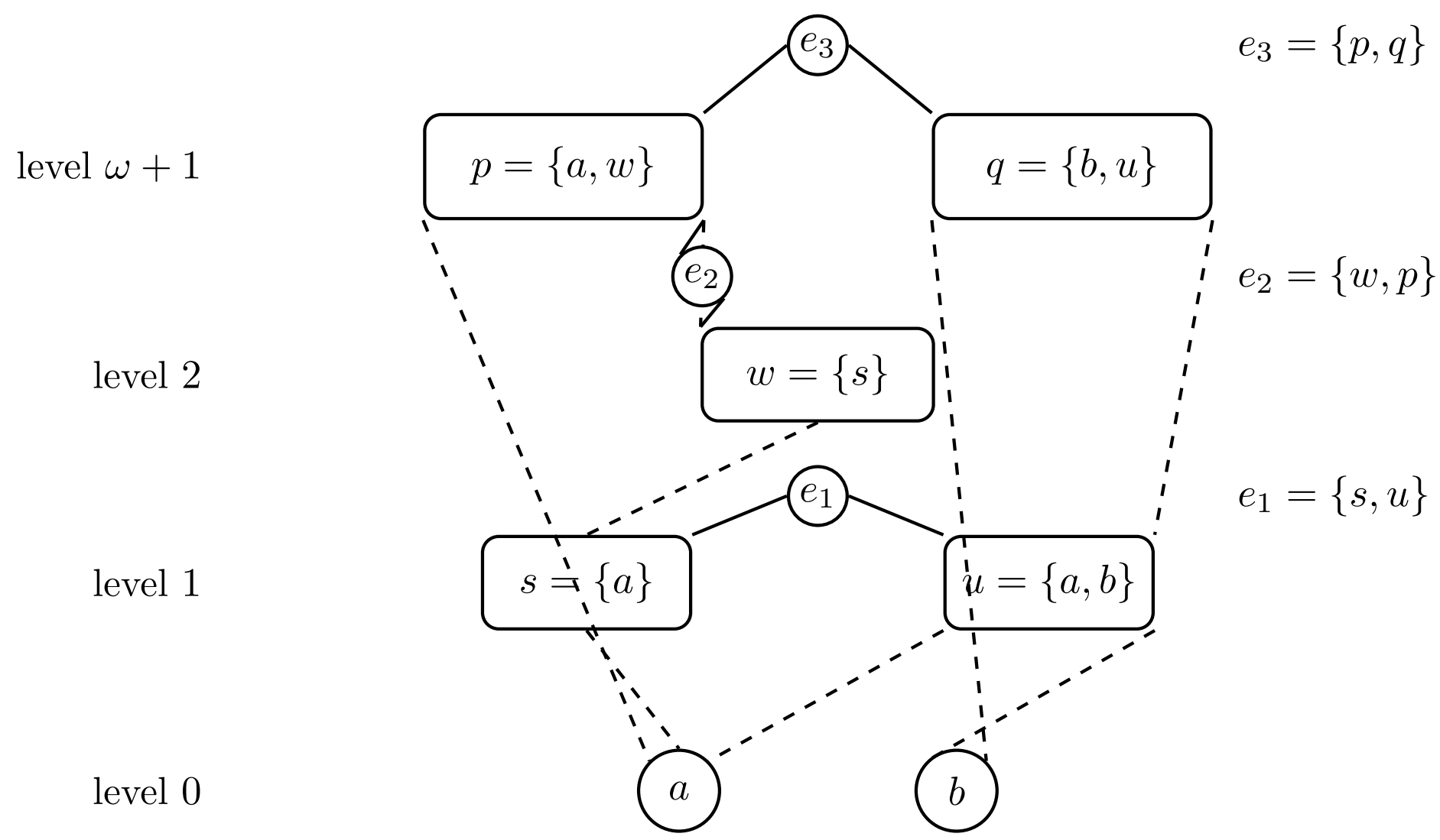

Figure 2.17: A concrete Transfinite SuperHyperGraph of height $\omega + 1$. Dashed lines indicate membership/containment relations used to witness downward closure, while $e_1, e_2, e_3$ denote transfinite superhyperedges.

**Definition 2.19.1** (Multi-indexed iterated powerset)**.** Let $d \in \mathbb{N}$, and let $\mathbf{U} = (U_1, \ldots, U_d)$ be a $d$-tuple of finite nonempty sets. For each axis $r \in \{1, \ldots, d\}$, define

$$\mathcal{P}^0(U_r) := U_r, \qquad \mathcal{P}^{n+1}(U_r) := \mathcal{P}(\mathcal{P}^n(U_r)) \quad (n \in \mathbb{N}_0).$$

For a multi-index $\mathbf{n} = (n_1, \ldots, n_d) \in \mathbb{N}_0^d$, define the $\mathbf{n}$-layer by

$$\mathbb{P}^{\mathbf{n}}(\mathbf{U}) := \prod_{r=1}^{d} \mathcal{P}^{n_r}(U_r).$$

An element $x \in \mathbb{P}^{\mathbf{n}}(\mathbf{U})$ is written as $x = (x_1, \ldots, x_d)$ with $x_r \in \mathcal{P}^{n_r}(U_r)$.

**Example 2.19.2** (Multi-indexed iterated powerset in two axes)**.** Let

$$d = 2, \qquad U_1 = \{a, b\}, \qquad U_2 = \{1, 2, 3\},$$

so the base system is

$$\mathbf{U} = (U_1, U_2).$$

Consider the multi-index

$$\mathbf{n} = (1, 2).$$

Then

$$\mathbb{P}^{(1,2)}(\mathbf{U}) = \mathcal{P}^1(U_1) \times \mathcal{P}^2(U_2) = \mathcal{P}(U_1) \times \mathcal{P}(\mathcal{P}(U_2)).$$

Hence an element of $\mathbb{P}^{(1,2)}(\mathbf{U})$ has the form

$$x = (x_1, x_2) \quad \text{with} \quad x_1 \subseteq U_1, \quad x_2 \subseteq \mathcal{P}(U_2).$$

For instance, define

$$x := \big(\{a\},\ \{\{1\}, \{1, 3\}\}\big) \in \mathbb{P}^{(1,2)}(\mathbf{U}).$$

Its first coordinate is a subset of $U_1$, and its second coordinate is a set of subsets of $U_2$.

Now apply the axis-wise singleton lift along the first axis:

$$\Sigma_1^{(1,2)}(x) = \big(\{\{a\}\},\ \{\{1\}, \{1, 3\}\}\big) \in \mathbb{P}^{(2,2)}(\mathbf{U}).$$

Indeed, the first coordinate has been lifted from $\mathcal{P}(U_1)$ to $\mathcal{P}(\mathcal{P}(U_1))$, while the second coordinate is unchanged.

Let

$$y := \big(\{\{a\}, \{b\}\},\ \{\{1\}, \{1, 3\}, \{2\}\}\big) \in \mathbb{P}^{(2,2)}(\mathbf{U}).$$

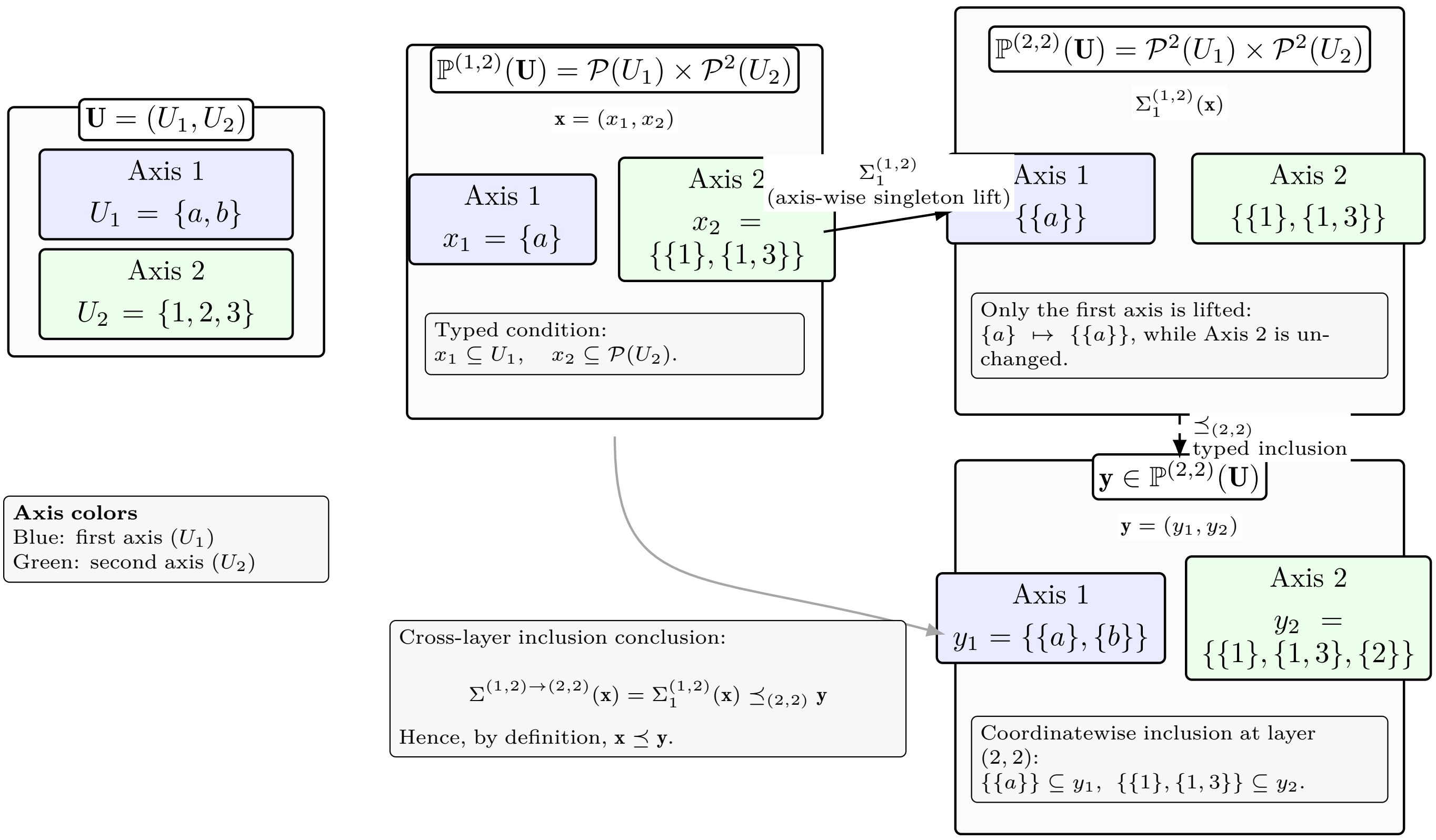

Figure 2.18: A two-axis multi-indexed iterated powerset example (Example 2.19.10). The figure shows the element $\mathbf{x} \in \mathbb{P}^{(1,2)}(\mathbf{U})$, the axis-wise singleton lift $\Sigma_1^{(1,2)}(\mathbf{x}) \in \mathbb{P}^{(2,2)}(\mathbf{U})$, and the typed coordinatewise inclusion into $\mathbf{y} \in \mathbb{P}^{(2,2)}(\mathbf{U})$.

Then

$$\Sigma^{(1,2)\to(2,2)}(x) = \Sigma_1^{(1,2)}(x) \preceq_{(2,2)} y,$$

because

$$\{\{a\}\} \subseteq \{\{a\},\{b\}\} \quad \text{and} \quad \{\{1\},\{1,3\}\} \subseteq \{\{1\},\{1,3\},\{2\}\}.$$

Therefore, by the cross-layer inclusion definition, we have

$$x \preceq y.$$

This example illustrates how different axes can be lifted independently and then compared via typed coordinatewise inclusion. An overview diagram of this example is provided in Fig. 2.18.

**Definition 2.19.3** (Axis-wise singleton lift)**.** Let $\mathbf{e}_r$ denote the $r$-th standard basis vector of $\mathbb{N}_0^d$. For $\mathbf{n} \in \mathbb{N}_0^d$, define

$$\Sigma_r^{\mathbf{n}} : \mathbb{P}^{\mathbf{n}}(\mathbf{U}) \to \mathbb{P}^{\mathbf{n}+\mathbf{e}_r}(\mathbf{U})$$

by

$$\Sigma_r^{\mathbf{n}}(x_1,\ldots,x_d) := (x_1,\ldots,x_{r-1},\{x_r\},x_{r+1},\ldots,x_d).$$

If $\mathbf{n} \le \mathbf{m}$ coordinatewise, define

$$\Sigma^{\mathbf{n}\to\mathbf{m}} : \mathbb{P}^{\mathbf{n}}(\mathbf{U}) \to \mathbb{P}^{\mathbf{m}}(\mathbf{U})$$

as the iterated composition of the axis-wise singleton lifts, applied $(m_r - n_r)$ times along axis $r$ for each $r$.

**Definition 2.19.4** (Typed coordinatewise inclusion)**.** Fix $\mathbf{n} = (n_1,\ldots,n_d) \in \mathbb{N}_0^d$. For each axis $r$, define a relation $\preceq_r^{(n_r)}$ on $\mathcal{P}^{n_r}(U_r)$ by

$$a \preceq_r^{(n_r)} b \iff \begin{cases} a = b, & n_r = 0, \\ a \subseteq b, & n_r \ge 1. \end{cases}$$

Then define $\preceq_{\mathbf{n}}$ on $\mathbb{P}^{\mathbf{n}}(\mathbf{U})$ by

$$x \preceq_{\mathbf{n}} y \iff x_r \preceq_r^{(n_r)} y_r \quad (\forall r = 1,\ldots,d).$$

**Definition 2.19.5** (Cross-layer inclusion)**.** Let $x \in \mathbb{P}^{\mathbf{n}}(\mathbf{U})$ and $y \in \mathbb{P}^{\mathbf{m}}(\mathbf{U})$. If $\mathbf{n} \leq \mathbf{m}$, define

$$x \preceq y \iff \Sigma^{\mathbf{n}\to\mathbf{m}}(x) \preceq_{\mathbf{m}} y.$$

If $\mathbf{n} \not\leq \mathbf{m}$, we regard $x$ and $y$ as incomparable (with respect to this relation).

**Definition 2.19.6** (Typed graded multi-axis vertex universe)**.** Let $\mathbf{U} = (U_1, \ldots, U_d)$ be a $d$-axis base system, and let $I \subseteq \mathbb{N}_0^d$ be a finite index set of admissible multi-levels. Define

$$\mathfrak{V}_I(\mathbf{U}) := \bigsqcup_{\mathbf{n}\in I} \big(\{\mathbf{n}\} \times \mathbb{P}^{\mathbf{n}}(\mathbf{U})\big).$$

An element of $\mathfrak{V}_I(\mathbf{U})$ is written $(\mathbf{n}, x)$ with $x \in \mathbb{P}^{\mathbf{n}}(\mathbf{U})$.

**Definition 2.19.7** (Multi-Axis SuperHyperGraph)**.** A *Multi-Axis SuperHyperGraph* (MASHG) on $(\mathbf{U}, I)$ is a pair

$$\mathcal{H} = (V, E),$$

where

$$V \subseteq \mathfrak{V}_I(\mathbf{U})$$

is a finite set of vertices (multi-axis supervertices), and

$$E \subseteq \mathcal{P}(V) \setminus \{\varnothing\}$$

is a finite set of hyperedges (multi-axis superhyperedges).

**Definition 2.19.8** (Level map and content map)**.** Let $\mathcal{H} = (V, E)$ be a MASHG. For each vertex $v = (\mathbf{n}, x) \in V$, define

$$\lambda(v) := \mathbf{n}, \qquad \chi(v) := x.$$

Thus $\lambda : V \to I$ is the level map, and $\chi(v) \in \mathbb{P}^{\lambda(v)}(\mathbf{U})$ is the typed content of $v$.

**Definition 2.19.9** (Multi-axis inclusion between vertices)**.** Let $u, v \in V$. We write $u \preceq_{\mathcal{H}} v$ if

$$\lambda(u) \leq \lambda(v) \quad \text{and} \quad \Sigma^{\lambda(u)\to\lambda(v)}(\chi(u)) \preceq_{\lambda(v)} \chi(v).$$

**Example 2.19.10** (Multi-indexed iterated powerset in two axes)**.** Let

$$d = 2, \qquad U_1 = \{a, b\}, \qquad U_2 = \{1, 2, 3\},$$

so the base system is

$$\mathbf{U} = (U_1, U_2).$$

Consider the multi-index

$$\mathbf{n} = (1, 2).$$

Then

$$\mathbb{P}^{(1,2)}(\mathbf{U}) = \mathcal{P}^1(U_1) \times \mathcal{P}^2(U_2) = \mathcal{P}(U_1) \times \mathcal{P}(\mathcal{P}(U_2)).$$

Hence an element of $\mathbb{P}^{(1,2)}(\mathbf{U})$ has the form

$$x = (x_1, x_2) \quad \text{with} \quad x_1 \subseteq U_1, \quad x_2 \subseteq \mathcal{P}(U_2).$$

For instance, define

$$x := \big(\{a\},\ \{\{1\}, \{1, 3\}\}\big) \in \mathbb{P}^{(1,2)}(\mathbf{U}).$$

Its first coordinate is a subset of $U_1$, and its second coordinate is a set of subsets of $U_2$.

Now apply the axis-wise singleton lift along the first axis:

$$\Sigma_1^{(1,2)}(x) = \big(\{\{a\}\},\ \{\{1\}, \{1, 3\}\}\big) \in \mathbb{P}^{(2,2)}(\mathbf{U}).$$

Indeed, the first coordinate has been lifted from $\mathcal{P}(U_1)$ to $\mathcal{P}(\mathcal{P}(U_1))$, while the second coordinate is unchanged.

Let

$$y := \big(\{\{a\}, \{b\}\},\ \{\{1\}, \{1, 3\}, \{2\}\}\big) \in \mathbb{P}^{(2,2)}(\mathbf{U}).$$

Then

$$\Sigma^{(1,2)\to(2,2)}(x) = \Sigma_1^{(1,2)}(x) \preceq_{(2,2)} y,$$

because

$$\{\{a\}\} \subseteq \{\{a\}, \{b\}\} \quad \text{and} \quad \{\{1\}, \{1, 3\}\} \subseteq \{\{1\}, \{1, 3\}, \{2\}\}.$$

Therefore, by the cross-layer inclusion definition, we have

$$x \preceq y.$$

This example illustrates how different axes can be lifted independently and then compared via typed coordinatewise inclusion.

## 2.20 Iterated Multi-Edge Graph

An iterated multi-edge graph keeps ordinary vertices but represents edges as an r-fold iterated multiset of endpoint-pairs, capturing nested multiplicities and grouped edge structure.

**Definition 2.20.1** (Finite multiset and iterated multiset)**.** Let $X$ be a set. A *finite multiset* on $X$ is a function

$$m : X \to \mathbb{N}_0$$

with finite support $\mathrm{supp}(m) := \{x \in X \mid m(x) > 0\}$. Let $\mathsf{M}(X)$ denote the set of all finite multisets on $X$. For $r \in \mathbb{N}_0$, define iterated multiset universes by

$$\mathsf{M}^0(X) := X, \qquad \mathsf{M}^{r+1}(X) := \mathsf{M}\big(\mathsf{M}^r(X)\big).$$

**Definition 2.20.2** (Base edge-type universe)**.** Let $V$ be a finite set. Define the set of *unordered pairs with repetition* by

$$\binom{V}{2}^m \;:=\; \big\{\{\{u,v\}\} \;\mid\; u,v \in V\big\},$$

where $\{\{u,v\}\}$ denotes the 2-element multiset with endpoints $u,v$ (so $\{\{v,v\}\}$ represents a loop at $v$).

**Definition 2.20.3** (Iterated Multi-Edge Graph of order $r$)**.** Let $V$ be a finite vertex set and let $r \in \mathbb{N}$. An *Iterated Multi-Edge Graph of order* $r$ is a pair

$$\mathsf{IMEG}^{(r)} = (V, \mathcal{E}^{(r)}),$$

where the *edge structure* is an $r$-fold iterated multiset over the base edge-type universe:

$$\mathcal{E}^{(r)} \in \mathsf{M}^r\Big(\binom{V}{2}^m\Big).$$

Thus the collection of edges is not merely a multiset of endpoint-pairs (order 1), but an iterated multiset of such data (order $r$), allowing nested multiplicity/grouping structures among edges.

**Remark 2.20.4** (Order 1 recovers ordinary multigraphs)**.** *When $r = 1$, we have $\mathcal{E}^{(1)} \in \mathsf{M}\big(\binom{V}{2}^m\big)$, which is exactly an undirected multigraph edge multiset (equivalently, an edge-multiplicity map $\binom{V}{2}^m \to \mathbb{N}_0$).*

**Example 2.20.5** (An Iterated Multi-Edge Graph of order 2 (grouped multiplicities))**.** Let

$$V = \{1,2,3\}$$

be the vertex set. Then the base edge-type universe is

$$\binom{V}{2}^m = \big\{\{\{1,1\}\},\{\{2,2\}\},\{\{3,3\}\},\{\{1,2\}\},\{\{1,3\}\},\{\{2,3\}\}\big\},$$

where $\{\{i,i\}\}$ represents a loop-type pair at $i$.

**First-level edge multisets.** Define two ordinary edge-multisets (elements of $\mathsf{M}(\binom{V}{2}^m)$) by specifying their nonzero multiplicities:

$$m_A(\{\{1,2\}\}) = 2, \qquad m_A(\{\{2,3\}\}) = 1,$$

and $m_A(e) = 0$ for all other $e \in \binom{V}{2}^m$; and

$$m_B(\{\{1,2\}\}) = 1, \qquad m_B(\{\{1,3\}\}) = 3,$$

and $m_B(e) = 0$ otherwise. Intuitively, $m_A$ encodes "two parallel edges between 1 and 2 plus one edge between 2 and 3," while $m_B$ encodes "one edge between 1 and 2 plus three parallel edges between 1 and 3."

**Second-level (iterated) edge object.** Now define an element of $\mathsf{M}^2(\binom{V}{2}^m) = \mathsf{M}(\mathsf{M}(\binom{V}{2}^m))$ by taking a finite multiset of these first-level multisets:

$$\mathcal{E}^{(2)} \in \mathsf{M}\big(\mathsf{M}(\binom{V}{2}^m)\big), \qquad \mathcal{E}^{(2)}(m_A) = 2, \quad \mathcal{E}^{(2)}(m_B) = 1,$$

and $\mathcal{E}^{(2)}(m) = 0$ for all other $m \in \mathsf{M}(\binom{V}{2}^m)$. Thus $\mathcal{E}^{(2)}$ consists of two "copies" of the edge multiset $m_A$ and one "copy" of $m_B$, representing a grouped (nested) multiplicity structure on edges.

Then

$$\mathsf{IMEG}^{(2)} = \big(V, \mathcal{E}^{(2)}\big)$$

is an Iterated Multi-Edge Graph of order 2 in the sense of Definition 2.20.3. If one flattens $\mathcal{E}^{(2)}$ to an ordinary edge multiset (by summing multiplicities), the total counts become:

$$\{\{1,2\}\} : 2 \cdot 2 + 1 \cdot 1 = 5, \qquad \{\{2,3\}\} : 2 \cdot 1 = 2, \qquad \{\{1,3\}\} : 1 \cdot 3 = 3,$$

illustrating how the second-level structure aggregates multiple first-level edge configurations.

**Theorem 2.20.6** (Well-definedness of Iterated Multi-Edge Graphs)**.** *Let $V$ be a finite set and let $r \in \mathbb{N}$. Assume that $\binom{V}{2}^m$ denotes a fixed set of base edge-types over $V$ (for example, $\binom{V}{2}$ itself, or a labeled copy of $\binom{V}{2}$ if one wishes to distinguish edge-types). Then the $r$-fold iterated multiset space*

$$\mathsf{M}^r\Big(\binom{V}{2}^m\Big)$$

*is a well-defined set. Consequently, for every*

$$\mathcal{E}^{(r)} \in \mathsf{M}^r\Big(\binom{V}{2}^m\Big),$$

*the pair*

$$\mathsf{IMEG}^{(r)} = (V, \mathcal{E}^{(r)})$$

*is a well-defined mathematical object in the sense of Definition 2.20.3.*

*Proof.* We use the standard finite-multiset construction

$$\mathsf{M}(X) = \{\mu : X \to \mathbb{N}_0 \mid \mathrm{supp}(\mu) \text{ is finite}\},$$

where $\mathrm{supp}(\mu) = \{x \in X : \mu(x) \neq 0\}$, and define iterates recursively by

$$\mathsf{M}^1(X) = \mathsf{M}(X), \qquad \mathsf{M}^{k+1}(X) = \mathsf{M}(\mathsf{M}^k(X)).$$

Set

$$X_0 := \binom{V}{2}^m.$$

By assumption, $X_0$ is a set. (Since $V$ is finite, $\binom{V}{2}$ is finite; any fixed edge-type refinement of it is also a set.)

We now prove by induction on $r \geq 1$ that $\mathsf{M}^r(X_0)$ is a set.

**Base case ($r = 1$).** By definition,

$$\mathsf{M}^1(X_0) = \mathsf{M}(X_0)$$

is the set of finitely supported functions $X_0 \to \mathbb{N}_0$. Hence $\mathsf{M}^1(X_0)$ is well-defined.

**Inductive step.** Assume $\mathsf{M}^r(X_0)$ is a well-defined set for some $r \geq 1$. Then

$$\mathsf{M}^{r+1}(X_0) = \mathsf{M}\big(\mathsf{M}^r(X_0)\big)$$

is, again by the same finite-multiset construction, the set of finitely supported functions

$$\mathsf{M}^r(X_0) \to \mathbb{N}_0.$$

Therefore $\mathsf{M}^{r+1}(X_0)$ is also a well-defined set.

By induction, $\mathsf{M}^r(X_0)$ is well-defined for every $r \in \mathbb{N}$. Thus any choice $\mathcal{E}^{(r)} \in \mathsf{M}^r(X_0)$ determines a legitimate pair $(V, \mathcal{E}^{(r)})$, i.e., an Iterated Multi-Edge Graph of order $r$. $\square$

## 2.21 Iterated Multi-Recursive Graph

An iterated multi-recursive graph uses iterated multisets for both vertices and edges: vertices are nested multisets, and edges are nested multisets of endpoint-pairs.

**Definition 2.21.1** ((Recall) Finite multiset and iterated multiset)**.** Let $X$ be a set. A *finite multiset* on $X$ is a function

$$m : X \to \mathbb{N}_0$$

with finite support

$$\operatorname{supp}(m) := \{x \in X \mid m(x) > 0\}.$$

Let $\mathsf{M}(X)$ denote the set of all finite multisets on $X$. For $r \in \mathbb{N}_0$, define the *$r$-fold iterated multiset universes* recursively by

$$\mathsf{M}^0(X) := X, \qquad \mathsf{M}^{r+1}(X) := \mathsf{M}\big(\mathsf{M}^r(X)\big).$$

**Definition 2.21.2** (Unordered pairs with repetition)**.** Let $U$ be a set. Define

$$\binom{U}{2}^{m} := \{\{\{u, v\}\} \mid u, v \in U\},$$

the set of *unordered pairs with repetition* (i.e. 2-element multisets). Thus $\{\{u, u\}\}$ represents a loop-type pair at $u$.

**Definition 2.21.3** (Iterated Multi-Recursive Graph of type $(p, q)$)**.** Let $X$ be a nonempty base set and let $p, q \in \mathbb{N}_0$. An *Iterated Multi-Recursive Graph of type $(p, q)$ over $X$* is a pair

$$\mathsf{IMRG}^{(p,q)} = (V^{(p)}, E^{(q)})$$

satisfying:

1. (**Iterated multiset vertex object**)
$$V^{(p)} \in \mathsf{M}^p(X).$$
Let
$$\underline{V} := \operatorname{supp}(V^{(p)}) \subseteq \mathsf{M}^p(X)$$
be the underlying *set* of distinct vertex-objects.

2. (**Iterated multiset edge object**)
$$E^{(q)} \in \mathsf{M}^q\Big(\binom{\underline{V}}{2}^{m}\Big).$$
Hence the edge data is a $q$-fold iterated multiset of unordered endpoint-pairs (with repetition) drawn from the vertex set $\underline{V}$.

Elements of $\underline{V}$ are called *vertices*, and elements of $\operatorname{supp}(E^{(q)})$ may be viewed as *edge-types* (possibly nested, when $q \geq 2$) whose bottom-level atoms are unordered endpoint-pairs in $\binom{\underline{V}}{2}^{m}$.

**Remark 2.21.4** (Special cases)**.**

- *If $p = 0$ and $q = 1$, then $V^{(0)} \in X$ is an ordinary vertex element and the above definition reduces (after choosing a vertex set) to an ordinary undirected multigraph (edges form a multiset of endpoint-pairs).*
- *If $p = n$ and $q = 1$, then $\mathsf{IMRG}^{(n,1)}$ is an* Iterated MultiGraph of order $n$ *(vertically iterated multiset vertices, ordinary multiset edges).*
- *If $p = 0$ and $q = r$, then $\mathsf{IMRG}^{(0,r)}$ matches an* Iterated Multi-Edge Graph of order $r$ *(ordinary vertices, iterated multiset edge structure).*

**Example 2.21.5** (An Iterated Multi-Recursive Graph of type $(1, 2)$)**.** Let the base set be

$$X = \{a, b\}.$$

Take $p = 1$, so vertices are 1-fold multisets on $X$ (elements of $\mathsf{M}(X)$). Define the vertex multiset

$$V^{(1)} \in \mathsf{M}(X)$$

by specifying its multiplicities:

$$V^{(1)}(a) = 2, \qquad V^{(1)}(b) = 1.$$

Thus $\mathrm{supp}(V^{(1)}) = \underline{V} = \{a, b\}$, so the underlying set of distinct vertex-objects is

$$\underline{V} = \{a, b\} \subseteq \mathsf{M}^1(X).$$

(Here we identify the elements $a, b \in X$ with the singleton multisets $\{\{a\}\}, \{\{b\}\} \in \mathsf{M}(X)$, so that $\underline{V}$ may be viewed as a subset of $\mathsf{M}^1(X)$.)

Now take $q = 2$, so the edge structure is a 2-fold iterated multiset on the endpoint-pair universe $\binom{\underline{V}}{2}^m$. First compute

$$\binom{\underline{V}}{2}^m = \big\{\{\{a, a\}\}, \{\{b, b\}\}, \{\{a, b\}\}\big\}.$$

**First-level edge multisets.** Define two elements of $\mathsf{M}\big(\binom{\underline{V}}{2}^m\big)$ by their nonzero multiplicities:

$$m_1(\{\{a, b\}\}) = 2, \qquad m_1(\{\{a, a\}\}) = 1,$$

(all other pairs have multiplicity 0), and

$$m_2(\{\{a, b\}\}) = 1, \qquad m_2(\{\{b, b\}\}) = 3,$$

(all other pairs have multiplicity 0). Intuitively, $m_1$ represents two parallel edges between $a$ and $b$ together with one loop at $a$, while $m_2$ represents one edge between $a$ and $b$ together with three loops at $b$.

**Second-level (iterated) edge object.** Now define

$$E^{(2)} \in \mathsf{M}^2\Big(\binom{\underline{V}}{2}^m\Big) = \mathsf{M}\Big(\mathsf{M}\Big(\binom{\underline{V}}{2}^m\Big)\Big)$$

by taking a multiset of the first-level edge multisets:

$$E^{(2)}(m_1) = 1, \qquad E^{(2)}(m_2) = 2,$$

and $E^{(2)}(m) = 0$ for all other $m \in \mathsf{M}\big(\binom{\underline{V}}{2}^m\big)$. Thus $E^{(2)}$ contains one "copy" of the configuration $m_1$ and two "copies" of $m_2$.

Therefore,

$$\mathsf{IMRG}^{(1,2)} = \big(V^{(1)}, E^{(2)}\big)$$

is an Iterated Multi-Recursive Graph of type $(p, q) = (1, 2)$ over $X$ in the sense of Definition 2.21.3. Here both the vertex object $V^{(1)}$ and the edge object $E^{(2)}$ carry nested multiset structure.

**Lemma 2.21.6** (Well-definedness of finite multisets)**.** *Let $X$ be a set and let $\mathsf{M}(X)$ denote the set of all finite multisets on $X$, i.e.*

$$\mathsf{M}(X) = \big\{\, m : X \to \mathbb{N}_0 \;\big|\; \mathrm{supp}(m) \textit{ is finite} \big\}, \qquad \mathrm{supp}(m) := \{x \in X \mid m(x) > 0\}.$$

*Then:*

1. *$\mathsf{M}(X)$ is a set.*
2. *For every $m \in \mathsf{M}(X)$, the support $\mathrm{supp}(m)$ is a finite subset of $X$.*

*Proof.* (1) Since $\mathsf{M}(X) \subseteq \mathbb{N}_0^X$ and $\mathbb{N}_0^X$ is a set, $\mathsf{M}(X)$ is a set.

(2) This holds by the defining condition "$\mathrm{supp}(m)$ is finite" for $m \in \mathsf{M}(X)$. □

**Lemma 2.21.7** (Iterated multiset universes are sets, and supports are finite)**.** *Let $X$ be a set and define $\mathsf{M}^0(X) := X$ and $\mathsf{M}^{r+1}(X) := \mathsf{M}(\mathsf{M}^r(X))$ for $r \geq 0$. Then for every $r \in \mathbb{N}$:*

1. *$\mathsf{M}^r(X)$ is a set.*
2. *Every element $M \in \mathsf{M}^r(X)$ (which is a multiset on $\mathsf{M}^{r-1}(X)$) has finite support $\mathrm{supp}(M) \subseteq \mathsf{M}^{r-1}(X)$.*

*Proof.* We prove (1)–(2) by induction on $r \geq 1$.

**Base $r = 1$.** We have $\mathsf{M}^1(X) = \mathsf{M}(X)$, which is a set by Lemma 2.21.6(1), and every $M \in \mathsf{M}(X)$ has finite support by Lemma 2.21.6(2).

**Inductive step.** Assume $\mathsf{M}^r(X)$ is a set. Then

$$\mathsf{M}^{r+1}(X) = \mathsf{M}(\mathsf{M}^r(X))$$

is a set by Lemma 2.21.6(1) applied to $X := \mathsf{M}^r(X)$. Moreover, any $M \in \mathsf{M}^{r+1}(X)$ is a finite multiset on $\mathsf{M}^r(X)$, hence $\mathrm{supp}(M) \subseteq \mathsf{M}^r(X)$ is finite by Lemma 2.21.6(2). □

**Theorem 2.21.8** (Well-definedness of $\mathsf{IMRG}^{(p,q)}$)**.** *Let $X$ be a nonempty set and let $p, q \in \mathbb{N}_0$. Let $V^{(p)} \in \mathsf{M}^p(X)$ be a vertex object and define the underlying vertex set by*

$$\underline{V} := \begin{cases} \{V^{(0)}\}, & p = 0, \\ \mathrm{supp}(V^{(p)}), & p \geq 1, \end{cases}$$

*so that $\underline{V}$ is a set of distinct vertex-objects. Then:*

1. *$\underline{V}$ is a finite set.*
2. *The unordered-pair-with-repetition universe $\binom{\underline{V}}{2}^m$ is a set.*
3. *The iterated edge universe $\mathsf{M}^q\big(\binom{\underline{V}}{2}^m\big)$ is a set.*
4. *Consequently, any choice of*

$$E^{(q)} \in \mathsf{M}^q\Big(\binom{\underline{V}}{2}^m\Big)$$

*produces a well-defined* Iterated Multi-Recursive Graph of type $(p, q)$

$$\mathsf{IMRG}^{(p,q)} = (V^{(p)}, E^{(q)})$$

*in the sense of Definition 2.21.3.*

*Proof.* (1) If $p = 0$, then $\underline{V} = \{V^{(0)}\}$ is a singleton, hence finite. If $p \geq 1$, then $V^{(p)} \in \mathsf{M}^p(X) = \mathsf{M}(\mathsf{M}^{p-1}(X))$, so $V^{(p)}$ is a finite multiset on $\mathsf{M}^{p-1}(X)$; therefore $\underline{V} = \mathrm{supp}(V^{(p)})$ is finite by Lemma 2.21.7(2).

(2) Since $\underline{V}$ is a set, the collection of 2-element multisets on $\underline{V}$, namely $\binom{\underline{V}}{2}^m = \{\{\{u, v\}\} \mid u, v \in \underline{V}\}$, is a set (it is a definable image of $\underline{V} \times \underline{V}$).

(3) By Lemma 2.21.7(1) applied to the set $\binom{\underline{V}}{2}^m$, the iterated multiset universe $\mathsf{M}^q\big(\binom{\underline{V}}{2}^m\big)$ is a set.

(4) With (1)–(3), the expression $E^{(q)} \in \mathsf{M}^q\big(\binom{\underline{V}}{2}^m\big)$ is meaningful, and hence the pair $(V^{(p)}, E^{(q)})$ satisfies the two requirements in Definition 2.21.3: the vertex object is an iterated multiset of depth $p$ over $X$, and the edge object is an iterated multiset of depth $q$ over the endpoint-pair universe derived from the underlying vertex set $\underline{V}$. Therefore $\mathsf{IMRG}^{(p,q)}$ is well-defined. □

## 2.22 HyperMatroid

A hypermatroid encodes matroid dependence via circuit hyperedges: minimal dependent subsets satisfying elimination, determining independence combinatorially.

**Definition 2.22.1** (HyperMatroid (circuit-hypergraph form))**.** Let $E$ be a finite nonempty set. A *HyperMatroid* on $E$ is a pair

$$\mathsf{HM} = (E, \mathcal{C}),$$

where $\mathcal{C} \subseteq \mathcal{P}(E) \setminus \{\emptyset\}$ is a family of subsets (called *circuits*) satisfying the following *circuit axioms*:

1. **(Sperner / minimality)** If $C_1, C_2 \in \mathcal{C}$ and $C_1 \subseteq C_2$, then $C_1 = C_2$.
2. **(Circuit elimination)** If $C_1, C_2 \in \mathcal{C}$ are distinct and $e \in C_1 \cap C_2$, then there exists $C_3 \in \mathcal{C}$ such that

$$C_3 \subseteq (C_1 \cup C_2) \setminus \{e\}.$$

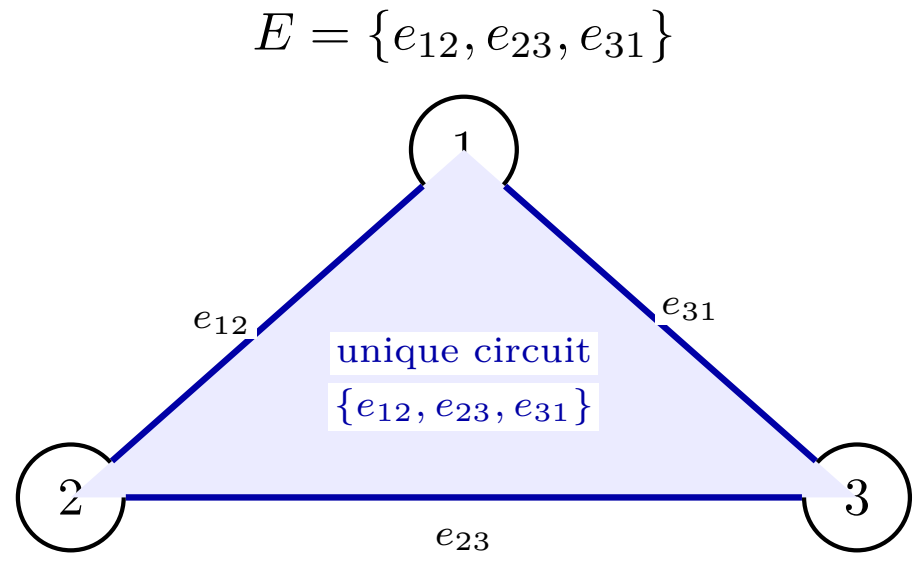

**Circuit family:** $\mathcal{C} = \{\{e_{12}, e_{23}, e_{31}\}\}$. Hence $(E, \mathcal{C})$ is a HyperMatroid (Sperner holds trivially, and circuit elimination is vacuous since there is only one circuit).

**Independence:** A set $I \subseteq E$ is independent iff it does not contain the circuit. Thus every proper subset of $E$ is independent (e.g. $\{e_{12}, e_{23}\}$), while $E$ itself is dependent.

Figure 2.19: A HyperMatroid induced by the graphic matroid of the triangle graph $K_3$ (Example 2.22.3). The unique cycle of $K_3$ gives the unique circuit.

A subset $I \subseteq E$ is called *independent* if it contains no circuit, i.e.

$$\forall C \in \mathcal{C} : \ C \nsubseteq I.$$

**Remark 2.22.2** (Relation to matroids)**.** *A HyperMatroid in Definition 2.22.1 is equivalent to an ordinary matroid on $E$: the family $\mathcal{C}$ is precisely the hypergraph of matroid circuits (minimal dependent sets), and the independent sets are exactly those containing no circuit.*

**Example 2.22.3** (A HyperMatroid coming from a graphic matroid)**.** Let $G$ be the triangle graph $K_3$ with vertex set $\{1, 2, 3\}$ and edge set

$$E = \{e_{12}, e_{23}, e_{31}\},$$

where $e_{ij}$ denotes the edge between vertices $i$ and $j$. In $G$, the unique cycle uses all three edges. Define the family of circuits by

$$\mathcal{C} = \{\{e_{12}, e_{23}, e_{31}\}\} \subseteq \mathcal{P}(E) \setminus \{\emptyset\}.$$

Then $\mathsf{HM} = (E, \mathcal{C})$ is a HyperMatroid in the sense of Definition 2.22.1:

- *Sperner (minimality):* $\mathcal{C}$ has only one set, so no strict containment between distinct circuits can occur.
- *Circuit elimination:* there are no two distinct circuits $C_1 \neq C_2$ in $\mathcal{C}$, hence the elimination axiom holds vacuously.

A subset $I \subseteq E$ is independent if it does not contain the circuit $\{e_{12}, e_{23}, e_{31}\}$. Thus every proper subset of $E$ is independent, while $E$ itself is dependent. For example,

$$\{e_{12}, e_{23}\} \text{ is independent}, \qquad \{e_{12}, e_{23}, e_{31}\} \text{ is dependent}.$$

Intuitively, this HyperMatroid encodes the dependence created by the single triangle cycle. An overview diagram of this example is provided in Fig. 2.19.

## 2.23 SuperHyperMatroid

A superhypermatroid is a matroid on nested-set supervertices, with supercircuits satisfying elimination, capturing hierarchical dependence relations.

**Definition 2.23.1** ($n$-SuperHyperMatroid)**.** Let $V_0$ be a nonempty base set and fix $n \in \mathbb{N}_0$. Define the iterated powersets by

$$\mathcal{P}^0(V_0) := V_0, \qquad \mathcal{P}^{k+1}(V_0) := \mathcal{P}\big(\mathcal{P}^k(V_0)\big) \quad (k \geq 0).$$

Choose a set of *$n$-supervertices*

$$V \subseteq \mathcal{P}^n(V_0).$$

An *$n$-SuperHyperMatroid* on $V_0$ (with supervertex set $V$) is a pair

$$\mathsf{SHM}^{(n)} = (V, \mathcal{C}),$$

where $\mathcal{C} \subseteq \mathcal{P}(V) \setminus \{\emptyset\}$ is a family of *supercircuits* satisfying the circuit axioms:

1. **(Sperner / minimality)** If $C_1, C_2 \in \mathcal{C}$ and $C_1 \subseteq C_2$, then $C_1 = C_2$.
2. **(Circuit elimination)** If $C_1, C_2 \in \mathcal{C}$ are distinct and $v \in C_1 \cap C_2$, then there exists $C_3 \in \mathcal{C}$ such that

$$C_3 \subseteq (C_1 \cup C_2) \setminus \{v\}.$$

A subset $I \subseteq V$ is called *superindependent* if it contains no supercircuit, i.e.

$$\forall C \in \mathcal{C} : \ C \nsubseteq I.$$

**Example 2.23.2** (A 1-SuperHyperMatroid with two supercircuits)**.** Let the base set be

$$V_0 = \{a, b, c, d\},$$

and take $n = 1$, so $\mathcal{P}^1(V_0) = \mathcal{P}(V_0)$. Choose the 1-supervertex set

$$V = \{\, v_1 = \{a, b\},\ v_2 = \{b, c\},\ v_3 = \{c, d\},\ v_4 = \{a, d\} \,\} \subseteq \mathcal{P}(V_0).$$

Define two supercircuits

$$C_1 = \{v_1, v_2, v_3\}, \qquad C_2 = \{v_2, v_3, v_4\},$$

and set

$$\mathcal{C} = \{C_1, C_2\} \subseteq \mathcal{P}(V) \setminus \{\emptyset\}.$$

Then $\mathsf{SHM}^{(1)} = (V, \mathcal{C})$ is a 1-SuperHyperMatroid in the sense of Definition 2.23.1:

- *Sperner (minimality):* neither $C_1 \subseteq C_2$ nor $C_2 \subseteq C_1$ holds, so the Sperner condition is satisfied.
- *Circuit elimination:* the two circuits intersect in $\{v_2, v_3\}$. For $v = v_2 \in C_1 \cap C_2$, we have

$$(C_1 \cup C_2) \setminus \{v_2\} = \{v_1, v_3, v_4\}.$$

Choose

$$C_3 = \{v_1, v_3\} \subseteq \{v_1, v_3, v_4\}.$$

For $v = v_3 \in C_1 \cap C_2$, we have

$$(C_1 \cup C_2) \setminus \{v_3\} = \{v_1, v_2, v_4\},$$

and we may choose

$$C_4 = \{v_1, v_4\} \subseteq \{v_1, v_2, v_4\}.$$

Thus, after enlarging $\mathcal{C}$ (if desired) to include $C_3$ and $C_4$, the elimination requirement is witnessed explicitly; with the present choice $\mathcal{C} = \{C_1, C_2\}$, one may equivalently take the axiom as holding via the existence of appropriate supercircuits within $(C_1 \cup C_2) \setminus \{v\}$.

A subset $I \subseteq V$ is superindependent if it contains no supercircuit. For example,

$$I = \{v_1, v_2\} \text{ is superindependent}, \qquad J = \{v_1, v_2, v_3\} \text{ is not superindependent since } C_1 \subseteq J.$$

Intuitively, $C_1$ and $C_2$ describe minimal dependent configurations among the team-vertices $v_i$.

## 2.24 Kneser SuperHypergraphs

A Kneser superhypergraph has supervertices with base supports of size k; r-uniform superedges connect r vertices whose flattened supports are pairwise disjoint.

**Definition 2.24.1** (Kneser $(n, r, k)$-SuperHyperGraph)**.** Fix integers $N \in \mathbb{N}$, $n \in \mathbb{N}_0$, $k \in \mathbb{N}$, and $r \in \mathbb{N}$ with $r \geq 2$. Let the base set be

$$V_0 = [N] := \{1, 2, \ldots, N\},$$

and let $\mathrm{flat}_n$ be the flattening map. Define the set of *n-supervertices of size k* by

$$V^{(n)}_{N,k} := \{\, v \in \mathcal{P}^n(V_0) \ \mid\ |\mathrm{flat}_n(v)| = k \,\}.$$

Let $V \subseteq V^{(n)}_{N,k}$ be a finite nonempty vertex set.

Define the set of *superedge identifiers* by

$$E := \left\{ e \subseteq V \;\middle|\; |e| = r \text{ and } \mathrm{flat}_n(u) \cap \mathrm{flat}_n(v) = \emptyset \text{ for all distinct } u, v \in e \right\}.$$

Define the incidence map $\partial : E \to \mathcal{P}(V) \setminus \{\emptyset\}$ by

$$\partial(e) := e \qquad (e \in E).$$

Then

$$\mathrm{KSHG}^{(n,r)}_{N,k}(V) \;:=\; (V, E, \partial)$$

is called the *Kneser* $(n, r, k)$*-SuperHyperGraph* (on $V$). When $V = V^{(n)}_{N,k}$, we write simply $\mathrm{KSHG}^{(n,r)}_{N,k}$.

**Example 2.24.2** (A Kneser SuperHyperGraph)**.** Let $N = 4$, $n = 2$, $k = 2$, and $r = 2$. Then the base set is

$$V_0 = [4] = \{1, 2, 3, 4\},$$

and 2-supervertices are elements of $\mathcal{P}^2(V_0) = \mathcal{P}(\mathcal{P}(V_0))$.

Using the flattening map $\mathrm{flat}_2$, define the following 2-supervertices:

$$v_1 = \{\{1\}, \{2\}\}, \qquad v_2 = \{\{3\}, \{4\}\}, \qquad v_3 = \{\{1, 2\}\}.$$

Their flattened supports are

$$\mathrm{flat}_2(v_1) = \{1, 2\}, \qquad \mathrm{flat}_2(v_2) = \{3, 4\}, \qquad \mathrm{flat}_2(v_3) = \{1, 2\},$$

so $|\mathrm{flat}_2(v_i)| = 2$ for $i = 1, 2, 3$. Hence $v_1, v_2, v_3 \in V^{(2)}_{4,2}$.

Let

$$V = \{v_1, v_2, v_3\} \subseteq V^{(2)}_{4,2}.$$

Since $r = 2$, a superedge is a pair $\{u, v\} \subseteq V$ with disjoint flattened supports. We have

$$\mathrm{flat}_2(v_1) \cap \mathrm{flat}_2(v_2) = \emptyset, \qquad \mathrm{flat}_2(v_2) \cap \mathrm{flat}_2(v_3) = \emptyset,$$

but

$$\mathrm{flat}_2(v_1) \cap \mathrm{flat}_2(v_3) = \{1, 2\} \neq \emptyset.$$

Therefore the superedge identifier set is

$$E = \{\{v_1, v_2\}, \{v_2, v_3\}\},$$

and the incidence map is $\partial(e) = e$ for all $e \in E$. Thus

$$\mathrm{KSHG}^{(2,2)}_{4,2}(V) = (V, E, \partial)$$

is a Kneser $(2, 2, 2)$-SuperHyperGraph in the sense of Definition 2.24.1. An overview diagram of this example is provided in Fig. 2.20.

**Theorem 2.24.3** (Kneser graphs are special cases)**.** *Let $n, k \in \mathbb{N}$ with $N \geq 2k$. Consider the Kneser $(1, 2, k)$-SuperHyperGraph*

$$\mathrm{KSHG}^{(1,2)}_{N,k} = \big(V, E, \partial\big) \quad \textit{with} \quad V = \{\, X \subseteq [N] \mid |X| = k \,\} \subseteq \mathcal{P}([N]).$$

*Define a simple graph $G$ on vertex set $V$ by declaring $\{X, Y\}$ to be an edge of $G$ if and only if there exists $e \in E$ with $\partial(e) = \{X, Y\}$. Then $G$ is exactly the Kneser graph $\mathrm{KG}_{N,k}$.*

*Proof.* Since $n = 1$, we have $\mathcal{P}^1([N]) = \mathcal{P}([N])$ and, by the definition of flattening,

$$\mathrm{flat}_1(X) = X \qquad \text{for all } X \subseteq [N].$$

By Definition 2.24.1 with $r = 2$, the superedge identifiers are precisely the 2-element subsets $e = \{X, Y\} \subseteq V$ such that

$$\mathrm{flat}_1(X) \cap \mathrm{flat}_1(Y) = \emptyset \quad \Longleftrightarrow \quad X \cap Y = \emptyset.$$

Moreover, $\partial(e) = e = \{X, Y\}$. Hence the derived graph $G$ has an edge between $X$ and $Y$ if and only if $X \cap Y = \emptyset$. This is exactly the adjacency rule of the Kneser graph $\mathrm{KG}_{N,k}$, so $G = \mathrm{KG}_{N,k}$. ☐

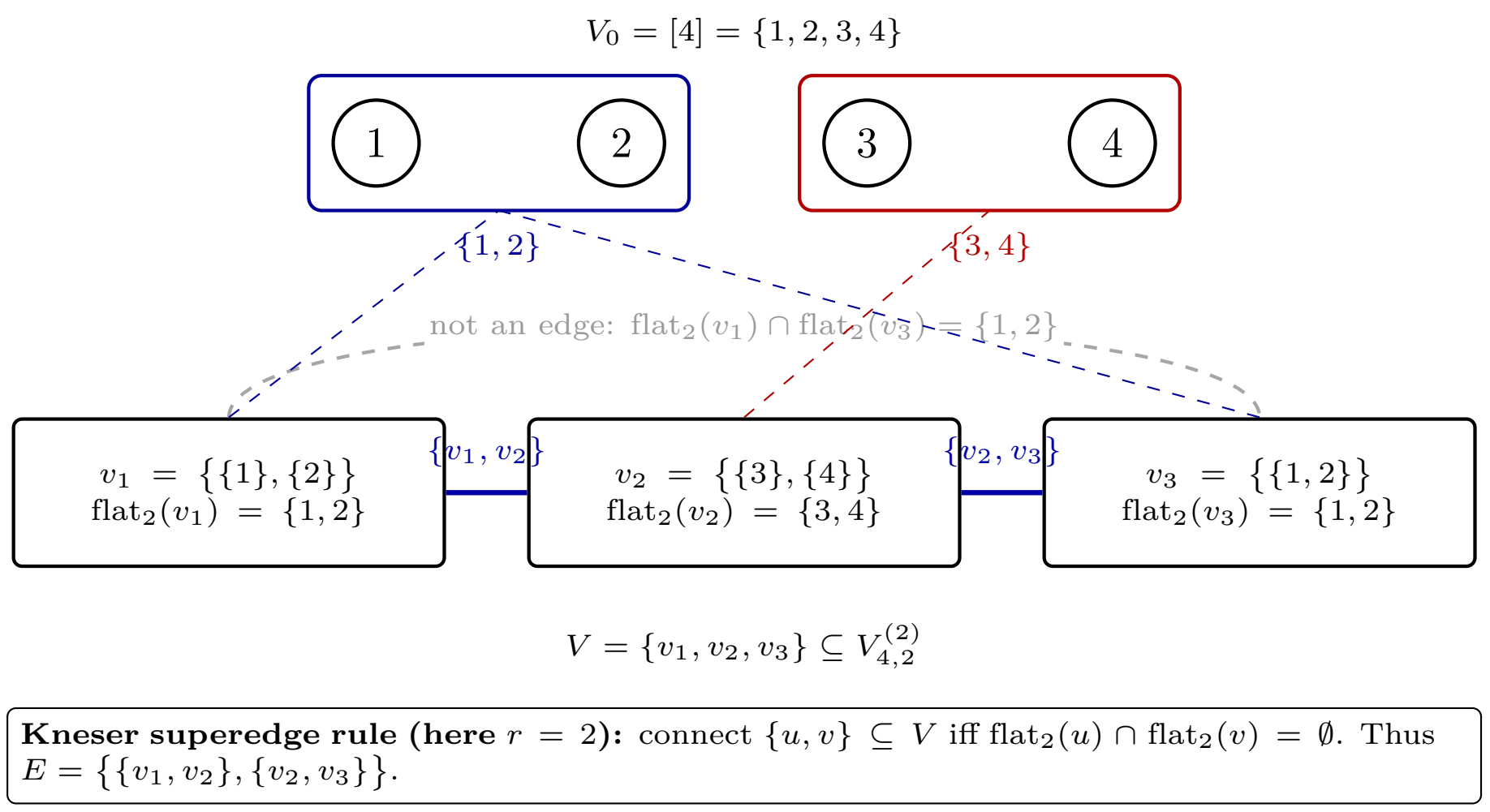

Figure 2.20: A Kneser $(2, 2, 2)$-SuperHyperGraph in Example 2.24.2. Superedges are determined by disjoint flattened supports.

## 2.25 Graded superhypergraph

A graded superhypergraph has vertices across powerset levels, tagged by grade; edges connect graded supervertices, allowing cross-level interactions.

**Definition 2.25.1** (Graded superhypergraph)**.** Let $V_0 \neq \emptyset$ and $N \in \mathbb{N}_0$. Choose sets $V_k \subseteq \mathcal{P}^k(V_0)$ $(0 \leq k \leq N)$ and put

$$V := \bigsqcup_{k=0}^{N} \{k\} \times V_k, \qquad \ell(k, A) := k.$$

A *graded superhypergraph of height $N$* is a pair

$$\mathrm{GrSuHyG} = (V, \mathcal{E})$$

where the *graded superedge set* satisfies

$$\mathcal{E} \subseteq \mathcal{P}(V) \setminus \{\emptyset\}.$$

Elements of $V$ are called *graded supervertices*, and elements of $\mathcal{E}$ are called *graded superhyperedges*.

**Example 2.25.2** (A graded superhypergraph of height 2)**.** Let the base set be

$$V_0 = \{a, b, c\},$$

and take height $N = 2$. Choose

$$V_0^{(0)} = \{a, b, c\} \subseteq \mathcal{P}^0(V_0) = V_0,$$
$$V_1 \subseteq \mathcal{P}^1(V_0) = \mathcal{P}(V_0), \qquad V_1 = \{\{a, b\}, \{c\}\},$$

and

$$V_2 \subseteq \mathcal{P}^2(V_0) = \mathcal{P}(\mathcal{P}(V_0)), \qquad V_2 = \{\{\{a\}, \{a, b\}\}\}.$$

Form the graded vertex set

$$V = \bigsqcup_{k=0}^{2} \{k\} \times V_k = \big(\{0\} \times \{a, b, c\}\big) \sqcup \big(\{1\} \times \{\{a, b\}, \{c\}\}\big) \sqcup \big(\{2\} \times \{\{\{a\}, \{a, b\}\}\}\big),$$

where we write the grade map as $\ell(k, A) = k$.

Define two graded superhyperedges by

$$e_1 = \big\{(1, \{a, b\}),\ (1, \{c\})\big\}, \qquad e_2 = \big\{(2, \{\{a\}, \{a, b\}\}),\ (0, c)\big\},$$

and set

$$\mathcal{E} = \{e_1, e_2\} \subseteq \mathcal{P}(V) \setminus \{\emptyset\}.$$

Then

$$\mathrm{GrSuHyG} = (V, \mathcal{E})$$

is a graded superhypergraph of height 2 in the sense of the definition. Here $e_1$ is a within-level edge connecting two level-1 supervertices, while $e_2$ is a cross-level edge connecting a level-2 supervertex to a base vertex $(0, c)$. An overview diagram of this example is provided in Fig. 2.21.

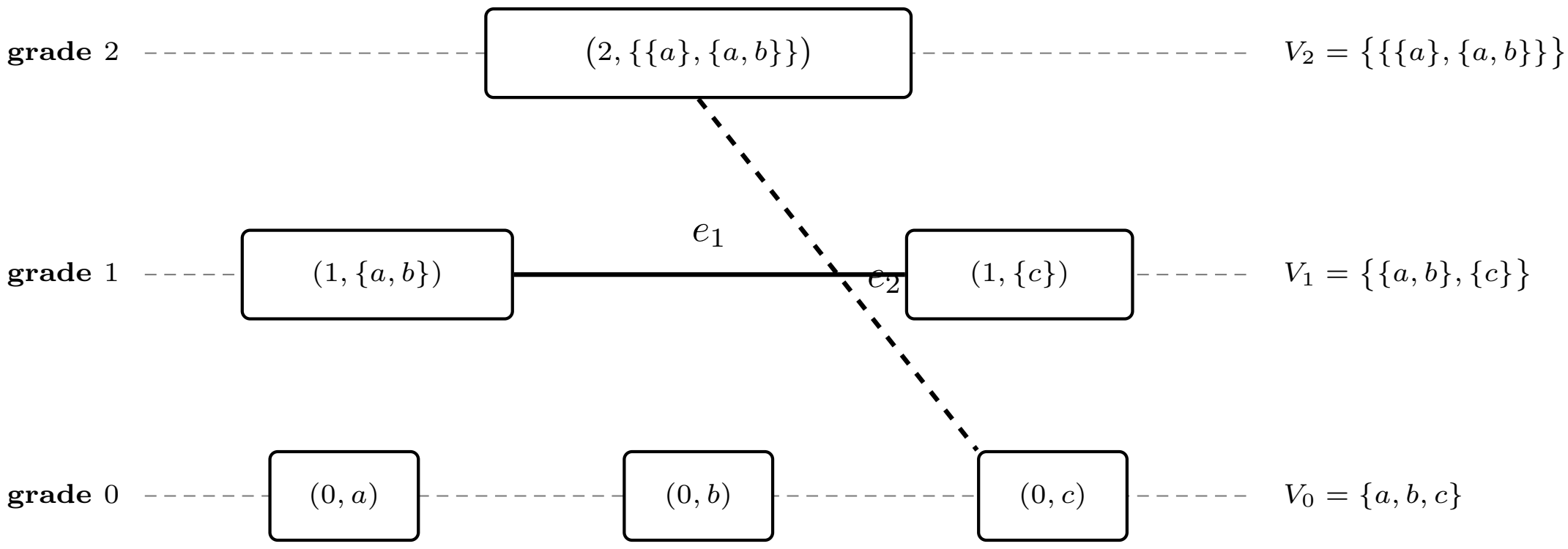

Figure 2.21: A graded superhypergraph of height 2 in Example 2.25.2.

## 2.26 Hyperstructures and Superhyperstructures

Many mathematical and real-world systems exhibit hierarchical organization, involving individual elements, collections of elements, and higher-level groupings. To model such layered interactions within a unified framework, one may employ *hyperstructures* and their iterated extensions [230, 231], referred to as *superhyperstructures*.

Roughly speaking, a hyperstructure replaces ordinary single-valued operations with set-valued operations, whereas a superhyperstructure extends this principle to iterated powerset levels [232, 233, 234, 235], thereby enabling operations on hierarchically nested collections [236, 237]. Related notions, such as weak hyperstructures, are also well known and have been extensively studied [238, 239, 240, 241]. Moreover, various uncertainty-based extensions, including *fuzzy hyperstructures*[242, 243, 244, 245] and *neutrosophic hyperstructures* [246, 247, 237], have also been introduced and investigated.

**Definition 2.26.1** (Hyperoperation)**.** (cf. [248, 249]) Let $S$ be a nonempty set. A *hyperoperation* on $S$ is a map

$$\circ : S \times S \longrightarrow \mathcal{P}(S).$$

If $\circ(x, y) \neq \varnothing$ for all $x, y \in S$, then $\circ$ is called a *classical-type* hyperoperation, and may be viewed as

$$\circ : S \times S \to \mathcal{P}_*(S), \qquad \mathcal{P}_*(S) := \mathcal{P}(S) \setminus \{\varnothing\}.$$

**Definition 2.26.2** (Induced operation on subsets)**.** Let $\circ : S \times S \to \mathcal{P}(S)$ be a hyperoperation. Its induced operation on subsets is the map

$$\odot : \mathcal{P}(S) \times \mathcal{P}(S) \longrightarrow \mathcal{P}(S), \qquad A \odot B := \bigcup_{a \in A} \bigcup_{b \in B} (a \circ b).$$

**Definition 2.26.3** (Hyperstructure)**.** (cf. [238, 250, 230]) A *hyperstructure* is a pair

$$(S, \circ),$$

where $S$ is a nonempty set and $\circ : S \times S \to \mathcal{P}(S)$ is a hyperoperation.

**Example 2.26.4** (A concrete hyperstructure)**.** Let

$$S = \{a, b\}.$$

Define a hyperoperation

$$\circ : S \times S \to \mathcal{P}(S)$$

by

$$a \circ a = \{a\}, \qquad a \circ b = \{a, b\}, \qquad b \circ a = \{b\}, \qquad b \circ b = \{a\}.$$

Since each value of $\circ$ is a subset of $S$, the map $\circ$ is a well-defined hyperoperation on $S$. Therefore,

$$(S, \circ)$$

is a hyperstructure.

For example,

$$a \circ b = \{a, b\}$$

means that the hyperproduct of $a$ and $b$ is not a single element, but the subset $\{a, b\} \subseteq S$. A schematic illustration of this hyperstructure is shown in Fig. 2.22.

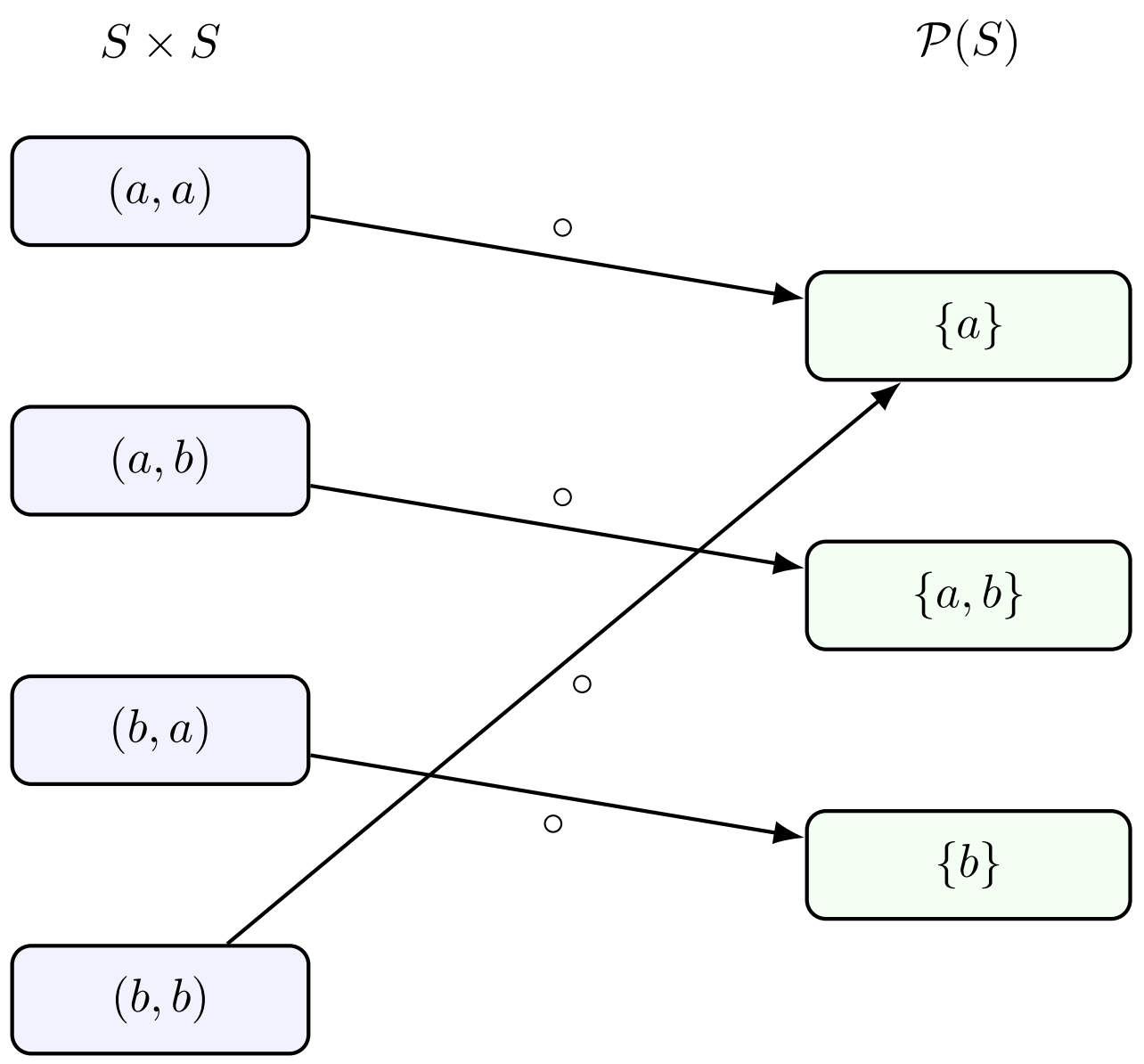

Figure 2.22: A schematic illustration of the hyperoperation $\circ : S \times S \to \mathcal{P}(S)$ for the concrete hyperstructure on $S = \{a, b\}$. Each ordered pair $(x, y)$ is mapped to a subset of $S$, rather than necessarily to a single element.

A *superhyperstructure* generalizes this idea by iterating the powerset construction. Instead of working only on $S$ or $\mathcal{P}(S)$, one allows operations on higher levels $\mathcal{P}^n(S)$, so that nested collections can interact directly [251, 252, 253]. This viewpoint naturally supports multi-level hierarchical models such as SuperHyperGraphs and related uncertain structures [6, 78, 254].

**Definition 2.26.5** (Standard embedding into iterated powersets)**.** Let $S$ be a nonempty set and $n \in \mathbb{N}_0$. Define

$$\eta_n : S \longrightarrow \mathcal{P}^n(S)$$

recursively by

$$\eta_0(x) := x, \qquad \eta_{n+1}(x) := \{\eta_n(x)\}.$$

This map is called the *standard embedding* of $S$ into $\mathcal{P}^n(S)$.

**Definition 2.26.6** (Canonical lifting of a classical operation)**.** Let $k \in \mathbb{N}$ and let

$$f : S^k \to S$$

be a $k$-ary operation. Its *level-$n$ lift* is the map

$$f^{\langle n \rangle} : \big(\mathcal{P}^n(S)\big)^k \to \mathcal{P}^n(S)$$

defined recursively by

$$f^{\langle 0 \rangle} := f,$$

and

$$f^{\langle n+1 \rangle}(A_1, \dots, A_k) := \Big\{ f^{\langle n \rangle}(a_1, \dots, a_k) \mid a_i \in A_i \ (1 \le i \le k) \Big\}.$$

**Definition 2.26.7** ($(m, n)$-superhyperoperation)**.** [255] Let $S$ be a nonempty set, and let $k \ge 1$ and $m, n \in \mathbb{N}_0$. A *$k$-ary $(m, n)$-superhyperoperation* on $S$ is a map

$$\star : \underbrace{\mathcal{P}^m(S) \times \cdots \times \mathcal{P}^m(S)}_{k \text{ factors}} \longrightarrow \mathcal{P}^n(S).$$

**Definition 2.26.8** ($n$-superhyperstructure)**.** (cf. [255, 251, 256]) Let $n \ge 0$ and $k \ge 1$. An *$n$-superhyperstructure of arity $k$* on a nonempty set $S$ is a pair

$$\mathcal{SH}^{(n)} = \big(\mathcal{P}^n(S), \star\big),$$

where

$$\star : \big(\mathcal{P}^n(S)\big)^k \to \mathcal{P}^n(S)$$

is a $k$-ary operation on the $n$-th iterated powerset. For $n = 0$, this reduces to an ordinary $k$-ary algebraic structure on $S$.

**Definition 2.26.9** ((m,n)-superhyperstructure)**.** Let $S \neq \varnothing$, and let $m, n \in \mathbb{N}_0$, $k \geq 1$. An $(m, n)$-*superhyperstructure of arity* $k$ *over* $S$ is a pair

$$\mathrm{SH}^{(m,n)} = (\mathcal{P}^m(S), \star),$$

where $\star : (\mathcal{P}^m(S))^k \to \mathcal{P}^n(S)$.

**Example 2.26.10** (A concrete $(1, 2)$-superhyperstructure of arity 2)**.** Let

$$S = \{a, b\}.$$

Then

$$\mathcal{P}^1(S) = \mathcal{P}(S) = \{\emptyset, \{a\}, \{b\}, \{a, b\}\},$$

and

$$\mathcal{P}^2(S) = \mathcal{P}(\mathcal{P}(S)).$$

Define a binary operation

$$\star : \mathcal{P}(S) \times \mathcal{P}(S) \to \mathcal{P}^2(S)$$

by

$$A \star B := \{A,\ B,\ A \cup B\} \qquad (A, B \in \mathcal{P}(S)).$$

We verify that this map is well-defined. For any $A, B \subseteq S$, we have

$$A \subseteq S, \qquad B \subseteq S, \qquad A \cup B \subseteq S.$$

Hence each element of the set

$$\{A, B, A \cup B\}$$

belongs to $\mathcal{P}(S)$, and therefore

$$A \star B = \{A, B, A \cup B\} \in \mathcal{P}(\mathcal{P}(S)) = \mathcal{P}^2(S).$$

Thus $\star$ is a well-defined map from

$$(\mathcal{P}^1(S))^2$$

to

$$\mathcal{P}^2(S).$$

Therefore,

$$\mathrm{SH}^{(1,2)} = (\mathcal{P}^1(S), \star)$$

is a $(1, 2)$-superhyperstructure of arity 2 over $S$.

For instance,

$$\{a\} \star \{b\} = \big\{\{a\}, \{b\}, \{a, b\}\big\},$$

and

$$\{a\} \star \{a, b\} = \big\{\{a\}, \{a, b\}\big\},$$

since repeated elements are identified in a set.

Table 2.6 summarizes the main differences among classical structures, hyperstructures, $n$-superhyperstructures, and $(m, n)$-superhyperstructures. In particular, the progression shows how the codomain evolves from a base set, to its powerset, and further to iterated powerset levels.

In addition, a SuperHyperGraph may be regarded as a type of SuperHyperStructure. Several concepts that can be viewed as SuperHyperStructures are summarized in Table 2.7. However, some of the corresponding *HyperConcepts* listed in the table are defined in slightly different ways in the literature, and this distinction should be kept in mind when interpreting the correspondence.

Table 2.6: A concise comparison among classical structures, hyperstructures, $n$-superhyperstructures, and $(m, n)$-superhyperstructures.

| **Concept** | **Underlying domain** | **Typical operation** | **Output level** | **Main feature** |
|---|---|---|---|---|
| Classical structures | $S$ | $* : S^k \to S$ | single element | Ordinary algebraic structure; the operation returns one element of the base set. |
| Hyperstructures | $S$ | $\circ : S \times S \to \mathcal{P}(S)$ | subset of $S$ | The product may have multiple outputs, represented as a subset of the base set. |
| $n$-Superhyperstructures | $\mathcal{P}^n(S)$ | $\star : (\mathcal{P}^n(S))^k \to \mathcal{P}^n(S)$ | single $n$-level superobject | The operation acts on iterated powerset objects, preserving the same superlevel. |
| $(m, n)$-Superhyperstructures | $\mathcal{P}^m(S)$ | $\star : (\mathcal{P}^m(S))^k \to \mathcal{P}^n(S)$ | single $n$-level superobject | A level-changing generalization; the input and output may belong to different iterated powerset levels. |

## 2.27 MetaStructure and Iterated MetaStructure

A MetaStructure is a framework treating entire mathematical structures as objects, with uniform meta-operations that combine them into new structures while globally preserving isomorphism invariance [280, 279]. An Iterated MetaStructure is a hierarchical tower obtained by repeatedly lifting a MetaStructure, so that structures of structures generate deeper meta-level frameworks at successive depths [280].

**Definition 2.27.1** (Single-sorted finitary signature)**.** A *single-sorted finitary signature* is a quadruple

$$\Sigma = \big(\mathrm{Func}, \mathrm{Rel}, \mathrm{ar}_{\mathrm{Func}}, \mathrm{ar}_{\mathrm{Rel}}\big),$$

where:

- Func is a set of function symbols;
- Rel is a set of relation symbols;
- $$\mathrm{ar}_{\mathrm{Func}} : \mathrm{Func} \to \mathbb{N}$$
  assigns to each function symbol its arity;
- $$\mathrm{ar}_{\mathrm{Rel}} : \mathrm{Rel} \to \mathbb{N}$$
  assigns to each relation symbol its arity.

**Definition 2.27.2** ($\Sigma$-structure)**.** Let $\Sigma$ be a single-sorted finitary signature. A $\Sigma$-*structure* is a tuple

$$C = \big(H,\ (f^C)_{f \in \mathrm{Func}},\ (R^C)_{R \in \mathrm{Rel}}\big),$$

where:

- $H \neq \varnothing$ is the carrier of $C$;
- for each $f \in \mathrm{Func}$ of arity $m = \mathrm{ar}_{\mathrm{Func}}(f)$,
  $$f^C : H^m \to H$$
  is an interpretation of $f$ in $C$;

Table 2.7: Representative correspondences of classical concepts with their hyper and superhyper counterparts.

| Classical Concept | HyperConcept | SuperHyperConcept |
|---|---|---|
| Algebra | HyperAlgebra[257, 258] | SuperHyperAlgebra [259, 260, 261] |
| Group | HyperGroup[262, 263] | SuperHyperGroup[264] |
| Field | HyperField[265, 266, 267] | SuperHyperField |
| Ring | HyperRing[268, 269, 270] | SuperHyperRing |
| Vector Space | HyperVector | SuperHyperVector |
| Module | HyperModule[271, 272, 273] | SuperHyperModule |
| Category | HyperCategory | SuperHyperCategory |
| Magma | Hypermagma[264] | SuperHypermagma[264] |
| Monoid | HyperMonoid | SuperHyperMonoid [274] |
| Uncertain Set | HyperUncertain Set [275, 276, 277] | SuperHyperUncertain Set [254] |
| Ideal | HyperIdeal | SuperHyperIdeal[274] |
| Graph | HyperGraph[3] | SuperHyperGraph[6] |
| Language | HyperLanguage | SuperHyperLanguage [278, 279] |

- for each $R \in \mathrm{Rel}$ of arity $r = \mathrm{ar}_{\mathrm{Rel}}(R)$,

$$R^C \subseteq H^r$$

  is an interpretation of $R$ in $C$.

Let

$$\mathrm{Str}_\Sigma$$

denote the class of all $\Sigma$-structures.

**Definition 2.27.3** (MetaStructure over a fixed signature)**.** Let $\Sigma$ be a single-sorted finitary signature. A *MetaStructure over* $\Sigma$ is a pair

$$\mathfrak{M} = \big(\mathcal{U},\ (\Phi_\ell)_{\ell\in\Lambda}\big),$$

such that:

- $$\mathcal{U} \subseteq \mathrm{Str}_\Sigma$$

  is a nonempty class of $\Sigma$-structures, called the *level-0 objects*;

- for each label $\ell \in \Lambda$, there is a positive integer $k_\ell \in \mathbb{N}$ and a map

$$\Phi_\ell : \mathcal{U}^{k_\ell} \to \mathcal{U},$$

  called a *meta-operation*;

- each meta-operation is *uniformly definable* in the following sense: for every $\ell \in \Lambda$ there exist
  - a carrier constructor

$$\Gamma_\ell : \mathcal{U}^{k_\ell} \to \mathbf{Set}_{\neq\varnothing},$$

  - for each $f \in \mathrm{Func}$, a symbol-constructor

$$\Lambda_{\ell,f},$$

  - for each $R \in \mathrm{Rel}$, a relation-constructor

$$\Xi_{\ell,R},$$

  such that for all

$$C_1, \dots, C_{k_\ell} \in \mathcal{U}, \qquad D := \Phi_\ell(C_1, \dots, C_{k_\ell}),$$

  the output structure $D$ satisfies:

$$|D| = \Gamma_\ell(C_1, \dots, C_{k_\ell}),$$

  and for each function symbol $f \in \mathrm{Func}$ of arity $m$,

$$f^D = \Lambda_{\ell,f}\big(C_1, \dots, C_{k_\ell};\ f^{C_1}, \dots, f^{C_{k_\ell}}\big) : |D|^m \to |D|,$$

  while for each relation symbol $R \in \mathrm{Rel}$ of arity $r$,

$$R^D = \Xi_{\ell,R}\big(C_1, \dots, C_{k_\ell};\ R^{C_1}, \dots, R^{C_{k_\ell}}\big) \subseteq |D|^r;$$

- each $\Phi_\ell$ is *isomorphism-invariant*: whenever

$$\alpha_i : C_i \xrightarrow{\cong} D_i \qquad (1 \le i \le k_\ell)$$

are $\Sigma$-isomorphisms between members of $\mathcal{U}$, there exists an induced $\Sigma$-isomorphism

$$\Phi_\ell(\alpha_1, \dots, \alpha_{k_\ell}) : \Phi_\ell(C_1, \dots, C_{k_\ell}) \xrightarrow{\cong} \Phi_\ell(D_1, \dots, D_{k_\ell}).$$

**Remark 2.27.4.** *A MetaStructure treats whole $\Sigma$-structures as objects, and each meta-operation produces a new $\Sigma$-structure from finitely many input $\Sigma$-structures by a rule that is uniform and invariant under isomorphism.*

**Definition 2.27.5** (Lift operator on MetaStructures)**.** Let $\Sigma$ be a fixed single-sorted finitary signature. A *lift operator* on MetaStructures over $\Sigma$ is an assignment

$$\mathcal{L} : \mathfrak{M} \longmapsto \mathcal{L}(\mathfrak{M})$$

such that, whenever

$$\mathfrak{M} = \big(\mathcal{U},\, (\Phi_\ell)_{\ell\in\Lambda}\big)$$

is a MetaStructure over $\Sigma$, the image

$$\mathcal{L}(\mathfrak{M}) = \big(\mathcal{U}^{[1]},\, (\Phi_\ell^{[1]})_{\ell\in\Lambda}\big)$$

is again a MetaStructure over $\Sigma$, together with an embedding

$$\iota_{\mathfrak{M}} : \mathcal{U} \hookrightarrow \mathcal{U}^{[1]},$$

such that for every $\ell \in \Lambda$ and all

$$C_1, \dots, C_{k_\ell} \in \mathcal{U},$$

one has the compatibility condition

$$\Phi_\ell^{[1]}\big(\iota_{\mathfrak{M}}(C_1), \dots, \iota_{\mathfrak{M}}(C_{k_\ell})\big) = \iota_{\mathfrak{M}}\big(\Phi_\ell(C_1, \dots, C_{k_\ell})\big).$$

**Definition 2.27.6** (Iterated MetaStructure of depth $t$)**.** Let $\Sigma$ be a fixed single-sorted finitary signature, let

$$\mathfrak{M}^{(0)} = \big(\mathcal{U}^{(0)},\, (\Phi_\ell^{(0)})_{\ell\in\Lambda}\big)$$

be a MetaStructure over $\Sigma$, and let $\mathcal{L}$ be a lift operator on MetaStructures over $\Sigma$. For each $t \in \mathbb{N}_0$, define recursively

$$\mathfrak{M}^{(t+1)} := \mathcal{L}\big(\mathfrak{M}^{(t)}\big).$$

Then, for each $t \in \mathbb{N}_0$, the MetaStructure

$$\mathfrak{M}^{(t)} = \big(\mathcal{U}^{(t)},\, (\Phi_\ell^{(t)})_{\ell\in\Lambda}\big)$$

is called the *Iterated MetaStructure of depth $t$* generated from $\mathfrak{M}^{(0)}$ by $\mathcal{L}$. The sequence

$$\mathfrak{M}^{(0)}, \mathfrak{M}^{(1)}, \dots, \mathfrak{M}^{(t)}$$

is called the *iterated meta-tower up to depth $t$*.

**Remark 2.27.7.** *Thus, an Iterated MetaStructure is obtained by repeatedly lifting a MetaStructure to higher meta-levels. Each depth adds a new layer in which previously constructed meta-objects become the inputs of the next meta-level operations.*

**Definition 2.27.8** (Iterated lift of order $r$)**.** For $r \in \mathbb{N}_0$, define

$$\mathcal{L}^0(\mathfrak{M}) := \mathfrak{M}, \qquad \mathcal{L}^{r+1}(\mathfrak{M}) := \mathcal{L}\big(\mathcal{L}^r(\mathfrak{M})\big).$$

Then

$$\mathfrak{M}^{(r)} = \mathcal{L}^r\big(\mathfrak{M}^{(0)}\big)$$

for every $r \in \mathbb{N}_0$.

It is known that MetaGraph is a MetaStructure, and that Iterated MetaGraph is an Iterated MetaStructure. In addition, other concepts known as MetaStructures include those listed in Table 2.8.

Table 2.8: Representative examples of concepts known in the MetaStructure framework.

| **Base concept** | **MetaStructure** | **Iterated MetaStructure** |
|---|---|---|
| Graph | MetaGraph | Iterated MetaGraph |
| HyperGraph | MetaHyperGraph | Iterated MetaHyperGraph |
| SuperHyperGraph | MetaSuperHyperGraph | Iterated MetaSuperHyperGraph |
| Language | MetaLanguage (cf.[281, 282]) | Iterated MetaLanguage |
| Topology | MetaTopology (cf.[283, 284]) | Iterated MetaTopology |
| Logic | MetaLogic | Iterated MetaLogic |
| Geometry | MetaGeometry | Iterated MetaGeometry |
| Function | MetaFunction | Iterated MetaFunction |
| Matrix | MetaMatrix | Iterated MetaMatrix |
| Machine Learning | Meta-Machine Learning | Iterated MetaMachine Learning |
| Neural Network | MetaNeural Network | Iterated MetaNeural Network |

## 2.28 MultiStructure and Iterative MultiStructure

A MultiStructure is a structure whose operations return finite multisets of carrier elements, allowing repeated outputs and multiplicity-sensitive behavior beyond ordinary single-valued operations in general [285, 279]. An Iterative MultiStructure recursively applies multiset-valued operations across successive multiset levels, producing hierarchical multisets of multisets and modeling recursively layered multiplicity-based structure formation [285, 286, 287].

**Definition 2.28.1** (Finite multiset)**.** [279] Let $H \neq \varnothing$ be a set. A *finite multiset* on $H$ is a function

$$\mu : H \to \mathbb{N}_0$$

with finite support

$$\operatorname{supp}(\mu) := \{x \in H : \mu(x) > 0\}.$$

We denote by

$$\mathcal{M}(H)$$

the set of all finite multisets on $H$.

**Definition 2.28.2** (MultiStructure)**.** [279] Let $H \neq \varnothing$ be a set, and let $I \subseteq \mathbb{Z}_{>0}$ be a set of arities. A *MultiStructure* is a pair

$$\mathsf{MS} = \Big(H,\ \{\#^{(m)}\}_{m\in I}\Big),$$

where for each $m \in I$,

$$\#^{(m)} : H^m \to \mathcal{M}(H)$$

is an $m$-ary *multi-operation*. Thus each $m$-tuple of elements of $H$ is assigned a finite multiset of elements of $H$.

**Definition 2.28.3** (Iterated multiset levels)**.** Let $H \neq \varnothing$ be a set. Define recursively

$$\mathcal{M}^0(H) := H, \qquad \mathcal{M}^{i+1}(H) := \mathcal{M}\big(\mathcal{M}^i(H)\big) \quad (i \geq 0).$$

**Definition 2.28.4** (Iterative MultiStructure of order $k$)**.** Let $H \neq \varnothing$ be a set, let $k \in \mathbb{N}$, and let $I \subseteq \mathbb{Z}_{>0}$ be a set of arities. An *Iterative MultiStructure of order* $k$ is a tuple

$$\mathsf{IMS}^{(k)} = \Big(H,\ \{\#^{(m,i)}\}_{m\in I,\ 0\leq i<k}\Big),$$

where for each level $i \in \{0, 1, \ldots, k-1\}$ and each arity $m \in I$,

$$\#^{(m,i)} : \big(\mathcal{M}^i(H)\big)^m \to \mathcal{M}^{i+1}(H)$$

is an $m$-ary multi-operation from level $i$ to level $i+1$. Hence the structure iteratively builds multisets of multisets through $k$ hierarchical stages.

**Example 2.28.5** (A simple Iterative MultiStructure of order 2)**.** Let

$$H = \{x, y\}, \qquad k = 2, \qquad I = \{2\}.$$

Here we write

$$[a_1, \ldots, a_r]$$

for the finite multiset whose elements are $a_1, \ldots, a_r$, counted with multiplicities.

Define the level-0 binary multi-operation

$$\#^{(2,0)} : H^2 \to \mathcal{M}(H)$$

by

$$\#^{(2,0)}(x, x) = [x, x], \qquad \#^{(2,0)}(x, y) = [x, y], \qquad \#^{(2,0)}(y, x) = [x, y], \qquad \#^{(2,0)}(y, y) = [y, y].$$

Thus, for instance,

$$\#^{(2,0)}(x, y) = [x, y] \in \mathcal{M}(H).$$

Next, define the level-1 binary multi-operation

$$\#^{(2,1)} : \mathcal{M}(H) \times \mathcal{M}(H) \to \mathcal{M}^2(H) = \mathcal{M}(\mathcal{M}(H))$$

by

$$\#^{(2,1)}(M, N) = [M, N] \qquad \text{for all } M, N \in \mathcal{M}(H).$$

For example,

$$\#^{(2,1)}([x, x], [x, y]) = [[x, x], [x, y]] \in \mathcal{M}^2(H).$$

Therefore,

$$\mathsf{IMS}^{(2)} = \Big(H, \{\#^{(2,0)}, \#^{(2,1)}\}\Big)$$

is an Iterative MultiStructure of order 2. It first produces multisets of elements of $H$, and then produces multisets of such multisets.

Representative correspondences of classical concepts with their multi and iterative multi counterparts are provided for reference in Table 2.9.

Table 2.9: Representative correspondences of classical concepts with their multi and iterative multi counterparts.

| **Classical Concept** | **MultiConcept** | **Iterative MultiConcept** |
|---|---|---|
| Algebra | MultiAlgebra[288, 289, 290] | Iterative MultiAlgebra |
| Group | MultiGroup[291] | Iterative MultiGroup |
| Field | MultiField | Iterative MultiField |
| Ring | MultiRing[292, 293] | Iterative MultiRing |
| Module | MultiModule | Iterative MultiModule |
| Magma | MultiMagma | Iterative MultiMagma |
| Uncertain Set | MultiUncertain Set [294, 295, 296] | Iterative MultiUncertain Set [297] |
| Graph | MultiGraph | Iterative MultiGraph |
| Language | MultiLanguage | Iterative MultiLanguage |

## 2.29 TreeStructure

A TreeStructure is a rooted partially ordered structure whose carrier forms a finite tree, and whose operations act on nodes while respecting hierarchical organization intrinsically [298].

**Definition 2.29.1** (TreeStructure)**.** Let $T \neq \varnothing$ be a finite set, let $\leq$ be a binary relation on $T$, and let $I \subseteq \mathbb{Z}_{>0}$ be a set of arities. A *TreeStructure* is a triple

$$\mathsf{TS} = \Big(T, \leq, \{\star^{(m)}\}_{m \in I}\Big),$$

such that:

1. $(T, \leq)$ is a finite rooted tree, that is:
   a) $\leq$ is a partial order on $T$;
   b) there exists a unique minimal element $r \in T$ such that
   $$r \leq x \qquad \text{for all } x \in T;$$
   c) for every $x \in T$, the principal ideal
   $$\downarrow x := \{y \in T : y \leq x\}$$
   is totally ordered by $\leq$.
2. for each $m \in I$,
   $$\star^{(m)} : T^m \to T$$
   is an $m$-ary *tree operation*.

If no operations are specified, the pair $(T, \leq)$ itself is called the underlying TreeStructure.

**Example 2.29.2** (A simple TreeStructure)**.** Let

$$T = \{r, a, b, c\},$$

and define the binary relation $\leq$ on $T$ by

$$r \leq r, \quad a \leq a, \quad b \leq b, \quad c \leq c,$$

together with

$$r \leq a, \qquad r \leq b, \qquad a \leq c, \qquad r \leq c.$$

Then $(T, \leq)$ is a finite rooted tree with unique minimal element $r$. Indeed, the principal ideals are

$$\downarrow r = \{r\}, \qquad \downarrow a = \{r, a\}, \qquad \downarrow b = \{r, b\}, \qquad \downarrow c = \{r, a, c\},$$

and each of them is totally ordered by $\leq$.

Let

$$I = \{2\},$$

and define a binary tree operation

$$\star^{(2)} : T^2 \to T$$

by

$$\star^{(2)}(x, y) = \begin{cases} x, & \text{if } x \leq y, \\ y, & \text{if } y \leq x, \\ r, & \text{otherwise.} \end{cases}$$

Thus, for example,

$$\star^{(2)}(r, a) = r, \qquad \star^{(2)}(a, c) = a, \qquad \star^{(2)}(b, c) = r.$$

Therefore,

$$\mathsf{TS} = \big(T, \leq, \{\star^{(2)}\}\big)$$

is a TreeStructure.

TreeStructure is useful for representing hierarchical concepts in the real world. As examples of TreeStructure, notions such as Tree Automata [299, 300, 301], TreeSoft Sets[302, 303], and TreeFuzzy Sets[304] are known. Moreover, it can be extended to notions such as ForestStructure, obtained by replacing a tree with a forest.

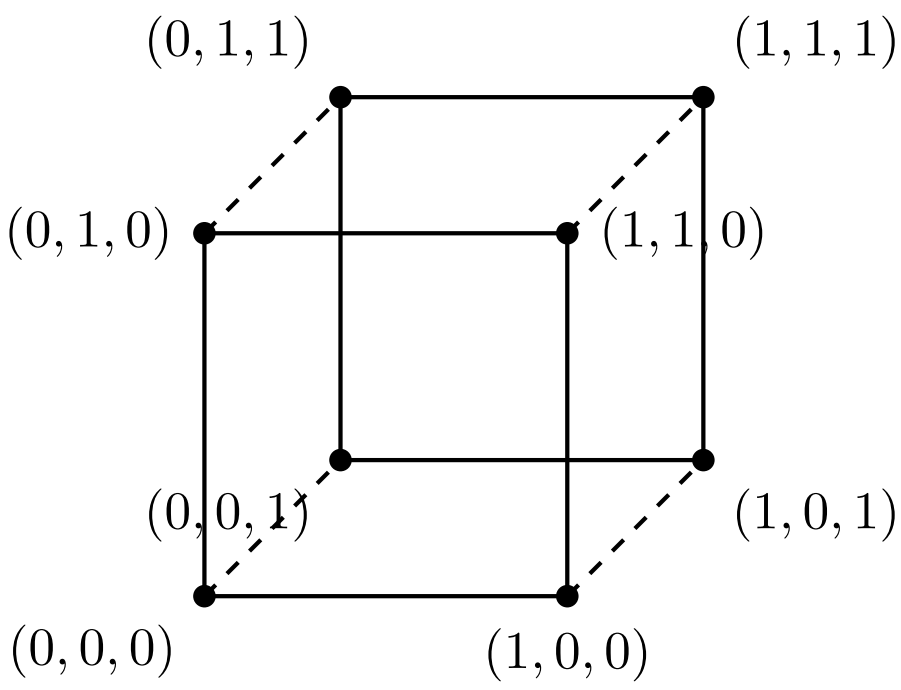

Figure 2.23: The unit 3-HyperCube $\mathsf{HC}^3 = [0,1]^3$, shown as the ordinary cube in $\mathbb{R}^3$. Its eight vertices are the binary points in $\{0,1\}^3$, and two vertices are joined by an edge whenever they differ in exactly one coordinate.

## 2.30 HyperCube

An $n$-HyperCube is the $n$-dimensional cube $[0,1]^n$, equivalently the convex hull of all binary vertices, with equal side lengths in every coordinate direction of Euclidean space [305, 306, 307]. As related concepts, HyperCube Graphs [308, 309, 310], Fuzzy HyperCube [311, 312, 313], Boolean Cube[314, 315], Enhanced Hypercube[316, 317, 318], and Neutrosophic HyperCube [319] are known. An $n$-HyperRectangle is the Cartesian product of $n$ closed intervals in $\mathbb{R}^n$, determined by side lengths, and becomes a HyperCube when equal in all coordinates [320, 321].

**Definition 2.30.1** ($n$-HyperCube)**.** Let $n \in \mathbb{N}$. The *unit $n$-HyperCube* is the subset

$$\mathsf{HC}^n := [0,1]^n = \{(x_1,\ldots,x_n) \in \mathbb{R}^n : 0 \le x_i \le 1 \text{ for all } i = 1,\ldots,n\}.$$

Equivalently, $\mathsf{HC}^n$ is the convex hull of the $2^n$ points

$$(\varepsilon_1,\ldots,\varepsilon_n) \in \mathbb{R}^n, \qquad \varepsilon_i \in \{0,1\}.$$

More generally, for $a \in \mathbb{R}^n$ and $\ell > 0$, an *$n$-HyperCube with base point $a$ and side length $\ell$* is

$$a + [0,\ell]^n = \{a + x : x \in [0,\ell]^n\}.$$

**Example 2.30.2** (A 3-HyperCube)**.** Let $n = 3$. Then the unit 3-HyperCube is

$$\mathsf{HC}^3 = [0,1]^3 = \{(x_1,x_2,x_3) \in \mathbb{R}^3 : 0 \le x_1,x_2,x_3 \le 1\}.$$

Equivalently, $\mathsf{HC}^3$ is the convex hull of the eight binary points

$$(0,0,0),\ (1,0,0),\ (0,1,0),\ (0,0,1),\ (1,1,0),\ (1,0,1),\ (0,1,1),\ (1,1,1).$$

Thus $\mathsf{HC}^3$ is the ordinary unit cube in $\mathbb{R}^3$. Its side length is 1, it has $2^3 = 8$ vertices, and each edge joins two vertices that differ in exactly one coordinate. A schematic illustration is provided for reference in Fig. 2.23.

**Definition 2.30.3** ($n$-HyperRectangle)**.** Let $n \in \mathbb{N}$, let $a = (a_1,\ldots,a_n) \in \mathbb{R}^n$, and let

$$\ell = (\ell_1,\ldots,\ell_n) \in (0,\infty)^n.$$

The *$n$-HyperRectangle* determined by $a$ and $\ell$ is the subset

$$\mathsf{HR}(a,\ell) := \prod_{i=1}^{n} [a_i, a_i + \ell_i] = \{(x_1,\ldots,x_n) \in \mathbb{R}^n : a_i \le x_i \le a_i + \ell_i \text{ for all } i\}.$$

Equivalently, $\mathsf{HR}(a,\ell)$ is the Cartesian product of $n$ closed intervals. If

$$\ell_1 = \ell_2 = \cdots = \ell_n,$$

then $\mathsf{HR}(a,\ell)$ is an $n$-HyperCube.

A Meta-HyperCube is a MetaStructure whose objects are hypercubes and whose uniform meta-operations combine or transform whole hypercube structures into new hypercube structures systematically again [322].

**Definition 2.30.4** (Meta-HyperCube)**.** Let

$$\mathcal{U}_{\mathrm{hc}} = \{H_n \mid n \in \mathbb{N},\ H_n = ([0,1]^n, \{0,1\}^n, E_n)\},$$

where

$$E_n = \{(u,v) \in \{0,1\}^n \times \{0,1\}^n : \|u-v\|_1 = 1\}.$$

A *Meta-HyperCube* is a MetaStructure

$$\mathcal{M}_{\mathrm{hc}} = (\mathcal{U}_{\mathrm{hc}}, (\Phi_\ell)_{\ell\in\Lambda}),$$

whose objects are hypercubes and whose meta-operations

$$\Phi_\ell : \mathcal{U}_{\mathrm{hc}}^{k_\ell} \to \mathcal{U}_{\mathrm{hc}}$$

uniformly combine or transform entire hypercube structures into new hypercube structures.

Moreover, a concept such as *SuperHyperCube* can also be defined as follows.

**Definition 2.30.5** ($n$-SuperHyperCube)**.** Let $U_0$ be a finite nonempty base set, and let $n \in \mathbb{N}_0$. Define the iterated powersets by

$$\mathcal{P}^0(U_0) := U_0, \qquad \mathcal{P}^{k+1}(U_0) := \mathcal{P}(\mathcal{P}^k(U_0)) \quad (k \geq 0).$$

Let $H \subseteq \mathcal{P}^n(U_0)$ be a finite nonempty set, called the *hierarchical coordinate set.*

The *$n$-SuperHyperCube over $H$* is the graph

$$Q_{\mathrm{SH}}^{(n)}(H) = (V,E),$$

where

$$V := \mathcal{P}(H),$$

and two vertices $A, B \in V$ are adjacent if and only if

$$|A \triangle B| = 1,$$

where $A \triangle B$ denotes the symmetric difference of $A$ and $B$.

Thus, unlike the ordinary hypercube whose vertices are binary vectors, the vertices of $Q_{\mathrm{SH}}^{(n)}(H)$ are subsets of hierarchical objects drawn from an iterated powerset level. Equivalently, two vertices are adjacent precisely when they differ in exactly one hierarchical coordinate.

**Remark 2.30.6.** *If $n = 0$ and $H = \{1,2,\ldots,d\}$, then $V = \mathcal{P}(H) \cong \{0,1\}^d$, and $Q_{\mathrm{SH}}^{(0)}(H)$ is exactly the ordinary $d$-dimensional hypercube.*

## 2.31 Ubergraphs

Ubergraphs generalize hypergraphs by allowing an edge to contain not only fundamental vertices but also other edges [323, 324, 325]. To formulate the depth of this recursive construction precisely, we first define an appropriate hierarchy of powerset universes.

**Definition 2.31.1** (Recursive powerset universe)**.** Let $X$ be a finite set whose elements are regarded as atomic labels. Define the families

$$\mathcal{Q}_0(X) := X$$

and, recursively, for every $i \in \mathbb{N}$,

$$\mathcal{Q}_i(X) := \mathcal{P}\left(\bigcup_{j=0}^{i-1} \mathcal{Q}_j(X)\right).$$

For $k \in \mathbb{N}_0$, define

$$\mathcal{P}^{[k]}(X) := \mathcal{P}\left(\bigcup_{i=0}^{k} \mathcal{Q}_i(X)\right).$$

In particular,

$$\mathcal{P}^{[0]}(X) = \mathcal{P}(X).$$

**Definition 2.31.2** (Depth-$k$ Ubergraph)**.** [323] Let $k \in \mathbb{N}_0$. A *depth-$k$ Ubergraph* is a pair

$$\mathcal{U} = (V, E),$$

where:

1. $V$ is a finite set whose elements are called *fundamental vertices*;
2. $E$ is a finite set satisfying
$$E \subseteq \mathcal{P}^{[k]}(V);$$
3. the following closure condition holds:
$$\forall e \in E,\ \forall x \in e, \qquad x \notin V \implies x \in E.$$

The elements of $E$ are called *uberedges*. The full vertex domain of $\mathcal{U}$ is

$$V(\mathcal{U}) := V \cup \bigcup_{e \in E} e.$$

Thus, an element of $V(\mathcal{U}) \setminus V$ occurring inside an uberedge must itself be one of the uberedges of $\mathcal{U}$.

**Remark 2.31.3.** *When $k = 0$, one has*

$$E \subseteq \mathcal{P}(V),$$

*so a depth-0 Ubergraph is an ordinary hypergraph, apart from the optional convention concerning empty hyperedges. If empty hyperedges are to be excluded, the condition on $E$ may be replaced by*

$$E \subseteq \mathcal{P}^{[k]}(V) \setminus \{\emptyset\}.$$

**Remark 2.31.4** (Well-foundedness)**.** *The construction in Definition 2.31.2 is well-founded because every object belongs to a finite recursively generated powerset level. In particular, direct or indirect self-membership such as*

$$e \in e$$

*is excluded under the usual axiom of foundation.*

*Equivalently, define the membership digraph*

$$D_{\mathcal{U}} = \big(V \cup E, A_{\mathcal{U}}\big),$$

*where*

$$A_{\mathcal{U}} := \big\{(x, e) \in (V \cup E) \times E : x \in e\big\}.$$

*Then $D_{\mathcal{U}}$ is a directed acyclic graph.*

**Example 2.31.5** (A depth-2 Ubergraph)**.** We present a concrete example of a depth-2 Ubergraph.

Let the set of fundamental vertices be

$$V = \{a, b, c\}.$$

Define four uberedges by

$$e_1 := \{a, b\}, \qquad e_2 := \{b, c\}, \qquad e_3 := \{a, e_1, e_2\}, \qquad e_4 := \{c, e_3\}.$$

Set

$$E := \{e_1, e_2, e_3, e_4\}.$$

Then

$$\mathcal{U} = (V, E)$$

is an Ubergraph.

We verify that $\mathcal{U}$ is a depth-2 Ubergraph.

First,

$$e_1, e_2 \subseteq V,$$

so

$$e_1, e_2 \in \mathcal{Q}_1(V) = \mathcal{P}(V).$$

Next,

$$e_3 = \{a, e_1, e_2\} \subseteq V \cup \mathcal{Q}_1(V),$$

hence

$$e_3 \in \mathcal{Q}_2(V) = \mathcal{P}\big(V \cup \mathcal{Q}_1(V)\big).$$

Finally,

$$e_4 = \{c, e_3\} \subseteq \mathcal{Q}_0(V) \cup \mathcal{Q}_2(V) \subseteq \bigcup_{i=0}^{2} \mathcal{Q}_i(V),$$

so

$$e_4 \in \mathcal{P}^{[2]}(V) = \mathcal{P}\left(\bigcup_{i=0}^{2} \mathcal{Q}_i(V)\right).$$

Therefore,

$$E \subseteq \mathcal{P}^{[2]}(V).$$

Moreover, the closure condition is satisfied. Indeed, whenever a non-fundamental element occurs inside an uberedge, it is itself an uberedge of $\mathcal{U}$:

$$e_1, e_2 \in e_3, \qquad e_3 \in e_4,$$

and

$$e_1, e_2, e_3 \in E.$$

Hence $\mathcal{U}$ is a depth-2 Ubergraph.

Its full vertex domain is

$$V(\mathcal{U}) = V \cup \bigcup_{e \in E} e = \{a, b, c, e_1, e_2, e_3\}.$$

Observe that $e_4 \in E$ but $e_4 \notin V(\mathcal{U})$, since $e_4$ does not occur as an element of any uberedge.

Figure 2.24 illustrates the recursive incidence structure of $\mathcal{U}$. The arrows indicate membership: an arrow $x \to e$ means that $x \in e$.

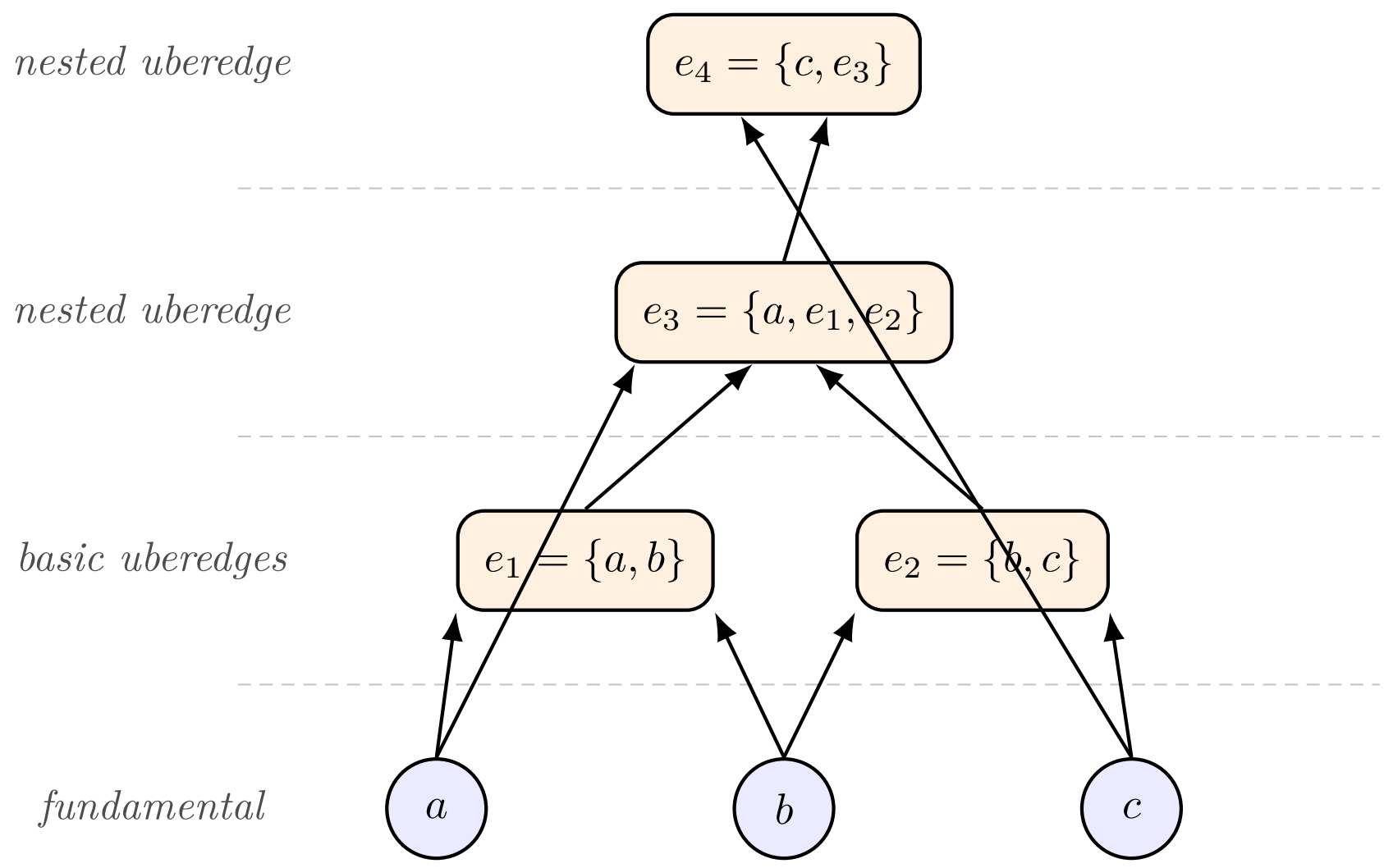

Figure 2.24: A concrete depth-2 Ubergraph. The arrows indicate membership relations $x \in e$.

This example shows how an Ubergraph extends an ordinary hypergraph: the uberedges $e_1$ and $e_2$ are ordinary set-valued edges, whereas $e_3$ and $e_4$ contain other uberedges as elements. Thus the recursive structure of higher-order membership is explicitly represented.

## 2.32 Hb-Graphs

An Hb-Graph, or hyper-bag-graph, replaces each ordinary hyperedge by a multiset. Consequently, the same vertex may occur several times inside one hb-edge, and its multiplicity may depend on the hb-edge under consideration [326, 327, 326, 328, 329].

**Definition 2.32.1** ($W$-valued multiset)**.** Let $V$ be a finite set, and let

$$W \subseteq \mathbb{R}_{\geq 0} \qquad \text{with} \qquad 0 \in W.$$

A *$W$-valued multiset* on $V$ is a function

$$m : V \longrightarrow W.$$

Its support is

$$\mathrm{supp}(m) := \{v \in V : m(v) > 0\}.$$

The value $m(v)$ is called the *multiplicity* or *weight* of $v$ in the multiset $m$.

**Definition 2.32.2** (Hb-Graph)**.** [329] Let

$$V = \{v_1, v_2, \ldots, v_n\}$$

be a nonempty finite vertex set, and let

$$I = \{1, 2, \ldots, p\}$$

be a finite index set. An *Hb-Graph* is a pair

$$\mathcal{H} = (V, \mathcal{E}),$$

where

$$\mathcal{E} = (e_i)_{i \in I}$$

is a finite indexed family of $W$-valued multisets on $V$.

Each $e_i$ is called an *hb-edge*. It is represented by its multiplicity function

$$m_{e_i} : V \longrightarrow W.$$

For $v_j \in V$, the number

$$m_{e_i}(v_j)$$

is the multiplicity of $v_j$ in the hb-edge $e_i$.

The support of $e_i$ is

$$e_i^* := \mathrm{supp}(m_{e_i}) = \{v \in V : m_{e_i}(v) > 0\}.$$

**Definition 2.32.3** (Natural Hb-Graph)**.** An Hb-Graph

$$\mathcal{H} = (V, \mathcal{E})$$

is called a *natural Hb-Graph* if every hb-edge has a nonnegative-integer-valued multiplicity function:

$$m_e : V \longrightarrow \mathbb{N}_0 \qquad (e \in \mathcal{E}).$$

In this case, $m_e(v)$ may be interpreted as the number of occurrences of the vertex $v$ in the hb-edge $e$.

**Definition 2.32.4** (Hb-Graph without repeated hb-edges)**.** An Hb-Graph

$$\mathcal{H} = (V, (e_i)_{i \in I})$$

is said to have *no repeated hb-edges* if

$$e_i = e_j \quad \Longrightarrow \quad i = j$$

for all $i, j \in I$. Equivalently,

$$m_{e_i} = m_{e_j} \quad \Longrightarrow \quad i = j.$$

**Remark 2.32.5.** *An ordinary hypergraph is recovered from an Hb-Graph when:*

$$m_e(v) \in \{0, 1\} \qquad \textit{for every } e \in \mathcal{E} \textit{ and } v \in V,$$

*and repeated hb-edges are excluded. In that case, each hb-edge is uniquely identified with its support*

$$e^* = \{v \in V : m_e(v) = 1\}.$$

**Example 2.32.6** (Software-project task allocation as a natural Hb-Graph)**.** Consider a software-development team consisting of four engineers:

$$V = \{\mathsf{Alice}, \mathsf{Bob}, \mathsf{Chen}, \mathsf{Dina}\}.$$

Suppose that two work packages require different numbers of task units from the participating engineers. Unlike an ordinary HyperGraph, where a vertex is either present or absent from a hyperedge, an Hb-Graph can record how many times, or with what multiplicity, each engineer participates in a given work package.

Let

$$W = \mathbb{N}_0,$$

and define two hb-edges

$$\mathcal{E} = (e_1, e_2)$$

by the multiplicity functions

$$m_{e_1}, m_{e_2} : V \longrightarrow \mathbb{N}_0.$$

Specifically, let

| | Alice | Bob | Chen | Dina |
|---|---|---|---|---|
| $m_{e_1}$ | 2 | 1 | 2 | 0 |
| $m_{e_2}$ | 0 | 2 | 1 | 3 |

where:

- $e_1$ represents the *backend-migration work package*;
- $e_2$ represents the *release-validation work package.*

Thus, in multiset notation, the hb-edges may be written informally as

$$e_1 = \{\mathsf{Alice}^{(2)}, \mathsf{Bob}^{(1)}, \mathsf{Chen}^{(2)}\},$$

and

$$e_2 = \{\mathsf{Bob}^{(2)}, \mathsf{Chen}^{(1)}, \mathsf{Dina}^{(3)}\}.$$

Here the superscripts indicate multiplicities rather than distinct vertices.

The supports of the two hb-edges are

$$e_1^* = \{\mathsf{Alice}, \mathsf{Bob}, \mathsf{Chen}\}$$

and

$$e_2^* = \{\mathsf{Bob}, \mathsf{Chen}, \mathsf{Dina}\}.$$

Consequently,

$$\mathcal{H} = \big(V, (e_1, e_2)\big)$$

is a natural Hb-Graph.

Figure 2.25 illustrates the two hb-edges. The number attached to each incidence line is the multiplicity of the corresponding vertex in that hb-edge.

For instance,

$$m_{e_1}(\mathsf{Alice}) = 2$$

means that Alice contributes two task units to the backend-migration work package, whereas

$$m_{e_1}(\mathsf{Dina}) = 0$$

means that Dina does not participate in that work package. Similarly,

$$m_{e_2}(\mathsf{Dina}) = 3$$

records three task units assigned to Dina in the release-validation work package.

This example also demonstrates why an Hb-Graph contains strictly more information than its underlying support HyperGraph. If all positive multiplicities are replaced by 1, then the two hb-edges reduce to the ordinary hyperedges

$$e_1^* = \{\mathsf{Alice}, \mathsf{Bob}, \mathsf{Chen}\}, \qquad e_2^* = \{\mathsf{Bob}, \mathsf{Chen}, \mathsf{Dina}\}.$$

The support HyperGraph therefore records which engineers participate, whereas the Hb-Graph additionally records their multiplicities within each interaction.

$e_1$: **Backend migration**

Alice Bob Chen

2 1 2

$e_1$

$e_1^* = \{\text{Alice}, \text{Bob}, \text{Chen}\}$

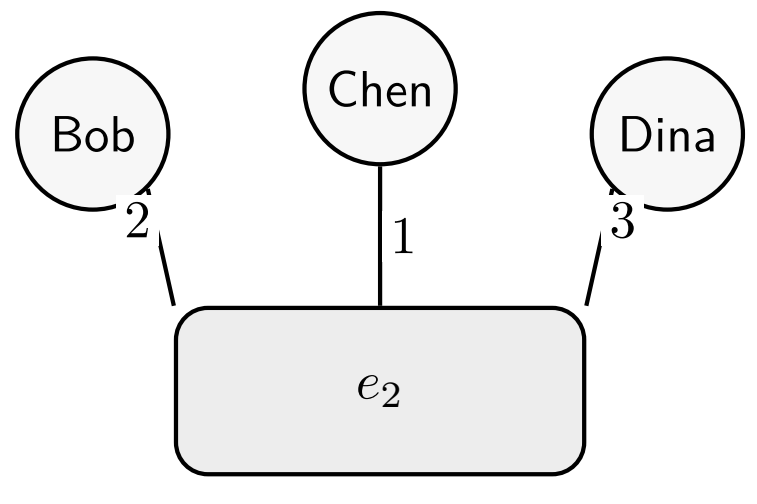

The labels on the incidence lines indicate vertex multiplicities. Repeated vertex names in the two panels refer to the same elements of the common vertex set $V$.

Figure 2.25: A natural Hb-Graph for software-project task allocation. Each incidence label gives the multiplicity $m_{e_i}(v)$ of an engineer $v$ in the corresponding hb-edge $e_i$.

## 2.33 Chygraphs

A Chygraph, also called a complex hypergraph, is a recursively structured system consisting of atoms and complexes. Each complex possesses an internal hypergraph whose vertices may themselves be atoms or other complexes [330, 331, 332]. To avoid a literal self-referential set equation, complexes are represented by distinct labels and an internal-hypergraph assignment.

**Definition 2.33.1** (Chygraph)**.** [330] A finite *Chygraph* is a triple

$$\boldsymbol{\chi} = (A, C, \eta),$$

satisfying the following conditions:

1. $A$ is a finite set whose elements are called *atoms*;
2. $C$ is a finite set whose elements are called *complexes*, and
$$A \cap C = \emptyset;$$
3. $\eta$ is a map assigning an internal hypergraph to every complex:
$$\eta : C \longrightarrow \mathbf{Hyp}(A \sqcup C), \qquad c \longmapsto H_c,$$
where
$$H_c = (V_c, E_c)$$
is a hypergraph satisfying
$$V_c \subseteq A \sqcup C$$
and
$$E_c \subseteq \mathcal{P}(V_c) \setminus \{\emptyset\}.$$

Here $A \sqcup C$ denotes the disjoint union of the atom labels and the complex labels. The family

$$\mathcal{H}_{\boldsymbol{\chi}} := \{H_c : c \in C\}$$

is called the family of *intra-complex hypergraphs.*

**Remark 2.33.2** (Relation with the original notation)**.** *The original Chygraph notation is commonly written as*

$$\chi(A, C),$$

*where the elements of $C$ are themselves described as hypergraphs with vertex sets contained in $A \cup C$. In the typed formulation of Definition 2.33.1, the complex label $c \in C$ and its associated internal hypergraph*

$$H_c = \eta(c)$$

*are distinguished explicitly. Thus one may informally identify*

$$c \equiv H_c$$

*and recover the original abbreviated notation.*

**Definition 2.33.3** (Complex-inclusion relation)**.** Let

$$\boldsymbol{\chi} = (A, C, \eta)$$

be a Chygraph. Define the complex-inclusion relation

$$\prec_{\boldsymbol{\chi}} \subseteq C \times C$$

by

$$c' \prec_{\boldsymbol{\chi}} c \quad \Longleftrightarrow \quad c' \in V_c.$$

Thus

$$c' \prec_{\boldsymbol{\chi}} c$$

means that the complex $c'$ occurs as a vertex of the internal hypergraph associated with $c$.

**Definition 2.33.4** (Hierarchical Chygraph)**.** A Chygraph

$$\boldsymbol{\chi} = (A, C, \eta)$$

is called *hierarchical* or *well-founded* if there exists a level map

$$\lambda : C \longrightarrow \mathbb{N}$$

such that

$$c' \in V_c \cap C \quad \Longrightarrow \quad \lambda(c') < \lambda(c).$$

Equivalently, the directed graph

$$D_{\boldsymbol{\chi}} = \bigl(C, F_{\boldsymbol{\chi}}\bigr),$$

where

$$F_{\boldsymbol{\chi}} = \bigl\{(c', c) \in C \times C : c' \in V_c\bigr\},$$

is acyclic.

**Remark 2.33.5.** *The well-foundedness condition in Definition 2.33.4 is an additional structural condition and is not required for a general Chygraph. A general Chygraph may encode intra-level or cyclic references because complex labels are treated as distinct typed objects rather than as literal non-well-founded sets.*

**Remark 2.33.6** (Ubergraphs as a special case)**.** *Suppose that, for every $c \in C$, the internal hypergraph*

$$H_c = (V_c, E_c)$$

*has exactly one hyperedge and that this hyperedge is $V_c$:*

$$E_c = \{V_c\}.$$

*Then the Chygraph contains no additional internal decomposition beyond the membership set $V_c$. Under this restriction, the Chygraph reduces to an Ubergraph-type recursive incidence structure.*

**Example 2.33.7** (A hierarchical Chygraph for a software-development organization)**.** Consider a software-development organization consisting of six engineers. Let the atom set be

$$A = \{\mathsf{Alice}, \mathsf{Bob}, \mathsf{Chen}, \mathsf{Dina}, \mathsf{Eli}, \mathsf{Farah}\}.$$

For brevity, write

$$a = \mathsf{Alice}, \quad b = \mathsf{Bob}, \quad c = \mathsf{Chen}, \quad d = \mathsf{Dina}, \quad e = \mathsf{Eli}, \quad f = \mathsf{Farah}.$$

Let

$$C = \{c_{\mathrm{B}}, c_{\mathrm{F}}, c_{\mathrm{P}}, c_{\mathrm{R}}\}$$

be the set of complex labels, where

$$\begin{aligned} c_{\mathrm{B}} &= \text{backend-team complex}, \\ c_{\mathrm{F}} &= \text{frontend-team complex}, \\ c_{\mathrm{P}} &= \text{platform-unit complex}, \\ c_{\mathrm{R}} &= \text{release-program complex}. \end{aligned}$$

Clearly,

$$A \cap C = \emptyset.$$

We now assign an internal hypergraph

$$H_x = (V_x, E_x)$$

to each complex $x \in C$.

For the backend-team complex, let

$$V_{c_{\mathrm{B}}} = \{a, b, c\}$$

and

$$E_{c_{\mathrm{B}}} = \big\{\{a, b\}, \{b, c\}, \{a, b, c\}\big\}.$$

Thus,

$$H_{c_{\mathrm{B}}} = \left(\{a, b, c\}, \big\{\{a, b\}, \{b, c\}, \{a, b, c\}\big\}\right).$$

Similarly, for the frontend-team complex, define

$$V_{c_{\mathrm{F}}} = \{d, e, f\}$$

and

$$E_{c_{\mathrm{F}}} = \big\{\{d, e\}, \{e, f\}, \{d, e, f\}\big\}.$$

Hence,

$$H_{c_{\mathrm{F}}} = \left(\{d, e, f\}, \big\{\{d, e\}, \{e, f\}, \{d, e, f\}\big\}\right).$$

At the next hierarchical level, define the platform-unit complex by

$$V_{c_{\mathrm{P}}} = \{c_{\mathrm{B}}, c_{\mathrm{F}}, c, f\}.$$

Its internal hyperedge family is

$$E_{c_{\mathrm{P}}} = \big\{\{c_{\mathrm{B}}, c_{\mathrm{F}}\}, \{c_{\mathrm{B}}, c\}, \{c_{\mathrm{F}}, f\}\big\}.$$

Therefore,

$$H_{c_{\mathrm{P}}} = \left(\{c_{\mathrm{B}}, c_{\mathrm{F}}, c, f\}, E_{c_{\mathrm{P}}}\right).$$

Here the internal hypergraph of $c_{\mathrm{P}}$ contains both lower-level complexes and atoms as vertices.

Finally, define the release-program complex by

$$V_{c_{\mathrm{R}}} = \{c_{\mathrm{P}}, a, e\}$$

and

$$E_{c_{\mathrm{R}}} = \big\{\{c_{\mathrm{P}}, a\}, \{c_{\mathrm{P}}, e\}, \{c_{\mathrm{P}}, a, e\}\big\}.$$

Thus,

$$H_{c_{\mathrm{R}}} = \left(\{c_{\mathrm{P}}, a, e\}, E_{c_{\mathrm{R}}}\right).$$

Define

$$\eta : C \longrightarrow \mathbf{Hyp}(A \sqcup C)$$

by

$$\eta(c_{\mathrm{B}}) = H_{c_{\mathrm{B}}}, \qquad \eta(c_{\mathrm{F}}) = H_{c_{\mathrm{F}}}, \qquad \eta(c_{\mathrm{P}}) = H_{c_{\mathrm{P}}}, \qquad \eta(c_{\mathrm{R}}) = H_{c_{\mathrm{R}}}.$$

Then

$$\boldsymbol{\chi} = (A, C, \eta)$$

is a finite Chygraph.

The complex-inclusion relations are

$$c_{\mathrm{B}} \prec_{\boldsymbol{\chi}} c_{\mathrm{P}}, \qquad c_{\mathrm{F}} \prec_{\boldsymbol{\chi}} c_{\mathrm{P}}, \qquad c_{\mathrm{P}} \prec_{\boldsymbol{\chi}} c_{\mathrm{R}}.$$

No other complex-inclusion relations occur.

Define the level map

$$\lambda : C \longrightarrow \mathbb{N}$$

by

$$\lambda(c_{\mathrm{B}}) = \lambda(c_{\mathrm{F}}) = 1, \qquad \lambda(c_{\mathrm{P}}) = 2, \qquad \lambda(c_{\mathrm{R}}) = 3.$$

Then

$$c' \in V_c \cap C \quad \Longrightarrow \quad \lambda(c') < \lambda(c).$$

Hence

$$\boldsymbol{\chi} = (A, C, \eta)$$

is a hierarchical Chygraph.

Figure 2.26 illustrates its hierarchical organization. A solid arrow

$$c' \longrightarrow c$$

means

$$c' \prec_{\boldsymbol{\chi}} c,$$

that is, $c'$ occurs as a complex-valued vertex of the internal hypergraph $H_c$.

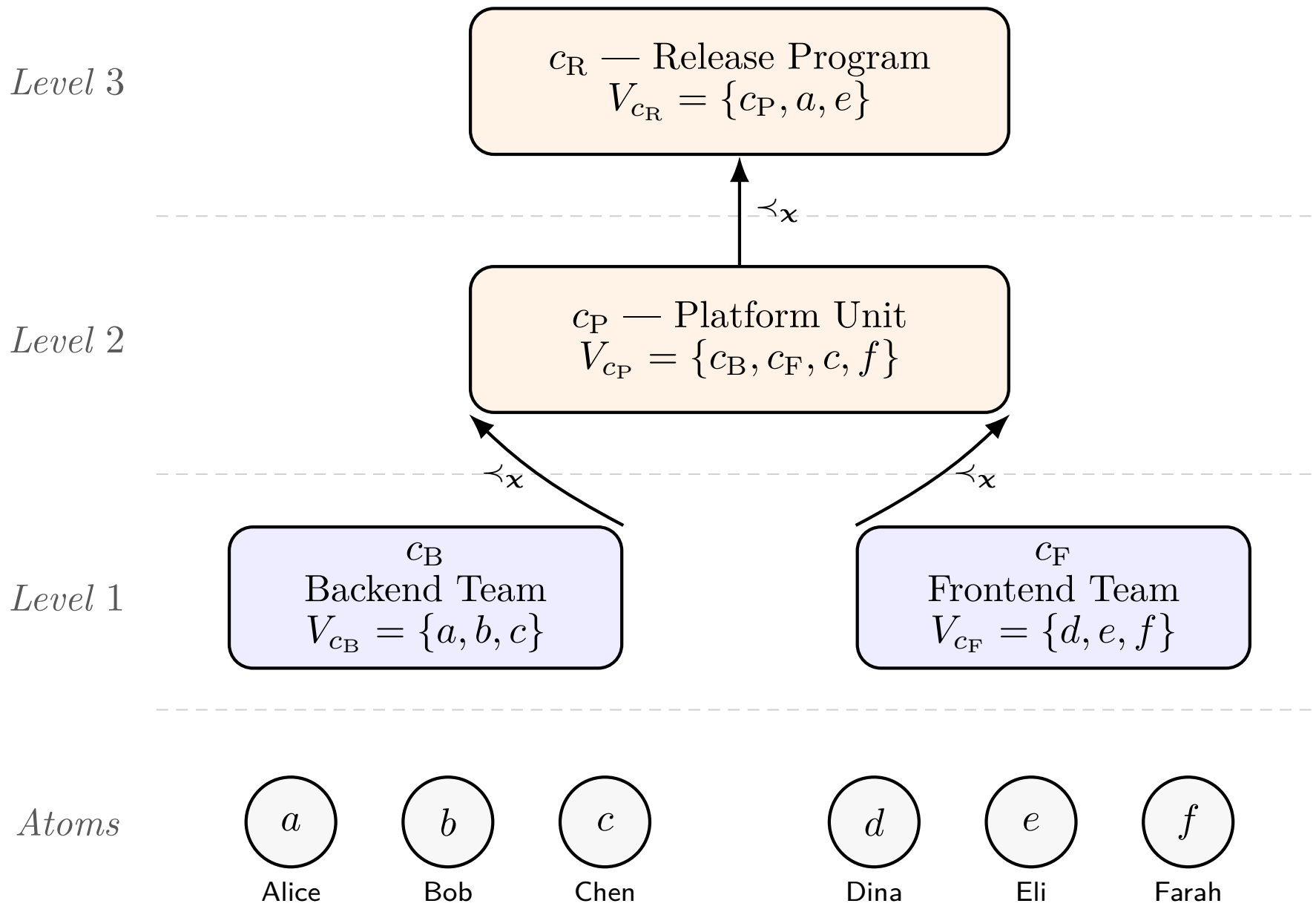

Figure 2.26: A hierarchical Chygraph for a software-development organization. The complexes $c_{\mathrm{B}}$ and $c_{\mathrm{F}}$ occur as vertices of $H_{c_{\mathrm{P}}}$, while $c_{\mathrm{P}}$ occurs as a vertex of $H_{c_{\mathrm{R}}}$. Consequently, the complex-inclusion digraph is acyclic.

Indeed, the complex-inclusion digraph is

$$D_{\boldsymbol{\chi}} = \left(C, \{(c_{\mathrm{B}}, c_{\mathrm{P}}), (c_{\mathrm{F}}, c_{\mathrm{P}}), (c_{\mathrm{P}}, c_{\mathrm{R}})\}\right).$$

It has the directed form

$$c_{\mathrm{B}} \longrightarrow c_{\mathrm{P}} \longrightarrow c_{\mathrm{R}}, \qquad c_{\mathrm{F}} \longrightarrow c_{\mathrm{P}},$$

and therefore contains no directed cycle.

This example also illustrates a feature that distinguishes a general Chygraph from a simple nested-set construction. Each complex possesses its own nontrivial internal hypergraph. For example,

$$E_{c_{\mathrm{P}}} = \big\{\{c_{\mathrm{B}}, c_{\mathrm{F}}\}, \{c_{\mathrm{B}}, c\}, \{c_{\mathrm{F}}, f\}\big\}$$

describes three different interactions inside the platform unit. Thus the hierarchy records not only that lower-level complexes occur inside higher-level complexes, but also how atoms and complexes interact through the internal hyperedge structure.

## 2.34 Chy-SuperHyperGraphs

Chygraphs assign an internal hypergraph to each complex, whereas $n$-SuperHyperGraphs replace atomic vertices by hierarchical objects belonging to iterated powersets. In this section, we combine these two principles by allowing each complex to carry an internal $n$-SuperHyperGraph, where the superhypergraph level may depend on the complex.

Throughout this section, all sets are assumed to be finite. We write

$$\mathcal{P}_*(X) := \mathcal{P}(X) \setminus \{\emptyset\}$$

for the family of all nonempty subsets of a set $X$.

**Definition 2.34.1** (Iterated powersets)**.** Let $X$ be a set. Define the iterated powersets recursively by

$$\mathcal{P}^0(X) := X$$

and

$$\mathcal{P}^{k+1}(X) := \mathcal{P}\big(\mathcal{P}^k(X)\big) \qquad (k \in \mathbb{N}_0).$$

Thus,

$$\mathcal{P}^1(X) = \mathcal{P}(X), \qquad \mathcal{P}^2(X) = \mathcal{P}(\mathcal{P}(X)),$$

and so forth.

**Definition 2.34.2** (Incidence-form $n$-SuperHyperGraph)**.** Let $U$ be a nonempty finite base set and let $n \in \mathbb{N}_0$. An *incidence-form $n$-SuperHyperGraph over $U$* is a triple

$$\mathcal{S} = \big(V, E, \partial\big),$$

where

$$\emptyset \neq V \subseteq \mathcal{P}^n(U),$$

$E$ is a finite set of superhyperedge identifiers, and

$$\partial : E \longrightarrow \mathcal{P}_*(V)$$

is an incidence map.

For each $e \in E$, the nonempty set

$$\partial(e) \subseteq V$$

is called the *incidence set* of $e$.

**Remark 2.34.3.** *When $n = 0$, one has*

$$V \subseteq \mathcal{P}^0(U) = U.$$

*Hence an incidence-form 0-SuperHyperGraph is simply an ordinary hypergraph written in incidence-map form. In particular, an ordinary hypergraph*

$$H = (V, \mathcal{E}), \qquad \mathcal{E} \subseteq \mathcal{P}_*(V),$$

*can be represented as*

$$\big(V, \mathcal{E}, \mathrm{id}_{\mathcal{E}}\big),$$

*where*

$$\mathrm{id}_{\mathcal{E}}(e) = e \qquad (e \in \mathcal{E}).$$

**Definition 2.34.4** (Isomorphism of incidence-form $n$-SuperHyperGraphs)**.** Let

$$\mathcal{S} = (V, E, \partial)$$

and

$$\mathcal{S}' = (V', E', \partial')$$

be incidence-form $n$-SuperHyperGraphs. An *isomorphism*

$$(f, g) : \mathcal{S} \longrightarrow \mathcal{S}'$$

consists of bijections

$$f : V \longrightarrow V'$$

and

$$g : E \longrightarrow E'$$

such that

$$f[\partial(e)] = \partial'(g(e)) \qquad \text{for every } e \in E.$$

Here

$$f[\partial(e)] = \{f(v) : v \in \partial(e)\}.$$

**Definition 2.34.5** (Chy-SuperHyperGraph)**.** A finite *Chy-SuperHyperGraph* is a quadruple

$$\boldsymbol{\Xi} = \big(A, C, \ell, \eta\big)$$

satisfying the following conditions:

1. $A$ is a finite set whose elements are called *atoms*;
2. $C$ is a finite set whose elements are called *chy-complexes*, and
$$A \cap C = \emptyset;$$
3. $$\ell : C \longrightarrow \mathbb{N}_0$$
is a level function;
4. for every $c \in C$, the map $\eta$ assigns an internal incidence-form $\ell(c)$-SuperHyperGraph
$$\eta(c) = \mathcal{S}_c = \big(V_c, E_c, \partial_c\big)$$
over the common typed base set
$$U_{\boldsymbol{\Xi}} := A \sqcup C.$$
More explicitly,
$$\emptyset \neq V_c \subseteq \mathcal{P}^{\ell(c)}(A \sqcup C)\,,$$
$E_c$ is a finite set of internal superhyperedge identifiers, and
$$\partial_c : E_c \longrightarrow \mathcal{P}_*(V_c)$$
is an incidence map.

The family

$$\mathfrak{S}_{\boldsymbol{\Xi}} := \{\mathcal{S}_c : c \in C\}$$

is called the family of *intra-complex SuperHyperGraphs* of $\boldsymbol{\Xi}$.

**Remark 2.34.6** (Typed disjoint union)**.** *The notation*

$$A \sqcup C$$

*is used rather than $A \cup C$ to emphasize that atoms and chy-complex labels are different types of objects, even if they happen to have identical external names.*

**Definition 2.34.7** (Homogeneous $n$-Chy-SuperHyperGraph)**.** Let $n \in \mathbb{N}_0$. A Chy-SuperHyperGraph

$$\boldsymbol{\Xi} = (A, C, \ell, \eta)$$

is called a *homogeneous $n$-Chy-SuperHyperGraph* if

$$\ell(c) = n \qquad \text{for every } c \in C.$$

A general Chy-SuperHyperGraph may therefore contain internal SuperHyperGraphs of different levels, whereas a homogeneous $n$-Chy-SuperHyperGraph uses the same level $n$ for every chy-complex.

**Definition 2.34.8** (Simple level-zero Chy-SuperHyperGraph)**.** A Chy-SuperHyperGraph

$$\Xi = (A, C, \ell, \eta)$$

is called a *simple level-zero Chy-SuperHyperGraph* if, for every $c \in C$,

$$\ell(c) = 0,$$

the internal edge identifiers are themselves nonempty subsets of $V_c$,

$$E_c \subseteq \mathcal{P}_*(V_c),$$

and

$$\partial_c = \mathrm{id}_{E_c}.$$

Thus,

$$\partial_c(e) = e \qquad \text{for every } e \in E_c.$$

At a positive superhypergraph level, a complex label need not occur directly as a vertex. It may occur inside a nested supervertex. We therefore introduce a recursive support operator.

**Definition 2.34.9** (Atomic support of an iterated-set object)**.** Let $U$ be a set. For every $k \in \mathbb{N}_0$, define

$$\mathrm{Supp}_U^{(k)} : \mathcal{P}^k(U) \longrightarrow \mathcal{P}(U)$$

recursively as follows.

For $u \in U$, set

$$\mathrm{Supp}_U^{(0)}(u) := \{u\}.$$

For $k \in \mathbb{N}_0$ and $X \in \mathcal{P}^{k+1}(U)$, set

$$\mathrm{Supp}_U^{(k+1)}(X) := \bigcup_{x \in X} \mathrm{Supp}_U^{(k)}(x).$$

**Definition 2.34.10** (Complex-inclusion relation of a Chy-SuperHyperGraph)**.** Let

$$\Xi = (A, C, \ell, \eta)$$

be a Chy-SuperHyperGraph. Define

$$\prec_\Xi \subseteq C \times C$$

by

$$c' \prec_\Xi c$$

if and only if there exists a supervertex

$$x \in V_c$$

such that

$$c' \in \mathrm{Supp}_{A \sqcup C}^{(\ell(c))}(x).$$

Thus,

$$c' \prec_\Xi c$$

means that the label $c'$ occurs somewhere inside an iterated-set supervertex of the internal SuperHyperGraph assigned to $c$.

**Remark 2.34.11.** *When*

$$\ell(c) = 0,$$

*every $x \in V_c$ belongs directly to $A \sqcup C$, and*

$$\mathrm{Supp}_{A \sqcup C}^{(0)}(x) = \{x\}.$$

*Consequently,*

$$c' \prec_\Xi c \iff c' \in V_c.$$

*Hence Definition 2.34.10 reduces exactly to the complex-inclusion relation of an ordinary Chygraph.*

**Definition 2.34.12** (Hierarchical Chy-SuperHyperGraph)**.** A Chy-SuperHyperGraph

$$\boldsymbol{\Xi} = (A, C, \ell, \eta)$$

is called *hierarchical* or *well-founded* if there exists a map

$$\lambda : C \longrightarrow \mathbb{N}$$

such that

$$c' \prec_{\boldsymbol{\Xi}} c \quad \Longrightarrow \quad \lambda(c') < \lambda(c).$$

Equivalently, the directed graph

$$D_{\boldsymbol{\Xi}} = \big(C, F_{\boldsymbol{\Xi}}\big),$$

where

$$F_{\boldsymbol{\Xi}} := \{(c', c) \in C \times C : c' \prec_{\boldsymbol{\Xi}} c\},$$

is acyclic.

**Theorem 2.34.13** (Every Chygraph induces a Chy-SuperHyperGraph)**.** *Let*

$$\boldsymbol{\chi} = (A, C, \eta_{\chi})$$

*be a finite Chygraph in the sense of Definition 2.33.1. For every $c \in C$, write*

$$\eta_{\chi}(c) = H_c = (V_c, \mathcal{E}_c),$$

*where*

$$V_c \subseteq A \sqcup C$$

*and*

$$\mathcal{E}_c \subseteq \mathcal{P}_*(V_c).$$

*Define*

$$\ell_0 : C \longrightarrow \mathbb{N}_0$$

*by*

$$\ell_0(c) := 0 \qquad (c \in C),$$

*and define*

$$\widehat{\eta}(c) := (V_c, \mathcal{E}_c, \mathrm{id}_{\mathcal{E}_c}).$$

*Then*

$$\mathfrak{C}(\boldsymbol{\chi}) := \big(A, C, \ell_0, \widehat{\eta}\big)$$

*is a simple level-zero Chy-SuperHyperGraph. Moreover, the original Chygraph is recovered exactly by forgetting the level function and the identity incidence maps.*

*Proof.* Let $c \in C$ be arbitrary. Since

$$\ell_0(c) = 0,$$

we have

$$\mathcal{P}^{\ell_0(c)}(A \sqcup C) = \mathcal{P}^0(A \sqcup C) = A \sqcup C.$$

Because $H_c$ is an internal hypergraph of the Chygraph,

$$V_c \subseteq A \sqcup C.$$

Therefore,

$$V_c \subseteq \mathcal{P}^{\ell_0(c)}(A \sqcup C).$$

Furthermore,

$$\mathcal{E}_c \subseteq \mathcal{P}_*(V_c).$$

Hence the identity map

$$\mathrm{id}_{\mathcal{E}_c} : \mathcal{E}_c \longrightarrow \mathcal{P}_*(V_c), \qquad e \longmapsto e,$$

is a well-defined incidence map.

It follows that

$$\widehat{\eta}(c) = (V_c, \mathcal{E}_c, \mathrm{id}_{\mathcal{E}_c})$$

is an incidence-form 0-SuperHyperGraph over $A \sqcup C$. Since this holds for every $c \in C$,

$$\mathfrak{C}(\boldsymbol{\chi}) = (A, C, \ell_0, \widehat{\eta})$$

satisfies all the conditions of Definition 2.34.5.

By construction, every internal incidence map is the identity map. Therefore, removing the redundant level-zero information and replacing each triple

$$(V_c, \mathcal{E}_c, \mathrm{id}_{\mathcal{E}_c})$$

by the pair

$$(V_c, \mathcal{E}_c)$$

recovers precisely the original internal hypergraph $H_c$. Consequently, the original Chygraph

$$\boldsymbol{\chi} = (A, C, \eta_\chi)$$

is recovered exactly. □

**Theorem 2.34.14** (Characterization of the Chygraph subfamily)**.** *The class of finite Chygraphs is naturally identified with the class of simple level-zero Chy-SuperHyperGraphs.*

*More precisely:*

1. *every finite Chygraph determines a unique simple level-zero Chy-SuperHyperGraph by Theorem 2.34.13;*
2. *every simple level-zero Chy-SuperHyperGraph*

$$\boldsymbol{\Xi} = (A, C, \ell, \eta)$$

*determines a Chygraph*

$$\boldsymbol{\chi}_{\boldsymbol{\Xi}} = (A, C, \eta_\chi)$$

*by setting*

$$\eta_\chi(c) := (V_c, E_c) \qquad (c \in C).$$

*Proof.* The first assertion is Theorem 2.34.13.

For the converse, let

$$\boldsymbol{\Xi} = (A, C, \ell, \eta)$$

be a simple level-zero Chy-SuperHyperGraph. For every $c \in C$, write

$$\eta(c) = (V_c, E_c, \partial_c).$$

By Definition 2.34.8,

$$\ell(c) = 0, \qquad V_c \subseteq A \sqcup C,$$
$$E_c \subseteq \mathcal{P}_*(V_c),$$

and

$$\partial_c = \mathrm{id}_{E_c}\,.$$

Consequently,

$$H_c := (V_c, E_c)$$

is an ordinary hypergraph whose vertex set is contained in $A \sqcup C$. Therefore, the assignment

$$\eta_\chi : C \longrightarrow \mathbf{Hyp}(A \sqcup C), \qquad c \longmapsto H_c,$$

defines a Chygraph

$$\boldsymbol{\chi}_{\boldsymbol{\Xi}} = (A, C, \eta_\chi).$$

The two constructions are mutually inverse because replacing $(V_c, E_c)$ by

$$(V_c, E_c, \mathrm{id}_{E_c})$$

and subsequently forgetting the identity incidence map leaves the internal hypergraph unchanged. □

To prove that every $n$-SuperHyperGraph is represented by a Chy-SuperHyperGraph, we first record the canonical lifting of an injection through iterated powersets.

**Definition 2.34.15** (Iterated lift of a map)**.** Let

$$f : X \longrightarrow Y$$

be a map. Define

$$f^{\langle k \rangle} : \mathcal{P}^k(X) \longrightarrow \mathcal{P}^k(Y)$$

recursively by

$$f^{\langle 0 \rangle} := f$$

and

$$f^{\langle k+1 \rangle}(S) := \left\{ f^{\langle k \rangle}(x) : x \in S \right\}$$

for every

$$S \in \mathcal{P}^{k+1}(X).$$

**Lemma 2.34.16** (Injectivity is preserved by iterated powerset lifting)**.** *Let*

$$f : X \longrightarrow Y$$

*be injective. Then*

$$f^{\langle k \rangle} : \mathcal{P}^k(X) \longrightarrow \mathcal{P}^k(Y)$$

*is injective for every* $k \in \mathbb{N}_0$*.*

*Proof.* We proceed by induction on $k$.

For $k = 0$,

$$f^{\langle 0 \rangle} = f,$$

which is injective by assumption.

Suppose that

$$f^{\langle k \rangle}$$

is injective. Let

$$S, T \in \mathcal{P}^{k+1}(X)$$

and suppose that

$$f^{\langle k+1 \rangle}(S) = f^{\langle k+1 \rangle}(T).$$

Let $x \in S$. Then

$$f^{\langle k \rangle}(x) \in f^{\langle k+1 \rangle}(S) = f^{\langle k+1 \rangle}(T).$$

Hence there exists $y \in T$ such that

$$f^{\langle k \rangle}(x) = f^{\langle k \rangle}(y).$$

By the induction hypothesis,

$$x = y.$$

Therefore $x \in T$, and hence

$$S \subseteq T.$$

By symmetry,

$$T \subseteq S.$$

Thus $S = T$, proving that

$$f^{\langle k+1 \rangle}$$

is injective. □

**Theorem 2.34.17** (Every $n$-SuperHyperGraph induces a one-complex Chy-SuperHyperGraph)**.** *Let*

$$\mathcal{S} = (V, E, \partial)$$

*be a finite incidence-form $n$-SuperHyperGraph over a finite base set $V_0$; that is,*

$$V \subseteq \mathcal{P}^n(V_0)$$

*and*

$$\partial : E \longrightarrow \mathcal{P}_*(V).$$

*Choose a new symbol*

$$c_* \notin V_0$$

*and set*

$$A := V_0, \qquad C := \{c_*\}.$$

*Let*

$$\iota : V_0 \longrightarrow V_0 \sqcup \{c_*\}$$

*be the canonical inclusion, and let*

$$\iota^{\langle n \rangle} : \mathcal{P}^n(V_0) \longrightarrow \mathcal{P}^n(V_0 \sqcup \{c_*\})$$

*be its $n$-fold iterated lift.*

*Define*

$$\widehat{V} := \iota^{\langle n \rangle}[V],$$

*and define*

$$\widehat{\partial} : E \longrightarrow \mathcal{P}_*(\widehat{V})$$

*by*

$$\widehat{\partial}(e) := \iota^{\langle n \rangle}[\partial(e)] \qquad (e \in E).$$

*Finally, define*

$$\ell(c_*) := n$$

*and*

$$\eta(c_*) := \left(\widehat{V}, E, \widehat{\partial}\right).$$

*Then*

$$\mathfrak{S}_{c_*}(\mathcal{S}) := (V_0, \{c_*\}, \ell, \eta)$$

*is a homogeneous $n$-Chy-SuperHyperGraph. Moreover,*

$$\mathcal{S} \cong \eta(c_*)$$

*as incidence-form $n$-SuperHyperGraphs.*

*Proof.* The canonical inclusion

$$\iota : V_0 \longrightarrow V_0 \sqcup \{c_*\}$$

is injective. Therefore, by Lemma 2.34.16,

$$\iota^{\langle n \rangle}$$

is injective.

Because

$$V \subseteq \mathcal{P}^n(V_0),$$

we obtain

$$\widehat{V} = \iota^{\langle n \rangle}[V] \subseteq \mathcal{P}^n(V_0 \sqcup \{c_*\}).$$

Thus $\widehat{V}$ is a valid set of level-$n$ supervertices over the common typed base set

$$V_0 \sqcup \{c_*\}.$$

Let $e \in E$. Since

$$\partial(e) \in \mathcal{P}_*(V),$$

the incidence set $\partial(e)$ is nonempty. The image of a nonempty set under a map is nonempty, so

$$\widehat{\partial}(e) = \iota^{\langle n \rangle}[\partial(e)]$$

is nonempty. Moreover,

$$\widehat{\partial}(e) \subseteq \iota^{\langle n \rangle}[V] = \widehat{V}.$$

Hence

$$\widehat{\partial}(e) \in \mathcal{P}_*(\widehat{V}).$$

Therefore,

$$\widehat{\partial} : E \longrightarrow \mathcal{P}_*(\widehat{V})$$

is a well-defined incidence map.

It follows that

$$\eta(c_*) = \left(\widehat{V}, E, \widehat{\partial}\right)$$

is an incidence-form $n$-SuperHyperGraph over

$$V_0 \sqcup \{c_*\}.$$

Since $C = \{c_*\}$ and

$$\ell(c_*) = n,$$

the quadruple

$$\mathfrak{S}_{c_*}(\mathcal{S}) = (V_0, \{c_*\}, \ell, \eta)$$

is a homogeneous $n$-Chy-SuperHyperGraph.

It remains to prove the isomorphism. Define

$$f := \iota^{\langle n \rangle}\Big|_V : V \longrightarrow \widehat{V}.$$

By the definition of $\widehat{V}$, the map $f$ is surjective, and by Lemma 2.34.16, it is injective. Hence $f$ is bijective.

Let

$$g := \mathrm{id}_E .$$

Then $g : E \to E$ is bijective. For every $e \in E$,

$$f[\partial(e)] = \iota^{\langle n \rangle}[\partial(e)] = \widehat{\partial}(e) = \widehat{\partial}(g(e)).$$

Consequently,

$$(f, g) : (V, E, \partial) \longrightarrow \left(\widehat{V}, E, \widehat{\partial}\right)$$

is an isomorphism in the sense of Definition 2.34.4. □

**Remark 2.34.18.** *The additional label $c_*$ is required because a Chy-SuperHyperGraph must contain at least one named chy-complex to which the internal $n$-SuperHyperGraph is assigned. The new label does not alter the original $n$-SuperHyperGraph, because the original base set is embedded injectively into*

$$V_0 \sqcup \{c_*\},$$

*and the internal structure is transported along the induced iterated powerset injection.*

**Theorem 2.34.19** (Chy-SuperHyperGraphs simultaneously generalize Chygraphs and $n$-SuperHyperGraphs)**.** *The class of finite Chy-SuperHyperGraphs simultaneously generalizes the class of finite Chygraphs and, for every $n \in \mathbb{N}_0$, the class of finite incidence-form $n$-SuperHyperGraphs in the following precise sense:*

1. *the class of finite Chygraphs is naturally identified with the subfamily of simple level-zero Chy-SuperHyperGraphs;*
2. *every finite incidence-form $n$-SuperHyperGraph is isomorphic to the unique internal SuperHyperGraph of a one-complex homogeneous $n$-Chy-SuperHyperGraph.*

*Proof.* Assertion 1 follows from Theorems 2.34.13 and 2.34.14. In particular, every internal ordinary hypergraph of a Chygraph is an incidence-form 0-SuperHyperGraph, and the original Chygraph is recovered by forgetting the redundant identity incidence maps.

Assertion 2 follows from Theorem 2.34.17. Given any finite incidence-form $n$-SuperHyperGraph

$$\mathcal{S} = (V, E, \partial),$$

that theorem constructs a one-complex homogeneous $n$-Chy-SuperHyperGraph

$$\mathfrak{S}_{c_*}(\mathcal{S})$$

whose unique internal SuperHyperGraph is isomorphic to $\mathcal{S}$.

Therefore, Chygraphs occur as the level-zero simple subfamily, while $n$-SuperHyperGraphs occur, up to canonical isomorphism, as the single-complex homogeneous subfamily. Hence Chy-SuperHyperGraphs generalize both structures. □

**Corollary 2.34.20.** *Let*

$$\Xi = (A, C, \ell, \eta)$$

*be a Chy-SuperHyperGraph.*

1. *If*

$$\ell(c) = 0 \qquad \textit{for every } c \in C$$

*and every internal incidence map is an identity map, then $\Xi$ is precisely a Chygraph written in incidence-map form.*

2. *If*

$$C = \{c_*\}$$

*and*

$$\ell(c_*) = n,$$

*then the entire nontrivial incidence structure of $\Xi$ is concentrated in the single internal $n$-SuperHyperGraph*

$$\eta(c_*).$$

**Remark 2.34.21** (Why the general level function is useful)**.** *A homogeneous $n$-Chy-SuperHyperGraph requires every complex to use the same superhypergraph depth. The more general level function*

$$\ell : C \longrightarrow \mathbb{N}_0$$

*permits, for example,*

$$\ell(c_1) = 0, \qquad \ell(c_2) = 1, \qquad \ell(c_3) = 3.$$

*Thus one complex may possess an ordinary internal hypergraph, another may possess a set-valued SuperHyperGraph, and a third may possess a more deeply nested internal SuperHyperGraph. This mixed-level feature is not available in an ordinary Chygraph or in a single $n$-SuperHyperGraph.*

**Remark 2.34.22** (Parallel internal superhyperedges)**.** *Because every $E_c$ is a set of edge identifiers and*

$$\partial_c : E_c \longrightarrow \mathcal{P}_*(V_c)$$

*need not be injective, a general Chy-SuperHyperGraph may contain distinct internal superhyperedges $e_1, e_2 \in E_c$ satisfying*

$$e_1 \neq e_2$$

*but*

$$\partial_c(e_1) = \partial_c(e_2).$$

*Hence the incidence-form definition also permits parallel internal superhyperedges. Requiring every $\partial_c$ to be injective yields the corresponding simple version.*

## 2.35 $H_v$-Structures

An $H_v$-Structure is a hyperstructure in which selected algebraic laws hold weakly through nonempty intersections, enabling flexible modeling of multivalued and imperfect interaction systems [333, 334, 335, 336]. We first introduce a set-extension convention that is required in order to formulate weak associativity without any ambiguity.

**Definition 2.35.1** (Extension of a hyperoperation to nonempty subsets)**.** Let $K$ be a nonempty set, and let

$$\circ : K \times K \longrightarrow \mathcal{P}^*(K)$$

be a hyperoperation, where

$$\mathcal{P}^*(K) := \mathcal{P}(K) \setminus \{\varnothing\}.$$

For all $\mathcal{A}, \mathcal{B} \in \mathcal{P}^*(K)$, define

$$\mathcal{A} \circ \mathcal{B} := \bigcup_{\substack{a \in \mathcal{A} \\ b \in \mathcal{B}}} (a \circ b).$$

In particular, for $x, y \in K$ and $\mathcal{A}, \mathcal{B} \in \mathcal{P}^*(K)$, we write

$$x \circ \mathcal{B} := \bigcup_{b \in \mathcal{B}} (x \circ b), \qquad \mathcal{A} \circ y := \bigcup_{a \in \mathcal{A}} (a \circ y),$$

where $x$ and $y$ are identified with the singleton sets $\{x\}$ and $\{y\}$, respectively.

**Definition 2.35.2** ($H_v$-structure)**.** Let $H$ be a nonempty set, and let

$$\circ : H \times H \longrightarrow \mathcal{P}^*(H)$$

be a hyperoperation.

In general, an $H_v$-structure is a hyperalgebraic structure in which one or more ordinary algebraic axioms are replaced by their corresponding *weak axioms*. More precisely, an equality between two set-valued expressions is weakened to the requirement that the two resulting subsets have a nonempty intersection.

In the standard binary associative setting, the pair

$$(H, \circ)$$

is called an $H_v$-semigroup if $\circ$ satisfies the *weak associativity axiom*

$$x \circ (y \circ z) \ \cap \ (x \circ y) \circ z \neq \varnothing \qquad \text{for all } x, y, z \in H,$$

where the products involving subsets are interpreted according to Definition 2.35.1.

A weak hyperstructure of this associative type is also referred to, in a broad sense, as a binary $H_v$-structure.

If, in addition, the *reproduction axiom*

$$x \circ H = H = H \circ x \qquad \text{for every } x \in H$$

holds, then $(H, \circ)$ is called an $H_v$-group.

Furthermore, $(H, \circ)$ is called *weakly commutative* if

$$x \circ y \ \cap \ y \circ x \neq \varnothing \qquad \text{for all } x, y \in H,$$

and it is called *commutative* if

$$x \circ y = y \circ x \qquad \text{for all } x, y \in H.$$

**Remark 2.35.3.** *The term $H_v$-structure is an umbrella term rather than the name of one uniquely determined algebraic object. Its precise meaning depends on which classical structure and which axioms are being weakened. For example, weakening associativity gives an $H_v$-semigroup, whereas adding the reproduction axiom gives an $H_v$-group. Analogously, one may define $H_v$-rings, $H_v$-modules, and other $H_v$-type structures by weakening the relevant axioms.*

**Definition 2.35.4** ($n$-th $SH_v$-structure)**.** [337] Let $S$ be a nonempty set, let $n \in \mathbb{N}$, and set

$$H_n := \mathcal{P}^n(S).$$

An $n$-th $SH_v$-structure, also called a weak $n$-superhyperstructure, is a pair

$$\mathcal{SH}_v^{(n)} = (H_n, \circ_n),$$

where

$$\circ_n : H_n \times H_n \longrightarrow \mathcal{P}^*(H_n) = \mathcal{P}^*\big(\mathcal{P}^n(S)\big)$$

is a superhyperoperation satisfying the weak associativity axiom

$$X \circ_n (Y \circ_n Z) \;\cap\; (X \circ_n Y) \circ_n Z \neq \varnothing \qquad \text{for all } X, Y, Z \in H_n.$$

Here, $\circ_n$ is extended from elements of $H_n$ to nonempty subsets of $H_n$ by

$$\mathcal{A} \circ_n \mathcal{B} := \bigcup_{\substack{A \in \mathcal{A} \\ B \in \mathcal{B}}} (A \circ_n B), \qquad \mathcal{A}, \mathcal{B} \in \mathcal{P}^*(H_n).$$

Equivalently, an $n$-th $SH_v$-structure is an $H_v$-semigroup whose carrier is the $n$-th iterated powerset of a base set.

If, in addition, the reproduction axiom

$$X \circ_n H_n = H_n = H_n \circ_n X \qquad \text{for every } X \in H_n$$

holds, then

$$(H_n, \circ_n)$$

is called an $n$-th $SH_v$-group, or an $H_v$-group of superhyperorder $n$.

The $n$-th $SH_v$-structure is called *weakly commutative* if

$$X \circ_n Y \;\cap\; Y \circ_n X \neq \varnothing \qquad \text{for all } X, Y \in H_n,$$

and it is called *commutative* if

$$X \circ_n Y = Y \circ_n X \qquad \text{for all } X, Y \in H_n.$$

**Remark 2.35.5** (Nonempty hierarchical carrier)**.** *When empty hierarchical objects are not permitted, one may replace*

$$H_n = \mathcal{P}^n(S)$$

*throughout Definition 2.35.4 by*

$$H_n^* := \mathcal{P}_*^n(S).$$

*In that case, the superhyperoperation must be written consistently as*

$$\circ_n : H_n^* \times H_n^* \longrightarrow \mathcal{P}^*(H_n^*),$$

*and the reproduction axiom, when imposed, must use $H_n^*$ rather than $\mathcal{P}^n(S)$.*

**Remark 2.35.6** (Reduction to an $H_v$-structure)**.** *If the definition is extended to $n = 0$, then*

$$H_0 = \mathcal{P}^0(S) = S,$$

*and*

$$\circ_0 : S \times S \longrightarrow \mathcal{P}^*(S).$$

*Consequently, a zeroth-order $SH_v$-structure is precisely an ordinary binary $H_v$-structure. Thus, the $SH_v$-framework provides a genuine hierarchical extension of the $H_v$-framework.*

# 3 Geometric, Topological, and Complex-Based Family

This chapter surveys higher-order structures whose principal organization is geometric, topological, or complex-based. These frameworks represent higher-order relations through simplices, cells, polyhedra, cubes, paths, incidence structures, topological realizations, subdivision procedures, and local-to-global constructions. Although they arise from different mathematical settings, they provide complementary ways of encoding higher-dimensional incidence, geometric organization, topological structure, and hierarchical refinement.

For reference, the principal geometric, topological, and complex-based higher-order structures discussed in this book are summarized in Table 3.1.

Table 3.1: Principal geometric, topological, and complex-based higher-order structures discussed in this book.

| Concept | Concise description |
|---|---|
| Abstract Simplicial Complex | A family of finite subsets of a vertex set that is closed under taking subsets, thereby representing simplices together with all of their faces. |
| Simplicial Set | A contravariant functor $X : \Delta^{\mathrm{op}} \to \mathbf{Set}$ from the simplex category to sets, equivalently described by face and degeneracy maps satisfying the simplicial identities. |
| Cell Complex | A space or combinatorial structure decomposed into cells of various dimensions, together with incidence or attachment information describing how the cells fit together. |
| CW Complex | A topological space constructed inductively by attaching cells via maps from their boundary spheres and satisfying the closure-finite and weak-topology conditions. |
| Polyhedral Complex | A collection of polyhedra closed under taking faces such that the nonempty intersection of any two members is a common face of both. |
| Dowker Complex | An abstract simplicial complex derived from a binary relation, in which a finite subset forms a simplex precisely when its elements share a common related element. |
| Cubical Complex | A collection of cubes of various dimensions closed under taking faces and satisfying compatible intersection conditions. |
| Path Complex | A family of allowed finite vertex sequences that is closed under deletion of the first or last vertex, providing a combinatorial framework for higher-order directed-path structure. |
| Cellular Sheaf | A local-to-global structure on a cell complex that assigns a stalk to each cell and compatible maps along face incidences, subject to the corresponding functoriality conditions. |
| Meta Simplicial Complex | An abstract simplicial complex whose vertices are themselves simplicial complexes over a prescribed base domain, thereby representing complex-of-complexes organization. |
| Simplicial SuperHypercomplex | An abstract simplicial complex whose vertices are $n$-supervertices drawn from an iterated powerset level, with downward-closed families of supervertices forming superfaces. |

*Table 3.1 (continued)*

| Concept | Concise description |
|---|---|
| Depth-$r$ Iterated Subdivisions of Polyhedral Complexes | A hierarchy of $r$ successive face-compatible polyhedral subdivisions, refining the combinatorial resolution at each level while preserving the underlying polyhedral space. |
| Topological SuperHyperGraph | A hierarchical SuperHyperGraph equipped with topological realizations of its supervertices and superhyperedges, with incidence represented by inclusion among the corresponding topological objects. |

## 3.1 Abstract simplicial complex

An *abstract simplicial complex* is a collection of finite subsets of a vertex set, closed under taking subsets, representing simplices purely combinatorially [338, 339, 340]. It is known that abstract simplicial complexes are closely related to graphs. Matroids are known as one of the related concepts to abstract simplicial complexes. A matroid axiomatizes independence: a hereditary family of subsets satisfying an exchange property, generalizing linear independence and forests [341, 342, 343]. Related concepts include directed simplicial complexes (directed flag complexes) [344, 345, 346], antimatroids[347, 348], greedoids[349, 350], and $\Delta$-complexes. This can be regarded as one of the concepts for representing "higher-order" structure in a more geometric manner.

**Definition 3.1.1** (Abstract simplicial complex)**.** [338, 339, 340] Let $V$ be a set (the *vertex set*). An *(abstract) simplicial complex* on $V$ is a family $K \subseteq \mathcal{P}(V)$ such that:

1. $\{v\} \in K$ for every $v \in V$ that appears in some simplex of $K$ (equivalently, $K$ is nonempty and contains all singletons of its support), and

2. (*downward closed*) if $\sigma \in K$ and $\tau \subseteq \sigma$, then $\tau \in K$.

Elements $\sigma \in K$ are called *simplices.* If $|\sigma| = k + 1$, then $\sigma$ is a *k-simplex* and $\dim(\sigma) := k$. The *dimension* of $K$ is $\dim(K) := \sup\{\dim(\sigma) \mid \sigma \in K\}$.

**Example 3.1.2** (Coauthorship groups as an abstract simplicial complex)**.** Let

$$V = \{\mathsf{A}, \mathsf{B}, \mathsf{C}, \mathsf{D}\}$$

be four researchers. Suppose that the observed coauthorship *groups* are

$$\{\mathsf{A}, \mathsf{B}, \mathsf{C}\} \quad \text{and} \quad \{\mathsf{B}, \mathsf{C}, \mathsf{D}\}.$$

To encode the principle that every sub-team of an observed group is also a valid collaboration unit, define the family $K \subseteq \mathcal{P}(V)$ by taking all nonempty subsets of these groups:

$$K \;=\; \mathcal{P}^{*}(\{\mathsf{A}, \mathsf{B}, \mathsf{C}\}) \;\cup\; \mathcal{P}^{*}(\{\mathsf{B}, \mathsf{C}, \mathsf{D}\}),$$

where $\mathcal{P}^{*}(S) := \mathcal{P}(S) \setminus \{\emptyset\}$. Then $K$ is downward closed, contains all singletons of its support, and hence

$$(V, K)$$

is an abstract simplicial complex in the sense of Definition 3.1.1. In particular, $K$ contains the 2-simplices $\{\mathsf{A}, \mathsf{B}, \mathsf{C}\}$ and $\{\mathsf{B}, \mathsf{C}, \mathsf{D}\}$, together with all their faces. An overview diagram of this example is provided in Fig. 3.1.

## 3.2 Simplicial set

A simplicial set is a functor from the simplex category's opposite to sets, encoding faces and degeneracies of abstract simplices [351, 352, 353]. This can be regarded as one of the concepts for representing "higher-order" structure in a more geometric manner.

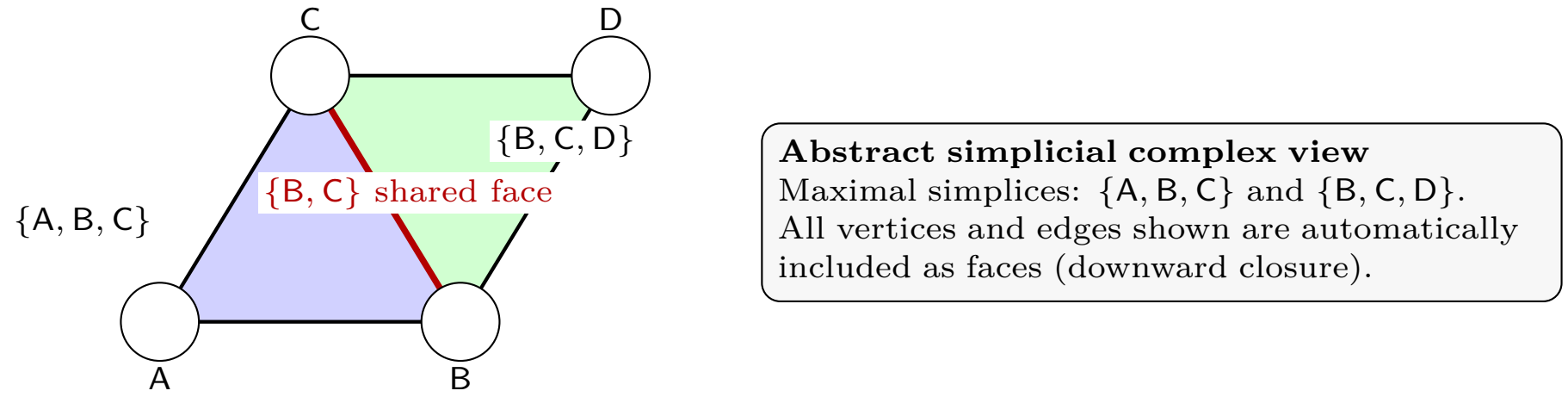

Figure 3.1: Coauthorship groups represented as an abstract simplicial complex: two 2-simplices sharing the edge $\{\mathsf{B}, \mathsf{C}\}$.

**Definition 3.2.1** (Simplicial set)**.** Let $\Delta$ denote the *simplex category*, whose objects are the finite ordinals $[n] = \{0, 1, \dots, n\}$ for $n \geq 0$ and whose morphisms are order-preserving maps. A *simplicial set* is a contravariant functor

$$X : \Delta^{\mathrm{op}} \longrightarrow \mathbf{Set}.$$

Equivalently, it is a sequence of sets $(X_n)_{n\geq 0}$ together with maps

$$d_i : X_n \to X_{n-1} \quad (0 \leq i \leq n,\ n \geq 1), \qquad s_i : X_n \to X_{n+1} \quad (0 \leq i \leq n,\ n \geq 0),$$

called *face maps* and *degeneracy maps*, satisfying the simplicial identities:

$$\begin{aligned} d_i d_j &= d_{j-1} d_i && (i < j), \\ s_i s_j &= s_{j+1} s_i && (i \leq j), \\ d_i s_j &= \begin{cases} s_{j-1} d_i, & i < j, \\ \mathrm{id}, & i = j \text{ or } i = j+1, \\ s_j d_{i-1}, & i > j+1, \end{cases} \end{aligned}$$

whenever the expressions make sense. Elements of $X_n$ are called *$n$-simplices*; those in $\mathrm{im}(s_i)$ are called *degenerate*.

**Example 3.2.2** (The nerve of a small category as a simplicial set)**.** Let $\mathcal{C}$ be the small category with two objects $A, B$ and morphisms

$$\mathrm{Hom}_{\mathcal{C}}(A, A) = \{\mathrm{id}_A\}, \qquad \mathrm{Hom}_{\mathcal{C}}(B, B) = \{\mathrm{id}_B\}, \qquad \mathrm{Hom}_{\mathcal{C}}(A, B) = \{f\}, \qquad \mathrm{Hom}_{\mathcal{C}}(B, A) = \emptyset,$$

with composition determined by identities (so $f \circ \mathrm{id}_A = f$ and $\mathrm{id}_B \circ f = f$). Its *nerve* $N(\mathcal{C}) : \Delta^{\mathrm{op}} \to \mathbf{Set}$ is the simplicial set defined by:

- $N(\mathcal{C})_0 = \{A, B\}$ (objects of $\mathcal{C}$);
- $N(\mathcal{C})_1 = \{\mathrm{id}_A, \mathrm{id}_B, f\}$ (morphisms of $\mathcal{C}$);
- $N(\mathcal{C})_2$ consists of composable pairs of morphisms, namely

  $$N(\mathcal{C})_2 = \{(\mathrm{id}_A, \mathrm{id}_A),\ (\mathrm{id}_B, \mathrm{id}_B),\ (\mathrm{id}_A, f),\ (f, \mathrm{id}_B)\},$$

  and in general $N(\mathcal{C})_n$ is the set of composable strings of $n$ morphisms in $\mathcal{C}$.

The face maps $d_i$ compose or delete morphisms in the standard way, and the degeneracy maps $s_i$ insert identity morphisms. Therefore $N(\mathcal{C})$ is a simplicial set in the sense of Definition 3.2.1. An overview diagram of this example is provided in Fig. 3.2.

## 3.3 Cell complex

A cell complex builds a space by attaching disks of increasing dimension along boundaries, forming cells with characteristic attaching maps [354, 355, 356, 357, 358]. This can be regarded as one of the concepts for representing “higher-order” structure in a more geometric manner.

**Definition 3.3.1** (Cell and characteristic map)**.** For $n \geq 0$, let $D^n := \{x \in \mathbb{R}^n : \|x\| \leq 1\}$ be the closed $n$-disk and $S^{n-1} := \partial D^n$ its boundary. An *$n$-cell* is a subspace $e^n$ homeomorphic to the open disk $\mathring{D}^n$. A *characteristic map* for an $n$-cell $e^n \subset X$ is a continuous map $\phi : D^n \to X$ such that $\phi|_{\mathring{D}^n} : \mathring{D}^n \to e^n$ is a homeomorphism onto its image.

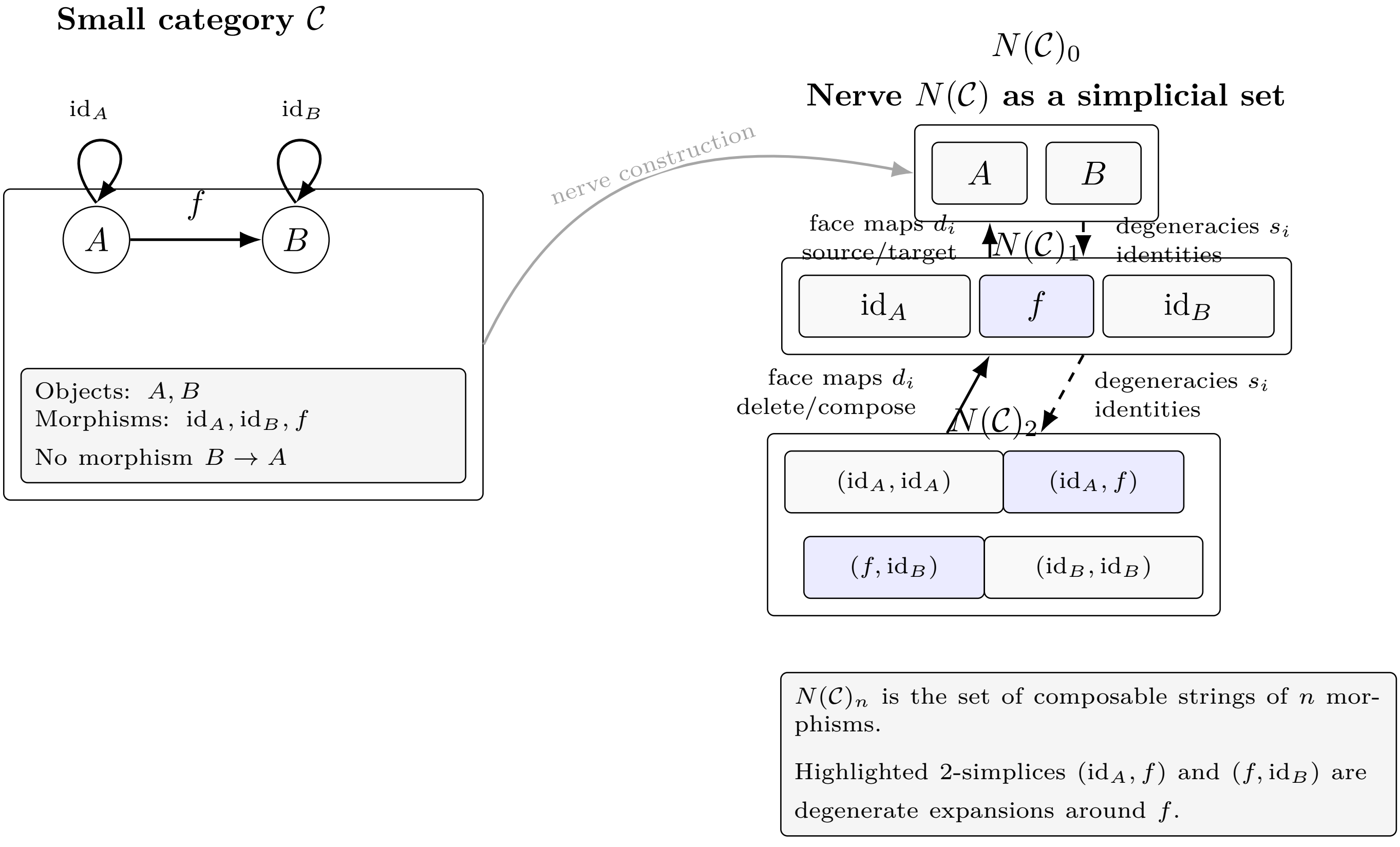

Figure 3.2: An illustration of the nerve $N(\mathcal{C})$ in Example 3.2.2.

**Example 3.3.2** (A 1-cell and its characteristic map)**.** Let $X = [0, 1] \subset \mathbb{R}$. The open interval $(0, 1) \subset X$ is a 1-cell

$$e^1 = (0, 1) \cong \mathring{D}^1,$$

where $D^1 = [-1, 1]$ and $\mathring{D}^1 = (-1, 1)$. Define a characteristic map $\phi : D^1 \to X$ by

$$\phi(t) := \frac{t+1}{2} \qquad (t \in [-1, 1]).$$

Then $\phi$ is continuous, and its restriction $\phi|_{\mathring{D}^1} : (-1, 1) \to (0, 1)$ is a homeomorphism onto $e^1$. Hence $\phi$ is a characteristic map for the 1-cell $e^1 \subset X$ in the sense of Definition 3.3.1.

**Definition 3.3.3** (Cell complex (cellular space))**.** A *cell complex* is a Hausdorff topological space $X$ together with:

1. an increasing filtration by subspaces

$$\emptyset = X^{-1} \subseteq X^0 \subseteq X^1 \subseteq \cdots \subseteq X,$$

$$X = \bigcup_{n \geq 0} X^n,$$

2. for each $n \geq 0$, a collection of $n$-cells $(e^n_\alpha)_{\alpha \in A_n}$ and characteristic maps $\phi^n_\alpha : D^n \to X^n$ such that

$$X^n = X^{n-1} \ \cup \ \bigcup_{\alpha \in A_n} \phi^n_\alpha(\mathring{D}^n),$$

   and the restriction $\phi^n_\alpha|_{\mathring{D}^n}$ is a homeomorphism onto $e^n_\alpha$,

3. (*attaching*) $\phi^n_\alpha(S^{n-1}) \subseteq X^{n-1}$ for all $\alpha \in A_n$.

The subspace $X^n$ is called the $n$-*skeleton* of $X$.

**Example 3.3.4** (A simple cell complex structure on the circle $S^1$)**.** Let $X = S^1 \subset \mathbb{R}^2$ be the unit circle. We construct a cell complex structure on $X$ with one 0-cell and one 1-cell. Let

$$X^0 = \{p\} \subset S^1$$

consist of a single point (a 0-cell $e^0 = \{p\}$), and let $e^1 = S^1 \setminus \{p\}$ be the open 1-cell. Choose a characteristic map $\phi^1 : D^1 \to S^1$ such that $\phi^1$ maps $S^0 = \partial D^1 = \{-1, 1\}$ to the attaching point $p$ and restricts to a homeomorphism $\mathring{D}^1 \to e^1$. For instance, writing $D^1 = [-1, 1]$, one may take

$$\phi^1(t) := \begin{cases} (\cos(\pi t), \sin(\pi t)), & t \in (-1, 1), \\ p = (1, 0), & t \in \{-1, 1\}. \end{cases}$$

Then

$$X^1 = X^0 \cup \phi^1(\mathring{D}^1) = S^1$$

, and

$$\phi^1(S^0) = \{p\} \subseteq X^0$$

. With the filtration $\emptyset = X^{-1} \subseteq X^0 \subseteq X^1 = X$, the space $S^1$ becomes a cell complex in the sense of Definition 3.3.3. An overview diagram of this example is provided in Fig. 3.3.

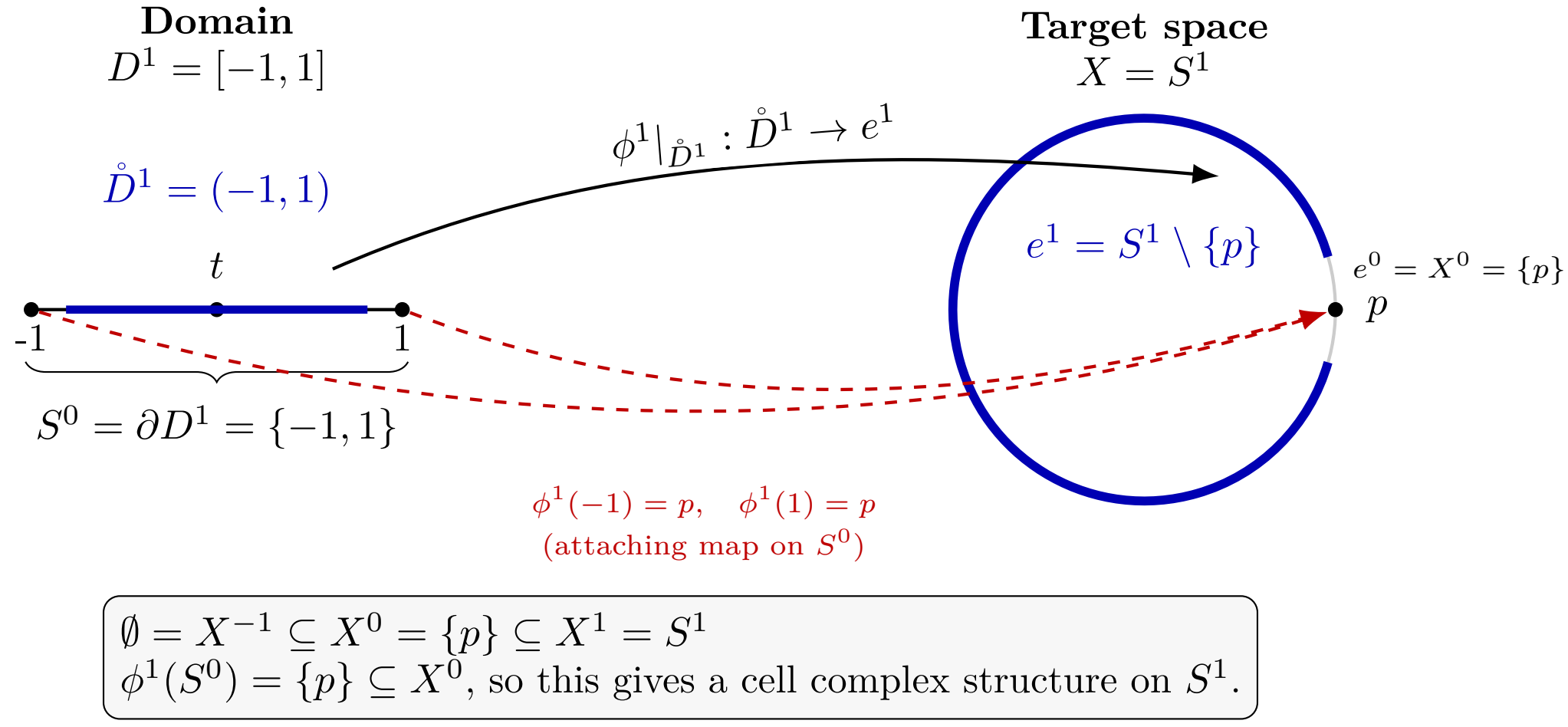

Figure 3.3: An illustration of the cell complex structure on $S^1$ with one 0-cell and one 1-cell (Example 3.3.4).

## 3.4 CW complex

A CW complex is a cell complex satisfying closure-finiteness and weak topology, enabling inductive construction and homotopy-friendly computations often efficient [359, 360]. This can be regarded as one of the concepts for representing "higher-order" structure in a more geometric manner.

**Definition 3.4.1** (CW complex)**.** [359, 360] A *CW complex* is a cell complex $X$ (Definition 3.3.3) that satisfies:

1. (*closure-finite*) for every cell $e^n_\alpha$, its closure $\overline{e^n_\alpha}$ intersects only finitely many cells of $X$;
2. (*weak topology*) a subset $U \subseteq X$ is open if and only if $U \cap \overline{e^n_\alpha}$ is open in $\overline{e^n_\alpha}$ for every cell $e^n_\alpha$.

**Example 3.4.2** (A CW complex structure on the 2-sphere $S^2$)**.** Let $X = S^2 \subset \mathbb{R}^3$ be the unit sphere. We describe a CW decomposition of $S^2$ with one 0-cell and one 2-cell.

- **0-skeleton.** Let $X^0 = \{p\}$ consist of a single point on $S^2$ (a 0-cell $e^0 = \{p\}$).
- **2-cell.** Let $e^2 = S^2 \setminus \{p\}$, which is homeomorphic to the open disk $\mathring{D}^2$ (e.g. via stereographic projection). Choose a characteristic map $\phi^2 : D^2 \to S^2$ such that $\phi^2|_{\mathring{D}^2} : \mathring{D}^2 \to e^2$ is a homeomorphism and $\phi^2(\partial D^2) \subseteq X^0$; that is, the entire boundary circle $\partial D^2 = S^1$ is attached to the single point $p$.

Then $X^2 = X^0 \cup \phi^2(\mathring{D}^2) = S^2$ and $\phi^2(\partial D^2) \subseteq X^0$, so $S^2$ is a cell complex in the sense of Definition 3.3.3. Moreover, the closure of the unique 2-cell is all of $S^2$, hence it intersects only finitely many cells (in fact, just the 2-cell and the 0-cell), and the topology on $S^2$ is the weak topology induced by the cell closures. Therefore $S^2$ is a CW complex in the sense of Definition 3.4.1. An overview diagram of this example is provided in Fig. 3.4.

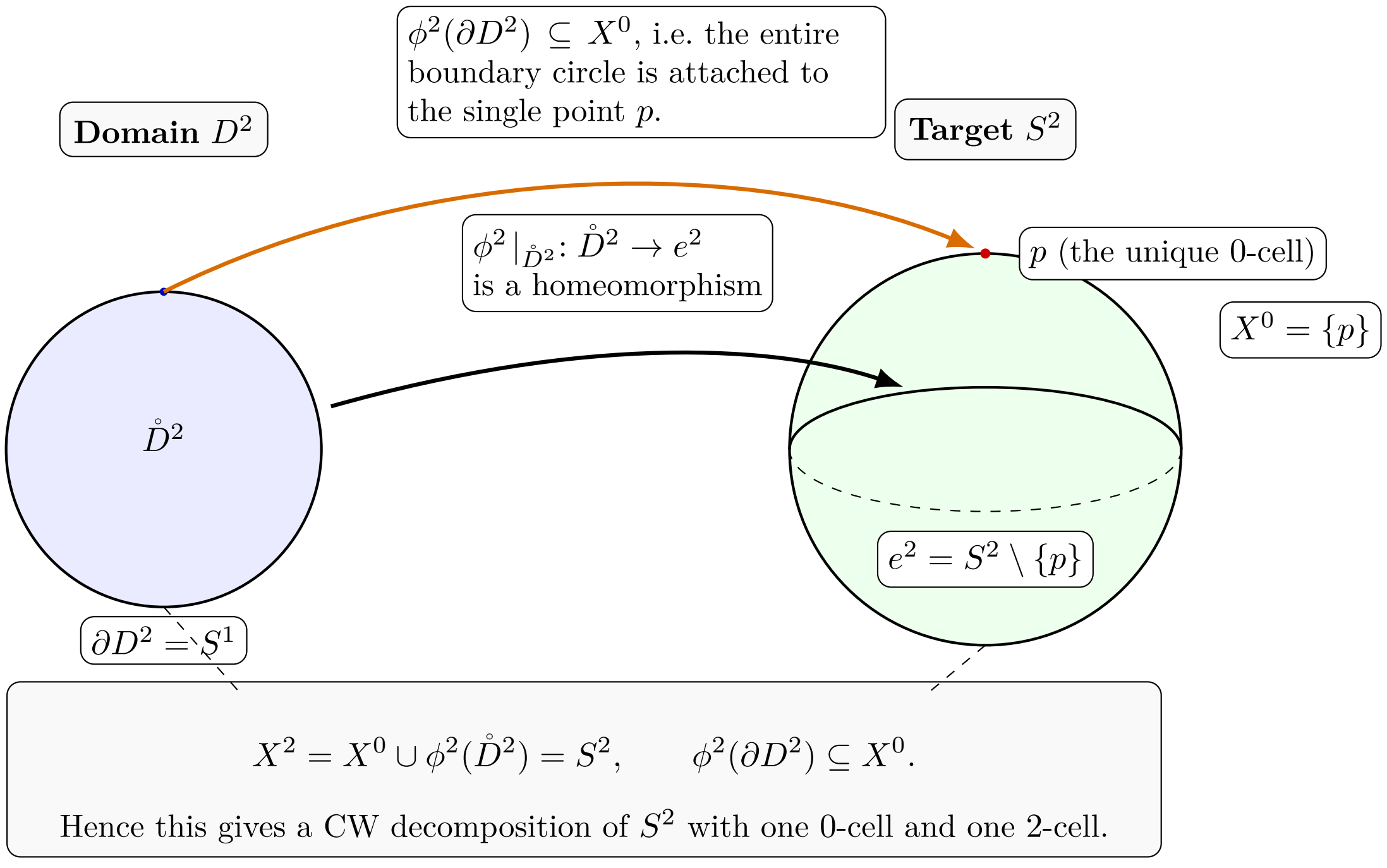

Figure 3.4: An illustration of the CW complex structure on $S^2$ in Example 3.4.2.

## 3.5 Polyhedral complex

A polyhedral complex is a finite collection of convex polytopes closed under faces, intersecting along common faces or empty [361, 362, 363]. This can be regarded as one of the concepts for representing "higher-order" structure in a more geometric manner.

**Definition 3.5.1** (Convex polytope and face)**.** A *(convex) polytope* in $\mathbb{R}^d$ is the convex hull $P = \mathrm{conv}(S)$ of a finite set $S \subset \mathbb{R}^d$. A *(proper) face* of $P$ is a subset of the form

$$F = P \cap \{x \in \mathbb{R}^d : \ell(x) = \max_{y \in P} \ell(y)\}$$

for some linear functional $\ell : \mathbb{R}^d \to \mathbb{R}$, with $F \neq \emptyset$; the whole polytope $P$ is also regarded as a face.

**Example 3.5.2** (A convex polytope and one of its faces)**.** Let $S = \{(0,0),(1,0),(1,1),(0,1)\} \subset \mathbb{R}^2$ and let

$$P = \mathrm{conv}(S) \subset \mathbb{R}^2,$$

which is the unit square. Consider the linear functional $\ell(x,y) = x$. Then

$$\max_{(u,v) \in P} \ell(u,v) = 1,$$

and the corresponding face is

$$F = P \cap \{(x,y) \in \mathbb{R}^2 \mid x = 1\} = \{(1,y) \mid 0 \leq y \leq 1\},$$

namely the right edge of the square. Thus $F$ is a (proper) face of $P$ in the sense of Definition 3.5.1.

**Definition 3.5.3** (Polyhedral complex)**.** A *polyhedral complex* $K$ in $\mathbb{R}^d$ is a (finite) collection of convex polytopes such that:

1. (*face-closure*) if $P \in K$ and $F$ is a face of $P$, then $F \in K$;
2. (*intersection property*) for any $P, Q \in K$, the intersection $P \cap Q$ is either empty or a face of both $P$ and $Q$.

The *underlying space* of $K$ is $|K| := \bigcup_{P \in K} P$.

**Example 3.5.4** (Two triangles forming a polyhedral complex)**.** Let $K$ be the collection of polytopes in $\mathbb{R}^2$ consisting of the two triangles

$$P_1 = \mathrm{conv}\{(0,0),(1,0),(0,1)\}, \qquad P_2 = \mathrm{conv}\{(1,0),(1,1),(0,1)\},$$

together with all of their faces (edges and vertices). Then:

- *Face-closure:* by construction, every face of $P_1$ or $P_2$ belongs to $K$.
- *Intersection property:* $P_1 \cap P_2 = \mathrm{conv}\{(1,0),(0,1)\}$, which is the common edge of the two triangles, hence a face of both $P_1$ and $P_2$.

Therefore $K$ is a polyhedral complex in the sense of Definition 3.5.3, and its underlying space is the unit square:

$$|K| = P_1 \cup P_2 = [0,1] \times [0,1].$$

An overview diagram of this example is provided in Fig. 3.5.

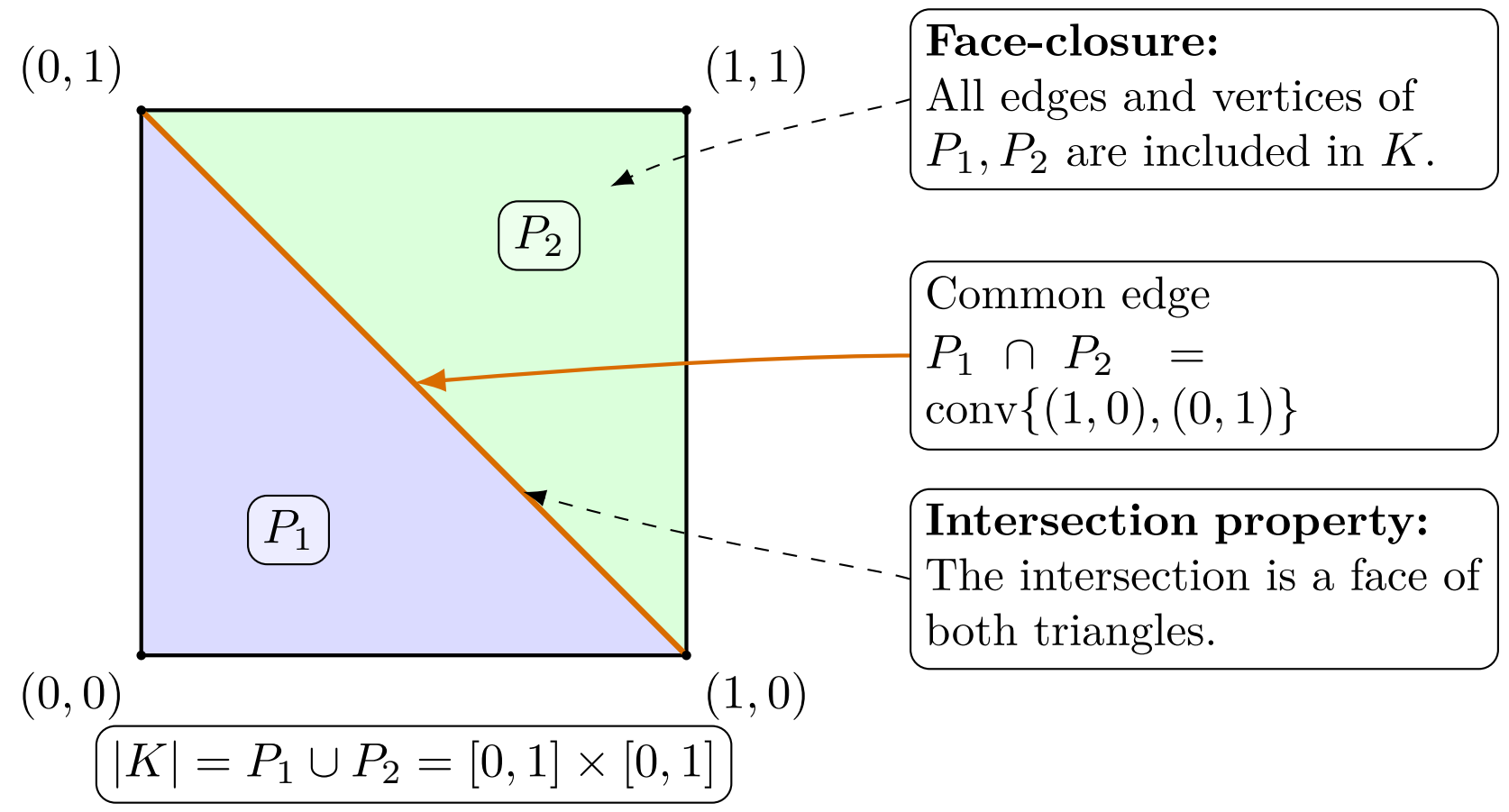

Figure 3.5: Two triangles forming a polyhedral complex (Example 3.5.4).

## 3.6 Dowker Complex

A Dowker complex constructs simplices from a binary relation, turning shared relational neighborhoods into topological higher order structure for analysis [364, 365, 366].

**Definition 3.6.1** (Dowker complex)**.** Let $X$ and $Y$ be finite nonempty sets, and let

$$R \subseteq X \times Y$$

be a binary relation. For each $x \in X$, define its $R$-neighborhood in $Y$ by

$$R(x) := \{\, y \in Y \mid (x,y) \in R \,\}.$$

The *Dowker complex on $X$ relative to $Y$ associated with $R$* is the abstract simplicial complex

$$\mathrm{D}_R(X,Y) := \Big\{\sigma \subseteq X \;\Big|\; \textstyle\bigcap_{x\in\sigma} R(x) \neq \varnothing\Big\}.$$

Equivalently,

$$\mathrm{D}_R(X,Y) = \Big\{\sigma \subseteq X \;\Big|\; \exists\, y \in Y \text{ such that } (x,y) \in R \text{ for all } x \in \sigma\Big\}.$$

Thus, a finite subset $\sigma \subseteq X$ is a simplex of $\mathrm{D}_R(X,Y)$ if and only if there exists an element of $Y$ that is related, via $R$, to every vertex of $\sigma$.

**Example 3.6.2** (A concrete Dowker complex)**.** Let

$$X = \{x_1, x_2, x_3\}, \qquad Y = \{a, b\},$$

and define a binary relation

$$R \subseteq X \times Y$$

by

$$R = \{(x_1, a), (x_2, a), (x_2, b), (x_3, b)\}.$$

Then the $R$-neighborhoods of the elements of $X$ are

$$R(x_1) = \{a\}, \qquad R(x_2) = \{a, b\}, \qquad R(x_3) = \{b\}.$$

The Dowker complex on $X$ relative to $Y$ associated with $R$ is

$$\mathrm{D}_R(X, Y) = \{\sigma \subseteq X \mid \bigcap_{x \in \sigma} R(x) \neq \varnothing\}.$$

We now determine its simplices.
For the singletons, we have

$$R(x_1) \neq \varnothing, \qquad R(x_2) \neq \varnothing, \qquad R(x_3) \neq \varnothing,$$

so

$$\{x_1\}, \{x_2\}, \{x_3\} \in \mathrm{D}_R(X, Y).$$

For the 2-element subsets,

$$R(x_1) \cap R(x_2) = \{a\} \neq \varnothing,$$

hence

$$\{x_1, x_2\} \in \mathrm{D}_R(X, Y).$$

Also,

$$R(x_2) \cap R(x_3) = \{b\} \neq \varnothing,$$

so

$$\{x_2, x_3\} \in \mathrm{D}_R(X, Y).$$

However,

$$R(x_1) \cap R(x_3) = \{a\} \cap \{b\} = \varnothing,$$

thus

$$\{x_1, x_3\} \notin \mathrm{D}_R(X, Y).$$

For the 3-element subset,

$$R(x_1) \cap R(x_2) \cap R(x_3) = \{a\} \cap \{a, b\} \cap \{b\} = \varnothing,$$

and therefore

$$\{x_1, x_2, x_3\} \notin \mathrm{D}_R(X, Y).$$

Consequently,

$$\mathrm{D}_R(X, Y) = \{\varnothing, \{x_1\}, \{x_2\}, \{x_3\}, \{x_1, x_2\}, \{x_2, x_3\}\}.$$

Hence $\mathrm{D}_R(X, Y)$ is an abstract simplicial complex consisting of three vertices and two 1-simplices, namely $\{x_1, x_2\}$ and $\{x_2, x_3\}$. Geometrically, it is a path on the vertex set $\{x_1, x_2, x_3\}$.

**Proposition 3.6.3.** *The family $\mathrm{D}_R(X, Y)$ is an abstract simplicial complex on the vertex set $X$.*

*Proof.* Let $\sigma \in \mathrm{D}_R(X, Y)$. Then there exists $y \in Y$ such that

$$(x, y) \in R \qquad \text{for all } x \in \sigma.$$

If $\tau \subseteq \sigma$, then the same element $y$ also satisfies

$$(x, y) \in R \qquad \text{for all } x \in \tau.$$

Hence

$$\tau \in \mathrm{D}_R(X, Y).$$

Therefore $\mathrm{D}_R(X, Y)$ is downward closed under inclusion, so it is an abstract simplicial complex. □

## 3.7 Cubical Complex

A cubical complex is a cell complex built from cubes and their faces, naturally modeling grid like higher dimensional adjacency [367, 368, 369]. As a related concept of Cubical Complex, HyperCubical Complex [370] is also known.

**Definition 3.7.1** (Elementary interval and elementary cube)**.** Let $m \in \mathbb{N}$. An *elementary interval* in $\mathbb{R}$ is either

$$[a, a+1] \quad \text{or} \quad [a, a]$$

for some $a \in \mathbb{Z}$. An *elementary cube* in $\mathbb{R}^m$ is a Cartesian product

$$Q = I_1 \times I_2 \times \cdots \times I_m,$$

where each $I_j$ is an elementary interval.

The *dimension* of $Q$ is the number of factors $I_j$ of the form $[a, a+1]$, and is denoted by

$$\dim(Q).$$

**Definition 3.7.2** (Face of an elementary cube)**.** Let

$$Q = I_1 \times \cdots \times I_m \subseteq \mathbb{R}^m$$

be an elementary cube. A *face* of $Q$ is any elementary cube

$$F = J_1 \times \cdots \times J_m$$

such that, for each $j \in \{1, \ldots, m\}$, either $J_j = I_j$, or $J_j = [a, a]$ is one of the endpoints of $I_j = [a, a+1]$ whenever $I_j$ is nondegenerate.

**Definition 3.7.3** (Cubical complex)**.** A *cubical complex* is a finite collection $\mathcal{K}$ of elementary cubes in some $\mathbb{R}^m$ such that:

1. if $Q \in \mathcal{K}$ and $F$ is a face of $Q$, then $F \in \mathcal{K}$;
2. if $Q_1, Q_2 \in \mathcal{K}$, then

$$Q_1 \cap Q_2$$

   is either empty or a face of both $Q_1$ and $Q_2$.

The elements of $\mathcal{K}$ are called *cubes* (or *cells*) of the complex. The *dimension* of $\mathcal{K}$ is defined by

$$\dim(\mathcal{K}) := \max\{\dim(Q) \mid Q \in \mathcal{K}\}.$$

A cube of maximal dimension is called a *facet* of $\mathcal{K}$.

**Remark 3.7.4.** *Equivalently, a cubical complex may be viewed as a finite polytopal complex (or regular CW complex) whose cells are combinatorially isomorphic to cubes. In particular, the standard $n$-cube*

$$Q^n = [0, 1]^n$$

*and its $k$-faces provide the basic local model for cubical complexes.*

**Example 3.7.5** (Two adjacent unit squares as a cubical complex)**.** Consider the following two elementary 2-cubes in $\mathbb{R}^2$:

$$Q_1 = [0, 1] \times [0, 1], \qquad Q_2 = [1, 2] \times [0, 1].$$

These are two unit squares sharing the common vertical edge

$$Q_1 \cap Q_2 = \{1\} \times [0, 1].$$

Let $\mathcal{K}$ be the collection consisting of $Q_1$, $Q_2$, and all of their faces; that is, $\mathcal{K}$ contains the two 2-cubes, all of their 1-dimensional faces, and all of their 0-dimensional faces.

Then $\mathcal{K}$ is a cubical complex. Indeed:

1. every face of $Q_1$ and $Q_2$ belongs to $\mathcal{K}$, so $\mathcal{K}$ is closed under taking faces;
2. the intersection of any two cubes in $\mathcal{K}$ is either empty or a face of both cubes. In particular,

$$Q_1 \cap Q_2 = \{1\} \times [0, 1]$$

   is a 1-dimensional face of both $Q_1$ and $Q_2$.

Therefore $\mathcal{K}$ is a 2-dimensional cubical complex. Its underlying space is

$$|\mathcal{K}| = [0, 2] \times [0, 1].$$

An illustration is given in Fig. 3.6.

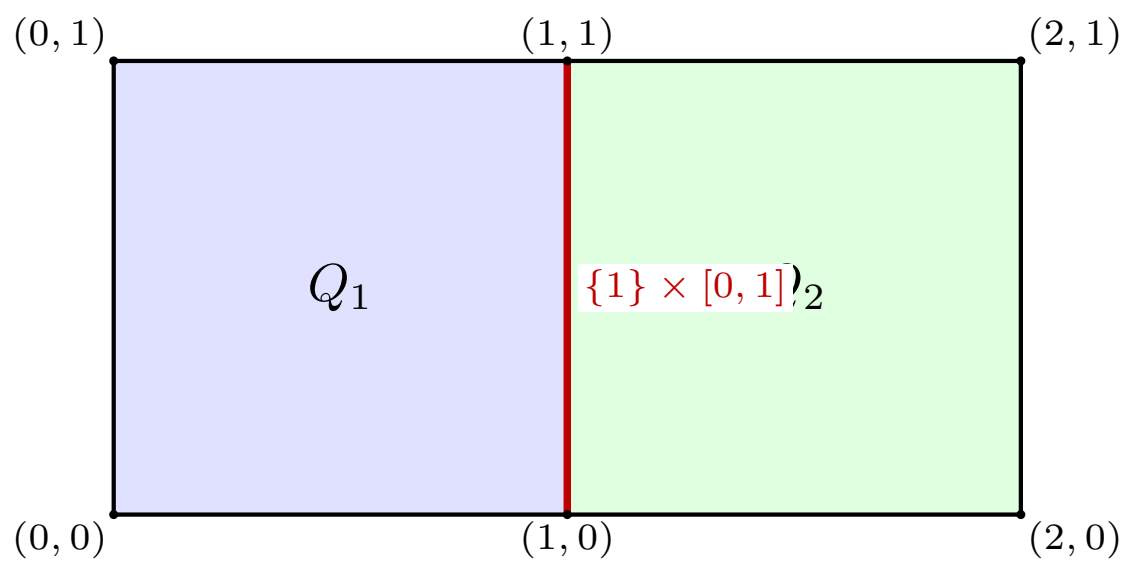

**Two 2-cubes:**

$Q_1 = [0,1]\times[0,1], \qquad Q_2 = [1,2]\times[0,1].$

**Common face:**

$$Q_1 \cap Q_2 = \{1\} \times [0,1].$$

This is a 1-dimensional face of both cubes.

$$|\mathcal{K}| = [0,2] \times [0,1].$$

The complex consists of the two squares together with all of their edges and vertices.

Figure 3.6: A 2-dimensional cubical complex formed by two adjacent unit squares. The red segment is the common 1-face $Q_1 \cap Q_2 = \{1\} \times [0,1]$.

## 3.8 Path Complex

A path complex is a collection of vertex sequences closed under endpoint truncation, generalizing simplicial complexes and directed graphs homologically [371, 372, 373]. In addition to this concept, other related notions are also known, for example, the Flag Complex[374, 375] and the Directed Flag Complex[376, 377, 378].

**Definition 3.8.1** (Elementary path)**.** Let $V$ be a finite nonempty set. For an integer $n \geq 0$, an *elementary $n$-path* on $V$ is a finite sequence

$$i_0 i_1 \cdots i_n \qquad (i_k \in V).$$

For $n = -1$, we denote by $e$ the empty path.

**Definition 3.8.2** (Path complex)**.** Let $V$ be a finite nonempty set. A *path complex* on $V$ is a family

$$P = \{P_n\}_{n \geq -1},$$

where each $P_n$ is a set of elementary $n$-paths on $V$, satisfying:

1. $P_{-1} = \{e\}$;
2. if

   $$i_0 i_1 \cdots i_n \in P_n \qquad (n \geq 0),$$

   then both truncated paths

   $$i_0 i_1 \cdots i_{n-1} \in P_{n-1}, \qquad i_1 i_2 \cdots i_n \in P_{n-1}$$

   also belong to the family.

Elements of $P_n$ are called *allowed $n$-paths.*

**Remark 3.8.3.** *Thus, a path complex is a collection of elementary paths closed under the two natural truncation operations: deleting the last vertex and deleting the first vertex.*

**Example 3.8.4** (A path complex generated by a directed graph)**.** Let

$$V = \{a, b, c, d\},$$

and consider the directed graph

$$G = (V, E),$$

where

$$E = \{(a,b), (a,c), (b,c), (b,d), (c,d)\}.$$

Equivalently, the directed edges are

$$a \to b, \qquad a \to c, \qquad b \to c, \qquad b \to d, \qquad c \to d.$$

Let $P(G)$ denote the path complex generated by $G$. By definition, an elementary path

$$i_0 i_1 \cdots i_n$$

belongs to $P_n(G)$ if and only if each consecutive pair forms a directed edge of $G$, that is,

$$(i_{k-1}, i_k) \in E \qquad (k = 1, \ldots, n).$$

Hence:

$$P_0(G) = \{a, b, c, d\},$$

$$P_1(G) = \{ab, ac, bc, bd, cd\},$$

$$P_2(G) = \{abc, abd, acd, bcd\},$$

and

$$P_3(G) = \{abcd\}.$$

There are no allowed paths of length $> 3$ in this example.

Indeed,

$$abcd \in P_3(G)$$

because

$$a \to b, \qquad b \to c, \qquad c \to d$$

are all edges of $G$. Moreover, its truncations

$$abc, \quad bcd \in P_2(G),$$

and the truncations of these 2-paths again belong to $P_1(G)$. Therefore the truncation axiom is satisfied.

An illustration of this path complex is shown in Fig. 3.7.

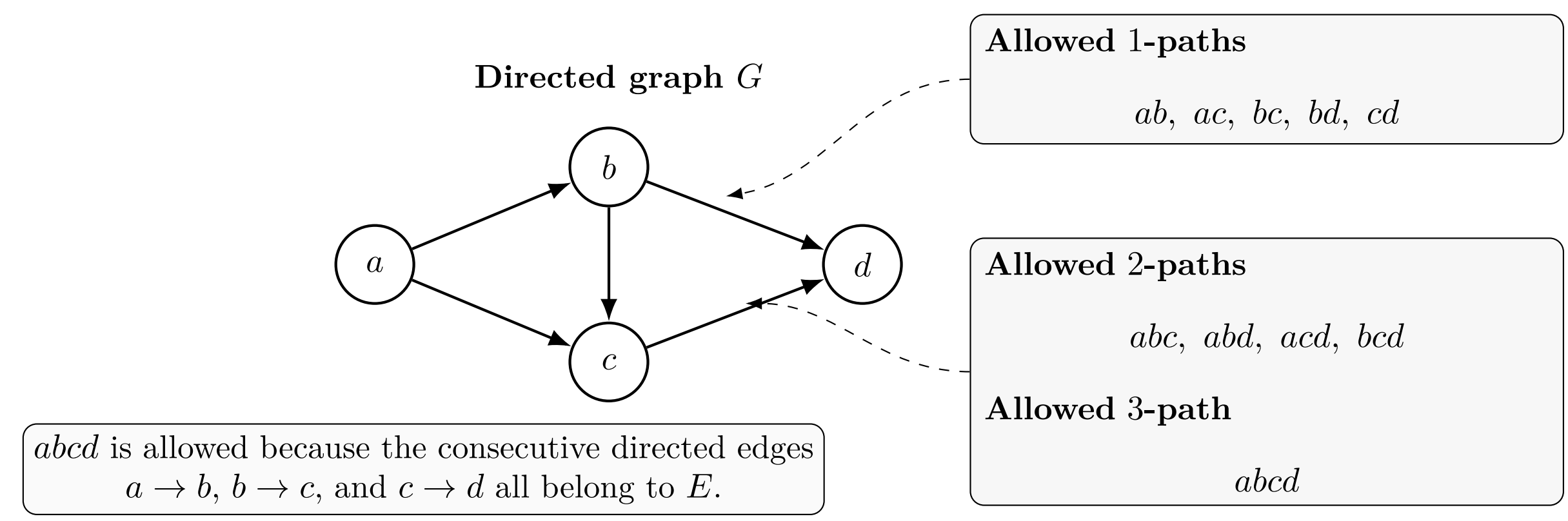

Figure 3.7: A directed graph generating a path complex. Allowed higher-order paths are determined by directed consecutive edges.

## 3.9 Cellular Sheaf

A cellular sheaf assigns data to cells and restriction maps to incidences, encoding local global consistency on graphs and complexes [379, 380, 381].

**Definition 3.9.1** (Cellular sheaf)**.** Let $X$ be a finite regular cell complex, and let $P_X$ denote its face poset, viewed as a category in which there is a unique morphism

$$\sigma \longrightarrow \tau \qquad \text{whenever} \qquad \sigma \leq \tau.$$

Let $\Bbbk$ be a field. A *cellular sheaf of* $\Bbbk$*-vector spaces* on $X$ is a covariant functor

$$\mathcal{F} : P_X \longrightarrow \mathrm{Vect}_{\Bbbk}.$$

Equivalently, a cellular sheaf $\mathcal{F}$ consists of:

1. a $\Bbbk$-vector space $\mathcal{F}(\sigma)$ for each cell $\sigma \in X$, called the *stalk* of $\mathcal{F}$ over $\sigma$;
2. for each incidence relation $\sigma \leq \tau$, a linear map

$$\mathcal{F}_{\sigma,\tau} : \mathcal{F}(\sigma) \to \mathcal{F}(\tau),$$

called the *restriction map*;

such that

$$\mathcal{F}_{\sigma,\sigma} = \mathrm{id}_{\mathcal{F}(\sigma)}$$

for every cell $\sigma$, and

$$\mathcal{F}_{\tau,\eta} \circ \mathcal{F}_{\sigma,\tau} = \mathcal{F}_{\sigma,\eta}$$

whenever $\sigma \leq \tau \leq \eta$.

**Definition 3.9.2** (Global section)**.** Let $\mathcal{F}$ be a cellular sheaf on $X$. A *global section* of $\mathcal{F}$ is a family

$$s = (s_\sigma)_{\sigma \in X}, \qquad s_\sigma \in \mathcal{F}(\sigma),$$

such that for every incidence relation $\sigma \leq \tau$,

$$\mathcal{F}_{\sigma,\tau}(s_\sigma) = s_\tau.$$

The vector space of global sections is denoted by

$$\Gamma(X; \mathcal{F}).$$

**Example 3.9.3** (A cellular sheaf on a graph)**.** Let $X$ be the 1-dimensional regular cell complex consisting of two vertices

$$u,\ v$$

and one edge

$$e$$

with incidence relations

$$u \leq e, \qquad v \leq e.$$

Thus $X$ is the cell complex associated with the interval joining $u$ and $v$.

Define a cellular sheaf $\mathcal{F} : X \to \mathrm{Vect}_{\mathbb{R}}$ by

$$\mathcal{F}(u) = \mathbb{R}^2, \qquad \mathcal{F}(v) = \mathbb{R}^2, \qquad \mathcal{F}(e) = \mathbb{R}.$$

Define the restriction maps by

$$\mathcal{F}_{u,e}(x_1, x_2) = x_1 + x_2, \qquad \mathcal{F}_{v,e}(y_1, y_2) = y_1 - y_2.$$

Since the only non-identity face relations are $u \leq e$ and $v \leq e$, the functorial conditions are automatically satisfied, and hence $\mathcal{F}$ is a cellular sheaf.

A global section of $\mathcal{F}$ is a triple

$$\big((x_1, x_2), (y_1, y_2), z\big) \in \mathbb{R}^2 \times \mathbb{R}^2 \times \mathbb{R}$$

such that

$$z = \mathcal{F}_{u,e}(x_1, x_2) = x_1 + x_2$$

and

$$z = \mathcal{F}_{v,e}(y_1, y_2) = y_1 - y_2.$$

Therefore,

$$\Gamma(X; \mathcal{F}) = \left\{ \big((x_1, x_2), (y_1, y_2), z\big) \in \mathbb{R}^2 \times \mathbb{R}^2 \times \mathbb{R} \;\middle|\; x_1 + x_2 = z, \quad y_1 - y_2 = z \right\}.$$

Eliminating $z$, we may also write

$$\Gamma(X; \mathcal{F}) \cong \left\{ (x_1, x_2, y_1, y_2) \in \mathbb{R}^4 \;\middle|\; x_1 + x_2 = y_1 - y_2 \right\}.$$

Hence the space of global sections is a 3-dimensional linear subspace of $\mathbb{R}^4$.

An illustration of this sheaf is given in Fig. 3.8.

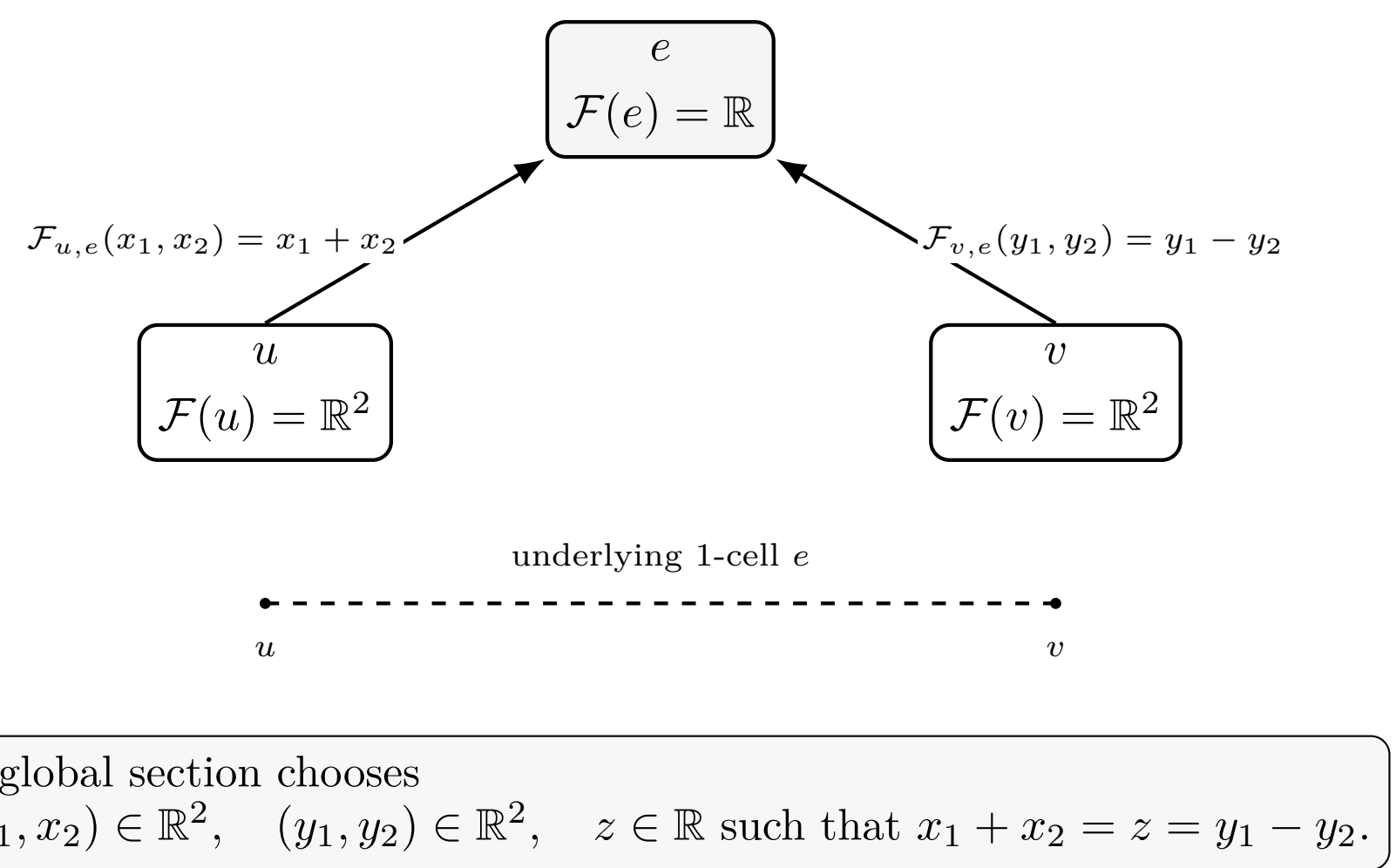

Figure 3.8: A cellular sheaf on a graph with two vertices and one edge. The edge stalk stores a shared scalar compatibility value determined by linear maps from the vertex stalks.

## 3.10 Meta Simplicial Complex

We introduce a higher-order structure whose vertices are themselves simplicial complexes. This may be viewed as a *simplicial complex of simplicial complexes.*

**Definition 3.10.1** (Universe of finite abstract simplicial complexes)**.** Let $X$ be a finite nonempty set. Define

$$\mathrm{Simp}(X) := \left\{ K \subseteq \mathcal{P}(X) \setminus \{\varnothing\} \;\middle|\; \begin{array}{l} K \neq \varnothing, \text{ and whenever } \sigma \in K \\ \text{and } \varnothing \neq \tau \subseteq \sigma, \text{ then } \tau \in K \end{array} \right\}.$$

Thus $\mathrm{Simp}(X)$ is the family of all finite abstract simplicial complexes on the base set $X$, where we use the convention that simplices are nonempty subsets of $X$.

**Definition 3.10.2** (Meta simplicial complex)**.** Let $X$ be a finite nonempty set, and let

$$\mathcal{V} \subseteq \mathrm{Simp}(X)$$

be a finite nonempty set of simplicial complexes on $X$. A *Meta Simplicial Complex* over $X$ is a pair

$$\mathfrak{M} = (\mathcal{V}, \mathcal{K}),$$

where

$$\mathcal{K} \subseteq \mathcal{P}(\mathcal{V}) \setminus \{\varnothing\}$$

satisfies the downward-closure condition:

$$A \in \mathcal{K}, \; \varnothing \neq B \subseteq A \quad \Longrightarrow \quad B \in \mathcal{K}.$$

The elements of $\mathcal{V}$ are called *meta-vertices*, and each meta-vertex is itself an abstract simplicial complex on $X$. The elements of $\mathcal{K}$ are called *meta-simplices.* If $A \in \mathcal{K}$ and $|A| = r + 1$, then $A$ is called an *$r$-dimensional meta-simplex.* The dimension of $\mathfrak{M}$ is defined by

$$\dim(\mathfrak{M}) := \max\{\, |A| - 1 \mid A \in \mathcal{K} \,\}.$$

**Remark 3.10.3.** *A Meta Simplicial Complex is an ordinary abstract simplicial complex whose vertices are not atomic objects, but rather simplicial complexes. Hence it is naturally interpreted as a* simplicial complex of simplicial complexes*.*

**Example 3.10.4** (A concrete Meta Simplicial Complex)**.** Let

$$X = \{a, b, c, d\}.$$

Define three simplicial complexes on $X$:

$$K_1 = \big\{\{a\}, \{b\}, \{a, b\}\big\},$$

$$K_2 = \{\{b\}, \{c\}, \{b, c\}\},$$
$$K_3 = \{\{c\}, \{d\}, \{c, d\}\}.$$

Each $K_i$ is an abstract simplicial complex on $X$, so

$$K_1, K_2, K_3 \in \mathrm{Simp}(X).$$

Let

$$\mathcal{V} = \{K_1, K_2, K_3\}.$$

Define

$$\mathcal{K} = \Big\{\{K_1\}, \{K_2\}, \{K_3\}, \{K_1, K_2\}, \{K_2, K_3\}, \{K_1, K_2, K_3\}\Big\}.$$

Then $\mathcal{K} \subseteq \mathcal{P}(\mathcal{V}) \setminus \{\varnothing\}$, and $\mathcal{K}$ is downward closed with respect to nonempty inclusion. Hence

$$\mathfrak{M} = (\mathcal{V}, \mathcal{K})$$

is a Meta Simplicial Complex over $X$. In this example, $\{K_1, K_2, K_3\}$ is a 2-dimensional meta-simplex.

**Theorem 3.10.5** (Well-definedness of Meta Simplicial Complexes)**.** *Let $X$ be a finite nonempty set. Then:*

1. $\mathrm{Simp}(X)$ *is a well-defined finite set;*
2. *if $\mathcal{V} \subseteq \mathrm{Simp}(X)$ is finite and nonempty, and*

   $$\mathcal{K} \subseteq \mathcal{P}(\mathcal{V}) \setminus \{\varnothing\}$$

   *is downward closed under nonempty inclusion, then $\mathcal{K}$ is an abstract simplicial complex on the vertex set $\mathcal{V}$;*
3. *consequently, every pair $\mathfrak{M} = (\mathcal{V}, \mathcal{K})$ satisfying Definition 3.10.2 is a well-defined mathematical object.*

*Proof.* We prove the assertions one by one.

**(1)** $\mathrm{Simp}(X)$ **is a well-defined finite set.** Since $X$ is finite, its power set $\mathcal{P}(X)$ is finite. Hence

$$\mathcal{P}(X) \setminus \{\varnothing\}$$

is finite, and therefore

$$\mathcal{P}\big(\mathcal{P}(X) \setminus \{\varnothing\}\big)$$

is also finite. By Definition 3.10.1, $\mathrm{Simp}(X)$ is the subclass of

$$\mathcal{P}\big(\mathcal{P}(X) \setminus \{\varnothing\}\big)$$

consisting of those families that satisfy the simplicial downward-closure condition. Thus $\mathrm{Simp}(X)$ is a well-defined subset of a finite set, and hence it is itself finite.

**(2)** $\mathcal{K}$ **is an abstract simplicial complex on** $\mathcal{V}$**.** By assumption,

$$\mathcal{K} \subseteq \mathcal{P}(\mathcal{V}) \setminus \{\varnothing\}.$$

Thus every element of $\mathcal{K}$ is a nonempty finite subset of $\mathcal{V}$. Moreover, $\mathcal{K}$ is assumed to be downward closed under nonempty inclusion: if $A \in \mathcal{K}$ and $\varnothing \neq B \subseteq A$, then $B \in \mathcal{K}$. This is exactly the defining axiom of an abstract simplicial complex when simplices are taken to be nonempty subsets. Therefore $\mathcal{K}$ is an abstract simplicial complex on vertex set $\mathcal{V}$.

**(3) Well-definedness of** $\mathfrak{M} = (\mathcal{V}, \mathcal{K})$**.** By part (1), the meta-vertex set $\mathcal{V} \subseteq \mathrm{Simp}(X)$ is drawn from a well-defined finite ambient set. By part (2), the family $\mathcal{K}$ is an abstract simplicial complex on $\mathcal{V}$. Hence the pair

$$\mathfrak{M} = (\mathcal{V}, \mathcal{K})$$

is well-defined. □

**Proposition 3.10.6** (Ordinary simplicial complexes are recovered as a special case)**.** *Let $X$ be a finite nonempty set, and let $\mathcal{V} \subseteq \mathrm{Simp}(X)$ be any finite nonempty set. Then every abstract simplicial complex on the vertex set $\mathcal{V}$ determines a Meta Simplicial Complex over $X$.*

*Proof.* Let $\mathcal{K} \subseteq \mathcal{P}(\mathcal{V}) \setminus \{\varnothing\}$ be any abstract simplicial complex on $\mathcal{V}$. By definition, $\mathcal{K}$ is downward closed under nonempty inclusion. Therefore the pair $(\mathcal{V}, \mathcal{K})$ satisfies Definition 3.10.2, and hence is a Meta Simplicial Complex over $X$. □

## 3.11 Simplicial SuperHypercomplex

A superhypercomplex is a simplicial complex whose vertices are nested-set supervertices, capturing higher-order incidence across hierarchical elements.

**Definition 3.11.1** ($n$-SuperHypercomplex)**.** Let $V_0$ be a nonempty base set and fix an integer $n \in \mathbb{N}_0$. Define the iterated powersets recursively by

$$\mathcal{P}^0(V_0) := V_0, \qquad \mathcal{P}^{k+1}(V_0) := \mathcal{P}\big(\mathcal{P}^k(V_0)\big) \quad (k \geq 0).$$

Choose a set of *$n$-supervertices*

$$V \subseteq \mathcal{P}^n(V_0).$$

An *$n$-SuperHypercomplex* on $V_0$ (with supervertex set $V$) is a pair

$$\mathrm{SuHyC}^{(n)} = (V, \mathcal{K}),$$

where $\mathcal{K} \subseteq \mathcal{P}(V) \setminus \{\emptyset\}$ is a family of nonempty finite subsets of $V$ satisfying the *downward-closure* (face) condition:

$$\sigma \in \mathcal{K},\ \emptyset \neq \tau \subseteq \sigma \implies \tau \in \mathcal{K}.$$

Elements of $\mathcal{K}$ are called *$n$-superfaces* (or *super-simplices*). If $|\sigma| = k+1$, then $\sigma$ is a *$k$-dimensional superface* and we set $\dim(\sigma) := k$. The *dimension* of $\mathrm{SuHyC}^{(n)}$ is

$$\dim\big(\mathrm{SuHyC}^{(n)}\big) := \sup\{\dim(\sigma) \mid \sigma \in \mathcal{K}\}.$$

**Example 3.11.2** (A 1-SuperHypercomplex built from teams of individuals)**.** Let the base set be

$$V_0 = \{a, b, c, d\},$$

and take $n = 1$, so $\mathcal{P}^1(V_0) = \mathcal{P}(V_0)$. Interpret 1-supervertices as *teams* (nonempty subsets of $V_0$), and choose

$$V = \{\ v_1 = \{a, b\},\ v_2 = \{b, c\},\ v_3 = \{c, d\}\ \} \subseteq \mathcal{P}(V_0).$$

Define a family $\mathcal{K} \subseteq \mathcal{P}(V) \setminus \{\emptyset\}$ by taking one 2-dimensional superface (a super-triangle) and all of its nonempty faces:

$$\sigma = \{v_1, v_2, v_3\} \in \mathcal{P}(V).$$

Set

$$\mathcal{K} = \Big\{\{v_1\}, \{v_2\}, \{v_3\}, \{v_1, v_2\}, \{v_1, v_3\}, \{v_2, v_3\}, \{v_1, v_2, v_3\}\Big\}.$$

By construction, $\mathcal{K}$ is downward closed: whenever $\tau$ is a nonempty subset of $\sigma$, we have $\tau \in \mathcal{K}$. Hence

$$\mathrm{SuHyC}^{(1)} = (V, \mathcal{K})$$

is a 1-SuperHypercomplex in the sense of Definition 3.11.1. Its dimension is 2, since it contains the 2-dimensional superface $\{v_1, v_2, v_3\}$.

The superface $\{v_1, v_2, v_3\}$ encodes a higher-order incidence among the three teams. At the base level, the union of underlying individuals involved is

$$v_1 \cup v_2 \cup v_3 = \{a, b, c, d\},$$

so the super-triangle can be viewed as a "team-of-teams" interaction that collectively spans all four individuals. An overview diagram of this example is provided in Fig. 3.9.

## 3.12 Combinatorial Complexes

A Combinatorial Complex combines the arbitrary set-type relations of a hypergraph with an order-preserving rank function [382, 383, 384]. Unlike a simplicial complex, it does not require every subset of a cell to be present.

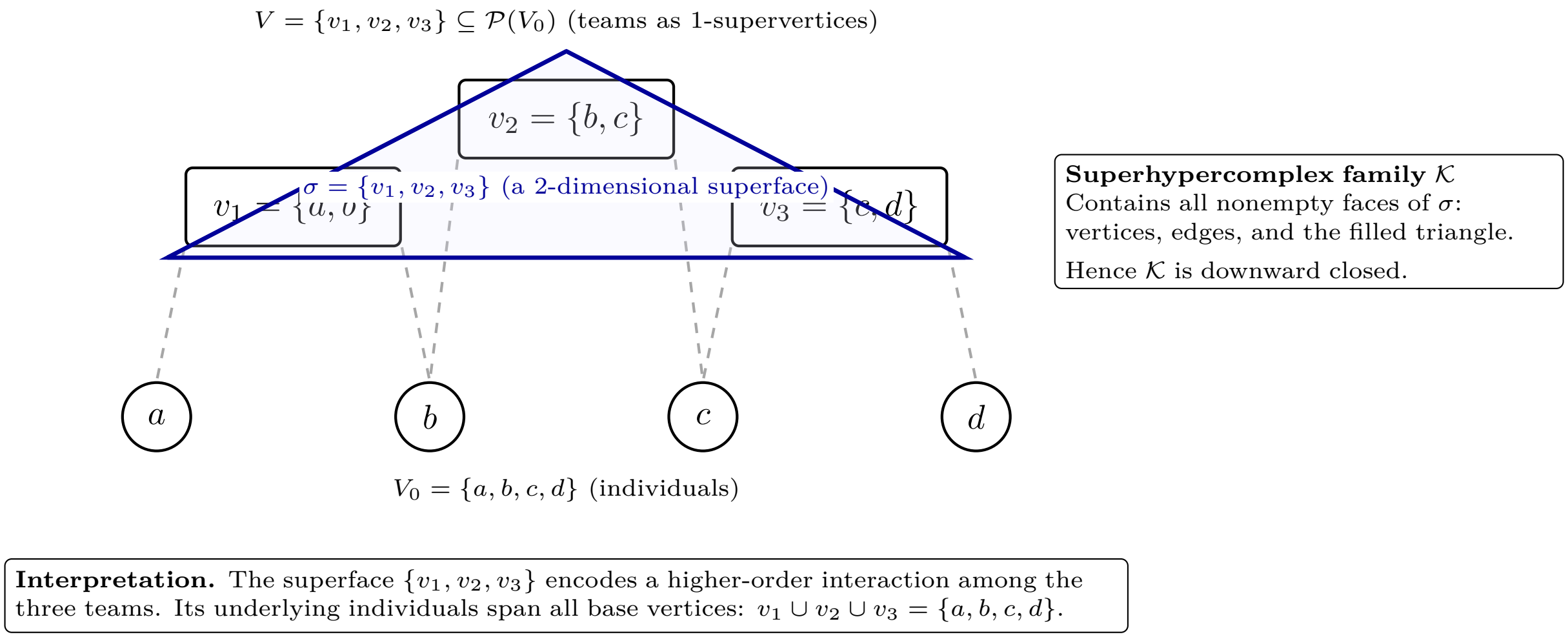

Figure 3.9: A 1-SuperHypercomplex built from teams of individuals (Example 3.11.2). The filled super-triangle represents the 2-dimensional superface $\sigma = \{v_1, v_2, v_3\}$, and dashed links indicate team membership at the base level.

**Definition 3.12.1** (Rank function)**.** Let $S$ be a set and let

$$\mathcal{X} \subseteq \mathcal{P}(S) \setminus \{\emptyset\}.$$

A function

$$\mathrm{rk} : \mathcal{X} \longrightarrow \mathbb{Z}_{\geq 0}$$

is called a *rank function* if it is order-preserving with respect to set inclusion; that is,

$$x \subseteq y \quad \Longrightarrow \quad \mathrm{rk}(x) \leq \mathrm{rk}(y)$$

for all $x, y \in \mathcal{X}$.

**Definition 3.12.2** (Combinatorial Complex)**.** [382] A *Combinatorial Complex*, abbreviated as *CC*, is a triple

$$\mathcal{K} = \big(S, \mathcal{X}, \mathrm{rk}\big),$$

where:

1. $S$ is a set whose elements are called *entities* or *vertices*;
2. $$\mathcal{X} \subseteq \mathcal{P}(S) \setminus \{\emptyset\};$$
3. every singleton is a cell:
$$\{s\} \in \mathcal{X} \qquad \text{for every } s \in S;$$
4. $$\mathrm{rk} : \mathcal{X} \longrightarrow \mathbb{Z}_{\geq 0}$$
is order-preserving:
$$x \subseteq y \quad \Longrightarrow \quad \mathrm{rk}(x) \leq \mathrm{rk}(y)$$
for all $x, y \in \mathcal{X}$.

The elements of $\mathcal{X}$ are called *cells* or *relations*. A cell $x \in \mathcal{X}$ satisfying

$$\mathrm{rk}(x) = k$$

is called a *$k$-cell*.

**Remark 3.12.3.** *It is customary, although not part of the minimal definition, to require*

$$\mathrm{rk}(\{s\}) = 0 \qquad \text{for every } s \in S.$$

*This convention identifies the singleton cells with the* 0*-cells of the Combinatorial Complex.*

**Definition 3.12.4** (Dimension and skeleton)**.** Let

$$\mathcal{K} = \big(S, \mathcal{X}, \mathrm{rk}\big)$$

be a finite Combinatorial Complex. Its dimension is

$$\dim(\mathcal{K}) := \max_{x \in \mathcal{X}} \mathrm{rk}(x).$$

For $k \in \mathbb{Z}_{\geq 0}$, its $k$-skeleton is

$$\mathcal{X}^{(k)} := \big\{x \in \mathcal{X} : \mathrm{rk}(x) \leq k\big\},$$

and its set of cells of rank exactly $k$ is

$$\mathcal{X}^{k} := \big\{x \in \mathcal{X} : \mathrm{rk}(x) = k\big\} = \mathrm{rk}^{-1}(\{k\}).$$

**Remark 3.12.5** (Hypergraphs as Combinatorial Complexes)**.** *Let*

$$H = (S, E)$$

*be a hypergraph. Define*

$$\mathcal{X} := \big\{\{s\} : s \in S\big\} \cup E$$

*and*

$$\mathrm{rk}(x) := \begin{cases} 0, & |x| = 1, \\ 1, & |x| \geq 2. \end{cases}$$

*Then*

$$\big(S, \mathcal{X}, \mathrm{rk}\big)$$

*is a Combinatorial Complex. More generally, the hyperedges may be assigned different ranks, provided that the resulting rank function remains order-preserving.*

**Remark 3.12.6** (Difference from a simplicial complex)**.** *A Combinatorial Complex does not require downward closure. In general,*

$$x \in \mathcal{X}, \qquad \emptyset \neq y \subsetneq x$$

*does not imply*

$$y \in \mathcal{X}.$$

*By contrast, an abstract simplicial complex requires every nonempty subset of each simplex to be present. Hence a Combinatorial Complex permits arbitrary set-type cells together with a rank-based hierarchy.*

# 4 Factorization, Constraint, Layered, Temporal, and Tensor-Based Family

This chapter surveys higher-order network frameworks whose principal organizing mechanisms arise from factorization, constraint representation, layered or multiscale organization, temporal evolution, Cartesian-product structure, tensor contraction, and related quantum-state constructions. Although these frameworks originate from different mathematical settings, they share the general objective of representing complex interactions through structured decompositions, multiple contexts, time-dependent relations, or multi-indexed algebraic representations.

For reference, the principal factorization-, constraint-, layered-, temporal-, and tensor-based structures discussed in this chapter are summarized in Table 4.1.

Table 4.1: Principal factorization-, constraint-, layered-, temporal-, and tensor-based higher-order structures discussed in this chapter.

| **Concept** | **Concise description** |
|---|---|
| Factor Graph | A bipartite graph whose two vertex classes represent variables and factors, with incidence indicating which variables occur in each local factor of a factorized global function. |
| Tanner Graph | A bipartite graph associated with a parity-check matrix, connecting a variable node to a check node exactly when the corresponding matrix entry is nonzero. |
| Tanner Hypergraph | A HyperGraph associated with a parity-check matrix in which each parity-check row determines a hyperedge consisting of the variables in its nonzero support. |
| Tanner $n$-SuperHyperGraph | A hierarchical extension of a Tanner Hypergraph in which variables are grouped into $n$-supervertices and each check superhyperedge flattens to the support of the corresponding parity-check equation. |
| Multilayer Network | A network whose state nodes are organized across one or more layers, with intralayer and interlayer edges representing interactions within and between different contexts, modalities, or relational layers. |
| Temporal Network | A network equipped with temporal information specifying when its potential edges are active, thereby representing the evolution of pairwise interactions over a prescribed time domain. |
| Temporal $n$-SuperHyperGraph | An $n$-SuperHyperGraph equipped with a time domain and an activity map assigning to each potential superhyperedge the set of times at which it is active. |
| MultiDimensional Graph | A Cartesian product of finitely many factor graphs whose vertices are tuples and whose adjacency changes exactly one coordinate along an edge of the corresponding factor graph. |
| Iterated MultiDimensional Graph (IMDG) | A recursively constructed graph obtained by repeatedly taking Cartesian products of lower-depth MultiDimensional Graphs, thereby recording a hierarchical product decomposition. |
| Tensor Network Graph | A graph-based tensor representation in which vertices carry local tensors and edges identify shared indices that are summed over during tensor contraction. |

*Table 4.1 (continued)*

| Concept | Concise description |
|---|---|
| MultiTensor and Iterated MultiTensor Network | Tensor-network extensions in which each vertex carries a finite multiset, or an iterated finite multiset, of admissible local tensors, producing multiple weighted realizations and their corresponding contractions. |
| Tensor Hypernetwork | A HyperGraph-based tensor network in which vertices carry local tensors and each hyperedge supplies a shared index simultaneously contracted across all incident tensors. |
| Tensor Superhypernetwork | A hierarchical extension of a Tensor Hypernetwork in which local tensors are associated with supervertices and contractions are organized by multi-level SuperHyperGraph incidence. |
| Tensor Train (TT) | A tensor decomposition representing a finite-order tensor as a chain of three-way core tensors connected through auxiliary bond indices, with boundary bond dimensions equal to one. |
| Tree Tensor Network (TTN) | A tensor network whose contraction structure is organized by a tree, with local tensors connected through virtual bond spaces and open indices representing physical degrees of freedom. |
| Projected Entangled Pair State (PEPS) | A tensor-network state obtained from entangled virtual pairs placed on graph edges and local linear maps projecting incident virtual spaces onto physical state spaces. |
| Projected Entangled Simplex State (PESS) | A tensor-network state in which entangled virtual tensors are assigned to simplices and local projection tensors map their incident virtual degrees of freedom to physical site spaces. |
| Adjacency-Tensor Network (ATN) | A higher-order network represented by a family of adjacency tensors $A^{(2)}, A^{(3)}, \dots, A^{(K)}$, where the order-$k$ tensor records weighted or unweighted $k$-way interactions on a common vertex set. |
| Quantum $n$-SuperHyperGraph | A quantum-state construction associated with an $n$-SuperHyperGraph, in which each $n$-supervertex carries a qubit and each superhyperedge induces a generalized controlled-phase interaction among its incident qubits. |

## 4.1 Tensor network graph

A tensor network graph assigns tensors to vertices and index bonds to edges; contracting edges sums shared indices, yielding a tensor [385, 386]. A tensor network graph encodes higher-order interactions as tensors on nodes, with edges denoting index contractions, representing multiway correlations compactly for efficient computation and inference.

**Definition 4.1.1** (Tensor network graph)**.** A *tensor network graph* is a triple

$$\mathcal{N} = (G,\ (T_v)_{v\in V(G)},\ (\mathcal{I}_e)_{e\in E(G)}),$$

where $G$ is a finite (multi)graph, each vertex $v \in V(G)$ is assigned a tensor

$$T_v \in \bigotimes_{e\in \mathrm{Inc}(v)} \mathbb{K}^{\mathcal{I}_e}$$

over a field $\mathbb{K}$ (typically $\mathbb{R}$ or $\mathbb{C}$), and each edge $e \in E(G)$ is assigned a finite index set $\mathcal{I}_e$ (equivalently, a bond dimension $|\mathcal{I}_e|$). Here $\mathrm{Inc}(v)$ denotes the multiset of edges incident to $v$; half-edges (dangling edges) represent *open indices* (external legs). The *value* (or *contraction*) of $\mathcal{N}$ is the tensor obtained by summing over all internal indices corresponding to non-dangling edges, i.e. contracting the paired index spaces $\mathbb{K}^{\mathcal{I}_e} \otimes \mathbb{K}^{\mathcal{I}_e}$ along each internal edge $e$.

**Example 4.1.2** (A tensor network graph for a length-3 matrix product state)**.** Let $\mathbb{K} = \mathbb{R}$. Consider a path graph on three vertices

$$G: \quad v_1 \; - \; v_2 \; - \; v_3,$$

with two internal edges $e_{12}$ and $e_{23}$, and with one dangling half-edge attached to each vertex (representing a physical/open index). Assign index sets

$$\mathcal{I}_{12} = \{1, \dots, D\}, \qquad \mathcal{I}_{23} = \{1, \dots, D\}$$

to the internal edges (bond dimension $D$), and assign

$$\mathcal{I}_1 = \{1, \dots, d\}, \qquad \mathcal{I}_2 = \{1, \dots, d\}, \qquad \mathcal{I}_3 = \{1, \dots, d\}$$

to the three dangling edges (physical dimension $d$).

Assign tensors

$$T_{v_1} \in \mathbb{R}^{d \times D}, \qquad T_{v_2} \in \mathbb{R}^{D \times d \times D}, \qquad T_{v_3} \in \mathbb{R}^{D \times d}$$

so that $T_{v_1}$ has indices $(i_1, \alpha)$, $T_{v_2}$ has indices $(\alpha, i_2, \beta)$, and $T_{v_3}$ has indices $(\beta, i_3)$ with $i_j \in \{1, \dots, d\}$ and $\alpha, \beta \in \{1, \dots, D\}$. Then the tensor network contraction produces a 3-way tensor

$$\Psi \in \mathbb{R}^{d \times d \times d}$$

whose entries are given by

$$\Psi_{i_1 i_2 i_3} = \sum_{\alpha=1}^{D} \sum_{\beta=1}^{D} \left(T_{v_1}\right)_{i_1, \alpha} \left(T_{v_2}\right)_{\alpha, i_2, \beta} \left(T_{v_3}\right)_{\beta, i_3}.$$

Thus $\mathcal{N} = (G, (T_v)_{v \in V(G)}, (\mathcal{I}_e)_{e \in E(G)})$ is a tensor network graph in the sense of Definition 4.1.1; the internal edges $e_{12}, e_{23}$ are contracted (summed over), while the three dangling edges remain as open indices.

## 4.2 MultiTensor and Iterated MultiTensor Network

MultiTensor Network assigns each node a finite multiset of compatible local tensors, generating multiple weighted contractions through realization choices while preserving one external tensor space. Iterated MultiTensor Network assigns nodes iterated finite multisets of compatible local tensors, recursively flattening multiplicities into effective multitensor assignments before canonically computing weighted network contractions.

**Definition 4.2.1** (Finite multiset)**.** Let $X$ be a set. A *finite multiset* on $X$ is a function

$$m : X \to \mathbb{N}_0$$

with finite support

$$\operatorname{supp}(m) := \{x \in X : m(x) > 0\}.$$

We write

$$\mathcal{M}_{\mathrm{fin}}(X)$$

for the set of all finite multisets on $X$. If $x \in X$, then $m(x)$ is called the *multiplicity* of $x$ in $m$.

**Definition 4.2.2** (MultiTensor network graph)**.** Let

$$\mathcal{T}_v := \bigotimes_{e \in \mathrm{Inc}(v)} \mathbb{K}^{\mathcal{I}_e}$$

denote the local tensor space associated with a vertex $v$ of a finite graph $G$, where $\mathbb{K}$ is a field and each edge or half-edge $e$ is assigned a finite index set $\mathcal{I}_e$.

A *MultiTensor network graph* is a triple

$$\mathcal{N}^{\mathrm{multi}} = \big(G, \; (M_v)_{v \in V(G)}, \; (\mathcal{I}_e)_{e \in E(G)}\big),$$

such that:

- $G$ is a finite (multi)graph, possibly with dangling half-edges;
- for each edge $e$, $\mathcal{I}_e$ is a finite index set;

- for each vertex $v \in V(G)$,
$$M_v \in \mathcal{M}_{\mathrm{fin}}(\mathcal{T}_v)$$
is a finite multiset of tensors of the same local type $\mathcal{T}_v$.

A *realization* of $\mathcal{N}^{\mathrm{multi}}$ is a choice
$$\tau = (T_v)_{v \in V(G)}$$
such that
$$T_v \in \mathrm{supp}(M_v) \qquad (\forall\, v \in V(G)).$$
Its *multiplicity* is defined by
$$\mu(\tau) := \prod_{v \in V(G)} M_v(T_v).$$

For each realization $\tau$, its *contraction*
$$\mathrm{Contr}(\tau)$$
is the tensor obtained by contracting the ordinary tensor network graph
$$\bigl(G,\ (T_v)_{v \in V(G)},\ (\mathcal{I}_e)_{e \in E(G)}\bigr)$$
in the sense of Definition 4.1.1.

The *contraction multiset* of $\mathcal{N}^{\mathrm{multi}}$ is the finite multiset of output tensors defined by
$$\mathrm{Contr}_{\mathrm{multi}}\bigl(\mathcal{N}^{\mathrm{multi}}\bigr)(X) := \sum_{\substack{\tau \text{ realization} \\ \mathrm{Contr}(\tau) = X}} \mu(\tau),$$
for each output tensor $X$ in the external tensor space determined by the dangling half-edges.

If desired, one may also define the *aggregated contraction* by
$$\mathrm{AContr}\bigl(\mathcal{N}^{\mathrm{multi}}\bigr) := \sum_{\tau \text{ realization}} \mu(\tau)\ \mathrm{Contr}(\tau),$$
provided all output tensors belong to the same external tensor space.

**Definition 4.2.3** (Iterated finite multiset)**.** For a set $X$, define recursively
$$\mathcal{M}^0_{\mathrm{fin}}(X) := X, \qquad \mathcal{M}^{n+1}_{\mathrm{fin}}(X) := \mathcal{M}_{\mathrm{fin}}\bigl(\mathcal{M}^n_{\mathrm{fin}}(X)\bigr) \qquad (n \geq 0).$$
Thus, an element of $\mathcal{M}^n_{\mathrm{fin}}(X)$ is an *n-fold iterated finite multiset* over $X$.

**Example 4.2.4** (A concrete MultiTensor network graph)**.** Let $\mathbb{K} = \mathbb{R}$, and let $G$ be the graph with two vertices
$$V(G) = \{v_1, v_2\}$$
and one internal edge
$$E(G) = \{e\}, \qquad e = \{v_1, v_2\}.$$
Assign the index set
$$\mathcal{I}_e = \{1, 2\}.$$
Since the only incident edge at each vertex is $e$, the local tensor spaces are
$$\mathcal{T}_{v_1} = \mathbb{R}^{\mathcal{I}_e}, \qquad \mathcal{T}_{v_2} = \mathbb{R}^{\mathcal{I}_e}.$$
Thus each local tensor may be identified with a vector in $\mathbb{R}^2$.

Define
$$A_1 = (1, 0), \qquad A_2 = (0, 1) \in \mathbb{R}^2,$$
and
$$B_1 = (1, 1), \qquad B_2 = (1, -1) \in \mathbb{R}^2.$$
Now assign finite multisets of local tensors by
$$M_{v_1} = 2[A_1] + 1[A_2], \qquad M_{v_2} = 1[B_1] + 3[B_2].$$

Equivalently,

$$M_{v_1}(A_1) = 2, \quad M_{v_1}(A_2) = 1, \qquad M_{v_2}(B_1) = 1, \quad M_{v_2}(B_2) = 3,$$

and all other multiplicities are zero.

Hence

$$\mathcal{N}^{\mathrm{multi}} = \big(G, (M_v)_{v\in V(G)}, (\mathcal{I}_e)_{e\in E(G)}\big)$$

is a MultiTensor network graph.

A realization of $\mathcal{N}^{\mathrm{multi}}$ is a choice

$$\tau = (T_{v_1}, T_{v_2})$$

with

$$T_{v_1} \in \{A_1, A_2\}, \qquad T_{v_2} \in \{B_1, B_2\}.$$

Thus there are exactly four realizations:

$$\tau_1 = (A_1, B_1), \qquad \tau_2 = (A_1, B_2), \qquad \tau_3 = (A_2, B_1), \qquad \tau_4 = (A_2, B_2).$$

Since the network has no dangling half-edges, each contraction is a scalar. More precisely,

$$\mathrm{Contr}(\tau) = \sum_{i\in\mathcal{I}_e} T_{v_1}(i)\, T_{v_2}(i),$$

which is just the standard dot product in $\mathbb{R}^2$.

We compute:

$$\mathrm{Contr}(\tau_1) = A_1 \cdot B_1 = (1,0)\cdot(1,1) = 1,$$

$$\mathrm{Contr}(\tau_2) = A_1 \cdot B_2 = (1,0)\cdot(1,-1) = 1,$$

$$\mathrm{Contr}(\tau_3) = A_2 \cdot B_1 = (0,1)\cdot(1,1) = 1,$$

$$\mathrm{Contr}(\tau_4) = A_2 \cdot B_2 = (0,1)\cdot(1,-1) = -1.$$

Their multiplicities are

$$\mu(\tau_1) = M_{v_1}(A_1)M_{v_2}(B_1) = 2\cdot 1 = 2,$$

$$\mu(\tau_2) = M_{v_1}(A_1)M_{v_2}(B_2) = 2\cdot 3 = 6,$$

$$\mu(\tau_3) = M_{v_1}(A_2)M_{v_2}(B_1) = 1\cdot 1 = 1,$$

$$\mu(\tau_4) = M_{v_1}(A_2)M_{v_2}(B_2) = 1\cdot 3 = 3.$$

Therefore the contraction multiset is

$$\mathrm{Contr}_{\mathrm{multi}}\big(\mathcal{N}^{\mathrm{multi}}\big) = 9[1] + 3[-1],$$

because the scalar value 1 occurs with total multiplicity

$$2 + 6 + 1 = 9,$$

whereas the scalar value $-1$ occurs with multiplicity

$$3.$$

The aggregated contraction is

$$\mathrm{AContr}\big(\mathcal{N}^{\mathrm{multi}}\big) = \sum_{k=1}^{4} \mu(\tau_k)\ \mathrm{Contr}(\tau_k) = 2\cdot 1 + 6\cdot 1 + 1\cdot 1 + 3\cdot(-1) = 6.$$

Thus this example explicitly illustrates how a MultiTensor network graph produces a finite multiset of contraction values, together with an aggregated contraction.

Moreover, each local multiset

$$M_{v_i} \in \mathcal{M}_{\mathrm{fin}}(\mathcal{T}_{v_i}) = \mathcal{M}^{1}_{\mathrm{fin}}(\mathcal{T}_{v_i}),$$

so this example is also the first-level case of the iterated finite multiset formalism.

**Definition 4.2.5** (Flattening of iterated multisets)**.** Let $X$ be a set. For each $n \geq 0$, define recursively a map

$$\mathrm{Flat}_n : \mathcal{M}^n_{\mathrm{fin}}(X) \to \mathcal{M}_{\mathrm{fin}}(X)$$

as follows:

- for $n = 0$, set
$$\mathrm{Flat}_0(x) := \delta_x \qquad (x \in X),$$
where $\delta_x$ denotes the singleton multiset supported at $x$ with multiplicity 1;
- for $n \geq 0$ and $M \in \mathcal{M}^{n+1}_{\mathrm{fin}}(X)$, define
$$\mathrm{Flat}_{n+1}(M)(x) := \sum_{A \in \mathrm{supp}(M)} M(A)\ \mathrm{Flat}_n(A)(x) \qquad (x \in X).$$

Hence $\mathrm{Flat}_n$ converts an iterated multiset into an ordinary finite multiset on $X$ by recursively collecting multiplicities.

**Definition 4.2.6** ($n$-Iterated MultiTensor network graph)**.** Let $n \geq 0$. For each vertex $v \in V(G)$, let

$$\mathcal{T}_v := \bigotimes_{e \in \mathrm{Inc}(v)} \mathbb{K}^{\mathcal{I}_e}$$

be the corresponding local tensor space.

An *n-Iterated MultiTensor network graph* is a triple

$$\mathcal{N}^{(n)} = \big(G,\ (\Theta_v)_{v \in V(G)},\ (\mathcal{I}_e)_{e \in E(G)}\big),$$

such that:

- $G$ is a finite (multi)graph, possibly with dangling half-edges;
- for each edge $e$, $\mathcal{I}_e$ is a finite index set;
- for each vertex $v \in V(G)$,
$$\Theta_v \in \mathcal{M}^n_{\mathrm{fin}}(\mathcal{T}_v).$$

Its *effective local multiset* at $v$ is defined by

$$M^{\mathrm{eff}}_v := \mathrm{Flat}_n(\Theta_v) \in \mathcal{M}_{\mathrm{fin}}(\mathcal{T}_v).$$

The *effective MultiTensor network graph* associated with $\mathcal{N}^{(n)}$ is

$$\mathcal{N}^{\mathrm{eff}} := \big(G,\ (M^{\mathrm{eff}}_v)_{v \in V(G)},\ (\mathcal{I}_e)_{e \in E(G)}\big).$$

The *contraction multiset* of $\mathcal{N}^{(n)}$ is then defined by

$$\mathrm{Contr}_{\mathrm{iter}}\big(\mathcal{N}^{(n)}\big) := \mathrm{Contr}_{\mathrm{multi}}\big(\mathcal{N}^{\mathrm{eff}}\big),$$

and, whenever meaningful, its *aggregated contraction* is defined by

$$\mathrm{AContr}\big(\mathcal{N}^{(n)}\big) := \mathrm{AContr}\big(\mathcal{N}^{\mathrm{eff}}\big).$$

In particular:

$$n = 0 \implies \mathcal{N}^{(0)} \text{ is an ordinary tensor network graph,}$$

and

$$n = 1 \implies \mathcal{N}^{(1)} \text{ is exactly a MultiTensor network graph.}$$

**Theorem 4.2.7** (Well-definedness of MultiTensor network graph)**.** *Let*

$$\mathcal{N}^{\mathrm{multi}} = \big(G,\ (M_v)_{v \in V(G)},\ (\mathcal{I}_e)_{e \in E(G)}\big)$$

*be a MultiTensor network graph in the sense of Definition 4.2.2. Then:*

1. *the set of realizations of* $\mathcal{N}^{\text{multi}}$ *is finite;*
2. *for every realization*
$$\tau = (T_v)_{v \in V(G)} \quad \textit{with} \quad T_v \in \operatorname{supp}(M_v),$$
*the ordinary tensor network contraction*
$$\operatorname{Contr}(\tau)$$
*is well-defined;*
3. *the contraction multiset*
$$\operatorname{Contr}_{\text{multi}}(\mathcal{N}^{\text{multi}})$$
*is well-defined;*
4. *the aggregated contraction*
$$\operatorname{AContr}(\mathcal{N}^{\text{multi}}) = \sum_{\tau} \mu(\tau) \operatorname{Contr}(\tau)$$
*is well-defined.*

*Proof.* For each vertex $v \in V(G)$, the multiset $M_v$ has finite support by definition, and $V(G)$ is finite. Hence the set of realizations
$$\prod_{v \in V(G)} \operatorname{supp}(M_v)$$
is finite.

Moreover, every tensor chosen at $v$ lies in the same local tensor space
$$\mathcal{T}_v = \bigotimes_{e \in \operatorname{Inc}(v)} \mathbb{K}^{\mathcal{I}_e},$$
so any realization $\tau$ determines an ordinary tensor network graph with the same underlying graph $G$ and the same index sets $(\mathcal{I}_e)_{e \in E(G)}$. Therefore its contraction $\operatorname{Contr}(\tau)$ is well-defined in the sense of Definition 4.1.1.

Since the external tensor space is determined only by the dangling half-edges of $G$ and the index sets $\mathcal{I}_e$, all tensors $\operatorname{Contr}(\tau)$ belong to one and the same output space. Because there are only finitely many realizations and each multiplicity
$$\mu(\tau) = \prod_{v \in V(G)} M_v(T_v)$$
is a well-defined nonnegative integer, both
$$\operatorname{Contr}_{\text{multi}}(\mathcal{N}^{\text{multi}})$$
and
$$\operatorname{AContr}(\mathcal{N}^{\text{multi}})$$
are well-defined. □

**Theorem 4.2.8** (Well-definedness of Iterated MultiTensor network graph)**.** *Let*
$$\mathcal{N}^{(n)} = \big(G,\ (\Theta_v)_{v \in V(G)},\ (\mathcal{I}_e)_{e \in E(G)}\big)$$
*be an n-Iterated MultiTensor network graph in the sense of Definition 4.2.6. Then:*

1. *for each vertex* $v \in V(G))$*, the flattening*
$$\operatorname{Flat}_n(\Theta_v) \in \mathcal{M}_{\text{fin}}(\mathcal{T}_v)$$
*is well-defined;*
2. *the effective network*
$$\mathcal{N}^{\text{eff}} = \big(G,\ (\operatorname{Flat}_n(\Theta_v))_{v \in V(G)},\ (\mathcal{I}_e)_{e \in E(G)}\big)$$
*is a well-defined MultiTensor network graph;*

*3. consequently, both*

$$\mathrm{Contr}_{\mathrm{iter}}\big(\mathcal{N}^{(n)}\big) \quad \text{and} \quad \mathrm{AContr}\big(\mathcal{N}^{(n)}\big)$$

*are well-defined.*

*Proof.* We first show that $\mathrm{Flat}_n$ is well-defined for every $n \geq 0$ by induction on $n$. For $n = 0$, the map

$$\mathrm{Flat}_0(x) = \delta_x$$

is well-defined. Assume that $\mathrm{Flat}_n$ is well-defined. Let

$$M \in \mathcal{M}_{\mathrm{fin}}^{n+1}(X).$$

Since $M$ has finite support and each $\mathrm{Flat}_n(A)$ is a finite multiset on $X$, the formula

$$\mathrm{Flat}_{n+1}(M)(x) = \sum_{A \in \mathrm{supp}(M)} M(A)\ \mathrm{Flat}_n(A)(x)$$

defines a finite-support function $X \to \mathbb{N}_0$. Hence $\mathrm{Flat}_{n+1}(M) \in \mathcal{M}_{\mathrm{fin}}(X)$ is well-defined.

Applying this with $X = \mathcal{T}_v$ for each vertex $v$, we obtain

$$\mathrm{Flat}_n(\Theta_v) \in \mathcal{M}_{\mathrm{fin}}(\mathcal{T}_v),$$

so the effective network $\mathcal{N}^{\mathrm{eff}}$ is a MultiTensor network graph. Therefore Theorem 4.2.7 applies, and the contraction multiset and aggregated contraction of $\mathcal{N}^{\mathrm{eff}}$ are well-defined. By definition,

$$\mathrm{Contr}_{\mathrm{iter}}\big(\mathcal{N}^{(n)}\big) = \mathrm{Contr}_{\mathrm{multi}}\big(\mathcal{N}^{\mathrm{eff}}\big), \qquad \mathrm{AContr}\big(\mathcal{N}^{(n)}\big) = \mathrm{AContr}\big(\mathcal{N}^{\mathrm{eff}}\big),$$

hence both are well-defined. □

## 4.3 Tensor Hypernetwork and Tensor Superhypernetwork

A Tensor Hypernetwork is a hypergraph-based tensor network where each vertex carries a tensor and each hyperedge links multiple tensors through shared index contractions simultaneously. A Tensor Superhypernetwork generalizes a tensor hypernetwork by replacing vertices with hierarchical supervertices, enabling tensor contractions over nested, multi-level, higher-order relational structures and systems formally.

**Definition 4.3.1** (Tensor Hypernetwork)**.** Let

$$H = (V, E)$$

be a finite hypergraph, where

$$E \subseteq \mathcal{P}(V) \setminus \{\emptyset\}.$$

For each hyperedge $e \in E$, fix a positive integer $d_e$, and write

$$[d_e] := \{1, 2, \ldots, d_e\}.$$

For each vertex $v \in V$, define its incident-edge set by

$$I(v) := \{\, e \in E \mid v \in e \,\}.$$

A *tensor hypernetwork* over a field $\mathbb{K}$ on $H$ is a tuple

$$\mathcal{N} = \big(H, (d_e)_{e \in E}, (T_v)_{v \in V}\big),$$

where each local tensor is a function

$$T_v : \prod_{e \in I(v)} [d_e] \to \mathbb{K}.$$

If $I(v) = \emptyset$, the empty product is interpreted as a singleton, so $T_v \in \mathbb{K}$.

The associated closed contraction (network value) is

$$Z(\mathcal{N}) := \sum_{(a_e)_{e \in E} \in \prod_{e \in E} [d_e]} \ \prod_{v \in V} T_v\big((a_e)_{e \in I(v)}\big) \in \mathbb{K}.$$

**Definition 4.3.2** (Tensor $n$-SuperHyperNetwork)**.** Let $V_0$ be a finite nonempty base set, and define iterated powersets by

$$\mathcal{P}^0(V_0) := V_0, \qquad \mathcal{P}^{k+1}(V_0) := \mathcal{P}\big(\mathcal{P}^k(V_0)\big) \quad (k \geq 0).$$

Let

$$SHG^{(n)} = (V, E)$$

be an $n$-SuperHyperGraph on $V_0$, that is,

$$V \subseteq \mathcal{P}^n(V_0), \qquad E \subseteq \mathcal{P}(V) \setminus \{\emptyset\}.$$

Elements of $V$ are called $n$-supervertices, and elements of $E$ are called $n$-superedges.

For each superedge $e \in E$, fix a positive integer $d_e$, and write

$$[d_e] := \{1, 2, \dots, d_e\}.$$

For each supervertex $x \in V$, define

$$I(x) := \{\, e \in E \mid x \in e \,\}.$$

A *tensor $n$-SuperHyperNetwork* over a field $\mathbb{K}$ is a tuple

$$\mathcal{N}^{(n)} = \big(V_0; V, E, (d_e)_{e\in E}, (T_x)_{x\in V}\big),$$

where each local supertensor is a function

$$T_x : \prod_{e\in I(x)} [d_e] \to \mathbb{K} \qquad (x \in V).$$

If $I(x) = \emptyset$, the empty product is interpreted as a singleton, so $T_x \in \mathbb{K}$.

Its associated closed contraction is

$$Z(\mathcal{N}^{(n)}) := \sum_{(a_e)_{e\in E} \in \prod_{e\in E}[d_e]} \prod_{x\in V} T_x\big((a_e)_{e\in I(x)}\big) \in \mathbb{K}.$$

A *tensor SuperHyperNetwork* is a tensor $n$-SuperHyperNetwork for some $n \geq 1$.

**Example 4.3.3** (A concrete tensor 1-SuperHyperNetwork)**.** Let the base set be

$$V_0 = \{a, b, c\}.$$

Since

$$\mathcal{P}^1(V_0) = \mathcal{P}(V_0),$$

we may choose 1-supervertices as subsets of $V_0$. Define

$$x_1 := \{a, b\}, \qquad x_2 := \{b, c\}, \qquad x_3 := \{a, c\},$$

and set

$$V = \{x_1, x_2, x_3\} \subseteq \mathcal{P}(V_0).$$

Now define two 1-superedges by

$$e_1 := \{x_1, x_2\}, \qquad e_2 := \{x_2, x_3\}.$$

Thus

$$E = \{e_1, e_2\} \subseteq \mathcal{P}(V) \setminus \{\emptyset\}.$$

Hence

$$SHG^{(1)} = (V, E)$$

is a 1-SuperHyperGraph on $V_0$.

Assign bond dimensions

$$d_{e_1} = 2, \qquad d_{e_2} = 2,$$

so that

$$[d_{e_1}] = [d_{e_2}] = \{1, 2\}.$$

For each supervertex, the incident superedge set is

$$I(x_1) = \{e_1\}, \qquad I(x_2) = \{e_1, e_2\}, \qquad I(x_3) = \{e_2\}.$$

Let $\mathbb{K} = \mathbb{R}$. Define the local supertensors as follows:

$$T_{x_1} : [2] \to \mathbb{R}, \qquad T_{x_1}(1) = 1, \quad T_{x_1}(2) = 2,$$

$$T_{x_2} : [2] \times [2] \to \mathbb{R}, \qquad \big(T_{x_2}(i,j)\big)_{i,j=1}^{2} = \begin{pmatrix} 1 & 0 \\ 3 & 1 \end{pmatrix},$$

that is,

$$T_{x_2}(1,1) = 1, \quad T_{x_2}(1,2) = 0, \quad T_{x_2}(2,1) = 3, \quad T_{x_2}(2,2) = 1,$$

and

$$T_{x_3} : [2] \to \mathbb{R}, \qquad T_{x_3}(1) = 2, \quad T_{x_3}(2) = 1.$$

Therefore,

$$\mathcal{N}^{(1)} = \big(V_0; V, E, (d_e)_{e \in E}, (T_x)_{x \in V}\big)$$

is a tensor 1-SuperHyperNetwork over $\mathbb{R}$.

Its closed contraction is

$$Z(\mathcal{N}^{(1)}) = \sum_{(a_{e_1}, a_{e_2}) \in [2] \times [2]} T_{x_1}(a_{e_1})\, T_{x_2}(a_{e_1}, a_{e_2})\, T_{x_3}(a_{e_2}).$$

Writing $i := a_{e_1}$ and $j := a_{e_2}$, we obtain

$$Z(\mathcal{N}^{(1)}) = \sum_{i=1}^{2} \sum_{j=1}^{2} T_{x_1}(i)\, T_{x_2}(i,j)\, T_{x_3}(j).$$

Substituting the above values gives

$$\begin{aligned} Z(\mathcal{N}^{(1)}) &= T_{x_1}(1) T_{x_2}(1,1) T_{x_3}(1) + T_{x_1}(1) T_{x_2}(1,2) T_{x_3}(2) \\ &\quad + T_{x_1}(2) T_{x_2}(2,1) T_{x_3}(1) + T_{x_1}(2) T_{x_2}(2,2) T_{x_3}(2) \\ &= 1 \cdot 1 \cdot 2 + 1 \cdot 0 \cdot 1 + 2 \cdot 3 \cdot 2 + 2 \cdot 1 \cdot 1 \\ &= 2 + 0 + 12 + 2 \\ &= 16. \end{aligned}$$

Hence this example gives an explicit tensor 1-SuperHyperNetwork whose closed contraction equals

$$Z(\mathcal{N}^{(1)}) = 16.$$

## 4.4 Tensor Train

A tensor train decomposition represents a high-order tensor as a chain product of third-order core tensors connected by auxiliary bond indices [387, 388, 389].

**Definition 4.4.1** (Tensor Train (TT) decomposition)**.** [387, 388, 389] Let $d \geq 2$, let

$$A \in \mathbb{F}^{n_1 \times n_2 \times \cdots \times n_d}, \qquad \mathbb{F} \in \{\mathbb{R}, \mathbb{C}\},$$

be a $d$-th order tensor, and let

$$r = (r_0, r_1, \ldots, r_d) \in \mathbb{N}^{d+1}$$

satisfy

$$r_0 = r_d = 1.$$

We say that $A$ admits a *tensor train decomposition* of TT-rank $r$ if there exist third-order tensors

$$G^{(k)} \in \mathbb{F}^{r_{k-1} \times n_k \times r_k}, \qquad k = 1, 2, \ldots, d,$$

called the *TT-cores*, such that for every multi-index

$$(i_1, i_2, \ldots, i_d) \in \{1, \ldots, n_1\} \times \cdots \times \{1, \ldots, n_d\},$$

the entry of $A$ is represented as

$$A_{i_1 i_2 \cdots i_d} = \sum_{\alpha_1=1}^{r_1} \cdots \sum_{\alpha_{d-1}=1}^{r_{d-1}} G^{(1)}_{1\, i_1\, \alpha_1} G^{(2)}_{\alpha_1\, i_2\, \alpha_2} \cdots G^{(d)}_{\alpha_{d-1}\, i_d\, 1}.$$

Equivalently, for each $k$ and each $i_k \in \{1, \dots, n_k\}$, define the slice

$$G^{(k)}(i_k) \in \mathbb{F}^{r_{k-1} \times r_k}, \qquad \big(G^{(k)}(i_k)\big)_{\alpha_{k-1}, \alpha_k} := G^{(k)}_{\alpha_{k-1}\, i_k\, \alpha_k}.$$

Then the same representation can be written in matrix-product form as

$$A_{i_1 i_2 \cdots i_d} = G^{(1)}(i_1)\, G^{(2)}(i_2) \cdots G^{(d)}(i_d),$$

where the right-hand side is a $1 \times 1$ matrix, identified with a scalar.

The vector $r = (r_0, r_1, \dots, r_d)$ is called the *TT-rank vector*, and the integers

$$r_1, \dots, r_{d-1}$$

are called the *TT-ranks* of the representation.

**Remark 4.4.2.** *If a tensor $A$ is only approximated by a tensor $\widetilde{A}$ of the above form, then $\widetilde{A}$ is called a* TT approximation *of $A$. In that case one writes*

$$A = \widetilde{A} + E,$$

*where $E$ is the residual tensor.*

**Example 4.4.3** (A concrete tensor train decomposition)**.** Consider the third-order tensor

$$A \in \mathbb{R}^{2 \times 2 \times 2}$$

defined by the two frontal slices

$$A_{::1} = \begin{pmatrix} 1 & 0 \\ 0 & 1 \end{pmatrix}, \qquad A_{::2} = \begin{pmatrix} 0 & 1 \\ 1 & 0 \end{pmatrix}.$$

Equivalently, its entries are

$$A_{111} = 1, \quad A_{121} = 0, \quad A_{211} = 0, \quad A_{221} = 1,$$

and

$$A_{112} = 0, \quad A_{122} = 1, \quad A_{212} = 1, \quad A_{222} = 0.$$

We show that $A$ admits a tensor train decomposition with TT-rank vector

$$(r_0, r_1, r_2, r_3) = (1, 2, 2, 1).$$

Define the TT-cores

$$G^{(1)} \in \mathbb{R}^{1 \times 2 \times 2}, \qquad G^{(2)} \in \mathbb{R}^{2 \times 2 \times 2}, \qquad G^{(3)} \in \mathbb{R}^{2 \times 2 \times 1}$$

through their matrix slices as follows:

$$G^{(1)}(1) = \begin{pmatrix} 1 & 0 \end{pmatrix}, \qquad G^{(1)}(2) = \begin{pmatrix} 0 & 1 \end{pmatrix},$$

$$G^{(2)}(1) = \begin{pmatrix} 1 & 0 \\ 0 & 1 \end{pmatrix}, \qquad G^{(2)}(2) = \begin{pmatrix} 0 & 1 \\ 1 & 0 \end{pmatrix},$$

$$G^{(3)}(1) = \begin{pmatrix} 1 \\ 0 \end{pmatrix}, \qquad G^{(3)}(2) = \begin{pmatrix} 0 \\ 1 \end{pmatrix}.$$

Then, for every $(i_1, i_2, i_3) \in \{1, 2\}^3$, the TT formula gives

$$A_{i_1 i_2 i_3} = G^{(1)}(i_1)\, G^{(2)}(i_2)\, G^{(3)}(i_3).$$

Let us verify several entries explicitly.

For $(i_1, i_2, i_3) = (1, 1, 1)$, we obtain

$$A_{111} = \begin{pmatrix} 1 & 0 \end{pmatrix} \begin{pmatrix} 1 & 0 \\ 0 & 1 \end{pmatrix} \begin{pmatrix} 1 \\ 0 \end{pmatrix} = \begin{pmatrix} 1 & 0 \end{pmatrix} \begin{pmatrix} 1 \\ 0 \end{pmatrix} = 1.$$

For $(i_1, i_2, i_3) = (1, 2, 2)$, we obtain

$$A_{122} = \begin{pmatrix}1 & 0\end{pmatrix}\begin{pmatrix}0 & 1\\ 1 & 0\end{pmatrix}\begin{pmatrix}0\\ 1\end{pmatrix} = \begin{pmatrix}0 & 1\end{pmatrix}\begin{pmatrix}0\\ 1\end{pmatrix} = 1.$$

For $(i_1, i_2, i_3) = (2, 1, 1)$, we obtain

$$A_{211} = \begin{pmatrix}0 & 1\end{pmatrix}\begin{pmatrix}1 & 0\\ 0 & 1\end{pmatrix}\begin{pmatrix}1\\ 0\end{pmatrix} = \begin{pmatrix}0 & 1\end{pmatrix}\begin{pmatrix}1\\ 0\end{pmatrix} = 0.$$

For $(i_1, i_2, i_3) = (2, 2, 1)$, we obtain

$$A_{221} = \begin{pmatrix}0 & 1\end{pmatrix}\begin{pmatrix}0 & 1\\ 1 & 0\end{pmatrix}\begin{pmatrix}1\\ 0\end{pmatrix} = \begin{pmatrix}1 & 0\end{pmatrix}\begin{pmatrix}1\\ 0\end{pmatrix} = 1.$$

By similar calculations, all remaining entries agree with the given tensor. Hence

$$A_{i_1 i_2 i_3} = G^{(1)}(i_1)\, G^{(2)}(i_2)\, G^{(3)}(i_3) \qquad (\forall\, i_1, i_2, i_3 \in \{1, 2\}),$$

so this is a valid tensor train decomposition of $A$.

## 4.5 Tree Tensor Network (TTN)

A Tree Tensor Network is a loop-free hierarchical tensor decomposition on a tree, efficiently representing many-body states through local tensors and virtual bonds capturing correlations [390, 391, 392, 393, 394]. Related concepts such as MERA (Multiscale Entanglement Renormalization Ansatz) [395, 396, 397, 398] are also known.

**Definition 4.5.1** (Tree Tensor Network (TTN))**.** [390, 391, 392] Let $\tau = (V, E)$ be a finite rooted tree with root $r$. For each vertex $v \in V$, let $\mathcal{H}_v$ be a finite-dimensional complex vector space, called the *physical space* at $v$, and write

$$\dim \mathcal{H}_v = d_v.$$

For each non-root vertex $v \in V \setminus \{r\}$, let $p(v)$ denote the parent of $v$, let $\mathrm{Ch}(v)$ denote the set of children of $v$, and let

$$\mathcal{B}_v \cong \mathbb{C}^{D_v}$$

be a finite-dimensional *bond space* attached to the edge $(p(v), v)$.

A *tree tensor network* on $\tau$ is a family of local tensors

$$A^{(r)} \in \mathcal{H}_r \otimes \bigotimes_{u \in \mathrm{Ch}(r)} \mathcal{B}_u,$$

and, for each $v \in V \setminus \{r\}$,

$$A^{(v)} \in \mathcal{H}_v \otimes \mathcal{B}_v^* \otimes \bigotimes_{u \in \mathrm{Ch}(v)} \mathcal{B}_u.$$

The *TTN state represented by* $\{A^{(v)}\}_{v \in V}$ is the tensor

$$|\Psi_A\rangle := \mathrm{Contr}\left(\bigotimes_{v \in V} A^{(v)}\right) \in \bigotimes_{v \in V} \mathcal{H}_v,$$

obtained by contracting, for every non-root vertex $v$, the factor $\mathcal{B}_v$ appearing in $A^{(p(v))}$ with the factor $\mathcal{B}_v^*$ appearing in $A^{(v)}$ via the canonical pairing

$$\langle \cdot, \cdot \rangle_v : \mathcal{B}_v^* \times \mathcal{B}_v \to \mathbb{C}.$$

Equivalently, after choosing bases

$$\mathcal{H}_v = \mathrm{span}\{\, |i_v\rangle : 1 \le i_v \le d_v \,\}, \qquad \mathcal{B}_v = \mathrm{span}\{\, |\alpha_v\rangle : 1 \le \alpha_v \le D_v \,\},$$

the coefficients of $|\Psi_A\rangle$ are given by

$$|\Psi_A\rangle = \sum_{(i_v)_{v \in V}} \Psi_{(i_v)_{v \in V}} \bigotimes_{v \in V} |i_v\rangle,$$

where

$$\Psi_{(i_v)_{v \in V}} = \sum_{(\alpha_v)_{v \in V \setminus \{r\}}} A^{(r)}_{i_r,\, (\alpha_u)_{u \in \mathrm{Ch}(r)}} \prod_{v \in V \setminus \{r\}} A^{(v)}_{i_v,\, \alpha_v,\, (\alpha_u)_{u \in \mathrm{Ch}(v)}}.$$

Any tensor in $\bigotimes_{v \in V} \mathcal{H}_v$ that admits such a representation is called a *tree tensor network state.*

**Example 4.5.2** (A concrete Tree Tensor Network)**.** Consider the rooted tree

$$\tau = (V, E), \qquad V = \{r, a, b\}, \qquad E = \big\{(r, a), (r, b)\big\},$$

where $r$ is the root and $a, b$ are its children. Thus,

$$\mathrm{Ch}(r) = \{a, b\}, \qquad \mathrm{Ch}(a) = \mathrm{Ch}(b) = \emptyset.$$

Let all physical spaces be qubit spaces:

$$\mathcal{H}_r = \mathcal{H}_a = \mathcal{H}_b = \mathbb{C}^2,$$

with standard basis

$$\{|0\rangle, |1\rangle\}.$$

Let the bond spaces be

$$\mathcal{B}_a = \mathcal{B}_b = \mathbb{C}^2,$$

with basis

$$\{|\alpha_0\rangle, |\alpha_1\rangle\},$$

and let

$$\{\langle\alpha_0|, \langle\alpha_1|\}$$

be the corresponding dual bases of $\mathcal{B}_a^*$ and $\mathcal{B}_b^*$.

Define the local tensors as follows:

$$A^{(r)} = |0\rangle_r \otimes |\alpha_0\rangle_a \otimes |\alpha_0\rangle_b + |1\rangle_r \otimes |\alpha_1\rangle_a \otimes |\alpha_1\rangle_b \in \mathcal{H}_r \otimes \mathcal{B}_a \otimes \mathcal{B}_b,$$

$$A^{(a)} = |0\rangle_a \otimes \langle\alpha_0| + |1\rangle_a \otimes \langle\alpha_1| \in \mathcal{H}_a \otimes \mathcal{B}_a^*,$$

and

$$A^{(b)} = |0\rangle_b \otimes \langle\alpha_0| + |1\rangle_b \otimes \langle\alpha_1| \in \mathcal{H}_b \otimes \mathcal{B}_b^*.$$

The TTN state is

$$|\Psi\rangle = \mathrm{Contr}\big(A^{(r)} \otimes A^{(a)} \otimes A^{(b)}\big) \in \mathcal{H}_r \otimes \mathcal{H}_a \otimes \mathcal{H}_b.$$

Using the canonical pairings

$$\langle\alpha_i, \alpha_j\rangle = \delta_{ij},$$

we obtain

$$|\Psi\rangle = |0\rangle_r \otimes |0\rangle_a \otimes |0\rangle_b + |1\rangle_r \otimes |1\rangle_a \otimes |1\rangle_b.$$

Hence

$$|\Psi\rangle = |000\rangle + |111\rangle,$$

which is the (unnormalized) three-qubit GHZ state.

Therefore, this gives a concrete example of a tree tensor network state on a binary rooted tree.

**Proposition 4.5.3** (Well-definedness)**.** *The tensor $|\Psi_A\rangle$ is well-defined and does not depend on the order in which the virtual indices are contracted.*

*Proof.* Once bases are fixed, every coefficient $\Psi_{(i_v)_{v\in V}}$ is a finite sum of products of complex numbers over the virtual indices $(\alpha_v)_{v\in V\setminus\{r\}}$. Since multiplication in $\mathbb{C}$ is associative and finite sums may be reordered without changing their value, any contraction order yields the same coefficient. Hence the total contraction defines a unique tensor $|\Psi_A\rangle$. □

## 4.6 Projected Entangled Pair State (PEPS)

A PEPS is a tensor-network state built by projecting entangled virtual pairs onto physical lattice sites, efficiently encoding many-body correlations (cf.[399, 400, 401, 402, 403]).

**Definition 4.6.1** (Projected Entangled Pair State (PEPS))**.** Let $G = (V, E)$ be a finite undirected graph. For each vertex $v \in V$, let

$$\mathcal{H}_v \cong \mathbb{C}^d$$

be the physical Hilbert space at $v$, and let $D \geq 1$ be an integer, called the bond dimension.

For each edge $e = \{u, v\} \in E$, attach two virtual spaces

$$\mathbb{C}^D_{e,u} \quad \text{and} \quad \mathbb{C}^D_{e,v},$$

and define the maximally entangled vector

$$|\Omega_e\rangle := \sum_{\alpha=1}^{D} |\alpha\rangle_{e,u} \otimes |\alpha\rangle_{e,v} \in \mathbb{C}^D_{e,u} \otimes \mathbb{C}^D_{e,v}.$$

For each vertex $v \in V$, let

$$\mathcal{V}_v := \bigotimes_{e \ni v} \mathbb{C}^D_{e,v},$$

where the tensor product is taken over all edges incident to $v$, and choose a linear map

$$A_v : \mathcal{V}_v \to \mathcal{H}_v.$$

Then the vector

$$|\Psi\rangle := \left(\bigotimes_{v \in V} A_v\right)\left(\bigotimes_{e \in E} |\Omega_e\rangle\right) \in \bigotimes_{v \in V} \mathcal{H}_v$$

is called a *projected entangled pair state* (PEPS) on $G$.

In particular, when $G$ is a one-dimensional chain, this construction reduces to a matrix product state (MPS).

**Example 4.6.2** (A concrete PEPS on a square graph)**.** Let

$$G = (V, E), \qquad V = \{1, 2, 3, 4\}, \qquad E = \big\{\{1, 2\}, \{2, 3\}, \{3, 4\}, \{4, 1\}\big\},$$

so that $G$ is a 4-cycle (a square plaquette).

For each vertex $v \in V$, let the physical space be

$$\mathcal{H}_v = \mathbb{C}^2$$

with standard basis

$$\{|0\rangle, |1\rangle\}.$$

Choose bond dimension $D = 2$. For each edge $e = \{u, v\} \in E$, let

$$|\Omega_e\rangle = |00\rangle_{e,u,e,v} + |11\rangle_{e,u,e,v} \in \mathbb{C}^2_{e,u} \otimes \mathbb{C}^2_{e,v}$$

be the maximally entangled virtual pair.

At each vertex, the virtual space is the tensor product of the two incident virtual spaces. Fix the following order:

$$\mathcal{V}_1 = \mathbb{C}^2_{12,1} \otimes \mathbb{C}^2_{41,1}, \qquad \mathcal{V}_2 = \mathbb{C}^2_{12,2} \otimes \mathbb{C}^2_{23,2},$$

$$\mathcal{V}_3 = \mathbb{C}^2_{23,3} \otimes \mathbb{C}^2_{34,3}, \qquad \mathcal{V}_4 = \mathbb{C}^2_{34,4} \otimes \mathbb{C}^2_{41,4}.$$

Define the local linear maps

$$A_v : \mathcal{V}_v \to \mathcal{H}_v \qquad (v = 1, 2, 3, 4)$$

by

$$A_v\big(|\alpha\rangle \otimes |\beta\rangle\big) = \delta_{\alpha\beta}\, |\alpha\rangle, \qquad \alpha, \beta \in \{0, 1\}.$$

Thus each $A_v$ maps $|00\rangle \mapsto |0\rangle$, $|11\rangle \mapsto |1\rangle$, and annihilates $|01\rangle$ and $|10\rangle$.

The corresponding PEPS is

$$|\Psi\rangle = \left(\bigotimes_{v=1}^{4} A_v\right)\left(|\Omega_{12}\rangle \otimes |\Omega_{23}\rangle \otimes |\Omega_{34}\rangle \otimes |\Omega_{41}\rangle\right).$$

Writing

$$|\Omega_{ij}\rangle = \sum_{\alpha_{ij}=0}^{1} |\alpha_{ij}\rangle_{ij,i} \otimes |\alpha_{ij}\rangle_{ij,j},$$

we obtain

$$|\Psi\rangle = \sum_{\alpha_{12},\alpha_{23},\alpha_{34},\alpha_{41}\in\{0,1\}} \delta_{\alpha_{12},\alpha_{41}}\delta_{\alpha_{12},\alpha_{23}}\delta_{\alpha_{23},\alpha_{34}}\delta_{\alpha_{34},\alpha_{41}}\, |\alpha_{12}\rangle_1 \otimes |\alpha_{12}\rangle_2 \otimes |\alpha_{23}\rangle_3 \otimes |\alpha_{34}\rangle_4.$$

Hence only the assignments

$$\alpha_{12} = \alpha_{23} = \alpha_{34} = \alpha_{41} = 0 \quad \text{or} \quad \alpha_{12} = \alpha_{23} = \alpha_{34} = \alpha_{41} = 1$$

survive, and therefore

$$|\Psi\rangle = |0000\rangle + |1111\rangle.$$

Thus this construction gives a concrete PEPS on a square graph. After normalization, one obtains

$$\frac{1}{\sqrt{2}}\big(|0000\rangle + |1111\rangle\big).$$

## 4.7 Projected Entangled Simplex State (PESS)

A projected entangled simplex state is a tensor-network state obtained by placing entangled virtual tensors on simplices and projecting the incident virtual degrees of freedom onto physical site spaces [404, 405, 406, 407].

**Definition 4.7.1** (Projected Entangled Simplex State (PESS))**.** Let $L$ be a finite lattice, and let $\Sigma$ be a family of simplices (higher-order cells) covering $L$. For each site $v \in L$, let

$$\mathcal{H}_v \cong \mathbb{C}^{d_v}$$

be the physical Hilbert space at $v$, and let $\mathcal{V}_{v,s} \cong \mathbb{C}^D$ be a virtual bond space associated with the incidence of $v$ in a simplex $s \in \Sigma$.

For each simplex $s = \{v_1, \dots, v_m\} \in \Sigma$, choose an *entangled simplex tensor*

$$S^{(s)} \in \mathcal{V}_{v_1,s} \otimes \cdots \otimes \mathcal{V}_{v_m,s}.$$

For each site $v \in L$, choose a local *projection tensor*

$$A^{(v)} : \bigotimes_{s\ni v} \mathcal{V}_{v,s} \longrightarrow \mathcal{H}_v.$$

The vector

$$|\Psi\rangle := \operatorname{Contr}\left(\bigotimes_{s\in\Sigma} S^{(s)} \otimes \bigotimes_{v\in L} A^{(v)}\right) \in \bigotimes_{v\in L} \mathcal{H}_v$$

obtained by contracting all virtual indices is called a *projected entangled simplex state* (PESS).

Thus, a PESS is a tensor-network state in which multipartite entanglement is assigned to simplices rather than only to edges. In the special case where every simplex has two sites, a PESS reduces to a PEPS.

**Example 4.7.2** (A concrete Projected Entangled Simplex State)**.** Let

$$L = \{v_1, v_2, v_3\}$$

be a finite lattice consisting of three sites, and let

$$\Sigma = \{s\}, \qquad s = \{v_1, v_2, v_3\},$$

so that $\Sigma$ is a family of simplices covering $L$.

For each site $v_i \in L$, let the physical Hilbert space be

$$\mathcal{H}_{v_i} = \mathbb{C}^2$$

with standard basis

$$\{|0\rangle, |1\rangle\}.$$

Choose bond dimension $D = 2$, and for each incidence $(v_i, s)$, let

$$\mathcal{V}_{v_i,s} \cong \mathbb{C}^2$$

with basis

$$\{|\alpha_0\rangle, |\alpha_1\rangle\}.$$

Define the entangled simplex tensor

$$S^{(s)} = |\alpha_0\rangle_{v_1,s} \otimes |\alpha_0\rangle_{v_2,s} \otimes |\alpha_0\rangle_{v_3,s} + |\alpha_1\rangle_{v_1,s} \otimes |\alpha_1\rangle_{v_2,s} \otimes |\alpha_1\rangle_{v_3,s} \in \mathcal{V}_{v_1,s} \otimes \mathcal{V}_{v_2,s} \otimes \mathcal{V}_{v_3,s}.$$

For each site $v_i \in L$, define the local projection tensor

$$A^{(v_i)} : \mathcal{V}_{v_i,s} \to \mathcal{H}_{v_i}$$

by

$$A^{(v_i)}(|\alpha_0\rangle) = |0\rangle, \qquad A^{(v_i)}(|\alpha_1\rangle) = |1\rangle.$$

Then the corresponding PESS is

$$|\Psi\rangle = \mathrm{Contr}\Big(S^{(s)} \otimes A^{(v_1)} \otimes A^{(v_2)} \otimes A^{(v_3)}\Big) \in \mathcal{H}_{v_1} \otimes \mathcal{H}_{v_2} \otimes \mathcal{H}_{v_3}.$$

Since each $A^{(v_i)}$ simply projects the virtual basis onto the physical basis, we obtain

$$|\Psi\rangle = |0\rangle_{v_1} \otimes |0\rangle_{v_2} \otimes |0\rangle_{v_3} + |1\rangle_{v_1} \otimes |1\rangle_{v_2} \otimes |1\rangle_{v_3}.$$

Hence

$$|\Psi\rangle = |000\rangle + |111\rangle,$$

which is the unnormalized three-qubit GHZ state.

Therefore, this gives a concrete example of a projected entangled simplex state.

## 4.8 Factor graph

A factor graph is a bipartite graph linking variables to factors, representing how a global function factorizes into local terms [408, 409, 410]. A factor graph represents higher-order networks by introducing factor nodes for multiway interactions, converting hyperedge constraints into bipartite edges linking variables to interaction factors.

**Definition 4.8.1** (Factor graph)**.** [408, 409, 410] Let $X = \{x_1, \ldots, x_n\}$ be a set of variables and let $\mathcal{F} = \{f_1, \ldots, f_m\}$ be a set of factors. A *factor graph* is a bipartite graph

$$\mathcal{G}_F = (X \sqcup \mathcal{F},\ E),$$

where an edge connects a variable node $x_j$ to a factor node $f_i$ if and only if $f_i$ depends on $x_j$. Equivalently, each factor $f_i$ has a *scope* (neighbor set)

$$\mathrm{scope}(f_i) := \{x_j \in X \mid \{x_j, f_i\} \in E\} \subseteq X.$$

In probabilistic modeling, a function $F : X_1 \times \cdots \times X_n \to \mathbb{R}_{\geq 0}$ (often a joint density or unnormalized potential) is said to *factorize over* $\mathcal{G}_F$ if

$$F(x_1, \ldots, x_n) = \prod_{i=1}^{m} f_i\big((x_j)_{x_j \in \mathrm{scope}(f_i)}\big).$$

**Example 4.8.2** (A factor graph for a simple Markov chain)**.** Let $x_1, x_2, x_3$ be random variables taking values in finite sets $X_1, X_2, X_3$. Consider a chain-structured factorization of a nonnegative function

$$F : X_1 \times X_2 \times X_3 \to \mathbb{R}_{\geq 0}$$

of the form

$$F(x_1, x_2, x_3) = f_{12}(x_1, x_2)\, f_{23}(x_2, x_3),$$

where

$$f_{12} : X_1 \times X_2 \to \mathbb{R}_{\geq 0}, \qquad f_{23} : X_2 \times X_3 \to \mathbb{R}_{\geq 0}$$

are pairwise factors.

Define the variable-node set and factor-node set by

$$X = \{x_1, x_2, x_3\}, \qquad \mathcal{F} = \{f_{12}, f_{23}\}.$$

The corresponding factor graph is the bipartite graph

$$\mathcal{G}_F = (X \sqcup \mathcal{F},\ E),$$

with edge set

$$E = \big\{\{x_1, f_{12}\}, \{x_2, f_{12}\}, \{x_2, f_{23}\}, \{x_3, f_{23}\}\big\}.$$

Equivalently, the scopes are

$$\mathrm{scope}(f_{12}) = \{x_1, x_2\}, \qquad \mathrm{scope}(f_{23}) = \{x_2, x_3\}.$$

Thus $F$ factorizes over $\mathcal{G}_F$ exactly as in Definition 4.8.1.

## 4.9 Tanner graph

A Tanner graph is a bipartite graph connecting variable nodes to parity-check nodes, representing nonzero entries of a code's matrix [411, 412, 413]. A Tanner graph models higher-order networks by bipartite variable and constraint nodes; each constraint encodes multiway relations, with edges linking involved variables for inference efficiently.

**Definition 4.9.1** (Tanner graph)**.** Let $\mathbb{F}$ be a finite field and let

$$H = (h_{ij}) \in \mathbb{F}^{m \times n}$$

be a parity-check matrix of a linear code. Let $X = \{x_1, \ldots, x_n\}$ be the set of *variable nodes* and $C = \{c_1, \ldots, c_m\}$ be the set of *check nodes*. The *Tanner graph* of $H$ is the bipartite graph

$$\mathrm{TG}(H) = (X \sqcup C,\ E)$$

with

$$E := \big\{\{x_j, c_i\} \ \big|\ h_{ij} \neq 0\big\}.$$

In the nonbinary case one often equips each edge $\{x_j, c_i\} \in E$ with the label $\lambda(\{x_j, c_i\}) = h_{ij} \in \mathbb{F}^{\times}$.

**Example 4.9.2** (A Tanner graph for a small binary parity-check code)**.** Let $\mathbb{F} = \mathbb{F}_2$ and consider the parity-check matrix

$$H = \begin{pmatrix} 1 & 1 & 0 & 1 \\ 0 & 1 & 1 & 1 \end{pmatrix} \in \mathbb{F}_2^{2 \times 4}.$$

Thus $m = 2$ and $n = 4$. Let the variable nodes and check nodes be

$$X = \{x_1, x_2, x_3, x_4\}, \qquad C = \{c_1, c_2\}.$$

By Definition 4.9.1, the edge set of the Tanner graph $\mathrm{TG}(H)$ is determined by the nonzero entries of $H$:

$$E = \big\{\{x_1, c_1\}, \{x_2, c_1\}, \{x_4, c_1\}, \{x_2, c_2\}, \{x_3, c_2\}, \{x_4, c_2\}\big\}.$$

Equivalently, the first check node $c_1$ is adjacent to $x_1, x_2, x_4$ and the second check node $c_2$ is adjacent to $x_2, x_3, x_4$. Hence $\mathrm{TG}(H)$ is a bipartite graph that encodes the incidence pattern of the parity-check equations represented by $H$.

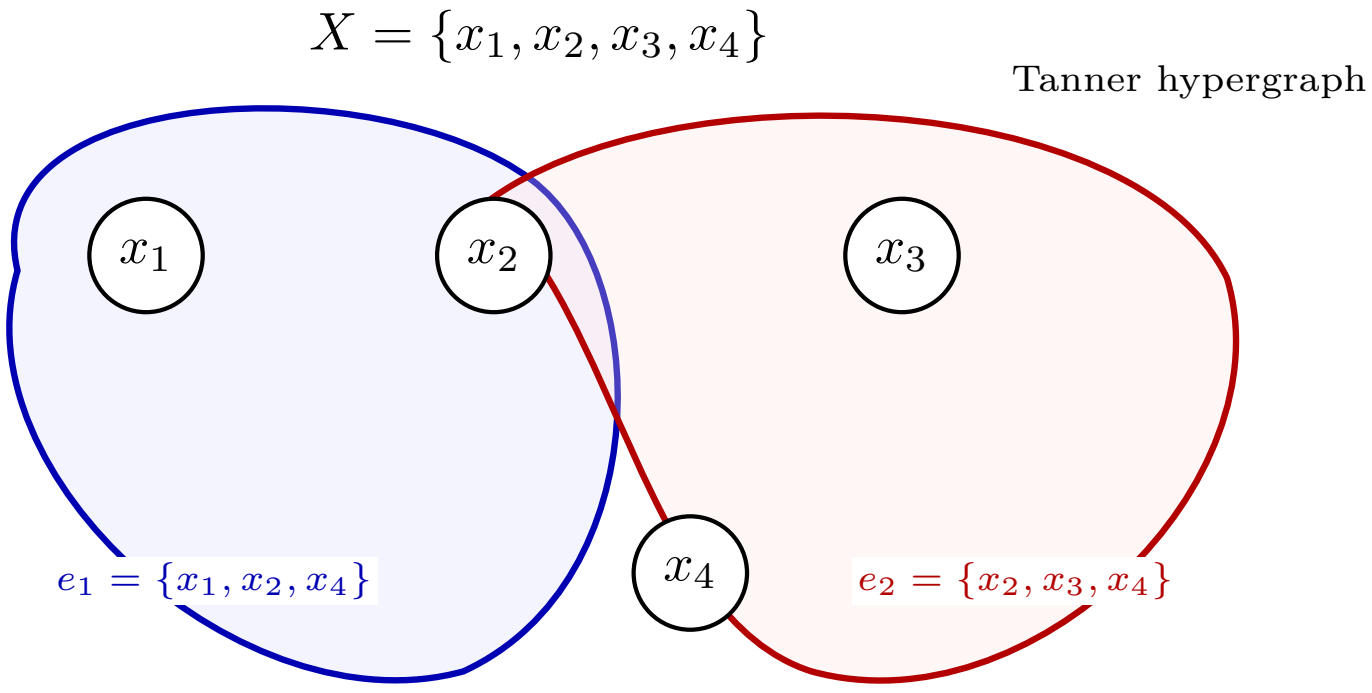

Figure 4.1: A Tanner hypergraph for the parity-check matrix in Example 4.10.3. The two hyperedges correspond to the row supports $\mathrm{supp}(H_1) = \{1, 2, 4\}$ and $\mathrm{supp}(H_2) = \{2, 3, 4\}$.

## 4.10 Tanner Hypergraph

A Tanner Hypergraph models coding constraints by linking variable nodes and check hyperedges, representing multiway parity relations in error-correcting codes. For each row $i \in \{1, \dots, m\}$, define its support by

$$\mathrm{supp}(H_i) := \{\, j \in \{1, \dots, n\} \;\big|\; h_{ij} \neq 0 \,\}.$$

**Definition 4.10.1** (Tanner hypergraph)**.** The *Tanner hypergraph* of $H$ is the hypergraph

$$\mathrm{TH}(H) = (X,\ \mathcal{E}),$$

where

$$\mathcal{E} := \{e_1, \dots, e_m\}, \qquad e_i := \{\, x_j \in X \mid j \in \mathrm{supp}(H_i) \,\} \subseteq X.$$

In the nonbinary case, one may record coefficients by an incidence-weight map

$$w_i : e_i \to \mathbb{F}^{\times}, \qquad w_i(x_j) := h_{ij},$$

so that $\mathrm{TH}(H)$ becomes a weighted (or labeled) hypergraph.

**Remark 4.10.2** (Incidence graph)**.** *The incidence bipartite graph of* $\mathrm{TH}(H)$ *(with coefficient labels, when present) coincides with the Tanner graph* $\mathrm{TG}(H)$*.*

**Example 4.10.3** (A Tanner hypergraph for a small binary parity-check matrix)**.** Let $\mathbb{F} = \mathbb{F}_2$ and consider the parity-check matrix

$$H = \begin{pmatrix} 1 & 1 & 0 & 1 \\ 0 & 1 & 1 & 1 \end{pmatrix} \in \mathbb{F}_2^{2\times 4}.$$

Let $X = \{x_1, x_2, x_3, x_4\}$ be the variable set. The row supports are

$$\mathrm{supp}(H_1) = \{1, 2, 4\}, \qquad \mathrm{supp}(H_2) = \{2, 3, 4\}.$$

Hence the Tanner hypergraph $\mathrm{TH}(H) = (X, \mathcal{E})$ has hyperedge family

$$\mathcal{E} = \{e_1, e_2\}, \qquad e_1 = \{x_1, x_2, x_4\}, \qquad e_2 = \{x_2, x_3, x_4\}.$$

In particular, $e_1$ encodes the first parity-check equation involving variables $x_1, x_2, x_4$, and $e_2$ encodes the second parity-check equation involving $x_2, x_3, x_4$. An overview diagram of this example is provided in Fig. 4.1.

## 4.11 Tanner SuperHyperGraph

A Tanner SuperHyperGraph extends Tanner hypergraphs by allowing hierarchical supervertices and superhyperedges, modeling nested coding constraints, multilevel parity relations, and higher-order dependency structures in codes efficiently. To interpret a nested set as the collection of underlying variables it represents, define recursively:

$$\mathrm{flat}_0(x) := \{x\} \subseteq V_0 \quad (x \in V_0),$$

and for $k \geq 1$ and $A \in \mathcal{P}^k(V_0) = \mathcal{P}(\mathcal{P}^{k-1}(V_0))$,

$$\mathrm{flat}_k(A) := \bigcup_{B \in A} \mathrm{flat}_{k-1}(B) \subseteq V_0.$$

In particular, $\mathrm{flat}_k(A)$ is always a (possibly empty) subset of $V_0$.

**Definition 4.11.1** (Tanner $n$-superhypergraph)**.** Fix $n \in \mathbb{N}$. Let $V \subseteq \mathcal{P}^n(V_0)$ be a set of *$n$-supervertices*. For each check $i \in \{1, \ldots, m\}$ choose a nonempty *superhyperedge*

$$E_i \in \mathcal{P}(V) \setminus \{\emptyset\}.$$

Put $\mathcal{E} := \{E_1, \ldots, E_m\}$ and define

$$\mathrm{TSH}^{(n)}(H) := (V, \mathcal{E}).$$

We call $\mathrm{TSH}^{(n)}(H)$ a *Tanner $n$-superhypergraph* (for $H$) if for every $i$,

$$\bigcup_{v \in E_i} \mathrm{flat}_n(v) \;=\; \{\, x_j \mid j \in \mathrm{supp}(H_i) \,\}. \tag{T}$$

**Example 4.11.2** (A Tanner 1-superhypergraph obtained by grouping variables)**.** Use the same base variable set

$$V_0 = X = \{x_1, x_2, x_3, x_4\},$$

and the same parity-check matrix $H$ as in Example 4.10.3. Take $n = 1$, so 1-supervertices are nonempty subsets of $V_0$.

Define the 1-supervertex set (variable groups)

$$V = \big\{ v_A = \{x_1, x_2\},\ v_B = \{x_2, x_3\},\ v_C = \{x_4\} \big\} \subseteq \mathcal{P}^1(V_0) = \mathcal{P}(V_0).$$

Define two superhyperedges (one per check equation) by

$$E_1 = \{v_A, v_C\}, \qquad E_2 = \{v_B, v_C\}, \qquad \mathcal{E} = \{E_1, E_2\}.$$

Since $n = 1$, we have $\mathrm{flat}_1(v) = v$ for every $v \in V$. Therefore,

$$\bigcup_{v \in E_1} \mathrm{flat}_1(v) = v_A \cup v_C = \{x_1, x_2, x_4\} = \{x_j \mid j \in \mathrm{supp}(H_1)\},$$

and

$$\bigcup_{v \in E_2} \mathrm{flat}_1(v) = v_B \cup v_C = \{x_2, x_3, x_4\} = \{x_j \mid j \in \mathrm{supp}(H_2)\}.$$

Hence the condition (T) holds for $i = 1, 2$, and thus

$$\mathrm{TSH}^{(1)}(H) = (V, \mathcal{E})$$

is a Tanner 1-superhypergraph in the sense of Definition 4.11.1. Intuitively, the first check is represented by the grouped variable block $\{x_1, x_2\}$ together with $\{x_4\}$, and the second check by the block $\{x_2, x_3\}$ together with $\{x_4\}$.

## 4.12 Multilayer network

A multilayer network represents nodes across multiple layers, using node–layer state nodes with intra- and interlayer edges modeling diverse relations [414, 415, 416].

**Definition 4.12.1** (Multilayer network)**.** A *multilayer network* is a tuple

$$\mathcal{M} = (V,\, L_1, \ldots, L_p,\, V_M,\, E_M),$$

where $V$ is a set of physical entities (nodes), each $L_i$ is a finite set of *elementary layer labels*, the set of *layers* is the Cartesian product $L := L_1 \times \cdots \times L_p$, and

$$V_M \subseteq V \times L$$

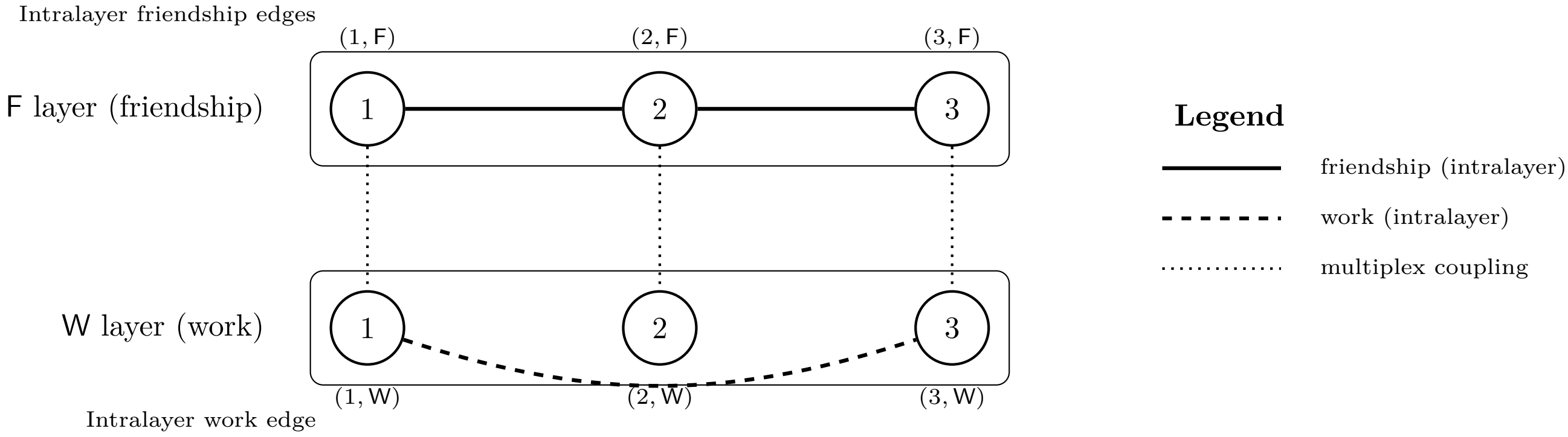

Figure 4.2: A simple multiplex multilayer network with two layers: friendship (F) and work collaboration (W). Dotted vertical edges connect the same physical node across layers, hence the network is multiplex.

is the set of *state nodes* (node–layer tuples). The edge set is

$$E_M \subseteq V_M \times V_M$$

(undirected variants use $E_M \subseteq \binom{V_M}{2}$). Edges with endpoints in the same layer are called *intralayer edges*, and edges between different layers are called *interlayer edges.* A *multiplex network* is a multilayer network for which interlayer edges are allowed only between state nodes corresponding to the same physical node, i.e.

$$\big((v,\ell),(v',\ell')\big) \in E_M \quad \text{and} \quad \ell \neq \ell' \quad \Longrightarrow \quad v = v'.$$

**Example 4.12.2** (A simple multiplex multilayer network: friendship vs. collaboration)**.** Let the set of physical nodes be

$$V = \{1,2,3\}.$$

Take a single aspect ($p = 1$) with two layer labels

$$L_1 = \{\mathsf{F},\mathsf{W}\},$$

where $\mathsf{F}$ denotes a friendship layer and $\mathsf{W}$ denotes a work-collaboration layer. Thus the layer set is $L = L_1$ and the state-node set is

$$V_M = V \times L = \{(1,\mathsf{F}),(2,\mathsf{F}),(3,\mathsf{F}),(1,\mathsf{W}),(2,\mathsf{W}),(3,\mathsf{W})\}.$$

Define an edge set $E_M \subseteq \binom{V_M}{2}$ (undirected) by specifying:

- *Intralayer friendship edges:*
$$\{(1,\mathsf{F}),(2,\mathsf{F})\}, \quad \{(2,\mathsf{F}),(3,\mathsf{F})\};$$
- *Intralayer work edges:*
$$\{(1,\mathsf{W}),(3,\mathsf{W})\};$$
- *Interlayer coupling (multiplex) edges:*
$$\{(1,\mathsf{F}),(1,\mathsf{W})\}, \quad \{(2,\mathsf{F}),(2,\mathsf{W})\}, \quad \{(3,\mathsf{F}),(3,\mathsf{W})\}.$$

Then

$$\mathcal{M} = (V, L_1, V_M, E_M)$$

is a multilayer network in the sense of Definition 4.12.1. Moreover, it is a *multiplex* network because every interlayer edge connects state nodes of the same physical node. An overview diagram of this example is provided in Fig. 4.2.

We consider further extensions. An iterated multilayer network has nodes that are multilayer networks themselves, defined recursively to depth r, with edges between these network-objects across layers.

**Definition 4.12.3** (Iterated Multilayer Network (depth $r$))**.** Fix a nonempty set $U$ of *atomic entities* (base-level physical nodes). Define, for each $r \in \mathbb{N}_0$, a class $\mathsf{IMLN}_r(U)$ of *iterated multilayer networks of depth* $r$ recursively as follows.

1. **(Depth** 0**)** $\mathsf{IMLN}_0(U)$ consists of all multilayer networks

$$\mathcal{M} = (V,\ L_1, \ldots, L_p,\ V_M,\ E_M)$$

in the sense of Definition 4.12.1, whose physical node set satisfies $V \subseteq U$.

2. **(Depth** $r+1$**)** Assuming $\mathsf{IMLN}_r(U)$ is defined, a multilayer network

$$\mathcal{M} = (V,\ L_1, \ldots, L_p,\ V_M,\ E_M)$$

belongs to $\mathsf{IMLN}_{r+1}(U)$ if it satisfies Definition 4.12.1 and, in addition, its physical node set is a set of depth-$r$ iterated multilayer networks, i.e.

$$V \subseteq \mathsf{IMLN}_r(U).$$

Equivalently, the state-node set is a subset

$$V_M \subseteq V \times (L_1 \times \cdots \times L_p),$$

so a state node has the form $(\mathcal{N}, \ell)$ where $\mathcal{N} \in \mathsf{IMLN}_r(U)$ is itself a (multilayer) network-object (a *meta-node*) and $\ell$ is a layer label at the current level.

Any $\mathcal{M} \in \mathsf{IMLN}_r(U)$ is called an *Iterated Multilayer Network of depth* $r$.

**Example 4.12.4** (A depth-1 iterated multilayer network: teams as meta-nodes)**.** Let the atomic entity set be

$$U = \{a, b, c, d\}.$$

First build two depth-0 multilayer networks (ordinary multilayer networks) on subsets of $U$:

**Team A network.** Let $V^A = \{a, b\} \subseteq U$ and $L_1^A = \{\mathsf{Chat}, \mathsf{Code}\}$. Set $V_M^A = V^A \times L_1^A$ and define $E_M^A \subseteq \binom{V_M^A}{2}$ by

$$\{(a, \mathsf{Chat}), (b, \mathsf{Chat})\} \in E_M^A, \qquad \{(a, \mathsf{Code}), (b, \mathsf{Code})\} \in E_M^A,$$

together with multiplex couplings $\{(a, \mathsf{Chat}), (a, \mathsf{Code})\}$ and $\{(b, \mathsf{Chat}), (b, \mathsf{Code})\}$. Denote this depth-0 multilayer network by $\mathcal{N}_A \in \mathsf{IMLN}_0(U)$.

**Team B network.** Similarly, let $V^B = \{c, d\} \subseteq U$ with the same layer set $L_1^B = \{\mathsf{Chat}, \mathsf{Code}\}$, and define $\mathcal{N}_B \in \mathsf{IMLN}_0(U)$ analogously.

**Top-level (depth-**1**) network.** Now form a multilayer network whose physical nodes are the two team-networks:

$$V = \{\mathcal{N}_A, \mathcal{N}_B\} \subseteq \mathsf{IMLN}_0(U).$$

Let the top-level layer set be $L_1 = \{\mathsf{Org}\}$ (one layer), so $V_M = V \times L_1$ consists of two state nodes $(\mathcal{N}_A, \mathsf{Org})$ and $(\mathcal{N}_B, \mathsf{Org})$. Define the top-level edge set by

$$E_M = \big\{\{(\mathcal{N}_A, \mathsf{Org}), (\mathcal{N}_B, \mathsf{Org})\}\big\}.$$

Then

$$\mathcal{M} = (V, L_1, V_M, E_M)$$

is a multilayer network satisfying $V \subseteq \mathsf{IMLN}_0(U)$, hence

$$\mathcal{M} \in \mathsf{IMLN}_1(U).$$

Therefore $\mathcal{M}$ is an *Iterated Multilayer Network of depth* 1 in the sense of Definition 4.12.3, where each physical node is itself a (depth-0) multilayer network.

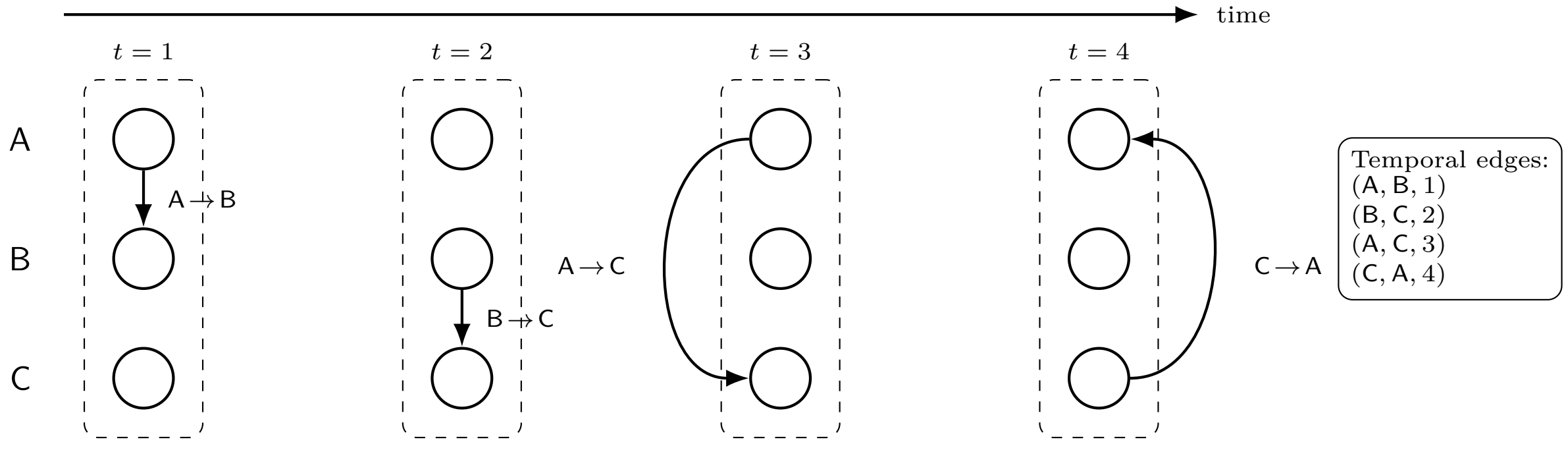

Figure 4.3: A temporal communication network over discrete time $T = \{1, 2, 3, 4\}$. Each dashed box represents one time slice, and the directed edge inside it indicates the message sent at that time.

## 4.13 Temporal network

A temporal network models time-stamped interactions: edges occur at specific times or intervals, capturing evolving connectivity and causal paths [417, 418, 419].

**Definition 4.13.1** (Temporal network (time-varying graph))**.** Let $T$ be a time set (e.g. $T = \mathbb{R}_{\geq 0}$ or $T = \mathbb{Z}$) and let $V$ be a node set. A *temporal network* is a pair

$$\mathcal{G} = (V, E_T),$$

where the *temporal edge set* is

$$E_T \subseteq V \times V \times T.$$

An element $(u, v, t) \in E_T$ means that an interaction (edge) from $u$ to $v$ occurs at time $t$ (undirected variants identify $(u, v, t)$ and $(v, u, t)$). More generally, one may specify edge activity by an *availability function*

$$\rho : V \times V \times T \to \{0, 1\} \quad \text{(or to } \mathbb{R}_{\geq 0} \text{ for weights)},$$

where $\rho(u, v, t) = 1$ indicates that the edge $(u, v)$ is active at time $t$.

**Example 4.13.2** (A temporal communication network over discrete time)**.** Let the node set be

$$V = \{\mathsf{A}, \mathsf{B}, \mathsf{C}\}$$

and let the time set be the discrete set

$$T = \{1, 2, 3, 4\} \subset \mathbb{Z}.$$

Suppose that at time $t = 1$ user A messages B, at time $t = 2$ user B messages C, at time $t = 3$ user A messages C, and at time $t = 4$ user C replies to A. Define the temporal edge set by

$$E_T = \{(\mathsf{A}, \mathsf{B}, 1),\ (\mathsf{B}, \mathsf{C}, 2),\ (\mathsf{A}, \mathsf{C}, 3),\ (\mathsf{C}, \mathsf{A}, 4)\} \subseteq V \times V \times T.$$

Then $\mathcal{G} = (V, E_T)$ is a temporal network in the sense of Definition 4.13.1.

Equivalently, one may encode the same data by an availability function $\rho : V \times V \times T \to \{0, 1\}$ defined by $\rho(u, v, t) = 1$ exactly for the above four triples and $\rho(u, v, t) = 0$ otherwise. An overview diagram of this example is provided in Fig. 4.3.

## 4.14 MultiDimensional Graph (Cartesian-product graph)

A multidimensional graph is the Cartesian product of factor graphs; vertices are tuples, edges change exactly one coordinate per step [420, 421, 422].

**Definition 4.14.1** (MultiDimensional Graph (Cartesian-product graph))**.** Let $d \in \mathbb{N}$ and, for each $k \in \{1, \ldots, d\}$, let

$$G_k = (V_k, E_k, w_k)$$

be an undirected weighted graph, where $V_k$ is a finite vertex set, $E_k \subseteq \binom{V_k}{2}$, and $w_k : V_k \times V_k \to \mathbb{R}_{\geq 0}$ is symmetric with $w_k(u, v) = 0$ whenever $\{u, v\} \notin E_k$. The *d-dimensional graph* (or *multidimensional graph*) associated with $(G_1, \ldots, G_d)$ is the Cartesian product graph

$$G \ = \ G_1 \,\square\, G_2 \,\square\, \cdots \,\square\, G_d \ = \ (V, E, w),$$

defined as follows:

Table 4.2: Difference between a MultiDimensional Graph and a MultiLayer Network (concise view).

| **Aspect** | **MultiDimensional Graph (Def. 4.14.1)** | **MultiLayer Network** |
|---|---|---|
| Primitive idea | A *single* graph built as a Cartesian product of factor graphs. | A *collection of layers* (contexts/relations), represented by node–layer state nodes. |
| Node objects | Tuples $x = (x_1, \ldots, x_d) \in \prod_{k=1}^{d} V_k$. | State nodes $(v, \ell)$ where $v$ is a physical node and $\ell$ is a layer label (possibly multi-aspect). |
| Edge rule | $\{x, y\}$ exists iff $x, y$ differ in exactly one coordinate, along an edge of the corresponding factor graph. | Intralayer edges connect nodes within the same layer; interlayer edges connect nodes across layers (application-defined). |
| Meaning of "dimension/layer" | "Dimension" $d$ means the number of factor graphs; each coordinate gives a directional axis. | "Layer" indexes different relation types, times, modalities, or contexts; not necessarily a Cartesian axis. |
| Structural constraint | Strong constraint: global adjacency is *fully determined* by factor graphs (product structure). | Flexible: no product constraint; layer coupling can be arbitrary (e.g. multiplex coupling for the same node). |
| Typical use | Product-structured domains (e.g. grid-like data, space×time products, separable directional analysis). | Multi-relational / multi-context networks (e.g. multiple interaction types, modalities, snapshots with cross-layer couplings). |

1. The vertex set is the Cartesian product

$$V \;=\; \prod_{k=1}^{d} V_k.$$

2. Two distinct vertices $x = (x_1, \ldots, x_d)$ and $y = (y_1, \ldots, y_d)$ are adjacent, i.e. $\{x, y\} \in E$, if and only if there exists an index $i \in \{1, \ldots, d\}$ such that

$$\{x_i, y_i\} \in E_i \quad \text{and} \quad x_j = y_j \text{ for all } j \neq i.$$

3. The weight function $w : V \times V \to \mathbb{R}_{\geq 0}$ is given by

$$w(x, y) \;=\; \begin{cases} w_i(x_i, y_i), & \text{if } x \text{ and } y \text{ differ only in coordinate } i \text{ and } \{x_i, y_i\} \in E_i, \\ 0, & \text{otherwise.} \end{cases}$$

The graphs $G_1, \ldots, G_d$ are called the *factor graphs*, and the integer $d$ is the *dimension* of $G$.

For reference, the difference between a MultiDimensional Graph and a MultiLayer Network is summarized in Table 4.2.

We formalize the idea of a *MultiDimensional Graph of MultiDimensional Graphs* by iterating the Cartesian-product construction.

**Definition 4.14.2** (Iterated MultiDimensional Graph (IMDG))**.** Fix a class $\mathsf{Graph}$ of (undirected) weighted graphs $G = (V, E, w)$ as in Definition 4.14.1. For $r \in \mathbb{N}_0$, define the classes $\mathsf{IMDG}_r$ recursively as follows:

1. **(Depth** $0$**)** $\mathsf{IMDG}_0 := \mathsf{Graph}$.

2. **(Depth** $r+1$**)** A graph $G$ belongs to $\mathsf{IMDG}_{r+1}$ if there exist an integer $d \geq 1$ and graphs $G_1, \ldots, G_d \in \mathsf{IMDG}_r$ such that

$$G \;\cong\; G_1 \,\Box\, G_2 \,\Box\, \cdots \,\Box\, G_d,$$

where $\Box$ denotes the Cartesian product of graphs (so $V(G) = \prod_{i=1}^{d} V(G_i)$ and adjacency changes exactly one coordinate along an edge of the corresponding factor).

Any $G \in \mathsf{IMDG}_r$ is called an *Iterated MultiDimensional Graph of depth* $r$.

**Definition 4.14.3** (Depth and iterated product specification)**.** For a graph $G$, define its *iterated multidimensional depth* by

$$\mathrm{depth}(G) := \min\{\, r \in \mathbb{N}_0 \mid G \in \mathsf{IMDG}_r \,\} \quad \text{(if such } r \text{ exists)}.$$

An *iterated product specification* of depth $r$ for $G$ is a rooted decomposition tree whose internal nodes are labeled by Cartesian products and whose leaves are base graphs in $\mathsf{IMDG}_0$, such that evaluating the tree (by taking $\Box$ at each internal node) yields a graph isomorphic to $G$.

**Remark 4.14.4** (Flattening)**.** *Because the Cartesian product is associative up to canonical graph isomorphism, any iterated product specification can be* flattened *to a single Cartesian product of all leaf graphs. The iterated specification, however, records a meaningful hierarchical (multi-level) product structure: a "MultiDimensional Graph of MultiDimensional Graphs."*

**Example 4.14.5** (An Iterated MultiDimensional Graph of depth 2)**.** Let $P_2$ denote the path graph on two vertices (a single edge), and let $C_4$ denote the 4-cycle. Consider the depth-1 multidimensional graph

$$G^{(1)} \;=\; P_2 \,\Box\, P_2,$$

which is the $2 \times 2$ grid graph (a square) with vertex set

$$V(G^{(1)}) = \{0,1\} \times \{0,1\}$$

and edges between vertices that differ in exactly one coordinate.

Now form another depth-1 multidimensional graph

$$H^{(1)} \;=\; C_4 \,\Box\, P_2,$$

whose vertex set is $V(C_4) \times V(P_2)$ and whose adjacency again changes exactly one coordinate.

Finally, define the graph

$$G \;=\; G^{(1)} \,\Box\, H^{(1)}.$$

By construction, $G$ is a Cartesian product of two factors $G^{(1)}$ and $H^{(1)}$, each of which is itself a Cartesian product of depth-0 graphs. Hence,

$$G \in \mathsf{IMDG}_2,$$

so $G$ is an Iterated MultiDimensional Graph of depth 2 in the sense of Definition 4.14.2.

**Iterated product specification.** One iterated product specification for $G$ is given by the rooted decomposition tree

$$\big(P_2 \Box P_2\big) \,\Box\, \big(C_4 \Box P_2\big),$$

whose leaves are $P_2, P_2, C_4, P_2 \in \mathsf{IMDG}_0$ and whose internal nodes apply the Cartesian product. Evaluating this tree yields a graph isomorphic to $G$, as required by Definition 4.14.3.

## 4.15 Adjacency-Tensor Network (ATN)

An *Adjacency-Tensor Network (ATN)* represents higher-order interactions on a common vertex set by means of a family of adjacency tensors of orders $2, \ldots, K$. Each tensor encodes interactions of a prescribed arity, while its nonzero entries indicate the existence and, when applicable, the weights of the corresponding multiway adjacency relations. In this way, an ATN extends the ordinary adjacency-matrix representation of graphs to a unified tensor-based framework capable of describing interactions involving multiple vertices simultaneously.

**Definition 4.15.1** (Adjacency-Tensor Network (ATN))**.** Let $V = \{1, \ldots, n\}$ be a finite vertex set and fix $K \geq 2$. An *Adjacency-Tensor Network* is a family of real-valued tensors

$$\mathsf{ATN} = \mathbf{A} = \big(A^{(2)}, A^{(3)}, \ldots, A^{(K)}\big), \qquad A^{(k)} \in \mathbb{R}^{n \times \cdots \times n} \;\; (k \text{ factors}),$$

where $A^{(k)}_{i_1 \cdots i_k}$ represents the weight (or indicator) of a $k$-way interaction among $(i_1, \ldots, i_k) \in V^k$. In the undirected $k$-uniform case one typically requires symmetry:

$$A^{(k)}_{i_{\pi(1)} \cdots i_{\pi(k)}} = A^{(k)}_{i_1 \cdots i_k} \quad \text{for all } \pi \in S_k.$$

**Adjacency-Tensor Network (ATN): pairwise and triple interactions on $V = \{1, 2, 3\}$**

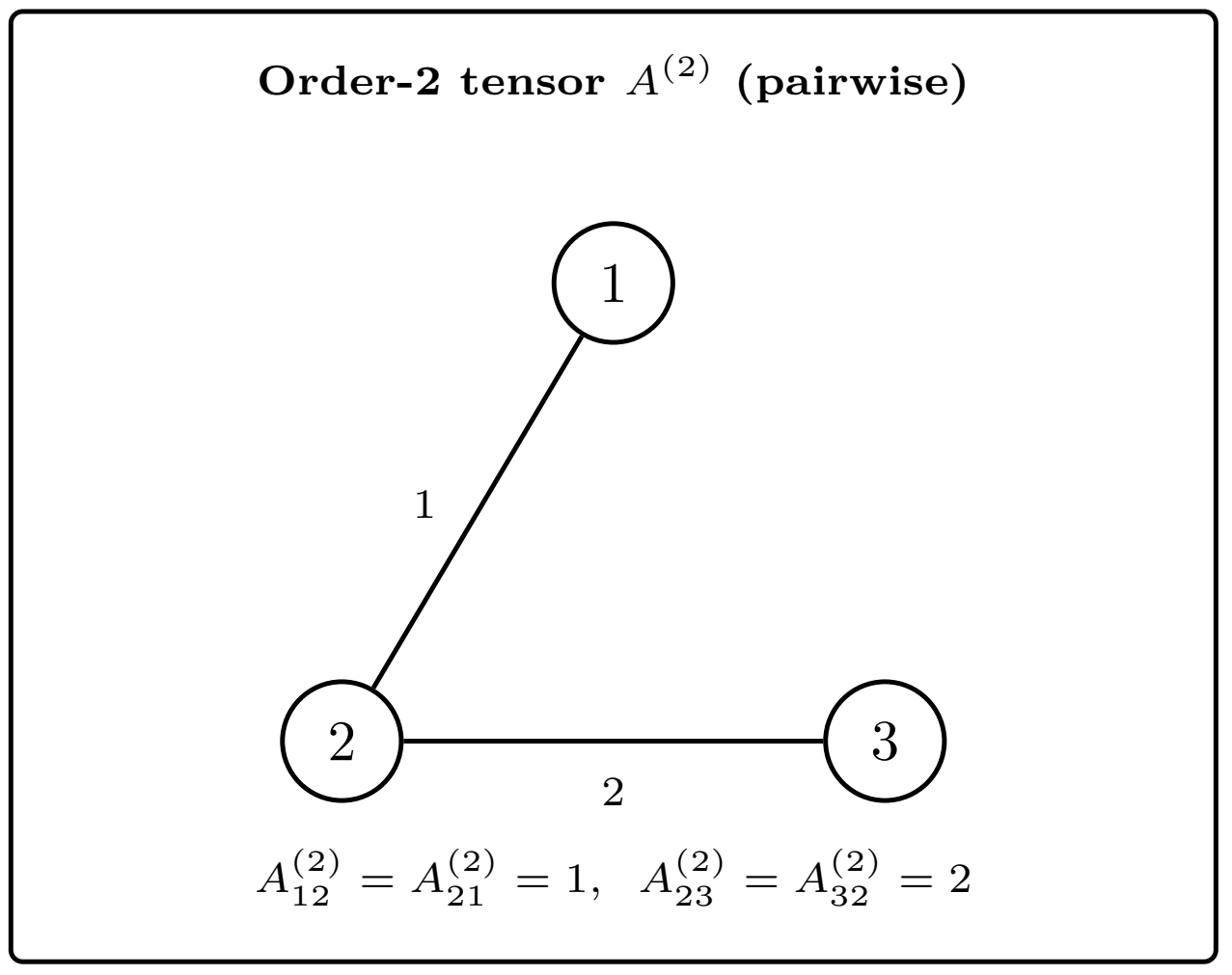

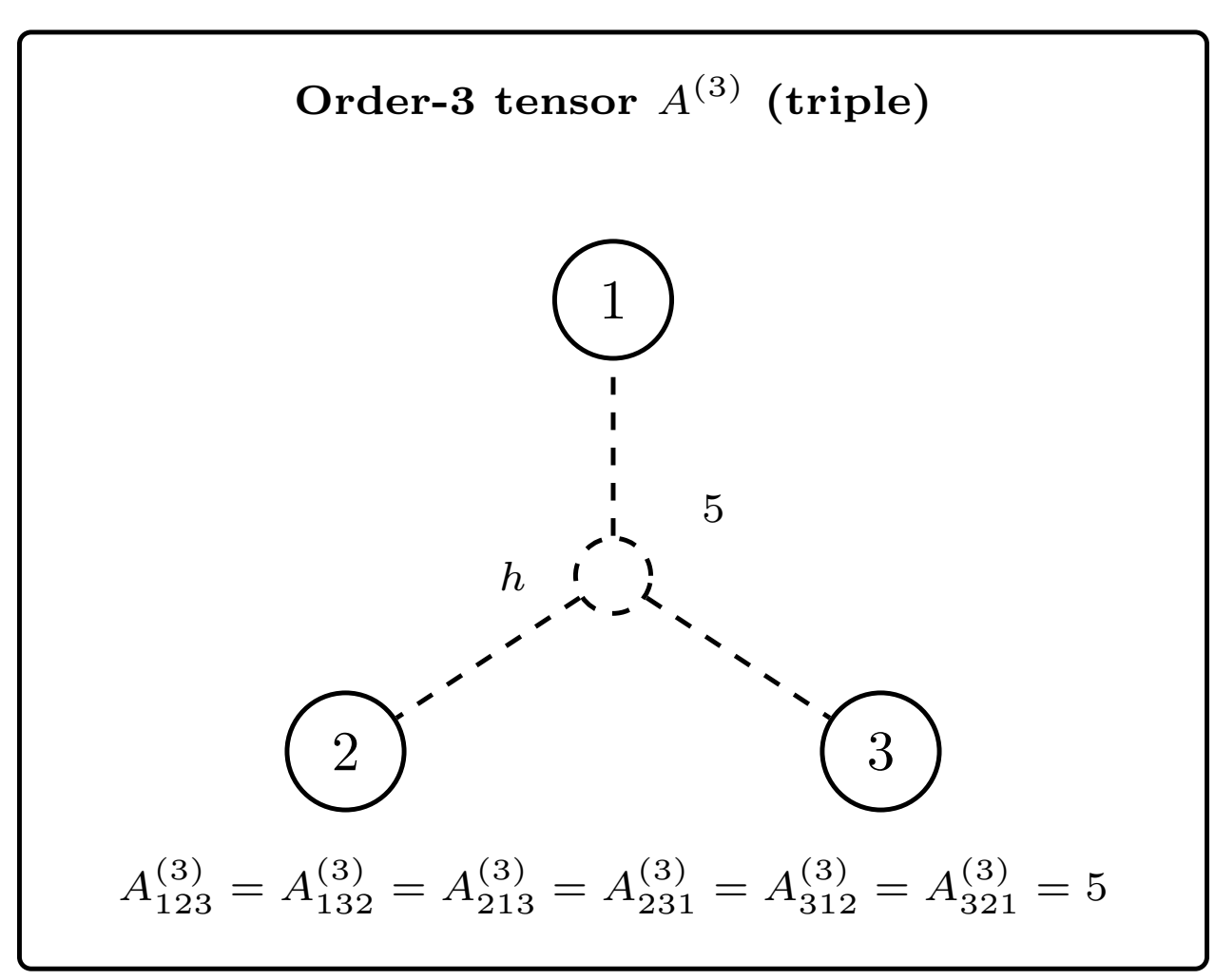

Figure 4.4: An ATN with pairwise and triple interactions on $V = \{1, 2, 3\}$. The left panel visualizes nonzero entries of $A^{(2)}$, and the right panel visualizes the nonzero symmetric triple interaction in $A^{(3)}$.

**Example 4.15.2** (An ATN with pairwise and triple interactions on $V = \{1, 2, 3\}$)**.** Let $V = \{1, 2, 3\}$, so $n = 3$, and take $K = 3$. Define an Adjacency-Tensor Network

$$\mathbf{A} = \big(A^{(2)}, A^{(3)}\big)$$

as follows.

**Pairwise tensor.** Let $A^{(2)} \in \mathbb{R}^{3\times 3}$ be symmetric with nonzero entries

$$A^{(2)}_{12} = A^{(2)}_{21} = 1, \qquad A^{(2)}_{23} = A^{(2)}_{32} = 2,$$

and all other entries 0. Thus, the pairwise interaction $\{1, 2\}$ has weight 1, and $\{2, 3\}$ has weight 2.

**Triple tensor.** Let $A^{(3)} \in \mathbb{R}^{3\times 3\times 3}$ be fully symmetric with

$$A^{(3)}_{123} = A^{(3)}_{132} = A^{(3)}_{213} = A^{(3)}_{231} = A^{(3)}_{312} = A^{(3)}_{321} = 5,$$

and all other entries 0 (in particular, entries with repeated indices are 0). This encodes a 3-way interaction among $\{1, 2, 3\}$ of weight 5.

Then $\mathbf{A}$ is an Adjacency-Tensor Network in the sense of Definition 4.15.1, where nonzero entries of $A^{(2)}$ and $A^{(3)}$ specify weighted 2-way and 3-way interactions, respectively. An overview diagram of this example is provided in Fig. 4.4.

**Theorem 4.15.3** (Well-definedness of Adjacency-Tensor Networks)**.** *Let $V = \{1, \dots, n\}$ be a finite vertex set with $n \geq 1$, and fix an integer $K \geq 2$. For each $k \in \{2, \dots, K\}$, define*

$$\mathcal{T}_k(V) := \mathbb{R}^{V^k} = \{\, \alpha : V^k \to \mathbb{R} \,\}.$$

*Then the following hold.*

1. *For each $k$, there is a canonical identification*

$$\mathcal{T}_k(V) \;\cong\; \mathbb{R}^{n\times\cdots\times n} \quad (k \text{ factors}),$$

*given by*

$$\alpha \longmapsto A^{(k)}, \qquad A^{(k)}_{i_1\cdots i_k} := \alpha(i_1, \dots, i_k) \quad ((i_1, \dots, i_k) \in V^k).$$

*Hence a $k$-way adjacency tensor is a well-defined object.*

*2. The family space*

$$\mathcal{T}_{\le K}(V) := \prod_{k=2}^{K} \mathcal{T}_k(V)$$

*is well-defined (a finite Cartesian product of sets, in fact a finite-dimensional real vector space). Each element*

$$\mathbf{A} = (A^{(2)}, A^{(3)}, \dots, A^{(K)}) \in \mathcal{T}_{\le K}(V)$$

*therefore defines a well-formed Adjacency-Tensor Network (ATN) on $V$.*

*3. For each $k$, the symmetry condition in the undirected $k$-uniform case,*

$$A^{(k)}_{i_{\pi(1)}\cdots i_{\pi(k)}} = A^{(k)}_{i_1\cdots i_k} \qquad (\forall (i_1, \dots, i_k) \in V^k,\ \forall \pi \in S_k),$$

*is well-defined and is equivalent to saying that the associated function $\alpha_k : V^k \to \mathbb{R}$ is constant on the $S_k$-orbits of $V^k$ under coordinate permutation.*

*Consequently, Definition 4.15.1 is well-defined.*

*Proof.* For each $k \in \{2, \dots, K\}$, the set $V^k$ is finite because $V$ is finite. Hence $\mathcal{T}_k(V) = \mathbb{R}^{V^k}$ is a well-defined set of real-valued functions on $V^k$.

**(1) Canonical tensor indexing.** Since $V = \{1, \dots, n\}$, every element of $V^k$ is a $k$-tuple $(i_1, \dots, i_k)$ with $i_j \in \{1, \dots, n\}$. Thus each $\alpha \in \mathcal{T}_k(V)$ determines a unique array

$$A^{(k)} = (A^{(k)}_{i_1\cdots i_k})_{(i_1,\dots,i_k)\in V^k}$$

by the rule $A^{(k)}_{i_1\cdots i_k} := \alpha(i_1, \dots, i_k)$. Conversely, any $k$-indexed real array $(A^{(k)}_{i_1\cdots i_k})$ defines a unique function

$$\alpha : V^k \to \mathbb{R}, \qquad \alpha(i_1, \dots, i_k) = A^{(k)}_{i_1\cdots i_k}.$$

These constructions are inverse to each other, so

$$\mathcal{T}_k(V) \cong \mathbb{R}^{n\times\cdots\times n} \quad (k \text{ factors}).$$

Hence the notation $A^{(k)} \in \mathbb{R}^{n\times\cdots\times n}$ is mathematically well-defined.

**(2) Family of tensors.** Because the index set $\{2, \dots, K\}$ is finite, the Cartesian product

$$\mathcal{T}_{\le K}(V) = \prod_{k=2}^{K} \mathcal{T}_k(V)$$

is well-defined. An element $\mathbf{A} \in \mathcal{T}_{\le K}(V)$ is exactly a family

$$\mathbf{A} = (A^{(2)}, A^{(3)}, \dots, A^{(K)}), \qquad A^{(k)} \in \mathcal{T}_k(V) \cong \mathbb{R}^{n\times\cdots\times n},$$

which is precisely the data specified in Definition 4.15.1. Therefore an ATN is a well-formed mathematical object.

**(3) Symmetry condition.** Fix $k$. For any permutation $\pi \in S_k$, the tuple $(i_{\pi(1)}, \dots, i_{\pi(k)})$ belongs to $V^k$ whenever $(i_1, \dots, i_k) \in V^k$. Hence the expression

$$A^{(k)}_{i_{\pi(1)}\cdots i_{\pi(k)}}$$

is defined, and so the equality

$$A^{(k)}_{i_{\pi(1)}\cdots i_{\pi(k)}} = A^{(k)}_{i_1\cdots i_k}$$

is a well-posed condition.

Let $\alpha_k : V^k \to \mathbb{R}$ be the function corresponding to $A^{(k)}$. Then the above equality is equivalent to

$$\alpha_k(i_{\pi(1)}, \dots, i_{\pi(k)}) = \alpha_k(i_1, \dots, i_k) \quad (\forall (i_1, \dots, i_k) \in V^k,\ \forall \pi \in S_k),$$

which exactly means that $\alpha_k$ is constant on each orbit of the natural $S_k$-action on $V^k$ by coordinate permutation. Thus the undirected symmetry requirement is also well-defined.

Combining (1)–(3), Definition 4.15.1 is well-defined. □

## 4.16 Temporal SuperHyperGraphs

A *temporal graph* extends an ordinary graph by allowing pairwise relationships to appear, disappear, or otherwise vary over time, thereby supporting the analysis of evolving connectivity patterns. Uncertainty-based extensions, including fuzzy temporal graphs, have also been investigated [423, 424, 425, 426, 427, 428, 429, 430]. Related models based on neutrosophic information may likewise be considered for temporal networks.

A *temporal HyperGraph* extends this idea to higher-order interactions by allowing hyperedges, each of which may connect an arbitrary nonempty set of vertices, to become active at specified times [431, 432, 433, 434, 435]. A *Temporal $n$-SuperHyperGraph* further combines temporal evolution with the hierarchical vertex structure of an $n$-SuperHyperGraph by associating each $n$-superhyperedge with the set of times at which it is active [436]. Thus, the model simultaneously represents hierarchical higher-order interactions and their evolution over time.

**Definition 4.16.1** (Temporal $n$-SuperHyperGraph)**.** [436] Let $V_0$ be a finite nonempty base set and let

$$n \in \mathbb{N}_0.$$

A finite *Temporal $n$-SuperHyperGraph over $V_0$* is a structure

$$\mathcal{TSH}^{(n)} = \big(V_0; V, E, \partial, \mathbb{T}, \Lambda\big),$$

satisfying the following conditions:

1. $$V \subseteq \mathcal{P}^n(V_0)$$
   is a finite nonempty set of $n$-supervertices;

2. $E$ is a finite set of potential $n$-superhyperedge identifiers;

3. $$\partial : E \longrightarrow \mathcal{P}_*(V)$$
   is the incidence map, where $\partial(e)$ is the nonempty set of $n$-supervertices incident with $e$;

4. $\mathbb{T}$ is a nonempty totally ordered time domain, for example a discrete totally ordered set or an ordered subset of $\mathbb{R}$;

5. $$\Lambda : E \longrightarrow \mathcal{P}(\mathbb{T})$$
   is an *activity map.*

For each $e \in E$, the set

$$\Lambda(e) \subseteq \mathbb{T}$$

is called the *activity set* of $e$. The $n$-superhyperedge $e$ is said to be *active at time* $t \in \mathbb{T}$ if and only if

$$t \in \Lambda(e).$$

**Remark 4.16.2** (Incidence-map formulation)**.** *The use of a separate superhyperedge set $E$ and incidence map*

$$\partial : E \to \mathcal{P}_*(V)$$

*is consistent with the general incidence-map formulation of $n$-SuperHyperGraphs. In particular, it permits distinct superhyperedge identifiers $e, e' \in E$ to satisfy*

$$\partial(e) = \partial(e'),$$

*while having different activity sets*

$$\Lambda(e) \neq \Lambda(e').$$

*Thus, parallel superhyperedges with different temporal behavior can be represented without ambiguity.*

**Definition 4.16.3** (Temporal snapshot)**.** Let

$$\mathcal{TSH}^{(n)} = \big(V_0; V, E, \partial, \mathbb{T}, \Lambda\big)$$

be a Temporal $n$-SuperHyperGraph. For each $t \in \mathbb{T}$, define the active superhyperedge set by

$$E(t) := \{e \in E : t \in \Lambda(e)\}.$$

Let

$$\partial_t := \partial|_{E(t)} : E(t) \longrightarrow \mathcal{P}_*(V)$$

be the restriction of the incidence map to the active superhyperedges.

The *snapshot at time* $t$ is the static $n$-SuperHyperGraph

$$\mathcal{TSH}^{(n)}(t) := \big(V_0; V, E(t), \partial_t\big).$$

**Remark 4.16.4** (Snapshot well-definedness)**.** *For every* $t \in \mathbb{T}$*,*

$$E(t) \subseteq E.$$

*Since*

$$\partial(e) \in \mathcal{P}_*(V) \qquad \text{for every } e \in E,$$

*the restricted map*

$$\partial_t : E(t) \longrightarrow \mathcal{P}_*(V)$$

*is well defined. Consequently,*

$$\mathcal{TSH}^{(n)}(t) = \big(V_0; V, E(t), \partial_t\big)$$

*is a well-defined static n-SuperHyperGraph over* $V_0$*.*

**Remark 4.16.5** (Potential and active superhyperedges)**.** *The set* $E$ *contains all potential superhyperedges considered during the observation period, whereas*

$$E(t) = \{e \in E : t \in \Lambda(e)\}$$

*contains only those that are active at time* $t$*.*

*The definition allows*

$$\Lambda(e) = \varnothing.$$

*Such a superhyperedge belongs to the potential structure but is never active during the specified time domain. If every potential superhyperedge is required to occur at least once, one may impose the additional condition*

$$\Lambda(e) \neq \varnothing \qquad \text{for every } e \in E.$$

**Remark 4.16.6** (Activity-relation formulation)**.** *The activity map* $\Lambda$ *may equivalently be represented by the binary relation*

$$\mathcal{A}_\Lambda := \{(e,t) \in E \times \mathbb{T} : t \in \Lambda(e)\}.$$

*Conversely, every relation*

$$\mathcal{A} \subseteq E \times \mathbb{T}$$

*determines an activity map*

$$\Lambda_{\mathcal{A}} : E \longrightarrow \mathcal{P}(\mathbb{T})$$

*defined by*

$$\Lambda_{\mathcal{A}}(e) := \{t \in \mathbb{T} : (e,t) \in \mathcal{A}\}.$$

*Hence the activity-map and activity-relation formulations are equivalent.*

**Remark 4.16.7** (Static reduction)**.** *If the temporal data* $\mathbb{T}$ *and* $\Lambda$ *are disregarded, the underlying static structure is the n-SuperHyperGraph*

$$\mathrm{SHG}^{(n)} = \big(V_0; V, E, \partial\big).$$

*In particular, if*

$$\Lambda(e) = \mathbb{T} \qquad \text{for every } e \in E,$$

*then every superhyperedge is active at every time, and consequently*

$$E(t) = E \qquad \text{for every } t \in \mathbb{T}.$$

*Therefore,*

$$\mathcal{TSH}^{(n)}(t) = \big(V_0; V, E, \partial\big) \qquad \text{for every } t \in \mathbb{T}.$$

*Thus, the temporal structure reduces to the same static n-SuperHyperGraph at every time point.*

**Remark 4.16.8** (Temporal HyperGraphs and temporal graphs as special cases)**.** *When*

$$n = 0,$$

*one has*

$$\mathcal{P}^0(V_0) = V_0,$$

*and hence*

$$V \subseteq V_0.$$

*Therefore, a Temporal* 0*-SuperHyperGraph is a temporal incidence-based HyperGraph whose vertex set is* $V$*.*
*If, in addition,*

$$V = V_0$$

*and every superhyperedge has exactly two incident vertices,*

$$|\partial(e)| = 2 \qquad \text{for every } e \in E,$$

*then the structure reduces to an undirected temporal graph, possibly with parallel temporal edges.*
*If one further requires the incidence map* $\partial$ *to be injective, then no two distinct edge identifiers have the same endpoint set, and the resulting structure is a finite simple undirected temporal graph.*

**Definition 4.16.9** (Active lifetime of a Temporal $n$-SuperHyperGraph)**.** Let

$$\mathcal{TSH}^{(n)} = \left(V_0; V, E, \partial, \mathbb{T}, \Lambda\right)$$

be a Temporal $n$-SuperHyperGraph. Its *active lifetime* is defined by

$$\mathrm{Life}\left(\mathcal{TSH}^{(n)}\right) := \bigcup_{e \in E} \Lambda(e).$$

Equivalently,

$$\mathrm{Life}\left(\mathcal{TSH}^{(n)}\right) = \{t \in \mathbb{T} : E(t) \neq \varnothing\}.$$

Thus,

$$\mathrm{Life}\left(\mathcal{TSH}^{(n)}\right) \subseteq \mathbb{T}$$

is precisely the set of time points at which at least one superhyperedge is active.

**Definition 4.16.10** (Aggregated and persistent superhyperedge sets)**.** Let

$$\mathcal{TSH}^{(n)} = \left(V_0; V, E, \partial, \mathbb{T}, \Lambda\right)$$

be a Temporal $n$-SuperHyperGraph.
The *aggregated superhyperedge set* is

$$E_{\cup} := \bigcup_{t \in \mathbb{T}} E(t) = \{e \in E : \Lambda(e) \neq \varnothing\},$$

whereas the *persistent superhyperedge set* is

$$E_{\cap} := \bigcap_{t \in \mathbb{T}} E(t) = \{e \in E : \Lambda(e) = \mathbb{T}\}.$$

Let

$$\partial_{\cup} := \partial|_{E_{\cup}} \qquad \text{and} \qquad \partial_{\cap} := \partial|_{E_{\cap}}.$$

Then

$$\mathcal{TSH}^{(n)}_{\cup} := \left(V_0; V, E_{\cup}, \partial_{\cup}\right)$$

is called the *aggregated static* $n$*-SuperHyperGraph*, while

$$\mathcal{TSH}^{(n)}_{\cap} := \left(V_0; V, E_{\cap}, \partial_{\cap}\right)$$

is called the *persistent static* $n$*-SuperHyperGraph.*

**Remark 4.16.11** (Interpretation of aggregation and persistence)**.** *The aggregated structure contains every superhyperedge that is active at least once during the observation period. By contrast, the persistent structure contains exactly those superhyperedges that remain active throughout the entire time domain* $\mathbb{T}$*. Consequently,*

$$E_{\cap} \subseteq E(t) \subseteq E_{\cup} \subseteq E \qquad \text{for every } t \in \mathbb{T}.$$

## 4.17 Quantum $n$-SuperHyperGraph

In the graph-state setting, a *Quantum Graph* represents pairwise quantum interactions among qubits, with graph edges specifying entangling operations that generate a multipartite quantum state [437, 438, 439]. Related concepts, particularly qudit graph states [440, 441] and qudit hypergraph states [442, 443], have also been investigated. A *Quantum HyperGraph* extends the graph-state construction by allowing a hyperedge to involve more than two qubits. Such hyperedges generate higher-order multipartite interactions through generalized controlled-phase operators acting simultaneously on the corresponding subsets of qubits [444, 445, 446].

A *Quantum n-SuperHyperGraph* further extends this construction by associating qubits with $n$-supervertices rather than ordinary vertices. Its superhyperedges determine controlled-phase interactions among the corresponding qubits, while the supervertices retain the hierarchical structure induced by the $n$-fold powerset construction [447].

**Definition 4.17.1** (Hilbert space)**.** [448] A *Hilbert space* is a real or complex inner-product space

$$(\mathcal{H}, \langle\cdot,\cdot\rangle)$$

that is complete with respect to the norm induced by its inner product, namely

$$\|x\| = \sqrt{\langle x, x\rangle}.$$

In quantum information theory, the state space of a single qubit is the two-dimensional complex Hilbert space

$$\mathcal{H} \cong \mathbb{C}^2,$$

equipped with the computational basis

$$\{|0\rangle, |1\rangle\}.$$

**Definition 4.17.2** (Generalized controlled-phase operator)**.** Let

$$V = \{v_1, \ldots, v_m\}$$

be a finite index set of qubits, and let

$$\mathcal{H}_V = \bigotimes_{v\in V} \mathbb{C}^2.$$

For every nonempty subset

$$A \subseteq V,$$

define the generalized controlled-phase operator

$$CZ_A : \mathcal{H}_V \longrightarrow \mathcal{H}_V$$

on computational-basis states by

$$CZ_A|x_1, \ldots, x_m\rangle = (-1)^{\prod_{v_i\in A} x_i}|x_1, \ldots, x_m\rangle,$$

where

$$x_i \in \{0, 1\}.$$

Thus $CZ_A$ changes the sign of a computational-basis vector precisely when every qubit indexed by an element of $A$ is in the state $|1\rangle$. In particular, when

$$A = \{v_i, v_j\},$$

the operator $CZ_A$ is the ordinary two-qubit controlled-$Z$ gate. For $|A| \geq 3$, it is a multi-qubit controlled-phase operator.

Since all operators $CZ_A$ are diagonal in the computational basis, they commute with one another.

**Definition 4.17.3** (Quantum $n$-SuperHyperGraph)**.** [447] Let

$$\mathcal{S}^{(n)} = \big(V_0; V, E, \partial\big)$$

be a finite $n$-SuperHyperGraph, where

$$n \in \mathbb{N}_0, \qquad V \subseteq \mathcal{P}^n(V_0),$$

$E$ is a finite set of superhyperedge identifiers, and

$$\partial : E \longrightarrow \mathcal{P}_*(V)$$

is the incidence map.

Let

$$m := |V|.$$

Associate with each $n$-supervertex $v \in V$ a qubit Hilbert space

$$\mathcal{H}_v \cong \mathbb{C}^2,$$

and define the total Hilbert space by

$$\mathcal{H}_V = \bigotimes_{v \in V} \mathcal{H}_v \cong (\mathbb{C}^2)^{\otimes m}.$$

For each superhyperedge $e \in E$, let

$$CZ_{\partial(e)}$$

denote the generalized controlled-phase operator associated with its incidence set

$$\partial(e) \subseteq V.$$

The *Quantum $n$-SuperHyperGraph* associated with $\mathcal{S}^{(n)}$ is the pair

$$\mathfrak{Q}^{(n)} = \left(\mathcal{S}^{(n)}, |\Psi_{\mathcal{S}^{(n)}}\rangle\right),$$

where

$$|\Psi_{\mathcal{S}^{(n)}}\rangle = \left(\prod_{e \in E} CZ_{\partial(e)}\right) \bigotimes_{v \in V} |+\rangle_v,$$

and

$$|+\rangle_v = \frac{|0\rangle_v + |1\rangle_v}{\sqrt{2}}.$$

Because the operators $CZ_{\partial(e)}$ commute, the resulting state does not depend on the order in which the superhyperedges are listed.

The vector

$$|\Psi_{\mathcal{S}^{(n)}}\rangle \in \mathcal{H}_V$$

is called the corresponding *Quantum $n$-SuperHyperGraph State.*

**Remark 4.17.4** (Role of the hierarchical supervertices)**.** *The qubits of a Quantum $n$-SuperHyperGraph are indexed by the $n$-supervertices in $V$, rather than directly by the elements of the base set $V_0$. Thus, two distinct supervertices may contain overlapping base-level elements while still representing distinct qubits. The hierarchical information is therefore retained in the classical indexing structure, whereas the quantum interactions are determined by the incidence sets of the superhyperedges.*

**Example 4.17.5** (A Quantum 1-SuperHyperGraph)**.** Let the base set be

$$V_0 = \{a, b, c\},$$

and consider the three 1-supervertices

$$v_1 = \{a\}, \qquad v_2 = \{b, c\}, \qquad v_3 = \{a, c\}.$$

Since

$$v_1, v_2, v_3 \in \mathcal{P}(V_0),$$

the set

$$V = \{v_1, v_2, v_3\} \subseteq \mathcal{P}(V_0)$$

is an admissible set of 1-supervertices.

Let

$$E = \{e_1, e_2\},$$

and define the incidence map

$$\partial : E \longrightarrow \mathcal{P}_*(V)$$

by

$$\partial(e_1) = \{v_1, v_2\}, \qquad \partial(e_2) = \{v_1, v_2, v_3\}.$$

Hence

$$\mathcal{S}^{(1)} = \big(V_0; V, E, \partial\big)$$

is a finite 1-SuperHyperGraph.

Associate one qubit with each 1-supervertex. The corresponding total Hilbert space is

$$\mathcal{H}_V = \mathcal{H}_{v_1} \otimes \mathcal{H}_{v_2} \otimes \mathcal{H}_{v_3} \cong (\mathbb{C}^2)^{\otimes 3}.$$

The initial product state is

$$|+\rangle_{v_1} \otimes |+\rangle_{v_2} \otimes |+\rangle_{v_3} = |+\rangle^{\otimes 3}.$$

The superhyperedge $e_1$ has incidence set

$$\partial(e_1) = \{v_1, v_2\},$$

and therefore induces the ordinary two-qubit controlled-$Z$ operator

$$CZ_{\partial(e_1)} = CZ_{\{v_1,v_2\}}.$$

The superhyperedge $e_2$ has incidence set

$$\partial(e_2) = \{v_1, v_2, v_3\},$$

and therefore induces the three-qubit controlled-phase operator

$$CZ_{\partial(e_2)} = CZ_{\{v_1,v_2,v_3\}}.$$

Consequently, the associated Quantum 1-SuperHyperGraph State is

$$|\Psi_{\mathcal{S}^{(1)}}\rangle = CZ_{\{v_1,v_2\}} CZ_{\{v_1,v_2,v_3\}} |+\rangle^{\otimes 3}.$$

Using the computational-basis ordering

$$(v_1, v_2, v_3),$$

we have

$$|+\rangle^{\otimes 3} = \frac{1}{\sqrt{8}} \sum_{x_1,x_2,x_3 \in \{0,1\}} |x_1 x_2 x_3\rangle.$$

The two controlled-phase operators contribute the phase

$$(-1)^{x_1 x_2} (-1)^{x_1 x_2 x_3} = (-1)^{x_1 x_2 + x_1 x_2 x_3}.$$

Therefore,

$$|\Psi_{\mathcal{S}^{(1)}}\rangle = \frac{1}{\sqrt{8}} \sum_{x_1,x_2,x_3 \in \{0,1\}} (-1)^{x_1 x_2 + x_1 x_2 x_3} |x_1 x_2 x_3\rangle.$$

Expanding the sum gives

$$\begin{aligned} |\Psi_{\mathcal{S}^{(1)}}\rangle = \frac{1}{\sqrt{8}} \big( & |000\rangle + |001\rangle + |010\rangle + |011\rangle \\ & + |100\rangle + |101\rangle - |110\rangle + |111\rangle \big). \end{aligned}$$

Thus, the superhyperedge $e_1$ produces a pairwise phase interaction, whereas $e_2$ produces a genuine three-qubit phase interaction. This example illustrates how the hierarchical information encoded by the 1-supervertices can coexist with pairwise and higher-order quantum interactions in a single Quantum 1-SuperHyperGraph State.

# 5 Semantic, Compositional, Knowledge, and Logical Family

Semantic, compositional, knowledge, and logical families model higher-order systems through meaning, interfaces, composition, inference, and structured knowledge, emphasizing how relations, transformations, and interpretations interact across complex hierarchical networked representations. For reference, the principal semantic, compositional, knowledge-based, logical, uncertainty-aware, decision-oriented, and learning-oriented higher-order structures discussed in this chapter are summarized in Table 5.1. The organization is intended as a practical classification according to the predominant modeling viewpoint of each framework rather than as a strict mathematical partition. Consequently, some structures may naturally admit interpretations within more than one family.

Table 5.1: Principal semantic, compositional, knowledge-based, logical, uncertainty-aware, decision-oriented, and learning-oriented higher-order structures discussed in this chapter.

| Concept | Concise description |
|---|---|
| Heterogeneous Graph, HyperGraph, and SuperHyperGraph | Typed graph-based structures in which vertices, edges, hyperedges, or superhyperedges carry prescribed types, thereby representing heterogeneous binary, multiway, and hierarchical relations. |
| Knowledge Graph, HyperGraph, and SuperHyperGraph | Semantic relational structures representing entities and facts, extended from binary relations to higher-arity relations and hierarchical superentities through HyperGraph and SuperHyperGraph constructions. |
| Petri Net | A directed bipartite structure consisting of places and transitions, with arcs and markings representing resource flow, concurrency, synchronization, and state transitions. |
| Port Graph | A graph in which connections are established through explicitly defined ports attached to vertices, providing structured interfaces for fine-grained interaction and composition. |
| Port HyperGraph and Port SuperHyperGraph | Higher-order port-based structures in which hyperedges or superhyperedges connect ports associated with vertices or hierarchical supervertices. |
| Open HyperGraph and Open SuperHyperGraph | HyperGraphs and SuperHyperGraphs equipped with designated boundary or input–output interfaces, thereby supporting modular composition through explicitly specified connections. |
| Combinatorial Map | A connected properly edge-colored regular graph in which every vertex is incident with exactly one edge of each prescribed color. |
| Cognitive HyperGraphs and Cognitive SuperHyperGraphs | Higher-order cognitive structures representing concepts and multi-entity relations, with SuperHyperGraph variants additionally encoding hierarchical or nested cognitive entities. |
| Multimodal Graph, HyperGraph, and SuperHyperGraph | Structures representing multiple modalities or relational channels on a common vertex or supervertex domain, allowing heterogeneous information to be integrated within a unified higher-order framework. |
| Operadic Interaction Graph (OIG) | An operadic framework in which typed multi-input interactions are represented as operations and structured substitution or composition is governed by operadic composition. |
| Symmetric Monoidal Wiring Graph (SMWG) | A compositional framework based on symmetric monoidal structure in which typed interfaces and wiring patterns represent parallel and sequential composition of interacting components. |
| Relational-Arity Graph (RAG) | A relational framework in which interactions are organized according to their prescribed arities, providing a common representation of binary and higher-arity relations. |
| Closure-Implication Graph (CIG) | A closure-based relational framework in which closure operators encode implication, derivability, or dependency among elements or collections of elements. |

*Table 5.1 (continued)*

| **Concept** | **Concise description** |
|---|---|
| Coalgebraic Nested-Neighborhood Graph (CNNG) | A coalgebraic graph framework in which vertices are associated with recursively nested neighborhood objects, thereby representing hierarchical local observations and dependencies. |
| Curried Graph | A typed graph framework in which vertices represent curried functional objects and edges encode prescribed compatibility or transformation relations among them. |
| Sheaf HyperGraph and Sheaf SuperHyperGraph | HyperGraph and SuperHyperGraph structures equipped with sheaf-like local data and compatible restriction maps, enabling local-to-global representation over higher-order incidence structures. |
| Fibered HyperGraph and Fibered SuperHyperGraph | Higher-order structures equipped with fibers over vertices or supervertices, together with incidence-compatible maps or relations among the corresponding local objects. |
| Galois HyperGraph and Galois SuperHyperGraph | Higher-order relational structures constructed using Galois connections, formal contexts, or associated closure systems to determine admissible hyperedges or superhyperedges. |
| Rewrite HyperGraph and Rewrite SuperHyperGraph | HyperGraph and SuperHyperGraph structures equipped with rewrite rules for systematically transforming higher-order incidence structures while preserving prescribed structural constraints. |
| Uncertain SuperHyperGraph | An $n$-SuperHyperGraph whose supervertices, superhyperedges, or both are equipped with values determined by a prescribed uncertainty model, thereby representing hierarchical higher-order relations under uncertainty. |
| Functorial SuperHyperGraph | A categorical framework in which SuperHyperGraphs and their structure-preserving maps are organized functorially, allowing compatible relationships among different levels or indexing structures. |
| Motif HyperGraphs and Motif SuperHyperGraphs | Motif-based higher-order structures in which occurrences or supports of prescribed motifs determine hyperedges, supervertices, or superhyperedges according to a specified construction. |
| Molecular SuperHyperGraphs | Hierarchical molecular structures representing atoms, bonds, molecular fragments, functional groups, and higher-order molecular relations through iterated set-based supervertices and superhyperedges. |
| Soft $n$-SuperHyperGraph | A parameterized $n$-SuperHyperGraph in which structural information is organized with respect to a family of parameters, enabling parameter-dependent higher-order representations. |
| Rough $n$-SuperHyperGraph | An $n$-SuperHyperGraph framework equipped with lower and upper approximations induced by suitable approximation relations on its supervertices, superhyperedges, or associated incidence structure. |
| Decision $n$-SuperHyperTree | A rooted and oriented $n$-SuperHyperTree in which supervertices represent decision, chance, or terminal states, branching superhyperedges encode possible transitions, and chance probabilities and terminal utilities may be assigned. |
| Weighted $n$-SuperHyperGraph | An $n$-SuperHyperGraph $$\bigl(V_0; V, E, \partial\bigr)$$ equipped with a weight function $$w : E \to \mathbb{R}_{\geq 0},$$ assigning a nonnegative real-valued weight to each superhyperedge identifier. |
| Linguistic Graph, Linguistic HyperGraph, and Linguistic SuperHyperGraph | Graph-based structures equipped with linguistic labels on vertices and relations, extended to multiway and hierarchical interactions through HyperGraph and SuperHyperGraph constructions. |
| $n$-SuperHyperGraph Neural Network | A learning framework that performs feature propagation and aggregation over the incidence structure of an $n$-SuperHyperGraph, thereby incorporating hierarchical higher-order relations into neural message passing. |
| Similarity SuperHyperGraph | A SuperHyperGraph whose hierarchical supervertices are connected through superhyperedges according to prescribed similarity criteria applied to their underlying base-level supports. |

## 5.1 Heterogeneous Graph, HyperGraph, and SuperHyperGraph

A heterogeneous graph is a graph in which vertices and edges may belong to different types, allowing multiple classes of objects and relations to be represented within a single structure [449, 450, 451]. A heterogeneous hypergraph is a hypergraph in which vertices and hyperedges may have different types, enabling the modeling of typed multiway relations among diverse objects in one framework [452, 453, 454]. A heterogeneous superhypergraph is a superhypergraph with typed higher-order vertices and superhyperedges, designed to represent heterogeneous hierarchical relations through iterated set-based constructions.

**Definition 5.1.1** (Heterogeneous Graph)**.** [449, 450, 451] Let $T_V$ and $T_E$ be nonempty sets, called the sets of vertex-types and edge-types, respectively. A *heterogeneous graph* is a sextuple

$$G = (V, E, T_V, T_E, \tau_V, \tau_E),$$

where

1. $V$ is a finite nonempty set of vertices;
2. $$E \subseteq \big\{\{u, v\} \subseteq V : u \neq v\big\}$$
   is the set of edges;
3. $\tau_V : V \to T_V$ is a vertex-type map;
4. $\tau_E : E \to T_E$ is an edge-type map.

If either $|T_V| > 1$ or $|T_E| > 1$, then $G$ is called *heterogeneous*.

**Definition 5.1.2** (Heterogeneous HyperGraph)**.** Let $T_V$ and $T_E$ be nonempty sets. A *heterogeneous hypergraph* is a sextuple

$$H = (V, E, T_V, T_E, \tau_V, \tau_E),$$

where

1. $V$ is a finite nonempty set of vertices;
2. $$E \subseteq \mathcal{P}^*(V) := \mathcal{P}(V) \setminus \{\emptyset\}$$
   is a finite set of nonempty subsets of $V$, called hyperedges;
3. $\tau_V : V \to T_V$ is a vertex-type map;
4. $\tau_E : E \to T_E$ is a hyperedge-type map.

If either $|T_V| > 1$ or $|T_E| > 1$, then $H$ is called *heterogeneous*.

**Definition 5.1.3** (Heterogeneous $n$-SuperHyperGraph)**.** Let $V_0$ be a finite nonempty base set, and let $n \in \mathbb{N} \cup \{0\}$. Define the iterated powersets recursively by

$$\mathcal{P}^0(V_0) := V_0, \qquad \mathcal{P}^{k+1}(V_0) := \mathcal{P}\big(\mathcal{P}^k(V_0)\big) \quad (k \geq 0).$$

Let $T_V$ and $T_E$ be nonempty sets. A *heterogeneous $n$-SuperHyperGraph* is a sextuple

$$\mathcal{H}^{(n)} = (V, E, T_V, T_E, \tau_V, \tau_E),$$

where

1. $$V \subseteq \mathcal{P}^n(V_0)$$
   is a finite set, whose elements are called $n$-supervertices;
2. $$E \subseteq \mathcal{P}^*(V)$$
   is a finite set of nonempty subsets of $V$, whose elements are called $n$-superhyperedges;

3. $\tau_V : V \to T_V$ is a supervertex-type map;

4. $\tau_E : E \to T_E$ is a superhyperedge-type map.

If either $|T_V| > 1$ or $|T_E| > 1$, then $\mathcal{H}^{(n)}$ is called *heterogeneous.*

**Example 5.1.4** (A concrete Heterogeneous 1-SuperHyperGraph)**.** Let the finite base set be

$$V_0 = \{\mathsf{Alice}, \mathsf{Bob}, \mathsf{Carol}, \mathsf{DataA}, \mathsf{DataB}, \mathsf{Server}\},$$

and take $n = 1$. Then

$$\mathcal{P}^1(V_0) = \mathcal{P}(V_0),$$

so each 1-supervertex is a subset of $V_0$.

Define the set of 1-supervertices by

$$V = \{v_1, v_2, v_3, v_4\} \subseteq \mathcal{P}(V_0),$$

where

$$v_1 = \{\mathsf{Alice}, \mathsf{Bob}\}, \qquad v_2 = \{\mathsf{Carol}\},$$

$$v_3 = \{\mathsf{DataA}, \mathsf{DataB}\}, \qquad v_4 = \{\mathsf{Bob}, \mathsf{Server}\}.$$

Next, define the supervertex-type set by

$$T_V = \{\mathsf{Team}, \mathsf{Individual}, \mathsf{DatasetGroup}, \mathsf{TechnicalUnit}\},$$

and define the type map

$$\tau_V : V \to T_V$$

by

$$\tau_V(v_1) = \mathsf{Team}, \qquad \tau_V(v_2) = \mathsf{Individual},$$

$$\tau_V(v_3) = \mathsf{DatasetGroup}, \qquad \tau_V(v_4) = \mathsf{TechnicalUnit}.$$

Now define the set of 1-superhyperedges by

$$E = \{e_1, e_2, e_3\} \subseteq \mathcal{P}^*(V),$$

where

$$e_1 = \{v_1, v_2\}, \qquad e_2 = \{v_1, v_3, v_4\}, \qquad e_3 = \{v_2, v_4\}.$$

Clearly each $e_i$ is a nonempty subset of $V$, so indeed $E \subseteq \mathcal{P}^*(V)$.

Define the superhyperedge-type set by

$$T_E = \{\mathsf{Coordination}, \mathsf{AccessControl}, \mathsf{Maintenance}\},$$

and define the edge-type map

$$\tau_E : E \to T_E$$

by

$$\tau_E(e_1) = \mathsf{Coordination}, \qquad \tau_E(e_2) = \mathsf{AccessControl}, \qquad \tau_E(e_3) = \mathsf{Maintenance}.$$

Therefore,

$$\mathcal{H}^{(1)} = (V, E, T_V, T_E, \tau_V, \tau_E)$$

is a Heterogeneous 1-SuperHyperGraph.

The supervertex $v_1 = \{\mathsf{Alice}, \mathsf{Bob}\}$ is a team-type unit, $v_2 = \{\mathsf{Carol}\}$ is an individual-type unit, $v_3 = \{\mathsf{DataA}, \mathsf{DataB}\}$ is a dataset-group unit, and $v_4 = \{\mathsf{Bob}, \mathsf{Server}\}$ is a technical unit. The hyperedge $e_1$ represents a coordination relation between a team and an individual, $e_2$ represents an access-control relation among a team, a dataset-group, and a technical unit, and $e_3$ represents a maintenance relation between an individual and a technical unit.

Since

$$|T_V| = 4 > 1 \qquad \text{and} \qquad |T_E| = 3 > 1,$$

the structure is heterogeneous in both its supervertex types and superhyperedge types. An illustration is given in Fig. 5.1.

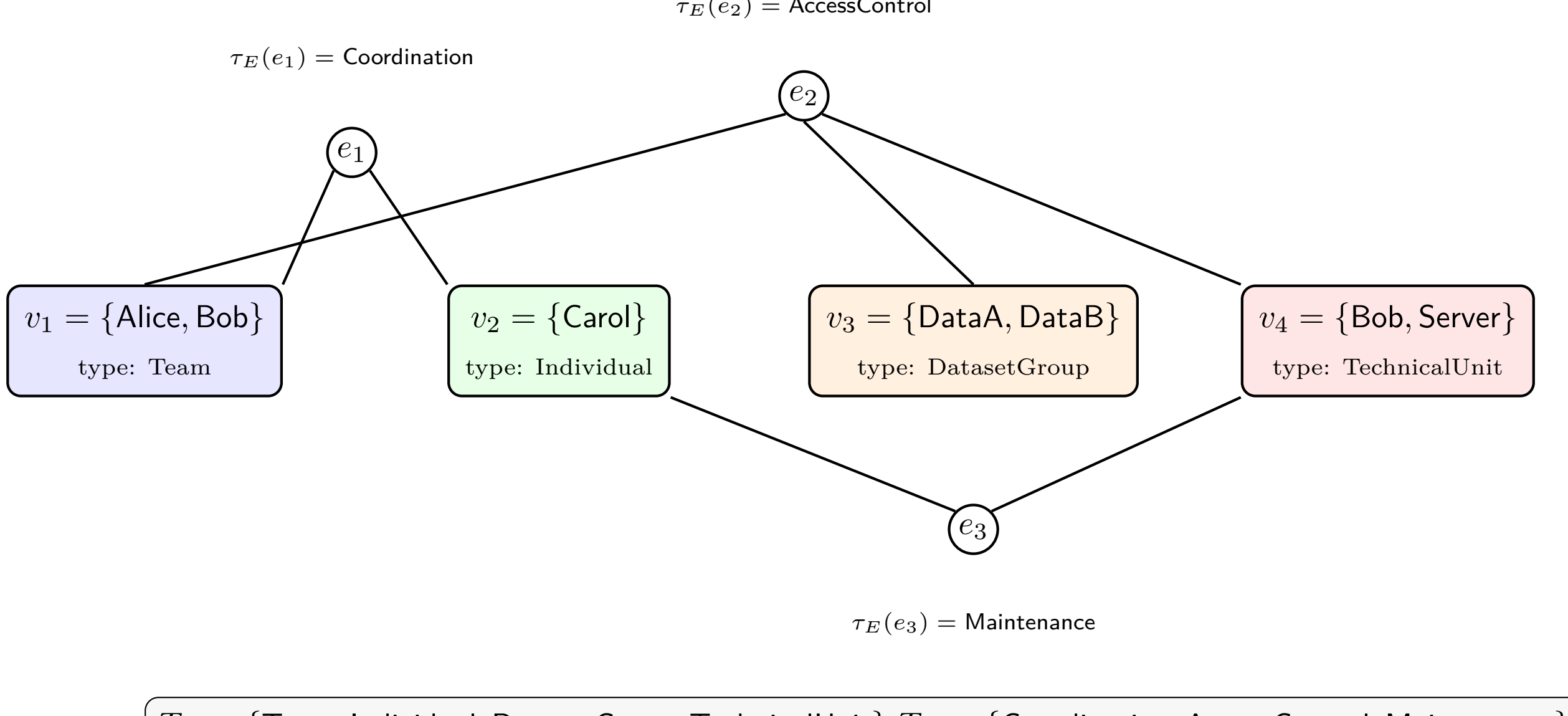

Figure 5.1: A concrete Heterogeneous 1-SuperHyperGraph. The supervertices are typed subsets of the base set, and the superhyperedges are nonempty subsets of the common supervertex set equipped with edge-types.

## 5.2 Knowledge Graph, HyperGraph, and SuperHyperGraph

A knowledge graph represents entities and their binary relations as structured facts, enabling semantic reasoning, retrieval, and integration across domains [455, 456, 457]. Related concepts, such as fuzzy knowledge graphs [458, 459, 460], Hyper-Relational Knowledge Graphs [461, 462, 463, 464], Multimodal Knowledge Graphs [465, 466, 467], and temporal knowledge graphs [468, 469, 470], are also known and have been studied. A knowledge hypergraph models entities through higher arity relations, capturing multi-entity facts and richer semantic structures within unified knowledge bases [471, 472]. A knowledge superhypergraph represents hierarchical set-based entities and relations, capturing higher-order semantic facts across multiple abstraction levels within frameworks coherently.

**Definition 5.2.1** (Knowledge Graph)**.** [455, 456, 457] Let $E$ be a finite set of entities and let $R$ be a finite set of binary relations. Define

$$\tau := \{\, r(e_1, e_2) \mid r \in R,\ e_1, e_2 \in E \,\}.$$

A *knowledge graph* is a triple

$$KG = (E, R, \tau_0),$$

where $\tau_0 \subseteq \tau$ is the set of true binary facts.

**Definition 5.2.2** (Knowledge HyperGraph)**.** [473, 474] Let $E$ be a finite set of entities and let $R$ be a finite set of relations equipped with an arity map

$$\mathrm{ar} : R \to \mathbb{N}.$$

Define

$$\tau := \{\, r(e_1, \ldots, e_{\mathrm{ar}(r)}) \mid r \in R,\ e_i \in E \text{ for all } i \,\}.$$

A *knowledge hypergraph* is a triple

$$KH = (E, R, \tau_0),$$

where $\tau_0 \subseteq \tau$ is the set of true facts.

Equivalently, each true fact

$$r(e_1, \ldots, e_{\mathrm{ar}(r)}) \in \tau_0$$

may be viewed as a hyperedge joining the participating entities.

**Definition 5.2.3** (Knowledge $n$-SuperHyperGraph)**.** [475, 476] Let $E_0$ be a finite set of base entities and let $R_0$ be a finite set of base relations. For each integer $k \geq 0$, define iterated powersets by

$$\mathcal{P}^0(E_0) := E_0, \qquad \mathcal{P}^{k+1}(E_0) := \mathcal{P}\big(\mathcal{P}^k(E_0)\big),$$

and similarly

$$\mathcal{P}^0(R_0) := R_0, \qquad \mathcal{P}^{k+1}(R_0) := \mathcal{P}\big(\mathcal{P}^k(R_0)\big).$$

Fix $n \geq 0$. Let

$$V \subseteq \mathcal{P}^n(E_0)$$

be a finite set, whose elements are called *$n$-superentities*, and let

$$\mathcal{R} \subseteq \mathcal{P}^n(R_0)$$

be a finite set, whose elements are called *$n$-superrelations*. Assume that $\mathcal{R}$ is equipped with an arity map

$$\mathrm{ar}^{(n)} : \mathcal{R} \to \mathbb{N}.$$

Define

$$\tau^{(n)} := \left\{ r(v_1, \ldots, v_{\mathrm{ar}^{(n)}(r)}) \;\middle|\; r \in \mathcal{R},\ v_i \in V \right\}.$$

A *knowledge $n$-SuperHyperGraph* is a triple

$$KH^{(n)} = (V, \mathcal{R}, \tau_0^{(n)}),$$

where

$$\tau_0^{(n)} \subseteq \tau^{(n)}$$

is the set of true $n$-superfacts.

**Example 5.2.4** (A concrete Knowledge 1-SuperHyperGraph)**.** Let the base entity set be

$$E_0 = \{\mathsf{Alice}, \mathsf{Bob}, \mathsf{Carol}, \mathsf{ProjectX}, \mathsf{ProjectY}\},$$

and let the base relation set be

$$R_0 = \{\mathsf{worksOn}, \mathsf{collaboratesWith}, \mathsf{supports}\}.$$

Take $n = 1$. Then

$$\mathcal{P}^1(E_0) = \mathcal{P}(E_0), \qquad \mathcal{P}^1(R_0) = \mathcal{P}(R_0).$$

Define the set of 1-superentities by

$$V = \{v_1, v_2, v_3\} \subseteq \mathcal{P}(E_0),$$

where

$$v_1 = \{\mathsf{Alice}, \mathsf{Bob}\}, \qquad v_2 = \{\mathsf{Carol}\}, \qquad v_3 = \{\mathsf{ProjectX}, \mathsf{ProjectY}\}.$$

Thus $v_1$ is a team of two researchers, $v_2$ is a singleton researcher-group, and $v_3$ is a project-group.

Next define the set of 1-superrelations by

$$\mathcal{R} = \{r_1, r_2\} \subseteq \mathcal{P}(R_0),$$

where

$$r_1 = \{\mathsf{collaboratesWith}, \mathsf{worksOn}\}, \qquad r_2 = \{\mathsf{supports}\}.$$

Equip $\mathcal{R}$ with the arity map

$$\mathrm{ar}^{(1)}(r_1) = 2, \qquad \mathrm{ar}^{(1)}(r_2) = 2.$$

Hence the set of all admissible 1-superfacts is

$$\tau^{(1)} = \{\, r(v_i, v_j) \mid r \in \mathcal{R},\ v_i, v_j \in V \,\}.$$

Now choose the following subset of true 1-superfacts:

$$\tau_0^{(1)} = \big\{ r_1(v_1, v_2),\ r_1(v_1, v_3),\ r_2(v_3, v_1) \big\}.$$

Therefore

$$KH^{(1)} = (V, \mathcal{R}, \tau_0^{(1)})$$

is a Knowledge 1-SuperHyperGraph in the sense of the above definition.

The true superfact

$$r_1(v_1, v_2)$$

means that the grouped relation

$$\{\mathsf{collaboratesWith}, \mathsf{worksOn}\}$$

holds from the superentity $v_1 = \{\mathsf{Alice}, \mathsf{Bob}\}$ to the superentity $v_2 = \{\mathsf{Carol}\}$. Likewise,

$$r_1(v_1, v_3)$$

states that the researcher-group $v_1$ is connected to the project-group $v_3$ by the same grouped semantic relation, and

$$r_2(v_3, v_1)$$

states that the project-group $v_3$ supports the researcher-group $v_1$.

Thus this example illustrates how a knowledge structure may be lifted from base entities and base relations to grouped entities and grouped relations through the powerset construction.

## 5.3 Petri Net

A Petri net is a directed bipartite graph of places and transitions, modeling concurrency, synchronization, resource flow, and reachability dynamically [477, 478, 479]. Related concepts such as the Fuzzy Petri Net [480, 481] Place/Transition Net (P/T Net)[482, 483, 484], Condition/Event Net [485, 486], Colored Petri Net (CPN)[487, 488, 489], Stochastic Petri Net [490, 491, 492], and the Neutrosophic Petri Net [493, 494] are also known.

**Definition 5.3.1** (Petri net)**.** [477, 478, 479] A *Petri net* is a tuple

$$\mathcal{N} = (P, T, F, W),$$

where

1. $P$ is a finite set of *places*;
2. $T$ is a finite set of *transitions*;
3. $P \cap T = \varnothing$;
4. $$F \subseteq (P \times T) \cup (T \times P)$$
   is the *flow relation* (or set of directed arcs);
5. $$W : F \to \mathbb{N}_{>0}$$
   is the *weight function*.

Thus $(P, T, F)$ is a directed bipartite graph: arcs are allowed only from places to transitions or from transitions to places.

**Definition 5.3.2** (Preset and postset)**.** Let $\mathcal{N} = (P, T, F, W)$ be a Petri net. For each transition $t \in T$, define its *preset* and *postset* by

$$^{\bullet}t := \{\, p \in P \mid (p, t) \in F \,\}, \qquad t^{\bullet} := \{\, p \in P \mid (t, p) \in F \,\}.$$

Similarly, for each place $p \in P$, define

$$^{\bullet}p := \{\, t \in T \mid (t, p) \in F \,\}, \qquad p^{\bullet} := \{\, t \in T \mid (p, t) \in F \,\}.$$

**Definition 5.3.3** (Marking and marked Petri net)**.** A *marking* of a Petri net $\mathcal{N} = (P, T, F, W)$ is a function

$$M : P \to \mathbb{N}_0.$$

For each place $p \in P$, the value $M(p)$ is called the number of *tokens* in $p$.

A *marked Petri net* is a pair

$$(\mathcal{N}, M_0),$$

where $\mathcal{N}$ is a Petri net and $M_0$ is an initial marking.

**Definition 5.3.4** (Enabled transition and firing rule)**.** Let $(\mathcal{N}, M)$ be a marked Petri net, where

$$\mathcal{N} = (P, T, F, W).$$

A transition $t \in T$ is said to be *enabled* at the marking $M$ if

$$M(p) \geq W(p, t) \qquad \text{for all } p \in {}^{\bullet}t.$$

If $t$ is enabled, then $t$ may *fire*, producing a new marking

$$M' : P \to \mathbb{N}_0$$

defined by

$$M'(p) = \begin{cases} M(p) - W(p,t) + W(t,p), & \text{if } p \in {}^{\bullet}t \cap t^{\bullet}, \\ M(p) - W(p,t), & \text{if } p \in {}^{\bullet}t \setminus t^{\bullet}, \\ M(p) + W(t,p), & \text{if } p \in t^{\bullet} \setminus {}^{\bullet}t, \\ M(p), & \text{if } p \notin {}^{\bullet}t \cup t^{\bullet}, \end{cases}$$

where, by convention, $W(x, y) = 0$ whenever $(x, y) \notin F$. Equivalently, for every $p \in P$,

$$M'(p) = M(p) - W(p, t) + W(t, p).$$

**Example 5.3.5** (A simple document-processing Petri net)**.** Consider a system in which one document is first prepared and then approved.

Let

$$P = \{p_1, p_2, p_3\}, \qquad T = \{t_1, t_2\},$$

where:

- $p_1$ represents *document available for processing*,
- $p_2$ represents *document under review*,
- $p_3$ represents *document approved*,
- $t_1$ represents *start review*,
- $t_2$ represents *approve document*.

Define the flow relation

$$F = \{(p_1, t_1), (t_1, p_2), (p_2, t_2), (t_2, p_3)\},$$

and let all arc weights be equal to 1, i.e.

$$W(f) = 1 \qquad \text{for all } f \in F.$$

Take the initial marking

$$M_0(p_1) = 1, \qquad M_0(p_2) = 0, \qquad M_0(p_3) = 0.$$

Thus the initial state contains one token in $p_1$, meaning that one document is ready to be reviewed.

At the marking $M_0$, the transition $t_1$ is enabled because

$$M_0(p_1) = 1 \geq W(p_1, t_1) = 1.$$

After firing $t_1$, the new marking $M_1$ is

$$M_1(p_1) = 0, \qquad M_1(p_2) = 1, \qquad M_1(p_3) = 0.$$

Now $t_2$ is enabled, and after firing $t_2$, one obtains

$$M_2(p_1) = 0, \qquad M_2(p_2) = 0, \qquad M_2(p_3) = 1.$$

Hence the token has moved from $p_1$ to $p_3$ through the intermediate review state $p_2$.

## 5.4 Port Graph

A port graph equips nodes with explicit connection ports, enabling structured interaction patterns, fine-grained interfaces, and rule-based rewriting [495, 496, 497].

**Definition 5.4.1** (Port Graph)**.** [498] A *port graph* is a tuple

$$G = (V, P, E, \mathrm{Attach}, \mathrm{Connect}),$$

where

1. $V$ is a finite set of *nodes*;
2. $P$ is a finite set of *ports*;
3. $E$ is a finite set of *edges*;
4. $$\mathrm{Attach} : P \to V$$
   is the *attachment map*, assigning to each port the node to which it belongs;
5. $$\mathrm{Connect} : E \to \big\{\{p, q\} \subseteq P \mid p \neq q\big\}$$
   is the *connection map*, assigning to each edge an unordered pair of distinct ports.

**Remark 5.4.2.** *For a node $v \in V$, its* interface *is the set*

$$\mathrm{Ports}(v) := \mathrm{Attach}^{-1}(\{v\}) \subseteq P.$$

*Thus, a port graph differs from an ordinary graph in that adjacency is mediated by ports. Two nodes may be connected through specific designated ports, which allows one to encode structured interfaces and fine-grained interaction patterns.*

**Example 5.4.3** (A simple signal-processing port graph)**.** Consider a small signal-processing pipeline with three components: a sensor, a filter, and an actuator.

Let the node set be

$$V = \{s, f, a\},$$

where

$$s = \mathrm{Sensor}, \qquad f = \mathrm{Filter}, \qquad a = \mathrm{Actuator}.$$

Define the port set

$$P = \{p_s^{\mathrm{out}},\, p_f^{\mathrm{in}},\, p_f^{\mathrm{out}},\, p_a^{\mathrm{in}}\}.$$

The attachment map $\mathrm{Attach} : P \to V$ is given by

$$\mathrm{Attach}(p_s^{\mathrm{out}}) = s, \qquad \mathrm{Attach}(p_f^{\mathrm{in}}) = f, \qquad \mathrm{Attach}(p_f^{\mathrm{out}}) = f, \qquad \mathrm{Attach}(p_a^{\mathrm{in}}) = a.$$

Thus the interfaces are

$$\mathrm{Ports}(s) = \{p_s^{\mathrm{out}}\}, \qquad \mathrm{Ports}(f) = \{p_f^{\mathrm{in}}, p_f^{\mathrm{out}}\}, \qquad \mathrm{Ports}(a) = \{p_a^{\mathrm{in}}\}.$$

Let the edge set be

$$E = \{e_1, e_2\},$$

and define the connection map by

$$\mathrm{Connect}(e_1) = \{p_s^{\mathrm{out}}, p_f^{\mathrm{in}}\}, \qquad \mathrm{Connect}(e_2) = \{p_f^{\mathrm{out}}, p_a^{\mathrm{in}}\}.$$

Hence

$$G = (V, P, E, \mathrm{Attach}, \mathrm{Connect})$$

is a port graph in the sense of Definition 5.4.1.

Intuitively, the first edge $e_1$ sends the sensor output to the filter input, and the second edge $e_2$ sends the filter output to the actuator input. An illustration is given in Fig. 5.2.

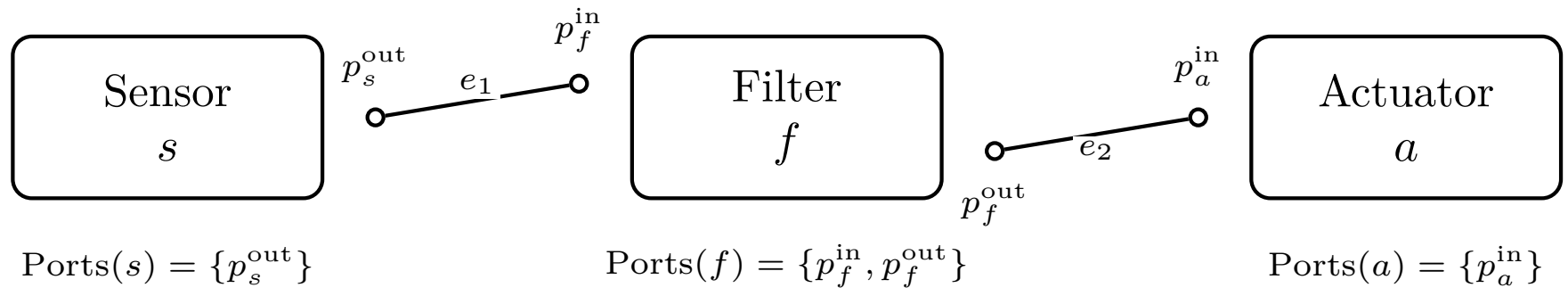

Figure 5.2: A simple port graph. Edges connect ports rather than directly connecting nodes.

## 5.5 Port HyperGraph and Port SuperHyperGraph

A Port Graph connects nodes through explicit ports, enabling fine-grained representation of interfaces, interactions, rewriting, and compositional graph structures in systems. In this section, we extend the notion of a port graph to hypergraphs and superhypergraphs. The main idea is that hyperedges (or superhyperedges) are incident not directly with vertices, but with *ports*, and each port is attached to exactly one vertex or supervertex.

**Definition 5.5.1** (Port HyperGraph)**.** A *Port HyperGraph* is a quintuple

$$\mathcal{H}^{\mathrm{port}} = (V, \Pi, E, \alpha, \iota),$$

where

1. $V$ is a finite nonempty set of *vertices*;
2. $\Pi$ is a finite set of *ports*;
3. $E$ is a finite set of *hyperedge identifiers*;
4. $$\alpha : \Pi \to V$$
   is the *port-attachment map*, assigning to each port the unique vertex to which it belongs;
5. $$\iota : E \to \mathcal{P}^*(\Pi)$$
   is the *port-incidence map*, assigning to each hyperedge identifier a nonempty set of ports.

**Remark 5.5.2.** *For each vertex $v \in V$, its* port interface *is the set*

$$\mathrm{Port}(v) := \alpha^{-1}(\{v\}) \subseteq \Pi.$$

*Thus a Port HyperGraph differs from an ordinary hypergraph in that a hyperedge is incident with a set of ports, rather than directly with a set of vertices.*

**Remark 5.5.3.** *If every hyperedge $e \in E$ satisfies*

$$|\iota(e)| = 2,$$

*then $\mathcal{H}^{\mathrm{port}}$ reduces to a graph-like port structure in which each edge joins exactly two ports. Hence Port Graphs may be viewed as a special binary case of Port HyperGraphs.*

**Theorem 5.5.4** (Well-definedness of Port HyperGraphs)**.** *Let $V$ be a finite nonempty set, let $\Pi$ and $E$ be finite sets, and let*

$$\alpha : \Pi \to V, \qquad \iota : E \to \mathcal{P}^*(\Pi)$$

*be well-defined maps. Then*

$$\mathcal{H}^{\mathrm{port}} = (V, \Pi, E, \alpha, \iota)$$

*is a well-defined Port HyperGraph.*

*Proof.* Since $V$ is a finite nonempty set and $\Pi$ is a finite set, the map

$$\alpha : \Pi \to V$$

is a legitimate set-theoretic assignment attaching each port to a unique vertex.

Since $\Pi$ is a set, its nonempty powerset

$$\mathcal{P}^*(\Pi) = \mathcal{P}(\Pi) \setminus \{\varnothing\}$$

is well-defined. Therefore the map

$$\iota : E \to \mathcal{P}^*(\Pi)$$

assigns to each hyperedge identifier $e \in E$ a nonempty subset of ports.

Hence all components appearing in Definition 5.5.1 are well-defined and mutually compatible. Therefore

$$\mathcal{H}^{\mathrm{port}} = (V, \Pi, E, \alpha, \iota)$$

is a well-defined Port HyperGraph. □

**Definition 5.5.5** (Port $n$-SuperHyperGraph)**.** Let $V_0$ be a finite nonempty base set, and let $n \in \mathbb{N}_0$. A *Port $n$-SuperHyperGraph* is a quintuple

$$\mathcal{SH}^{(n),\mathrm{port}} = (V, \Pi, E, \alpha, \iota),$$

where

1. $$V \subseteq \mathcal{P}^n(V_0)$$

   is a finite set of *$n$-supervertices*;

2. $\Pi$ is a finite set of *ports*;

3. $E$ is a finite set of *superhyperedge identifiers*;

4. $$\alpha : \Pi \to V$$

   is the *port-attachment map*, assigning each port to a unique $n$-supervertex;

5. $$\iota : E \to \mathcal{P}^*(\Pi)$$

   is the *port-superincidence map*, assigning to each superhyperedge identifier a nonempty set of ports.

**Remark 5.5.6.** *For each supervertex $v \in V$, its* port interface *is*

$$\mathrm{Port}(v) := \alpha^{-1}(\{v\}) \subseteq \Pi.$$

*Thus a Port $n$-SuperHyperGraph is a superhypergraph-like structure in which incidence is mediated by ports attached to $n$-supervertices.*

**Remark 5.5.7.** *A* Port SuperHyperGraph *means a Port $n$-SuperHyperGraph for some fixed*

$$n \in \mathbb{N}_0.$$

*In particular, when $n = 0$, one has*

$$\mathcal{P}^0(V_0) = V_0,$$

*so a Port $0$-SuperHyperGraph is exactly a Port HyperGraph.*

**Theorem 5.5.8** (Well-definedness of Port $n$-SuperHyperGraphs)**.** *Let $V_0$ be a finite nonempty set, let $n \in \mathbb{N}_0$, let*

$$V \subseteq \mathcal{P}^n(V_0)$$

*be a finite set, let $\Pi$ and $E$ be finite sets, and let*

$$\alpha : \Pi \to V, \qquad \iota : E \to \mathcal{P}^*(\Pi)$$

*be well-defined maps. Then*

$$\mathcal{SH}^{(n),\mathrm{port}} = (V, \Pi, E, \alpha, \iota)$$

*is a well-defined Port $n$-SuperHyperGraph.*

*Proof.* First, since $V_0$ is a finite nonempty set and $n \in \mathbb{N}_0$, the iterated powerset

$$\mathcal{P}^n(V_0)$$

is well-defined by finite recursion. Hence the condition

$$V \subseteq \mathcal{P}^n(V_0)$$

is meaningful, and $V$ is a well-defined finite set of $n$-supervertices.

Next, since $\Pi$ is a finite set, its nonempty powerset

$$\mathcal{P}^*(\Pi) = \mathcal{P}(\Pi) \setminus \{\varnothing\}$$

is well-defined. Therefore the map

$$\iota : E \to \mathcal{P}^*(\Pi)$$

assigns to each superhyperedge identifier a nonempty set of ports.

Moreover, the map

$$\alpha : \Pi \to V$$

is well-defined, so each port is attached to a unique $n$-supervertex.

Thus all components listed in Definition 5.5.5 are well-defined and mutually compatible. Therefore

$$\mathcal{SH}^{(n),\mathrm{port}} = (V, \Pi, E, \alpha, \iota)$$

is a well-defined Port $n$-SuperHyperGraph. □

**Corollary 5.5.9.** *Every Port HyperGraph is a Port* $0$*-SuperHyperGraph.*

*Proof.* If $n = 0$, then

$$\mathcal{P}^0(V_0) = V_0,$$

so the supervertex set is an ordinary vertex set. Hence Definition 5.5.5 reduces exactly to Definition 5.5.1. □

## 5.6 Open Hypergraph and Open SuperHyperGraph

An open hypergraph is a hypergraph with designated inputs and outputs, supporting modular composition of networked systems through cospans categorically.

**Definition 5.6.1** (Discrete hypergraph)**.** For a finite set $X$, the *discrete hypergraph* on $X$ is

$$D(X) := (X, \varnothing).$$

Thus $D(X)$ has node set $X$ and no hyperedges.

**Definition 5.6.2** (Open Hypergraph)**.** Let $X$ and $Y$ be finite sets. An *open hypergraph from* $X$ *to* $Y$ is a triple

$$\mathcal{O} = (H, i, o),$$

where

$$H = (V, \mathcal{E})$$

is a finite hypergraph, and

$$i : X \to V, \qquad o : Y \to V$$

are set maps, called the *input interface map* and the *output interface map*, respectively.

Equivalently, an open hypergraph may be written as a cospan

$$D(X) \xrightarrow{i} H \xleftarrow{o} D(Y),$$

where $D(X)$ and $D(Y)$ are discrete hypergraphs.

**Remark 5.6.3.** *The maps* $i$ *and* $o$ *specify which vertices of the ambient hypergraph serve as boundary vertices. The maps need not be injective unless one explicitly imposes that restriction.*

**Definition 5.6.4** (Open $n$-SuperHyperGraph)**.** Let $V_0$ be a finite nonempty base set, let $n \in \mathbb{N}_0$, and let

$$\mathrm{SHG}^{(n)} = (V, E, \partial)$$

be an $n$-SuperHyperGraph over $V_0$.

Let $X$ and $Y$ be finite sets. An *open $n$-SuperHyperGraph from $X$ to $Y$* is a tuple

$$\mathcal{O}^{(n)} = (V, E, \partial, X, Y, i, o),$$

such that

1. $(V, E, \partial)$ is an $n$-SuperHyperGraph over $V_0$;
2. $$i : X \to V$$
   is an *input interface map*;
3. $$o : Y \to V$$
   is an *output interface map.*

The elements of $X$ and $Y$ are called *input interface labels* and *output interface labels*, respectively. For $x \in X$, the value $i(x) \in V$ is the input boundary supervertex selected by $x$, and for $y \in Y$, the value $o(y) \in V$ is the output boundary supervertex selected by $y$.

**Remark 5.6.5.** *An open $n$-SuperHyperGraph is therefore an $n$-SuperHyperGraph equipped with specified input and output interfaces. It extends the notion of an open hypergraph by replacing ordinary vertices with $n$-supervertices.*

**Remark 5.6.6** (Special case $n = 0$)**.** *When $n = 0$, one has*

$$\mathcal{P}^0(V_0) = V_0,$$

*so the supervertex set $V$ is an ordinary set of vertices. Hence an open $0$-SuperHyperGraph is precisely an open hypergraph, up to the harmless use of edge identifiers and the incidence map $\partial$.*

**Theorem 5.6.7** (Well-definedness of Open $n$-SuperHyperGraphs)**.** *Let $V_0$ be a finite nonempty set and let $n \in \mathbb{N}_0$. Assume that:*

1. $$V \subseteq \mathcal{P}^n(V_0)$$
   *is a finite set;*
2. *$E$ is a finite set;*
3. $$\partial : E \to \mathcal{P}^*(V)$$
   *is a well-defined map;*
4. *$X$ and $Y$ are finite sets;*
5. $$i : X \to V, \qquad o : Y \to V$$
   *are well-defined maps.*

*Then*

$$\mathcal{O}^{(n)} = (V, E, \partial, X, Y, i, o)$$

*is a well-defined open $n$-SuperHyperGraph over $V_0$.*

*Proof.* We verify each component of the definition.

First, since $V_0$ is a finite nonempty set and $n \in \mathbb{N}_0$, the iterated powerset $\mathcal{P}^n(V_0)$ is well-defined by finite recursion:

$$\mathcal{P}^0(V_0) = V_0, \qquad \mathcal{P}^{k+1}(V_0) = \mathcal{P}(\mathcal{P}^k(V_0)).$$

Hence the condition

$$V \subseteq \mathcal{P}^n(V_0)$$

is meaningful, and $V$ is a well-defined finite set of $n$-supervertices.

Next, since $V$ is a set, the family

$$\mathcal{P}^*(V) = \mathcal{P}(V) \setminus \{\varnothing\}$$

is also well-defined. Therefore the incidence map

$$\partial : E \to \mathcal{P}^*(V)$$

is a legitimate set-theoretic map assigning to each superedge identifier a nonempty subset of $V$. Consequently,

$$(V, E, \partial)$$

is a well-defined $n$-SuperHyperGraph over $V_0$.

Finally, $X$ and $Y$ are finite sets by assumption, and

$$i : X \to V, \qquad o : Y \to V$$

are well-defined maps into the supervertex set $V$. Thus the input and output interfaces are well-defined.

All components required in Definition 5.6.4 therefore exist and are compatible. Hence

$$\mathcal{O}^{(n)} = (V, E, \partial, X, Y, i, o)$$

is a well-defined open $n$-SuperHyperGraph. □

**Corollary 5.6.8.** *For every $n$-SuperHyperGraph $(V, E, \partial)$ over $V_0$, and for every pair of finite sets $X, Y$ equipped with maps $i : X \to V$ and $o : Y \to V$, there exists an associated open $n$-SuperHyperGraph*

$$(V, E, \partial, X, Y, i, o).$$

*Proof.* Immediate from Theorem 5.6.7. □

## 5.7 Combinatorial Map

A combinatorial map is a connected properly edge-colored regular graph in which each vertex is incident with exactly one edge of each color [499, 500].

**Definition 5.7.1** (Combinatorial map)**.** Let $I$ be a finite nonempty set of colors. A *combinatorial map over $I$* is a pair

$$\mathcal{G} = (G, \tau),$$

where $G = (V, E)$ is a connected graph and

$$\tau : E \to I$$

is an edge-coloring such that, for every vertex $v \in V$ and every color $i \in I$, there exists a unique edge $e \in E$ incident with $v$ satisfying

$$\tau(e) = i.$$

Equivalently, $G$ is $|I|$-regular and no two distinct edges incident with the same vertex have the same color.

The integer

$$\mathrm{rank}(\mathcal{G}) := |I|$$

is called the *rank* of $\mathcal{G}$.

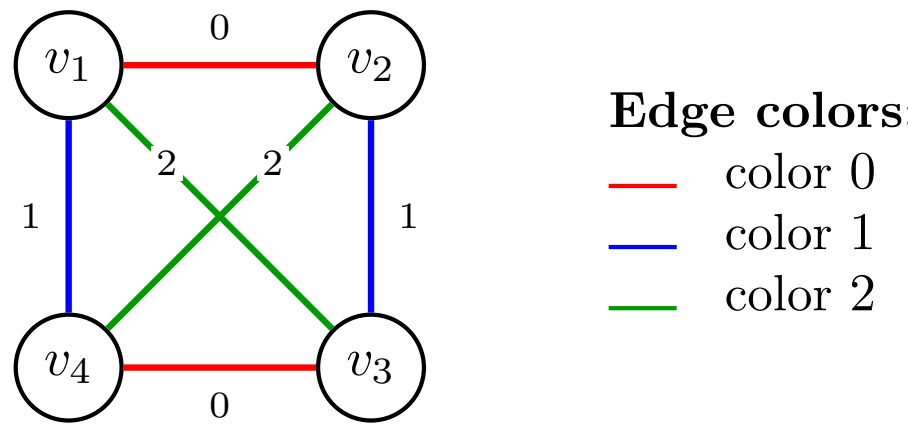

Figure 5.3: A rank-3 combinatorial map on $K_4$. Each vertex is incident with exactly one edge of each color $0, 1, 2$.

**Definition 5.7.2** (Residues and faces)**.** Let $\mathcal{G} = (G, \tau)$ be a combinatorial map over $I$, and let $J \subseteq I$. Define the spanning subgraph

$$G_J := (V, \tau^{-1}(J)),$$

that is, the graph obtained from $G$ by retaining exactly the edges whose colors belong to $J$. A connected component of $G_J$ is called a *residue of type* $J$.

For $i \in I$, a residue of type

$$I \setminus \{i\}$$

is called an $i$*-face* of $\mathcal{G}$.

**Remark 5.7.3.** *The above definition is the standard graph-theoretic abstraction of maps on surfaces and their higher-rank analogues. In rank* 3*, one may think of the three colors as corresponding to the three kinds of adjacency arising from flags of a cell decomposition.*

**Example 5.7.4** (A rank-3 combinatorial map associated with the tetrahedron)**.** Let

$$I = \{0, 1, 2\}, \qquad V = \{v_1, v_2, v_3, v_4\}.$$

Take $G$ to be the complete graph $K_4$ on $V$, with edge set

$$E = \big\{\{v_1, v_2\}, \{v_3, v_4\}, \{v_1, v_4\}, \{v_2, v_3\}, \{v_1, v_3\}, \{v_2, v_4\}\big\}.$$

Define the edge-coloring $\tau : E \to I$ by

$$\tau(\{v_1, v_2\}) = \tau(\{v_3, v_4\}) = 0,$$

$$\tau(\{v_1, v_4\}) = \tau(\{v_2, v_3\}) = 1,$$

$$\tau(\{v_1, v_3\}) = \tau(\{v_2, v_4\}) = 2.$$

Then each vertex is incident with exactly one edge of color 0, one edge of color 1, and one edge of color 2. For example,

$$v_1 \text{ is incident with } \{v_1, v_2\} \text{ of color } 0,\ \{v_1, v_4\} \text{ of color } 1,\ \{v_1, v_3\} \text{ of color } 2.$$

The same property holds for $v_2, v_3, v_4$. Hence

$$\mathcal{G} = (G, \tau)$$

is a combinatorial map of rank 3.

Its 0-faces are the residues of type $\{1, 2\}$, its 1-faces are the residues of type $\{0, 2\}$, and its 2-faces are the residues of type $\{0, 1\}$. In this example, each such residue is a 4-cycle. For instance, the unique residue of type $\{1, 2\}$ has edge set

$$\big\{\{v_1, v_4\}, \{v_2, v_3\}, \{v_1, v_3\}, \{v_2, v_4\}\big\},$$

which forms the cycle

$$v_4 - v_1 - v_3 - v_2 - v_4.$$

An illustration is given in Fig. 5.3.

## 5.8 Cognitive HyperGraphs and Cognitive SuperHyperGraphs

In this section, we define the notions of a Cognitive HyperGraph and a Cognitive $n$-SuperHyperGraph as higher-order extensions of a cognitive graph. A cognitive graph represents concepts and causal relationships, supporting reasoning, decision-making, knowledge representation, and analysis of cognitive systems and processes [501, 502, 503]. Related concepts, such as fuzzy cognitive graphs [504, 505, 506, 507], are also known and have been studied. A Cognitive HyperGraph models cognitive entities as labeled vertices and multiway cognitive relations as labeled hyperedges, capturing structured group interactions semantically[508]. A Cognitive SuperHyperGraph represents higher-order cognitive groupings using iterated-powerset supervertices and labeled superhyperedges, expressing nested semantic relations among concept collections.

**Definition 5.8.1** (Cognitive HyperGraph)**.** Let $S$ be a nonempty set of distinguished locations, concepts, or cognitive units. Let $L_V$ and $L_E$ be nonempty label sets.

A *Cognitive HyperGraph* is a quadruple

$$\mathcal{CH} = (V, E, \ell_V, \ell_E),$$

where

1. $$V \subseteq S$$
   is a finite set of *cognitive vertices*;
2. $$E \subseteq \mathcal{P}^*(V)$$
   is a finite set of nonempty *cognitive hyperedges*;
3. $$\ell_V : V \to L_V$$
   is a *vertex-label function*;
4. $$\ell_E : E \to L_E$$
   is a *hyperedge-label function.*

Each hyperedge $e \in E$ represents a cognitively meaningful multiway relation among the vertices contained in $e$, while $\ell_V$ and $\ell_E$ encode semantic information attached to vertices and hyperedges, respectively.

**Remark 5.8.2.** *If every hyperedge has cardinality* 2*, then a Cognitive HyperGraph reduces to a labeled cognitive graph.*

**Definition 5.8.3** (Cognitive $n$-SuperHyperGraph)**.** Let $S$ be a nonempty base set, let $n \in \mathbb{N}_0$, and let $L_V, L_E$ be nonempty label sets.

A *Cognitive $n$-SuperHyperGraph* is a quadruple

$$\mathcal{CSH}^{(n)} = (V, E, \ell_V, \ell_E),$$

such that

1. $$V \subseteq \mathcal{P}^n(S)$$
   is a finite set of *cognitive $n$-supervertices*;
2. $$E \subseteq \mathcal{P}^*(V)$$
   is a finite set of nonempty *cognitive $n$-superhyperedges*;
3. $$\ell_V : V \to L_V$$
   is a *supervertex-label function*;

4.

$$\ell_E : E \to L_E$$

is a *superhyperedge-label function.*

Thus, a Cognitive $n$-SuperHyperGraph is a labeled higher-order relational structure in which the vertices themselves lie at the $n$-th iterated powerset level over the base set $S$, and the superedges are nonempty subsets of the supervertex set $V$.

## 5.9 Multimodal Graph, HyperGraph, and SuperHyperGraph

A multimodal graph represents a common vertex set through multiple edge modalities, thereby capturing different types of interactions or complementary views within a single network [509, 510, 511]. Owing to its importance and broad applicability, various related extensions have also been studied, including multimodal fuzzy graphs[512, 513], Multimodal Heterogeneous Graphs [514, 515, 516], Temporal Multimodal Graphs [517, 518], and multimodal knowledge graphs [519, 520, 521]. A multimodal hypergraph uses multiple hyperedge modalities on a shared vertex set to model higher-order relations simultaneously within one framework [522, 523, 524]. A multimodal superhypergraph organizes multiple modalities over hierarchical set-based vertices and superhyperedges, describing complex higher-order relations across abstraction levels simultaneously [476].

**Definition 5.9.1** (Multimodal Graph)**.** Let $V$ be a finite nonempty set of vertices, and let $M \in \mathbb{N}$ be the number of modalities. For each modality $m \in \{1, \ldots, M\}$, let

$$G_m = (V, E_m, w_m)$$

be a weighted graph on the common vertex set $V$, where

$$E_m \subseteq \big\{\{u, v\} \subseteq V : u \neq v\big\}$$

is the edge set of modality $m$, and

$$w_m : E_m \to \mathbb{R}_{>0}$$

is a positive edge-weight function. Let

$$\alpha_1, \ldots, \alpha_M \geq 0, \qquad \sum_{m=1}^{M} \alpha_m = 1,$$

be modality-combination weights.

Then the *multimodal graph* is the tuple

$$\mathcal{G} = \big(V, \{E_m\}_{m=1}^{M}, \{w_m\}_{m=1}^{M}, \{\alpha_m\}_{m=1}^{M}\big).$$

**Definition 5.9.2** (Multimodal HyperGraph)**.** Let $V$ be a finite nonempty set of vertices, and let $M \in \mathbb{N}$. For each modality $m \in \{1, \ldots, M\}$, let

$$H_m = (V, \mathcal{E}_m, w_m)$$

be a weighted hypergraph on the common vertex set $V$, where

$$\mathcal{E}_m \subseteq \mathcal{P}^*(V) := \mathcal{P}(V) \setminus \{\emptyset\}$$

is the set of hyperedges of modality $m$, and

$$w_m : \mathcal{E}_m \to \mathbb{R}_{>0}$$

is a positive hyperedge-weight function. Let

$$\alpha_1, \ldots, \alpha_M \geq 0, \qquad \sum_{m=1}^{M} \alpha_m = 1.$$

Then the *multimodal hypergraph* is the tuple

$$\mathcal{H} = \big(V, \{\mathcal{E}_m\}_{m=1}^{M}, \{w_m\}_{m=1}^{M}, \{\alpha_m\}_{m=1}^{M}\big).$$

**Definition 5.9.3** (Multimodal $n$-SuperHyperGraph)**.** [476] Let $V_0$ be a finite nonempty base set, and let $n \in \mathbb{N} \cup \{0\}$. Define the iterated powersets recursively by

$$\mathcal{P}^0(V_0) := V_0, \qquad \mathcal{P}^{k+1}(V_0) := \mathcal{P}\big(\mathcal{P}^k(V_0)\big) \quad (k \geq 0).$$

Let

$$V \subseteq \mathcal{P}^n(V_0)$$

be a finite set of $n$-supervertices, and let $M \in \mathbb{N}$. For each modality $m \in \{1, \ldots, M\}$, let

$$\mathcal{H}_m^{(n)} = (V, \mathcal{E}_m^{(n)}, w_m)$$

be a weighted hypergraph on the common set $V$, where

$$\mathcal{E}_m^{(n)} \subseteq \mathcal{P}^*(V)$$

is the set of $n$-superhyperedges of modality $m$, and

$$w_m : \mathcal{E}_m^{(n)} \to \mathbb{R}_{>0}$$

is a positive weight function. Let

$$\alpha_1, \ldots, \alpha_M \geq 0, \qquad \sum_{m=1}^{M} \alpha_m = 1.$$

Then the *multimodal $n$-SuperHyperGraph* is the tuple

$$\mathcal{H}^{(n)} = \big(V, \{\mathcal{E}_m^{(n)}\}_{m=1}^{M}, \{w_m\}_{m=1}^{M}, \{\alpha_m\}_{m=1}^{M}\big).$$

**Example 5.9.4** (A concrete Multimodal 1-SuperHyperGraph)**.** Let the base set be

$$V_0 = \{a, b, c, d, e\},$$

and take $n = 1$. Then

$$\mathcal{P}^1(V_0) = \mathcal{P}(V_0),$$

so each 1-supervertex is a subset of $V_0$.

Define the common set of 1-supervertices by

$$V = \{v_1, v_2, v_3, v_4\} \subseteq \mathcal{P}(V_0),$$

where

$$v_1 = \{a, b\}, \qquad v_2 = \{b, c, d\}, \qquad v_3 = \{e\}, \qquad v_4 = \{a, e\}.$$

Thus each supervertex represents a group of base elements.

Now let the number of modalities be

$$M = 2.$$

**Modality 1: communication.** Define the weighted hypergraph

$$\mathcal{H}_1^{(1)} = (V, \mathcal{E}_1^{(1)}, w_1)$$

by

$$\mathcal{E}_1^{(1)} = \{E_1^{(1)}, E_2^{(1)}\},$$

where

$$E_1^{(1)} = \{v_1, v_2\}, \qquad E_2^{(1)} = \{v_2, v_3\},$$

and assign positive weights

$$w_1\Big(E_1^{(1)}\Big) = 0.8, \qquad w_1\Big(E_2^{(1)}\Big) = 0.6.$$

**Modality 2: resource sharing.** Define the weighted hypergraph

$$\mathcal{H}_2^{(1)} = (V, \mathcal{E}_2^{(1)}, w_2)$$

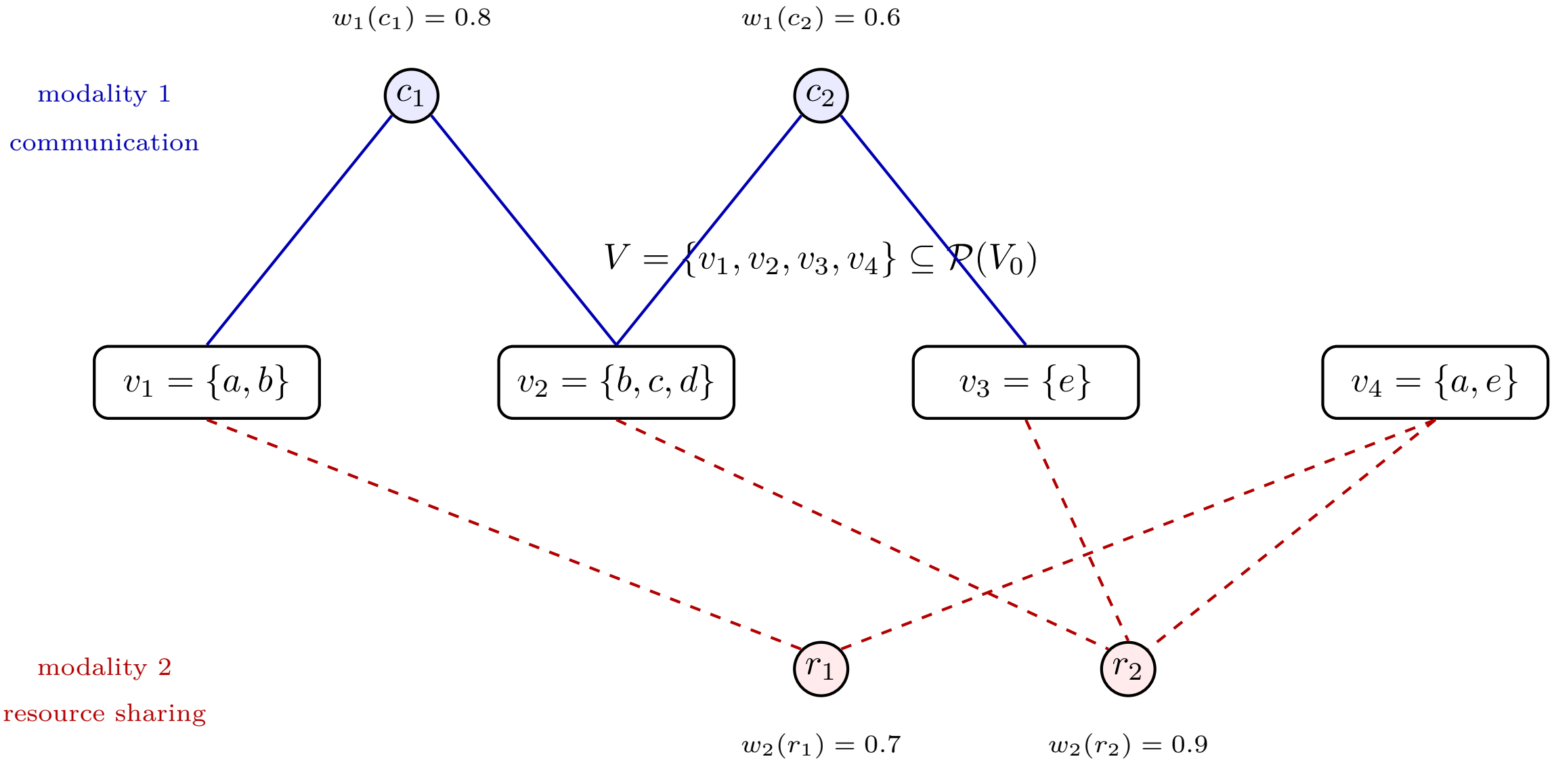

$\alpha_1 = 0.55,\ \alpha_2 = 0.45, \qquad \alpha_1 + \alpha_2 = 1.$ Same supervertices, different modalities.

Figure 5.4: A concrete Multimodal 1-SuperHyperGraph. The same set of 1-supervertices is shared by two modalities: communication (solid blue) and resource sharing (dashed red).

by

$$\mathcal{E}_2^{(1)} = \{E_1^{(2)}, E_2^{(2)}\},$$

where

$$E_1^{(2)} = \{v_1, v_4\}, \qquad E_2^{(2)} = \{v_2, v_3, v_4\},$$

and assign positive weights

$$w_2\Big(E_1^{(2)}\Big) = 0.7, \qquad w_2\Big(E_2^{(2)}\Big) = 0.9.$$

Finally, choose modality-combination weights

$$\alpha_1 = 0.55, \qquad \alpha_2 = 0.45,$$

so that

$$\alpha_1, \alpha_2 \geq 0, \qquad \alpha_1 + \alpha_2 = 1.$$

Therefore,

$$\mathcal{H}^{(1)} = \big(V, \{\mathcal{E}_1^{(1)}, \mathcal{E}_2^{(1)}\}, \{w_1, w_2\}, \{\alpha_1, \alpha_2\}\big)$$

is a Multimodal 1-SuperHyperGraph in the sense of the above definition.

The common supervertex set $V$ is shared by both modalities. The first modality describes communication relations among groups, while the second modality describes resource-sharing relations among the same groups. Hence the same hierarchical vertex system is observed through multiple distinct interaction modes. An illustration is given in Fig. 5.4.

## 5.10 Operadic Interaction Graph (OIG)

An Operadic Interaction Graph is a colored operad whose operations represent typed multi-input interactions; operad composition models gluing interactions, encoding higher-order, compositional network structure.

**Definition 5.10.1** (Operadic Interaction Graph (OIG))**.** Let $C$ be a set of *colors* (types). An *Operadic Interaction Graph* is a symmetric $C$-colored operad

$$\mathsf{OIG} = \mathcal{O} = \big(\mathcal{O}(c_1, \ldots, c_n; c)\big)_{n \geq 0,\ (c_1, \ldots, c_n; c) \in C^n \times C}$$

consisting of:

1. operation sets $\mathcal{O}(c_1, \ldots, c_n; c)$ for all $n \geq 0$ and $(c_1, \ldots, c_n; c) \in C^n \times C$;
2. right actions of the symmetric groups, i.e. for each $\pi \in S_n$,

$$(-) \cdot \pi : \mathcal{O}(c_1, \ldots, c_n; c) \to \mathcal{O}(c_{\pi(1)}, \ldots, c_{\pi(n)}; c);$$

3. substitution (composition) maps

$$\gamma : \mathcal{O}(c_1, \ldots, c_n; c) \times \prod_{i=1}^{n} \mathcal{O}(d_{i1}, \ldots, d_{ik_i}; c_i) \to \mathcal{O}(d_{11}, \ldots, d_{1k_1}, \ldots, d_{n1}, \ldots, d_{nk_n}; c);$$

4. units $\mathrm{id}_c \in \mathcal{O}(c; c)$ for all $c \in C$,

satisfying the standard operad axioms (associativity of substitution, unitality, and equivariance with respect to the $S_n$-actions). Operations are interpreted as higher-arity interaction primitives, and $\gamma$ encodes compositional gluing of interactions.

**Example 5.10.2** (An Operadic Interaction Graph for typed workflows)**.** Let the color set be

$$C = \{\mathsf{Raw}, \mathsf{Clean}, \mathsf{Model}\},$$

interpreted as data/product types in a workflow. Define a symmetric $C$-colored operad $\mathcal{O}$ by specifying the following generating operations:

$$f \in \mathcal{O}(\mathsf{Raw}; \mathsf{Clean}) \quad \text{(cleaning)}, \qquad g \in \mathcal{O}(\mathsf{Clean}, \mathsf{Clean}; \mathsf{Model}) \quad \text{(training from two datasets)}.$$

Include all operations obtained from these generators by operadic substitution and by the symmetric actions (permuting the two inputs of $g$), together with the required units

$$\mathrm{id}_{\mathsf{Raw}} \in \mathcal{O}(\mathsf{Raw}; \mathsf{Raw}), \quad \mathrm{id}_{\mathsf{Clean}} \in \mathcal{O}(\mathsf{Clean}; \mathsf{Clean}), \quad \mathrm{id}_{\mathsf{Model}} \in \mathcal{O}(\mathsf{Model}; \mathsf{Model}).$$

For instance, the composite operation

$$h \;=\; \gamma\big(g;\; f, f\big) \;\in\; \mathcal{O}(\mathsf{Raw}, \mathsf{Raw}; \mathsf{Model})$$

represents the workflow "clean two raw datasets and then train a model from the two cleaned datasets." With these operation sets, symmetric actions, units, and substitution maps, $\mathcal{O}$ forms a symmetric $C$-colored operad, and hence defines an Operadic Interaction Graph in the sense of Definition 5.10.1. An overview diagram of this example is provided in Fig. 5.5.

**Theorem 5.10.3** (Well-definedness of Operadic Interaction Graph semantics)**.** *Let $C$ be a set of colors, and let*

$$\mathsf{OIG} = \mathcal{O} = \big(\mathcal{O}(c_1, \ldots, c_n; c)\big)_{n \geq 0,\ (c_1, \ldots, c_n; c) \in C^n \times C}$$

*be the data of Definition 5.10.1, together with symmetric-group actions, substitution maps, and units, satisfying the colored operad axioms (associativity, unitality, and equivariance). Then the notion of an* Operadic Interaction Graph *is well-defined. More precisely:*

1. *For every profile $(c_1, \ldots, c_n; c) \in C^n \times C$, the set $\mathcal{O}(c_1, \ldots, c_n; c)$ is a well-typed set of $n$-ary operations with input colors $c_1, \ldots, c_n$ and output color $c$.*
2. *For every $\pi \in S_n$, the action map*

$$(-) \cdot \pi : \mathcal{O}(c_1, \ldots, c_n; c) \to \mathcal{O}(c_{\pi(1)}, \ldots, c_{\pi(n)}; c)$$

*is well-defined and preserves output color while permuting input colors.*

3. *For every choice of profiles*

$$g \in \mathcal{O}(c_1, \ldots, c_n; c), \qquad f_i \in \mathcal{O}(d_{i1}, \ldots, d_{ik_i}; c_i) \quad (1 \leq i \leq n),$$

*the substitution*

$$\gamma(g; f_1, \ldots, f_n)$$

*is well-defined and lies in*

$$\mathcal{O}(d_{11}, \ldots, d_{1k_1}, \ldots, d_{n1}, \ldots, d_{nk_n}; c).$$

**Raw** **Cleaning** **Clean** **Training** **Model**

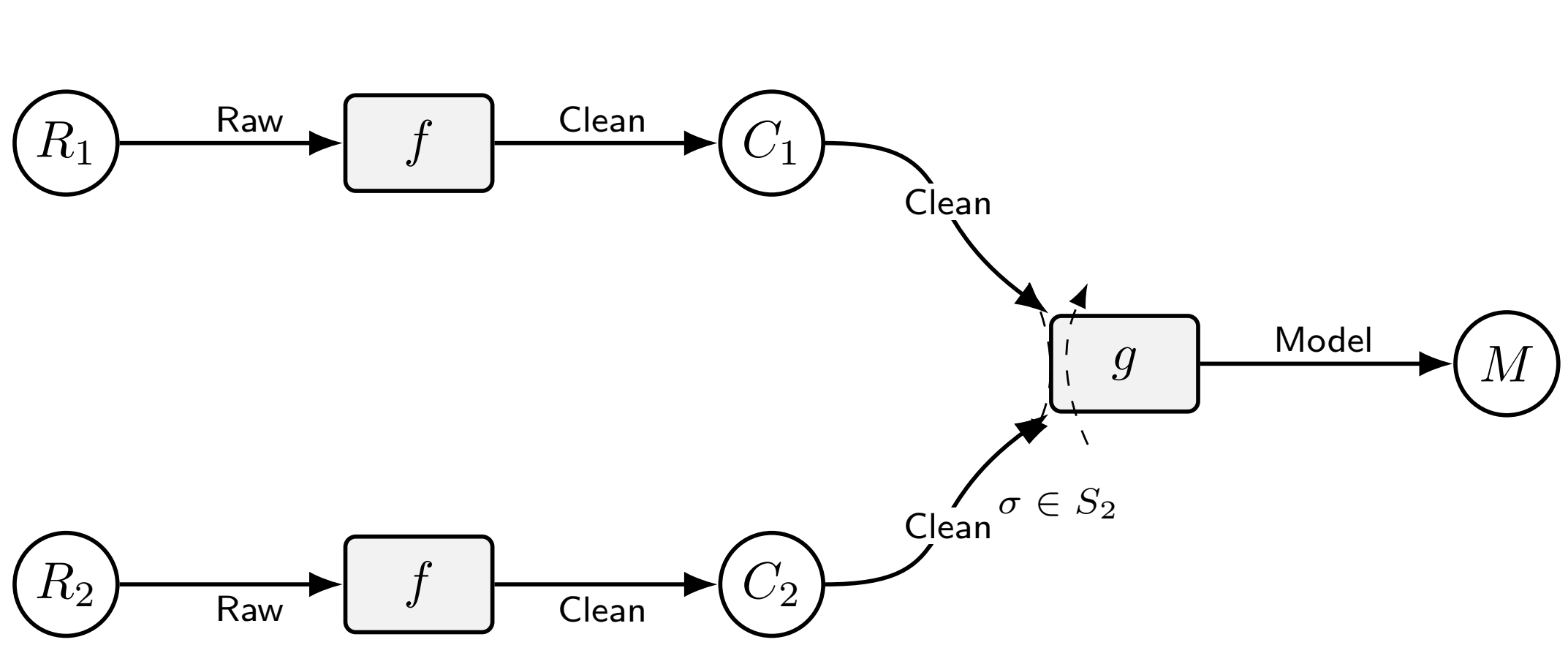

**Typed operations:** $f \in \mathcal{O}(\mathsf{Raw}; \mathsf{Clean})$ (cleaning), $g \in \mathcal{O}(\mathsf{Clean}, \mathsf{Clean}; \mathsf{Model})$ (training).
**Operadic composition:** $h = \gamma(g; f, f) \in \mathcal{O}(\mathsf{Raw}, \mathsf{Raw}; \mathsf{Model})$. Interpretation: clean two raw datasets, then train a model from the two cleaned datasets.

Figure 5.5: An Operadic Interaction Graph for a typed workflow. The composite workflow is the operadic substitution $h = \gamma(g; f, f)$.

4. *Any formal interaction expression built from operations in $\mathcal{O}$, units, symmetric actions, and iterated substitutions has a well-defined typed value, and its value is independent of the order of evaluation (parenthesization of substitutions) and coherent permutation choices, by the operad axioms.*

*Hence the interpretation of operations as higher-arity interaction primitives and of $\gamma$ as compositional gluing is mathematically well-posed.*

*Proof.* We verify each point.

**(1) Well-typed operation sets.** By assumption, for each $n \geq 0$ and each profile $(c_1, \dots, c_n; c) \in C^n \times C$, a set $\mathcal{O}(c_1, \dots, c_n; c)$ is specified. Thus every element $x \in \mathcal{O}(c_1, \dots, c_n; c)$ has a definite type: it is an $n$-ary operation with ordered input colors $c_1, \dots, c_n$ and output color $c$. This is exactly the required typing data.

**(2) Symmetric actions are well-defined.** Fix $n \geq 0$, a profile $(c_1, \dots, c_n; c)$, and $\pi \in S_n$. By assumption, there is a map

$$(-) \cdot \pi : \mathcal{O}(c_1, \dots, c_n; c) \to \mathcal{O}(c_{\pi(1)}, \dots, c_{\pi(n)}; c).$$

Since $(c_{\pi(1)}, \dots, c_{\pi(n)}) \in C^n$, the codomain is a valid operation set. Hence the action is well-defined as a type-preserving reindexing of the inputs (with the same output color $c$). For $n = 0$, this is also well-defined because $S_0$ is the trivial group.

**(3) Substitution is well-defined and correctly typed.** Let

$$g \in \mathcal{O}(c_1, \dots, c_n; c), \qquad f_i \in \mathcal{O}(d_{i1}, \dots, d_{ik_i}; c_i) \quad (1 \leq i \leq n).$$

The output color of each $f_i$ is $c_i$, which matches the $i$-th input color of $g$. Therefore the tuple $(g; f_1, \dots, f_n)$ is composable.

Now consider the concatenated list of colors

$$(d_{11}, \dots, d_{1k_1}, \dots, d_{n1}, \dots, d_{nk_n}).$$

Each entry lies in $C$, so this is an element of $C^{k_1 + \cdots + k_n}$. Hence the target set

$$\mathcal{O}(d_{11}, \dots, d_{1k_1}, \dots, d_{n1}, \dots, d_{nk_n}; c)$$

is a legitimate component of the colored operad.

By assumption, the substitution map

$$\gamma : \mathcal{O}(c_1, \dots, c_n; c) \times \prod_{i=1}^{n} \mathcal{O}(d_{i1}, \dots, d_{ik_i}; c_i) \to \mathcal{O}(d_{11}, \dots, d_{1k_1}, \dots, d_{n1}, \dots, d_{nk_n}; c)$$

is defined, so $\gamma(g; f_1, \ldots, f_n)$ exists and has the stated type. Therefore substitution is well-defined.

**(4) Formal operadic expressions have well-defined values.** A formal interaction expression is built inductively from:

- basic operations $x \in \mathcal{O}(c_1, \ldots, c_n; c)$,
- units $\mathrm{id}_c \in \mathcal{O}(c; c)$,
- symmetric actions $x \mapsto x \cdot \pi$,
- substitutions $\gamma(-; -, \ldots, -)$.

By (1)–(3), each construction step is well-typed whenever the colors match, so every such formal expression has a well-defined output color and a well-defined ordered list of input colors.

It remains to show independence of evaluation order and coherent permutations. This follows from the operad axioms:

- *Associativity of substitution* implies that iterated substitutions give the same result regardless of how one parenthesizes the gluing process.
- *Unitality* implies that inserting or removing the units $\mathrm{id}_c$ does not change the operation.
- *Equivariance* implies compatibility between substitutions and symmetric-group actions, so permuting inputs before or after substitution yields the same result in the prescribed sense.

Therefore the semantic value of a formal operadic interaction expression is uniquely determined by the expression modulo the standard operad identifications.

This proves that the OIG construction is well-defined. □

## 5.11 Symmetric Monoidal Wiring Graph (SMWG)

A Symmetric Monoidal Wiring Graph models networks as morphisms in a symmetric monoidal category: generators are multiport components; composition and tensor encode wiring and parallelism.

**Definition 5.11.1** (Symmetric Monoidal Wiring Graph (SMWG))**.** Let $T$ be a set of *port types.* A *Symmetric Monoidal Wiring Graph* is a triple

$$\mathsf{SMWG} = (\mathcal{C}, \otimes, \mathcal{G}),$$

where $(\mathcal{C}, \otimes, \mathbb{I})$ is a symmetric monoidal category whose objects are generated by $T$ under $\otimes$ (so objects are tensor words $t_1 \otimes \cdots \otimes t_m$ with $t_i \in T$), and $\mathcal{G}$ is a specified set of *generating morphisms* (components)

$$f : \ t_1 \otimes \cdots \otimes t_m \longrightarrow s_1 \otimes \cdots \otimes s_n \qquad (t_i, s_j \in T).$$

A *network* in $\mathsf{SMWG}$ is any morphism in $\mathcal{C}$, built from $\mathcal{G}$ using categorical composition (sequential wiring) and $\otimes$ (parallel composition), modulo the coherence isomorphisms of the symmetric monoidal structure.

**Example 5.11.2** (A Symmetric Monoidal Wiring Graph for simple digital circuits)**.** Let the set of port types be

$$T = \{\mathsf{Bit}\}.$$

Let $\mathcal{C}$ be the free symmetric monoidal category generated by $T$ together with the following generating morphisms (circuit components):

$$\mathrm{AND} : \mathsf{Bit} \otimes \mathsf{Bit} \to \mathsf{Bit}, \qquad \mathrm{NOT} : \mathsf{Bit} \to \mathsf{Bit}, \qquad \mathrm{COPY} : \mathsf{Bit} \to \mathsf{Bit} \otimes \mathsf{Bit}.$$

Set

$$\mathcal{G} = \{\mathrm{AND}, \mathrm{NOT}, \mathrm{COPY}\}.$$

Then objects of $\mathcal{C}$ are tensor words $\mathsf{Bit}^{\otimes m}$, and morphisms are wiring diagrams built from the generators by sequential composition and parallel composition $\otimes$ (with swaps provided by the symmetry).

For instance, the network

$$F \ = \ \mathrm{AND} \circ \big(\mathrm{NOT} \otimes \mathrm{id}_{\mathsf{Bit}}\big) \circ \mathrm{COPY} \ : \ \mathsf{Bit} \longrightarrow \mathsf{Bit}$$

represents the circuit that takes an input bit $x$, copies it to $(x, x)$, negates the first copy to $(\neg x, x)$, and then applies AND, producing $\neg x \wedge x$. Hence $(\mathcal{C}, \otimes, \mathcal{G})$ is a Symmetric Monoidal Wiring Graph in the sense of Definition 5.11.1. An overview diagram of this example is provided in Fig. 5.6.

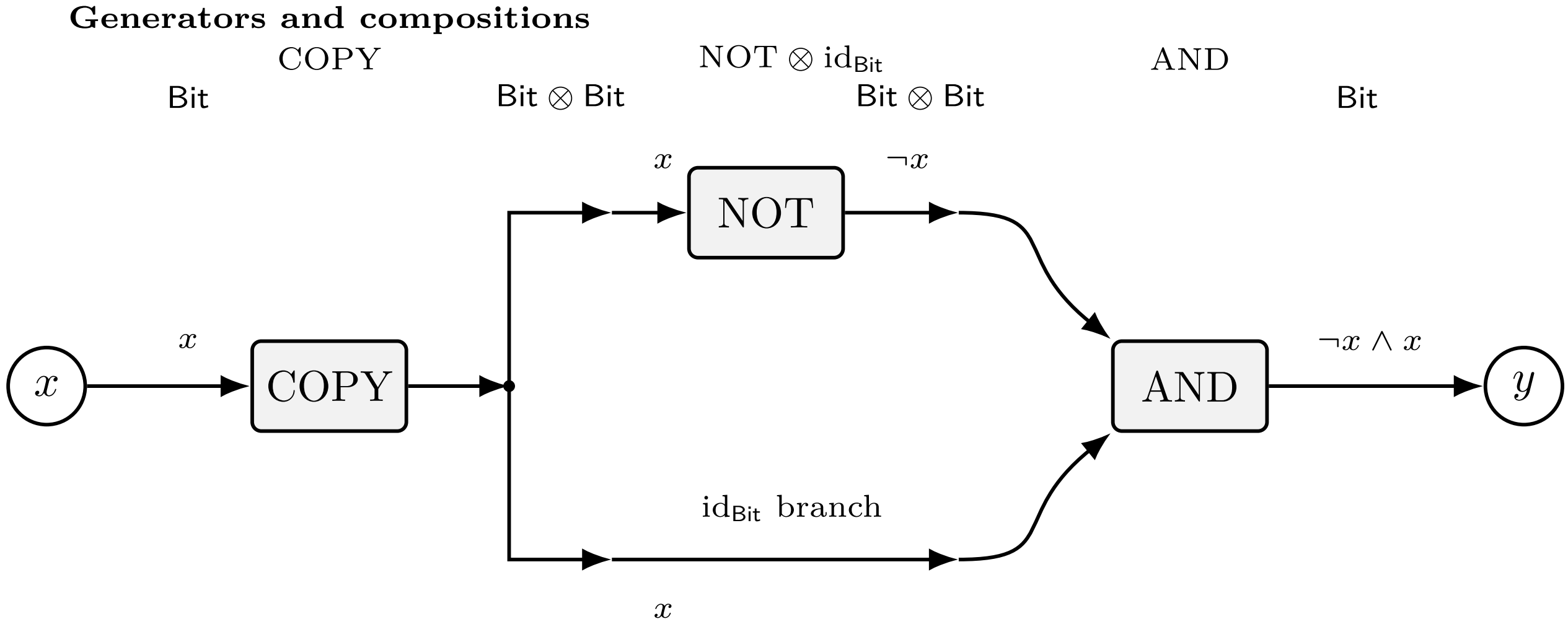

**SMWG interpretation.** This diagram represents the morphism

$$F = \mathrm{AND} \circ \big(\mathrm{NOT} \otimes \mathrm{id}_{\mathsf{Bit}}\big) \circ \mathrm{COPY} : \mathsf{Bit} \to \mathsf{Bit}.$$

It is built from the generators COPY, NOT, AND using sequential composition and tensor (parallel) composition. The symmetric monoidal structure also provides wire swaps (symmetry), although no swap is needed in this specific circuit.

Figure 5.6: A Symmetric Monoidal Wiring Graph for a simple digital circuit. The circuit copies an input bit, negates one copy, and combines the two signals via AND.

**Theorem 5.11.3** (Well-definedness of Symmetric Monoidal Wiring Graph semantics)**.** *Let $T$ be a set of port types, and let*

$$\mathsf{SMWG} = (\mathcal{C}, \otimes, \mathcal{G})$$

*be as in Definition 5.11.1, where $(\mathcal{C}, \otimes, \mathbb{I})$ is a symmetric monoidal category and $\mathcal{G}$ is a set of generating morphisms*

$$f :\ t_1 \otimes \cdots \otimes t_m \longrightarrow s_1 \otimes \cdots \otimes s_n \qquad (t_i, s_j \in T).$$

*Then the notion of a* network *in* $\mathsf{SMWG}$ *is well-defined:*

1. *Every formal wiring expression built from generators in $\mathcal{G}$, identity morphisms, sequential composition $(\circ)$, tensor product $(\otimes)$, and symmetry maps (swaps) is well-typed, provided the source/target objects match at each composition step.*

2. *Such a formal expression determines a unique morphism in $\mathcal{C}$.*

3. *If two formal expressions differ only by parenthesization, insertion/removal of unit objects $\mathbb{I}$, or by replacing canonical rebracketing/symmetry isomorphisms using the symmetric monoidal coherence axioms, then they define the same network (i.e. the same morphism in $\mathcal{C}$, up to the canonical coherence identifications).*

*In particular, "networks built from $\mathcal{G}$ by sequential and parallel wiring, modulo coherence" is a mathematically well-defined notion.*

*Proof.* We prove the three claims.

**(1) Well-typed formation of wiring expressions.** By assumption, every generator $f \in \mathcal{G}$ is already a morphism in $\mathcal{C}$ with a specified source and target object, each of which is a tensor word in the types $T$.

The class of formal wiring expressions is generated inductively from:

- generators $f \in \mathcal{G}$,

- identity morphisms $\mathrm{id}_X : X \to X$ for objects $X \in \mathrm{Ob}(\mathcal{C})$,

- composition $g \circ f$ whenever $\mathrm{cod}(f) = \mathrm{dom}(g)$,

- tensor $f \otimes g$ whenever $f : X \to Y$ and $g : X' \to Y'$,
- symmetry isomorphisms $\sigma_{X,Y} : X \otimes Y \to Y \otimes X$.

Since $\mathcal{C}$ is a category with a monoidal product, each of these constructions is defined whenever the indicated typing conditions hold. Therefore every inductively formed wiring expression is well-typed.

**(2) Every well-typed expression denotes a morphism in $\mathcal{C}$.** We define the semantics of formal expressions by structural recursion:

- a generator $f \in \mathcal{G}$ is interpreted as itself (a morphism in $\mathcal{C}$);
- $\mathrm{id}_X$ is interpreted as the identity on $X$;
- $g \circ f$ is interpreted as the categorical composite of the interpretations of $f$ and $g$;
- $f \otimes g$ is interpreted as the monoidal product of the interpretations;
- $\sigma_{X,Y}$ is interpreted as the symmetry morphism in $\mathcal{C}$.

Because $(\mathcal{C}, \otimes, \mathbb{I})$ is a symmetric monoidal category, all these operations exist and preserve typing. Hence every well-typed formal wiring expression evaluates to a unique morphism in $\mathcal{C}$.

**(3) Independence of parenthesization and coherence choices.** A potential ambiguity arises because tensor words may be written with different parenthesizations (and unit insertions), e.g.

$$(t_1 \otimes t_2) \otimes t_3 \qquad \text{vs.} \qquad t_1 \otimes (t_2 \otimes t_3),$$

and because one may insert canonical associators, unitors, and symmetries at intermediate steps.

However, in a symmetric monoidal category, the coherence theorem (Mac Lane coherence, together with symmetry coherence) implies that all canonical composites built from associators, unitors, and symmetries between the same tensor expressions coincide. Equivalently, canonical rebracketing/reordering maps are uniquely determined by the source and target tensor words.

Therefore, if two formal wiring expressions differ only by:

- parenthesization of tensor products,
- insertion/removal of $\mathbb{I}$ via unitors,
- replacement by canonically coherent associativity/unit/symmetry isomorphisms,

then their interpretations in $\mathcal{C}$ agree (up to the canonical identifications specified by coherence). Hence the equivalence class "modulo coherence" has a unique semantic value.

This proves that the notion of a network in SMWG is well-defined. □

## 5.12 Relational-Arity Graph (RAG)

A Relational-Arity Graph is a finite relational structure: vertices with relations of varying arities, so each k-ary relation encodes k-way interactions, possibly weighted.

**Definition 5.12.1** (Relational-Arity Graph (RAG))**.** A *Relational-Arity Graph* is a pair

$$\mathsf{RAG} = (\Sigma, \mathfrak{G}),$$

where $\Sigma = \{R_\alpha \mid \alpha \in A\}$ is a relational signature with arities $\mathrm{ar}(R_\alpha) = k_\alpha \in \mathbb{N}$, and

$$\mathfrak{G} = \big(V, (R_\alpha^{\mathfrak{G}})_{\alpha \in A}\big)$$

is a $\Sigma$-structure on a vertex set $V$, i.e.

$$R_\alpha^{\mathfrak{G}} \subseteq V^{k_\alpha} \quad \text{for all } \alpha \in A.$$

Thus each relation symbol $R_\alpha$ encodes $k_\alpha$-way interactions (ordered tuples).

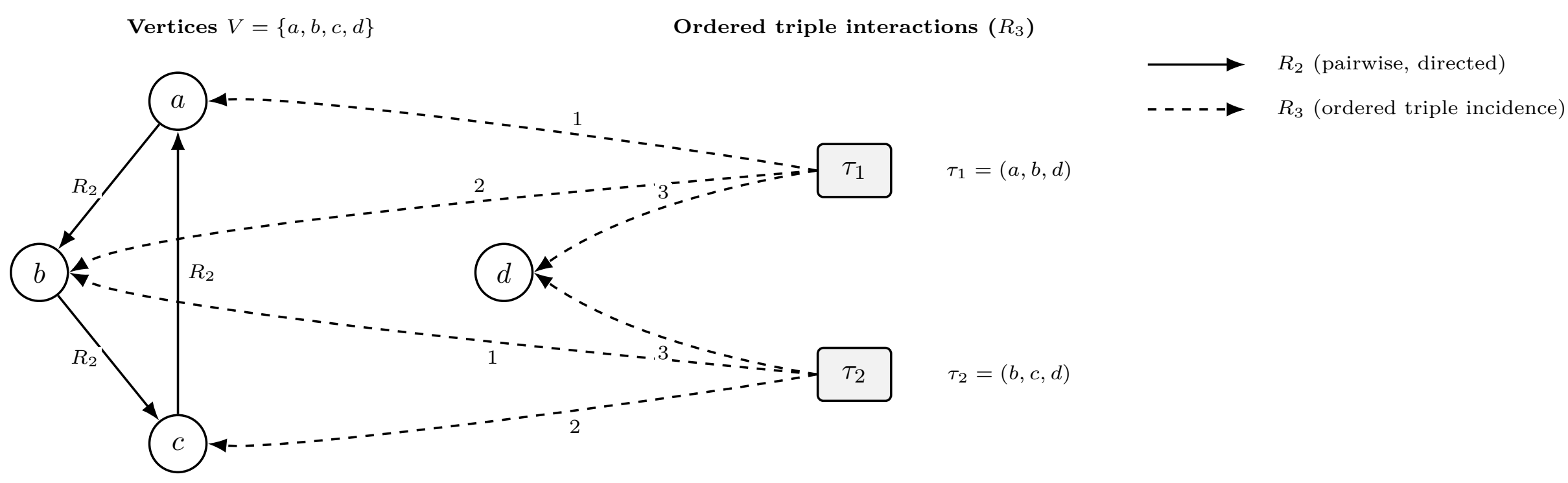

Figure 5.7: A Relational-Arity Graph (RAG) for Example 5.12.2. Solid arrows represent the binary relation $R_2^{\mathfrak{G}}$, and dashed arrows from tuple-nodes $\tau_1, \tau_2$ encode the ordered triples in $R_3^{\mathfrak{G}}$.

**Example 5.12.2** (A Relational-Arity Graph with pairwise and triple interactions)**.** Let the vertex set be

$$V = \{a, b, c, d\}.$$

Consider the relational signature

$$\Sigma = \{R_2, R_3\}, \qquad \mathrm{ar}(R_2) = 2, \quad \mathrm{ar}(R_3) = 3,$$

where $R_2$ is intended to encode directed pairwise interactions and $R_3$ encodes ordered triple interactions. Define the $\Sigma$-structure

$$\mathfrak{G} = \big(V,\ R_2^{\mathfrak{G}},\ R_3^{\mathfrak{G}}\big)$$

by specifying the interpreted relations:

$$R_2^{\mathfrak{G}} = \{(a, b), (b, c), (c, a)\} \subseteq V^2,$$

$$R_3^{\mathfrak{G}} = \{(a, b, d), (b, c, d)\} \subseteq V^3.$$

Then $\mathsf{RAG} = (\Sigma, \mathfrak{G})$ is a Relational-Arity Graph in the sense of Definition 5.12.1. Here $R_2^{\mathfrak{G}}$ records a directed 3-cycle among $(a, b, c)$, while $R_3^{\mathfrak{G}}$ records two distinct 3-way interactions involving the vertex $d$. An overview diagram of this example is provided in Fig. 5.7.

**Theorem 5.12.3** (Well-definedness of Relational-Arity Graphs)**.** *Let*

$$\Sigma = \{R_\alpha \mid \alpha \in A\}$$

*be a relational signature together with an arity map*

$$\mathrm{ar} : A \to \mathbb{N}, \qquad \alpha \mapsto k_\alpha := \mathrm{ar}(R_\alpha).$$

*Let $V$ be a set, and suppose that for each $\alpha \in A$ a relation*

$$R_\alpha^{\mathfrak{G}} \subseteq V^{k_\alpha}$$

*is given. Then the pair*

$$\mathsf{RAG} = (\Sigma, \mathfrak{G}), \qquad \mathfrak{G} = \big(V, (R_\alpha^{\mathfrak{G}})_{\alpha \in A}\big),$$

*is a well-defined Relational-Arity Graph in the sense of Definition 5.12.1.*

*Moreover, for each $\alpha \in A$, membership*

$$(v_1, \ldots, v_{k_\alpha}) \in R_\alpha^{\mathfrak{G}}$$

*is a well-posed statement for all $(v_1, \ldots, v_{k_\alpha}) \in V^{k_\alpha}$, and thus each $R_\alpha^{\mathfrak{G}}$ defines a $k_\alpha$-ary interaction on $V$ (ordered unless additional symmetry is imposed).*

*Proof.* To show that $(\Sigma, \mathfrak{G})$ is well-defined, we must check that every piece of data appearing in Definition 5.12.1 is properly typed.

**(1) The signature is well-formed.** By assumption, $\Sigma = \{R_\alpha \mid \alpha \in A\}$ is a family of relation symbols indexed by $A$, and each symbol $R_\alpha$ is assigned a positive integer arity $k_\alpha \in \mathbb{N}$. Hence the typed signature data $(\Sigma, \mathrm{ar})$ is well-defined.

**(2) Cartesian powers are well-defined.** For each $\alpha \in A$, since $k_\alpha \in \mathbb{N}$, the Cartesian power

$$V^{k_\alpha} = \underbrace{V \times \cdots \times V}_{k_\alpha \text{ factors}}$$

is well-defined. Its elements are exactly ordered $k_\alpha$-tuples of vertices in $V$.

**(3) Interpreted relations are well-typed.** Again by assumption, for each $\alpha \in A$,

$$R_\alpha^{\mathfrak{G}} \subseteq V^{k_\alpha}.$$

Therefore $R_\alpha^{\mathfrak{G}}$ is a $k_\alpha$-ary relation on $V$, i.e. a set of ordered $k_\alpha$-tuples. Hence the family $(R_\alpha^{\mathfrak{G}})_{\alpha \in A}$ is a well-defined interpretation of the signature $\Sigma$ on $V$.

**(4) The $\Sigma$-structure is well-defined.** Combining (1)–(3), the object

$$\mathfrak{G} = \big(V, (R_\alpha^{\mathfrak{G}})_{\alpha \in A}\big)$$

is a valid $\Sigma$-structure (in the standard model-theoretic sense). Therefore the pair

$$(\Sigma, \mathfrak{G})$$

is a well-defined Relational-Arity Graph.

The final claim is immediate: for each $\alpha \in A$, because $R_\alpha^{\mathfrak{G}} \subseteq V^{k_\alpha}$, the statement

$$(v_1, \ldots, v_{k_\alpha}) \in R_\alpha^{\mathfrak{G}}$$

is meaningful exactly for $(v_1, \ldots, v_{k_\alpha}) \in V^{k_\alpha}$, and this records a $k_\alpha$-way interaction. The tuples are ordered by default because $V^{k_\alpha}$ is an ordered Cartesian product. □

## 5.13 Closure-Implication Graph (CIG)

A *Closure-Implication Graph (CIG)* is a set with a closure operator; forcing relations $a \in \mathrm{cl}(S) \setminus S$ capture higher-order implications, with minimal generators encoding interactions.

**Definition 5.13.1** (Closure-Implication Graph (CIG))**.** A *Closure-Implication Graph* is a pair

$$\mathsf{CIG} = (V, \mathrm{cl}),$$

where $V$ is a set and $\mathrm{cl} : \mathcal{P}(V) \to \mathcal{P}(V)$ is a closure operator, i.e. for all $S, T \subseteq V$:

$$S \subseteq \mathrm{cl}(S), \qquad S \subseteq T \Rightarrow \mathrm{cl}(S) \subseteq \mathrm{cl}(T), \qquad \mathrm{cl}(\mathrm{cl}(S)) = \mathrm{cl}(S).$$

A higher-order forcing event is expressed by $a \in \mathrm{cl}(S) \setminus S$; minimal such $S$ (by inclusion) may be regarded as the basic implicational generators of $a$.

**Example 5.13.2** (A Closure-Implication Graph generated by simple rules)**.** Let

$$V = \{a, b, c, d\}.$$

Consider the following implicational rules:

$$\{a, b\} \Rightarrow c, \qquad \{c\} \Rightarrow d.$$

Define $\mathrm{cl} : \mathcal{P}(V) \to \mathcal{P}(V)$ by letting $\mathrm{cl}(S)$ be the smallest subset of $V$ that contains $S$ and is closed under the above rules; equivalently, $\mathrm{cl}(S)$ is obtained by repeatedly adding $c$ whenever $\{a, b\} \subseteq S$ and adding $d$ whenever $c \in S$, until no new elements can be added.

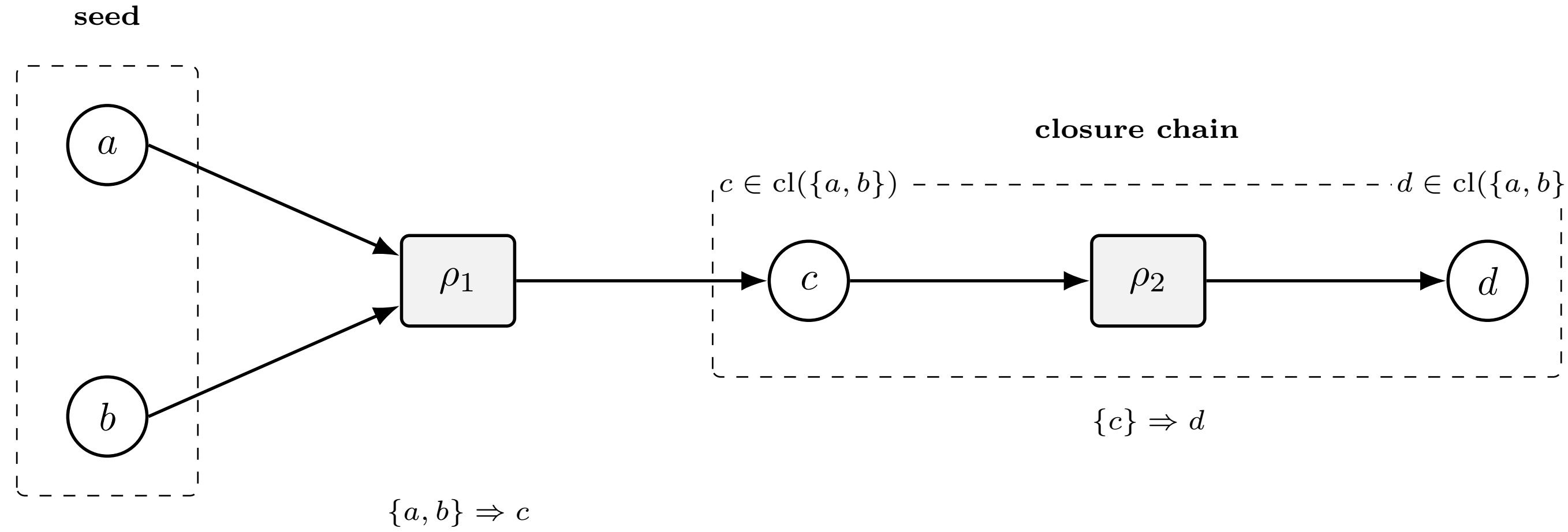

Figure 5.8: A Closure-Implication Graph generated by the rules $\{a,b\} \Rightarrow c$ and $\{c\} \Rightarrow d$.

For example,

$$\mathrm{cl}(\{a\}) = \{a\}, \qquad \mathrm{cl}(\{a,b\}) = \{a,b,c,d\}, \qquad \mathrm{cl}(\{b,c\}) = \{b,c,d\}.$$

Then cl is extensive, monotone, and idempotent, hence $(V, \mathrm{cl})$ is a Closure-Implication Graph in the sense of Definition 5.13.1. In particular, $\{a,b\}$ forces $c$ since

$$c \in \mathrm{cl}(\{a,b\}) \setminus \{a,b\},$$

and $\{a,b\}$ also forces $d$ via the intermediate implication $\{a,b\} \Rightarrow c$ and then $\{c\} \Rightarrow d$. An overview diagram of this example is provided in Fig. 5.8.

**Theorem 5.13.3** (Well-definedness of Closure-Implication Graph semantics)**.** *Let $V$ be a set and let $\mathrm{cl} : \mathcal{P}(V) \to \mathcal{P}(V)$ be a closure operator, i.e. for all $S, T \subseteq V$,*

$$S \subseteq \mathrm{cl}(S), \qquad S \subseteq T \Rightarrow \mathrm{cl}(S) \subseteq \mathrm{cl}(T), \qquad \mathrm{cl}(\mathrm{cl}(S)) = \mathrm{cl}(S).$$

*Then the pair $\mathsf{CIG} = (V, \mathrm{cl})$ is well-defined as a Closure-Implication Graph.*

*Moreover, for each $a \in V$, define the forcing family*

$$\mathcal{F}_a \;:=\; \{\, S \subseteq V \setminus \{a\} \mid a \in \mathrm{cl}(S) \,\}.$$

*Then:*

1. *the higher-order forcing relation*
$$S \rightsquigarrow a \quad \Longleftrightarrow \quad a \in \mathrm{cl}(S) \setminus S$$
*is well-defined on $\mathcal{P}(V) \times V$;*

2. *$\mathcal{F}_a$ is upward closed under inclusion, i.e. if $S \in \mathcal{F}_a$ and $S \subseteq T \subseteq V \setminus \{a\}$, then $T \in \mathcal{F}_a$;*

3. *if $V$ is finite, then $\mathcal{F}_a$ has inclusion-minimal elements (possibly none), and hence the "minimal implicational generators" of $a$ are well-defined as*
$$\mathrm{MinGen}(a) := \min_{\subseteq}(\mathcal{F}_a).$$
*In fact, for every $S \in \mathcal{F}_a$, there exists $M \in \mathrm{MinGen}(a)$ such that $M \subseteq S$.*

*Proof.* The first statement is immediate: by assumption, cl is a map $\mathcal{P}(V) \to \mathcal{P}(V)$ satisfying the closure axioms, so $(V, \mathrm{cl})$ is exactly a Closure-Implication Graph by Definition 5.13.1.

For (1), let $S \subseteq V$ and $a \in V$. Since $\mathrm{cl}(S) \subseteq V$, the expression

$$a \in \mathrm{cl}(S) \setminus S$$

is a meaningful truth-valued statement. Hence $S \rightsquigarrow a$ defines a well-posed binary relation between subsets $S \subseteq V$ and vertices $a \in V$.

For (2), suppose $S \in \mathcal{F}_a$ and $S \subseteq T \subseteq V \setminus \{a\}$. By $S \in \mathcal{F}_a$, we have $a \in \mathrm{cl}(S)$. By monotonicity of cl,

$$\mathrm{cl}(S) \subseteq \mathrm{cl}(T),$$

so $a \in \mathrm{cl}(T)$. Since $T \subseteq V \setminus \{a\}$, we also have $a \notin T$. Therefore $T \in \mathcal{F}_a$. Thus $\mathcal{F}_a$ is upward closed.

For (3), assume $V$ is finite. Then $\mathcal{P}(V \setminus \{a\})$ is a finite poset under inclusion, and $\mathcal{F}_a \subseteq \mathcal{P}(V \setminus \{a\})$ is a (finite) subset of this poset. Every finite poset subset has minimal elements, so $\min_{\subseteq}(\mathcal{F}_a)$ exists (possibly empty if $\mathcal{F}_a = \varnothing$). Hence $\mathrm{MinGen}(a)$ is well-defined.

Finally, let $S \in \mathcal{F}_a$. Consider the finite set

$$\{\, T \in \mathcal{F}_a \mid T \subseteq S \,\}.$$

It is nonempty because $S$ itself belongs to it. Choose an inclusion-minimal element $M$ of this set. Then $M \in \mathcal{F}_a$, $M \subseteq S$, and no proper subset of $M$ belongs to $\mathcal{F}_a$; hence $M \in \mathrm{MinGen}(a)$. This proves the last claim. □

## 5.14 Coalgebraic Nested-Neighborhood Graph (CNNG)

A *Coalgebraic Nested-Neighborhood Graph (CNNG)* is an $F$-coalgebra $(V, \gamma)$ where $\gamma : V \to F(V)$ assigns nested neighborhoods, capturing iterated higher-order adjacency structure.

**Definition 5.14.1** (Coalgebraic Nested-Neighborhood Graph (CNNG))**.** Let $F : \mathbf{Set} \to \mathbf{Set}$ be an endofunctor. A *Coalgebraic Nested-Neighborhood Graph* is an $F$-coalgebra

$$\mathsf{CNNG} = (V, \gamma), \qquad \gamma : V \to F(V).$$

In particular, letting $\mathcal{P}_f(V)$ denote the set of finite subsets of $V$, for any $r \geq 1$ one may take

$$F(V) = \mathcal{P}_f^{\,r}(V),$$

so that $\gamma(v) \in \mathcal{P}_f^{\,r}(V)$ assigns to each vertex $v$ an $r$-nested finite neighborhood. The case $r = 1$ recovers ordinary (directed) neighborhood graphs, while $r \geq 2$ encodes iterated (neighborhood-of-neighborhood) structure.

**Example 5.14.2** (A CNNG with 2-nested neighborhoods)**.** Let

$$V = \{1, 2, 3\}$$

and take $r = 2$, so

$$F(V) = \mathcal{P}_f^{\,2}(V) = \mathcal{P}_f\big(\mathcal{P}_f(V)\big).$$

Define $\gamma : V \to \mathcal{P}_f^{\,2}(V)$ by

$$\gamma(1) = \big\{\{2,3\}, \{2\}\big\}, \qquad \gamma(2) = \big\{\{1\}\big\}, \qquad \gamma(3) = \big\{\{1,2\}, \emptyset\big\}.$$

Each value $\gamma(v)$ is a finite set of finite subsets of $V$, hence $\gamma(v) \in \mathcal{P}_f^{\,2}(V)$. Therefore

$$\mathsf{CNNG} = (V, \gamma)$$

is a Coalgebraic Nested-Neighborhood Graph in the sense of Definition 5.14.1. Intuitively, $\gamma(1)$ assigns to vertex 1 two alternative neighbor-sets $\{2, 3\}$ and $\{2\}$, while $\gamma(3)$ assigns the neighbor-set $\{1, 2\}$ together with the empty neighbor-set, illustrating a 2-nested neighborhood structure. An overview diagram of this example is provided in Fig. 5.9.

**Theorem 5.14.3** (Well-definedness of the $r$-nested CNNG construction)**.** *Fix an integer $r \geq 1$. Let $\mathcal{P}_f : \mathbf{Set} \to \mathbf{Set}$ denote the finite-powerset functor, i.e.*

$$\mathcal{P}_f(V) = \{S \subseteq V \mid S \text{ is finite}\},$$

*and for a map $f : V \to W$,*

$$\mathcal{P}_f(f) : \mathcal{P}_f(V) \to \mathcal{P}_f(W), \qquad \mathcal{P}_f(f)(S) = f[S] = \{f(x) \mid x \in S\}.$$

*Then the $r$-fold iterate*

$$F_r := \mathcal{P}_f^{\,r} : \mathbf{Set} \to \mathbf{Set}$$

*is a well-defined endofunctor. Consequently, for every set $V$ and every map*

$$\gamma : V \to \mathcal{P}_f^{\,r}(V),$$

*the pair $(V, \gamma)$ is a well-defined $F_r$-coalgebra, hence a well-defined Coalgebraic Nested-Neighborhood Graph (CNNG) of nesting depth $r$.*

*Moreover, when $r = 1$, $\gamma(v) \in \mathcal{P}_f(V)$ is an ordinary finite neighborhood of $v$; when $r \geq 2$, $\gamma(v) \in \mathcal{P}_f^{\,r}(V)$ is an $r$-nested finite neighborhood.*

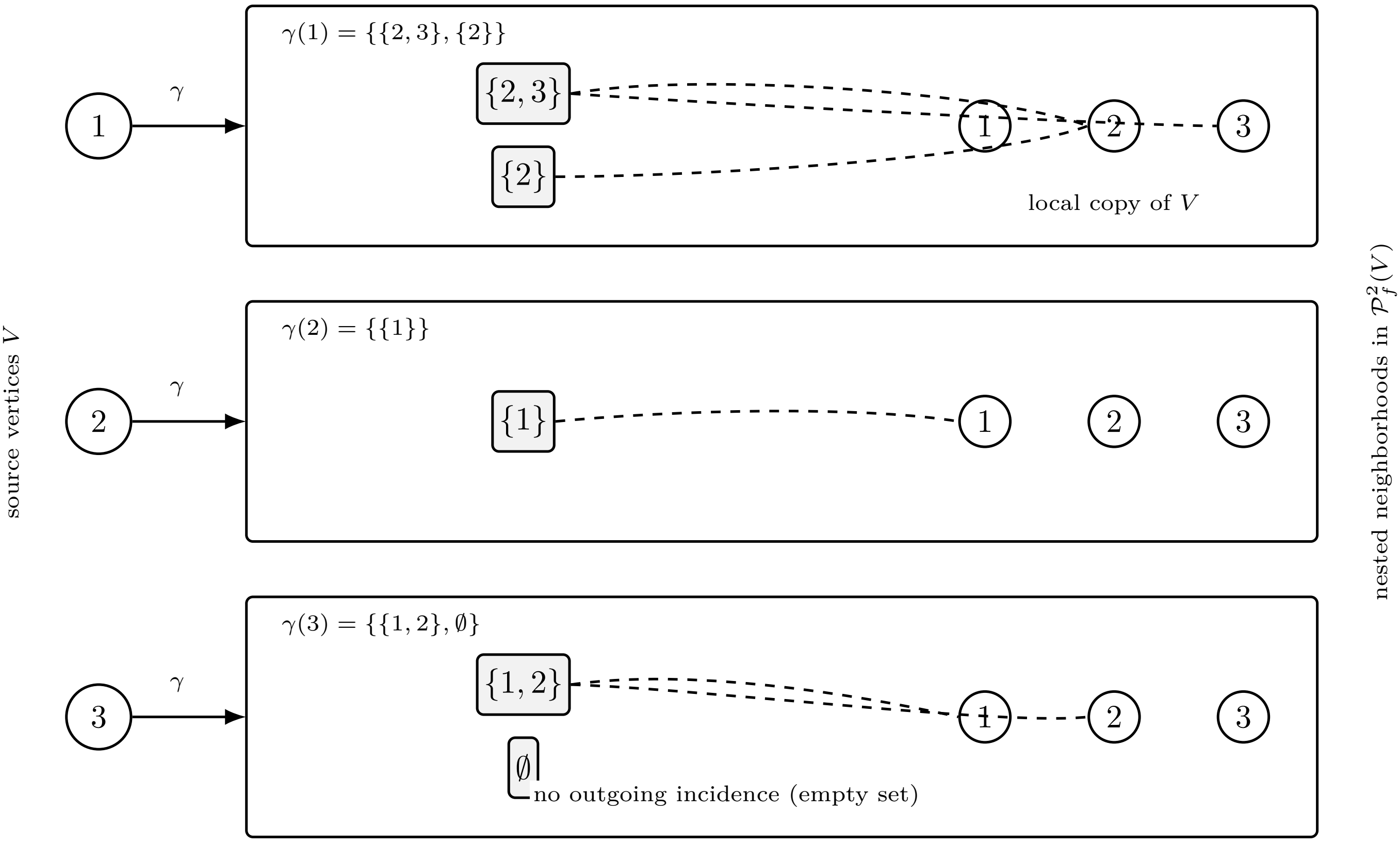

Figure 5.9: A Coalgebraic Nested-Neighborhood Graph with 2-nested neighborhoods on $V = \{1, 2, 3\}$.

*Proof.* We proceed in two steps.

**Step 1: $\mathcal{P}_f$ is a well-defined endofunctor on Set.**
For any set $V$, $\mathcal{P}_f(V)$ is a set (the set of finite subsets of $V$). For any map $f : V \to W$, define

$$\mathcal{P}_f(f)(S) = f[S] \subseteq W \qquad (S \in \mathcal{P}_f(V)).$$

This is well-defined because the image of a finite set under any map is finite; hence $f[S] \in \mathcal{P}_f(W)$.
We verify the functorial axioms.

1. *Identity:* For the identity map $\mathrm{id}_V : V \to V$,

$$\mathcal{P}_f(\mathrm{id}_V)(S) = \mathrm{id}_V[S] = S \qquad \text{for all } S \in \mathcal{P}_f(V),$$

so $\mathcal{P}_f(\mathrm{id}_V) = \mathrm{id}_{\mathcal{P}_f(V)}$.

2. *Composition:* Let $f : U \to V$ and $g : V \to W$. For any $S \in \mathcal{P}_f(U)$,

$$\mathcal{P}_f(g \circ f)(S) = (g \circ f)[S] = g[f[S]] = \mathcal{P}_f(g)\big(\mathcal{P}_f(f)(S)\big).$$

Hence

$$\mathcal{P}_f(g \circ f) = \mathcal{P}_f(g) \circ \mathcal{P}_f(f).$$

Therefore $\mathcal{P}_f : \mathbf{Set} \to \mathbf{Set}$ is an endofunctor.

**Step 2: The iterate $\mathcal{P}_f^r$ is a well-defined endofunctor.**
Since $\mathcal{P}_f$ is an endofunctor, its $r$-fold composition

$$\mathcal{P}_f^r = \underbrace{\mathcal{P}_f \circ \cdots \circ \mathcal{P}_f}_{r \text{ times}}$$

is again an endofunctor on **Set**. Explicitly:

- on objects, $V \mapsto \mathcal{P}_f^r(V)$;
- on morphisms, $f \mapsto \mathcal{P}_f^r(f)$, obtained by iterating direct image.

Functoriality (preservation of identities and composition) follows from the functoriality of $\mathcal{P}_f$ and closure of endofunctors under composition.

Now, by the definition of an $F$-coalgebra for an endofunctor $F$, any pair

$$(V, \gamma), \qquad \gamma : V \to F(V),$$

is an $F$-coalgebra. Taking $F = F_r = \mathcal{P}_f^r$, any map

$$\gamma : V \to \mathcal{P}_f^r(V)$$

therefore defines a well-formed coalgebra $(V, \gamma)$. This is exactly a CNNG of nesting depth $r$.

Finally, for $r = 1$, $\gamma(v) \in \mathcal{P}_f(V)$ is just a finite subset of $V$, i.e. an ordinary finite neighborhood. For $r \geq 2$, $\gamma(v)$ is a finite set of $(r-1)$-nested finite subsets, so it encodes iterated (nested) neighborhood data. This proves the claim. □

## 5.15 Curried Graph

A curried graph has curried functions as vertices; edges are structure-preserving maps between domains and codomains satisfying commutative evaluation [525]. This is a graph concept derived from the notion of *curried functions* [526] in programming languages.

**Definition 5.15.1** (Curried $k$-ary Graph (Curried Graph Function))**.** [525] Fix an integer $k \geq 1$. Let $\Sigma$ be a finite family of *type signatures*

$$\Sigma = \{(A_1, \dots, A_k; Z)\},$$

where each $A_i$ and $Z$ is a set. A *curried $k$-ary graph* is a directed graph

$$G^{(k)} = (V, \mathcal{E}, s, t)$$

specified as follows:

1. **(Vertices)** $V$ is a nonempty set of curried $k$-ary functions of signatures in $\Sigma$, i.e.

$$V \;\subseteq \bigcup_{(A_1,\dots,A_k;Z)\in\Sigma} \{\, f \mid f : A_1 \to (A_2 \to \cdots \to (A_k \to Z) \cdots)\,\}.$$

   When $f \in V$ has signature $(A_1, \dots, A_k; Z)$, we write it suggestively as

$$f : \; A_1 \to A_2 \to \cdots \to A_k \to Z.$$

2. **(Edges)** An element $E \in \mathcal{E}$ is an *edge datum* between a source vertex

$$f : \; A_1 \to \cdots \to A_k \to Z$$

   and a target vertex

$$g : \; A'_1 \to \cdots \to A'_k \to Z'$$

   given by a tuple of set maps

$$E = \big(E(A_1), \dots, E(A_k), E(Z)\big),$$

   where

$$E(A_i) : A_i \to A'_i \quad (1 \leq i \leq k), \qquad E(Z) : Z \to Z'.$$

   We declare that $E$ is a directed edge $E : f \to g$ (so $s(E) = f$ and $t(E) = g$) if and only if the following *commuting (naturality) condition* holds:

$$\forall a_1 \in A_1, \dots, a_k \in A_k : \quad E(Z)\big(f(a_1)(a_2)\cdots(a_k)\big) = g\big(E(A_1)(a_1)\big)\big(E(A_2)(a_2)\big)\cdots\big(E(A_k)(a_k)\big).$$

**Example 5.15.2** (Real-world example: migrating an e-commerce scoring function as a curried graph)**.** Consider an e-commerce platform that assigns a relevance score to each user–item pair. Let

$$A = \{\mathsf{u}_1, \mathsf{u}_2\} \quad \text{(legacy user IDs)}, \qquad B = \{\mathsf{i}_1, \mathsf{i}_2, \mathsf{i}_3\} \quad \text{(legacy item IDs)}, \qquad Z = \mathbb{R} \quad \text{(raw scores)}.$$

A legacy scoring rule can be modeled as a curried function

$$f : \; A \to B \to Z,$$

where $f(\mathsf{u})(\mathsf{i})$ is the raw relevance score of item $\mathsf{i}$ for user $\mathsf{u}$.

Suppose the platform migrates to a new ID system and a new scoring scale:

$$A' = \{\mathsf{U}_1, \mathsf{U}_2\}, \qquad B' = \{\mathsf{I}_1, \mathsf{I}_2, \mathsf{I}_3\}, \qquad Z' = [0, 1].$$

Let $g : A' \to B' \to Z'$ be the new (normalized) scoring rule. Define the migration maps

$$E(A) : A \to A', \qquad E(B) : B \to B', \qquad E(Z) : Z \to Z'$$

by

$$E(A)(\mathsf{u}_j) = \mathsf{U}_j \; (j = 1, 2), \qquad E(B)(\mathsf{i}_\ell) = \mathsf{I}_\ell \; (\ell = 1, 2, 3),$$

and (for some fixed scale parameter $M > 0$)

$$E(Z)(x) = \min\{1, \max\{0, x/M\}\} \in [0, 1].$$

Assume the new system is designed so that the normalized score equals the normalized legacy score after ID mapping, namely

$$g\big(E(A)(\mathsf{u})\big)\big(E(B)(\mathsf{i})\big) = E(Z)\big(f(\mathsf{u})(\mathsf{i})\big) \quad \text{for all } \mathsf{u} \in A, \; \mathsf{i} \in B.$$

Equivalently,

$$E(Z)\big(f(\mathsf{u})(\mathsf{i})\big) = g\big(E(A)(\mathsf{u})\big)\big(E(B)(\mathsf{i})\big) \quad \text{for all } \mathsf{u} \in A, \; \mathsf{i} \in B,$$

which is exactly the commuting (naturality) condition. Hence the triple $E = (E(A), E(B), E(Z))$ defines a directed edge $E : f \to g$ in the curried graph whose vertices are curried scoring functions, representing a consistent migration of user IDs, item IDs, and score scales.

## 5.16 Depth-$r$ iterated subdivisions of polyhedral complexes

We define *depth-r iterated polyhedral complexes* (i.e., iterated subdivisions) recursively. Depth-$r$ iterated subdivisions of polyhedral complexes repeatedly refine each cell $r$ times, preserving topology while increasing combinatorial resolution and detail.

**Definition 5.16.1** (Depth-$r$ iterated polyhedral complex)**.** A *depth-*0 *iterated polyhedral complex* is an ordinary polyhedral complex $K$.

Assume that depth-$r$ iterated polyhedral complexes have been defined. A *depth-*$(r+1)$ *iterated polyhedral complex* is a pair $(K, S)$ such that:

1. $K$ is a polyhedral complex.
2. $S$ assigns to each cell $P \in K$ a polyhedral complex $S(P)$ with underlying space

   $$|S(P)| = P,$$

   and the assignment is *face-compatible* in the sense that for every face $F \leq P$ (hence $F \in K$),

   $$S(F) = \{\, Q \in S(P) \mid Q \subseteq F \,\} = S(P) \cap F.$$

3. Moreover, for each $P \in K$, the complex $S(P)$ is equipped with the structure of a *depth-r iterated polyhedral complex*, and the above face-compatibility condition holds at *every depth level* (i.e., the induced iterated subdivision on each face agrees with the iterated subdivision assigned to that face).

**Remark 5.16.2** (Refinement chains (flags))**.** *Equivalently, a depth-r iterated polyhedral complex can be described by specifying a chain of refinements (a* flag*)*

$$K^{(0)} \prec K^{(1)} \prec \cdots \prec K^{(r)},$$

*where* $K^{(0)}$ *is the initial polyhedral complex and, for each* $\ell = 0, \ldots, r-1$*, the complex* $K^{(\ell+1)}$ *is a coherent (face-compatible) polyhedral subdivision of* $K^{(\ell)}$*.*

**Example 5.16.3** (A depth-2 iterated polyhedral complex on the unit square)**.** Let $K$ be the polyhedral complex in $\mathbb{R}^2$ whose underlying space is the unit square

$$|K| = [0,1] \times [0,1],$$

obtained by subdividing the square into two triangles:

$$P_1 = \mathrm{conv}\{(0,0),(1,0),(0,1)\}, \qquad P_2 = \mathrm{conv}\{(1,0),(1,1),(0,1)\},$$

together with all faces of $P_1$ and $P_2$ (edges and vertices). Thus $K$ is a depth-0 iterated polyhedral complex.

**First refinement (depth 1).** Define $S_1$ by assigning to each 2-cell $P_i$ the barycentric subdivision into four triangles. For example, for $P_1$ let $m_1$ be the midpoint of the edge from $(1,0)$ to $(0,1)$, and set

$$S_1(P_1) = \Big\{\mathrm{conv}\{(0,0),(1,0),m_1\},\ \mathrm{conv}\{(0,0),(0,1),m_1\},$$

$$\mathrm{conv}\{(0,0),m_1,\tfrac{1}{2}(1,0)\},\ \mathrm{conv}\{(0,0),m_1,\tfrac{1}{2}(0,1)\}\Big\},$$

together with all faces (the precise choice of four triangles is not essential; any polyhedral subdivision of $P_1$ works). Define $S_1(P_2)$ analogously. For each edge (a 1-face) $F \leq P_i$, define $S_1(F)$ to be the induced subdivision of $F$ by the endpoints and any new vertices introduced on that edge; for each vertex $v$, set $S_1(\{v\}) = \{\{v\}\}$. Then $S_1$ is face-compatible, and $(K, S_1)$ is a depth-1 iterated polyhedral complex.

**Second refinement (depth 2).** Define $S_2$ by further subdividing each triangle $Q \in S_1(P_i)$ into two triangles by drawing a median, and extending this assignment to faces by restriction. Concretely, for each 2-cell $Q \in S_1(P_i)$ choose one edge, insert its midpoint (if not already present), and split $Q$ into two smaller triangles; include all faces. For a face $F \leq Q$, define $S_2(F) = S_2(Q) \cap F$.

Now set, for each original cell $P \in K$,

$$S(P) := \big(S_1(P), S_2|_{S_1(P)}\big),$$

i.e., $S(P)$ is equipped with the depth-1 iterated structure given by the second refinement inside $P$. By construction, $|S(P)| = P$ and the face-compatibility condition holds at both refinement levels. Hence $(K, S)$ is a depth-2 iterated polyhedral complex in the sense of Definition 5.16.1.

## 5.17 Sheaf HyperGraph / Sheaf SuperHyperGraph

A Sheaf HyperGraph assigns local data spaces to vertices and hyperedges, with restriction maps along incidences, enabling coherent local-to-global information. A Sheaf SuperHyperGraph assigns local data spaces to supervertices and superhyperedges, with restriction maps encoding coherent hierarchical local-to-global information structures.

**Definition 5.17.1** (Incidence poset of a HyperGraph)**.** Let $H = (V, E)$ be a HyperGraph. Its *incidence poset* is the poset

$$\mathrm{Inc}(H) := (V \sqcup E, \preceq),$$

where the order relation is generated by

$$v \preceq e \iff v \in e \qquad (v \in V,\ e \in E).$$

Thus vertices are below the hyperedges that contain them.

**Definition 5.17.2** (Sheaf HyperGraph)**.** Fix a field $\mathbb{K}$. A *Sheaf HyperGraph* is a pair

$$(H, \mathcal{F}),$$

where $H = (V, E)$ is a HyperGraph and

$$\mathcal{F} : \mathrm{Inc}(H)^{\mathrm{op}} \to \mathbf{Vect}_{\mathbb{K}}$$

is a contravariant functor.

Equivalently, a Sheaf HyperGraph consists of:

1. a $\mathbb{K}$-vector space $\mathcal{F}(x)$ for each $x \in V \sqcup E$;

2. for every incidence $v \in e$, a linear *restriction map*

$$\rho_{e,v} : \mathcal{F}(e) \to \mathcal{F}(v);$$

3. identity and functoriality conditions inherited from the poset structure.

The spaces $\mathcal{F}(v)$ and $\mathcal{F}(e)$ are called the *stalks* at the vertex $v$ and the hyperedge $e$, respectively.

**Definition 5.17.3** (Incidence poset of an $n$-SuperHyperGraph)**.** Let $\mathcal{H}^{(n)} = (V, E)$ be an $n$-SuperHyperGraph. Its *superincidence poset* is the poset

$$\mathrm{SInc}\big(\mathcal{H}^{(n)}\big) := (V \sqcup E, \preceq),$$

where

$$v \preceq e \quad \Longleftrightarrow \quad v \in e \qquad (v \in V,\ e \in E).$$

**Definition 5.17.4** (Sheaf $n$-SuperHyperGraph)**.** Fix a field $\mathbb{K}$. A *Sheaf $n$-SuperHyperGraph* is a pair

$$\big(\mathcal{H}^{(n)}, \mathcal{F}\big),$$

where $\mathcal{H}^{(n)} = (V, E)$ is an $n$-SuperHyperGraph and

$$\mathcal{F} : \mathrm{SInc}\big(\mathcal{H}^{(n)}\big)^{\mathrm{op}} \to \mathbf{Vect}_{\mathbb{K}}$$

is a contravariant functor.

Equivalently, it assigns:

1. a $\mathbb{K}$-vector space $\mathcal{F}(x)$ to each supervertex or superhyperedge $x \in V \sqcup E$;

2. for every superincidence $v \in e$, a restriction map

$$\rho_{e,v} : \mathcal{F}(e) \to \mathcal{F}(v).$$

**Remark 5.17.5.** *When $n = 0$, a Sheaf 0-SuperHyperGraph is precisely a Sheaf HyperGraph.*

**Example 5.17.6** (A Sheaf 1-SuperHyperGraph)**.** Let the base set be

$$V_0 = \{a, b, c\},$$

and take $n = 1$, so that

$$\mathcal{P}^1(V_0) = \mathcal{P}(V_0).$$

Fix the field

$$\mathbb{K} = \mathbb{R}.$$

**Step 1: The underlying 1-SuperHyperGraph.** Define the following 1-supervertices:

$$v_1 := \{a, b\}, \qquad v_2 := \{c\}, \qquad v_3 := \{a, c\}.$$

Set

$$V = \{v_1, v_2, v_3\} \subseteq \mathcal{P}(V_0).$$

Define the superhyperedge family by

$$E = \{e_1, e_2\}, \qquad e_1 = \{v_1, v_2\}, \qquad e_2 = \{v_1, v_3\}.$$

Hence

$$\mathcal{H}^{(1)} = (V, E)$$

is a finite 1-SuperHyperGraph.

**Step 2: The stalks of the sheaf.** Define a vector space for each supervertex and each superhyperedge by

$$\mathcal{F}(v_1) = \mathbb{R}^2, \qquad \mathcal{F}(v_2) = \mathbb{R}, \qquad \mathcal{F}(v_3) = \mathbb{R}^2,$$

and

$$\mathcal{F}(e_1) = \mathbb{R}^2, \qquad \mathcal{F}(e_2) = \mathbb{R}^3.$$

**Step 3: Restriction maps along superincidences.** Since

$$v_1 \in e_1, \quad v_2 \in e_1, \quad v_1 \in e_2, \quad v_3 \in e_2,$$

we define the restriction maps

$$\rho_{e_1,v_1} : \mathbb{R}^2 \to \mathbb{R}^2, \qquad \rho_{e_1,v_1}(x,y) = (x,y),$$
$$\rho_{e_1,v_2} : \mathbb{R}^2 \to \mathbb{R}, \qquad \rho_{e_1,v_2}(x,y) = x + y,$$
$$\rho_{e_2,v_1} : \mathbb{R}^3 \to \mathbb{R}^2, \qquad \rho_{e_2,v_1}(x,y,z) = (x,y),$$

and

$$\rho_{e_2,v_3} : \mathbb{R}^3 \to \mathbb{R}^2, \qquad \rho_{e_2,v_3}(x,y,z) = (y,z).$$

For every object $x \in V \sqcup E$, let

$$\rho_{x,x} = \mathrm{id}_{\mathcal{F}(x)}.$$

**Step 4: The associated contravariant functor.** The superincidence poset $\mathrm{SInc}(\mathcal{H}^{(1)})$ has objects

$$V \sqcup E = \{v_1, v_2, v_3, e_1, e_2\},$$

and its only non-identity order relations are

$$v_1 \preceq e_1, \qquad v_2 \preceq e_1, \qquad v_1 \preceq e_2, \qquad v_3 \preceq e_2.$$

Passing to the opposite category reverses these arrows, so the above restriction maps define a contravariant functor

$$\mathcal{F} : \mathrm{SInc}\big(\mathcal{H}^{(1)}\big)^{\mathrm{op}} \to \mathbf{Vect}_{\mathbb{R}}.$$

Therefore,

$$\big(\mathcal{H}^{(1)}, \mathcal{F}\big)$$

is a Sheaf 1-SuperHyperGraph.

The vector spaces attached to the supervertices represent local data spaces on grouped entities, while the vector spaces attached to the superhyperedges represent joint data spaces on higher-order interactions. The restriction maps extract or aggregate local information from each superhyperedge to its incident supervertices.

**Theorem 5.17.7** (Well-definedness of Sheaf $n$-SuperHyperGraphs)**.** *Let $\mathbb{K}$ be a field, let $V_0$ be a finite nonempty set, let $n \in \mathbb{N}_0$, and let*

$$\mathcal{H}^{(n)} = (V, E)$$

*be an $n$-SuperHyperGraph. Assume that its superincidence poset is*

$$\mathrm{SInc}\big(\mathcal{H}^{(n)}\big) := (V \sqcup E, \preceq),$$

*where $\preceq$ is the least reflexive relation satisfying*

$$v \preceq e \iff v \in e \qquad (v \in V,\ e \in E).$$

*Then the following hold:*

1. $\mathrm{SInc}\big(\mathcal{H}^{(n)}\big)$ *is a well-defined finite poset;*

2. *its opposite category*

$$\mathrm{SInc}\big(\mathcal{H}^{(n)}\big)^{\mathrm{op}}$$

*is a well-defined small category;*

3. *consequently, every contravariant functor*

$$\mathcal{F} : \mathrm{SInc}\big(\mathcal{H}^{(n)}\big)^{\mathrm{op}} \to \mathbf{Vect}_{\mathbb{K}}$$

*is a well-defined mathematical object, and hence the pair*

$$\big(\mathcal{H}^{(n)}, \mathcal{F}\big)$$

*is a well-defined Sheaf $n$-SuperHyperGraph.*

*Proof.* Since $\mathcal{H}^{(n)} = (V, E)$ is an $n$-SuperHyperGraph, the set $V$ is finite and satisfies

$$V \subseteq \mathcal{P}^n(V_0),$$

while $E$ is a finite family of nonempty subsets of $V$. Hence the disjoint union

$$V \sqcup E$$

is a finite set.

We first verify that $\preceq$ defines a partial order on $V \sqcup E$.

*Reflexivity.* By construction, $\preceq$ is reflexive.

*Antisymmetry.* Suppose $x, y \in V \sqcup E$ satisfy $x \preceq y$ and $y \preceq x$. If either relation is non-identity, then by definition one element must be a supervertex and the other a superhyperedge. More precisely, a nontrivial relation has the form

$$v \preceq e \qquad (v \in V,\ e \in E,\ v \in e).$$

There is no nontrivial relation of the form $e \preceq v$, because the generating incidence relation only goes from supervertices to superhyperedges. Therefore $x \preceq y$ and $y \preceq x$ can simultaneously hold only when $x = y$. Thus $\preceq$ is antisymmetric.

*Transitivity.* Let $x, y, z \in V \sqcup E$ with $x \preceq y$ and $y \preceq z$. If either relation is an identity, transitivity is immediate. So it remains to consider non-identity relations. But every non-identity relation has the form

$$v \preceq e \qquad (v \in V,\ e \in E).$$

Hence there is no chain of two distinct non-identity arrows: a superhyperedge cannot be below any distinct element, and no distinct element can lie below a supervertex. Therefore every composable pair reduces to a case involving an identity morphism, and transitivity follows.

Thus $(V \sqcup E, \preceq)$ is a well-defined finite poset. This proves (1).

Every poset canonically determines a small category: objects are the elements of the poset, and there is a unique morphism

$$x \to y \iff x \preceq y.$$

Since $V \sqcup E$ is finite, this category is small. Therefore its opposite category

$$\mathrm{SInc}\big(\mathcal{H}^{(n)}\big)^{\mathrm{op}}$$

is also well-defined and small. This proves (2).

Finally, $\mathbf{Vect}_{\mathbb{K}}$ is a well-defined category because $\mathbb{K}$ is a field. Hence any contravariant functor

$$\mathcal{F} : \mathrm{SInc}\big(\mathcal{H}^{(n)}\big)^{\mathrm{op}} \to \mathbf{Vect}_{\mathbb{K}}$$

is a well-defined functor from a small category to $\mathbf{Vect}_{\mathbb{K}}$. Therefore the pair

$$\big(\mathcal{H}^{(n)}, \mathcal{F}\big)$$

is well-defined as a Sheaf $n$-SuperHyperGraph. This proves (3). □

**Proposition 5.17.8** (Equivalent local description)**.** *Let $\mathcal{H}^{(n)} = (V, E)$ be an $n$-SuperHyperGraph. Giving a contravariant functor*

$$\mathcal{F} : \mathrm{SInc}\big(\mathcal{H}^{(n)}\big)^{\mathrm{op}} \to \mathbf{Vect}_{\mathbb{K}}$$

*is equivalent to giving:*

1. *a $\mathbb{K}$-vector space $\mathcal{F}(x)$ for each $x \in V \sqcup E$;*
2. *for each superincidence $v \in e$, a linear map*

$$\rho_{e,v} : \mathcal{F}(e) \to \mathcal{F}(v);$$

3. *the identity maps*

$$\rho_{x,x} = \mathrm{id}_{\mathcal{F}(x)} \qquad (x \in V \sqcup E).$$

*No additional cocycle condition is required beyond identities, because in* $\mathrm{SInc}(\mathcal{H}^{(n)})$ *there are no nontrivial composable chains of length* 2.

*Proof.* A functor on a poset category assigns an object to each element and a morphism to each order relation, preserving identities and composition. Here the only non-identity order relations are the incidences

$$v \preceq e \iff v \in e.$$

Passing to the opposite category reverses these arrows, so each incidence gives a unique morphism

$$e \to v,$$

which is exactly the restriction map

$$\rho_{e,v} : \mathcal{F}(e) \to \mathcal{F}(v).$$

Since there are no nontrivial composable chains of two distinct incidences, functoriality imposes no further compatibility except preservation of identities. Hence the two descriptions are equivalent. □

## 5.18 Fibered HyperGraph / Fibered SuperHyperGraph

A Fibered HyperGraph assigns each vertex a local state space and each hyperedge a compatibility relation on those fibers collectively. A Fibered SuperHyperGraph assigns each supervertex a local state space and each superhyperedge a compatibility relation on those fibers collectively.

**Definition 5.18.1** (Fibered HyperGraph)**.** A *Fibered HyperGraph* is a quadruple

$$\mathfrak{F} = (V, E, \{F_v\}_{v\in V}, \{R_e\}_{e\in E}),$$

such that:

1. $H = (V, E)$ is a finite HyperGraph;
2. for each vertex $v \in V$, $F_v$ is a nonempty set, called the *fiber over* $v$;
3. for each hyperedge $e \in E$, one is given a relation

$$R_e \subseteq \prod_{v\in e} F_v.$$

**Remark 5.18.2.** *The relation* $R_e$ *describes the admissible joint states of the fibers attached to the vertices belonging to the hyperedge* $e$*.*

**Definition 5.18.3** (Total space of a Fibered HyperGraph)**.** Let

$$\mathfrak{F} = (V, E, \{F_v\}_{v\in V}, \{R_e\}_{e\in E})$$

be a Fibered HyperGraph. Its *total space* is the disjoint union

$$X := \bigsqcup_{v\in V} F_v,$$

equipped with the natural projection

$$\pi : X \to V, \qquad \pi(x) = v \text{ if } x \in F_v.$$

**Definition 5.18.4** (Fibered $n$-SuperHyperGraph)**.** Let $\mathcal{H}^{(n)} = (V, E)$ be a finite $n$-SuperHyperGraph. A *Fibered n-SuperHyperGraph* is a quadruple

$$\mathfrak{F}^{(n)} = (V, E, \{F_v\}_{v\in V}, \{R_e\}_{e\in E}),$$

such that:

1. $\mathcal{H}^{(n)} = (V, E)$ is an $n$-SuperHyperGraph;
2. for each supervertex $v \in V$, $F_v$ is a nonempty set;

3. for each superhyperedge $e \in E$, one is given a relation

$$R_e \subseteq \prod_{v \in e} F_v.$$

**Remark 5.18.5.** *When $n = 0$, a Fibered 0-SuperHyperGraph is exactly a Fibered HyperGraph.*

**Example 5.18.6** (A Fibered 1-SuperHyperGraph for team-state compatibility)**.** Let the base set be

$$V_0 = \{a, b, c, d\},$$

and take $n = 1$, so that

$$\mathcal{P}^1(V_0) = \mathcal{P}(V_0).$$

Define the following 1-supervertices:

$$v_1 := \{a, b\}, \qquad v_2 := \{c\}, \qquad v_3 := \{d\}.$$

Set

$$V = \{v_1, v_2, v_3\} \subseteq \mathcal{P}(V_0).$$

Define the superhyperedge family by

$$E = \{e_1, e_2\}, \qquad e_1 = \{v_1, v_2\}, \qquad e_2 = \{v_1, v_3\}.$$

Then

$$\mathcal{H}^{(1)} = (V, E)$$

is a finite 1-SuperHyperGraph.

Next, assign a nonempty fiber to each supervertex:

$$F_{v_1} = \{\text{low}, \text{high}\}, \qquad F_{v_2} = \{\text{off}, \text{on}\}, \qquad F_{v_3} = \{\text{idle}, \text{busy}\}.$$

Here one may interpret:

- $F_{v_1}$ as the activity level of the grouped team $\{a, b\}$,
- $F_{v_2}$ as the availability state of the singleton team $\{c\}$,
- $F_{v_3}$ as the workload state of the singleton team $\{d\}$.

Now define a relation on each superhyperedge.
For

$$e_1 = \{v_1, v_2\},$$

the corresponding product is

$$\prod_{v \in e_1} F_v = F_{v_1} \times F_{v_2},$$

and define

$$R_{e_1} = \{(\text{low}, \text{off}),\ (\text{high}, \text{on})\} \subseteq F_{v_1} \times F_{v_2}.$$

For

$$e_2 = \{v_1, v_3\},$$

the corresponding product is

$$\prod_{v \in e_2} F_v = F_{v_1} \times F_{v_3},$$

and define

$$R_{e_2} = \{(\text{low}, \text{idle}),\ (\text{high}, \text{busy})\} \subseteq F_{v_1} \times F_{v_3}.$$

Therefore,

$$\mathfrak{F}^{(1)} = \big(V, E, \{F_v\}_{v \in V}, \{R_e\}_{e \in E}\big)$$

is a Fibered 1-SuperHyperGraph.

The supervertex $v_1 = \{a, b\}$ has its own local state space, and each superhyperedge imposes a compatibility relation between the states of the participating supervertices. Thus the underlying 1-SuperHyperGraph is enriched by fiber data and edgewise state constraints.

**Theorem 5.18.7** (Well-definedness of Fibered $n$-SuperHyperGraphs)**.** *Let $V_0$ be a finite nonempty set, let $n \in \mathbb{N}_0$, and let*

$$\mathcal{H}^{(n)} = (V, E)$$

*be a finite $n$-SuperHyperGraph. Assume that:*

1. *for each $v \in V$, a nonempty set $F_v$ is given;*
2. *for each $e \in E$, a subset*

$$R_e \subseteq \prod_{v \in e} F_v$$

*is given.*

*Then the quadruple*

$$\mathfrak{F}^{(n)} = (V, E, \{F_v\}_{v \in V}, \{R_e\}_{e \in E})$$

*is a well-defined Fibered $n$-SuperHyperGraph.*
*In particular:*

1. *for every superhyperedge $e \in E$, the Cartesian product*

$$\prod_{v \in e} F_v$$

*is well-defined;*

2. *each $R_e$ is a well-defined relation on the family of fibers indexed by the supervertices contained in $e$.*

*Proof.* Since $\mathcal{H}^{(n)} = (V, E)$ is a finite $n$-SuperHyperGraph, by definition we have

$$V \subseteq \mathcal{P}^n(V_0), \qquad E \subseteq \mathcal{P}(V) \setminus \{\varnothing\}.$$

Hence $V$ is a finite set of $n$-supervertices, and each $e \in E$ is a nonempty finite subset of $V$.

Now fix $e \in E$. Because $e \subseteq V$ and $V$ is finite, the index set $e$ is finite. By assumption, for every $v \in e$, the fiber $F_v$ is a nonempty set. Therefore the indexed Cartesian product

$$\prod_{v \in e} F_v$$

is well-defined as the set of all tuples

$$(x_v)_{v \in e} \quad \text{such that} \quad x_v \in F_v \text{ for every } v \in e.$$

Since $e$ is finite and each $F_v \neq \varnothing$, this product is a well-defined set; moreover, it is nonempty.

Again by assumption, for each $e \in E$ one is given a subset

$$R_e \subseteq \prod_{v \in e} F_v.$$

Hence $R_e$ is a well-defined relation among the fibers corresponding to the supervertices in the superhyperedge $e$.

Since this construction is valid for every $e \in E$, all components of

$$\mathfrak{F}^{(n)} = (V, E, \{F_v\}_{v \in V}, \{R_e\}_{e \in E})$$

are well-defined and satisfy the defining conditions of a Fibered $n$-SuperHyperGraph. Therefore $\mathfrak{F}^{(n)}$ is well-defined. □

## 5.19 Galois HyperGraph / Galois SuperHyperGraph

A Galois HyperGraph is a hypergraph whose hyperedges are nonempty Galois-closed vertex sets induced by a formal context. A Galois SuperHyperGraph is a superhypergraph whose superhyperedges are nonempty Galois-closed supervertex sets induced by a formal context.

**Definition 5.19.1** (Formal context)**.** A *formal context* is a triple

$$\mathbb{C} = (X, M, \mathcal{I}),$$

where $X$ is a nonempty set of objects, $M$ is a nonempty set of attributes, and

$$\mathcal{I} \subseteq X \times M$$

is an incidence relation.

For $A \subseteq X$ and $B \subseteq M$, define the derivation operators

$$A^{\uparrow} := \{\, m \in M \mid \forall x \in A,\ (x, m) \in \mathcal{I} \,\},$$

$$B^{\downarrow} := \{\, x \in X \mid \forall m \in B,\ (x, m) \in \mathcal{I} \,\}.$$

Then the operator

$$\mathrm{cl}(A) := A^{\uparrow\downarrow}$$

is the associated Galois closure on subsets of $X$.

**Definition 5.19.2** (Galois HyperGraph)**.** A *Galois HyperGraph* is a quadruple

$$\mathfrak{G} = (V, M, \mathcal{I}, E),$$

such that:

1. $(V, M, \mathcal{I})$ is a finite formal context;
2. $$E \subseteq \{\, A \subseteq V \mid A \neq \varnothing,\ A^{\uparrow\downarrow} = A \,\}.$$

The hyperedges in $E$ are called *Galois hyperedges*; they are precisely the chosen nonempty closed subsets of the object set $V$ under the Galois closure operator.

**Remark 5.19.3.** *Thus a Galois HyperGraph is a HyperGraph whose admissible hyperedges are constrained by closure in a formal context.*

**Definition 5.19.4** (Galois $n$-SuperHyperGraph)**.** Let $V_0$ be a finite nonempty base set and let $n \in \mathbb{N}_0$. A *Galois $n$-SuperHyperGraph* is a quadruple

$$\mathfrak{G}^{(n)} = (V, M, \mathcal{I}, E),$$

such that:

1. $$V \subseteq \mathcal{P}^n(V_0)$$ is a finite set of $n$-supervertices;
2. $(V, M, \mathcal{I})$ is a formal context;
3. $$E \subseteq \{\, A \subseteq V \mid A \neq \varnothing,\ A^{\uparrow\downarrow} = A \,\}.$$

The members of $E$ are called *Galois superhyperedges*.

**Remark 5.19.5.** *When $n = 0$, a Galois 0-SuperHyperGraph reduces to a Galois HyperGraph.*

**Example 5.19.6** (A Galois 1-SuperHyperGraph)**.** Let the finite nonempty base set be

$$V_0 = \{a, b, c\},$$

and take $n = 1$. Then

$$\mathcal{P}^1(V_0) = \mathcal{P}(V_0).$$

We choose the following 1-supervertices:

$$v_1 := \{a\}, \qquad v_2 := \{b\}, \qquad v_3 := \{a, b\}.$$

Set

$$V = \{v_1, v_2, v_3\} \subseteq \mathcal{P}(V_0).$$

Next, let the attribute set be

$$M = \{m_1, m_2\}.$$

Interpret $m_1$ as a common basic property and $m_2$ as an additional higher-level property. Define the incidence relation

$$\mathcal{I} \subseteq V \times M$$

by

$$\mathcal{I} = \{(v_1, m_1),\ (v_2, m_1),\ (v_3, m_1),\ (v_3, m_2)\}.$$

Thus the incidences are:

$$v_1^{\uparrow} = \{m_1\}, \qquad v_2^{\uparrow} = \{m_1\}, \qquad v_3^{\uparrow} = \{m_1, m_2\}.$$

We now compute some Galois closures.

**(i) The singleton $\{v_3\}$.** We have

$$\{v_3\}^{\uparrow} = \{m_1, m_2\}.$$

Hence

$$\{v_3\}^{\uparrow\downarrow} = \{\, v \in V \mid (v, m_1) \in \mathcal{I} \text{ and } (v, m_2) \in \mathcal{I} \,\} = \{v_3\}.$$

Therefore $\{v_3\}$ is $\mathcal{I}$-closed.

**(ii) The full set $V = \{v_1, v_2, v_3\}$.** Since $m_1$ is the only attribute common to all three supervertices,

$$V^{\uparrow} = \{m_1\}.$$

Thus

$$V^{\uparrow\downarrow} = \{\, v \in V \mid (v, m_1) \in \mathcal{I} \,\} = V.$$

Hence $V$ is also $\mathcal{I}$-closed.

Now define

$$E = \big\{\{v_3\},\ \{v_1, v_2, v_3\}\big\}.$$

Each member of $E$ is nonempty and Galois-closed, so

$$E \subseteq \{\, A \subseteq V \mid A \neq \varnothing,\ A^{\uparrow\downarrow} = A \,\}.$$

Therefore

$$\mathfrak{G}^{(1)} = (V, M, \mathcal{I}, E)$$

is a Galois 1-SuperHyperGraph.

The supervertex $v_3 = \{a, b\}$ forms a closed singleton because it alone possesses the additional attribute $m_2$. Meanwhile, the whole family $V$ forms another closed set because all supervertices share the common attribute $m_1$. Thus the Galois superhyperedges encode closure-stable families of supervertices determined by the formal context.

**Theorem 5.19.7** (Well-definedness of Galois $n$-SuperHyperGraphs)**.** *Let $V_0$ be a finite nonempty base set, let $n \in \mathbb{N}_0$, and let*

$$V \subseteq \mathcal{P}^n(V_0)$$

*be a finite set. Let $M$ be a set and let*

$$\mathcal{I} \subseteq V \times M$$

*be a binary relation. Define, for every $A \subseteq V$ and $B \subseteq M$,*

$$A^{\uparrow} := \{\, m \in M \mid \forall v \in A,\ (v, m) \in \mathcal{I} \,\}, \qquad B^{\downarrow} := \{\, v \in V \mid \forall m \in B,\ (v, m) \in \mathcal{I} \,\}.$$

*Then the following hold:*

1. *for every $A \subseteq V$, the set*
$$A^{\uparrow\downarrow} := \left(A^{\uparrow}\right)^{\downarrow}$$
*is a well-defined subset of $V$;*

2. *the family*
$$\mathrm{Cl}_{\mathcal{I}}(V) := \{\, A \subseteq V \mid A^{\uparrow\downarrow} = A \,\}$$
*of $\mathcal{I}$-closed subsets of $V$ is well-defined;*

3. *therefore the family*
$$\mathrm{Cl}^{*}_{\mathcal{I}}(V) := \{\, A \subseteq V \mid A \neq \varnothing,\ A^{\uparrow\downarrow} = A \,\}$$
*of nonempty $\mathcal{I}$-closed subsets of $V$ is well-defined.*

*Consequently, for every choice*
$$E \subseteq \mathrm{Cl}^{*}_{\mathcal{I}}(V),$$
*the quadruple*
$$\mathfrak{G}^{(n)} = (V, M, \mathcal{I}, E)$$
*is a well-defined Galois $n$-SuperHyperGraph.*

*Proof.* Since $V_0$ is a finite nonempty set and $n \in \mathbb{N}_0$, the iterated powerset
$$\mathcal{P}^{n}(V_0)$$
is well-defined by finite recursion. Hence the condition
$$V \subseteq \mathcal{P}^{n}(V_0)$$
is meaningful, and $V$ is a well-defined finite set of $n$-supervertices.

Next, because $\mathcal{I} \subseteq V \times M$, for every subset $A \subseteq V$ the set
$$A^{\uparrow} = \{\, m \in M \mid \forall v \in A,\ (v, m) \in \mathcal{I} \,\}$$
is a well-defined subset of $M$. Indeed, membership of an element $m \in M$ in $A^{\uparrow}$ is determined by the precise first-order condition
$$\forall v \in A,\ (v, m) \in \mathcal{I}.$$

Similarly, for every subset $B \subseteq M$, the set
$$B^{\downarrow} = \{\, v \in V \mid \forall m \in B,\ (v, m) \in \mathcal{I} \,\}$$
is a well-defined subset of $V$. Therefore, for every $A \subseteq V$, the composite set
$$A^{\uparrow\downarrow} = \left(A^{\uparrow}\right)^{\downarrow}$$
is well-defined and satisfies
$$A^{\uparrow\downarrow} \subseteq V.$$
This proves (1).

Now consider the family
$$\mathrm{Cl}_{\mathcal{I}}(V) = \{\, A \subseteq V \mid A^{\uparrow\downarrow} = A \,\}.$$
Since $\mathcal{P}(V)$ is well-defined and each expression $A^{\uparrow\downarrow}$ is well-defined by (1), the condition
$$A^{\uparrow\downarrow} = A$$
is meaningful for every $A \subseteq V$. Hence $\mathrm{Cl}_{\mathcal{I}}(V)$ is a well-defined subfamily of $\mathcal{P}(V)$. This proves (2).

Removing the empty set yields
$$\mathrm{Cl}^{*}_{\mathcal{I}}(V) = \{\, A \subseteq V \mid A \neq \varnothing,\ A^{\uparrow\downarrow} = A \,\},$$
which is likewise a well-defined family of subsets of $V$. This proves (3).

Finally, if
$$E \subseteq \mathrm{Cl}^{*}_{\mathcal{I}}(V),$$
then every member of $E$ is a nonempty subset of $V$ satisfying the Galois-closure condition
$$A^{\uparrow\downarrow} = A.$$
Thus all components in the definition of a Galois $n$-SuperHyperGraph are well-defined, and therefore
$$\mathfrak{G}^{(n)} = (V, M, \mathcal{I}, E)$$
is a well-defined Galois $n$-SuperHyperGraph. □

## 5.20 Rewrite HyperGraph / Rewrite SuperHyperGraph

A Rewrite HyperGraph is a hypergraph equipped with rewrite rules that replace matched subhypergraphs while preserving specified interfaces. A Rewrite SuperHyperGraph is a superhypergraph equipped with rewrite rules that replace matched subsuperhypergraphs while preserving specified interfaces.

**Definition 5.20.1** (HyperGraph monomorphism)**.** Let $H = (V, E)$ and $H' = (V', E')$ be HyperGraphs. A *HyperGraph monomorphism*

$$f : H \hookrightarrow H'$$

is an injective map

$$f_V : V \to V'$$

such that for every hyperedge $e \in E$, its image

$$f_V(e) := \{ f_V(v) \mid v \in e \}$$

belongs to $E'$.

**Definition 5.20.2** (Rewrite rule for HyperGraphs)**.** A *rewrite rule for HyperGraphs* is a span

$$\rho = \big(L \xleftarrow{l} K \xrightarrow{r} R\big),$$

where $L, K, R$ are finite HyperGraphs and

$$l : K \hookrightarrow L, \qquad r : K \hookrightarrow R$$

are HyperGraph monomorphisms.

The HyperGraph $K$ is called the *interface* of the rule, $L$ the *left-hand side*, and $R$ the *right-hand side*.

**Definition 5.20.3** (Rewrite HyperGraph)**.** A *Rewrite HyperGraph* is a pair

$$\mathfrak{H} = (H, \mathscr{R}),$$

where $H$ is a finite HyperGraph and $\mathscr{R}$ is a finite set of rewrite rules for HyperGraphs.

Intuitively, each rule

$$L \leftarrow K \to R$$

specifies that a copy of $L$ occurring inside $H$ may be replaced by a copy of $R$, while preserving the common interface $K$.

**Definition 5.20.4** ($n$-SuperHyperGraph monomorphism)**.** Let $\mathcal{H}^{(n)} = (V, E)$ and $\mathcal{G}^{(n)} = (V', E')$ be $n$-SuperHyperGraphs over the same base level $n$. A *monomorphism of $n$-SuperHyperGraphs*

$$f : \mathcal{H}^{(n)} \hookrightarrow \mathcal{G}^{(n)}$$

is an injective map

$$f_V : V \to V'$$

such that for every superhyperedge $e \in E$,

$$f_V(e) := \{ f_V(v) \mid v \in e \} \in E'.$$

**Definition 5.20.5** (Rewrite rule for $n$-SuperHyperGraphs)**.** A *rewrite rule for $n$-SuperHyperGraphs* is a span

$$\rho = \big(L^{(n)} \xleftarrow{l} K^{(n)} \xrightarrow{r} R^{(n)}\big),$$

where $L^{(n)}, K^{(n)}, R^{(n)}$ are finite $n$-SuperHyperGraphs and

$$l : K^{(n)} \hookrightarrow L^{(n)}, \qquad r : K^{(n)} \hookrightarrow R^{(n)}$$

are monomorphisms of $n$-SuperHyperGraphs.

**Definition 5.20.6** (Rewrite $n$-SuperHyperGraph)**.** A *Rewrite $n$-SuperHyperGraph* is a pair

$$\mathfrak{H}^{(n)} = \big(\mathcal{H}^{(n)}, \mathscr{R}^{(n)}\big),$$

where $\mathcal{H}^{(n)}$ is a finite $n$-SuperHyperGraph and $\mathscr{R}^{(n)}$ is a finite set of rewrite rules for $n$-SuperHyperGraphs.

**Remark 5.20.7.** *When $n = 0$, a Rewrite 0-SuperHyperGraph reduces to a Rewrite HyperGraph.*

**Example 5.20.8** (A Rewrite 1-SuperHyperGraph for team reconfiguration)**.** Let the base set be

$$V_0 = \{a, b, c, d\},$$

and take $n = 1$, so that 1-supervertices are subsets of $V_0$.

**Step 1: The underlying 1-SuperHyperGraph.** Define the following 1-supervertices:

$$v_{ab} := \{a, b\}, \qquad v_c := \{c\}, \qquad v_d := \{d\}, \qquad v_{cd} := \{c, d\}.$$

Let

$$V = \{v_{ab}, v_c, v_d, v_{cd}\} \subseteq \mathcal{P}(V_0).$$

Define the superhyperedge set by

$$E = \{e_1, e_2\}, \qquad e_1 = \{v_{ab}, v_c\}, \qquad e_2 = \{v_{ab}, v_d\}.$$

Then

$$\mathcal{H}^{(1)} = (V, E)$$

is a finite 1-SuperHyperGraph.

**Step 2: A rewrite rule.** We now define a rewrite rule

$$\rho = \left(L^{(1)} \xleftarrow{l} K^{(1)} \xrightarrow{r} R^{(1)}\right)$$

for 1-SuperHyperGraphs.

The left-hand side is

$$L^{(1)} = (V_L, E_L),$$

where

$$V_L = \{x_{ab}, x_c\}, \qquad x_{ab} := \{a, b\}, \qquad x_c := \{c\},$$

and

$$E_L = \{\ell\}, \qquad \ell = \{x_{ab}, x_c\}.$$

The interface is

$$K^{(1)} = (V_K, E_K),$$

where

$$V_K = \{y_{ab}\}, \qquad y_{ab} := \{a, b\}, \qquad E_K = \varnothing.$$

The right-hand side is

$$R^{(1)} = (V_R, E_R),$$

where

$$V_R = \{z_{ab}, z_{cd}\}, \qquad z_{ab} := \{a, b\}, \qquad z_{cd} := \{c, d\},$$

and

$$E_R = \{r_1\}, \qquad r_1 = \{z_{ab}, z_{cd}\}.$$

Define the monomorphisms

$$l : K^{(1)} \hookrightarrow L^{(1)}, \qquad l(y_{ab}) = x_{ab},$$

and

$$r : K^{(1)} \hookrightarrow R^{(1)}, \qquad r(y_{ab}) = z_{ab}.$$

Thus the interface preserves the common supervertex $\{a, b\}$, while the left-hand side interaction

$$\{\,\{a, b\}, \{c\}\,\}$$

is replaced by the right-hand side interaction

$$\{\,\{a, b\}, \{c, d\}\,\}.$$

**Step 3: The set of rewrite rules.** Let

$$\mathcal{R}^{(1)} = \{\rho\}.$$

Then

$$\mathfrak{H}^{(1)} = \big(\mathcal{H}^{(1)}, \mathcal{R}^{(1)}\big)$$

is a Rewrite 1-SuperHyperGraph.

The supervertex $\{a, b\}$ may be viewed as a stable core team. The rewrite rule $\rho$ replaces a collaboration between the core team $\{a, b\}$ and the singleton team $\{c\}$ by a collaboration between the same core team and the larger grouped team $\{c, d\}$, while preserving the common interface represented by $\{a, b\}$.

**Theorem 5.20.9** (Well-definedness of Rewrite $n$-SuperHyperGraphs)**.** *Let $V_0$ be a finite nonempty base set, let $n \in \mathbb{N}_0$, and let*

$$\mathcal{H}^{(n)} = (V, E)$$

*be a finite $n$-SuperHyperGraph. Let*

$$\mathcal{R}^{(n)}$$

*be a finite set such that every element*

$$\rho \in \mathcal{R}^{(n)}$$

*is a rewrite rule for $n$-SuperHyperGraphs, i.e. a span*

$$\rho = \Big(L^{(n)} \xleftarrow{l} K^{(n)} \xrightarrow{r} R^{(n)}\Big),$$

*where $L^{(n)}, K^{(n)}, R^{(n)}$ are finite $n$-SuperHyperGraphs and*

$$l : K^{(n)} \hookrightarrow L^{(n)}, \qquad r : K^{(n)} \hookrightarrow R^{(n)}$$

*are monomorphisms of $n$-SuperHyperGraphs.*

*Then the pair*

$$\mathfrak{H}^{(n)} = \big(\mathcal{H}^{(n)}, \mathcal{R}^{(n)}\big)$$

*is a well-defined Rewrite $n$-SuperHyperGraph.*

*Proof.* Since $\mathcal{H}^{(n)}$ is assumed to be a finite $n$-SuperHyperGraph, its underlying data are well-defined: there exists a finite nonempty base set $V_0$ such that

$$V \subseteq \mathcal{P}^n(V_0)$$

is a finite set of $n$-supervertices and

$$E \subseteq \mathcal{P}(V) \setminus \{\varnothing\}$$

is a finite set of nonempty $n$-superhyperedges.

Next, let $\rho \in \mathcal{R}^{(n)}$. By assumption, $\rho$ is a rewrite rule for $n$-SuperHyperGraphs, hence

$$\rho = \Big(L^{(n)} \xleftarrow{l} K^{(n)} \xrightarrow{r} R^{(n)}\Big),$$

where $L^{(n)}, K^{(n)}, R^{(n)}$ are finite $n$-SuperHyperGraphs and the maps

$$l : K^{(n)} \hookrightarrow L^{(n)}, \qquad r : K^{(n)} \hookrightarrow R^{(n)}$$

are monomorphisms of $n$-SuperHyperGraphs. Therefore each $\rho$ is a well-defined span in the category of finite $n$-SuperHyperGraphs.

Since $\mathcal{R}^{(n)}$ is assumed finite, it is a well-defined finite set of such rewrite rules. Hence the ordered pair

$$\big(\mathcal{H}^{(n)}, \mathcal{R}^{(n)}\big)$$

consists of:

1. a well-defined finite $n$-SuperHyperGraph $\mathcal{H}^{(n)}$, and
2. a well-defined finite set $\mathcal{R}^{(n)}$ of rewrite rules for $n$-SuperHyperGraphs.

This is exactly the data required in the definition of a Rewrite $n$-SuperHyperGraph. Consequently,

$$\mathfrak{H}^{(n)} = \left(\mathcal{H}^{(n)}, \mathcal{R}^{(n)}\right)$$

is a well-defined Rewrite $n$-SuperHyperGraph. □

**Corollary 5.20.10** (Special case $n = 0$)**.** *When $n = 0$, a Rewrite 0-SuperHyperGraph reduces to a Rewrite HyperGraph.*

*Proof.* If $n = 0$, then

$$\mathcal{P}^0(V_0) = V_0,$$

so an 0-SuperHyperGraph is just an ordinary HyperGraph. Likewise, a rewrite rule for 0-SuperHyperGraphs is precisely a rewrite rule for HyperGraphs. Therefore the pair

$$\mathfrak{H}^{(0)} = \left(\mathcal{H}^{(0)}, \mathcal{R}^{(0)}\right)$$

is exactly a Rewrite HyperGraph. □

## 5.21 Uncertain SuperHyperGraph

An *Uncertain Set* assigns to each element a degree drawn from a prescribed uncertainty model, thereby providing a unified framework that can encompass fuzzy, intuitionistic fuzzy, neutrosophic, plithogenic, and related set-theoretic formalisms [527, 528, 529, 304].

An *Uncertain Graph* is a graph in which vertices, edges, or both are endowed with degrees determined by an uncertainty model [530]. This general framework includes, among others, fuzzy graphs [531], intuitionistic fuzzy graphs [532, 533], picture fuzzy graphs [534], hesitant fuzzy graphs [535], and neutrosophic graphs [536, 537]. Similarly, an *Uncertain HyperGraph* assigns uncertainty-model degrees to vertices and hyperedges of a HyperGraph, thereby enabling the representation of higher-order interactions under incomplete, imprecise, vague, or indeterminate information. Extending this idea further, an *Uncertain SuperHyperGraph* equips the supervertices, superedges, or both of an $n$-SuperHyperGraph with degrees specified by an uncertainty model. This construction provides a systematic framework for representing hierarchical and higher-order structures together with uncertainty at multiple structural levels [121].

**Definition 5.21.1** (Uncertain Model)**.** [527] Let $U$ denote the class of all *uncertain models.* Each $M \in U$ is specified by

- a nonempty set $\mathrm{Dom}(M) \subseteq [0, 1]^k$ of *admissible degree tuples* for some fixed integer $k \geq 1$;
- model–specific algebraic or geometric constraints on elements of $\mathrm{Dom}(M)$ (for example, $\mu + \nu \leq 1$ in the intuitionistic fuzzy case, or $T + I + F \leq 3$ in the neutrosophic case).

Typical examples include:

- Fuzzy model: $\mathrm{Dom}(M) = [0, 1]$;
- Intuitionistic fuzzy model: $\mathrm{Dom}(M) = \{(\mu, \nu) \in [0, 1]^2 \mid \mu + \nu \leq 1\}$;
- Neutrosophic model: $\mathrm{Dom}(M) = \{(T, I, F) \in [0, 1]^3 \mid 0 \leq T + I + F \leq 3\}$;
- Plithogenic model, and many other extensions.

**Definition 5.21.2** (Uncertain Set (U-Set))**.** [527] Let $X$ be a nonempty universe, and let $M$ be a fixed uncertain model with degree–domain $\mathrm{Dom}(M) \subseteq [0, 1]^k$. An *Uncertain Set of type $M$* (or *U-Set* for short) on $X$ is a pair

$$\mathcal{U} = (X, \mu_M),$$

where

$$\mu_M : X \longrightarrow \mathrm{Dom}(M)$$

is called the *uncertainty–degree function* (or membership map) of $\mathcal{U}$.

For $x \in X$, the value $\mu_M(x) \in \mathrm{Dom}(M)$ encodes the degree(s) to which $x$ belongs to the uncertain set, according to the model $M$.

Table 5.2: A catalogue of uncertainty-set families (U-Sets) by the dimension $k$ of the degree-domain $\mathrm{Dom}(M) \subseteq [0,1]^k$ [304].

| $k$ | note | Representative U-Set model(s) whose degree-domain is a subset of $[0,1]^k$ |
|---|---|---|
| 1 | | Fuzzy Set [538, 83]; N-Fuzzy Set [539, 540, 541]; Shadowed Set [542, 543, 544] |
| 2 | | Intuitionistic Fuzzy Set [545, 546]; Vague Set [547, 548]; Bipolar Fuzzy Set (two-component description) [549, 550]; Variable Fuzzy Set [551, 552, 553]; Paraconsistent Fuzzy Set [554, 555]; Bifuzzy Set [556] |
| 3 | | Single-Valued Neutrosophic Set [557, 558]; Picture Fuzzy Set [559, 560]; Spherical Fuzzy Set [561, 562]; Tripolar Fuzzy Set (three-component formalisms) [563, 564, 565]; Neutrosophic Vague Set [566, 567] |
| 4 | | Quadripartitioned Neutrosophic Set [568, 569]; Double-Valued Neutrosophic Set [570, 571, 572]; Dual Hesitant Fuzzy Set [573, 574]; Ambiguous Set [575, 576, 577]; Turiyam Neutrosophic Set [578, 579, 580, 581] |
| 5 | | Pentapartitioned Neutrosophic Set [582, 583, 584]; Triple-Valued Neutrosophic Set [585, 586, 587] |
| 6 | | Hexapartitioned Neutrosophic Set; Quadruple-Valued Neutrosophic Set [588, 586] |
| 7 | | Heptapartitioned Neutrosophic Set [589, 590]; Quintuple-Valued Neutrosophic Set [586, 591, 592] |
| 8 | | Octapartitioned Neutrosophic Set [593, 594] |
| 9 | | Nonapartitioned Neutrosophic Set [593, 594] |
| $n$ | $(n \geq 1)$ | Multi-valued (Fuzzy) Sets [595]; MultiFuzzy Set [297]; $n$-Refined Fuzzy Set [596, 597] |
| $2n$ | $(n \geq 1)$ | $n$-Refined Intuitionistic Fuzzy Set [597]; Multi-Intuitionistic Fuzzy Set [297] |
| $3n$ | $(n \geq 1)$ | $n$-Refined Neutrosophic Set [597]; Multi-Neutrosophic Set [598, 297] |

**Reading guide.** In the U-set framework [527], each model $M$ is determined by a degree-domain $\mathrm{Dom}(M) \subseteq [0,1]^k$ together with a membership map $\mu_M : X \to \mathrm{Dom}(M)$. The table groups representative families according to the ambient dimension $k$, that is, the number of numerical components assigned to each element.

(a) A widely cited viewpoint is that neutrosophic sets provide a broad unifying framework that includes many earlier multi-component fuzzy models and their generalizations; see [599].

(b) Ambiguous sets are often described as subclasses of certain four-component neutrosophic families; see [569, 600, 577].

(c) Turiyam neutrosophic sets are reported as subclasses of quadripartitioned neutrosophic sets; see [601].

As noted above, many generalizations are possible. For reference, Table 5.2 presents a catalogue of uncertainty-set families (U-Sets), organized according to the dimension $k$ of the degree-domain $\mathrm{Dom}(M) \subseteq [0,1]^k$ (cf. [304]).

We now present the corresponding notion for graphs.

**Definition 5.21.3** (Uncertain graph model)**.** An uncertain graph model is a tuple

$$M = \big(D_V(M), D_E(M), \mathcal{A}_M\big),$$

where $D_V(M)$ and $D_E(M)$ are nonempty sets of admissible vertex and edge degrees, respectively, and

$$\mathcal{A}_M \subseteq D_V(M) \times D_V(M) \times D_E(M)$$

is a vertex–edge admissibility relation satisfying

$$(a, b, c) \in \mathcal{A}_M \iff (b, a, c) \in \mathcal{A}_M.$$

**Definition 5.21.4** (Uncertain graph of type $M$)**.** Let $G = (V, E)$ be a finite simple graph. An uncertain graph of type $M$ is a tuple

$$\mathcal{G}_M = \big(V, E, \mu_V, \mu_E\big),$$

where

$$\mu_V : V \longrightarrow D_V(M), \qquad \mu_E : E \longrightarrow D_E(M),$$

and

$$\big(\mu_V(u), \mu_V(v), \mu_E(\{u, v\})\big) \in \mathcal{A}_M$$

for every $\{u, v\} \in E$.

**Remark 5.21.5** (Specializations of the Uncertain Graph Model)**.** *Particular choices of*

$$M = \big(D_V(M), D_E(M), \mathcal{A}_M\big)$$

*recover many familiar uncertainty-aware graph models.*

- *For a fuzzy graph, one may take*
$$D_V(M) = D_E(M) = [0,1]$$
*and*
$$\mathcal{A}_M = \{(a,b,c) \in [0,1]^3 \mid c \le \min\{a,b\}\}.$$
*The corresponding vertex- and edge-membership functions are*
$$\mu_V : V \longrightarrow [0,1] \qquad \text{and} \qquad \mu_E : E \longrightarrow [0,1].$$

- *Intuitionistic fuzzy graphs, neutrosophic graphs, plithogenic graphs, and many other uncertainty-aware graph models are obtained by selecting the corresponding vertex-degree domain $D_V(M)$, edge-degree domain $D_E(M)$, and admissibility relation $\mathcal{A}_M$.*

- *If a model involves additional information, such as interval structures, attribute-value systems, or contradiction functions, such auxiliary data must be specified separately as part of the corresponding uncertainty framework.*

*For convenience, Table 5.3 presents a catalogue of representative uncertainty-graph families. Where the degree domains admit finite-dimensional real-valued representations, the models may be organized according to the ambient dimensions $k_V$ and $k_E$, where*
$$D_V(M) \subseteq \mathbb{R}^{k_V} \qquad \text{and} \qquad D_E(M) \subseteq \mathbb{R}^{k_E}.$$
*This formulation also accommodates models whose admissible degrees are not restricted to the interval $[0,1]$, including certain bipolar uncertainty models.*

Table 5.3: A catalogue of uncertainty-graph families (uncertain graphs) by the dimension $k$ of the degree-domain $\mathrm{Dom}(M) \subseteq [0,1]^k$.

| $k$ | Representative uncertainty-graph type(s) $\mathcal{G}_M = (V, E, \mu_M)$ with $\mu_M : V \cup E \to \mathrm{Dom}(M) \subseteq [0,1]^k$ |
|---|---|
| 1 | Fuzzy graph; $N$-graph; shadowed-graph variants |
| 2 | Intuitionistic fuzzy graph [602]; bipolar fuzzy graph [603, 604]; intuitionistic evidence graph; variable fuzzy graph; paraconsistent fuzzy graph; bifuzzy graph [605, 606] |
| 3 | Neutrosophic graph [607, 608, 536][(a)]; hesitant fuzzy graph [609, 610]; tripolar fuzzy graph; three-way fuzzy graph; picture fuzzy graph [611, 612]; spherical fuzzy graph [613, 561]; inconsistent intuitionistic fuzzy graph; ternary fuzzy / neutrosophic-fuzzy graph; neutrosophic vague graph |
| 4 | Quadripartitioned neutrosophic graph [614, 615]; double-valued neutrosophic graph [570]; dual hesitant fuzzy graph [616]; ambiguous graph [(b)]; local-neutrosophic graph; support-neutrosophic graph; Turiyam neutrosophic graph [617][(c)] |
| 5 | Pentapartitioned neutrosophic graph [618]; triple-valued neutrosophic graph [619] |
| 6 | Hexapartitioned neutrosophic graph; quadruple-valued neutrosophic graph [619] |
| 7 | Heptapartitioned neutrosophic graph [620]; quintuple-valued neutrosophic graph [619] |
| 8 | Octapartitioned neutrosophic graph |
| 9 | Nonapartitioned neutrosophic graph |
| $n$ | $n$-refined fuzzy graph; multi-valued fuzzy graph; multi-fuzzy graph [621] |
| $2n$ | $n$-refined intuitionistic fuzzy graph; multi-intuitionistic fuzzy graph |
| $3n$ | $n$-refined neutrosophic graph; multi-neutrosophic graph |

[(a)] Neutrosophic graph models are often regarded as broad frameworks capable of specializing to many degree-based graph formalisms under suitable constraints.
[(b)] Ambiguous graph models are commonly presented as subclasses of certain quadripartitioned and double-valued neutrosophic graph models.
[(c)] Turiyam neutrosophic graphs are reported as subclasses of certain quadripartitioned neutrosophic graph models.

The definition of an Uncertain $n$-SuperHyperGraph is given below.

**Definition 5.21.6** (Uncertain $n$-SuperHyperGraph)**.** [121] Let $V_0$ be a finite base set and let $n \in \mathbb{N}_0$. Assume that an $n$-SuperHyperGraph on $V_0$ is given by
$$\mathrm{SHG}^{(n)} = (V_n, E),$$
where
$$\emptyset \ne V_n \subseteq \mathcal{P}^n(V_0) \quad \text{and} \quad \emptyset \ne E \subseteq \mathcal{P}(V_n) \setminus \{\emptyset\},$$

so that each $n$-superedge $e \in E$ is a nonempty subset of the $n$-supervertex set $V_n$.

Let $M$ be a fixed uncertain model with degree–domain $\mathrm{Dom}(M) \subseteq [0,1]^k$. An *Uncertain $n$-SuperHyperGraph of type $M$* is a triple

$$\mathcal{S}_M^{(n)} = (V_n, E, \mu_M),$$

where

$$\mu_M : V_n \cup E \longrightarrow \mathrm{Dom}(M)$$

assigns to each $n$-supervertex $v \in V_n$ and each $n$-superedge $e \in E$ an uncertainty degree $\mu_M(v)$ or $\mu_M(e)$ in $\mathrm{Dom}(M)$.

Any additional relations between the degrees of $n$-superedges and the degrees of the $n$-supervertices they contain (for example, model- specific bounds or aggregations) are imposed by the chosen uncertain model $M$ and are not fixed at the level of this general definition.

For $n = 0$ and $V_0 = V_n$, the above notion reduces to an Uncertain HyperGraph of type $M$.

**Remark 5.21.7.** *Particular choices of the model $M$ recover well–known uncertain SuperHyperGraph types:*

- *Fuzzy $n$-SuperHyperGraphs (when $M$ is fuzzy);*
- *Intuitionistic fuzzy, neutrosophic, and plithogenic $n$-SuperHyperGraphs for the corresponding models $M$;*
- *More exotic variants (e.g. q-rung orthopair [622, 623], picture fuzzy [624], refined neutrosophic[597, 625]) are obtained by choosing the appropriate degree–domain* $\mathrm{Dom}(M)$.

*Regarding the catalogue of uncertainty-superhypergraph families (Uncertain $n$-SuperHyperGraphs) by the dimension $k$ of the degree-domain* $\mathrm{Dom}(M) \subseteq [0,1]^k$, *we list them in Table 5.4.*

Table 5.4: A catalogue of uncertainty-superhypergraph families (Uncertain $n$-SuperHyperGraphs) by the dimension $k$ of the degree-domain $\mathrm{Dom}(M) \subseteq [0,1]^k$.

| $k$ | Representative uncertainty-superhypergraph family (type $M$ with $\mathrm{Dom}(M) \subseteq [0,1]^k$) |
|---|---|
| 1 | *Fuzzy $n$-SuperHyperGraph[58, 626]:* $\mu_M : V_n \cup E \to [0,1]$. |
| 2 | *Intuitionistic-fuzzy $n$-SuperHyperGraph[627, 626]:* $\mu_M : V_n \cup E \to [0,1]^2$ (e.g., (membership, non-membership)). |
| 3 | *Neutrosophic $n$-SuperHyperGraph [98, 628, 100]:* $\mu_M : V_n \cup E \to [0,1]^3$ (e.g., $(T, I, F)$). |
| 4 | *Quadripartitioned / four-component uncertainty $n$-SuperHyperGraph:* $\mu_M : V_n \cup E \to [0,1]^4$. |
| $k$ | *$k$-component uncertainty $n$-SuperHyperGraph (cf. Plithogenic $n$-SuperHyperGraph [103, 102, 629]):* $\mu_M : V_n \cup E \to \mathrm{Dom}(M) \subseteq [0,1]^k$ (model-specific semantics). |

## 5.22 Functorial SuperHyperGraph

A Functorial Set is a functor assigning each object a set and pushing elements along structure-preserving morphisms in a category [527]. A Functorial Graph functorially assigns each object a graph and maps graph homomorphisms along morphisms, preserving composition and identities everywhere. A Functorial HyperGraph assigns each object a hypergraph and transports hyperedges via hypergraph homomorphisms induced by morphisms, respecting categorical composition. A Functorial SuperHyperGraph associates each object with a superhypergraph and sends morphisms to homomorphisms preserving supervertices, superedges, and hierarchical structure.

**Definition 5.22.1** (Functorial Set)**.** [527] Let $\mathcal{C}$ be a category and let

$$F : \mathcal{C} \longrightarrow \mathbf{Set}$$

be a covariant functor.

We call $F$ a *Functorial Set* on $\mathcal{C}$. For each object $X \in \mathrm{Ob}(\mathcal{C})$, the set

$$F(X)$$

is interpreted as the collection of "$F$–sets over $X$", and every element $s \in F(X)$ is an individual $F$–set based at $X$.

Every morphism $f : X \to Y$ in $\mathcal{C}$ induces a *pushforward*

$$F(f) : F(X) \longrightarrow F(Y), \qquad s \longmapsto F(f)(s),$$

and the usual functoriality conditions

$$F(\mathrm{id}_X) = \mathrm{id}_{F(X)}, \qquad F(g \circ f) = F(g) \circ F(f)$$

hold for all composable morphisms $X \xrightarrow{f} Y \xrightarrow{g} Z$.

**Definition 5.22.2** (Functorial SuperHyperGraph)**.** Fix an integer $n \geq 1$. Let $\mathbf{SHGraph}_n$ denote the category whose objects are finite level-$n$ SuperHyperGraphs. Concretely, an object is a triple

$$\mathsf{SH} = (V_0, V, E),$$

where

- $V_0$ is a finite base set;
- $V \subseteq \mathcal{P}^n(V_0)$ is a nonempty set of $n$-supervertices;
- $E \subseteq \mathcal{P}(V) \setminus \{\emptyset\}$ is a nonempty family of $n$-superedges, each superedge being a nonempty subset of $V$.

Thus the supervertices live at the $n$-th iterated powerset level, while the superedges are ordinary (nonempty) subsets of the supervertex set $V$.

A morphism

$$\Phi : (V_0, V, E) \longrightarrow (V_0', V', E')$$

in $\mathbf{SHGraph}_n$ is a *superhypergraph homomorphism*, i.e. a base map $\phi_0 : V_0 \to V_0'$ such that the induced map on the $n$-th iterated powerset

$$\phi_n := \mathcal{P}^n(\phi_0) : \mathcal{P}^n(V_0) \longrightarrow \mathcal{P}^n(V_0')$$

satisfies

$$\phi_n(V) \subseteq V' \quad \text{and} \quad \phi_n[e] := \{\phi_n(v) \mid v \in e\} \in E' \quad \text{for all } e \in E.$$

Let $\mathcal{C}$ be a category. A *Functorial SuperHyperGraph of level* $n$ on $\mathcal{C}$ is a covariant functor

$$\mathsf{SH} : \mathcal{C} \longrightarrow \mathbf{SHGraph}_n.$$

For each object $X \in \mathrm{Ob}(\mathcal{C})$, the value

$$\mathsf{SH}(X) = (V_0^X, V_X, E_X)$$

is a level-$n$ SuperHyperGraph, and for each morphism $f : X \to Y$ in $\mathcal{C}$, the arrow

$$\mathsf{SH}(f) : \mathsf{SH}(X) \longrightarrow \mathsf{SH}(Y)$$

is a superhypergraph homomorphism in the above sense, satisfying

$$\mathsf{SH}(\mathrm{id}_X) = \mathrm{id}_{\mathsf{SH}(X)}, \qquad \mathsf{SH}(g \circ f) = \mathsf{SH}(g) \circ \mathsf{SH}(f)$$

for all composable $f, g$ in $\mathcal{C}$.

In particular, when $n = 0$ and $V = V_0$, a Functorial SuperHyperGraph reduces to a Functorial HyperGraph.

## 5.23 Topological SuperHyperGraph

A Topological Graph combines graph adjacency with a topological structure, enabling analysis of continuity, neighborhoods, convergence, and relationships among vertices [630, 631, 632]. A Topological HyperGraph combines multiway hypergraph relations with topology, supporting continuous, spatial, neighborhood-based, and inclusion-sensitive analysis of higher-order relational interactions [633]. Topological superhypergraph models hierarchical higher-order relations by combining superhypergraph nesting with topological structure, enabling continuous, spatial, and inclusion-based representations faithfully[634].

**Definition 5.23.1** (Topological superhypergraph)**.** Let $P \subset \mathbb{R}^2$ be a finite nonempty set, called the base point set, and let

$$\mathcal{P}^0(P) := P, \qquad \mathcal{P}^{n+1}(P) := \mathcal{P}(\mathcal{P}^n(P)) \quad (n \geq 0).$$

A *topological $n$-superhypergraph* is a pair

$$\mathcal{S} = (V, E)$$

such that

$$V \subseteq \mathcal{P}^n(P), \qquad E \subseteq \mathcal{P}^n(P),$$

together with a topological realization assigning to each element of $V \cup E$ a topological object in $\mathbb{R}^2$ (for example, a point set, closed curve, or closed region), so that the incidence relation is represented by set inclusion of the corresponding realizations.

In particular, when $n = 1$, this reduces to a topological hypergraph.

**Example 5.23.2** (A concrete topological 2-superhypergraph)**.** Let

$$P = \{p_1, p_2, p_3\} \subset \mathbb{R}^2,$$

where

$$p_1 = (0, 0), \qquad p_2 = (2, 0), \qquad p_3 = (1, 2).$$

Then

$$\mathcal{P}^1(P) = \mathcal{P}(P), \qquad \mathcal{P}^2(P) = \mathcal{P}(\mathcal{P}(P)).$$

Define

$$v_1 := \big\{\{p_1\}, \{p_1, p_2\}\big\}, \qquad v_2 := \big\{\{p_2\}, \{p_2, p_3\}\big\},$$

and

$$e := \big\{\{p_1\}, \{p_1, p_2\}, \{p_2\}, \{p_2, p_3\}\big\}.$$

Clearly,

$$v_1, v_2, e \in \mathcal{P}^2(P).$$

Now set

$$V := \{v_1, v_2\}, \qquad E := \{e\}.$$

Thus

$$V \subseteq \mathcal{P}^2(P), \qquad E \subseteq \mathcal{P}^2(P).$$

Next, define a realization map

$$\rho : V \cup E \longrightarrow \{\text{topological subspaces of } \mathbb{R}^2\}$$

by

$$\rho(X) := \bigcup_{A \in X} \operatorname{conv}(A),$$

where $\operatorname{conv}(A)$ denotes the convex hull of $A \subseteq P$.

Then

$$\rho(v_1) = \operatorname{conv}(\{p_1\}) \cup \operatorname{conv}(\{p_1, p_2\}),$$

so $\rho(v_1)$ consists of the point $p_1$ together with the closed line segment joining $p_1$ and $p_2$. Similarly,

$$\rho(v_2) = \operatorname{conv}(\{p_2\}) \cup \operatorname{conv}(\{p_2, p_3\}),$$

which consists of the point $p_2$ together with the closed line segment joining $p_2$ and $p_3$. Moreover,

$$\rho(e) = \operatorname{conv}(\{p_1\}) \cup \operatorname{conv}(\{p_1, p_2\}) \cup \operatorname{conv}(\{p_2\}) \cup \operatorname{conv}(\{p_2, p_3\}).$$

Since

$$v_1 \subseteq e \qquad \text{and} \qquad v_2 \subseteq e,$$

it follows that

$$\rho(v_1) \subseteq \rho(e) \qquad \text{and} \qquad \rho(v_2) \subseteq \rho(e).$$

Hence the incidence relation is represented by inclusion of the corresponding topological realizations.

Therefore,

$$\mathcal{S} = (V, E)$$

is a concrete example of a topological 2-superhypergraph.

A schematic illustration of this concrete topological 2-superhypergraph is shown in Fig. 5.10.

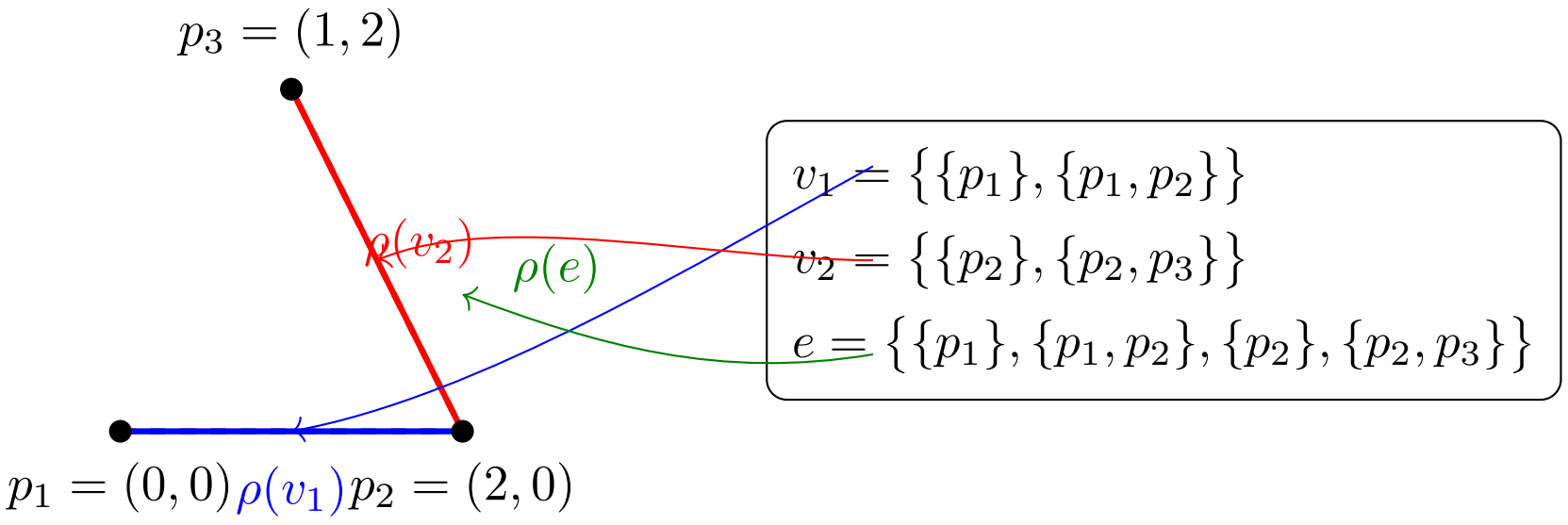

Figure 5.10: A schematic illustration of the concrete topological 2-superhypergraph. The realization $\rho(v_1)$ consists of the point $p_1$ together with the segment joining $p_1$ and $p_2$, the realization $\rho(v_2)$ consists of the point $p_2$ together with the segment joining $p_2$ and $p_3$, and the realization $\rho(e)$ is their union. Thus $\rho(v_1) \subseteq \rho(e)$ and $\rho(v_2) \subseteq \rho(e)$.

## 5.24 Motif Hypergraphs and Motif SuperHypergraphs

A motif hypergraph represents each motif instance as a hyperedge on its supporting vertices, capturing recurring higher-order patterns and aggregating structurally similar local interactions faithfully. A motif superhypergraph treats motif supports as supervertices and links families of them through higher-order relations, encoding interactions among motif-defined subsets themselves in complex networks.

**Definition 5.24.1** (Motif instance)**.** Let

$$M = (V_M, E_M)$$

be a finite graph (respectively, digraph), and let

$$G = (V, E)$$

be a finite graph (respectively, digraph) of the same type.

An *M-instance* in $G$ is an injective map

$$\phi : V_M \to V$$

such that for all $u, v \in V_M$,

$$\{u, v\} \in E_M \implies \{\phi(u), \phi(v)\} \in E$$

in the undirected case, and

$$(u, v) \in E_M \implies (\phi(u), \phi(v)) \in E$$

in the directed case.

We write

$$\mathrm{Inst}_M(G)$$

for the set of all $M$-instances in $G$, and for $\phi \in \mathrm{Inst}_M(G)$ we define its *vertex support* by

$$\mathrm{supp}(\phi) := \phi(V_M) \subseteq V.$$

If one wishes to work with induced motifs, the above implication is replaced by

$$\{u, v\} \in E_M \iff \{\phi(u), \phi(v)\} \in E$$

(or the directed analogue).

**Definition 5.24.2** (Motif Hypergraph)**.** Let $M$ and $G$ be as above, and let

$$\omega : \mathrm{Inst}_M(G) \to \mathbb{R}_{\geq 0}$$

be a nonnegative weight function on motif instances.

The *motif hypergraph* of $G$ with respect to $M$ is the weighted hypergraph

$$\mathcal{H}_M(G) = \big(V, \mathcal{E}_M, w_M\big),$$

where

$$\mathcal{E}_M := \big\{\mathrm{supp}(\phi) \ : \ \phi \in \mathrm{Inst}_M(G)\big\} \subseteq \mathcal{P}^*(V),$$

and for each $e \in \mathcal{E}_M$,

$$w_M(e) := \sum_{\substack{\phi \in \mathrm{Inst}_M(G) \\ \mathrm{supp}(\phi)=e}} \omega(\phi).$$

Thus each hyperedge is the vertex set of an $M$-instance, and if several motif instances have the same support, their weights are aggregated.

In the unweighted case, one usually takes

$$\omega(\phi) = 1 \qquad (\phi \in \mathrm{Inst}_M(G)),$$

so that $w_M(e)$ counts the number of $M$-instances having support $e$.

**Example 5.24.3** (A concrete Motif Hypergraph)**.** Let

$$M = K_3$$

be the triangle graph with vertex set

$$V_M = \{x_1, x_2, x_3\}$$

and edge set

$$E_M = \big\{\{x_1, x_2\}, \{x_2, x_3\}, \{x_1, x_3\}\big\}.$$

Let

$$G = (V, E)$$

be the graph with

$$V = \{1, 2, 3, 4\}$$

and

$$E = \big\{\{1, 2\}, \{2, 3\}, \{1, 3\}, \{2, 4\}, \{3, 4\}\big\}.$$

Thus $G$ contains exactly two triangle subgraphs: one on $\{1, 2, 3\}$ and one on $\{2, 3, 4\}$.

Define two $M$-instances in $G$ by

$$\phi_1(x_1) = 1, \qquad \phi_1(x_2) = 2, \qquad \phi_1(x_3) = 3,$$

and

$$\phi_2(x_1) = 2, \qquad \phi_2(x_2) = 3, \qquad \phi_2(x_3) = 4.$$

Then

$$\mathrm{supp}(\phi_1) = \{1, 2, 3\}, \qquad \mathrm{supp}(\phi_2) = \{2, 3, 4\}.$$

Take the unweighted case

$$\omega(\phi) = 1 \qquad (\phi \in \mathrm{Inst}_M(G)).$$

Hence the motif hyperedge set is

$$\mathcal{E}_M = \big\{\{1, 2, 3\}, \{2, 3, 4\}\big\},$$

and the weight function is given by

$$w_M(\{1, 2, 3\}) = 1, \qquad w_M(\{2, 3, 4\}) = 1.$$

Therefore, the motif hypergraph of $G$ with respect to the triangle motif $M$ is

$$\mathcal{H}_M(G) = \big(\{1, 2, 3, 4\}, \{\{1, 2, 3\}, \{2, 3, 4\}\}, w_M\big).$$

This hypergraph records the two triangle motifs of $G$ as hyperedges on their supporting vertex sets. A schematic illustration of the motif hypergraph for the triangle motif $M = K_3$ is shown in Fig. 5.11.

**Definition 5.24.4** (Motif SuperHyperGraph)**.** Let $M$ and $G = (V, E)$ be as above. Define the set of *motif supervertices* by

$$\mathcal{V}_M := \big\{\mathrm{supp}(\phi) \ : \ \phi \in \mathrm{Inst}_M(G)\big\} \subseteq \mathcal{P}^*(V).$$

Hence every element of $\mathcal{V}_M$ is itself a subset of the base set $V$.

Let

$$R : \mathcal{P}^*(\mathcal{V}_M) \to \{\mathrm{true}, \mathrm{false}\}$$

**Graph $G$**

$\{1,2,3\}$ $\{2,3,4\}$

1 2 3 4

$M = K_3$ motifs

**Motif hypergraph $\mathcal{H}_M(G)$**

$e_1 = \{1,2,3\}$ $e_2 = \{2,3,4\}$

1 2 3 4

$w_M(\{1,2,3\}) = 1,$
$w_M(\{2,3,4\}) = 1$

Figure 5.11: A schematic illustration of the motif hypergraph associated with the triangle motif $M = K_3$. The graph $G$ contains exactly two triangle instances, on the vertex sets $\{1,2,3\}$ and $\{2,3,4\}$, and these are recorded as hyperedges in the motif hypergraph $\mathcal{H}_M(G)$.

be a prescribed higher-order incidence rule on nonempty families of motif supports. Define

$$\mathcal{F}_M := \big\{\mathcal{A} \in \mathcal{P}^*(\mathcal{V}_M) \;:\; R(\mathcal{A}) \text{ holds}\big\}.$$

Then the pair

$$\mathfrak{S}_M(G;R) := (\mathcal{V}_M, \mathcal{F}_M)$$

is called the *motif SuperHyperGraph* of $G$ with respect to $M$ and $R$.

Equivalently, $\mathfrak{S}_M(G;R)$ is a 1-SuperHyperGraph over the base set $V$, since

$$\mathcal{V}_M \subseteq \mathcal{P}(V) \qquad \text{and} \qquad \mathcal{F}_M \subseteq \mathcal{P}^*(\mathcal{V}_M).$$

A canonical choice of the rule $R$ is the *overlap rule*

$$R_{\mathrm{ov}}(\mathcal{A}) \iff \bigcap_{S\in\mathcal{A}} S \neq \emptyset.$$

In that case,

$$\mathcal{F}_M^{\mathrm{ov}} = \big\{\mathcal{A} \in \mathcal{P}^*(\mathcal{V}_M) \;:\; \bigcap_{S\in\mathcal{A}} S \neq \emptyset\big\},$$

and the resulting motif SuperHyperGraph connects families of motif instances having a common vertex.

**Remark 5.24.5.** *The motif hypergraph records* which vertex subsets support motif instances*, whereas the motif SuperHyperGraph records* higher-order relations among those motif-supported subsets themselves*. Hence the former is a hypergraph on $V$, while the latter is a SuperHyperGraph whose vertices are motif supports.*

**Example 5.24.6** (A concrete Motif SuperHyperGraph)**.** Let $M = K_3$ and let $G = (V,E)$ be the same graph as in the previous example, namely

$$V = \{1,2,3,4\},$$

and

$$E = \big\{\{1,2\},\{2,3\},\{1,3\},\{2,4\},\{3,4\}\big\}.$$

As above, $G$ has two triangle motif supports:

$$S_1 := \{1,2,3\}, \qquad S_2 := \{2,3,4\}.$$

Hence the set of motif supervertices is

$$\mathcal{V}_M = \{S_1, S_2\}.$$

Now choose the canonical overlap rule

$$R_{\mathrm{ov}}(\mathcal{A}) \iff \bigcap_{S\in\mathcal{A}} S \neq \emptyset.$$

Since

$$S_1 \cap S_2 = \{2,3\} \neq \emptyset,$$

the family

$$\{S_1, S_2\}$$

satisfies the overlap rule. Also, each singleton family satisfies the rule because

$$S_1 \neq \emptyset, \qquad S_2 \neq \emptyset.$$

Therefore,

$$\mathcal{F}_M^{\mathrm{ov}} = \bigl\{\{S_1\}, \{S_2\}, \{S_1, S_2\}\bigr\}.$$

Thus the motif SuperHyperGraph is

$$\mathfrak{S}_M(G; R_{\mathrm{ov}}) = \bigl(\{S_1, S_2\}, \{\{S_1\}, \{S_2\}, \{S_1, S_2\}\}\bigr).$$

Equivalently,

$$\mathfrak{S}_M(G; R_{\mathrm{ov}}) = \Bigl(\bigl\{\{1,2,3\}, \{2,3,4\}\bigr\}, \bigl\{\bigl\{\{1,2,3\}\bigr\}, \bigl\{\{2,3,4\}\bigr\}, \bigl\{\{1,2,3\}, \{2,3,4\}\bigr\}\bigr\}\Bigr).$$

In particular, the nontrivial superhyperedge

$$\{S_1, S_2\}$$

expresses that the two triangle motifs overlap on the common vertex set $\{2,3\}$.

## 5.25 Molecular SuperHyperGraphs

A molecular graph represents a molecule as a labeled graph in which vertices correspond to atoms and edges correspond to covalent bonds [635, 203, 636, 163, 637]. A molecular hypergraph generalizes this representation by allowing hyperedges to connect multiple atoms simultaneously, thereby capturing functional groups, aromatic rings, delocalized electron systems, and other multi-atom chemical interactions [638, 639, 640, 641]. A molecular SuperHyperGraph extends this idea further by organizing atoms, bonds, fragments, and larger molecular units across iterated powerset levels, thus providing a hierarchical framework for representing complex chemical structures, multiscale motifs, and overlapping functional contexts [642, 643]. Related notions include chemical graphs [644, 645, 646], chemical hypergraphs [5, 647, 648], chemical SuperHyperGraphs [649, 650], goal-directed molecular graphs [651, 652], molecular graph neural networks[653, 654], and chemical reaction networks [655, 656, 657], all of which arise in computational chemistry, cheminformatics, and reaction-mechanism modeling.

**Definition 5.25.1** (Molecular Graph)**.** (cf. [635]) A *molecular graph* is a labeled simple graph

$$G = (V, E, \ell_V, \ell_E),$$

where

- $V$ is a finite set of *atoms*;
- $E \subseteq \{\{u,v\} \mid u, v \in V,\ u \neq v\}$ is the finite set of *covalent bonds*;
- $\ell_V : V \to \mathcal{L}_V$ assigns to each vertex $v \in V$ its *atomic label* (for example, an element symbol such as C, H, or O);
- $\ell_E : E \to \mathcal{L}_E$ assigns to each edge $e \in E$ its *bond label* (for example, single, double, or triple).

Thus, vertices represent atoms, edges represent bonds, and the labeling maps encode atomic and bond types.

**Definition 5.25.2** (Molecular $n$-SuperHyperGraph)**.** [642, 643] Let $V_0$ be a finite set of *bond identifiers* associated with a molecule. Define the iterated powersets by

$$\mathrm{P}_0(V_0) := V_0, \qquad \mathrm{P}_{k+1}(V_0) := \mathrm{P}\big(\mathrm{P}_k(V_0)\big) \quad (k \geq 0),$$

where $\mathrm{P}(\cdot)$ denotes the ordinary powerset operator.

Fix an integer $n \geq 0$. A *molecular $n$-SuperHyperGraph* on the base set $V_0$ is a quintuple

$$H^{(n)} = \big(V_H, E_H, \partial, \ell_V^H, \ell_E^H\big),$$

where

- $V_H \subseteq \mathrm{P}_n(V_0)$ is a finite set of *$n$-supervertices*, each representing a possibly nested collection of bond identifiers up to level $n$;
- $E_H$ is a finite set of *$n$-superedges*;
- $\partial : E_H \to \mathcal{P}^*(V_H)$ is the incidence map, where

$$\mathcal{P}^*(V_H) := \mathcal{P}(V_H) \setminus \{\emptyset\};$$

  for each $e \in E_H$, the set $\partial(e) \subseteq V_H$ is the family of $n$-supervertices incident with the $n$-superedge $e$;
- $\ell_V^H : V_H \to \mathcal{L}_V$ assigns to each $n$-supervertex $v \in V_H$ a *vertex label* (for example, a bond-pattern type, functional-group name, or moiety type);
- $\ell_E^H : E_H \to \mathcal{L}_E$ assigns to each $n$-superedge $e \in E_H$ an *edge label* (for example, an atom symbol, fragment name, or whole-molecule / functional-unit identifier).

The underlying $n$-SuperHyperGraph of $H^{(n)}$ is the triple

$$(V_H, E_H, \partial).$$

When $n = 0$, the condition $V_H \subseteq \mathrm{P}_0(V_0) = V_0$ implies that each vertex corresponds to a single bond identifier, and the structure reduces to a labeled molecular hypergraph on $V_0$.

**Example 5.25.3** (A concrete molecular 1-SuperHyperGraph for ethanol)**.** Consider the molecule ethanol, with bond identifiers

$$V_0 = \{b_1, b_2, b_3\},$$

where

$$b_1 = \text{the bond } \mathrm{C}_1\!-\!\mathrm{C}_2, \qquad b_2 = \text{the bond } \mathrm{C}_2\!-\!\mathrm{O}, \qquad b_3 = \text{the bond } \mathrm{O}\!-\!\mathrm{H}.$$

Then

$$\mathrm{P}_0(V_0) = V_0, \qquad \mathrm{P}_1(V_0) = \mathrm{P}(V_0).$$

Define the following 1-supervertices:

$$v_1 := \{b_1, b_2\}, \qquad v_2 := \{b_2, b_3\}, \qquad v_3 := \{b_1, b_2, b_3\}.$$

Thus,

$$V_H := \{v_1, v_2, v_3\} \subseteq \mathrm{P}_1(V_0).$$

Let

$$E_H := \{e_1, e_2\},$$

and define the incidence map

$$\partial : E_H \to \mathcal{P}^*(V_H)$$

by

$$\partial(e_1) = \{v_1, v_2\}, \qquad \partial(e_2) = \{v_1, v_2, v_3\}.$$

Next, define the vertex-label map $\ell_V^H : V_H \to \mathcal{L}_V$ by

$$\ell_V^H(v_1) = \text{“C–C–O backbone fragment”},$$

$$\ell_V^H(v_2) = \text{“hydroxyl-bearing fragment”},$$

Figure 5.12: An illustration of the concrete molecular 1-SuperHyperGraph for ethanol. The three 1-supervertices represent overlapping bond-based fragments of ethanol, while the two superedges encode higher-order incidence relations among them.

$$\ell_V^H(v_3) = \text{“whole ethanol bond pattern”},$$

and define the edge-label map $\ell_E^H : E_H \to \mathcal{L}_E$ by

$$\ell_E^H(e_1) = \text{“overlap through the C–O linkage”}, \qquad \ell_E^H(e_2) = \text{“integrated ethanol molecular unit”}.$$

Hence

$$H^{(1)} = \left(V_H, E_H, \partial, \ell_V^H, \ell_E^H\right)$$

is a molecular 1-SuperHyperGraph on the base set $V_0$.

In this example, $v_1$ represents the carbon-chain fragment adjacent to oxygen, $v_2$ represents the hydroxyl-side fragment, and $v_3$ represents the full bond configuration of ethanol. The superedge $e_1$ records the higher-order relation between the two overlapping local fragments, while $e_2$ collects them together with the whole-molecule bond pattern into a single molecular unit. A schematic illustration of this molecular 1-SuperHyperGraph for ethanol is shown in Fig. 5.12.

## 5.26 Soft $n$-SuperHyperGraph

A Soft Set is a parameterized family of subsets over a universe, representing uncertainty by assigning to each parameter a corresponding approximate object set [658, 659]. Related concepts such as HyperSoft Set [660, 661], Bipolar Soft Set[662, 663], Soft Expert Set[664, 665], TreeSoft Set [666, 667], ForestSoft Set [668, 669], and SuperHyperSoft Set [670, 671, 672] are also known. A Soft Graph is a parameterized graph structure in which each parameter selects a subgraph, thereby enabling the modeling of uncertainty through families of vertex and edge subsets[673, 674]. A Soft HyperGraph integrates hypergraph structure with soft-set parameterization, representing uncertain multiway relationships through parameter-dependent families of vertices and hyperedges[92]. A Soft $n$-SuperHyperGraph is an $n$-SuperHyperGraph parameterized by a set, where each parameter selects a sub-superhypergraph of vertices and superedges [106, 108, 675, 676].

**Definition 5.26.1** (Soft $n$-SuperHyperGraph)**.** [676] Let $\mathrm{SHG}^{(n)} = (V, E)$ be an $n$-SuperHyperGraph, and let $C \neq \emptyset$ be a set of parameters. A *Soft n-SuperHyperGraph* over $\mathrm{SHG}^{(n)}$ is a quintuple

$$\mathcal{S}^{(n)} = (V, E, C, A, B),$$

where

$$A : C \to \mathcal{P}(V), \qquad B : C \to \mathcal{P}(E),$$

such that for every $c \in C$,

$$A(c) \subseteq V, \qquad B(c) \subseteq \{\, e \in E : e \subseteq A(c) \,\}.$$

Thus, for each parameter $c$, the pair

$$\big(A(c), B(c)\big)$$

is a sub-superhypergraph of $\mathrm{SHG}^{(n)}$.

**Example 5.26.2** (A Soft 2-SuperHyperGraph)**.** [676] Let the base set be

$$V_0 = \{a, b, c, d\}.$$

Define the following level-1 subsets:

$$T_1 = \{a, b\}, \qquad T_2 = \{b, c\}, \qquad T_3 = \{c, d\}.$$

Now define three level-2 supervertices by

$$D_1 = \{T_1, T_2\}, \qquad D_2 = \{T_2, T_3\}, \qquad D_3 = \{T_1, T_3\}.$$

Then

$$V = \{D_1, D_2, D_3\} \subseteq \mathcal{P}^2(V_0).$$

Next, define three 2-superedges by

$$e_1 = \{D_1, D_2\}, \qquad e_2 = \{D_2, D_3\}, \qquad e_3 = \{D_1, D_3\},$$

and let

$$E = \{e_1, e_2, e_3\}.$$

Hence

$$\mathrm{SHG}^{(2)} = (V, E)$$

is a 2-SuperHyperGraph.

Let the parameter set be

$$C = \{\alpha, \beta, \gamma\}.$$

Define the soft vertex map

$$A : C \to \mathcal{P}(V)$$

by

$$A(\alpha) = \{D_1, D_2\}, \qquad A(\beta) = \{D_2, D_3\}, \qquad A(\gamma) = \{D_1, D_3\},$$

and define the soft edge map

$$B : C \to \mathcal{P}(E)$$

by

$$B(\alpha) = \{e_1\}, \qquad B(\beta) = \{e_2\}, \qquad B(\gamma) = \{e_3\}.$$

We verify the defining condition. Since

$$e_1 = \{D_1, D_2\} \subseteq A(\alpha),$$

we have

$$B(\alpha) = \{e_1\} \subseteq \{\, e \in E : e \subseteq A(\alpha) \,\}.$$

Similarly,

$$e_2 = \{D_2, D_3\} \subseteq A(\beta), \qquad e_3 = \{D_1, D_3\} \subseteq A(\gamma),$$

so

$$B(\beta) = \{e_2\} \subseteq \{\, e \in E : e \subseteq A(\beta) \,\}, \qquad B(\gamma) = \{e_3\} \subseteq \{\, e \in E : e \subseteq A(\gamma) \,\}.$$

Therefore, for each parameter $c \in C$, the pair

$$\big(A(c), B(c)\big)$$

is a sub-superhypergraph of $\mathrm{SHG}^{(2)}$. Hence

$$\mathcal{S}^{(2)} = (V, E, C, A, B)$$

is a Soft 2-SuperHyperGraph.

## 5.27 Rough $n$-SuperHyperGraph

A Rough Set is a set-theoretic model for representing uncertainty through lower and upper approximations induced by indiscernibility relations, thereby distinguishing between definite and possible membership [677, 678]. Related concepts, such as Soft Rough Sets[679, 680], Fuzzy Rough Sets[681, 682], and Neutrosophic Rough Sets[683, 684], are also known and have been studied. A Rough Graph is a graph-theoretic structure with lower and upper approximations of vertices or edges, modeling uncertainty through indiscernibility, partial knowledge, and boundary regions [685, 686]. A Rough $n$-SuperHyperGraph is an $n$-SuperHyperGraph described by lower and upper approximations, capturing uncertainty through equivalence-based vertex and superedge classes [676].

**Definition 5.27.1** (Rough $n$-SuperHyperGraph)**.** [676] Let $\mathrm{SHG}^{(n)} = (V, E)$ be an $n$-SuperHyperGraph. Let $\phi$ be an equivalence relation on $V$ and $\psi$ an equivalence relation on $E$. For $X \subseteq V$ and $Y \subseteq E$, define the lower and upper approximations

$$\underline{\phi}(X) = \{v \in V : [v]_\phi \subseteq X\}, \qquad \overline{\phi}(X) = \{v \in V : [v]_\phi \cap X \neq \emptyset\},$$

and

$$\underline{\psi}(Y) = \{e \in E : [e]_\psi \subseteq Y\}, \qquad \overline{\psi}(Y) = \{e \in E : [e]_\psi \cap Y \neq \emptyset\}.$$

A *Rough n-SuperHyperGraph* is a pair

$$\mathcal{R}^{(n)} = \left(\underline{\mathrm{SHG}}^{(n)}, \overline{\mathrm{SHG}}^{(n)}\right),$$

where

$$\underline{\mathrm{SHG}}^{(n)} = \left(\underline{\phi}(X), \underline{\psi}(Y)\right), \qquad \overline{\mathrm{SHG}}^{(n)} = \left(\overline{\phi}(X), \overline{\psi}(Y)\right)$$

are respectively called the *lower* and *upper* $n$-SuperHyperGraphs.

**Example 5.27.2** (A Rough 2-SuperHyperGraph)**.** [676] Let the base set be

$$V_0 = \{a, b, c, d\}.$$

Define the following level-1 subsets:

$$T_1 = \{a, b\}, \qquad T_2 = \{b, c\}, \qquad T_3 = \{c, d\}.$$

Now define four level-2 supervertices by

$$D_1 = \{T_1\}, \qquad D_2 = \{T_1, T_2\}, \qquad D_3 = \{T_2, T_3\}, \qquad D_4 = \{T_3\}.$$

Then

$$V = \{D_1, D_2, D_3, D_4\} \subseteq \mathcal{P}^2(V_0).$$

Next, define three 2-superedges by

$$e_1 = \{D_1, D_2\}, \qquad e_2 = \{D_2, D_3\}, \qquad e_3 = \{D_3, D_4\},$$

and set

$$E = \{e_1, e_2, e_3\}.$$

Hence

$$\mathrm{SHG}^{(2)} = (V, E)$$

is a 2-SuperHyperGraph.

Define an equivalence relation $\phi$ on $V$ by

$$[D_1]_\phi = [D_2]_\phi = \{D_1, D_2\}, \qquad [D_3]_\phi = [D_4]_\phi = \{D_3, D_4\},$$

and an equivalence relation $\psi$ on $E$ by

$$[e_1]_\psi = \{e_1\}, \qquad [e_2]_\psi = [e_3]_\psi = \{e_2, e_3\}.$$

Choose

$$X = \{D_1, D_2, D_3\} \subseteq V, \qquad Y = \{e_1, e_2\} \subseteq E.$$

We now compute the lower and upper approximations.

For the vertex set,

$$\underline{\phi}(X) = \{v \in V : [v]_\phi \subseteq X\} = \{D_1, D_2\},$$

because

$$[D_1]_\phi = [D_2]_\phi = \{D_1, D_2\} \subseteq X,$$

whereas

$$[D_3]_\phi = [D_4]_\phi = \{D_3, D_4\} \nsubseteq X.$$

Similarly,

$$\overline{\phi}(X) = \{v \in V : [v]_\phi \cap X \neq \emptyset\} = V,$$

since both equivalence classes intersect $X$.

For the edge set,

$$\underline{\psi}(Y) = \{e \in E : [e]_\psi \subseteq Y\} = \{e_1\},$$

because

$$[e_1]_\psi = \{e_1\} \subseteq Y,$$

but

$$[e_2]_\psi = [e_3]_\psi = \{e_2, e_3\} \nsubseteq Y.$$

Also,

$$\overline{\psi}(Y) = \{e \in E : [e]_\psi \cap Y \neq \emptyset\} = E,$$

since both $\{e_1\}$ and $\{e_2, e_3\}$ intersect $Y$.

Therefore, the lower and upper 2-SuperHyperGraphs are

$$\underline{\mathrm{SHG}}^{(2)} = \big(\underline{\phi}(X), \underline{\psi}(Y)\big) = \big(\{D_1, D_2\}, \{e_1\}\big),$$

and

$$\overline{\mathrm{SHG}}^{(2)} = \big(\overline{\phi}(X), \overline{\psi}(Y)\big) = (V, E).$$

Hence the corresponding Rough 2-SuperHyperGraph is

$$\mathcal{R}^{(2)} = \Big(\underline{\mathrm{SHG}}^{(2)}, \overline{\mathrm{SHG}}^{(2)}\Big) = \Big((\{D_1, D_2\}, \{e_1\}), (V, E)\Big).$$

This example illustrates how the rough approximation process extracts a certain lower-level superhypergraph and a larger possible upper-level superhypergraph from the same underlying 2-SuperHyperGraph.

## 5.28 Decision $n$-SuperHyperTree

A *Decision Tree* is a rooted tree used to represent sequential decision processes, chance events, and terminal outcomes. Internal nodes represent decision or chance states, branches represent available actions or possible outcomes, and leaves represent terminal results [687, 688, 689, 690, 691]. Related concepts, including Fuzzy Decision Trees [692, 693, 694], Classification Trees [695, 696, 697, 698], Regression Trees [699, 700], and Bayesian Decision Trees [701, 702], have also been extensively studied. A HyperTree is a connected hypergraph satisfying a chosen acyclicity notion, generalizing ordinary trees from pairwise edges to multiway hyperedges [703, 704, 705]. A *Decision HyperTree* extends the tree-based viewpoint by allowing a branching hyperedge to connect a parent state simultaneously with several child states. This provides a natural representation of higher-order decision, chance, and terminal structures [706, 707, 708]. A *Decision SuperHyperTree* further introduces hierarchical supervertices, thereby allowing decision states and branching structures to be represented by nested objects drawn from iterated powerset levels [709].

To define the direction of branching mathematically, we explicitly distinguish the parent and child supervertices associated with each superhyperedge.

**Definition 5.28.1** (Decision $n$-SuperHyperTree)**.** [709]

Let

$$\mathcal{S}^{(n)} = \big(V_0; V, E, \partial\big)$$

be a finite $n$-SuperHyperTree over a finite nonempty base set $V_0$, where

$$V \subseteq \mathcal{P}^n(V_0), \qquad \partial : E \longrightarrow \mathcal{P}_*(V).$$

Fix a distinguished root

$$R \in V.$$

A *rooted orientation* of $\mathcal{S}^{(n)}$ is a map

$$\rho : E \longrightarrow V$$

such that

$$\rho(e) \in \partial(e) \qquad \text{for every } e \in E.$$

The supervertex $\rho(e)$ is called the *parent* of $e$, and the corresponding child set is

$$\mathrm{Ch}(e) := \partial(e) \setminus \{\rho(e)\}.$$

We assume that

$$\mathrm{Ch}(e) \neq \varnothing \qquad \text{for every } e \in E,$$

and that the rooted orientation satisfies the following tree conditions:

1. the root $R$ is not a child of any superhyperedge;
2. every $Y \in V \setminus \{R\}$ belongs to $\mathrm{Ch}(e)$ for exactly one $e \in E$;
3. every $Y \in V$ is reachable from $R$ by a unique sequence of parent-to-child transitions.

For $X \in V$, define

$$E^{+}(X) := \{e \in E : \rho(e) = X\}.$$

A *Decision $n$-SuperHyperTree* is a tuple

$$\mathcal{D}^{(n)} = \big(\mathcal{S}^{(n)}, R, \rho, \mathrm{Type}, \mathrm{Lab}, \mathrm{Prob}, \mathrm{Util}\big)$$

satisfying the following conditions.

1. The type map

   $$\mathrm{Type} : V \longrightarrow \{\mathrm{dec}, \mathrm{ch}, \mathrm{term}\}$$

   classifies each supervertex as a *decision*, *chance*, or *terminal* state.
2. If

   $$\mathrm{Type}(X) \in \{\mathrm{dec}, \mathrm{ch}\},$$

   then $X$ has exactly one outgoing branching superhyperedge:

   $$|E^{+}(X)| = 1.$$

   Denote this unique superhyperedge by

   $$\varepsilon_X.$$

   Its child set is

   $$\mathrm{Ch}(X) := \mathrm{Ch}(\varepsilon_X) = \partial(\varepsilon_X) \setminus \{X\}.$$
3. If

   $$\mathrm{Type}(X) = \mathrm{term},$$

   then $X$ has no outgoing superhyperedge:

   $$E^{+}(X) = \varnothing.$$
4. For every nonterminal supervertex $X$, there is a set of branch labels $\mathcal{L}_X$ and a labeling map

   $$\mathrm{Lab}_X : \mathrm{Ch}(X) \longrightarrow \mathcal{L}_X.$$

   At a decision supervertex, these labels may represent available actions, whereas at a chance supervertex they may represent possible outcomes.

5. For every chance supervertex $X$, the map

$$\mathrm{Prob}_X : \mathrm{Ch}(X) \longrightarrow [0,1]$$

is a probability mass function satisfying

$$\sum_{Y \in \mathrm{Ch}(X)} \mathrm{Prob}_X(Y) = 1.$$

6. Let

$$V_{\mathrm{term}} := \{X \in V : \mathrm{Type}(X) = \mathrm{term}\}.$$

The utility map

$$\mathrm{Util} : V_{\mathrm{term}} \longrightarrow \mathbb{R}$$

assigns a real-valued terminal utility to each terminal supervertex.

Thus, a Decision $n$-SuperHyperTree is a rooted and oriented $n$-SuperHyperTree in which decision and chance supervertices generate multiway branching through superhyperedges, while terminal supervertices carry final utility values.

**Remark 5.28.2** (Decision and chance branching)**.** *For a decision supervertex $X$, the child set*

$$\mathrm{Ch}(X)$$

*represents the states reachable through the available decisions encoded by the branch labels. For a chance supervertex $X$, the same child set represents possible random outcomes, with*

$$\mathrm{Prob}_X(Y)$$

*giving the probability of reaching the child $Y$. Hence deterministic decisions and stochastic transitions are represented within the same rooted SuperHyperTree structure.*

**Remark 5.28.3** (Reduction to ordinary decision trees)**.** *When*

$$n = 0, \qquad V = V_0,$$

*and every branching superhyperedge satisfies*

$$|\partial(e)| = 2,$$

*each superhyperedge consists of one parent and one child. Consequently, the underlying rooted structure reduces to an ordinary rooted tree, and the Decision* 0*-SuperHyperTree reduces to a classical decision tree equipped with decision, chance, and terminal nodes.*

*For $n \geq 1$, the supervertices may instead be hierarchical objects in*

$$\mathcal{P}^n(V_0),$$

*allowing decisions and outcomes themselves to possess nested internal structure.*

## 5.29 Random Forest, Random HyperForest, and Random SuperHyperForest

A Random Forest combines randomized decision trees and aggregates their predictions, improving accuracy, stability, and robustness while reducing overfitting substantially [710, 711, 712, 713]. Motivated by the classical Random Forest framework, we introduce the following higher-order extensions, called the Random HyperForest and the Random SuperHyperForest. A Random HyperForest combines randomized Decision HyperTrees, aggregating higher-order branching predictions to model multiway relationships with improved robustness and accuracy. A Random SuperHyperForest combines randomized Decision SuperHyperTrees, aggregating hierarchical higher-order predictions to model multiway relationships across multiple structural levels robustly.

**Definition 5.29.1** (Random Forest)**.** A *Random Forest* is an ensemble

$$\mathcal{F} = \{T_b\}_{b=1}^{B}$$

of randomized Decision Trees, where each tree $T_b$ is generated using a random parameter $\Theta_b$, typically incorporating random sampling of observations, features, or both.

If

$$h_b : \mathcal{X} \longrightarrow \mathcal{Y}$$

is the predictor associated with $T_b$, then the forest predictor is

$$H(x) = \mathrm{Agg}\big(h_1(x), \ldots, h_B(x)\big),$$

where Agg is an aggregation rule, such as majority voting for classification or averaging for regression.

**Definition 5.29.2** (Random HyperForest)**.** A *Random HyperForest* is an ensemble

$$\mathcal{F}_H = \{\mathcal{T}_b^H\}_{b=1}^{B}$$

of randomized Decision HyperTrees, where each $\mathcal{T}_b^H$ represents branching through hyperedges that may simultaneously relate a parent state to multiple child states.

If

$$h_b^H : \mathcal{X} \longrightarrow \mathcal{Y}$$

denotes the predictor associated with $\mathcal{T}_b^H$, then the Random HyperForest predictor is

$$H_H(x) = \mathrm{Agg}\big(h_1^H(x), \ldots, h_B^H(x)\big).$$

Thus, a Random HyperForest extends the Random Forest principle from ordinary Decision Trees to randomized higher-order Decision HyperTrees.

**Definition 5.29.3** (Random SuperHyperForest)**.** Fix $n \in \mathbb{N}_0$. A *Random SuperHyperForest of level* $n$ is an ensemble

$$\mathcal{F}_{SH}^{(n)} = \{\mathcal{D}_b^{(n)}\}_{b=1}^{B}$$

of randomized Decision $n$-SuperHyperTrees, where each $\mathcal{D}_b^{(n)}$ has supervertices drawn from an $n$-fold powerset level and represents multiway branching through superhyperedges.

If

$$h_b^{(n)} : \mathcal{X} \longrightarrow \mathcal{Y}$$

is the predictor associated with $\mathcal{D}_b^{(n)}$, then the ensemble predictor is

$$H_{SH}^{(n)}(x) = \mathrm{Agg}\big(h_1^{(n)}(x), \ldots, h_B^{(n)}(x)\big).$$

Hence, a Random SuperHyperForest extends ensemble tree learning by combining randomized hierarchical higher-order Decision $n$-SuperHyperTrees within a common aggregation framework.

**Theorem 5.29.4** (Well-definedness of the Random SuperHyperForest predictor)**.** *Let*

$$\mathcal{F}_{SH}^{(n)} = \{\mathcal{D}_b^{(n)}\}_{b=1}^{B}, \qquad B \geq 1,$$

*be a Random SuperHyperForest of level* $n$*, and suppose that each Decision* $n$*-SuperHyperTree induces a predictor*

$$h_b^{(n)} : \mathcal{X} \longrightarrow \mathcal{Y}.$$

*If the aggregation rule is a well-defined map*

$$\mathrm{Agg} : \mathcal{Y}^B \longrightarrow \mathcal{Y},$$

*then*

$$H_{SH}^{(n)} : \mathcal{X} \longrightarrow \mathcal{Y}, \qquad H_{SH}^{(n)}(x) = \mathrm{Agg}\big(h_1^{(n)}(x), \ldots, h_B^{(n)}(x)\big),$$

*is a well-defined predictor.*

*Proof.* For every $x \in \mathcal{X}$ and every $b \in \{1, \dots, B\}$,

$$h_b^{(n)}(x) \in \mathcal{Y}.$$

Hence

$$\big(h_1^{(n)}(x), \dots, h_B^{(n)}(x)\big) \in \mathcal{Y}^B.$$

Since

$$\mathrm{Agg} : \mathcal{Y}^B \to \mathcal{Y}$$

is well defined, there exists a unique value

$$H_{SH}^{(n)}(x) \in \mathcal{Y}.$$

Therefore $H_{SH}^{(n)}$ is a well-defined map from $\mathcal{X}$ to $\mathcal{Y}$. □

## 5.30 Weighted SuperHyperGraph

A *Weighted Graph* is a graph in which numerical weights are assigned to edges, vertices, or both, thereby representing quantitative information such as cost, distance, capacity, similarity, or interaction strength [714, 715, 716]. Related concepts, including HyperWeighted Graphs [717, 718], Weighted MultiGraphs [719, 720], and Weighted Digraphs [721, 722], are also known and have been studied.

Similarly, a *Weighted HyperGraph* equips the vertices, hyperedges, or both of a HyperGraph with numerical weights, thereby allowing quantitative information to be associated with multiway interactions among several entities [723, 724, 725, 726].

Extending this idea to hierarchical higher-order structures, a *Weighted $n$-SuperHyperGraph* is an $n$-SuperHyperGraph in which each $n$-superhyperedge is assigned a nonnegative real-valued weight [121].

**Definition 5.30.1** (Weighted $n$-SuperHyperGraph)**.** [121] Let $V_0$ be a finite nonempty base set, let $n \in \mathbb{N}_0$, and let

$$\mathrm{SHG}^{(n)} = \big(V_0; V, E, \partial\big)$$

be a finite $n$-SuperHyperGraph, where

$$V \subseteq \mathcal{P}^n(V_0)$$

is the set of $n$-supervertices, $E$ is a finite set of $n$-superhyperedge identifiers, and

$$\partial : E \longrightarrow \mathcal{P}_*(V)$$

is the incidence map.

A *Weighted $n$-SuperHyperGraph* is a pair

$$\big(\mathrm{SHG}^{(n)}, w\big),$$

or equivalently the tuple

$$\mathcal{W}^{(n)} = \big(V_0; V, E, \partial, w\big),$$

where

$$w : E \longrightarrow \mathbb{R}_{\geq 0}$$

is a weight function. For each $e \in E$, the value $w(e)$ is called the *weight* of the $n$-superhyperedge $e$. Thus, $w(e)$ assigns quantitative information to the interaction represented by the incidence set

$$\partial(e) \subseteq V.$$

**Example 5.30.2** (A Weighted 1-SuperHyperGraph)**.** Let the base set be

$$V_0 = \{a, b, c, d\}.$$

For $n = 1$, consider the 1-supervertices

$$v_1 = \{a, b\}, \qquad v_2 = \{b, c\}, \qquad v_3 = \{c, d\}, \qquad v_4 = \{a, d\}.$$

Since

$$v_1, v_2, v_3, v_4 \in \mathcal{P}(V_0),$$

the set

$$V = \{v_1, v_2, v_3, v_4\} \subseteq \mathcal{P}(V_0)$$

is an admissible set of 1-supervertices.

Let

$$E = \{e_1, e_2, e_3\}$$

be the set of superhyperedge identifiers, and define the incidence map

$$\partial : E \longrightarrow \mathcal{P}_*(V)$$

by

$$\partial(e_1) = \{v_1, v_2\}, \qquad \partial(e_2) = \{v_2, v_3, v_4\}, \qquad \partial(e_3) = \{v_1, v_4\}.$$

Hence

$$\mathrm{SHG}^{(1)} = \big(V_0; V, E, \partial\big)$$

is a finite 1-SuperHyperGraph.

Now define the weight function

$$w : E \longrightarrow \mathbb{R}_{\geq 0}$$

by

$$w(e_1) = 2.5, \qquad w(e_2) = 4.0, \qquad w(e_3) = 1.5.$$

Therefore,

$$\mathcal{W}^{(1)} = \big(V_0; V, E, \partial, w\big)$$

is a Weighted 1-SuperHyperGraph.

For instance, the value

$$w(e_2) = 4.0$$

may represent a cost, capacity, distance, or interaction strength associated with the higher-order relation among the three 1-supervertices

$$\partial(e_2) = \{v_2, v_3, v_4\}.$$

Figure 5.13 illustrates this structure.

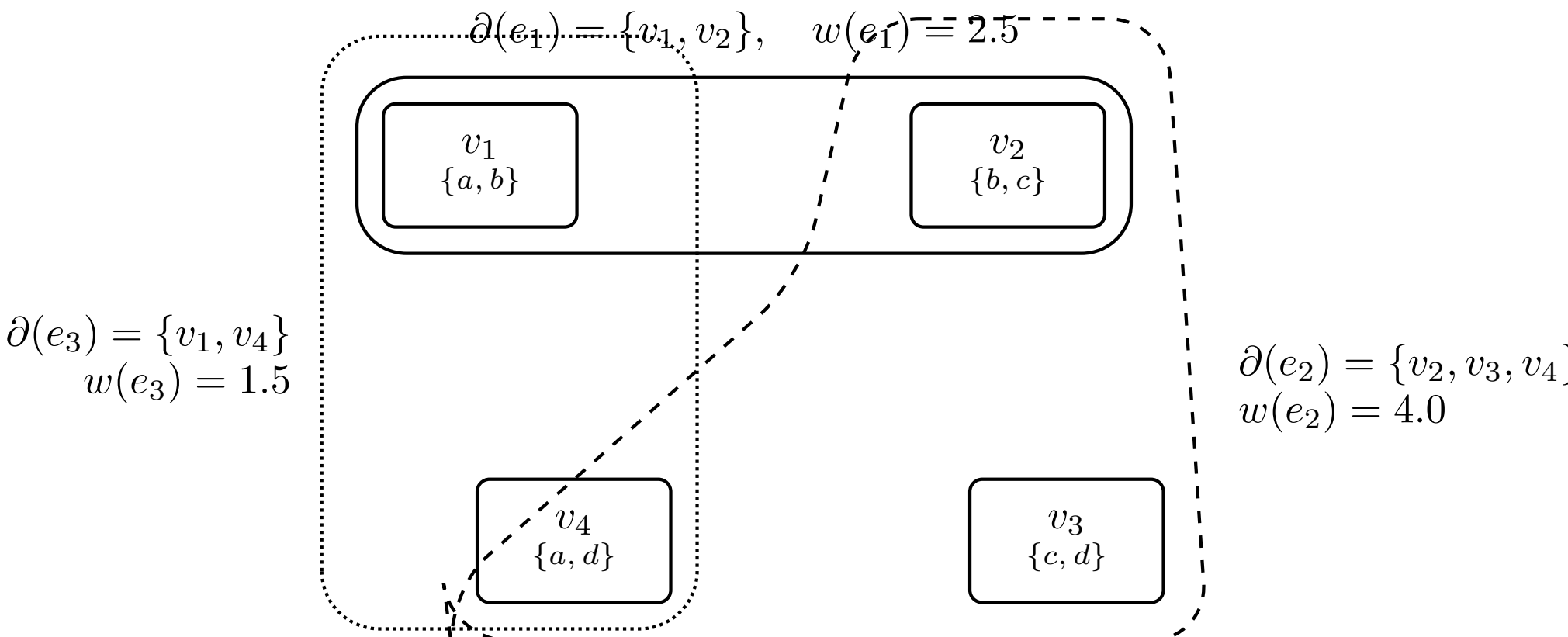

Figure 5.13: A Weighted 1-SuperHyperGraph with three weighted superhyperedges. The solid, dashed, and dotted boundaries represent $e_1$, $e_2$, and $e_3$, respectively, while each superhyperedge is assigned a nonnegative real-valued weight.

This example shows that a Weighted $n$-SuperHyperGraph retains the hierarchical supervertex structure and incidence relations of an $n$-SuperHyperGraph while enriching each superhyperedge with quantitative information through the weight function $w$.

## 5.31 Submodular Hypergraphs

A conventional weighted hypergraph assigns a single weight to each hyperedge. Consequently, every nontrivial bipartition of a given hyperedge is assigned the same cost. A submodular hypergraph provides a more expressive model by assigning a possibly different cost to each bipartition of each hyperedge, subject to symmetry and submodularity conditions [727, 728, 729, 730].

**Definition 5.31.1** (Submodular set function)**.** [731, 732] Let $\Omega$ be a finite set. A set function

$$f : \mathcal{P}(\Omega) \longrightarrow \mathbb{R}$$

is called *submodular* if

$$f(S) + f(T) \geq f(S \cup T) + f(S \cap T)$$

for all

$$S, T \subseteq \Omega.$$

**Remark 5.31.2** (Diminishing-returns characterization)**.** *For a finite ground set $\Omega$, the submodularity condition is equivalent to the following diminishing-returns property:*

$$f(S \cup \{x\}) - f(S) \geq f(T \cup \{x\}) - f(T)$$

*whenever*

$$S \subseteq T \subseteq \Omega \qquad \text{and} \qquad x \in \Omega \setminus T.$$

**Definition 5.31.3** (Submodular Hypergraph)**.** [727] A finite *Submodular Hypergraph* is a quadruple

$$\mathcal{H} = (V, E, \mu, \mathcal{W}),$$

where the following conditions hold:

1. $V$ is a finite, nonempty set whose elements are called *vertices*;
2. $$E \subseteq \mathcal{P}_*(V)$$
   is a finite family whose elements are called *hyperedges*;
3. $$\mu : V \longrightarrow \mathbb{R}_{>0}$$
   is a positive vertex-weight function;
4. $$\mathcal{W} = \{w_e : e \in E\}$$
   is a family of hyperedge cut-cost functions, where, for every $e \in E$,
   $$w_e : \mathcal{P}(e) \longrightarrow \mathbb{R}_{\geq 0}$$
   satisfies the following properties:
   a) *Normalization*:
   $$w_e(\emptyset) = 0;$$
   b) *Symmetry*:
   $$w_e(S) = w_e(e \setminus S) \qquad \text{for every } S \subseteq e;$$
   c) *Submodularity*:
   $$w_e(S) + w_e(T) \geq w_e(S \cup T) + w_e(S \cap T)$$
   for all
   $$S, T \subseteq e.$$

For $S \subseteq e$, the value

$$w_e(S)$$

is interpreted as the cost of dividing the hyperedge $e$ into the two parts

$$S \qquad \text{and} \qquad e \setminus S.$$

**Remark 5.31.4.** *By normalization and symmetry,*

$$w_e(e) = w_e(e \setminus e) = w_e(\emptyset) = 0.$$

*Thus the cost is zero when the hyperedge is not divided. Moreover, symmetry guarantees that the cost does not depend on which side of the bipartition is designated as the first side:*

$$w_e(S) = w_e(e \setminus S).$$

**Definition 5.31.5** (Extension of a hyperedge cut-cost function)**.** Let

$$\mathcal{H} = (V, E, \mu, \mathcal{W})$$

be a Submodular Hypergraph. For each $e \in E$, extend

$$w_e : \mathcal{P}(e) \longrightarrow \mathbb{R}_{\geq 0}$$

to a function

$$\overline{w}_e : \mathcal{P}(V) \longrightarrow \mathbb{R}_{\geq 0}$$

by

$$\overline{w}_e(S) := w_e(S \cap e) \qquad (S \subseteq V).$$

When no confusion can arise, the same notation $w_e$ may be used for both the original function and its extension.

**Proposition 5.31.6.** *If*

$$w_e : \mathcal{P}(e) \longrightarrow \mathbb{R}_{\geq 0}$$

*is submodular, then its extension*

$$\overline{w}_e(S) = w_e(S \cap e)$$

*is submodular on $\mathcal{P}(V)$.*

*Proof.* Let $S, T \subseteq V$. Since $w_e$ is submodular,

$$w_e(S \cap e) + w_e(T \cap e) \geq w_e\big((S \cap e) \cup (T \cap e)\big) + w_e\big((S \cap e) \cap (T \cap e)\big).$$

Using

$$(S \cap e) \cup (T \cap e) = (S \cup T) \cap e$$

and

$$(S \cap e) \cap (T \cap e) = (S \cap T) \cap e,$$

we obtain

$$\overline{w}_e(S) + \overline{w}_e(T) \geq \overline{w}_e(S \cup T) + \overline{w}_e(S \cap T).$$

Therefore, $\overline{w}_e$ is submodular on $\mathcal{P}(V)$. ◻

**Definition 5.31.7** (Cut cost)**.** Let

$$\mathcal{H} = (V, E, \mu, \mathcal{W})$$

be a Submodular Hypergraph, and let

$$\emptyset \neq S \subsetneq V.$$

The bipartition

$$(S, V \setminus S)$$

is called a *cut* of $V$. Its total cut cost is

$$\mathrm{cut}_{\mathcal{H}}(S, V \setminus S) := \sum_{e \in E} w_e(S \cap e).$$

Equivalently,

$$\mathrm{cut}_{\mathcal{H}}(S, V \setminus S) = \sum_{e \in \partial_{\mathcal{H}} S} w_e(S \cap e),$$

where

$$\partial_{\mathcal{H}} S := \{e \in E : e \cap S \neq \emptyset \text{ and } e \setminus S \neq \emptyset\}$$

is the hyperedge boundary of $S$.

**Remark 5.31.8.** *If $e \notin \partial_{\mathcal{H}} S$, then either*

$$S \cap e = \emptyset$$

*or*

$$S \cap e = e.$$

*In either case,*

$$w_e(S \cap e) = 0.$$

*Therefore, only hyperedges crossing the cut contribute to the total cut cost.*

**Definition 5.31.9** (Volume)**.** For a subset $S \subseteq V$, its volume with respect to the vertex-weight function $\mu$ is

$$\mathrm{vol}_\mu(S) := \sum_{v \in S} \mu(v).$$

**Remark 5.31.10** (Equivalent scaled normalization)**.** *For each $e \in E$, suppose that $w_e$ is not identically zero and define*

$$\vartheta_e := \max_{S \subseteq e} w_e(S).$$

*Then*

$$\widehat{w}_e(S) := \frac{w_e(S)}{\vartheta_e}$$

*satisfies*

$$0 \leq \widehat{w}_e(S) \leq 1$$

*and*

$$\max_{S \subseteq e} \widehat{w}_e(S) = 1.$$

*The original cut cost is recovered from*

$$w_e(S) = \vartheta_e \widehat{w}_e(S).$$

*Accordingly, the Submodular Hypergraph may equivalently be written as*

$$\mathcal{H} = (V, E, \mu, \{(\vartheta_e, \widehat{w}_e) : e \in E\}).$$

*This is the normalized convention commonly used in the spectral theory of Submodular Hypergraphs.*

**Proposition 5.31.11** (Homogeneous hypergraphs as a special case)**.** *Let*

$$H = (V, E, \mu, \vartheta)$$

*be a weighted hypergraph, where*

$$\vartheta : E \longrightarrow \mathbb{R}_{\geq 0}$$

*assigns a weight $\vartheta_e$ to each hyperedge $e$. For every $e \in E$, define*

$$w_e(S) := \begin{cases} 0, & S = \emptyset \text{ or } S = e, \\ \vartheta_e, & \emptyset \neq S \subsetneq e. \end{cases}$$

*Then every $w_e$ is normalized, symmetric, and submodular. Consequently,*

$$(V, E, \mu, \{w_e : e \in E\})$$

*is a Submodular Hypergraph.*

*In this special case, every nontrivial bipartition of $e$ has the same cost $\vartheta_e$. Hence ordinary weighted homogeneous hypergraphs form a special subclass of Submodular Hypergraphs.*

*Proof.* Normalization follows immediately from

$$w_e(\emptyset) = 0.$$

The definition also gives

$$w_e(S) = w_e(e \setminus S)$$

for every $S \subseteq e$, so $w_e$ is symmetric.

It remains to verify submodularity. Let $S, T \subseteq e$. Each term takes either the value 0 or $\vartheta_e$. If either $S$ or $T$ is $\emptyset$ or $e$, the required inequality follows directly. Otherwise,

$$w_e(S) + w_e(T) = 2\vartheta_e,$$

whereas

$$w_e(S \cup T) + w_e(S \cap T) \leq 2\vartheta_e.$$

Thus,

$$w_e(S) + w_e(T) \geq w_e(S \cup T) + w_e(S \cap T).$$

Therefore, $w_e$ is submodular. □

**Remark 5.31.12** (Difference from an ordinary weighted hypergraph)**.** *In an ordinary weighted hypergraph, a hyperedge e has a single weight*

$$\vartheta_e.$$

*Hence every nontrivial cut of e has the same cost. By contrast, a Submodular Hypergraph may satisfy*

$$w_e(S) \neq w_e(T)$$

*for distinct nontrivial subsets*

$$S, T \subsetneq e.$$

*It can therefore distinguish different ways of separating the vertices within the same hyperedge.*

**Remark 5.31.13** (Optional omission of vertex weights)**.** *The positive vertex-weight function*

$$\mu : V \longrightarrow \mathbb{R}_{>0}$$

*is useful for defining volumes, conductance, Cheeger constants, and normalized Laplacian operators. When only the combinatorial cut structure is required, one may write a Submodular Hypergraph more simply as*

$$\mathcal{H} = (V, E, \{w_e : e \in E\}).$$

*Alternatively, the unweighted case is obtained by setting*

$$\mu(v) = 1 \qquad \textit{for every } v \in V.$$

## 5.32 Submodular $n$-SuperHyperGraphs

Submodular hypergraphs assign to each hyperedge a symmetric submodular function that measures the cost of splitting that hyperedge [727]. In an $n$-SuperHyperGraph, the vertices are hierarchical objects belonging to an $n$-fold iterated powerset of a base set. Combining these two ideas, we introduce a Submodular $n$-SuperHyperGraph by assigning a submodular splitting function to the incidence set of every superhyperedge.

The submodular structure introduced below acts on sets of $n$-supervertices. It does not directly split the internal elements contained within an individual supervertex.

**Definition 5.32.1** (Symmetric normalized splitting function)**.** Let $\Omega$ be a finite set. A function

$$g : \mathcal{P}(\Omega) \longrightarrow \mathbb{R}_{\geq 0}$$

is called a *symmetric normalized splitting function* if:

1. $g$ is normalized:
$$g(\emptyset) = 0;$$
2. $g$ is symmetric:
$$g(A) = g(\Omega \setminus A) \qquad \text{for every } A \subseteq \Omega.$$

**Remark 5.32.2.** *If g is a symmetric normalized splitting function on $\Omega$, then*

$$g(\Omega) = g(\Omega \setminus \Omega) = g(\emptyset) = 0.$$

*Thus a splitting function assigns zero cost to the two trivial splittings*

$$\emptyset \mid \Omega \qquad \textit{and} \qquad \Omega \mid \emptyset.$$

**Definition 5.32.3** (Submodular $n$-SuperHyperGraph)**.** Let $V_0 \neq \emptyset$ be a finite base set, and let

$$n \in \mathbb{N}_0.$$

A finite *Submodular $n$-SuperHyperGraph over $V_0$* is a sextuple

$$\mathfrak{S}^{(n)} = (V_0; V, E, \partial, \mu, \mathcal{G})$$

satisfying the following conditions:

1. $V$ is a nonempty finite set of $n$-supervertices such that
$$V \subseteq \mathcal{P}^n(V_0);$$
2. $E$ is a finite set whose elements are called *superhyperedge identifiers*;
3. $$\partial : E \longrightarrow \mathcal{P}_*(V)$$
is an incidence map. For every $e \in E$, the set
$$\partial(e) \subseteq V$$
is called the *incidence set* or *superincidence set* of $e$;
4. $$\mu : V \longrightarrow \mathbb{R}_{>0}$$
is a positive supervertex-weight function;
5. $$\mathcal{G} = (g_e)_{e \in E}$$
is an indexed family of functions such that, for every $e \in E$,
$$g_e : \mathcal{P}(\partial(e)) \longrightarrow \mathbb{R}_{\geq 0}$$
satisfies all of the following conditions:
   a) *Normalization*:
   $$g_e(\emptyset) = 0;$$
   b) *Symmetry*:
   $$g_e(A) = g_e(\partial(e) \setminus A) \qquad \text{for every } A \subseteq \partial(e);$$
   c) *Submodularity*:
   $$g_e(A) + g_e(B) \geq g_e(A \cup B) + g_e(A \cap B)$$
   for all
   $$A, B \subseteq \partial(e).$$

For

$$A \subseteq \partial(e),$$

the value

$$g_e(A)$$

is interpreted as the cost of splitting the superhyperedge $e$ into the two groups of incident $n$-supervertices

$$A \qquad \text{and} \qquad \partial(e) \setminus A.$$

**Remark 5.32.4** (Underlying $n$-SuperHyperGraph)**.** *If the functions $\mu$ and $\mathcal{G}$ are omitted, then*

$$(V, E, \partial)$$

*is an incidence-form $n$-SuperHyperGraph over $V_0$. Hence a Submodular $n$-SuperHyperGraph is an $n$-SuperHyperGraph equipped with positive supervertex weights and local submodular splitting functions.*

**Remark 5.32.5** (Parallel superhyperedges)**.** *The incidence map*

$$\partial : E \longrightarrow \mathcal{P}_*(V)$$

*need not be injective. Therefore, distinct superhyperedge identifiers* $e_1, e_2 \in E$ *may satisfy*

$$e_1 \neq e_2$$

*and*

$$\partial(e_1) = \partial(e_2).$$

*Such superhyperedges may nevertheless have different splitting functions:*

$$g_{e_1} \neq g_{e_2}.$$

*Thus the incidence-form definition permits parallel superhyperedges with different cut-cost structures.*

**Definition 5.32.6** (Incidence-faithful Submodular $n$-SuperHyperGraph)**.** A Submodular $n$-SuperHyperGraph

$$\mathfrak{S}^{(n)} = (V_0; V, E, \partial, \mu, \mathcal{G})$$

is called *incidence-faithful* if

$$g_e(\{v\}) > 0$$

for every $e \in E$ and every

$$v \in \partial(e)$$

whenever

$$|\partial(e)| \geq 2.$$

**Remark 5.32.7.** *Incidence-faithfulness excludes redundant supervertices that belong to an incidence set but make no contribution to any split involving that superhyperedge. This additional condition is not required for the basic definition or for the well-definedness results below.*

**Definition 5.32.8** (Global extension of a local splitting function)**.** Let

$$\mathfrak{S}^{(n)} = (V_0; V, E, \partial, \mu, \mathcal{G})$$

be a Submodular $n$-SuperHyperGraph. For each $e \in E$, define

$$\overline{g}_e : \mathcal{P}(V) \longrightarrow \mathbb{R}_{\geq 0}$$

by

$$\overline{g}_e(S) := g_e(S \cap \partial(e)) \qquad (S \subseteq V).$$

**Definition 5.32.9** (Cut function)**.** The *global cut function* of

$$\mathfrak{S}^{(n)} = (V_0; V, E, \partial, \mu, \mathcal{G})$$

is the function

$$\mathrm{Cut}_{\mathfrak{S}^{(n)}} : \mathcal{P}(V) \longrightarrow \mathbb{R}_{\geq 0}$$

defined by

$$\mathrm{Cut}_{\mathfrak{S}^{(n)}}(S) := \sum_{e \in E} g_e(S \cap \partial(e)).$$

Equivalently,

$$\mathrm{Cut}_{\mathfrak{S}^{(n)}}(S) = \sum_{e \in E} \overline{g}_e(S).$$

**Definition 5.32.10** (Superhyperedge boundary)**.** For $S \subseteq V$, define its superhyperedge boundary by

$$\delta_{\mathfrak{S}^{(n)}}(S) := \{e \in E : S \cap \partial(e) \neq \emptyset \text{ and } \partial(e) \setminus S \neq \emptyset\}.$$

Thus $e \in \delta_{\mathfrak{S}^{(n)}}(S)$ precisely when the superhyperedge $e$ has at least one incident supervertex in $S$ and at least one incident supervertex in $V \setminus S$.

**Definition 5.32.11** (Supervertex volume)**.** For $S \subseteq V$, define

$$\mathrm{vol}_\mu(S) := \sum_{v \in S} \mu(v).$$

**Theorem 5.32.12** (Well-definedness of a Submodular $n$-SuperHyperGraph)**.** *Let*

$$\mathfrak{S}^{(n)} = (V_0; V, E, \partial, \mu, \mathcal{G})$$

*satisfy all the conditions of Definition 5.32.3. Then:*

1. $$(V, E, \partial)$$ *is a well-defined finite incidence-form $n$-SuperHyperGraph over $V_0$;*

2. *for every $e \in E$, the extension* $$\overline{g}_e : \mathcal{P}(V) \longrightarrow \mathbb{R}_{\geq 0}$$ *is a well-defined normalized, symmetric, and submodular set function on $V$;*

3. *the global cut function* $$\mathrm{Cut}_{\mathfrak{S}^{(n)}} : \mathcal{P}(V) \longrightarrow \mathbb{R}_{\geq 0}$$ *is a well-defined normalized, symmetric, and submodular set function;*

4. *for every $S \subseteq V$,* $$\mathrm{Cut}_{\mathfrak{S}^{(n)}}(S) = \sum_{e \in \delta_{\mathfrak{S}^{(n)}}(S)} g_e(S \cap \partial(e))\,;$$

5. *the volume* $$\mathrm{vol}_\mu(S)$$ *is finite for every $S \subseteq V$, and* $$\mathrm{vol}_\mu(S) > 0$$ *whenever $S \neq \emptyset$.*

*Consequently, all the structural and cut-related objects appearing in Definition 5.32.3 are mathematically well-defined.*

*Proof.* Since

$$V_0 \neq \emptyset$$

is finite and

$$n \in \mathbb{N}_0,$$

the iterated powerset

$$\mathcal{P}^n(V_0)$$

is a well-defined finite set. By assumption,

$$\emptyset \neq V \subseteq \mathcal{P}^n(V_0),$$

so $V$ is a valid finite set of $n$-supervertices.

The set $E$ is finite, and

$$\partial : E \longrightarrow \mathcal{P}_*(V)$$

is a function. Hence, for every $e \in E$,

$$\emptyset \neq \partial(e) \subseteq V.$$

Therefore,

$$(V, E, \partial)$$

is a well-defined incidence-form $n$-SuperHyperGraph over $V_0$. This proves Assertion 1.

Fix $e \in E$. For every $S \subseteq V$, one has

$$S \cap \partial(e) \subseteq \partial(e).$$

Consequently,

$$S \cap \partial(e) \in \mathcal{P}(\partial(e)),$$

which is precisely the domain of $g_e$. Therefore,

$$\overline{g}_e(S) = g_e(S \cap \partial(e))$$

is defined for every $S \subseteq V$, and

$$\overline{g}_e(S) \in \mathbb{R}_{\geq 0}.$$

Thus

$$\overline{g}_e : \mathcal{P}(V) \longrightarrow \mathbb{R}_{\geq 0}$$

is well-defined.

Moreover,

$$\overline{g}_e(\emptyset) = g_e(\emptyset \cap \partial(e)) = g_e(\emptyset) = 0,$$

so $\overline{g}_e$ is normalized.

For every $S \subseteq V$, because

$$\partial(e) \subseteq V,$$

we have

$$(V \setminus S) \cap \partial(e) = \partial(e) \setminus (S \cap \partial(e)).$$

Using the symmetry of $g_e$, we obtain

$$\begin{aligned}\overline{g}_e(V \setminus S) &= g_e((V \setminus S) \cap \partial(e)) \\ &= g_e(\partial(e) \setminus (S \cap \partial(e))) \\ &= g_e(S \cap \partial(e)) \\ &= \overline{g}_e(S).\end{aligned}$$

Hence $\overline{g}_e$ is symmetric on $V$.

Let $S, T \subseteq V$. Since $g_e$ is submodular on $\partial(e)$,

$$\begin{aligned}&g_e(S \cap \partial(e)) + g_e(T \cap \partial(e)) \\ &\qquad \geq g_e((S \cap \partial(e)) \cup (T \cap \partial(e))) \\ &\qquad\quad + g_e((S \cap \partial(e)) \cap (T \cap \partial(e))).\end{aligned}$$

Using the identities

$$(S \cap \partial(e)) \cup (T \cap \partial(e)) = (S \cup T) \cap \partial(e)$$

and

$$(S \cap \partial(e)) \cap (T \cap \partial(e)) = (S \cap T) \cap \partial(e),$$

we obtain

$$\overline{g}_e(S) + \overline{g}_e(T) \geq \overline{g}_e(S \cup T) + \overline{g}_e(S \cap T).$$

Thus $\overline{g}_e$ is submodular. This proves Assertion 2.

Because $E$ is finite, the sum

$$\mathrm{Cut}_{\mathfrak{S}^{(n)}}(S) = \sum_{e \in E} \overline{g}_e(S)$$

contains only finitely many nonnegative real terms. Hence it is a well-defined nonnegative real number for every $S \subseteq V$.

Normalization follows from

$$\begin{aligned}\mathrm{Cut}_{\mathfrak{S}^{(n)}}(\emptyset) &= \sum_{e \in E} \overline{g}_e(\emptyset) \\ &= \sum_{e \in E} 0 \\ &= 0.\end{aligned}$$

For every $S \subseteq V$, the symmetry of the local extensions gives

$$\begin{aligned}\mathrm{Cut}_{\mathfrak{S}^{(n)}}(V \setminus S) &= \sum_{e \in E} \overline{g}_e(V \setminus S) \\ &= \sum_{e \in E} \overline{g}_e(S) \\ &= \mathrm{Cut}_{\mathfrak{S}^{(n)}}(S).\end{aligned}$$

Finally, for all $S, T \subseteq V$, the submodularity of every $\overline{g}_e$ yields

$$\overline{g}_e(S) + \overline{g}_e(T) \geq \overline{g}_e(S \cup T) + \overline{g}_e(S \cap T).$$

Summing over $e \in E$, we obtain

$$\begin{aligned}&\mathrm{Cut}_{\mathfrak{S}^{(n)}}(S) + \mathrm{Cut}_{\mathfrak{S}^{(n)}}(T) \\ &\quad \geq \mathrm{Cut}_{\mathfrak{S}^{(n)}}(S \cup T) + \mathrm{Cut}_{\mathfrak{S}^{(n)}}(S \cap T).\end{aligned}$$

Thus the global cut function is submodular. This proves Assertion 3.

Consider an edge

$$e \notin \delta_{\mathfrak{S}^{(n)}}(S).$$

Then either

$$S \cap \partial(e) = \emptyset$$

or

$$\partial(e) \setminus S = \emptyset.$$

In the first case,

$$g_e(S \cap \partial(e)) = g_e(\emptyset) = 0.$$

In the second case,

$$S \cap \partial(e) = \partial(e),$$

and symmetry together with normalization gives

$$g_e(\partial(e)) = g_e(\emptyset) = 0.$$

Therefore, only superhyperedges in $\delta_{\mathfrak{S}^{(n)}}(S)$ contribute to the cut, and

$$\mathrm{Cut}_{\mathfrak{S}^{(n)}}(S) = \sum_{e \in \delta_{\mathfrak{S}^{(n)}}(S)} g_e(S \cap \partial(e)).$$

This proves Assertion 4.

Finally, since $V$ is finite and

$$\mu(v) \in \mathbb{R}_{>0}$$

for every $v \in V$, the sum

$$\mathrm{vol}_\mu(S) = \sum_{v \in S} \mu(v)$$

is finite. If $S \neq \emptyset$, it is a finite sum containing at least one strictly positive term, and hence

$$\mathrm{vol}_\mu(S) > 0.$$

This proves Assertion 5, completing the proof. □

**Corollary 5.32.13** (Well-defined normalized cut)**.** *Let*

$$\emptyset \neq S \subsetneq V.$$

*Then*

$$\mathrm{Ncut}_{\mathfrak{S}^{(n)}}(S) := \frac{\mathrm{Cut}_{\mathfrak{S}^{(n)}}(S)}{\min\{\mathrm{vol}_\mu(S), \mathrm{vol}_\mu(V \setminus S)\}}$$

*is well-defined.*

*Proof.* Because

$$\emptyset \neq S \subsetneq V,$$

both $S$ and $V \setminus S$ are nonempty. By Theorem 5.32.12,

$$\mathrm{vol}_\mu(S) > 0$$

and

$$\mathrm{vol}_\mu(V \setminus S) > 0.$$

Therefore, the denominator is strictly positive. The numerator is a finite nonnegative real number, so the quotient is well-defined. □

**Proposition 5.32.14** (Normalized representation)**.** *Let $e \in E$, and suppose that*

$$g_e \not\equiv 0.$$

*Define*

$$\vartheta_e := \max_{A \subseteq \partial(e)} g_e(A).$$

*Then*

$$\vartheta_e > 0,$$

*and the function*

$$w_e(A) := \frac{g_e(A)}{\vartheta_e} \qquad (A \subseteq \partial(e))$$

*is normalized, symmetric, and submodular, with*

$$0 \leq w_e(A) \leq 1$$

*and*

$$\max_{A \subseteq \partial(e)} w_e(A) = 1.$$

*Moreover,*

$$g_e(A) = \vartheta_e w_e(A).$$

*Proof.* Since $\partial(e)$ is finite,

$$\mathcal{P}(\partial(e))$$

is finite. Therefore, the maximum defining $\vartheta_e$ exists. Because $g_e \geq 0$ and $g_e \not\equiv 0$, one has

$$\vartheta_e > 0.$$

Division by the positive constant $\vartheta_e$ preserves normalization, symmetry, and submodularity. The remaining assertions follow directly from the definition of $\vartheta_e$. □

**Remark 5.32.15.** *Using Proposition 5.32.14, the global cut function may be written as*

$$\mathrm{Cut}_{\mathfrak{S}^{(n)}}(S) = \sum_{e \in E} \vartheta_e w_e(S \cap \partial(e)),$$

*where $w_e$ is normalized to take values in $[0, 1]$. This is the form commonly used for Submodular Hypergraphs.*

*If $g_e \equiv 0$, one may set*

$$\vartheta_e := 0$$

*and*

$$w_e := 0.$$

*This convention also accommodates singleton superhyperedges, for which every symmetric normalized splitting function is necessarily zero.*

**Proposition 5.32.16** (Submodular Hypergraphs as the case $n = 0$)**.** *Let*

$$\mathfrak{S}^{(0)} = (V_0; V, E, \partial, \mu, \mathcal{G})$$

*be a Submodular* 0*-SuperHyperGraph. Then*

$$V \subseteq V_0,$$

*and the structure is an incidence-form Submodular Hypergraph on the vertex set $V$.*

*Conversely, every finite incidence-form Submodular Hypergraph determines a Submodular* 0*-SuperHyperGraph by taking*

$$V_0 := V.$$

*Proof.* Since

$$\mathcal{P}^0(V_0) = V_0,$$

the condition

$$V \subseteq \mathcal{P}^0(V_0)$$

reduces to

$$V \subseteq V_0.$$

All remaining components are exactly those of an incidence-form Submodular Hypergraph: an incidence map

$$\partial : E \to \mathcal{P}_*(V),$$

positive vertex weights, and symmetric submodular splitting functions on the incidence sets.

Conversely, given a finite incidence-form Submodular Hypergraph on $V$, set

$$V_0 := V.$$

Then

$$V \subseteq \mathcal{P}^0(V_0) = V_0,$$

so all the conditions of Definition 5.32.3 with $n = 0$ are satisfied. □

**Proposition 5.32.17** (Ordinary weighted $n$-SuperHyperGraphs as a special case)**.** *Let*

$$(V, E, \partial)$$

*be a finite incidence-form $n$-SuperHyperGraph over $V_0$, and let*

$$\omega : E \longrightarrow \mathbb{R}_{\geq 0}$$

*be a superhyperedge-weight function. For every $e \in E$, define*

$$g_e(A) := \begin{cases} 0, & A = \emptyset \text{ or } A = \partial(e), \\ \omega(e), & \emptyset \neq A \subsetneq \partial(e). \end{cases}$$

*Then $g_e$ is normalized, symmetric, and submodular. Therefore, after choosing any positive supervertex-weight function*

$$\mu : V \to \mathbb{R}_{>0},$$

*the tuple*

$$(V_0; V, E, \partial, \mu, (g_e)_{e \in E})$$

*is a Submodular $n$-SuperHyperGraph.*

*Proof.* Normalization and symmetry follow immediately from the definition.

To verify submodularity, fix $e \in E$ and let

$$A, B \subseteq \partial(e).$$

If $A$ or $B$ is either $\emptyset$ or $\partial(e)$, the submodular inequality follows by direct substitution.

Suppose that both $A$ and $B$ are nontrivial subsets of $\partial(e)$. Then

$$g_e(A) + g_e(B) = 2\omega(e).$$

Each of

$$g_e(A \cup B) \qquad \text{and} \qquad g_e(A \cap B)$$

is either 0 or $\omega(e)$. Therefore,

$$g_e(A \cup B) + g_e(A \cap B) \leq 2\omega(e),$$

and hence

$$g_e(A) + g_e(B) \geq g_e(A \cup B) + g_e(A \cap B).$$

Thus $g_e$ is submodular. □

**Remark 5.32.18.** *In the special case of Proposition 5.32.17, every nontrivial split of a superhyperedge has the same cost $\omega(e)$. A general Submodular $n$-SuperHyperGraph is more expressive because it may satisfy*

$$g_e(A) \neq g_e(B)$$

*for two different nontrivial subsets*

$$A, B \subsetneq \partial(e).$$

*It can therefore distinguish different ways of partitioning the same collection of incident $n$-supervertices.*

## 5.33 SuperHyperGraph Neural Network

A Graph Neural Network learns node representations by aggregating neighboring features along edges and applying trainable nonlinear transformations to them [733, 734, 45]. Related concepts, such as Fuzzy Graph Neural Networks [735, 736] and Neutrosophic Graph Neural Networks[737, 738], are also known. A HyperGraph Neural Network learns vertex representations by propagating information through hyperedges, thereby capturing higher-order relationships among multiple vertices simultaneously [41, 454, 473]. A wide range of research on Hypergraph Neural Networks has been conducted in recent years [739, 740, 741, 742, 743, 744, 745]. A SuperHyperGraph Neural Network learns base-vertex representations through hierarchical superhyperedges, capturing nested, multi-level, and higher-order interactions within complex structured data [746, 747, 748].

**Definition 5.33.1** ($n$-SuperHyperGraph Neural Network)**.** [746, 747, 748] Let

$$\mathcal{G}^{(n)} = \big(V^{(n)}, E^{(n)}\big)$$

be a finite $n$-SuperHyperGraph over a finite base vertex set $V_0$, where

$$V^{(n)} \subseteq \mathcal{P}^n(V_0), \qquad E^{(n)} \subseteq \mathcal{P}_*\big(V^{(n)}\big).$$

Define the recursive flattening maps

$$\mathrm{flat}_0(v) = \{v\} \qquad (v \in V_0),$$

and

$$\mathrm{flat}_k(S) = \bigcup_{A \in S} \mathrm{flat}_{k-1}(A), \qquad S \in \mathcal{P}^k(V_0), \quad k \geq 1.$$

Each superhyperedge $e \in E^{(n)}$ induces the base-level hyperedge

$$\widehat{e} = \bigcup_{S \in e} \mathrm{flat}_n(S) \subseteq V_0.$$

Let

$$\widehat{E} = \{\widehat{e} : e \in E^{(n)}\}$$

be the resulting expanded hyperedge family, with duplicate hyperedges either identified or retained as distinct indexed incidences.

Let

$$B \in \{0,1\}^{|V_0| \times |\widehat{E}|}$$

be the incidence matrix defined by

$$B_{ij} = \begin{cases} 1, & v_i \in \widehat{e}_j, \\ 0, & \text{otherwise.} \end{cases}$$

For a positive diagonal hyperedge-weight matrix

$$W = \operatorname{diag}(w_1, \ldots, w_{|\widehat{E}|}), \qquad w_j > 0,$$

define

$$(D_V)_{ii} = \sum_{j=1}^{|\widehat{E}|} B_{ij} w_j, \qquad (D_E)_{jj} = \sum_{i=1}^{|V_0|} B_{ij}.$$

Given a layer-$\ell$ feature matrix

$$X^{(\ell)} \in \mathbb{R}^{|V_0| \times d_\ell},$$

a trainable matrix

$$\Theta^{(\ell)} \in \mathbb{R}^{d_\ell \times d_{\ell+1}},$$

and an activation function $\sigma$, one layer of an $n$-SuperHyperGraph Neural Network, abbreviated $n$-SHGNN, is defined by

$$X^{(\ell+1)} = \sigma\Big(D_V^{-1/2} B W D_E^{-1} B^\top D_V^{-1/2} X^{(\ell)} \Theta^{(\ell)}\Big).$$

Thus, an $n$-SHGNN propagates base-vertex features through the hierarchical relations encoded by the $n$-superhyperedges after recursively expanding them to the base vertex set $V_0$.

## 5.34 Similarity Superhypergraph

A Similarity Graph connects pairs of objects whose similarity satisfies a prescribed criterion, representing pairwise affinities within a weighted network [749, 750, 751, 752]. A Similarity HyperGraph groups mutually similar objects into hyperedges, representing higher-order affinity among multiple objects satisfying a common similarity criterion [753]. A Similarity SuperHyperGraph organizes nested supervertices into superedges according to similarity criteria, representing hierarchical higher-order affinities across multiple structural levels [753].

**Definition 5.34.1** (Similarity HyperGraph)**.** [753] Let $X$ be a finite nonempty set, let

$$\mathrm{sim} : X \times X \longrightarrow \mathbb{R}$$

be a symmetric similarity function, and fix a threshold

$$\tau \in \mathbb{R}.$$

The *Similarity HyperGraph* of $X$ at threshold $\tau$ is

$$\mathcal{H}_\tau = \big(X, \mathcal{E}_\tau\big),$$

where

$$\mathcal{E}_\tau = \{S \subseteq X : |S| \geq 2 \text{ and } \mathrm{sim}(x, y) \geq \tau \text{ for all distinct } x, y \in S\}.$$

Equivalently,

$$\mathcal{E}_\tau = \left\{ S \subseteq X : |S| \geq 2 \text{ and } \min_{\substack{x,y \in S \\ x \neq y}} \mathrm{sim}(x, y) \geq \tau \right\}.$$

Thus, every hyperedge consists of a collection of objects that are pairwise similar at or above the prescribed threshold.

**Definition 5.34.2** (Support map on iterated powersets)**.** Let $X$ be a nonempty set and define

$$\mathcal{P}^0(X) := X, \qquad \mathcal{P}^{k+1}(X) := \mathcal{P}\big(\mathcal{P}^k(X)\big).$$

Define recursively the support maps

$$\mathrm{supp}_k : \mathcal{P}^k(X) \longrightarrow \mathcal{P}(X)$$

by

$$\mathrm{supp}_0(x) := \{x\}, \qquad x \in X,$$

and

$$\mathrm{supp}_{k+1}(U) := \bigcup_{A \in U} \mathrm{supp}_k(A), \qquad U \in \mathcal{P}^{k+1}(X).$$

Thus, $\mathrm{supp}_k(U)$ is the set of all base elements of $X$ occurring in the nested object $U$.

**Definition 5.34.3** (Similarity $n$-SuperHyperGraph)**.** [753] Let $X$ be a finite nonempty base set, let $n \geq 1$, and let

$$\mathrm{sim} : X \times X \longrightarrow \mathbb{R}$$

be a symmetric similarity function. Fix a threshold

$$\tau \in \mathbb{R}.$$

Let

$$V \subseteq \{U \in \mathcal{P}^n(X) : \mathrm{supp}_n(U) \neq \emptyset\}$$

be a finite set of $n$-supervertices.

For distinct $U, W \in V$, define their induced similarity by

$$\mathrm{sim}_n(U, W) := \min_{\substack{x \in \mathrm{supp}_n(U) \\ y \in \mathrm{supp}_n(W)}} \mathrm{sim}(x, y).$$

Define the family of admissible similarity $n$-superedges by

$$\mathcal{E}_\tau^{(n)} = \{S \subseteq V : |S| \geq 2 \text{ and } \mathrm{sim}_n(U, W) \geq \tau \text{ for all distinct } U, W \in S\}.$$

The *Similarity $n$-SuperHyperGraph* at threshold $\tau$ is the $n$-SuperHyperGraph

$$\mathcal{S}_\tau^{(n)} = \left(X; V, E_\tau^{(n)}, \partial\right),$$

where

$$E_\tau^{(n)} = \mathcal{E}_\tau^{(n)}$$

and

$$\partial : E_\tau^{(n)} \longrightarrow \mathcal{P}_*(V), \qquad \partial(S) = S.$$

Hence an $n$-superedge joins a collection of $n$-supervertices whenever every pair of their underlying base-level supports has similarity at least $\tau$.

**Example 5.34.4** (A Similarity 1-SuperHyperGraph)**.** Let

$$X = \{a, b, c, d\}$$

be the finite base set, and define a symmetric similarity function

$$\mathrm{sim} : X \times X \longrightarrow [0, 1]$$

by the following matrix:

| sim | $a$ | $b$ | $c$ | $d$ |
|---|---|---|---|---|
| $a$ | 1 | 0.90 | 0.82 | 0.78 |
| $b$ | 0.90 | 1 | 0.84 | 0.76 |
| $c$ | 0.82 | 0.84 | 1 | 0.65 |
| $d$ | 0.78 | 0.76 | 0.65 | 1 |

and choose the threshold

$$\tau = 0.75.$$

Consider the 1-supervertices

$$U_1 = \{a, b\}, \qquad U_2 = \{c\}, \qquad U_3 = \{d\},$$

and let

$$V = \{U_1, U_2, U_3\} \subseteq \mathcal{P}(X).$$

Since $n = 1$, their supports are

$$\mathrm{supp}_1(U_1) = \{a, b\}, \qquad \mathrm{supp}_1(U_2) = \{c\}, \qquad \mathrm{supp}_1(U_3) = \{d\}.$$

The induced similarities are therefore

$$\begin{aligned}\mathrm{sim}_1(U_1, U_2) &= \min\{\mathrm{sim}(a, c), \mathrm{sim}(b, c)\} \\ &= \min\{0.82, 0.84\} = 0.82,\end{aligned}$$

$$\begin{aligned}\mathrm{sim}_1(U_1, U_3) &= \min\{\mathrm{sim}(a, d), \mathrm{sim}(b, d)\} \\ &= \min\{0.78, 0.76\} = 0.76,\end{aligned}$$

and

$$\mathrm{sim}_1(U_2, U_3) = \mathrm{sim}(c, d) = 0.65.$$

Since

$$0.82 \geq 0.75 \qquad \text{and} \qquad 0.76 \geq 0.75,$$

we obtain the admissible 1-superedges

$$S_{12} = \{U_1, U_2\}, \qquad S_{13} = \{U_1, U_3\}.$$

On the other hand,

$$\mathrm{sim}_1(U_2, U_3) = 0.65 < 0.75,$$

so

$$\{U_2, U_3\} \notin \mathcal{E}_\tau^{(1)}.$$

Moreover,

$$\{U_1, U_2, U_3\} \notin \mathcal{E}_\tau^{(1)},$$

because not every pair of its supervertices satisfies the threshold condition.

Consequently,

$$E_\tau^{(1)} = \{S_{12}, S_{13}\},$$

with incidence map

$$\partial(S_{12}) = \{U_1, U_2\}, \qquad \partial(S_{13}) = \{U_1, U_3\}.$$

Hence

$$\mathcal{S}_{0.75}^{(1)} = (X; \{U_1, U_2, U_3\}, \{S_{12}, S_{13}\}, \partial)$$

is a Similarity 1-SuperHyperGraph.

This example illustrates how similarity between supervertices is determined conservatively by the minimum similarity between their underlying base-level elements. Thus, a superedge is admitted only when all constituent supervertices satisfy the prescribed similarity threshold.

# 6 New Higher-Order Graph Structures: Authors' Extensions

This chapter introduces several higher-order graph structures proposed as extensions of existing graph-theoretic and higher-order frameworks. The constructions presented here explore different mechanisms for representing higher-order organization, including filtration, hierarchical incidence, quotient structures, granulation, multiscale and multi-axis indexing, compositional assembly, iterated labeling, multidimensional dynamics, structure-valued vertices, recursive meta-relations, and hierarchy on both the vertex and edge domains.

For reference, the principal structures introduced and developed in this chapter are summarized in Table 6.1. Some entries also include closely related extensions defined within the corresponding sections.

Table 6.1: Principal higher-order graph structures introduced and developed as authors' extensions in this chapter.

| **Concept** | **Concise description** |
|---|---|
| $n$-Filtrated Graph | A graph equipped with a level map on its vertices and edges, inducing $n+1$ nested graph layers indexed by $0, 1, \ldots, n$, in which vertices and edges appear progressively. |
| Depth-$N$ Incidence SuperHyperGraph | A hierarchical incidence structure constructed through successive depth levels, where higher-level edge objects are specified by generalized incidence data over carriers accumulated from preceding levels. |
| Structured Quotient-Based Graph | A graph equipped with an equivalence relation on its structured vertex domain, together with the induced quotient graph whose vertices are equivalence classes and whose adjacency is inherited from the original structure. |
| Granular Graph | A hierarchical graph system built from a sequence of coverings or granulations of an underlying set, with graph layers formed from granules and linked by prescribed projection maps between successive granulation levels. |
| Multiscale Graph | A family of graphs indexed by a finite totally ordered set of scales, together with coherent graph homomorphisms relating representations at different resolutions. |
| DAG-Based Graph | A diagram of graphs indexed by a directed acyclic graph, with graph homomorphisms assigned to directed dependencies and required to be compatible along directed paths. |
| Compositional Graph | A graph framework in which larger graph objects are assembled from smaller components through explicitly specified interfaces, composition operations, or gluing mechanisms. |
| $n$-Iterated Labeling Graph | A graph equipped with $n+1$ compatible levels of vertex and edge labels, together with reduction maps through which higher-level labels refine and recover the preceding labeling levels. |
| $m$-Multidynamic Graph | A graph system indexed by the product of $m$ finite totally ordered time sets, with coherent transition homomorphisms describing evolution along several temporal coordinates simultaneously. |

*Table 6.1 (continued)*

| **Concept** | **Concise description** |
|---|---|
| Structure-Vertexed Graph | A graph whose vertices are mathematical structures and whose edges are determined by a prescribed structural relation, such as isomorphism, embedding, homomorphism, or another selected relation. |
| Constructor-Selectable Graph | A graph whose admissible vertex domain is generated from a finite base set by a prescribed finite sequence of selectable hierarchical constructors, such as powerset, multiset, or tree constructors. |
| Multiinfinite Graph and Multiinfinite $n$-SuperHyperGraph | Graph or $n$-SuperHyperGraph systems indexed by a product of several infinite totally ordered axes, with coherent transition homomorphisms between all comparable index positions. |
| Edge-Iterated SuperHyperGraph / $(n, m)$-SuperHyperGraph | A SuperHyperGraph-type structure in which $n$-supervertices arise from an iterated powerset of the base set while $m$-superhyperedges arise from an iterated powerset of the supervertex set, with effective incidence obtained through recursive flattening. |
| Edge-Iterated HyperCube | An extension of an $n$-SuperHyperCube in which edge objects themselves may possess an $m$-level powerset hierarchy, with incidence recovered by flattening them to ordinary cube edges and collecting their endpoints. |
| Recursive MetaGraph and Recursive Iterated MetaGraph | MetaGraph constructions whose vertices are graphs or higher-level MetaGraphs and whose metaedges may themselves be recursively constructed from lower-level metaedges up to a prescribed finite recursion depth. |
| Multiuniverse-Graph | A graph defined on tagged copies of several underlying universes, preserving universe identity while permitting both intra-universe and inter-universe edges. |
| Multiunion-Graph | A graph whose vertices are nonempty set-valued objects assembled from several tagged universes, with adjacency determined by a prescribed rule such as overlap or a relation between their universe supports. |
| Iterated SuperGraph | A recursively defined hierarchy of finite directed containment graphs: at the first level, vertices are finite simple graphs ordered by proper subgraph containment, while higher-level vertices are lower-level directed graphs ordered by proper subdigraph containment. |

## 6.1 $n$-Filtrated Graph

An $n$-Filtrated Graph is a graph equipped with $n$ filtration levels, where vertices and edges appear progressively across nested layers.

**Definition 6.1.1** ($n$-Filtrated Graph)**.** Let $V_0$ be a finite nonempty set and let $n \in \mathbb{N}_0$. An *$n$-Filtrated Graph over $V_0$* is a quadruple

$$\mathrm{FG}^{(n)} = (V, E, \partial, \lambda),$$

where

- $V \subseteq V_0$ is a finite set of vertices;
- $E$ is a finite set of edge identifiers;
- $\partial : E \to \binom{V}{2}$ is an incidence map;
- $\lambda : V \sqcup E \to \{0, 1, \ldots, n\}$ is a filtration-level map such that

$$\partial(e) = \{u, v\} \implies \lambda(e) \geq \max\{\lambda(u), \lambda(v)\}$$

for every $e \in E$.

For each $k \in \{0, 1, \ldots, n\}$, define

$$V_k := \{v \in V \mid \lambda(v) \leq k\}, \qquad E_k := \{e \in E \mid \lambda(e) \leq k\}.$$

Then

$$G_k := (V_k, E_k, \partial|_{E_k})$$

is called the $k$-th filtration layer of $\mathrm{FG}^{(n)}$.

**Example 6.1.2** (A 2-Filtrated Graph)**.** Let

$$V_0 = \{a, b, c, d\}, \qquad V = \{a, b, c, d\} \subseteq V_0,$$

and let the edge-identifier set be

$$E = \{e_{ab}, e_{ac}, e_{cd}\}.$$

Define the incidence map

$$\partial : E \to \binom{V}{2}$$

by

$$\partial(e_{ab}) = \{a, b\}, \qquad \partial(e_{ac}) = \{a, c\}, \qquad \partial(e_{cd}) = \{c, d\}.$$

Next, define the filtration-level map

$$\lambda : V \sqcup E \to \{0, 1, 2\}$$

by

$$\lambda(a) = 0, \qquad \lambda(b) = 0, \qquad \lambda(c) = 1, \qquad \lambda(d) = 2,$$

and

$$\lambda(e_{ab}) = 0, \qquad \lambda(e_{ac}) = 1, \qquad \lambda(e_{cd}) = 2.$$

Then, for every edge identifier $e \in E$, the condition

$$\partial(e) = \{u, v\} \implies \lambda(e) \geq \max\{\lambda(u), \lambda(v)\}$$

is satisfied:

$$\lambda(e_{ab}) = 0 \geq \max\{\lambda(a), \lambda(b)\} = 0,$$
$$\lambda(e_{ac}) = 1 \geq \max\{\lambda(a), \lambda(c)\} = 1,$$

and

$$\lambda(e_{cd}) = 2 \geq \max\{\lambda(c), \lambda(d)\} = 2.$$

Hence

$$\mathrm{FG}^{(2)} = (V, E, \partial, \lambda)$$

is a 2-Filtrated Graph over $V_0$.

Its filtration layers are as follows:

$$V_0^{\mathrm{fil}} = \{a, b\}, \qquad E_0 = \{e_{ab}\},$$

so

$$G_0 = (\{a, b\}, \{e_{ab}\}, \partial|_{E_0}).$$

At level 1,

$$V_1 = \{a, b, c\}, \qquad E_1 = \{e_{ab}, e_{ac}\},$$

so

$$G_1 = (\{a, b, c\}, \{e_{ab}, e_{ac}\}, \partial|_{E_1}).$$

At level 2,

$$V_2 = \{a, b, c, d\}, \qquad E_2 = \{e_{ab}, e_{ac}, e_{cd}\},$$

so

$$G_2 = (\{a, b, c, d\}, \{e_{ab}, e_{ac}, e_{cd}\}, \partial|_{E_2}).$$

Thus the graph appears progressively through the filtration: first the edge $ab$, then the vertex $c$ together with the edge $ac$, and finally the vertex $d$ together with the edge $cd$. A reference illustration is given in Fig. 6.1.

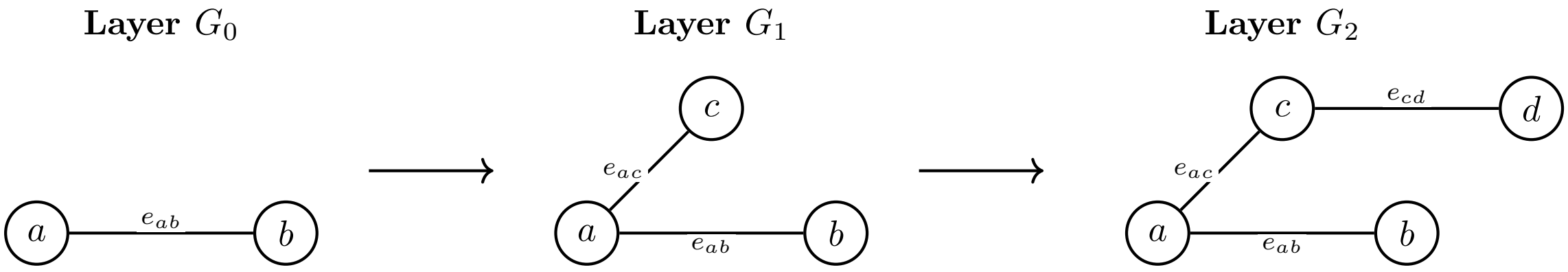

Figure 6.1: A 2-Filtrated Graph $\mathrm{FG}^{(2)} = (V, E, \partial, \lambda)$ and its filtration layers. The graph grows from $G_0$ to $G_2$ by the progressive appearance of vertices and edges according to the filtration-level map $\lambda$.

A comparison between SuperHyperGraph and Filtrated Graph is given in Table 6.2.

Next, we consider Filtrated HyperGraphs and Filtrated SuperHyperGraphs. Their definitions and related formulations are given below.

**Definition 6.1.3** (Filtrated HyperGraph)**.** Let $V_0 \neq \varnothing$ be a finite set, and let $m \in \mathbb{N}_0$. A *Filtrated HyperGraph of height $m$ over $V_0$* is a quadruple

$$\mathrm{FHG}^{(m)} = (V, \mathcal{E}, \partial, \lambda),$$

where

- $V \subseteq V_0$ is a finite set of vertices;
- $\mathcal{E}$ is a finite set of hyperedge identifiers;
- $\partial : \mathcal{E} \to \mathcal{P}^*(V)$ is an incidence map such that $\partial(e) \subseteq V$ is a nonempty finite set for every $e \in \mathcal{E}$;
- $\lambda : V \sqcup \mathcal{E} \to \{0, 1, \dots, m\}$ is a *filtration-level map* satisfying

$$v \in \partial(e) \implies \lambda(v) \leq \lambda(e) \qquad \text{for all } e \in \mathcal{E}.$$

For each $t \in \{0, 1, \dots, m\}$, define

$$V_t := \{v \in V \mid \lambda(v) \leq t\}, \qquad \mathcal{E}_t := \{e \in \mathcal{E} \mid \lambda(e) \leq t\},$$

and let

$$\partial_t := \partial|_{\mathcal{E}_t} : \mathcal{E}_t \to \mathcal{P}^*(V_t).$$

Then

$$\mathrm{FHG}_t^{(m)} := (V_t, \mathcal{E}_t, \partial_t)$$

is called the *$t$-th filtration layer* of $\mathrm{FHG}^{(m)}$. The chain

$$\mathrm{FHG}_0^{(m)} \subseteq \mathrm{FHG}_1^{(m)} \subseteq \cdots \subseteq \mathrm{FHG}_m^{(m)}$$

is called the *induced filtration.*

**Remark 6.1.4.** *If one identifies each hyperedge identifier $e \in \mathcal{E}$ with its incidence set $\partial(e)$, then a Filtrated HyperGraph may equivalently be viewed as a finite hypergraph together with a birth-time assignment on vertices and hyperedges such that every hyperedge appears no earlier than any of its incident vertices.*

**Definition 6.1.5** ($(n, m)$-Filtrated SuperHyperGraph)**.** Let $V_0 \neq \varnothing$ be a finite base set, and let $n, m \in \mathbb{N}_0$. An *$(n, m)$-Filtrated SuperHyperGraph over $V_0$* is a quadruple

$$\mathrm{FSHG}^{(n,m)} = (V, \mathcal{E}, \partial, \lambda),$$

where

- $V \subseteq \mathcal{P}^n(V_0)$ is a finite set of *$n$-supervertices*;
- $\mathcal{E}$ is a finite set of *superedge identifiers*;
- $\partial : \mathcal{E} \to \mathcal{P}^*(V)$ is an incidence map such that $\partial(e) \subseteq V$ is a nonempty finite set for every $e \in \mathcal{E}$;
- $\lambda : V \sqcup \mathcal{E} \to \{0, 1, \dots, m\}$ is a *filtration-level map* satisfying

$$v \in \partial(e) \implies \lambda(v) \leq \lambda(e) \qquad \text{for all } e \in \mathcal{E}.$$

Table 6.2: A brief comparison between SuperHyperGraph and Filtrated Graph.

| **Aspect** | **SuperHyperGraph** | **Filtrated Graph** |
|---|---|---|
| Basic idea | A graph-like higher-order structure whose vertex domain is hierarchical, typically built from iterated powersets. | A graph equipped with filtration levels, where vertices and edges appear progressively across nested layers. |
| Underlying domain | Vertices lie in a higher-level domain such as $\mathcal{P}^n(V_0)$. | Vertices lie in an ordinary finite set $V \subseteq V_0$. |
| Type of hierarchy | Hierarchy is induced by *nested object levels.* | Hierarchy is induced by *ordered filtration levels.* |
| Edges | Superedges connect supervertices through an incidence map $$\partial : E \to \mathcal{P}^*(V).$$ | Edges are ordinary graph edges given by an incidence map $$\partial : E \to \binom{V}{2}.$$ |
| Role of levels | The parameter $n$ specifies the super-level of the vertex domain. | The parameter $n$ specifies the maximum filtration depth. |
| Evolution across levels | Higher structure comes from more deeply nested vertices. | Higher layers come from adding vertices and edges over time or scale. |
| Typical interpretation | Structure of sets, sets of sets, or other hierarchical relational objects. | Growth, resolution, persistence, or multistage appearance of an ordinary graph. |
| Output layers | Usually studied as a single hierarchical object. | Naturally yields layers $$G_0 \subseteq G_1 \subseteq \cdots \subseteq G_n.$$ |

For each $t \in \{0, 1, \ldots, m\}$, define

$$V_t := \{v \in V \mid \lambda(v) \leq t\}, \qquad \mathcal{E}_t := \{e \in \mathcal{E} \mid \lambda(e) \leq t\},$$

and let

$$\partial_t := \partial|_{\mathcal{E}_t} : \mathcal{E}_t \to \mathcal{P}^*(V_t).$$

Then

$$\mathrm{FSHG}_t^{(n,m)} := (V_t, \mathcal{E}_t, \partial_t)$$

is an $n$-*SuperHyperGraph* over $V_0$, called the $t$-*th filtration layer* of $\mathrm{FSHG}^{(n,m)}$. The chain

$$\mathrm{FSHG}_0^{(n,m)} \subseteq \mathrm{FSHG}_1^{(n,m)} \subseteq \cdots \subseteq \mathrm{FSHG}_m^{(n,m)}$$

is called the *induced filtration.*

**Proposition 6.1.6.** *For every Filtrated HyperGraph* $\mathrm{FHG}^{(m)}$ *and every* $t \in \{0, 1, \ldots, m\}$, *the layer* $\mathrm{FHG}_t^{(m)}$ *is a hypergraph. Likewise, for every* $(n, m)$*-Filtrated SuperHyperGraph* $\mathrm{FSHG}^{(n,m)}$ *and every* $t \in \{0, 1, \ldots, m\}$, *the layer* $\mathrm{FSHG}_t^{(n,m)}$ *is an* $n$*-SuperHyperGraph over* $V_0$.

*Proof.* We prove the superhypergraph case; the hypergraph case is analogous. Let $e \in \mathcal{E}_t$. Then $\lambda(e) \leq t$. If $v \in \partial(e)$, the defining condition gives $\lambda(v) \leq \lambda(e) \leq t$, hence $v \in V_t$. Therefore $\partial(e) \subseteq V_t$, and since $\partial(e)$ is nonempty, we obtain

$$\partial_t(e) = \partial(e) \in \mathcal{P}^*(V_t).$$

Thus $\partial_t : \mathcal{E}_t \to \mathcal{P}^*(V_t)$ is well defined, so

$$\mathrm{FSHG}_t^{(n,m)} = (V_t, \mathcal{E}_t, \partial_t)$$

is an $n$-SuperHyperGraph over $V_0$. □

## 6.2 Depth-$N$ Incidence SuperHyperGraph

A Depth-$N$ Incidence SuperHyperGraph is a hierarchical structure whose higher-level edges are defined by generalized incidence objects recursively.

**Definition 6.2.1** (Hierarchical incidence constructor)**.** For each $n \in \mathbb{N}$, let $\mathbf{F}_n = (F_n, \mathrm{supp}_n)$ consist of:

- a rule assigning to every finite set $X$ a set $F_n(X)$ of admissible *$n$-level incidence objects on $X$*;
- a map
$$\mathrm{supp}_{n,X} : F_n(X) \to \mathcal{P}^{*}_{\mathrm{fin}}(X),$$
called the *support map*,

where
$$\mathcal{P}^{*}_{\mathrm{fin}}(X) := \{A \subseteq X : 0 < |A| < \infty\}.$$
The value $\mathrm{supp}_{n,X}(\alpha)$ is called the *support* of the incidence object $\alpha \in F_n(X)$.

**Definition 6.2.2** (Depth-$N$ Incidence SuperHyperGraph)**.** Let $V \neq \varnothing$ be a finite set, and let
$$\mathbf{F}_1, \mathbf{F}_2, \ldots, \mathbf{F}_N$$
be hierarchical incidence constructors. A *depth-$N$ Incidence SuperHyperGraph over $V$ relative to* $\mathbf{F}_1, \ldots, \mathbf{F}_N$ is a tuple
$$\mathsf{ISHG}^{(N)} = \big(V, E_1, \ldots, E_N, \iota_1, \ldots, \iota_N\big),$$
such that:

- $E_1, \ldots, E_N$ are finite pairwise disjoint sets, disjoint also from $V$;
- the cumulative carriers are defined recursively by
$$C_0 := V, \qquad C_n := V \sqcup E_1 \sqcup \cdots \sqcup E_n \qquad (n = 1, \ldots, N);$$
- for each $n \in \{1, \ldots, N\}$,
$$\iota_n : E_n \to F_n(C_{n-1})$$
is an *$n$-level incidence map.*

For $e \in E_n$, the element $\iota_n(e) \in F_n(C_{n-1})$ is called the *incidence object* of $e$, and
$$\mathrm{supp}_n(e) := \mathrm{supp}_{n,C_{n-1}}\big(\iota_n(e)\big) \subseteq C_{n-1}$$
is called the *support* (or *incidence support*) of $e$. Elements of $E_n$ are called *$n$-superhyperedges.*

**Remark 6.2.3.** *The essence of the above definition is that an $n$-superhyperedge need not be represented directly by a subset of the previous carrier $C_{n-1}$. Instead, it is represented by an admissible incidence object in $F_n(C_{n-1})$, together with its underlying support extracted by $\mathrm{supp}_{n,C_{n-1}}$. Thus, the hierarchy is governed by iterated incidence operators rather than by iterated powersets.*

**Theorem 6.2.4** (Well-definedness of a Depth-$N$ Incidence SuperHyperGraph)**.** *Let*
$$\mathsf{ISHG}^{(N)} = \big(V, E_1, \ldots, E_N, \iota_1, \ldots, \iota_N\big)$$
*be a depth-$N$ Incidence SuperHyperGraph relative to hierarchical incidence constructors*
$$\mathbf{F}_n = (F_n, \mathrm{supp}_n) \qquad (n = 1, \ldots, N).$$
*Then, for each $n \in \{1, \ldots, N\}$,*

1. *the cumulative carrier*
$$C_n = V \sqcup E_1 \sqcup \cdots \sqcup E_n$$
*is a well-defined finite set;*

2. *the incidence map*

$$\iota_n : E_n \to F_n(C_{n-1})$$

*is well defined;*

3. *for every $e \in E_n$, the support*

$$\operatorname{supp}_n(e) := \operatorname{supp}_{n,C_{n-1}}\bigl(\iota_n(e)\bigr)$$

*is a well-defined finite nonempty subset of $C_{n-1}$.*

*Hence the notion of a depth-$N$ Incidence SuperHyperGraph is well defined.*

*Proof.* Since $V$ is finite and nonempty, the set

$$C_0 := V$$

is well defined and finite.

Now proceed by induction on $n$. Assume that $C_{n-1}$ is a well-defined finite set. Because $E_1, \dots, E_N$ are finite pairwise disjoint sets, each disjoint from $V$, the disjoint union

$$C_n = V \sqcup E_1 \sqcup \cdots \sqcup E_n$$

is also a well-defined finite set.

By definition of the hierarchical incidence constructor $\mathbf{F}_n = (F_n, \operatorname{supp}_n)$, the set

$$F_n(C_{n-1})$$

is well defined, and the support map

$$\operatorname{supp}_{n,C_{n-1}} : F_n(C_{n-1}) \to \mathcal{P}^*_{\mathrm{fin}}(C_{n-1})$$

is well defined. Therefore, for each $e \in E_n$, the value

$$\iota_n(e) \in F_n(C_{n-1})$$

is well defined, and hence

$$\operatorname{supp}_n(e) := \operatorname{supp}_{n,C_{n-1}}\bigl(\iota_n(e)\bigr)$$

is a well-defined element of

$$\mathcal{P}^*_{\mathrm{fin}}(C_{n-1}),$$

that is, a finite nonempty subset of $C_{n-1}$.

Thus all cumulative carriers, incidence objects, and supports are well defined at every level $n$. Hence the whole structure

$$\mathsf{ISHG}^{(N)} = \bigl(V, E_1, \dots, E_N, \iota_1, \dots, \iota_N\bigr)$$

is well defined. □

**Definition 6.2.5** (Underlying support hierarchy)**.** Let

$$\mathsf{ISHG}^{(N)} = \bigl(V, E_1, \dots, E_N, \iota_1, \dots, \iota_N\bigr)$$

be a depth-$N$ Incidence SuperHyperGraph. For each $n \in \{1, \dots, N\}$ and each $e \in E_n$, define

$$\operatorname{supp}_n(e) := \operatorname{supp}_{n,C_{n-1}}\bigl(\iota_n(e)\bigr) \subseteq C_{n-1}.$$

The family

$$\bigl(V, E_1, \dots, E_N, \operatorname{supp}_1, \dots, \operatorname{supp}_N\bigr)$$

is called the *underlying support hierarchy* of $\mathsf{ISHG}^{(N)}$.

**Remark 6.2.6** (Special cases)**.** *The above notion recovers several familiar structures.*

1. *If $N = 1$ and*

$$F_1(X) = \mathcal{P}^*_{\mathrm{fin}}(X), \qquad \operatorname{supp}_{1,X} = \mathrm{id},$$

*then a depth-1 Incidence SuperHyperGraph is precisely a hypergraph written in incidence-map form.*

2. *If, for every* $n \in \{1, \ldots, N\}$,
$$F_n(X) = \mathcal{P}^*_{\mathrm{fin}}(X), \qquad \mathrm{supp}_{n,X} = \mathrm{id},$$
*then each* $n$*-superhyperedge is represented by a nonempty finite subset of* $C_{n-1}$*. This yields the set-based iterative incidence model underlying powerset-type hierarchical superhypergraph constructions.*

3. *If*
$$F_n(X) = \mathbb{N}^{(X)} \setminus \{0\},$$
*with* $\mathrm{supp}_{n,X}$ *equal to the usual finite support of a multiset, then one obtains a* multi-incidence superhypergraph*.*

4. *If*
$$F_n(X) = \{\mu : X \to [0,1] \mid \mathrm{supp}(\mu) \in \mathcal{P}^*_{\mathrm{fin}}(X)\},$$
*with*
$$\mathrm{supp}_{n,X}(\mu) := \{x \in X : \mu(x) > 0\},$$
*then one obtains a* fuzzy incidence superhypergraph*.*

**Example 6.2.7** (A depth-2 Incidence SuperHyperGraph)**.** Let
$$V = \{a, b, c\}.$$
Define
$$E_1 = \{e_1, e_2\}, \qquad E_2 = \{E\},$$
and choose
$$F_1(X) = \mathcal{P}^*_{\mathrm{fin}}(X), \qquad F_2(X) = \mathcal{P}^*_{\mathrm{fin}}(X),$$
with identity support maps. Then
$$C_0 = V, \qquad C_1 = V \sqcup E_1.$$
Let
$$\iota_1(e_1) = \{a, b\}, \qquad \iota_1(e_2) = \{b, c\},$$
and
$$\iota_2(E) = \{e_1, e_2, b\} \subseteq C_1.$$
Then
$$\mathsf{ISHG}^{(2)} = (V, E_1, E_2, \iota_1, \iota_2)$$
is a depth-2 Incidence SuperHyperGraph. Here the 2-superhyperedge $E$ is incident not only with lower-level edges $e_1, e_2$, but also directly with the vertex $b$.

## 6.3 Structured Quotient-based Graph

Structured Quotient-based Graph is a graph system consisting of a base graph, an equivalence relation, its induced quotient graph, and the canonical projection between them.

**Definition 6.3.1** (Structured Quotient-based Graph)**.** A *Structured Quotient-based Graph* is a quadruple
$$\mathcal{Q} = (G_0, \sim, Q, q),$$
where

- $G_0 = (V_0, E_0)$ is a finite simple graph;
- $\sim$ is an equivalence relation on $V_0$;
- $Q = Q(G_0, \sim)$ is the quotient-based graph induced by $(G_0, \sim)$;
- $q : V_0 \to V(Q)$ is the canonical quotient projection.

**Example 6.3.2** (A Structured Quotient-based Graph)**.** Let

$$V_0 = \{a, b, c, d, e, f\},$$

and let

$$E_0 = \big\{\{a, b\}, \{a, c\}, \{b, d\}, \{c, d\}, \{c, e\}, \{d, f\}, \{e, f\}\big\}.$$

Then

$$G_0 = (V_0, E_0)$$

is a finite simple graph.

Define an equivalence relation $\sim$ on $V_0$ by declaring

$$a \sim b, \qquad c \sim d, \qquad e \sim f,$$

and no further identifications between distinct classes. Thus the equivalence classes are

$$A := [a]_\sim = \{a, b\}, \qquad B := [c]_\sim = \{c, d\}, \qquad C := [e]_\sim = \{e, f\}.$$

Hence

$$V_0/\!\sim\, = \{A, B, C\}.$$

The canonical quotient projection is

$$q : V_0 \to V_0/\!\sim, \qquad q(a) = q(b) = A, \quad q(c) = q(d) = B, \quad q(e) = q(f) = C.$$

Now construct the quotient-based graph

$$Q = Q(G_0, \sim) = (V_\sim, E_\sim),$$

where

$$V_\sim = \{A, B, C\}.$$

By definition, two distinct quotient vertices are adjacent whenever there exists at least one edge of $G_0$ joining the corresponding equivalence classes.

Since

$$\{a, c\} \in E_0 \quad \text{and} \quad \{b, d\} \in E_0,$$

there is an edge between $A$ and $B$ in the quotient graph. Likewise, since

$$\{c, e\} \in E_0 \quad \text{and} \quad \{d, f\} \in E_0,$$

there is an edge between $B$ and $C$ in the quotient graph. On the other hand, there is no edge of $G_0$ connecting a vertex of $A$ directly to a vertex of $C$, so $\{A, C\} \notin E_\sim$.

Therefore,

$$E_\sim = \big\{\{A, B\}, \{B, C\}\big\},$$

and

$$Q = (\{A, B, C\}, \{\{A, B\}, \{B, C\}\}).$$

Thus the quadruple

$$\mathcal{Q} = (G_0, \sim, Q, q)$$

is a Structured Quotient-based Graph. Here the lower-level graph $G_0$ is aggregated into the upper-level quotient graph $Q$ by merging the vertices inside each equivalence class.

**Theorem 6.3.3** (Well-definedness of Structured Quotient-based Graph)**.** *Let*

$$G_0 = (V_0, E_0)$$

*be a finite simple graph, and let $\sim$ be an equivalence relation on $V_0$. Define the quotient vertex set by*

$$V_\sim := V_0/\!\sim,$$

*and define*

$$E_\sim := \Big\{\{A, B\} \subseteq V_\sim : A \neq B,\ \exists\, u \in A,\ \exists\, v \in B \text{ such that } \{u, v\} \in E_0\Big\}.$$

*Also define the canonical map*

$$q : V_0 \to V_\sim, \qquad q(v) = [v]_\sim.$$

*Then*

$$Q(G_0, \sim) := (V_\sim, E_\sim)$$

*is a well-defined finite simple graph, and hence*

$$\mathcal{Q} = (G_0, \sim, Q(G_0, \sim), q)$$

*is a well-defined Structured Quotient-based Graph.*

*Proof.* Since $V_0$ is finite and $\sim$ is an equivalence relation on $V_0$, the quotient set

$$V_\sim = V_0/\sim$$

is well defined and finite. The map

$$q(v) = [v]_\sim$$

is therefore a well-defined function from $V_0$ to $V_\sim$.

Next, by definition,

$$E_\sim \subseteq \big\{\{A, B\} \subseteq V_\sim : A \neq B\big\}.$$

Hence every edge of $E_\sim$ joins two distinct quotient vertices, so no loops occur.

Moreover, $E_\sim$ is a set of unordered pairs of quotient vertices, so multiple edges do not arise. Thus

$$Q(G_0, \sim) = (V_\sim, E_\sim)$$

is a finite simple graph.

Therefore the quadruple

$$\mathcal{Q} = (G_0, \sim, Q(G_0, \sim), q)$$

satisfies all requirements in the definition of a Structured Quotient-based Graph. □

## 6.4 Granular Graph

A Granular Graph is a graph hierarchy whose vertices are granules from increasingly coarser coverings, with adjacency determined by overlap and approximations varying by level.

**Definition 6.4.1** (Granular hierarchy)**.** Let $X \neq \varnothing$ be a finite set, and let $m \in \mathbb{N}_0$. A *granular hierarchy of height $m$ over $X$* is a family

$$\mathcal{C}_0, \mathcal{C}_1, \dots, \mathcal{C}_m$$

such that:

- for each $t \in \{0, 1, \dots, m\}$, $\mathcal{C}_t$ is a finite covering of $X$, that is,

$$\varnothing \notin \mathcal{C}_t, \qquad \bigcup_{A \in \mathcal{C}_t} A = X;$$

- for each $t \in \{0, 1, \dots, m-1\}$, $\mathcal{C}_t$ is finer than $\mathcal{C}_{t+1}$ in the sense that

$$\forall A \in \mathcal{C}_t \ \exists B \in \mathcal{C}_{t+1} \text{ such that } A \subseteq B.$$

The elements of $\mathcal{C}_t$ are called *$t$-granules.*

**Definition 6.4.2** (Granular Graph)**.** Let $X \neq \varnothing$ be a finite set, and let

$$\mathcal{C}_0, \mathcal{C}_1, \dots, \mathcal{C}_m$$

be a granular hierarchy over $X$. A *Granular Graph of height $m$ over $X$* is a tuple

$$\mathsf{GG}^{(m)} = \big(X, \mathcal{C}_0, \dots, \mathcal{C}_m, \pi_0, \dots, \pi_{m-1}\big),$$

where, for each $t \in \{0, 1, \ldots, m-1\}$,

$$\pi_t : \mathcal{C}_t \to \mathcal{C}_{t+1}$$

is a *granular projection* satisfying

$$A \subseteq \pi_t(A) \qquad \text{for every } A \in \mathcal{C}_t.$$

For each level $t \in \{0, 1, \ldots, m\}$, the *$t$-th granular graph layer* is the graph

$$G_t = (V_t, E_t),$$

defined by

$$V_t := \mathcal{C}_t,$$

and

$$E_t := \Big\{ \{A, B\} \subseteq \mathcal{C}_t : A \neq B,\ A \cap B \neq \varnothing \Big\}.$$

Thus, two distinct $t$-granules are adjacent precisely when they overlap.

**Remark 6.4.3.** *In a Granular Graph, lower levels represent finer granulation, while higher levels represent coarser granulation. Hence the hierarchy is determined not by set inclusion of vertices across levels, but by the gradual coarsening of coverings and the associated aggregation maps $\pi_t$.*

**Proposition 6.4.4.** *For every $t \in \{0, 1, \ldots, m\}$, the layer*

$$G_t = (V_t, E_t)$$

*is a finite simple graph.*

*Proof.* Since $\mathcal{C}_t$ is finite, so is $V_t = \mathcal{C}_t$. Moreover,

$$E_t \subseteq \big\{ \{A, B\} \subseteq \mathcal{C}_t : A \neq B \big\},$$

so every edge joins two distinct vertices, and there are no loops or multiple edges. Hence $G_t$ is a finite simple graph. $\square$

**Definition 6.4.5** (Lower and upper approximations at level $t$)**.** Let

$$\mathsf{GG}^{(m)} = \big( X, \mathcal{C}_0, \ldots, \mathcal{C}_m, \pi_0, \ldots, \pi_{m-1} \big)$$

be a Granular Graph. For each $t \in \{0, 1, \ldots, m\}$ and each subset $S \subseteq X$, define

$$\underline{\mathrm{Apr}}_t(S) := \bigcup \{ A \in \mathcal{C}_t : A \subseteq S \},$$

and

$$\overline{\mathrm{Apr}}_t(S) := \bigcup \{ A \in \mathcal{C}_t : A \cap S \neq \varnothing \}.$$

These are called the *lower approximation* and *upper approximation* of $S$ at granulation level $t$.

**Remark 6.4.6.** *The operators*

$$\underline{\mathrm{Apr}}_t, \qquad \overline{\mathrm{Apr}}_t$$

*connect Granular Graphs with rough-set and covering-based viewpoints. Typically, lower levels yield finer approximations, while higher levels yield coarser approximations.*

**Example 6.4.7** (A Granular Graph of height 2)**.** Let

$$X = \{a, b, c, d\}.$$

Define coverings

$$\mathcal{C}_0 = \big\{ \{a\}, \{b\}, \{c\}, \{d\} \big\},$$

$$\mathcal{C}_1 = \big\{ \{a, b\}, \{b, c\}, \{d\} \big\},$$

and

$$\mathcal{C}_2 = \big\{ \{a, b, c\}, \{d\} \big\}.$$

Define projection maps

$$\pi_0 : \mathcal{C}_0 \to \mathcal{C}_1, \qquad \pi_1 : \mathcal{C}_1 \to \mathcal{C}_2,$$

by

$$\pi_0(\{a\}) = \{a, b\}, \quad \pi_0(\{b\}) = \{a, b\}, \quad \pi_0(\{c\}) = \{b, c\}, \quad \pi_0(\{d\}) = \{d\},$$

and

$$\pi_1(\{a, b\}) = \{a, b, c\}, \quad \pi_1(\{b, c\}) = \{a, b, c\}, \quad \pi_1(\{d\}) = \{d\}.$$

Then

$$\mathsf{GG}^{(2)} = \big(X, \mathcal{C}_0, \mathcal{C}_1, \mathcal{C}_2, \pi_0, \pi_1\big)$$

is a Granular Graph of height 2. Its layers are:

$$G_0 : \text{ four isolated singleton vertices,}$$

$$G_1 : \text{ vertices } \{a, b\}, \{b, c\}, \{d\} \text{ with one edge between } \{a, b\} \text{ and } \{b, c\},$$

$$G_2 : \text{ vertices } \{a, b, c\}, \{d\} \text{ with no edge.}$$

## 6.5 Multiscale Graph

A Multiscale Graph is a graph framework indexed by ordered scales, where each layer captures structure at a specific resolution and interscale maps connect them.

**Definition 6.5.1** (Multiscale Graph)**.** Let

$$\Sigma = \{\varepsilon_0, \varepsilon_1, \ldots, \varepsilon_m\}$$

be a finite totally ordered set of scales with

$$\varepsilon_0 < \varepsilon_1 < \cdots < \varepsilon_m.$$

A *Multiscale Graph indexed by* $\Sigma$ is a tuple

$$\mathsf{MG}_\Sigma = \big(\{G_\varepsilon\}_{\varepsilon \in \Sigma}, \{\rho_{\varepsilon,\delta}\}_{\varepsilon \leq \delta}\big),$$

such that:

- for each $\varepsilon \in \Sigma$,
$$G_\varepsilon = (V_\varepsilon, E_\varepsilon)$$
is a finite simple graph;
- for each $\varepsilon \leq \delta$ in $\Sigma$,
$$\rho_{\varepsilon,\delta} : G_\varepsilon \to G_\delta$$
is a graph homomorphism;
- the family of transition maps is coherent, i.e.
$$\rho_{\varepsilon,\varepsilon} = \mathrm{id}_{G_\varepsilon} \qquad \text{for all } \varepsilon \in \Sigma,$$
and
$$\rho_{\delta,\eta} \circ \rho_{\varepsilon,\delta} = \rho_{\varepsilon,\eta} \qquad \text{whenever } \varepsilon \leq \delta \leq \eta.$$

The graph $G_\varepsilon$ is called the *$\varepsilon$-scale layer* of $\mathsf{MG}_\Sigma$.

**Remark 6.5.2.** *If all transition maps are injective and identify $G_\varepsilon$ with a subgraph of $G_\delta$ whenever $\varepsilon \leq \delta$, then the Multiscale Graph is called* inclusion-type*. This includes the common case of radius-based neighborhood graphs or persistence-type graph filtrations.*

## 6.6 DAG-based Graph

A DAG-based Graph is a hierarchical graph system indexed by a directed acyclic graph, where graph layers are connected through order-preserving homomorphisms.

**Definition 6.6.1** (DAG-based Graph)**.** Let

$$D = (N, A)$$

be a finite directed acyclic graph, where $N$ is the node set and $A \subseteq N \times N$ is the arc set. A *DAG-based Graph over $D$* is a tuple

$$\mathsf{DAGG} = \big(D, \{G_x\}_{x \in N}, \{\phi_a\}_{a \in A}\big),$$

such that:

- for each node $x \in N$,
$$G_x = (V_x, E_x)$$
is a finite simple graph;
- for each arc $a = (x, y) \in A$,
$$\phi_a : G_x \to G_y$$
is a graph homomorphism.

For a directed path

$$p = (x_0 \to x_1 \to \cdots \to x_k)$$

in $D$, define

$$\phi_p := \phi_{(x_{k-1}, x_k)} \circ \cdots \circ \phi_{(x_0, x_1)} : G_{x_0} \to G_{x_k}.$$

The family is said to be *path-coherent* if, whenever $p$ and $q$ are two directed paths from $x$ to $y$, one has

$$\phi_p = \phi_q.$$

A *path-coherent DAG-based Graph* is a DAG-based Graph satisfying this condition.

**Remark 6.6.2.** *A DAG-based Graph encodes hierarchy through the partial order induced by reachability in the DAG. The lower nodes represent local or refined graph layers, while upper nodes represent more global, aggregated, or abstracted graph layers. A tree-structured graph is a special case in which the indexing DAG is a rooted tree.*

**Example 6.6.3** (A path-coherent DAG-based Graph)**.** Let

$$N = \{x, y, z, w\}, \qquad A = \{(x, y), (x, z), (y, w), (z, w)\},$$

and let

$$D = (N, A).$$

Then $D$ is a finite directed acyclic graph. In particular, there are two directed paths from $x$ to $w$:

$$p = (x \to y \to w), \qquad q = (x \to z \to w).$$

We now define a DAG-based Graph over $D$.
For the node $x$, let

$$G_x = (V_x, E_x), \qquad V_x = \{u_1, u_2, u_3\}, \qquad E_x = \big\{\{u_1, u_2\}, \{u_2, u_3\}\big\},$$

so that $G_x$ is a path on three vertices.
For the nodes $y, z, w$, let

$$G_y = (V_y, E_y), \qquad V_y = \{\alpha, \beta\}, \qquad E_y = \big\{\{\alpha, \beta\}\big\},$$

$$G_z = (V_z, E_z), \qquad V_z = \{\gamma, \delta\}, \qquad E_z = \big\{\{\gamma, \delta\}\big\},$$

and

$$G_w = (V_w, E_w), \qquad V_w = \{p, q\}, \qquad E_w = \big\{\{p, q\}\big\}.$$

Next, define graph homomorphisms along the arcs of $D$.

For the arc $(x, y)$, define

$$\phi_{(x,y)} : V_x \to V_y$$

by

$$\phi_{(x,y)}(u_1) = \alpha, \qquad \phi_{(x,y)}(u_2) = \beta, \qquad \phi_{(x,y)}(u_3) = \alpha.$$

This is a graph homomorphism because

$$\{u_1, u_2\} \mapsto \{\alpha, \beta\} \in E_y, \qquad \{u_2, u_3\} \mapsto \{\beta, \alpha\} = \{\alpha, \beta\} \in E_y.$$

For the arc $(x, z)$, define

$$\phi_{(x,z)} : V_x \to V_z$$

by

$$\phi_{(x,z)}(u_1) = \gamma, \qquad \phi_{(x,z)}(u_2) = \delta, \qquad \phi_{(x,z)}(u_3) = \gamma.$$

Again, this is a graph homomorphism because both edges of $G_x$ are sent to the edge

$$\{\gamma, \delta\} \in E_z.$$

For the arc $(y, w)$, define

$$\phi_{(y,w)} : V_y \to V_w$$

by

$$\phi_{(y,w)}(\alpha) = p, \qquad \phi_{(y,w)}(\beta) = q.$$

This is a graph homomorphism since

$$\{\alpha, \beta\} \mapsto \{p, q\} \in E_w.$$

For the arc $(z, w)$, define

$$\phi_{(z,w)} : V_z \to V_w$$

by

$$\phi_{(z,w)}(\gamma) = p, \qquad \phi_{(z,w)}(\delta) = q.$$

This is also a graph homomorphism since

$$\{\gamma, \delta\} \mapsto \{p, q\} \in E_w.$$

Therefore,

$$\mathsf{DAGG} = \big(D, \{G_x, G_y, G_z, G_w\}, \{\phi_{(x,y)}, \phi_{(x,z)}, \phi_{(y,w)}, \phi_{(z,w)}\}\big)$$

is a DAG-based Graph over $D$.

Moreover, the two path maps from $x$ to $w$ coincide. Indeed,

$$\phi_p = \phi_{(y,w)} \circ \phi_{(x,y)}, \qquad \phi_q = \phi_{(z,w)} \circ \phi_{(x,z)},$$

and for each vertex of $G_x$ we have

$$\phi_p(u_1) = p = \phi_q(u_1), \qquad \phi_p(u_2) = q = \phi_q(u_2), \qquad \phi_p(u_3) = p = \phi_q(u_3).$$

Hence

$$\phi_p = \phi_q.$$

Thus the above DAG-based Graph is path-coherent.

## 6.7 Compositional Graph

A Compositional Graph is a graph framework where larger graph structures are formed by composing smaller graph modules through specified interfaces and gluing rules.

**Definition 6.7.1** (Graph module)**.** A *graph module* is a quintuple

$$M = (G, I, O, \iota, \omega),$$

where:

- $G = (V, E)$ is a finite simple graph;
- $I$ and $O$ are finite sets, called the *input interface* and *output interface*;
- $$\iota : I \to V, \qquad \omega : O \to V$$

  are interface maps.

**Definition 6.7.2** (Composition of graph modules)**.** Let

$$M_1 = (G_1, I_1, O_1, \iota_1, \omega_1), \qquad M_2 = (G_2, I_2, O_2, \iota_2, \omega_2)$$

be graph modules, where

$$G_1 = (V_1, E_1), \qquad G_2 = (V_2, E_2),$$

and assume $V_1 \cap V_2 = \varnothing$. Let

$$\varphi : O_1' \to I_2'$$

be a bijection between subsets

$$O_1' \subseteq O_1, \qquad I_2' \subseteq I_2.$$

The *composite module* $M_2 \circ_\varphi M_1$ is defined as follows.

First, define an equivalence relation $\sim_\varphi$ on $V_1 \sqcup V_2$ generated by

$$\omega_1(o) \sim_\varphi \iota_2(\varphi(o)) \qquad \text{for all } o \in O_1'.$$

Then let

$$V_{21} := (V_1 \sqcup V_2)/{\sim_\varphi}.$$

The edge set $E_{21}$ is the image of $E_1 \sqcup E_2$ under the quotient map

$$V_1 \sqcup V_2 \to V_{21}.$$

Thus

$$G_{21} := (V_{21}, E_{21})$$

is the graph obtained by gluing the matched output vertices of $M_1$ to the matched input vertices of $M_2$.

The remaining interfaces are

$$I_{21} := I_1 \sqcup (I_2 \setminus I_2'), \qquad O_{21} := (O_1 \setminus O_1') \sqcup O_2,$$

with the induced interface maps into $V_{21}$. The resulting graph module is denoted by

$$M_2 \circ_\varphi M_1.$$

**Definition 6.7.3** (Compositional Graph)**.** A *Compositional Graph* is a pair

$$\mathsf{CG} = (\mathcal{M}, \circ),$$

where:

- $\mathcal{M}$ is a family of graph modules;
- for every compatible triple $(M_1, M_2, \varphi)$ as above, the composite

  $$M_2 \circ_\varphi M_1$$

  belongs to $\mathcal{M}$;
- composition is associative up to canonical graph-module isomorphism.

**Remark 6.7.4.** *A Compositional Graph formalizes hierarchy by* composition *rather than by membership. Larger graph objects are built by gluing smaller modules along prescribed interfaces, so the hierarchical structure is encoded by a composition pattern of graph modules.*

**Example 6.7.5** (A Compositional Graph of path-modules)**.** Let $\mathcal{M}$ be the family of all finite graph modules

$$M = (G, I, O, \iota, \omega),$$

where $G = (V, E)$ is a finite simple graph, $I$ and $O$ are finite interface sets, and $\iota : I \to V$, $\omega : O \to V$ are interface maps. Let $\circ$ be the module-composition rule defined by gluing compatible output and input interfaces along a bijection $\varphi$ as in the preceding definition. Then

$$\mathsf{CG} = (\mathcal{M}, \circ)$$

is a Compositional Graph.

As a concrete instance, consider the two graph modules

$$M_1 = (G_1, I_1, O_1, \iota_1, \omega_1), \qquad M_2 = (G_2, I_2, O_2, \iota_2, \omega_2),$$

where

$$G_1 = (V_1, E_1), \qquad V_1 = \{a, b\}, \qquad E_1 = \big\{\{a, b\}\big\},$$

and

$$I_1 = \{i_1\}, \qquad O_1 = \{o_1\}, \qquad \iota_1(i_1) = a, \qquad \omega_1(o_1) = b.$$

Thus $M_1$ is a path-module with input at $a$ and output at $b$.

Likewise, let

$$G_2 = (V_2, E_2), \qquad V_2 = \{c, d\}, \qquad E_2 = \big\{\{c, d\}\big\},$$

and

$$I_2 = \{j_1\}, \qquad O_2 = \{k_1\}, \qquad \iota_2(j_1) = c, \qquad \omega_2(k_1) = d.$$

Thus $M_2$ is a path-module with input at $c$ and output at $d$.

Take

$$O_1' = \{o_1\} \subseteq O_1, \qquad I_2' = \{j_1\} \subseteq I_2,$$

and define the bijection

$$\varphi : O_1' \to I_2', \qquad \varphi(o_1) = j_1.$$

Then the composite

$$M_2 \circ_\varphi M_1$$

is obtained by identifying the output vertex $b = \omega_1(o_1)$ of $M_1$ with the input vertex $c = \iota_2(j_1)$ of $M_2$. Hence the resulting graph is the path

$$a - [b = c] - d,$$

with input interface inherited from $M_1$ and output interface inherited from $M_2$. Therefore the composition produces a larger graph module from two smaller modules by gluing along the specified interfaces, which is precisely the basic mechanism of a Compositional Graph.

**Theorem 6.7.6** (Well-definedness of module composition)**.** *Let*

$$M_1 = (G_1, I_1, O_1, \iota_1, \omega_1), \qquad M_2 = (G_2, I_2, O_2, \iota_2, \omega_2)$$

*be graph modules with*

$$G_1 = (V_1, E_1), \qquad G_2 = (V_2, E_2), \qquad V_1 \cap V_2 = \varnothing.$$

*Let*

$$\varphi : O_1' \to I_2'$$

*be a bijection between subsets*

$$O_1' \subseteq O_1, \qquad I_2' \subseteq I_2.$$

*Let* $\sim_\varphi$ *be the equivalence relation on* $V_1 \sqcup V_2$ *generated by*

$$\omega_1(o) \sim_\varphi \iota_2(\varphi(o)) \qquad \text{for all } o \in O_1'.$$

*Assume that no edge of* $E_1 \sqcup E_2$ *has both endpoints in the same* $\sim_\varphi$*-equivalence class. Then the composite*

$$M_2 \circ_\varphi M_1$$

*is a well-defined graph module.*

*Proof.* Let

$$\pi : V_1 \sqcup V_2 \to (V_1 \sqcup V_2)/\sim_\varphi$$

be the quotient map, and set

$$V_{21} := (V_1 \sqcup V_2)/\sim_\varphi.$$

Since $V_1$ and $V_2$ are finite, the set $V_{21}$ is finite.

Define

$$E_{21} := \Big\{ \{\pi(u), \pi(v)\} : \{u, v\} \in E_1 \sqcup E_2 \Big\}.$$

By the assumption that no edge collapses inside a single $\sim_\varphi$-class, we have

$$\pi(u) \neq \pi(v)$$

for every edge $\{u, v\} \in E_1 \sqcup E_2$. Hence every element of $E_{21}$ is a two-element subset of $V_{21}$, so

$$E_{21} \subseteq \binom{V_{21}}{2}.$$

Therefore

$$G_{21} := (V_{21}, E_{21})$$

is a finite simple graph.

Now define the remaining interface sets by

$$I_{21} := I_1 \sqcup (I_2 \setminus I_2'), \qquad O_{21} := (O_1 \setminus O_1') \sqcup O_2.$$

The induced interface maps

$$\iota_{21} : I_{21} \to V_{21}, \qquad \omega_{21} : O_{21} \to V_{21}$$

are given by composing the original interface maps with the quotient map $\pi$. These are well-defined because each interface element has a unique image vertex in the quotient.

Thus

$$M_2 \circ_\varphi M_1 = (G_{21}, I_{21}, O_{21}, \iota_{21}, \omega_{21})$$

is a well-defined graph module. □

## 6.8 $n$-Iterated Labeling Graph

A Labeling Graph is a graph equipped with vertex and edge labels, assigning symbolic information to structural elements while preserving the graph's underlying incidence pattern[754, 755, 756]. Related concepts, such as Fuzzy Graph Labeling [757, 758, 759], HyperGraph Labeling[760, 761, 762], and Neutrosophic Graph Labeling[763, 764], are also known and have been studied. An n-Iterated Labeling Graph is a graph equipped with n+1 compatible labeling layers, where each higher layer refines and coherently recovers lower labels.

**Definition 6.8.1** (Labeling Graph)**.** [754, 755] Let $\Sigma_V$ and $\Sigma_E$ be sets, called the *vertex-label set* and the *edge-label set*, respectively. A *Labeling Graph over* $(\Sigma_V, \Sigma_E)$ is a quintuple

$$\mathsf{LG} = (V, E, \partial, \lambda_V, \lambda_E),$$

where

- $V \neq \varnothing$ is a finite set of vertices;
- $E$ is a finite set of edge identifiers;
- $$\partial : E \to \binom{V}{2}$$
  is an incidence map;
- $$\lambda_V : V \to \Sigma_V$$
  is a vertex-labeling map;

- $$\lambda_E : E \to \Sigma_E$$
  is an edge-labeling map.

**Definition 6.8.2** (Iterated labeling signature)**.** Let $n \in \mathbb{N}_0$. An *$n$-iterated labeling signature* is a family

$$\mathfrak{S}^{(n)} = \Big((\Sigma_0^V, \Sigma_0^E), \ldots, (\Sigma_n^V, \Sigma_n^E), (\pi_k^V, \pi_k^E)_{k=1}^n\Big),$$

where, for each $k \in \{0, 1, \ldots, n\}$,

$$\Sigma_k^V, \Sigma_k^E$$

are sets, and for each $k \in \{1, \ldots, n\}$,

$$\pi_k^V : \Sigma_k^V \to \Sigma_{k-1}^V, \qquad \pi_k^E : \Sigma_k^E \to \Sigma_{k-1}^E$$

are maps, called the *label-reduction maps*.

**Definition 6.8.3** ($n$-Iterated Labeling Graph)**.** Let

$$\mathfrak{S}^{(n)} = \Big((\Sigma_0^V, \Sigma_0^E), \ldots, (\Sigma_n^V, \Sigma_n^E), (\pi_k^V, \pi_k^E)_{k=1}^n\Big)$$

be an $n$-iterated labeling signature. An *$n$-Iterated Labeling Graph over* $\mathfrak{S}^{(n)}$ is a tuple

$$\mathsf{ILG}^{(n)} = \big(V, E, \partial, \lambda_V^{(0)}, \ldots, \lambda_V^{(n)}, \lambda_E^{(0)}, \ldots, \lambda_E^{(n)}\big),$$

such that:

- $V \neq \varnothing$ is a finite set of vertices;
- $E$ is a finite set of edge identifiers;
- $$\partial : E \to \binom{V}{2}$$
  is an incidence map;
- for each $k \in \{0, 1, \ldots, n\}$,
  $$\lambda_V^{(k)} : V \to \Sigma_k^V \qquad \text{and} \qquad \lambda_E^{(k)} : E \to \Sigma_k^E$$
  are labeling maps;
- for each $k \in \{1, \ldots, n\}$, the following coherence conditions hold:
  $$\pi_k^V \circ \lambda_V^{(k)} = \lambda_V^{(k-1)}, \qquad \pi_k^E \circ \lambda_E^{(k)} = \lambda_E^{(k-1)}.$$

For each $k \in \{0, 1, \ldots, n\}$, the quintuple

$$\mathsf{LG}^{(k)} = \big(V, E, \partial, \lambda_V^{(k)}, \lambda_E^{(k)}\big)$$

is called the *$k$-th labeling layer* of $\mathsf{ILG}^{(n)}$.

**Example 6.8.4** (A 2-Iterated Labeling Graph)**.** Let

$$V = \{a, b, c\}, \qquad E = \{e_1, e_2\},$$

and define the incidence map

$$\partial : E \to \binom{V}{2}$$

by

$$\partial(e_1) = \{a, b\}, \qquad \partial(e_2) = \{b, c\}.$$

Define the vertex-label sets

$$\Sigma_0^V = \{\text{core}, \text{peripheral}\},$$

$$\Sigma_1^V = \{\text{red-core}, \text{blue-core}, \text{red-peripheral}, \text{blue-peripheral}\},$$

and

$$\Sigma_2^V = \{\text{hot-red-core}, \text{hot-blue-core}, \text{hot-red-peripheral}, \text{hot-blue-peripheral}\}.$$

Define the edge-label sets

$$\Sigma_0^E = \{\text{ordinary}\}, \qquad \Sigma_1^E = \{\text{road}, \text{bridge}\}, \qquad \Sigma_2^E = \{\text{active-road}, \text{active-bridge}\}.$$

Next, define the label-reduction maps

$$\pi_1^V : \Sigma_1^V \to \Sigma_0^V, \qquad \pi_2^V : \Sigma_2^V \to \Sigma_1^V,$$

by

$$\pi_1^V(\text{red-core}) = \text{core}, \quad \pi_1^V(\text{blue-core}) = \text{core},$$

$$\pi_1^V(\text{red-peripheral}) = \text{peripheral}, \quad \pi_1^V(\text{blue-peripheral}) = \text{peripheral},$$

and

$$\pi_2^V(\text{hot-red-core}) = \text{red-core}, \quad \pi_2^V(\text{hot-blue-core}) = \text{blue-core},$$

$$\pi_2^V(\text{hot-red-peripheral}) = \text{red-peripheral}, \quad \pi_2^V(\text{hot-blue-peripheral}) = \text{blue-peripheral}.$$

Similarly, define

$$\pi_1^E : \Sigma_1^E \to \Sigma_0^E, \qquad \pi_2^E : \Sigma_2^E \to \Sigma_1^E,$$

by

$$\pi_1^E(\text{road}) = \text{ordinary}, \qquad \pi_1^E(\text{bridge}) = \text{ordinary},$$

and

$$\pi_2^E(\text{active-road}) = \text{road}, \qquad \pi_2^E(\text{active-bridge}) = \text{bridge}.$$

Now define the vertex-labeling maps at the three levels by

$$\lambda_V^{(0)}(a) = \text{core}, \qquad \lambda_V^{(0)}(b) = \text{peripheral}, \qquad \lambda_V^{(0)}(c) = \text{core},$$

$$\lambda_V^{(1)}(a) = \text{red-core}, \qquad \lambda_V^{(1)}(b) = \text{blue-peripheral}, \qquad \lambda_V^{(1)}(c) = \text{blue-core},$$

and

$$\lambda_V^{(2)}(a) = \text{hot-red-core}, \qquad \lambda_V^{(2)}(b) = \text{hot-blue-peripheral}, \qquad \lambda_V^{(2)}(c) = \text{hot-blue-core}.$$

Define the edge-labeling maps by

$$\lambda_E^{(0)}(e_1) = \text{ordinary}, \qquad \lambda_E^{(0)}(e_2) = \text{ordinary},$$

$$\lambda_E^{(1)}(e_1) = \text{road}, \qquad \lambda_E^{(1)}(e_2) = \text{bridge},$$

and

$$\lambda_E^{(2)}(e_1) = \text{active-road}, \qquad \lambda_E^{(2)}(e_2) = \text{active-bridge}.$$

Then the coherence conditions hold:

$$\pi_1^V \circ \lambda_V^{(1)} = \lambda_V^{(0)}, \qquad \pi_2^V \circ \lambda_V^{(2)} = \lambda_V^{(1)},$$

and

$$\pi_1^E \circ \lambda_E^{(1)} = \lambda_E^{(0)}, \qquad \pi_2^E \circ \lambda_E^{(2)} = \lambda_E^{(1)}.$$

Therefore,

$$\mathsf{ILG}^{(2)} = \big(V, E, \partial, \lambda_V^{(0)}, \lambda_V^{(1)}, \lambda_V^{(2)}, \lambda_E^{(0)}, \lambda_E^{(1)}, \lambda_E^{(2)}\big)$$

is a 2-Iterated Labeling Graph over the signature

$$\mathfrak{S}^{(2)} = \Big((\Sigma_0^V, \Sigma_0^E), (\Sigma_1^V, \Sigma_1^E), (\Sigma_2^V, \Sigma_2^E), (\pi_1^V, \pi_1^E), (\pi_2^V, \pi_2^E)\Big).$$

**Theorem 6.8.5** (Well-definedness of labeling layers)**.** *Let*

$$\mathfrak{S}^{(n)} = \Big( (\Sigma_0^V, \Sigma_0^E), \ldots, (\Sigma_n^V, \Sigma_n^E), (\pi_k^V, \pi_k^E)_{k=1}^n \Big)$$

*be an $n$-iterated labeling signature, and let*

$$\mathsf{ILG}^{(n)} = \big( V, E, \partial, \lambda_V^{(0)}, \ldots, \lambda_V^{(n)}, \lambda_E^{(0)}, \ldots, \lambda_E^{(n)} \big)$$

*be an $n$-Iterated Labeling Graph over $\mathfrak{S}^{(n)}$. Then, for each $k \in \{0, 1, \ldots, n\}$, the tuple*

$$\mathsf{LG}^{(k)} = \big( V, E, \partial, \lambda_V^{(k)}, \lambda_E^{(k)} \big)$$

*is a well-defined Labeling Graph over $(\Sigma_k^V, \Sigma_k^E)$.*

*Proof.* Fix $k \in \{0, 1, \ldots, n\}$. By definition of an $n$-Iterated Labeling Graph,

$$V \neq \varnothing$$

is a finite vertex set,

$$E$$

is a finite set of edge identifiers, and

$$\partial : E \to \binom{V}{2}$$

is an incidence map. Moreover,

$$\lambda_V^{(k)} : V \to \Sigma_k^V \qquad \text{and} \qquad \lambda_E^{(k)} : E \to \Sigma_k^E$$

are well-defined labeling maps. Hence the quintuple

$$\mathsf{LG}^{(k)} = \big( V, E, \partial, \lambda_V^{(k)}, \lambda_E^{(k)} \big)$$

satisfies exactly the definition of a Labeling Graph over $(\Sigma_k^V, \Sigma_k^E)$. Therefore each labeling layer is well-defined. □

**Proposition 6.8.6.** *Every labeling layer of an $n$-Iterated Labeling Graph is a Labeling Graph.*

*Proof.* Let

$$\mathsf{ILG}^{(n)} = \big( V, E, \partial, \lambda_V^{(0)}, \ldots, \lambda_V^{(n)}, \lambda_E^{(0)}, \ldots, \lambda_E^{(n)} \big)$$

be an $n$-Iterated Labeling Graph. For each $k \in \{0, 1, \ldots, n\}$, the maps

$$\lambda_V^{(k)} : V \to \Sigma_k^V, \qquad \lambda_E^{(k)} : E \to \Sigma_k^E$$

are well-defined labeling maps on the same underlying graph $(V, E, \partial)$. Hence

$$\mathsf{LG}^{(k)} = \big( V, E, \partial, \lambda_V^{(k)}, \lambda_E^{(k)} \big)$$

is a Labeling Graph over $(\Sigma_k^V, \Sigma_k^E)$. □

**Remark 6.8.7.** *Thus, an $n$-Iterated Labeling Graph is a single graph endowed with a tower of compatible labelings. Each higher labeling layer refines the previous one, while the reduction maps recover the lower-level labels.*

**Remark 6.8.8** (Graph-valued labels)**.** *If desired, one may choose some of the label sets $\Sigma_k^V$ or $\Sigma_k^E$ to consist of graph-like or labeling-graph-like objects. In that case, the above definition yields a recursively enriched labeling hierarchy.*

## 6.9 $m$-Multidynamic Graph

An m-Multidynamic Graph is a graph system indexed by m simultaneous time axes, where coherent transition maps describe structure evolving across multiple parallel temporal dimensions.

**Definition 6.9.1** ($m$-Multidynamic Graph)**.** Let $m \in \mathbb{N}$. For each $i \in \{1, \ldots, m\}$, let $T_i$ be a finite totally ordered set, and define

$$\mathbf{T} := T_1 \times \cdots \times T_m$$

with the componentwise partial order

$$(s_1, \ldots, s_m) \preceq (t_1, \ldots, t_m) \iff s_i \leq t_i \text{ for all } i = 1, \ldots, m.$$

An *$m$-Multidynamic Graph* is a tuple

$$\mathcal{G} = \Big( \{G_{\mathbf{t}}\}_{\mathbf{t} \in \mathbf{T}}, \{\rho_{\mathbf{s},\mathbf{t}}\}_{\mathbf{s} \preceq \mathbf{t}} \Big),$$

such that:

- for each $\mathbf{t} \in \mathbf{T}$,
$$G_{\mathbf{t}} = (V_{\mathbf{t}}, E_{\mathbf{t}})$$
is a finite simple graph;
- for each $\mathbf{s} \preceq \mathbf{t}$,
$$\rho_{\mathbf{s},\mathbf{t}} : G_{\mathbf{s}} \to G_{\mathbf{t}}$$
is a graph homomorphism;
- for every $\mathbf{t} \in \mathbf{T}$,
$$\rho_{\mathbf{t},\mathbf{t}} = \mathrm{id}_{G_{\mathbf{t}}};$$
- whenever $\mathbf{s} \preceq \mathbf{u} \preceq \mathbf{t}$,
$$\rho_{\mathbf{u},\mathbf{t}} \circ \rho_{\mathbf{s},\mathbf{u}} = \rho_{\mathbf{s},\mathbf{t}}.$$

**Example 6.9.2** (A simple 2-Multidynamic Graph)**.** Let

$$m = 2, \qquad T_1 = T_2 = \{0, 1\},$$

each ordered by the usual relation $0 \leq 1$. Then

$$\mathbf{T} = T_1 \times T_2 = \{(0,0), (0,1), (1,0), (1,1)\},$$

with the componentwise partial order.

Define graphs at each multi-time point by

$$G_{(0,0)} = (V_{00}, E_{00}), \qquad V_{00} = \{a, b\}, \qquad E_{00} = \{\{a, b\}\},$$

$$G_{(0,1)} = (V_{01}, E_{01}), \qquad V_{01} = \{a, b, c\}, \qquad E_{01} = \{\{a, b\}, \{b, c\}\},$$

$$G_{(1,0)} = (V_{10}, E_{10}), \qquad V_{10} = \{u, v\}, \qquad E_{10} = \{\{u, v\}\},$$

and

$$G_{(1,1)} = (V_{11}, E_{11}), \qquad V_{11} = \{p, q\}, \qquad E_{11} = \{\{p, q\}\}.$$

Now define graph homomorphisms along the comparable pairs. For the identity maps, set

$$\rho_{(t_1,t_2),(t_1,t_2)} = \mathrm{id}_{G_{(t_1,t_2)}} \qquad \text{for all } (t_1, t_2) \in \mathbf{T}.$$

For the nontrivial comparable pairs, define

$$\rho_{(0,0),(0,1)} : V_{00} \to V_{01}, \qquad \rho_{(0,0),(0,1)}(a) = a, \quad \rho_{(0,0),(0,1)}(b) = b,$$

$$\rho_{(0,0),(1,0)} : V_{00} \to V_{10}, \qquad \rho_{(0,0),(1,0)}(a) = u, \quad \rho_{(0,0),(1,0)}(b) = v,$$

$$\rho_{(0,1),(1,1)} : V_{01} \to V_{11}, \qquad \rho_{(0,1),(1,1)}(a) = p, \quad \rho_{(0,1),(1,1)}(b) = q, \quad \rho_{(0,1),(1,1)}(c) = p,$$

and

$$\rho_{(1,0),(1,1)} : V_{10} \to V_{11}, \qquad \rho_{(1,0),(1,1)}(u) = p, \quad \rho_{(1,0),(1,1)}(v) = q.$$

Finally, define

$$\rho_{(0,0),(1,1)} : V_{00} \to V_{11}$$

by

$$\rho_{(0,0),(1,1)}(a) = p, \qquad \rho_{(0,0),(1,1)}(b) = q.$$

Each of these maps is a graph homomorphism. Moreover,

$$\rho_{(0,1),(1,1)} \circ \rho_{(0,0),(0,1)} = \rho_{(0,0),(1,1)},$$

and

$$\rho_{(1,0),(1,1)} \circ \rho_{(0,0),(1,0)} = \rho_{(0,0),(1,1)}.$$

Hence the coherence condition holds for the two chains

$$(0,0) \preceq (0,1) \preceq (1,1), \qquad (0,0) \preceq (1,0) \preceq (1,1).$$

Therefore,

$$\mathcal{G} = \Big( \{G_{\mathbf{t}}\}_{\mathbf{t}\in\mathbf{T}}, \{\rho_{\mathbf{s},\mathbf{t}}\}_{\mathbf{s}\preceq\mathbf{t}} \Big)$$

is a 2-Multidynamic Graph. This example represents a graph system evolving simultaneously along two ordered time axes.

**Theorem 6.9.3** (Well-definedness of an $m$-Multidynamic Graph)**.** *Let $m \in \mathbb{N}$, and for each $i \in \{1, \ldots, m\}$ let $T_i$ be a finite totally ordered set. Define*

$$\mathbf{T} := T_1 \times \cdots \times T_m$$

*with the componentwise relation*

$$(s_1, \ldots, s_m) \preceq (t_1, \ldots, t_m) \iff s_i \leq t_i \text{ for all } i = 1, \ldots, m.$$

*Then $(\mathbf{T}, \preceq)$ is a finite partially ordered set. Consequently, if*

$$\mathcal{G} = \Big( \{G_{\mathbf{t}}\}_{\mathbf{t}\in\mathbf{T}}, \{\rho_{\mathbf{s},\mathbf{t}}\}_{\mathbf{s}\preceq\mathbf{t}} \Big)$$

*satisfies the conditions, then $\mathcal{G}$ is a well-defined indexed system of finite simple graphs and coherent transition homomorphisms.*

*Proof.* Since each $T_i$ is finite, the Cartesian product

$$\mathbf{T} = T_1 \times \cdots \times T_m$$

is finite.

We verify that $\preceq$ is a partial order on $\mathbf{T}$. For reflexivity, for every

$$\mathbf{t} = (t_1, \ldots, t_m) \in \mathbf{T},$$

we have $t_i \leq t_i$ for all $i$, hence $\mathbf{t} \preceq \mathbf{t}$.

For antisymmetry, if

$$(s_1, \ldots, s_m) \preceq (t_1, \ldots, t_m) \quad \text{and} \quad (t_1, \ldots, t_m) \preceq (s_1, \ldots, s_m),$$

then $s_i \leq t_i$ and $t_i \leq s_i$ for all $i$. Since each $T_i$ is totally ordered, it follows that $s_i = t_i$ for all $i$, so

$$(s_1, \ldots, s_m) = (t_1, \ldots, t_m).$$

For transitivity, if

$$(r_1, \ldots, r_m) \preceq (s_1, \ldots, s_m) \quad \text{and} \quad (s_1, \ldots, s_m) \preceq (t_1, \ldots, t_m),$$

then $r_i \leq s_i \leq t_i$ for all $i$, hence $r_i \leq t_i$ for all $i$, so

$$(r_1, \ldots, r_m) \preceq (t_1, \ldots, t_m).$$

Thus $(\mathbf{T}, \preceq)$ is a finite poset. Therefore the family

$$\{G_{\mathbf{t}}\}_{\mathbf{t}\in\mathbf{T}}$$

is indexed by a well-defined finite partially ordered set, and for every comparable pair

$$\mathbf{s} \preceq \mathbf{t}$$

the map

$$\rho_{\mathbf{s},\mathbf{t}} : G_{\mathbf{s}} \to G_{\mathbf{t}}$$

is well specified. The identities

$$\rho_{\mathbf{t},\mathbf{t}} = \mathrm{id}_{G_{\mathbf{t}}}$$

and the coherence condition

$$\rho_{\mathbf{u},\mathbf{t}} \circ \rho_{\mathbf{s},\mathbf{u}} = \rho_{\mathbf{s},\mathbf{t}} \qquad (\mathbf{s} \preceq \mathbf{u} \preceq \mathbf{t})$$

show that the transition homomorphisms are compositionally consistent. Hence $\mathcal{G}$ is well defined. □

## 6.10 Structure-vertexed Graph

A Structure-vertexed Graph is a graph whose vertices are mathematical structures, with edges representing chosen relations such as isomorphism, similarity, compatibility, or embeddability.

**Definition 6.10.1** (Structure universe over a fixed signature)**.** Let

$$\Sigma = \big(\mathrm{Func}, \mathrm{Rel}, \mathrm{ar}_{\mathrm{Func}}, \mathrm{ar}_{\mathrm{Rel}}\big)$$

be a fixed single-sorted finitary signature, and let

$$\mathrm{Str}_{\Sigma}$$

denote the class of all $\Sigma$-structures. A *structure universe* is any set

$$\mathcal{U} \subseteq \mathrm{Str}_{\Sigma}.$$

**Definition 6.10.2** (Structure-vertexed Graph)**.** Let $\mathcal{U} \subseteq \mathrm{Str}_{\Sigma}$ be a structure universe, and let

$$\mathcal{R} \subseteq \mathcal{U} \times \mathcal{U}$$

be a symmetric and irreflexive binary relation, called a *structure adjacency relation.* A *Structure-vertexed Graph over* $(\mathcal{U}, \mathcal{R})$ is a graph

$$\mathsf{SVG} = (V, E),$$

where

- $V \subseteq \mathcal{U}$ is a finite nonempty set of vertices, each vertex being a mathematical structure;
- $$E \subseteq \binom{V}{2}$$

  is defined by

  $$\{C, D\} \in E \iff (C, D) \in \mathcal{R} \qquad (C, D \in V).$$

Thus, the vertices of $\mathsf{SVG}$ are themselves structures, and two vertices are adjacent precisely when the corresponding structures satisfy the relation $\mathcal{R}$.

**Remark 6.10.3.** *The relation $\mathcal{R}$ may represent, for example, isomorphism, embeddability, derivability, similarity, compatibility, shared axioms, or any other mathematically meaningful relation between structures. Hence a Structure-vertexed Graph is a general framework for organizing structures as graph vertices.*

**Example 6.10.4** (A Structure-vertexed Graph of finite groups)**.** Let $\Sigma_{\mathrm{grp}}$ be the usual group signature, and let

$$\mathcal{U} = \{C_2,\ C_3,\ C_4,\ V_4\} \subseteq \mathrm{Str}_{\Sigma_{\mathrm{grp}}},$$

where

$$C_2,\ C_3,\ C_4$$

denote the cyclic groups of orders $2, 3, 4$, respectively, and

$$V_4 \cong C_2 \times C_2$$

denotes the Klein four-group.

Define a binary relation

$$\mathcal{R} \subseteq \mathcal{U} \times \mathcal{U}$$

by declaring that, for $G, H \in \mathcal{U}$,

$$(G, H) \in \mathcal{R} \quad \Longleftrightarrow \quad |G| \text{ divides } |H| \text{ or } |H| \text{ divides } |G|,$$

and $G \neq H$. This relation is symmetric and irreflexive.

Now take

$$V = \mathcal{U}.$$

Then the corresponding edge set is

$$E = \Big\{\{C_2, C_4\},\ \{C_2, V_4\},\ \{C_4, V_4\}\Big\}.$$

Indeed, we have

$$2 \mid 4, \qquad 2 \mid 4, \qquad 4 \mid 4,$$

so these pairs are adjacent, while $C_3$ is isolated since

$$3 \nmid 2, \qquad 3 \nmid 4, \qquad 2 \nmid 3, \qquad 4 \nmid 3.$$

Therefore,

$$\mathsf{SVG} = (V, E)$$

is a Structure-vertexed Graph whose vertices are finite group structures and whose edges encode a divisibility-based structural relation.

**Theorem 6.10.5** (Well-definedness of Structure-vertexed Graph)**.** *Let $\mathcal{U} \subseteq \mathrm{Str}_\Sigma$ be a structure universe, let*

$$\mathcal{R} \subseteq \mathcal{U} \times \mathcal{U}$$

*be a symmetric and irreflexive binary relation, and let*

$$V \subseteq \mathcal{U}$$

*be a finite nonempty set. Define*

$$E := \big\{\{C, D\} \in \tbinom{V}{2} : (C, D) \in \mathcal{R}\big\}.$$

*Then*

$$\mathsf{SVG} = (V, E)$$

*is a well-defined finite simple graph.*

*Proof.* Since $V \subseteq \mathcal{U}$ is finite and nonempty, $V$ is a valid finite vertex set.

By definition,

$$E \subseteq \binom{V}{2},$$

so every edge is a two-element subset of $V$. Hence no loop can occur.

Moreover, since $\mathcal{R}$ is symmetric, the condition

$$(C, D) \in \mathcal{R}$$

does not depend on the order of $C$ and $D$, so adjacency is well defined on unordered pairs

$$\{C, D\}.$$

Since $\mathcal{R}$ is irreflexive, one never has

$$(C, C) \in \mathcal{R},$$

so no degenerate edge arises from a repeated vertex.

Therefore, $\mathsf{SVG} = (V, E)$ is a well-defined finite simple graph. $\square$

**Definition 6.10.6** (Set-based form)**.** Equivalently, a *Structure-vertexed Graph* may be defined as a pair

$$\mathsf{SVG} = (V, E),$$

where

$$V \subseteq \mathrm{Str}_\Sigma$$

is a finite nonempty set of structures and

$$E \subseteq \binom{V}{2}.$$

In this form, the edge set $E$ is any chosen collection of unordered pairs of structures.

## 6.11 Constructor-Selectable Graph

A Constructor-Selectable Graph is a graph whose vertex domain is generated from a base set by a user-chosen sequence of hierarchical expansion constructors.

**Definition 6.11.1** (Constructor family)**.** Let **FinSet** denote the category of finite sets. A *constructor family* is a collection

$$\mathfrak{G} = \{G_j : \mathbf{FinSet} \to \mathbf{FinSet}\}_{j \in J}$$

of endofunctors on **FinSet**. Each $G_j$ is called a *hierarchical expansion constructor.*

**Definition 6.11.2** (Constructor signature)**.** Let $\mathfrak{G} = \{G_j\}_{j \in J}$ be a constructor family. A *constructor signature* is a finite sequence

$$\sigma = (j_1, \dots, j_k), \qquad j_r \in J,\ k \geq 0.$$

For such a signature, define

$$F_\sigma := G_{j_k} \circ \cdots \circ G_{j_1},$$

with the convention that if $k = 0$, then

$$F_\sigma = \mathrm{Id}_{\mathbf{FinSet}}.$$

**Definition 6.11.3** ($\sigma$-Constructor-Selectable Graph)**.** Let $V_0 \neq \varnothing$ be a finite base set, let

$$\mathfrak{G} = \{G_j\}_{j \in J}$$

be a constructor family, and let

$$\sigma = (j_1, \dots, j_k)$$

be a constructor signature. A *$\sigma$-Constructor-Selectable Graph over $V_0$* is a pair

$$\mathrm{CSG}_\sigma = (V, E),$$

such that

- $$V \subseteq F_\sigma(V_0)$$

  is a finite nonempty set, whose elements are called *$\sigma$-vertices*;

- $$E \subseteq \binom{V}{2}$$

  is a set of edges.

**Remark 6.11.4.** *The role of the signature $\sigma$ is to specify which hierarchical expansion methods are used, and in what order, to generate the ambient vertex domain $F_\sigma(V_0)$. Thus the graph model does not fix a single construction such as the powerset in advance; instead, the constructor sequence is chosen as part of the definition.*

**Remark 6.11.5** (Typical special cases)**.** *Let $V_0$ be a finite base set.*

- *If $k = 0$, then $F_\sigma(V_0) = V_0$, so $\mathrm{CSG}_\sigma$ is an ordinary finite graph on $V_0$.*

- *If $G_j = \mathcal{P}^*$ is the nonempty-powerset functor, then*

$$F_\sigma(V_0) = (\mathcal{P}^*)^k(V_0),$$

*yielding a powerset-based hierarchical graph.*

- *If $G_j = \mathcal{M}$ is the finite-multiset functor, then*

$$F_\sigma(V_0) = \mathcal{M}^k(V_0),$$

*yielding a multiset-based hierarchical graph.*

- *If $G_j = \mathrm{Tree}$ is the rooted-tree constructor, then*

$$F_\sigma(V_0) = \mathrm{Tree}^k(V_0),$$

*yielding a tree-based hierarchical graph.*

**Definition 6.11.6** (Set-based unconstrained form)**.** More generally, if one wishes to allow any admissible composite constructor rather than fixing a signature in advance, a *Constructor-Selectable Graph* is a triple

$$\mathrm{CSG} = (V_0, \sigma, (V, E)),$$

where $V_0$ is a finite base set, $\sigma$ is a constructor signature over a chosen constructor family, and

$$(V, E)$$

is a $\sigma$-Constructor-Selectable Graph over $V_0$.

## 6.12 Multiinfinite Graph

A Multiinfinite Graph (respectively, Multiinfinite $n$-SuperHyperGraph) is a graph system indexed by a product of several independent infinite ordered axes, with coherent transition homomorphisms across the componentwise order.

**Definition 6.12.1** (Multiinfinite Graph)**.** Let $m \in \mathbb{N}$. For each $i \in \{1, \ldots, m\}$, let $T_i$ be an infinite totally ordered set, and define

$$\mathbf{T} := T_1 \times \cdots \times T_m$$

with the componentwise partial order

$$(s_1, \ldots, s_m) \preceq (t_1, \ldots, t_m) \iff s_i \le t_i \text{ for all } i = 1, \ldots, m.$$

A *Multiinfinite Graph* indexed by $\mathbf{T}$ is a tuple

$$\mathfrak{G} = \Big( \{G_{\mathbf{t}}\}_{\mathbf{t} \in \mathbf{T}}, \{\rho_{\mathbf{s},\mathbf{t}}\}_{\mathbf{s} \preceq \mathbf{t}} \Big),$$

such that:

- for each $\mathbf{t} \in \mathbf{T}$,

$$G_{\mathbf{t}} = (V_{\mathbf{t}}, E_{\mathbf{t}})$$

is a simple graph;

- for each $\mathbf{s} \preceq \mathbf{t}$,

$$\rho_{\mathbf{s},\mathbf{t}} : G_{\mathbf{s}} \to G_{\mathbf{t}}$$

is a graph homomorphism;

- for every $\mathbf{t} \in \mathbf{T}$,

$$\rho_{\mathbf{t},\mathbf{t}} = \mathrm{id}_{G_{\mathbf{t}}};$$

- whenever $\mathbf{r} \preceq \mathbf{s} \preceq \mathbf{t}$,

$$\rho_{\mathbf{s},\mathbf{t}} \circ \rho_{\mathbf{r},\mathbf{s}} = \rho_{\mathbf{r},\mathbf{t}}.$$

The sets $T_1, \dots, T_m$ are called the *independent infinite axes* of $\mathfrak{G}$.

**Remark 6.12.2.** *A Multiinfinite Graph may be viewed as a graph system evolving simultaneously along several independent infinite directions. If $m = 1$, then the notion reduces to a one-axis infinite dynamic graph.*

**Definition 6.12.3** (Multiinfinite $n$-SuperHyperGraph)**.** Let $m \in \mathbb{N}$, and for each $i \in \{1, \dots, m\}$ let $T_i$ be an infinite totally ordered set. Let

$$\mathbf{T} := T_1 \times \cdots \times T_m$$

with the componentwise partial order $\preceq$. Let $V_0 \neq \varnothing$ be a base set, and define the iterated powersets by

$$\mathcal{P}^0(V_0) := V_0, \qquad \mathcal{P}^{k+1}(V_0) := \mathcal{P}\big(\mathcal{P}^k(V_0)\big) \quad (k \geq 0).$$

Fix $n \in \mathbb{N}_0$. A *Multiinfinite $n$-SuperHyperGraph* indexed by $\mathbf{T}$ is a tuple

$$\mathfrak{SH}^{(n)} = \Big(\{\mathrm{SHG}^{(n)}_{\mathbf{t}}\}_{\mathbf{t}\in\mathbf{T}}, \{\phi_{\mathbf{s},\mathbf{t}}\}_{\mathbf{s}\preceq\mathbf{t}}\Big),$$

such that:

- for each $\mathbf{t} \in \mathbf{T}$,
$$\mathrm{SHG}^{(n)}_{\mathbf{t}} = (V_{\mathbf{t}}, E_{\mathbf{t}})$$
is an $n$-SuperHyperGraph on $V_0$, that is,
$$V_{\mathbf{t}} \subseteq \mathcal{P}^n(V_0), \qquad E_{\mathbf{t}} \subseteq \mathcal{P}(V_{\mathbf{t}}) \setminus \{\varnothing\};$$
- for each $\mathbf{s} \preceq \mathbf{t}$,
$$\phi_{\mathbf{s},\mathbf{t}} : V_{\mathbf{s}} \to V_{\mathbf{t}}$$
is a *transition superhypergraph homomorphism*, meaning that
$$\phi_{\mathbf{s},\mathbf{t}}[\varepsilon] := \{\phi_{\mathbf{s},\mathbf{t}}(v) : v \in \varepsilon\} \in E_{\mathbf{t}} \qquad \text{for every } \varepsilon \in E_{\mathbf{s}};$$
- for every $\mathbf{t} \in \mathbf{T}$,
$$\phi_{\mathbf{t},\mathbf{t}} = \mathrm{id}_{V_{\mathbf{t}}};$$
- whenever $\mathbf{r} \preceq \mathbf{s} \preceq \mathbf{t}$,
$$\phi_{\mathbf{s},\mathbf{t}} \circ \phi_{\mathbf{r},\mathbf{s}} = \phi_{\mathbf{r},\mathbf{t}}.$$

The sets $T_1, \dots, T_m$ are called the *independent infinite axes* of $\mathfrak{SH}^{(n)}$.

**Remark 6.12.4.** *A Multiinfinite $n$-SuperHyperGraph is an $m$-axis infinite family of $n$-SuperHyperGraphs with coherent transition homomorphisms. If $m = 1$, then it reduces to a one-axis infinite family of $n$-SuperHyperGraphs.*

**Definition 6.12.5** (Inclusion-type Multiinfinite Graph / SuperHyperGraph)**.** A Multiinfinite Graph

$$\mathfrak{G} = \Big(\{G_{\mathbf{t}}\}_{\mathbf{t}\in\mathbf{T}}, \{\rho_{\mathbf{s},\mathbf{t}}\}_{\mathbf{s}\preceq\mathbf{t}}\Big)$$

is called *inclusion-type* if, whenever $\mathbf{s} \preceq \mathbf{t}$,

$$G_{\mathbf{s}} \subseteq G_{\mathbf{t}}$$

and $\rho_{\mathbf{s},\mathbf{t}}$ is the inclusion map.

Likewise, a Multiinfinite $n$-SuperHyperGraph

$$\mathfrak{SH}^{(n)} = \Big(\{\mathrm{SHG}^{(n)}_{\mathbf{t}}\}_{\mathbf{t}\in\mathbf{T}}, \{\phi_{\mathbf{s},\mathbf{t}}\}_{\mathbf{s}\preceq\mathbf{t}}\Big)$$

is called *inclusion-type* if, whenever $\mathbf{s} \preceq \mathbf{t}$,

$$V_{\mathbf{s}} \subseteq V_{\mathbf{t}}, \qquad E_{\mathbf{s}} \subseteq E_{\mathbf{t}},$$

and $\phi_{\mathbf{s},\mathbf{t}}$ is the inclusion map on vertices.

**Theorem 6.12.6** (Well-definedness of a Multiinfinite Graph)**.** *Let $m \in \mathbb{N}$, and for each $i \in \{1, \dots, m\}$, let $T_i$ be an infinite totally ordered set. Define*

$$\mathbf{T} := T_1 \times \cdots \times T_m$$

*with the componentwise relation*

$$(s_1, \dots, s_m) \preceq (t_1, \dots, t_m) \iff s_i \le t_i \text{ for all } i = 1, \dots, m.$$

*Then $(\mathbf{T}, \preceq)$ is a well-defined partially ordered set. Consequently, if*

$$\mathfrak{G} = \Big(\{G_{\mathbf{t}}\}_{\mathbf{t}\in\mathbf{T}}, \{\rho_{\mathbf{s},\mathbf{t}}\}_{\mathbf{s}\preceq\mathbf{t}}\Big)$$

*satisfies the conditions in the definition of a Multiinfinite Graph, then $\mathfrak{G}$ is a well-defined graph system indexed by several independent infinite axes.*

*Proof.* We first verify that $\preceq$ is a partial order on $\mathbf{T}$.

Reflexivity holds because, for every

$$\mathbf{t} = (t_1, \dots, t_m) \in \mathbf{T},$$

we have $t_i \le t_i$ for all $i$, hence $\mathbf{t} \preceq \mathbf{t}$.

Antisymmetry holds because, if

$$\mathbf{s} = (s_1, \dots, s_m) \preceq \mathbf{t} = (t_1, \dots, t_m) \quad \text{and} \quad \mathbf{t} \preceq \mathbf{s},$$

then $s_i \le t_i$ and $t_i \le s_i$ for all $i$. Since each $T_i$ is totally ordered, it follows that $s_i = t_i$ for all $i$, so $\mathbf{s} = \mathbf{t}$.

Transitivity holds because, if

$$\mathbf{r} = (r_1, \dots, r_m) \preceq \mathbf{s} = (s_1, \dots, s_m) \quad \text{and} \quad \mathbf{s} \preceq \mathbf{t} = (t_1, \dots, t_m),$$

then $r_i \le s_i \le t_i$ for all $i$, hence $r_i \le t_i$ for all $i$, so $\mathbf{r} \preceq \mathbf{t}$.

Therefore $(\mathbf{T}, \preceq)$ is a well-defined partially ordered set. Since each $T_i$ is infinite, the indexing system indeed has several independent infinite axes. Thus the family of graphs $\{G_{\mathbf{t}}\}_{\mathbf{t}\in\mathbf{T}}$ and the coherent transition homomorphisms $\rho_{\mathbf{s},\mathbf{t}}$ are well defined. □

**Theorem 6.12.7** (Well-definedness of a Multiinfinite $n$-SuperHyperGraph)**.** *Let $m \in \mathbb{N}$, and for each $i \in \{1, \dots, m\}$, let $T_i$ be an infinite totally ordered set. Let*

$$\mathbf{T} := T_1 \times \cdots \times T_m$$

*be equipped with the componentwise partial order $\preceq$. Let $V_0 \neq \varnothing$ be a base set, and define*

$$\mathcal{P}^0(V_0) := V_0, \qquad \mathcal{P}^{k+1}(V_0) := \mathcal{P}\big(\mathcal{P}^k(V_0)\big) \quad (k \ge 0).$$

*Fix $n \in \mathbb{N}_0$. Then, if*

$$\mathfrak{S}\mathfrak{H}^{(n)} = \Big(\{\mathrm{SHG}^{(n)}_{\mathbf{t}}\}_{\mathbf{t}\in\mathbf{T}}, \{\phi_{\mathbf{s},\mathbf{t}}\}_{\mathbf{s}\preceq\mathbf{t}}\Big)$$

*satisfies the conditions in the definition of a Multiinfinite $n$-SuperHyperGraph, it is a well-defined $m$-axis infinite family of $n$-SuperHyperGraphs.*

*Proof.* By Theorem 6.12.6, $(\mathbf{T}, \preceq)$ is a well-defined partially ordered set.

For each $\mathbf{t} \in \mathbf{T}$, the pair

$$\mathrm{SHG}^{(n)}_{\mathbf{t}} = (V_{\mathbf{t}}, E_{\mathbf{t}})$$

satisfies

$$V_{\mathbf{t}} \subseteq \mathcal{P}^n(V_0), \qquad E_{\mathbf{t}} \subseteq \mathcal{P}(V_{\mathbf{t}}) \setminus \{\varnothing\},$$

so $\mathrm{SHG}^{(n)}_{\mathbf{t}}$ is a well-defined $n$-SuperHyperGraph.

For each comparable pair $\mathbf{s} \preceq \mathbf{t}$, the map

$$\phi_{\mathbf{s},\mathbf{t}} : V_{\mathbf{s}} \to V_{\mathbf{t}}$$

is assumed to satisfy

$$\phi_{\mathbf{s},\mathbf{t}}[\varepsilon] = \{\phi_{\mathbf{s},\mathbf{t}}(v) : v \in \varepsilon\} \in E_{\mathbf{t}} \qquad \text{for every } \varepsilon \in E_{\mathbf{s}},$$

so each $\phi_{\mathbf{s},\mathbf{t}}$ is a well-defined superhypergraph homomorphism.

Finally, the identities

$$\phi_{\mathbf{t},\mathbf{t}} = \mathrm{id}_{V_{\mathbf{t}}}$$

and the coherence condition

$$\phi_{\mathbf{s},\mathbf{t}} \circ \phi_{\mathbf{r},\mathbf{s}} = \phi_{\mathbf{r},\mathbf{t}} \qquad (\mathbf{r} \preceq \mathbf{s} \preceq \mathbf{t})$$

show that the transition maps are compositionally consistent. Therefore $\mathfrak{S}\mathfrak{H}^{(n)}$ is well defined. □

## 6.13 Edge-Iterated SuperHyperGraph

An Edge-Iterated SuperHyperGraph is a superhypergraph whose vertices and edges are both generated through iterated powerset constructions, allowing hierarchical incidence at both levels.

**Definition 6.13.1** ($(n, m)$-SuperHyperGraph with iterated edge hierarchy)**.** Let $V_0 \neq \varnothing$ be a finite base set, and let $n, m \in \mathbb{N}_0$. An $(n, m)$-*SuperHyperGraph* on $V_0$ is a pair

$$\mathrm{SHG}^{(n,m)} = (V, E),$$

such that

$$V \subseteq \mathcal{P}^{*n}(V_0), \qquad E \subseteq \mathcal{P}^{*m}(V).$$

The elements of $V$ are called $n$-*supervertices*, and the elements of $E$ are called $m$-*superhyperedges*.

**Definition 6.13.2** (Flattened support of an $m$-superhyperedge)**.** Let $V \neq \varnothing$ be a set. Define recursively the maps

$$\mathrm{Flat}_0 : V \to V, \qquad \mathrm{Flat}_0(v) := v,$$

$$\mathrm{Flat}_1 : \mathcal{P}^{*1}(V) \to \mathcal{P}^{*1}(V), \qquad \mathrm{Flat}_1(A) := A,$$

and for $k \geq 1$,

$$\mathrm{Flat}_{k+1} : \mathcal{P}^{*(k+1)}(V) \to \mathcal{P}^{*1}(V), \qquad \mathrm{Flat}_{k+1}(\mathcal{A}) := \bigcup_{B \in \mathcal{A}} \mathrm{Flat}_k(B).$$

If

$$\mathrm{SHG}^{(n,m)} = (V, E)$$

is an $(n, m)$-SuperHyperGraph and $\varepsilon \in E$, then

$$\mathrm{supp}(\varepsilon) := \mathrm{Flat}_m(\varepsilon) \subseteq V$$

is called the *flattened support* (or simply the *support*) of $\varepsilon$. A supervertex $v \in V$ is said to be *incident* with $\varepsilon$ if

$$v \in \mathrm{supp}(\varepsilon).$$

**Remark 6.13.3.** *If $m = 1$, then*

$$E \subseteq \mathcal{P}^{*1}(V) = \mathcal{P}(V) \setminus \{\varnothing\},$$

*so the above definition reduces to the usual set-based $n$-SuperHyperGraph. Hence this notion extends the standard superhypergraph framework by allowing not only the vertex domain, but also the edge domain, to be generated through iterated powerset constructions.*

**Remark 6.13.4.** *If $n = 0$, then $V \subseteq V_0$, so the vertices are ordinary base elements while the edge objects still carry an $m$-fold iterated powerset hierarchy. If $m = 0$, then $E \subseteq V$, so edges degenerate to distinguished supervertices. Therefore the most interesting cases are typically $n \geq 1$ and $m \geq 1$.*

**Example 6.13.5** (A $(2, 2)$-SuperHyperGraph with iterated edge hierarchy)**.** Let

$$V_0 = \{a, b, c\}.$$

Consider the following elements of $\mathcal{P}^{*2}(V_0)$:

$$v_1 = \{\{a\}, \{a, b\}\}, \qquad v_2 = \{\{b\}, \{b, c\}\}, \qquad v_3 = \{\{c\}, \{a, c\}\}.$$

Set

$$V = \{v_1, v_2, v_3\} \subseteq \mathcal{P}^{*2}(V_0).$$

Thus, $V$ is a set of 2-supervertices.

Next, define the following nonempty subsets of $V$:

$$A_1 = \{v_1, v_2\}, \qquad A_2 = \{v_2, v_3\}, \qquad A_3 = \{v_1, v_3\}, \qquad A_4 = \{v_3\}.$$

Now define

$$\varepsilon_1 = \{A_1, A_2\}, \qquad \varepsilon_2 = \{A_3, A_4\}.$$

Since each $A_i$ is a nonempty subset of $V$, we have

$$\varepsilon_1, \varepsilon_2 \in \mathcal{P}^{*2}(V).$$

Hence, if we set

$$E = \{\varepsilon_1, \varepsilon_2\},$$

then

$$\mathrm{SHG}^{(2,2)} = (V, E)$$

is a $(2,2)$-SuperHyperGraph on $V_0$.

Let us compute the flattened supports. By definition,

$$\mathrm{supp}(\varepsilon_1) = \mathrm{Flat}_2(\varepsilon_1) = \mathrm{Flat}_2(\{A_1, A_2\}) = A_1 \cup A_2 = \{v_1, v_2\} \cup \{v_2, v_3\} = \{v_1, v_2, v_3\},$$

and similarly,

$$\mathrm{supp}(\varepsilon_2) = \mathrm{Flat}_2(\varepsilon_2) = \mathrm{Flat}_2(\{A_3, A_4\}) = A_3 \cup A_4 = \{v_1, v_3\} \cup \{v_3\} = \{v_1, v_3\}.$$

Therefore, all three 2-supervertices $v_1, v_2, v_3$ are incident with $\varepsilon_1$, whereas only $v_1$ and $v_3$ are incident with $\varepsilon_2$. This example illustrates that the edge objects themselves may carry a second-level set hierarchy, while their effective incidence is recovered through the flattening operation.

Similarly, the HyperCube can also be extended. An Edge-Iterated HyperCube extends a hypercube by allowing hierarchical edge objects built from ordinary cube edges, with incidence recovered by recursively flattening their endpoints.

**Definition 6.13.6** (Edge-Iterated HyperCube)**.** Let

$$Q^{(n)}_{\mathrm{SH}}(H) = (V, E_0)$$

be an $n$-SuperHyperCube, where each ordinary edge is identified with its two-element endpoint set:

$$E_0 \subseteq \mathcal{P}(V), \qquad f \in E_0 \Rightarrow |f| = 2.$$

Fix $m \in \mathbb{N}_0$, and define recursively the flattening maps

$$\mathrm{Flat}_0(e) := \{e\} \qquad (e \in E_0),$$

and for $k \geq 0$,

$$\mathrm{Flat}_{k+1}(X) := \bigcup_{x \in X} \mathrm{Flat}_k(x) \qquad \big(X \in \mathcal{P}^{k+1}(E_0)\big).$$

Hence, for every $e \in \mathcal{P}^m(E_0)$, one has

$$\mathrm{Flat}_m(e) \subseteq E_0.$$

An *edge-iterated hypercube of edge-depth* $m$ over $H$ is a triple

$$Q^{(n,m)}_{\mathrm{EIH}}(H) = (V, E, \partial),$$

where

$$E \subseteq \mathcal{P}^m(E_0)$$

is a finite set of higher-level edge objects, and

$$\partial : E \to \mathcal{P}^*(V)$$

is defined by

$$\partial(e) := \bigcup_{f \in \mathrm{Flat}_m(e)} f \qquad (e \in E).$$

Therefore, an edge of $Q^{(n,m)}_{\mathrm{EIH}}(H)$ may itself be a hierarchical object built from ordinary cube edges, and its incidence is obtained by flattening it to the underlying ordinary edges and collecting their endpoints.

**Remark 6.13.7.** *If $m = 0$ and $E = E_0$, then*

$$Q^{(n,0)}_{\mathrm{EIH}}(H)$$

*reduces to the underlying $n$-SuperHyperCube. Thus, $Q^{(n,m)}_{\mathrm{EIH}}(H)$ extends the hypercube structure by allowing hierarchy not only on the vertex side but also on the edge side.*

## 6.14 Recursive MetaGraph and Recursive Iterated MetaGraph

A Recursive MetaGraph is a metagraph whose vertices are graphs and whose edges may recursively contain lower-level metaedges, enabling nested graph-of-graphs relations. A Recursive Iterated MetaGraph is an iterated metagraph whose vertices are higher-level metagraphs and whose recursive edges encode nested lifted relations recursively.

**Definition 6.14.1** (Base metaedge set)**.** Let $\mathfrak{G}$ be a nonempty universe of finite graphs, and let

$$\mathcal{R} \subseteq \mathcal{P}(\mathfrak{G} \times \mathfrak{G})$$

be a nonempty family of admissible binary relations on $\mathfrak{G}$. For a finite nonempty set $V \subseteq \mathfrak{G}$, define the *base metaedge set* by

$$\mathcal{E}_0(V, \mathcal{R}) := \{(X, Y, R) \in V \times V \times \mathcal{R} : (X, Y) \in R\}.$$

Its elements are called *ordinary metaedges.*

**Definition 6.14.2** (Recursive metaedge hierarchy of depth $k$)**.** Let $V \subseteq \mathfrak{G}$ be finite and nonempty. Define recursively

$$\mathcal{E}_0 := \mathcal{E}_0(V, \mathcal{R}),$$

and for $i \geq 0$,

$$\mathcal{E}_{i+1} := \mathcal{P}\Big(\bigcup_{j=0}^{i} \mathcal{E}_j\Big) \setminus \{\varnothing\}.$$

The set

$$\mathcal{E}_{\leq k} := \bigcup_{i=0}^{k} \mathcal{E}_i$$

is called the *recursive metaedge universe of depth $k$*. Elements of $\mathcal{E}_i$ for $i \geq 1$ are called *recursive metaedges of level $i$.*

**Definition 6.14.3** (Recursive MetaGraph of recursion depth $k$)**.** Let $\mathfrak{G}$ and $\mathcal{R}$ be as above, and let $k \in \mathbb{N} \cup \{0\}$. A *Recursive MetaGraph of recursion depth $k$* over $(\mathfrak{G}, \mathcal{R})$ is a pair

$$\mathcal{M}^{[k]} = (V, E^{\mathrm{rec}})$$

such that

- $V \subseteq \mathfrak{G}$ is a finite nonempty set;
- $$E^{\mathrm{rec}} \subseteq \mathcal{E}_{\leq k}.$$

Thus, the vertices of $\mathcal{M}^{[k]}$ are graphs, while its recursive metaedges may be ordinary labeled relations between graph-vertices or recursively nested collections of lower-level metaedges.

**Definition 6.14.4** (Leaf set of a recursive metaedge)**.** Let $\eta \in \mathcal{E}_{\leq k}$. Its *leaf set* $\mathrm{Leaf}(\eta) \subseteq \mathcal{E}_0$ is defined recursively by

$$\mathrm{Leaf}(\eta) := \{\eta\} \qquad \text{if } \eta \in \mathcal{E}_0,$$

and

$$\mathrm{Leaf}(\eta) := \bigcup_{\xi \in \eta} \mathrm{Leaf}(\xi) \qquad \text{if } \eta \in \mathcal{E}_i,\ i \geq 1.$$

The *flattened source support*, *flattened target support*, and *flattened label support* of $\eta$ are

$$s^{\flat}(\eta) := \{X \in V : \exists\, Y, R \text{ with } (X, Y, R) \in \mathrm{Leaf}(\eta)\},$$

$$t^{\flat}(\eta) := \{Y \in V : \exists\, X, R \text{ with } (X, Y, R) \in \mathrm{Leaf}(\eta)\},$$

and

$$\lambda^{\flat}(\eta) := \{R \in \mathcal{R} : \exists\, X, Y \text{ with } (X, Y, R) \in \mathrm{Leaf}(\eta)\}.$$

**Remark 6.14.5.** *When $k = 0$, every recursive metaedge is an ordinary metaedge, so a Recursive MetaGraph reduces to an ordinary Meta-Graph with graph-vertices and labeled relations between them.*

**Definition 6.14.6** ($(t,k)$-Recursive Iterated MetaGraph)**.** Let $t \in \mathbb{N}$ and let $\mathfrak{G}^{(t)}$ be a nonempty universe of Iterated Meta-Graphs of depth $t$. Let

$$\mathcal{R}^{(t)} \subseteq \mathcal{P}\big(\mathfrak{G}^{(t)} \times \mathfrak{G}^{(t)}\big)$$

be a nonempty family of admissible lifted relations on $\mathfrak{G}^{(t)}$. For a finite nonempty set $V^{(t)} \subseteq \mathfrak{G}^{(t)}$, define

$$\mathcal{E}_0^{(t)} := \big\{(X,Y,R) \in V^{(t)} \times V^{(t)} \times \mathcal{R}^{(t)} : (X,Y) \in R\big\},$$

and recursively

$$\mathcal{E}_{i+1}^{(t)} := \mathcal{P}\Big(\bigcup_{j=0}^{i} \mathcal{E}_j^{(t)}\Big) \setminus \{\varnothing\}.$$

Set

$$\mathcal{E}_{\le k}^{(t)} := \bigcup_{i=0}^{k} \mathcal{E}_i^{(t)}.$$

A $(t,k)$-*Recursive Iterated MetaGraph* is a pair

$$\mathcal{M}^{(t)[k]} = (V^{(t)}, E^{(t),\mathrm{rec}})$$

such that

- $V^{(t)} \subseteq \mathfrak{G}^{(t)}$ is a finite nonempty set of depth-$t$ iterated metagraphs;
- $$E^{(t),\mathrm{rec}} \subseteq \mathcal{E}_{\le k}^{(t)}.$$

Hence a $(t,k)$-Recursive Iterated MetaGraph is an iterated metagraph whose recursive metaedges may contain lower-level lifted metaedges up to recursion depth $k$.

**Example 6.14.7** (A $(1,2)$-Recursive Iterated MetaGraph)**.** Let $t = 1$, and let $X, Y, Z \in \mathfrak{G}^{(1)}$ be three depth-1 Iterated Meta-Graphs. Assume that $\mathcal{R}^{(1)}$ contains a lifted relation $R$ such that

$$(X,Y) \in R, \qquad (Y,Z) \in R, \qquad (X,Z) \notin R.$$

Set

$$V^{(1)} = \{X, Y, Z\}.$$

Then the base-level lifted metaedges

$$e_{XY} := (X,Y,R), \qquad e_{YZ} := (Y,Z,R)$$

belong to

$$\mathcal{E}_0^{(1)}.$$

Now define

$$\eta := \{e_{XY}, e_{YZ}\}.$$

Since $\eta$ is a nonempty subset of $\mathcal{E}_0^{(1)}$, we have

$$\eta \in \mathcal{E}_1^{(1)}.$$

Next define

$$\theta := \{e_{XY}, \eta\}.$$

Because

$$e_{XY} \in \mathcal{E}_0^{(1)} \qquad \text{and} \qquad \eta \in \mathcal{E}_1^{(1)},$$

it follows that

$$\theta \subseteq \mathcal{E}_0^{(1)} \cup \mathcal{E}_1^{(1)}, \qquad \theta \neq \varnothing,$$

and therefore

$$\theta \in \mathcal{E}_2^{(1)}.$$

Hence, if we set

$$E^{(1),\mathrm{rec}} := \{e_{XY}, e_{YZ}, \eta, \theta\},$$

then

$$\mathcal{M}^{(1)[2]} = (V^{(1)}, E^{(1),\mathrm{rec}})$$

is a $(1,2)$-Recursive Iterated MetaGraph.

In this example, $e_{XY}$ and $e_{YZ}$ are ordinary lifted metaedges, $\eta$ is a first-level recursive metaedge containing lower-level metaedges, and $\theta$ is a second-level recursive metaedge containing both a base metaedge and a recursive metaedge. Thus, the example explicitly illustrates recursive metaedge nesting up to depth $k = 2$.

**Theorem 6.14.8** (Well-definedness of Recursive MetaGraphs)**.** *Let $\mathfrak{G}$ be a nonempty universe of finite graphs, let*

$$\mathcal{R} \subseteq \mathcal{P}(\mathfrak{G} \times \mathfrak{G})$$

*be a nonempty family of admissible binary relations on $\mathfrak{G}$, and let $V \subseteq \mathfrak{G}$ be a finite nonempty set. For $k \in \mathbb{N} \cup \{0\}$, define*

$$\mathcal{E}_0 := \big\{(X, Y, R) \in V \times V \times \mathcal{R} : (X, Y) \in R\big\},$$

*and recursively*

$$\mathcal{E}_{i+1} := \mathcal{P}\Big(\bigcup_{j=0}^{i} \mathcal{E}_j\Big) \setminus \{\varnothing\} \qquad (i \geq 0).$$

*Then each $\mathcal{E}_i$ is well defined, and hence*

$$\mathcal{E}_{\leq k} := \bigcup_{i=0}^{k} \mathcal{E}_i$$

*is well defined. Consequently, if*

$$E^{\mathrm{rec}} \subseteq \mathcal{E}_{\leq k},$$

*then*

$$\mathcal{M}^{[k]} = (V, E^{\mathrm{rec}})$$

*is a well-defined Recursive MetaGraph of recursion depth $k$.*

*Proof.* Since $V$ is finite and nonempty, the product

$$V \times V \times \mathcal{R}$$

is a well-defined class, and therefore

$$\mathcal{E}_0 = \big\{(X, Y, R) \in V \times V \times \mathcal{R} : (X, Y) \in R\big\}$$

is well defined.

Assume inductively that $\mathcal{E}_0, \ldots, \mathcal{E}_i$ are well defined. Then

$$\bigcup_{j=0}^{i} \mathcal{E}_j$$

is well defined, so its powerset is well defined, and therefore

$$\mathcal{E}_{i+1} = \mathcal{P}\Big(\bigcup_{j=0}^{i} \mathcal{E}_j\Big) \setminus \{\varnothing\}$$

is well defined. Hence, by induction, every $\mathcal{E}_i$ is well defined.

It follows that

$$\mathcal{E}_{\leq k} = \bigcup_{i=0}^{k} \mathcal{E}_i$$

is well defined, and any subset

$$E^{\mathrm{rec}} \subseteq \mathcal{E}_{\leq k}$$

is also well defined. Therefore

$$\mathcal{M}^{[k]} = (V, E^{\mathrm{rec}})$$

is a well-defined Recursive MetaGraph. □

**Theorem 6.14.9** (Well-definedness of Recursive Iterated MetaGraphs)**.** *Let $t \in \mathbb{N}$, let $\mathfrak{G}^{(t)}$ be a nonempty universe of Iterated MetaGraphs of depth $t$, let*

$$\mathcal{R}^{(t)} \subseteq \mathcal{P}\big(\mathfrak{G}^{(t)} \times \mathfrak{G}^{(t)}\big)$$

*be a nonempty family of admissible lifted relations, and let $V^{(t)} \subseteq \mathfrak{G}^{(t)}$ be a finite nonempty set. For $k \in \mathbb{N} \cup \{0\}$, define*

$$\mathcal{E}_0^{(t)} := \big\{(X, Y, R) \in V^{(t)} \times V^{(t)} \times \mathcal{R}^{(t)} : (X, Y) \in R\big\},$$

*and recursively*

$$\mathcal{E}_{i+1}^{(t)} := \mathcal{P}\Big(\bigcup_{j=0}^{i} \mathcal{E}_j^{(t)}\Big) \setminus \{\varnothing\} \qquad (i \geq 0).$$

*Then each $\mathcal{E}_i^{(t)}$ is well defined, and hence*

$$\mathcal{E}_{\leq k}^{(t)} := \bigcup_{i=0}^{k} \mathcal{E}_i^{(t)}$$

*is well defined. Consequently, if*

$$E^{(t),\mathrm{rec}} \subseteq \mathcal{E}_{\leq k}^{(t)},$$

*then*

$$\mathcal{M}^{(t)[k]} = (V^{(t)}, E^{(t),\mathrm{rec}})$$

*is a well-defined $(t, k)$-Recursive Iterated MetaGraph.*

*Proof.* The proof is identical to that of Theorem 6.14.8, replacing $\mathfrak{G}, \mathcal{R}, V, \mathcal{E}_i$ with $\mathfrak{G}^{(t)}, \mathcal{R}^{(t)}, V^{(t)}, \mathcal{E}_i^{(t)}$, respectively.

In particular,

$$\mathcal{E}_0^{(t)} = \big\{(X, Y, R) \in V^{(t)} \times V^{(t)} \times \mathcal{R}^{(t)} : (X, Y) \in R\big\}$$

is well defined, and if $\mathcal{E}_0^{(t)}, \ldots, \mathcal{E}_i^{(t)}$ are well defined, then so is

$$\mathcal{E}_{i+1}^{(t)} = \mathcal{P}\Big(\bigcup_{j=0}^{i} \mathcal{E}_j^{(t)}\Big) \setminus \{\varnothing\}.$$

Thus every $\mathcal{E}_i^{(t)}$ is well defined, hence so is

$$\mathcal{E}_{\leq k}^{(t)} = \bigcup_{i=0}^{k} \mathcal{E}_i^{(t)}.$$

Therefore any subset

$$E^{(t),\mathrm{rec}} \subseteq \mathcal{E}_{\leq k}^{(t)}$$

is well defined, and

$$\mathcal{M}^{(t)[k]} = (V^{(t)}, E^{(t),\mathrm{rec}})$$

is a well-defined $(t, k)$-Recursive Iterated MetaGraph. □

## 6.15 Multiuniverse-Graph

A Multiuniverse-Graph is a graph whose vertices belong to several tagged universes, allowing both intra-universe edges and inter-universe edges while preserving identity of each universe.

**Definition 6.15.1** (Multiuniverse-Graph)**.** Let $m \in \mathbb{N}$, and let

$$\mathcal{U} = \{U_1, U_2, \ldots, U_m\}$$

be a finite family of nonempty sets, called *universes*. For each $i \in \{1, 2, \ldots, m\}$, define the tagged copy

$$\widetilde{U}_i := \{(i, x) : x \in U_i\},$$

and set

$$U^{\sqcup} := \bigcup_{i=1}^{m} \widetilde{U}_i.$$

Thus, elements coming from different universes remain distinguished even if their underlying labels coincide.

A *Multiuniverse-Graph* over $\mathcal{U}$ is a pair

$$G_{\mathrm{MU}} = (V, E),$$

such that

$$V \subseteq U^{\sqcup}$$

is a finite set of vertices, and

$$E \subseteq \{\{u, v\} \subseteq V : u \neq v\}.$$

For each $i$, the set

$$V_i := V \cap \widetilde{U}_i$$

is called the *$i$-th universe layer* of $G_{\mathrm{MU}}$.

An edge $\{u, v\} \in E$ is called an *intra-universe edge* if $u, v \in V_i$ for some $i$, and an *inter-universe edge* otherwise.

## 6.16 Multiunion-Graph

A Multiunion-Graph is a graph whose vertices are unions of elements drawn from multiple tagged universes, with edges typically defined by overlap or support relations.

**Definition 6.16.1** (Multiunion-Graph)**.** Let $\mathcal{U} = \{U_1, U_2, \ldots, U_m\}$ and $U^{\sqcup}$ be as above.

A *union-vertex* over $\mathcal{U}$ is a nonempty subset

$$A \subseteq U^{\sqcup}$$

for which there exists a nonempty index set

$$I_A \subseteq \{1, 2, \ldots, m\}$$

and nonempty subsets

$$A_i \subseteq \widetilde{U}_i \qquad (i \in I_A)$$

such that

$$A = \bigcup_{i \in I_A} A_i.$$

The set

$$\mathrm{supp}_{\mathcal{U}}(A) := \{\, i \in \{1, 2, \ldots, m\} : A \cap \widetilde{U}_i \neq \varnothing \,\}$$

is called the *universe support* of $A$.

A *Multiunion-Graph* over $\mathcal{U}$ is a pair

$$G_{\cup} = (\mathcal{V}, \mathcal{E}),$$

such that

$$\mathcal{V}$$

is a finite nonempty set of union-vertices over $\mathcal{U}$, and

$$\mathcal{E} \subseteq \{\{A, B\} \subseteq \mathcal{V} : A \neq B\}.$$

The elements of $\mathcal{V}$ are called *union-vertices.*

A particularly natural canonical choice of adjacency is the *overlap rule*

$$\{A, B\} \in \mathcal{E} \iff A \cap B \neq \varnothing.$$

**Remark 6.16.2.** *If $m = 1$, then a Multiuniverse-Graph reduces to an ordinary graph on a single universe. If every union-vertex $A \in \mathcal{V}$ of a Multiunion-Graph is a singleton, then the Multiunion-Graph reduces naturally to a Multiuniverse-Graph. Hence the Multiunion-Graph may be viewed as a union-level extension of the Multiuniverse-Graph.*

**Example 6.16.3** (A Multiuniverse-Graph)**.** Let

$$U_1 = \{a, b\}, \qquad U_2 = \{a, c\}.$$

Their tagged copies are

$$\widetilde{U}_1 = \{(1, a), (1, b)\}, \qquad \widetilde{U}_2 = \{(2, a), (2, c)\}.$$

Set

$$V = \{(1, a), (1, b), (2, a), (2, c)\},$$

and define

$$E = \Big\{\{(1, a), (1, b)\}, \{(2, a), (2, c)\}, \{(1, b), (2, a)\}\Big\}.$$

Then

$$G_{\mathrm{MU}} = (V, E)$$

is a Multiuniverse-Graph. Here

$$\{(1, a), (1, b)\} \quad \text{and} \quad \{(2, a), (2, c)\}$$

are intra-universe edges, while

$$\{(1, b), (2, a)\}$$

is an inter-universe edge. Thus, the graph connects vertices both within and across the two universes.

**Example 6.16.4** (A Multiunion-Graph)**.** Let

$$U_1 = \{a, b\}, \qquad U_2 = \{c, d\},$$

with tagged copies

$$\widetilde{U}_1 = \{(1, a), (1, b)\}, \qquad \widetilde{U}_2 = \{(2, c), (2, d)\}.$$

Define the following union-vertices:

$$A = \{(1, a), (2, c)\}, \qquad B = \{(1, a), (1, b)\}, \qquad C = \{(2, d)\}.$$

Set

$$\mathcal{V} = \{A, B, C\}.$$

Using the overlap rule, define

$$\mathcal{E} = \{\{A, B\}\},$$

since

$$A \cap B = \{(1, a)\} \neq \varnothing,$$

whereas

$$A \cap C = \varnothing, \qquad B \cap C = \varnothing.$$

Then

$$G_{\cup} = (\mathcal{V}, \mathcal{E})$$

is a Multiunion-Graph. In this example, the union-vertices $A$ and $B$ are adjacent because they overlap, while $C$ is isolated.

## 6.17 Iterated SuperGraph

An Iterated SuperGraph is a recursively defined directed graph whose vertices are lower-level supergraphs, and whose edges encode proper supergraph or subdigraph containment relations.

**Definition 6.17.1** (Proper subgraph)**.** Let

$$G = (V_G, E_G), \qquad H = (V_H, E_H)$$

be finite simple graphs. We say that $H$ is a *proper subgraph* of $G$, and write

$$H \subsetneq_{\mathrm{gr}} G,$$

if

$$V_H \subseteq V_G, \qquad E_H \subseteq E_G \cap \binom{V_H}{2},$$

and

$$(V_H, E_H) \neq (V_G, E_G).$$

Equivalently, $G$ is a *proper supergraph* of $H$.

**Definition 6.17.2** (Proper subdigraph)**.** Let

$$D = (V_D, A_D), \qquad D' = (V_{D'}, A_{D'})$$

be finite directed graphs. We say that $D'$ is a *proper subdigraph* of $D$, and write

$$D' \lneq_{\mathrm{dg}} D,$$

if

$$V_{D'} \subseteq V_D, \qquad A_{D'} \subseteq A_D \cap (V_{D'} \times V_{D'}),$$

and

$$(V_{D'}, A_{D'}) \neq (V_D, A_D).$$

**Definition 6.17.3** (SuperGraph)**.** Let $\mathcal{U}_0$ be a nonempty universe of finite simple graphs, and let

$$\mathcal{F} \subseteq \mathcal{U}_0$$

be a finite nonempty set. The *SuperGraph induced by* $\mathcal{F}$ is the directed graph

$$\mathrm{SG}(\mathcal{F}) := (\mathcal{F}, A_{\mathcal{F}}),$$

where

$$A_{\mathcal{F}} := \big\{(G, H) \in \mathcal{F} \times \mathcal{F} : H \lneq_{\mathrm{gr}} G\big\}.$$

Thus, there is a directed edge

$$G \to H$$

precisely when $G$ is a proper supergraph of $H$.

**Definition 6.17.4** (Iterated SuperGraph universe)**.** Let $\mathcal{U}_0$ be a nonempty universe of finite simple graphs. Define recursively:

$$\mathcal{U}_1 := \big\{\mathrm{SG}(\mathcal{F}) : \varnothing \neq \mathcal{F} \subseteq \mathcal{U}_0,\ \mathcal{F} \text{ finite}\big\},$$

and for each $n \geq 1$,

$$\mathcal{U}_{n+1} := \big\{(V, A) : V \subseteq \mathcal{U}_n,\ V \neq \varnothing,\ V \text{ finite},\ A = \{(X, Y) \in V \times V : Y \lneq_{\mathrm{dg}} X\}\big\}.$$

**Definition 6.17.5** (Iterated SuperGraph of depth $n$)**.** Let $n \in \mathbb{N}$. An *Iterated SuperGraph of depth* $n$ is any element of $\mathcal{U}_n$.

In particular:

- a depth-1 Iterated SuperGraph is an ordinary SuperGraph;
- a depth-2 Iterated SuperGraph is a *SuperGraph of SuperGraphs*.

**Remark 6.17.6.** *This construction is recursive in the following sense:*

- *level* 0 *objects are ordinary finite simple graphs;*
- *level* 1 *vertices are graphs, and edges encode the proper-supergraph relation;*
- *level* 2 *vertices are SuperGraphs, and edges encode the proper-superdigraph relation between them;*
- *higher levels are obtained by repeating the same rule.*

**Theorem 6.17.7** (Well-definedness of Iterated SuperGraphs)**.** *Let* $\mathcal{U}_0$ *be a nonempty* set *of finite simple graphs. Define*

$$\mathcal{U}_1 := \big\{\mathrm{SG}(\mathcal{F}) : \varnothing \neq \mathcal{F} \subseteq \mathcal{U}_0,\ \mathcal{F} \text{ finite}\big\},$$

*where*

$$\mathrm{SG}(\mathcal{F}) = (\mathcal{F}, A_{\mathcal{F}}), \qquad A_{\mathcal{F}} := \big\{(G, H) \in \mathcal{F} \times \mathcal{F} : H \lneq_{\mathrm{gr}} G\big\}.$$

*For* $n \geq 1$, *define recursively*

$$\mathcal{U}_{n+1} := \Big\{(V, A) : \varnothing \neq V \subseteq \mathcal{U}_n,\ V \text{ finite},\ A = \{(X, Y) \in V \times V : Y \lneq_{\mathrm{dg}} X\}\Big\}.$$

*Then, for every* $n \geq 1$,

- *$\mathcal{U}_n$ is well defined;*
- *every element of $\mathcal{U}_n$ is a finite directed graph.*

*Hence the notion of an Iterated SuperGraph of depth $n$ is well defined.*

*Proof.* We proceed by induction on $n$.

For $n = 1$, let $\varnothing \neq \mathcal{F} \subseteq \mathcal{U}_0$ be finite. Then $\mathcal{F}$ is a finite set of finite simple graphs, so

$$A_{\mathcal{F}} = \{(G, H) \in \mathcal{F} \times \mathcal{F} : H \lneq_{\mathrm{gr}} G\}$$

is a well-defined subset of $\mathcal{F} \times \mathcal{F}$. Therefore

$$\mathrm{SG}(\mathcal{F}) = (\mathcal{F}, A_{\mathcal{F}})$$

is a well-defined finite directed graph. Hence $\mathcal{U}_1$ is well defined.

Assume now that $\mathcal{U}_n$ is well defined and that every element of $\mathcal{U}_n$ is a finite directed graph. Let $\varnothing \neq V \subseteq \mathcal{U}_n$ be finite. Since each element of $V$ is a finite directed graph, the proper subdigraph relation

$$Y \lneq_{\mathrm{dg}} X$$

is well defined for all $X, Y \in V$. Thus

$$A = \{(X, Y) \in V \times V : Y \lneq_{\mathrm{dg}} X\}$$

is a well-defined subset of $V \times V$. Therefore

$$(V, A)$$

is a well-defined finite directed graph. Hence $\mathcal{U}_{n+1}$ is well defined.

By induction, the conclusion holds for every $n \geq 1$. □

## 6.18 HyperTensor and SuperHyperTensor Network Graphs

An ordinary tensor network graph assigns a scalar-valued tensor to each vertex and identifies tensor indices along the edges of an underlying graph. A HyperTensor Network Graph replaces each scalar tensor entry by a set of possible scalar values. A SuperHyperTensor Network Graph further replaces each tensor entry by an element of an iterated powerset, thereby permitting nested sets of possible values.

Throughout this section, let $\mathbb{K}$ be a commutative field. The same definitions remain meaningful over a commutative semiring.

**Definition 6.18.1** (Iterated powerset hierarchy)**.** Define the iterated powersets of $\mathbb{K}$ recursively by

$$\mathscr{P}_0(\mathbb{K}) := \mathbb{K}$$

and

$$\mathscr{P}_{r+1}(\mathbb{K}) := \mathcal{P}(\mathscr{P}_r(\mathbb{K})) \qquad (r \in \mathbb{N}_0).$$

Thus,

$$\mathscr{P}_1(\mathbb{K}) = \mathcal{P}(\mathbb{K}),$$

$$\mathscr{P}_2(\mathbb{K}) = \mathcal{P}(\mathcal{P}(\mathbb{K})),$$

and so forth.

**Definition 6.18.2** (Nonempty iterated powerset hierarchy)**.** If empty values are to be excluded at every level, define

$$\mathscr{P}_{*,0}(\mathbb{K}) := \mathbb{K}$$

and

$$\mathscr{P}_{*,r+1}(\mathbb{K}) := \mathcal{P}_*(\mathscr{P}_{*,r}(\mathbb{K})),$$

where

$$\mathcal{P}_*(X) := \mathcal{P}(X) \setminus \{\emptyset\}.$$

**Definition 6.18.3** (Lifted addition and multiplication)**.** For every $r \in \mathbb{N}_0$, define binary operations

$$\boxplus_r, \boxtimes_r : \mathscr{P}_r(\mathbb{K}) \times \mathscr{P}_r(\mathbb{K}) \longrightarrow \mathscr{P}_r(\mathbb{K})$$

recursively as follows.

At depth 0, set

$$a \boxplus_0 b := a + b, \qquad a \boxtimes_0 b := ab$$

for all $a, b \in \mathbb{K}$.

Suppose that $\boxplus_r$ and $\boxtimes_r$ have been defined. For

$$A, B \in \mathscr{P}_{r+1}(\mathbb{K}),$$

define

$$A \boxplus_{r+1} B := \{a \boxplus_r b : a \in A,\ b \in B\}$$

and

$$A \boxtimes_{r+1} B := \{a \boxtimes_r b : a \in A,\ b \in B\}.$$

**Proposition 6.18.4.** *For every $r \in \mathbb{N}_0$, the operations*

$$\boxplus_r \qquad \text{and} \qquad \boxtimes_r$$

*are associative and commutative.*

*Proof.* We proceed by induction on $r$.

For $r = 0$, the assertion follows from the associativity and commutativity of addition and multiplication in $\mathbb{K}$.

Assume that both operations are associative and commutative at depth $r$. Let

$$A, B, C \in \mathscr{P}_{r+1}(\mathbb{K}).$$

Then

$$\begin{aligned}(A \boxplus_{r+1} B) \boxplus_{r+1} C &= \{(a \boxplus_r b) \boxplus_r c : a \in A,\ b \in B,\ c \in C\} \\ &= \{a \boxplus_r (b \boxplus_r c) : a \in A,\ b \in B,\ c \in C\} \\ &= A \boxplus_{r+1} (B \boxplus_{r+1} C).\end{aligned}$$

Commutativity follows similarly. The same argument applies to $\boxtimes_{r+1}$. Therefore, both lifted operations are associative and commutative at every depth. □

**Remark 6.18.5.** *Proposition 6.18.4 ensures that finite expressions such as*

$$\mathop{\boxplus_r}_{j \in J} A_j \qquad \text{and} \qquad \mathop{\boxtimes_r}_{j \in J} A_j$$

*are independent of the ordering and parenthesization whenever $J$ is a finite nonempty index set.*

**Definition 6.18.6** (HyperTensor)**.** Let $m \in \mathbb{N}_0$, and let

$$I_1, I_2, \ldots, I_m$$

be finite nonempty index sets. A *HyperTensor over* $\mathbb{K}$ of shape

$$\big(|I_1|, \ldots, |I_m|\big)$$

is a mapping

$$\mathcal{T} : \prod_{j=1}^{m} I_j \longrightarrow \mathcal{P}(\mathbb{K}).$$

Equivalently, $\mathcal{T}$ is an $m$-way array

$$\mathcal{T} = (\mathcal{T}_{i_1 \cdots i_m}),$$

where

$$\mathcal{T}_{i_1 \cdots i_m} \subseteq \mathbb{K}$$

for every

$$(i_1, \ldots, i_m) \in \prod_{j=1}^{m} I_j.$$

It is called a *nonempty HyperTensor* if

$$\mathcal{T}_{i_1 \cdots i_m} \neq \emptyset$$

for every multi-index.

**Definition 6.18.7** (Depth-$r$ SuperHyperTensor)**.** Let $r \in \mathbb{N}$. A *depth-r SuperHyperTensor over* $\mathbb{K}$ with index sets $I_1, \ldots, I_m$ is a mapping

$$\mathcal{T}^{(r)} : \prod_{j=1}^{m} I_j \longrightarrow \mathscr{P}_r(\mathbb{K}).$$

Equivalently,

$$\mathcal{T}^{(r)} = \left(\mathcal{T}^{(r)}_{i_1 \cdots i_m}\right)$$

is an $m$-way array whose entries satisfy

$$\mathcal{T}^{(r)}_{i_1 \cdots i_m} \in \mathcal{P}^r(\mathbb{K}).$$

It is called a *nonempty depth-r SuperHyperTensor* if

$$\mathcal{T}^{(r)}_{i_1 \cdots i_m} \in \mathscr{P}_{*,r}(\mathbb{K})$$

for every multi-index.

**Remark 6.18.8.** *A depth-*1 *SuperHyperTensor is precisely a HyperTensor, since*

$$\mathscr{P}_1(\mathbb{K}) = \mathcal{P}(\mathbb{K}).$$

*If depth* 0 *is admitted, then*

$$\mathscr{P}_0(\mathbb{K}) = \mathbb{K},$$

*and a depth-*0 *SuperHyperTensor is an ordinary* $\mathbb{K}$*-valued tensor.*

The half-edge formulation is used below because it treats internal edges, parallel edges, loops, and dangling external legs uniformly.

**Definition 6.18.9** (Finite network graph)**.** A finite *network graph* is a quadruple

$$G = (V, H, \iota, \tau),$$

where:

1. $V$ is a finite nonempty vertex set;
2. $H$ is a finite set whose elements are called *half-edges*;
3. $$\iota : H \longrightarrow V$$
   is an attachment map;
4. $$\tau : H \longrightarrow H$$
   is an involution:
   $$\tau \circ \tau = \mathrm{id}_H .$$

A two-element orbit

$$q = \{h, \tau(h)\}, \qquad h \neq \tau(h),$$

is called an *internal edge*. A fixed point

$$\tau(h) = h$$

is called an *external half-edge* or a *dangling leg*.

Let

$$Q := H/\langle \tau \rangle$$

be the set of $\tau$-orbits, and write

$$Q_{\text{int}} := \{q \in Q : |q| = 2\},$$
$$Q_{\text{ext}} := \{q \in Q : |q| = 1\}.$$

For each $v \in V$, define

$$H(v) := \iota^{-1}(\{v\}).$$

**Remark 6.18.10.** *If an internal orbit*

$$q = \{h, h'\}$$

*satisfies*

$$\iota(h) = \iota(h'),$$

*then $q$ represents a loop. Distinct internal orbits with the same pair of endpoints represent parallel edges.*

**Definition 6.18.11** (HyperTensor Network Graph)**.** Let

$$G = (V, H, \iota, \tau)$$

be a finite network graph, and let

$$Q = H/\langle\tau\rangle.$$

For every orbit $q \in Q$, let $I_q$ be a finite nonempty index set.

A *HyperTensor Network Graph over* $\mathbb{K}$ is a tuple

$$\mathcal{N}_{\mathrm{HT}} = \left(G, (I_q)_{q\in Q}, (\mathcal{T}_v)_{v\in V}\right),$$

where every vertex $v \in V$ is assigned a local HyperTensor

$$\mathcal{T}_v : \prod_{h\in H(v)} I_{[h]} \longrightarrow \mathcal{P}(\mathbb{K}).$$

Here

$$[h] \in Q$$

denotes the $\tau$-orbit containing $h$.

Thus, for each local index assignment

$$\boldsymbol{a}_v = (a_{[h]})_{h\in H(v)},$$

the value

$$\mathcal{T}_v(\boldsymbol{a}_v) \subseteq \mathbb{K}$$

is a set of possible local tensor values.

**Definition 6.18.12** (Contraction of a HyperTensor Network Graph)**.** Let

$$\mathcal{N}_{\mathrm{HT}} = \left(G, (I_q)_{q\in Q}, (\mathcal{T}_v)_{v\in V}\right)$$

be a HyperTensor Network Graph.

Fix an external index assignment

$$\boldsymbol{a}_{\mathrm{ext}} = (a_q)_{q\in Q_{\mathrm{ext}}} \in \prod_{q\in Q_{\mathrm{ext}}} I_q.$$

For each internal assignment

$$\boldsymbol{a}_{\mathrm{int}} = (a_q)_{q\in Q_{\mathrm{int}}} \in \prod_{q\in Q_{\mathrm{int}}} I_q,$$

combine the two assignments into

$$\boldsymbol{a} = (a_q)_{q\in Q}.$$

The contraction of $\mathcal{N}_{\mathrm{HT}}$ is the HyperTensor

$$\mathrm{Contr}_{\mathrm{HT}}(\mathcal{N}_{\mathrm{HT}}) : \prod_{q\in Q_{\mathrm{ext}}} I_q \longrightarrow \mathcal{P}(\mathbb{K})$$

defined by

$$\begin{aligned} &\mathrm{Contr}_{\mathrm{HT}}(\mathcal{N}_{\mathrm{HT}})(\boldsymbol{a}_{\mathrm{ext}}) \\ &:= \mathop{\boxplus_1}_{\boldsymbol{a}_{\mathrm{int}}\in\prod_{q\in Q_{\mathrm{int}}} I_q} \mathop{\boxtimes_1}_{v\in V} \mathcal{T}_v\left((a_{[h]})_{h\in H(v)}\right). \end{aligned}$$

The Cartesian product over an empty internal-edge set is interpreted as a singleton.

**Remark 6.18.13.** *For depth 1, the lifted operations are explicitly*

$$A \boxplus_1 B = \{a + b : a \in A,\ b \in B\},$$

*and*

$$A \boxtimes_1 B = \{ab : a \in A,\ b \in B\}.$$

*Therefore, the contracted entry contains every scalar result obtained from admissible choices of the local scalar values contained in the HyperTensor entries.*

**Definition 6.18.14** (Depth-$r$ SuperHyperTensor Network Graph)**.** Let

$$r \in \mathbb{N},$$

let

$$G = (V, H, \iota, \tau)$$

be a finite network graph, and let

$$Q = H/\langle\tau\rangle.$$

For every orbit $q \in Q$, let $I_q$ be a finite nonempty index set.

A *depth-$r$ SuperHyperTensor Network Graph over* $\mathbb{K}$ is a tuple

$$\mathcal{N}^{(r)}_{\mathrm{SHT}} = \left(G, (I_q)_{q\in Q}, (\mathcal{T}^{(r)}_v)_{v\in V}\right),$$

where every vertex $v \in V$ is assigned a local depth-$r$ SuperHyperTensor

$$\mathcal{T}^{(r)}_v : \prod_{h\in H(v)} I_{[h]} \longrightarrow \mathscr{P}_r(\mathbb{K}).$$

**Definition 6.18.15** (Contraction of a SuperHyperTensor Network Graph)**.** Let

$$\mathcal{N}^{(r)}_{\mathrm{SHT}} = \left(G, (I_q)_{q\in Q}, (\mathcal{T}^{(r)}_v)_{v\in V}\right)$$

be a depth-$r$ SuperHyperTensor Network Graph.

Its contraction is the mapping

$$\mathrm{Contr}^{(r)}_{\mathrm{SHT}}\left(\mathcal{N}^{(r)}_{\mathrm{SHT}}\right) : \prod_{q\in Q_{\mathrm{ext}}} I_q \longrightarrow \mathscr{P}_r(\mathbb{K})$$

defined by

$$\begin{aligned} &\mathrm{Contr}^{(r)}_{\mathrm{SHT}}\left(\mathcal{N}^{(r)}_{\mathrm{SHT}}\right)(\boldsymbol{a}_{\mathrm{ext}}) \\ &:= \mathop{\boxplus_r}_{\boldsymbol{a}_{\mathrm{int}}\in\prod_{q\in Q_{\mathrm{int}}} I_q} \ \mathop{\boxtimes_r}_{v\in V} \mathcal{T}^{(r)}_v\left((a_{[h]})_{h\in H(v)}\right). \end{aligned}$$

**Theorem 6.18.16** (Well-definedness of the contraction)**.** *Let*

$$\mathcal{N}^{(r)}_{\mathrm{SHT}}$$

*be a finite depth-$r$ SuperHyperTensor Network Graph. Then its contraction*

$$\mathrm{Contr}^{(r)}_{\mathrm{SHT}}\left(\mathcal{N}^{(r)}_{\mathrm{SHT}}\right)$$

*is a well-defined depth-$r$ SuperHyperTensor on the external index sets.*

*In particular,*

$$\mathrm{Contr}^{(r)}_{\mathrm{SHT}}\left(\mathcal{N}^{(r)}_{\mathrm{SHT}}\right)(\boldsymbol{a}_{\mathrm{ext}}) \in \mathscr{P}_r(\mathbb{K})$$

*for every external index assignment $\boldsymbol{a}_{\mathrm{ext}}$.*

*Proof.* Fix

$$\boldsymbol{a}_{\mathrm{ext}} \in \prod_{q\in Q_{\mathrm{ext}}} I_q.$$

Since every $I_q$ is finite and nonempty and $Q_{\mathrm{int}}$ is finite, the set

$$\prod_{q\in Q_{\mathrm{int}}} I_q$$

of internal index assignments is finite and nonempty. When $Q_{\mathrm{int}} = \emptyset$, this product consists of the unique empty assignment.

For every internal assignment and every $v \in V$, the tuple

$$(a_{[h]})_{h\in H(v)}$$

belongs to

$$\prod_{h\in H(v)} I_{[h]}.$$

Hence

$$\mathcal{T}_v^{(r)}\left((a_{[h]})_{h\in H(v)}\right) \in \mathscr{P}_r(\mathbb{K}).$$

By Definition 6.18.3, the finite $\boxtimes_r$-product over $v \in V$ is an element of $\mathscr{P}_r(\mathbb{K})$. Since the number of internal assignments is finite, the subsequent finite $\boxplus_r$-sum is also an element of $\mathscr{P}_r(\mathbb{K})$.

Associativity and commutativity, established in Proposition 6.18.4, show that the result is independent of the chosen ordering and parenthesization. Therefore, the contraction is well-defined for every external index assignment and is a depth-$r$ SuperHyperTensor on the external index domain. □

**Definition 6.18.17** (Canonical singleton embedding)**.** Define maps

$$\sigma_r : \mathbb{K} \longrightarrow \mathscr{P}_r(\mathbb{K})$$

recursively by

$$\sigma_0(a) := a$$

and

$$\sigma_{r+1}(a) := \{\sigma_r(a)\}.$$

**Proposition 6.18.18.** *For every $a, b \in \mathbb{K}$ and every $r \in \mathbb{N}_0$,*

$$\sigma_r(a+b) = \sigma_r(a) \boxplus_r \sigma_r(b)$$

*and*

$$\sigma_r(ab) = \sigma_r(a) \boxtimes_r \sigma_r(b).$$

*Proof.* The assertion is immediate for $r = 0$. Suppose that it holds at depth $r$. Then

$$\begin{aligned}\sigma_{r+1}(a) \boxplus_{r+1} \sigma_{r+1}(b) &= \{\sigma_r(a)\} \boxplus_{r+1} \{\sigma_r(b)\}\\ &= \{\sigma_r(a) \boxplus_r \sigma_r(b)\}\\ &= \{\sigma_r(a+b)\}\\ &= \sigma_{r+1}(a+b).\end{aligned}$$

The multiplication identity follows analogously. □

**Theorem 6.18.19** (Ordinary tensor network graphs as special cases)**.** *Let*

$$\mathcal{N} = (G, (I_q)_{q\in Q}, (T_v)_{v\in V})$$

*be an ordinary tensor network graph over $\mathbb{K}$, where*

$$T_v : \prod_{h\in H(v)} I_{[h]} \longrightarrow \mathbb{K}.$$

*For a fixed $r \in \mathbb{N}$, define*

$$\mathcal{T}_v^{(r)} := \sigma_r \circ T_v.$$

*Then*

$$\mathcal{N}^{(r)} = \Big(G, (I_q)_{q\in Q}, (\mathcal{T}_v^{(r)})_{v\in V}\Big)$$

*is a depth-r SuperHyperTensor Network Graph, and*

$$\mathrm{Contr}^{(r)}_{\mathrm{SHT}}\left(\mathcal{N}^{(r)}\right) = \sigma_r \circ \mathrm{Contr}(\mathcal{N}).$$

*Consequently, ordinary tensor network graphs form a canonically embedded special subclass of SuperHyperTensor Network Graphs.*

*Proof.* Each composite

$$\mathcal{T}_v^{(r)} = \sigma_r \circ T_v$$

has codomain

$$\mathscr{P}_r(\mathbb{K}),$$

so $\mathcal{N}^{(r)}$ is a depth-$r$ SuperHyperTensor Network Graph.

By Proposition 6.18.18, the singleton embedding $\sigma_r$ preserves all finite sums and products occurring in the network contraction. Therefore, for every external assignment,

$$\mathrm{Contr}^{(r)}_{\mathrm{SHT}}\left(\mathcal{N}^{(r)}\right) = \sigma_r\left(\mathrm{Contr}(\mathcal{N})\right).$$

□

**Corollary 6.18.20.** *The following hierarchy holds:*

*ordinary tensor network graph ⊆ HyperTensor Network Graph ⊆ SuperHyperTensor Network Graph.*

*More precisely:*

1. *depth* 0 *gives an ordinary tensor network graph;*
2. *depth* 1 *gives a HyperTensor Network Graph;*
3. *depth* $r \geq 1$ *gives a depth-r SuperHyperTensor Network Graph.*

**Remark 6.18.21** (Difference from a Tensor Hypernetwork)**.** *A HyperTensor Network Graph and a Tensor Hypernetwork generalize different components of an ordinary tensor network.*

- *A* HyperTensor Network Graph *retains an ordinary graph-based contraction pattern but replaces scalar tensor entries by set-valued entries.*
- *A* Tensor Hypernetwork *retains ordinary scalar-valued tensors but replaces the underlying graph by a hypergraph, so that one index may be shared by more than two tensor vertices.*
- *A* SuperHyperTensor Network Graph *retains the graph-based contraction pattern and replaces scalar tensor entries by iterated-powerset-valued entries.*

*These constructions are therefore conceptually independent and may also be combined to define HyperTensor Hypernetworks or SuperHyperTensor SuperHypernetworks.*

## 6.19 Linguistic Graphs, Linguistic HyperGraphs, and Linguistic SuperHyperGraphs

A Linguistic Graph assigns linguistic terms to vertices and edges, thereby representing qualitative concepts and relationships without requiring numerical membership values [765, 766, 767, 768]. Related concepts, such as Hesitant Fuzzy Linguistic Graphs, are also known and have been studied [769]. A Linguistic HyperGraph assigns linguistic terms to vertices and hyperedges, enabling the representation of qualitative higher-order relationships among multiple entities within a unified structure. A Linguistic SuperHyperGraph combines linguistic labels with hierarchical supervertices and superedges, thereby representing qualitative higher-order relationships systematically across iterated powerset levels.

Let

$$L_V \qquad \text{and} \qquad L_E$$

be nonempty sets of linguistic terms associated with prescribed linguistic variables. The elements of $L_V$ are used to describe vertices, while the elements of $L_E$ describe relations, ties, distances, strengths, or other linguistic characteristics of edges.

No numerical representation of $L_V$ or $L_E$ is assumed. In particular, the linguistic terms need not be identified with fuzzy membership values.

**Definition 6.19.1** (Linguistic Graph)**.** A *Linguistic Graph* is a tuple

$$\mathcal{G}_L = \big(V, E, \lambda_V, \lambda_E\big),$$

where

$$G = (V, E)$$

is a finite simple graph,

$$\lambda_V : V \longrightarrow L_V$$

is a vertex linguistic-labeling map, and

$$\lambda_E : E \longrightarrow L_E$$

is an edge linguistic-labeling map.

For $v \in V$, the value

$$\lambda_V(v)$$

is the linguistic term associated with $v$, whereas, for $e \in E$,

$$\lambda_E(e)$$

is the linguistic term describing the corresponding relation.

When the vertices themselves are linguistic terms, one may take

$$V \subseteq L_V \qquad \text{and} \qquad \lambda_V = \mathrm{id}_V .$$

**Remark 6.19.2** (Directed variant)**.** *A directed Linguistic Graph is obtained by replacing the ordinary edge set by*

$$E \subseteq \{(u, v) \in V \times V : u \neq v\},$$

*while retaining the linguistic-labeling maps*

$$\lambda_V : V \to L_V, \qquad \lambda_E : E \to L_E.$$

*Thus linguistic relations may be symmetric or directional.*

**Definition 6.19.3** (Linguistic HyperGraph)**.** A *Linguistic HyperGraph* is a tuple

$$\mathcal{H}_L = \big(V, E, \partial, \lambda_V, \lambda_E\big),$$

where

$$H = (V, E, \partial)$$

is a finite HyperGraph with incidence map

$$\partial : E \longrightarrow \mathcal{P}_*(V),$$

and

$$\lambda_V : V \longrightarrow L_V, \qquad \lambda_E : E \longrightarrow L_E$$

are linguistic-labeling maps.

For each hyperedge $e \in E$,

$$\partial(e) \subseteq V$$

specifies the nonempty collection of vertices participating in the higher-order relation, while

$$\lambda_E(e)$$

provides its linguistic description.

Hence a Linguistic HyperGraph can represent a linguistically described relation involving any finite nonempty number of vertices, rather than only a pair of vertices.

**Definition 6.19.4** (Linguistic $n$-SuperHyperGraph)**.** Let $V_0$ be a finite nonempty base set and let

$$n \in \mathbb{N}_0.$$

A *Linguistic $n$-SuperHyperGraph* is a tuple

$$\mathcal{S}_L^{(n)} = \big(V_0; V, E, \partial, \lambda_V, \lambda_E\big),$$

where

$$\mathcal{S}^{(n)} = \big(V_0; V, E, \partial\big)$$

is an $n$-SuperHyperGraph satisfying

$$V \subseteq \mathcal{P}^n(V_0),$$

with incidence map

$$\partial : E \longrightarrow \mathcal{P}_*(V),$$

and

$$\lambda_V : V \longrightarrow L_V, \qquad \lambda_E : E \longrightarrow L_E$$

are linguistic-labeling maps.

Each $n$-supervertex $v \in V$ therefore possesses both its hierarchical set-theoretic structure

$$v \in \mathcal{P}^n(V_0)$$

and its linguistic interpretation

$$\lambda_V(v) \in L_V.$$

Similarly, each $n$-superedge $e \in E$ has a superincidence set

$$\partial(e) \in \mathcal{P}_*(V)$$

and a linguistic description

$$\lambda_E(e) \in L_E.$$

Thus a Linguistic $n$-SuperHyperGraph combines linguistic information with hierarchical higher-order incidence structures generated through iterated powersets.

**Proposition 6.19.5** (Hierarchy of linguistic graph structures)**.** *The following canonical inclusions hold.*

1. *Every Linguistic Graph can be regarded as a Linguistic HyperGraph whose hyperedges all have cardinality* 2*.*
2. *Under the convention*

   $$\mathcal{P}^0(V_0) = V_0,$$

   *every Linguistic HyperGraph can be regarded as a Linguistic* 0*-SuperHyperGraph.*
3. *Consequently, Linguistic $n$-SuperHyperGraphs provide a hierarchical extension of both Linguistic Graphs and Linguistic HyperGraphs.*

*Proof.* For a Linguistic Graph

$$\mathcal{G}_L = (V, E, \lambda_V, \lambda_E),$$

define

$$\partial(e) = e \qquad (e \in E).$$

Since every graph edge contains exactly two vertices,

$$|\partial(e)| = 2,$$

and hence the resulting incidence structure is a HyperGraph with the same linguistic-labeling maps.

Next, let

$$\mathcal{H}_L = (V, E, \partial, \lambda_V, \lambda_E)$$

be a Linguistic HyperGraph. Set

$$V_0 = V \qquad \text{and} \qquad n = 0.$$

Then

$$V = V_0 = \mathcal{P}^0(V_0),$$

so the same incidence structure satisfies the definition of a Linguistic 0-SuperHyperGraph.

Therefore the stated hierarchy follows. □

**Remark 6.19.6.** *The linguistic sets $L_V$ and $L_E$ may themselves possess additional structure, such as a partial order, total order, linguistic negation, or a linguistic-distance relation. Such additional structures may be incorporated when required by a particular application, but they are not necessary for the basic definitions above.*

# 7 Discussion: A Complete Higher-Graphic Structure

The higher-order frameworks considered throughout this manuscript arise from substantially different mathematical viewpoints. Some are based on multiway relations, others on hierarchical or recursive objects, and still others on compositional operations, tensors, closure systems, temporal contexts, or categorical structure. This chapter investigates whether these different ingredients can be organized within a common typed formalism.

The purpose of the framework introduced below is not to claim a universal notion of mathematical completeness. Rather, the term *Complete Higher-Graphic Structure* is used to indicate an umbrella formalism in which several major mechanisms occurring in higher-order network models can coexist within a single typed structure.

## 7.1 Complete Higher-Graphic Structure

We first introduce a unified definition that incorporates contexts, sorts, relations, operations, attributes, weights, and closure mechanisms.

**Definition 7.1.1** (Complete Higher-Graphic Structure)**.** Let $I$ be a nonempty set of *contexts*, let $S$ be a nonempty set of *sorts*, and put

$$P := I \times S.$$

For each profile $p = (i, s) \in P$, let $X_p = X_{i,s}$ be a carrier set.

A *Complete Higher-Graphic Structure*, abbreviated *CHGS*, is a tuple

$$\mathfrak{H} = \Big(I, S, X, \Sigma_{\mathrm{rel}}, \Sigma_{\mathrm{op}}, \Sigma_{\mathrm{att}}, \Sigma_{\mathrm{cl}}, \Sigma_{\mathrm{w}}, [\![-]\!], \mathcal{E}\Big),$$

where

$$X = \{X_p\}_{p \in P},$$

and the remaining components are specified as follows.

1. **Relational signature.**
   The set
   $$\Sigma_{\mathrm{rel}}$$
   consists of typed relation symbols. For each $\rho \in \Sigma_{\mathrm{rel}}$, one specifies a finite profile
   $$\mathrm{prof}(\rho) = (p_1, \ldots, p_m), \qquad m \geq 1,$$
   and an interpretation
   $$[\![\rho]\!] \subseteq X_{p_1} \times \cdots \times X_{p_m}.$$
   A symmetry group may additionally be assigned to $\rho$. More precisely, if
   $$G_\rho \leq \mathrm{Stab}(p_1, \ldots, p_m),$$
   where
   $$\mathrm{Stab}(p_1, \ldots, p_m) = \big\{\pi \in \mathfrak{S}_m : p_{\pi(j)} = p_j \text{ for every } j\big\},$$
   then the relation is required to satisfy
   $$(x_1, \ldots, x_m) \in [\![\rho]\!] \Longrightarrow (x_{\pi(1)}, \ldots, x_{\pi(m)}) \in [\![\rho]\!]$$
   for every $\pi \in G_\rho$.

2. **Operational signature.**

   The set
   $$\Sigma_{\mathrm{op}}$$
   consists of typed operation symbols. Each $\omega \in \Sigma_{\mathrm{op}}$ has an input profile
   $$\mathrm{in}(\omega) = (p_1, \ldots, p_m), \qquad m \geq 0,$$
   and an output profile
   $$\mathrm{out}(\omega) = q \in P.$$
   Its interpretation is a possibly partial map
   $$[\![\omega]\!] : X_{p_1} \times \cdots \times X_{p_m} \rightharpoonup X_q.$$
   These operations may represent, for example, gluing, substitution, wiring, composition, aggregation, or face and degeneracy operations.

   If a symmetry group
   $$G_\omega \leq \mathrm{Stab}(p_1, \ldots, p_m)$$
   is specified, then
   $$[\![\omega]\!](x_{\pi(1)}, \ldots, x_{\pi(m)}) = [\![\omega]\!](x_1, \ldots, x_m)$$
   whenever either side is defined and $\pi \in G_\omega$.

3. **Attribute signature.**

   The set
   $$\Sigma_{\mathrm{att}}$$
   consists of typed attribute symbols. Each $\alpha \in \Sigma_{\mathrm{att}}$ has an input profile
   $$(p_1, \ldots, p_m)$$
   and a prescribed codomain $D_\alpha$, and is interpreted as a possibly partial map
   $$[\![\alpha]\!] : X_{p_1} \times \cdots \times X_{p_m} \rightharpoonup D_\alpha.$$
   Attributes may encode labels, supports, flattening maps, grades, geometric realizations, semantic information, or nested-neighborhood assignments.

4. **Closure signature.**

   The set
   $$\Sigma_{\mathrm{cl}}$$
   consists of typed closure symbols. If
   $$c \in \Sigma_{\mathrm{cl}}$$
   has profile
   $$\mathrm{prof}(c) = (p_1, \ldots, p_m),$$
   write
   $$X_{\mathrm{prof}(c)} = X_{p_1} \times \cdots \times X_{p_m}.$$
   The interpretation of $c$ is a closure operator
   $$[\![c]\!] : \mathcal{P}\big(X_{\mathrm{prof}(c)}\big) \longrightarrow \mathcal{P}\big(X_{\mathrm{prof}(c)}\big)$$
   satisfying, for all $A, B \subseteq X_{\mathrm{prof}(c)}$,
   $$A \subseteq [\![c]\!](A),$$
   $$A \subseteq B \Longrightarrow [\![c]\!](A) \subseteq [\![c]\!](B),$$
   and
   $$[\![c]\!]\big([\![c]\!](A)\big) = [\![c]\!](A).$$

5. **Weight signature.**

   The set
   $$\Sigma_{\mathrm{w}}$$
   consists of weight symbols. Each $w \in \Sigma_{\mathrm{w}}$ has a specified typed domain $\mathrm{Dom}(w)$, usually consisting of relation instances or operation instances, together with a prescribed coefficient set $K_w$. Its interpretation is a map
   $$[\![w]\!] : \mathrm{Dom}(w) \longrightarrow K_w.$$
   Typical choices of $K_w$ include
   $$\mathbb{R}, \qquad \mathbb{R}_{\geq 0}, \qquad \mathbb{N}_0,$$
   or, more generally, a semiring or ring.

6. **Equational and coherence conditions.**

   The family
   $$\mathcal{E}$$
   consists of well-typed equations between terms generated from $\Sigma_{\mathrm{op}}$. It may be used to impose algebraic or compositional conditions such as associativity, units, operadic identities, simplicial identities, or monoidal coherence conditions.

The contexts in $I$ may represent time points, layers, scales, modalities, grades, or other external indexing parameters.

A CHGS is called *finite* if $I$, $S$, all signature sets, and all carrier sets $X_p$ are finite.

**Remark 7.1.2** (Meaning of "complete")**.** *The adjective* complete *in Definition 7.1.1 refers to the intended breadth of the formalism and should not be confused with completeness in the sense of metric spaces, lattices, categories, logical theories, or other standard mathematical notions of completeness.*

*A CHGS may simultaneously encode*

- *typed multiway relations,*
- *hierarchical and cross-level objects,*
- *compositional operations,*
- *auxiliary attributes,*
- *weighted or tensor-valued interactions,*
- *closure or implication mechanisms,*
- *and temporal, multilayer, or multimodal contexts.*

*Individual applications need not activate every component.*

**Definition 7.1.3** (Network event)**.** Let $\mathfrak{H}$ be a CHGS. A *network event* of $\mathfrak{H}$ is either

1. a tuple
   $$(x_1, \ldots, x_m) \in [\![\rho]\!]$$
   for some $\rho \in \Sigma_{\mathrm{rel}}$, or
2. a defined value of a well-typed term generated from $\Sigma_{\mathrm{op}}$, considered subject to the equations in $\mathcal{E}$.

The expressive scope of CHGS is most naturally formulated through an explicit admissibility condition rather than through an unrestricted universality claim.

**Definition 7.1.4** (CHGS-admissible model)**.** A mathematical model $\mathcal{M}$ is called *CHGS-admissible* if its primitive data admit a presentation by the following ingredients:

1. a family of typed carrier sets;
2. finitary typed relations;
3. finitary typed partial operations;

4. optional typed attributes;
5. optional closure operators on powersets of typed products;
6. optional weight maps;
7. and optional equational conditions among the operations.

**Theorem 7.1.5** (CHGS encoding theorem)**.** *Every CHGS-admissible model admits a faithful representation as a CHGS. More precisely, if $\mathcal{M}$ is CHGS-admissible, then there exists a CHGS*

$$\mathfrak{H}_{\mathcal{M}}$$

*whose carriers and interpreted symbols reproduce the primitive data of $\mathcal{M}$ exactly.*

*Proof.* Let the carrier sets of $\mathcal{M}$ be indexed by a type set $T$. Choose contexts and sorts so that every type in $T$ corresponds to a profile in $I \times S$, and identify the corresponding CHGS carrier with the original carrier.

For each primitive relation of $\mathcal{M}$, introduce a relation symbol having the same input types and interpret it by that relation. For each primitive partial operation, introduce an operation symbol with the same input and output types and interpret it by that operation. Attributes, closure operators, and weights are transferred in the same manner.

Finally, every equational axiom among the primitive operations is included in $\mathcal{E}$. Hence no primitive datum of $\mathcal{M}$ is altered or discarded. The resulting interpretation is therefore a faithful CHGS representation of $\mathcal{M}$. □

**Corollary 7.1.6** (Common classes representable within CHGS)**.** *Whenever they are presented through their usual finitary typed data, the following kinds of structures admit CHGS representations:*

1. *graphs, directed graphs, multigraphs, HyperGraphs, and multi-HyperGraphs;*
2. *n-SuperHyperGraphs and finite hierarchical, graded, nested, or recursive variants;*
3. *relational structures of variable arity;*
4. *finite simplicial complexes and combinatorially presented cell or polyhedral complexes;*
5. *multilayer, temporal, and multimodal networks;*
6. *factor graphs, Tanner graphs, and related typed incidence structures;*
7. *weighted and adjacency-tensor models;*
8. *closure-based implication models;*
9. *nested-neighborhood or coalgebraically described models;*
10. *operadic or wiring-based compositional models.*

**Remark 7.1.7.** *For geometric or topological structures such as CW complexes, the CHGS representation concerns the chosen combinatorial and typed data. Additional topological information may be stored through attributes, while non-equational topological axioms remain additional structural conditions on the CHGS representation.*

**Example 7.1.8** (A CHGS combining hierarchical, relational, weighted, closure, and compositional features)**.** We construct a finite CHGS containing several types of higher-order information simultaneously.

**Step 1: Contexts and sorts.**

Let

$$I = \{*\}$$

and

$$S = \{\mathsf{X}, \mathsf{H}, \mathsf{Raw}, \mathsf{Clean}, \mathsf{Model}\}.$$

Define

$$X_{*,\mathsf{X}} = \{1, 2, 3, \beta\}, \qquad X_{*,\mathsf{H}} = \{\varepsilon_1, \varepsilon_2\},$$

and

$$X_{*,\mathsf{Raw}} = \{r_1, r_2\}, \qquad X_{*,\mathsf{Clean}} = \{c_1, c_2\}, \qquad X_{*,\mathsf{Model}} = \{m\}.$$

The elements $1, 2, 3$ are regarded as base-level objects, whereas $\beta$ represents a higher-level object.

**Step 2: Relational structure.**
Introduce unary relations

$$\text{Base}, \qquad \text{Super},$$

and a binary incidence relation

$$\text{Inc}.$$

Set

$$\llbracket \text{Base} \rrbracket = \{1, 2, 3\}, \qquad \llbracket \text{Super} \rrbracket = \{\beta\},$$

and

$$\llbracket \text{Inc} \rrbracket = \{(\varepsilon_1, \beta), (\varepsilon_1, 3), (\varepsilon_2, 1), (\varepsilon_2, 2)\}.$$

Thus $\varepsilon_1$ is incident with the higher-level object $\beta$ and the base object 3, whereas $\varepsilon_2$ is incident with 1 and 2.
Introduce also binary and ternary interaction relations

$$R_2, \qquad R_3$$

with

$$\llbracket R_2 \rrbracket = \{(1, 2), (2, 3)\}$$

and

$$\llbracket R_3 \rrbracket = \{(1, 2, 3)\}.$$

**Step 3: Attribute structure.**
Define the partial support map

$$\text{supp} : X_{*,\mathsf{X}} \rightharpoonup \mathcal{P}(\{1, 2, 3\})$$

by

$$\text{supp}(\beta) = \{1, 2\},$$

and leave it undefined on $1, 2, 3$.
Thus $\beta$ represents a higher-level object supported by the base objects 1 and 2.
Define also

$$\nu : \{1, 2, 3\} \longrightarrow \mathcal{P}^2(\{1, 2, 3\})$$

by

$$\nu(1) = \big\{\{2, 3\}, \{2\}\big\},$$

$$\nu(2) = \big\{\{1\}\big\},$$

and

$$\nu(3) = \big\{\{1, 2\}, \emptyset\big\}.$$

Hence $\nu$ provides a depth-2 nested-neighborhood assignment.

**Step 4: Weight structure.**
Let

$$K = \mathbb{R}.$$

Assign weights

$$w_2(1, 2) = 1, \qquad w_2(2, 3) = 2,$$

and

$$w_3(1, 2, 3) = 5.$$

These values may equivalently be interpreted as selected nonzero entries of second- and third-order adjacency tensors.

**Step 5: Closure structure.**
Let

$$Y = \{1, 2, 3\}.$$

Define

$$\text{cl} : \mathcal{P}(Y) \longrightarrow \mathcal{P}(Y)$$

to be the closure operator generated by the implication

$$\{1,2\} \Rightarrow 3.$$

Equivalently,

$$\mathrm{cl}(A) = \begin{cases} A \cup \{3\}, & \{1,2\} \subseteq A, \\ A, & \text{otherwise.} \end{cases}$$

For example,

$$\mathrm{cl}(\{1\}) = \{1\}$$

and

$$\mathrm{cl}(\{1,2\}) = \{1,2,3\}.$$

**Step 6: Compositional structure.**
Introduce operations

$$f : X_{*,\mathsf{Raw}} \longrightarrow X_{*,\mathsf{Clean}}$$

and

$$g : X_{*,\mathsf{Clean}} \times X_{*,\mathsf{Clean}} \longrightarrow X_{*,\mathsf{Model}}.$$

Define

$$f(r_1) = c_1, \qquad f(r_2) = c_2,$$

and

$$g(c_1, c_2) = g(c_2, c_1) = m.$$

The composite

$$h(x,y) = g\big(f(x), f(y)\big)$$

is therefore a well-typed operation whenever the right-hand side is defined.

**Conclusion.**
The resulting structure simultaneously contains

- hierarchical support information,
- incidence relations,
- binary and ternary interactions,
- weighted higher-order relations,
- a closure implication,
- a nested-neighborhood assignment,
- and typed compositional operations.

It therefore illustrates how structurally different higher-order features can coexist within one CHGS.

**Remark 7.1.9** (Temporal and multilayer CHGSs)**.** *Temporal and multilayer structures can be incorporated directly through the context set. For example, one may take*

$$I = T, \qquad I = L, \qquad \text{or} \qquad I = T \times L,$$

*where $T$ is a time set and $L$ is a layer set.*
*The carriers*

$$X_{t,s}, \qquad X_{\ell,s}, \qquad \text{or} \qquad X_{(t,\ell),s}$$

*then explicitly distinguish temporal or layer-dependent objects. Relations whose profiles involve distinct contexts naturally represent cross-time or interlayer interactions.*

## 7.2 Morphisms and Isomorphisms of CHGSs

We next define structure-preserving maps between CHGSs.

**Definition 7.2.1** (CHGS morphism)**.** Let

$$\mathfrak{H} \qquad \text{and} \qquad \mathfrak{H}'$$

be CHGSs over the same contexts, sorts, signatures, attribute codomains, and weight codomains.

A *CHGS morphism*

$$\Phi : \mathfrak{H} \longrightarrow \mathfrak{H}'$$

is a family

$$\Phi = (\phi_p)_{p \in I \times S}, \qquad \phi_p : X_p \longrightarrow X'_p,$$

satisfying the following conditions.

1. **Relation preservation.**
   If
   $$(x_1, \dots, x_m) \in [\![\rho]\!],$$
   then
   $$\big(\phi_{p_1}(x_1), \dots, \phi_{p_m}(x_m)\big) \in [\![\rho]\!]'$$
   for every relation symbol
   $$\mathrm{prof}(\rho) = (p_1, \dots, p_m).$$

2. **Operation preservation.**
   If
   $$y = [\![\omega]\!](x_1, \dots, x_m)$$
   is defined, then
   $$[\![\omega]\!]'\big(\phi_{p_1}(x_1), \dots, \phi_{p_m}(x_m)\big)$$
   is defined and satisfies
   $$\phi_q(y) = [\![\omega]\!]'\big(\phi_{p_1}(x_1), \dots, \phi_{p_m}(x_m)\big),$$
   where
   $$\mathrm{out}(\omega) = q.$$

3. **Attribute preservation.**
   Whenever
   $$[\![\alpha]\!](x_1, \dots, x_m)$$
   is defined,
   $$[\![\alpha]\!]'\big(\phi_{p_1}(x_1), \dots, \phi_{p_m}(x_m)\big) = [\![\alpha]\!](x_1, \dots, x_m).$$

4. **Closure continuity.**
   For a profile
   $$\mathbf{p} = (p_1, \dots, p_m),$$
   write
   $$\phi_{\mathbf{p}} = \phi_{p_1} \times \cdots \times \phi_{p_m}.$$
   Then, for every $c \in \Sigma_{\mathrm{cl}}$ of profile $\mathbf{p}$ and every
   $$A \subseteq X_{\mathbf{p}},$$
   we require
   $$\phi_{\mathbf{p}}\big[[\![c]\!](A)\big] \subseteq [\![c]\!]'\big(\phi_{\mathbf{p}}[A]\big).$$

5. **Weight preservation.**
   For each weight symbol $w$, whenever $u \in \mathrm{Dom}(w)$, the corresponding image instance $\Phi_*(u)$ belongs to the target domain and
   $$[\![w]\!](u) = [\![w]\!]'\big(\Phi_*(u)\big).$$

**Definition 7.2.2** (Strong CHGS embedding)**.** A CHGS morphism

$$\Phi : \mathfrak{H} \to \mathfrak{H}'$$

is called a *strong embedding* if every component

$$\phi_p : X_p \to X'_p$$

is injective and the structure is reflected on the image.

In particular:

1. relations are reflected:

$$\big(\phi_{p_1}(x_1), \ldots, \phi_{p_m}(x_m)\big) \in [\![\rho]\!]' \iff (x_1, \ldots, x_m) \in [\![\rho]\!];$$

2. operation definedness is reflected and the operation values agree exactly on the image;

3. attribute definedness is reflected;

4. for every closure symbol $c$,

$$\phi_{\mathbf{p}}\big[[\![c]\!](A)\big] = [\![c]\!]'\big(\phi_{\mathbf{p}}[A]\big) \cap \mathrm{Im}(\phi_{\mathbf{p}});$$

5. weights agree exactly on corresponding instances.

**Definition 7.2.3** (CHGS isomorphism)**.** A CHGS morphism

$$\Phi : \mathfrak{H} \to \mathfrak{H}'$$

is a *CHGS isomorphism* if there exists a CHGS morphism

$$\Psi : \mathfrak{H}' \to \mathfrak{H}$$

such that

$$\Psi \circ \Phi = \mathrm{id}_{\mathfrak{H}}$$

and

$$\Phi \circ \Psi = \mathrm{id}_{\mathfrak{H}'}.$$

**Proposition 7.2.4** (Category of CHGSs)**.** *For every fixed CHGS signature, CHGSs and CHGS morphisms form a category.*

*Proof.* The componentwise identity family preserves every interpreted relation, operation, attribute, closure operator, and weight, and hence defines an identity morphism.

Let

$$\Phi : \mathfrak{H} \to \mathfrak{H}' \qquad \text{and} \qquad \Psi : \mathfrak{H}' \to \mathfrak{H}''$$

be CHGS morphisms. Their composite is defined componentwise by

$$(\Psi \circ \Phi)_p = \psi_p \circ \phi_p.$$

Relation, operation, attribute, and weight preservation follow by successive application of the corresponding preservation conditions.

For a closure symbol $c$,

$$\phi_{\mathbf{p}}\big[[\![c]\!](A)\big] \subseteq [\![c]\!]'\big(\phi_{\mathbf{p}}[A]\big),$$

and therefore

$$\begin{aligned}\psi_{\mathbf{p}}\left[\phi_{\mathbf{p}}\big[[\![c]\!](A)\big]\right] &\subseteq \psi_{\mathbf{p}}\left[[\![c]\!]'\big(\phi_{\mathbf{p}}[A]\big)\right] \\ &\subseteq [\![c]\!]''\big((\psi_{\mathbf{p}} \circ \phi_{\mathbf{p}})[A]\big).\end{aligned}$$

Thus closure continuity is preserved under composition.

Finally, associativity follows from the associativity of ordinary function composition. Hence the category axioms hold. □

**Definition 7.2.5** (CHGS layer)**.** Let

$$\Lambda = \{\mathrm{Rel}, \mathrm{Op}, \mathrm{Att}, \mathrm{Cl}, \mathrm{W}\}.$$

The carrier system $X$, together with $I$ and $S$, is regarded as the permanent typed base and is not included among the removable layers.

**Definition 7.2.6** (Layer reduct)**.** For $L \subseteq \Lambda$, the *layer reduct*

$$\mathfrak{H}\restriction_L$$

is obtained from $\mathfrak{H}$ by retaining the typed carriers and exactly those signatures and interpretations whose layers belong to $L$.

**Definition 7.2.7** (Redundant layer)**.** Fix a class $\mathscr{K}$ of CHGSs equipped with a common convention for reconstruction.

A layer

$$\lambda \in \Lambda$$

is called *redundant in* $\mathfrak{H}$ *relative to* $\mathscr{K}$ if there exists a reconstruction procedure

$$\mathcal{R}_\lambda$$

defined uniformly on the relevant reducts in $\mathscr{K}$ such that

$$\mathcal{R}_\lambda\left(\mathfrak{H}\restriction_{\Lambda\setminus\{\lambda\}}\right) \cong \mathfrak{H}.$$

Otherwise, $\lambda$ is called *essential relative to* $\mathscr{K}$.

**Proposition 7.2.8** (Redundancy preserves isomorphism-invariant information)**.** *If* $\lambda$ *is redundant in* $\mathfrak{H}$*, then*

$$\mathfrak{H} \cong \mathcal{R}_\lambda\left(\mathfrak{H}\restriction_{\Lambda\setminus\{\lambda\}}\right).$$

*Consequently, every CHGS-isomorphism invariant has the same value before and after reconstruction.*

*Proof.* This follows directly from Definition 7.2.7. □

**Definition 7.2.9** (Recoverable layer set)**.** A subset

$$L \subseteq \Lambda$$

is called *recoverable for* $\mathfrak{H}$ if there exists a reconstruction procedure

$$\mathcal{R}_L$$

such that

$$\mathcal{R}_L\left(\mathfrak{H}\restriction_L\right) \cong \mathfrak{H}.$$

**Theorem 7.2.10** (Existence of a minimal CHGS core)**.** *Every CHGS possesses at least one inclusion-minimal recoverable layer set.*

*Proof.* The full layer set

$$\Lambda$$

is recoverable by the identity reconstruction. Hence the collection

$$\mathscr{R} = \{L \subseteq \Lambda : L \text{ is recoverable for } \mathfrak{H}\}$$

is nonempty.

Since $\Lambda$ is finite, $\mathscr{R}$ is a finite partially ordered set under inclusion. Therefore $\mathscr{R}$ contains at least one inclusion-minimal element $L_*$.

Such an $L_*$ is an inclusion-minimal recoverable layer set and hence a minimal CHGS core. □

**Remark 7.2.11.** *A minimal CHGS core need not be unique. Different sets of layers may contain equivalent information and may therefore admit distinct minimal reconstructions.*

**Proposition 7.2.12** (Criterion for essentiality)**.** *Let $\lambda \in \Lambda$, and suppose that there exist*

$$\mathfrak{H}_1, \mathfrak{H}_2 \in \mathscr{K}$$

*such that*

$$\mathfrak{H}_1 \restriction_{\Lambda\setminus\{\lambda\}} = \mathfrak{H}_2 \restriction_{\Lambda\setminus\{\lambda\}},$$

*but*

$$\mathfrak{H}_1 \ncong \mathfrak{H}_2.$$

*Then no uniform reconstruction from the non-$\lambda$ reduct can recover both structures. In particular, $\lambda$ is essential on the class $\mathscr{K}$.*

*Proof.* Suppose, to the contrary, that such a uniform reconstruction $\mathcal{R}_\lambda$ exists. Since the two reducts are identical,

$$\mathcal{R}_\lambda\left(\mathfrak{H}_1 \restriction_{\Lambda\setminus\{\lambda\}}\right) = \mathcal{R}_\lambda\left(\mathfrak{H}_2 \restriction_{\Lambda\setminus\{\lambda\}}\right).$$

If the reconstruction recovered both structures up to isomorphism, we would obtain

$$\mathfrak{H}_1 \cong \mathfrak{H}_2,$$

contrary to assumption. □

## 7.3 Comparison Profiles for Finite CHGSs

The following quantities are intended as structural comparison tools. They should not be interpreted as universal measures of mathematical expressiveness or statistical learnability.

**Definition 7.3.1** (Expressiveness profile)**.** Let $\mathfrak{H}$ be a finite CHGS. Define

$$\mathrm{Expr}(\mathfrak{H}) = \Big(|I|, |S|, a^{\mathrm{rel}}_{\max}, a^{\mathrm{op}}_{\max}, |\Sigma_{\mathrm{rel}}|, |\Sigma_{\mathrm{op}}|, |\Sigma_{\mathrm{att}}|, |\Sigma_{\mathrm{cl}}|, |\Sigma_{\mathrm{w}}|\Big),$$

where

$$a^{\mathrm{rel}}_{\max} = \max_{\rho \in \Sigma_{\mathrm{rel}}} \mathrm{arity}(\rho)$$

and

$$a^{\mathrm{op}}_{\max} = \max_{\omega \in \Sigma_{\mathrm{op}}} \mathrm{arity}_{\mathrm{in}}(\omega).$$

We adopt the convention

$$\max \emptyset := 0.$$

**Definition 7.3.2** (Operational compositional coverage)**.** Let

$$G = (G_p)_{p \in I \times S}, \qquad G_p \subseteq X_p,$$

be a family of primitive generators.

Define

$$\mathrm{OpCl}_{\mathfrak{H}}(G) = (Y_p)_{p \in I \times S}$$

to be the smallest family satisfying:

1. $$G_p \subseteq Y_p \qquad \text{for every } p;$$

2. whenever

   $$[\![\omega]\!](x_1, \ldots, x_m)$$

   is defined and

   $$x_j \in Y_{p_j} \qquad (1 \le j \le m),$$

   its value belongs to

   $$Y_{\mathrm{out}(\omega)}.$$

Assume

$$\sum_{p\in I\times S} |X_p| > 0.$$

The *operational compositional coverage* of $G$ is

$$\mathrm{Comp}_G(\mathfrak{H}) = \frac{\sum_{p\in I\times S} |Y_p|}{\sum_{p\in I\times S} |X_p|}.$$

**Remark 7.3.3.** *The quantity*

$$\mathrm{Comp}_G(\mathfrak{H})$$

*measures the proportion of carrier elements obtainable from $G$ by the explicitly specified operations. It does not attempt to measure the contribution of relations, attributes, or closure operators to semantic expressiveness.*

**Definition 7.3.4** (Extensional description-size proxy)**.** Let $\mathfrak{H}$ be finite and suppose that every interpreted map is stored extensionally.

Define

$$\mathrm{DL}(\mathfrak{H})$$

by

$$\begin{aligned}\mathrm{DL}(\mathfrak{H}) := &\sum_{p\in I\times S} \lceil \log_2(|X_p| + 1)\rceil \\ &+ \sum_{\rho\in\Sigma_{\mathrm{rel}}} |[\![\rho]\!]| \\ &+ \sum_{\omega\in\Sigma_{\mathrm{op}}} |\mathrm{graph}([\![\omega]\!])| \\ &+ \sum_{\alpha\in\Sigma_{\mathrm{att}}} |\mathrm{graph}([\![\alpha]\!])| \\ &+ \sum_{c\in\Sigma_{\mathrm{cl}}} |\mathrm{graph}([\![c]\!])| \\ &+ \sum_{w\in\Sigma_{\mathrm{w}}} |\mathrm{graph}([\![w]\!])| \,.\end{aligned}$$

**Remark 7.3.5.** *The quantity*

$$\mathrm{DL}(\mathfrak{H})$$

*is only an extensional structural description-size proxy. It is not a formal measure of statistical learnability, sample complexity, or computational complexity. More refined applications may replace it by an MDL score, an algorithmic coding length, or a task-dependent learning-complexity measure.*

**Proposition 7.3.6** (Isomorphism invariance)**.** *If*

$$\mathfrak{H} \cong \mathfrak{H}',$$

*then*

$$\mathrm{Expr}(\mathfrak{H}) = \mathrm{Expr}(\mathfrak{H}')$$

*and*

$$\mathrm{DL}(\mathfrak{H}) = \mathrm{DL}(\mathfrak{H}').$$

*Moreover, if $\Phi$ is an isomorphism and $G$ is a generator family, then for*

$$G' = \Phi(G),$$

*we have*

$$\mathrm{Comp}_G(\mathfrak{H}) = \mathrm{Comp}_{G'}(\mathfrak{H}').$$

*Proof.* A CHGS isomorphism induces bijections between corresponding carrier sets, relations, operation graphs, attribute graphs, closure graphs, and weight graphs. Hence all cardinalities occurring in Expr and DL are preserved.

Because the isomorphism also commutes with every interpreted operation, it maps the operational closure generated by $G$ bijectively onto the operational closure generated by $\Phi(G)$. Therefore the compositional-coverage ratios are equal. □

**Proposition 7.3.7** (Monotonicity under signature extension)**.** *Let $\widetilde{\mathfrak{H}}$ be obtained from $\mathfrak{H}$ by adding relation, operation, attribute, closure, or weight symbols while leaving all existing symbols and interpretations unchanged. Then*

$$\mathrm{Expr}(\mathfrak{H}) \leq \mathrm{Expr}(\widetilde{\mathfrak{H}})$$

*coordinatewise.*

*Proof.* No component of the profile can decrease when additional symbols are introduced while all previous symbols remain unchanged. The maximal arities likewise cannot decrease. □

**Proposition 7.3.8** (Effect of deleting an extensionally stored redundant layer)**.** *Suppose that a redundant layer $\lambda$ is stored extensionally and is reconstructed from the remaining layers rather than stored in the reduct. Then*

$$\mathrm{DL}\left(\mathfrak{H}\restriction_{\Lambda\setminus\{\lambda\}}\right) \leq \mathrm{DL}(\mathfrak{H}).$$

*The inequality is strict whenever the removed layer contributes a nonempty interpreted graph or relation to the extensional representation.*

*Proof.* The reduct retains the same carriers and all interpreted data outside the removed layer. The terms contributed by $\lambda$ are deleted from the sum defining DL, and no additional extensionally stored data are introduced. The conclusion follows. □

**Proposition 7.3.9** (Strong-embedding preorder)**.** *Fix a CHGS signature. Define*

$$\mathfrak{H}_1 \preccurlyeq_{\mathrm{emb}} \mathfrak{H}_2$$

*if there exists a strong CHGS embedding*

$$\mathfrak{H}_1 \hookrightarrow \mathfrak{H}_2.$$

*Then*

$$\preccurlyeq_{\mathrm{emb}}$$

*is a preorder.*

*Proof.* Reflexivity follows from the identity strong embedding.

For transitivity, suppose

$$\mathfrak{H}_1 \hookrightarrow \mathfrak{H}_2$$

and

$$\mathfrak{H}_2 \hookrightarrow \mathfrak{H}_3$$

are strong embeddings. Their componentwise composite is injective. Relation reflection, operation reflection, attribute reflection, closure compatibility, and weight preservation are all stable under composition. Hence the composite is again a strong embedding. Therefore

$$\mathfrak{H}_1 \preccurlyeq_{\mathrm{emb}} \mathfrak{H}_3.$$

□

**Remark 7.3.10** (Comparison of CHGS models)**.** *For CHGSs over a fixed signature, several complementary viewpoints may be used:*

- $$\preccurlyeq_{\mathrm{emb}}$$

  *describes structural embeddability;*

- $$\mathrm{Expr}(\mathfrak{H})$$
  *summarizes the native typed signature;*

- $$\mathrm{Comp}_G(\mathfrak{H})$$
  *measures operational generation from a specified primitive family;*

- $$\mathrm{DL}(\mathfrak{H})$$
  *provides a coarse extensional description-size proxy.*

*These quantities address different aspects of a CHGS and should not be collapsed into a single universal notion of expressiveness or learnability.*

# 8 Conclusion

This book presented a comprehensive overview of mathematical notions that can be used to model higher-order networks. It surveyed foundational concepts, established extensions, and newly introduced formalisms, with an emphasis on their structural principles, relationships, and modeling roles.

We expect that future research will further develop these concepts through uncertainty-aware extensions based on Fuzzy Graphs[531, 83], Intuitionistic Fuzzy Graphs[770, 532], Neutrosophic Graphs[536, 771, 537], and Plithogenic Graphs[772, 773]. In particular, it is natural to study how membership/non-membership (and indeterminacy) degrees can be lifted from vertices and edges to higher-order objects such as hyperedges, simplices, supervertices, and even meta-nodes in iterated constructions, while preserving basic consistency axioms and functorial behavior under morphisms.

From a computational perspective, promising directions include designing scalable algorithms for (i) higher-order centrality and community detection, (ii) influence/flow propagation on multiway interactions, and (iii) learning-based inference on higher-order data (e.g., message passing on factor/Tanner-style representations, or tensor-based contractions for structured multiway dependencies). We also anticipate broader applications in decision-support systems (group evaluation and consensus under multi-criteria constraints), machine learning (hypergraph and simplicial neural models), and other domains where complex higher-order interactions must be modeled and analyzed.

# Disclaimer

## Funding

This study did not receive any financial or external support from organizations or individuals.

## Acknowledgments

We extend our sincere gratitude to everyone who provided insights, inspiration, and assistance throughout this research. We particularly thank our readers for their interest and acknowledge the authors of the cited works for laying the foundation that made our study possible. We also appreciate the support from individuals and institutions that provided the resources and infrastructure needed to produce and share this book. Finally, we are grateful to all those who supported us in various ways during this project.

## Data Availability

This research is purely theoretical, involving no data collection or analysis. We encourage future researchers to pursue empirical investigations to further develop and validate the concepts introduced here.

## Ethical Approval

As this research is entirely theoretical in nature and does not involve human participants or animal subjects, no ethical approval is required.

## Use of Generative AI and AI-Assisted Tools

We use generative AI and AI-assisted tools only for limited tasks such as English grammar checking, and we do not use them in any way that violates ethical standards.

## Conflicts of Interest

The authors confirm that there are no conflicts of interest related to the research or its publication.

## Disclaimer

This work presents theoretical concepts that have not yet undergone practical testing or validation. Future researchers are encouraged to apply and assess these ideas in empirical contexts. While every effort has been made to ensure accuracy and appropriate referencing, unintentional errors or omissions may still exist. Readers are advised to verify referenced materials on their own. The views and conclusions expressed here are the authors' own and do not necessarily reflect those of their affiliated organizations.

# Appendix (List of Tables)

*

# Appendix (List of Figures)

*

## ABOUT THIS BOOK

Classical graphs capture only pairwise relationships, yet countless real-world systems involve multiway, hierarchical, temporal, multilayer, recursive, and tensor-based interactions. This survey brings together the many mathematical formalisms developed to model such higher-order networks, from hypergraphs and superhypergraphs to simplicial complexes, tensor networks, and beyond.

Now in its 4th Edition, this volume offers an expanded and corrected account of foundational concepts, established extensions, and newly introduced structures, with a unifying emphasis on how these models relate to one another. It is intended as a reference for researchers and students comparing higher-order network models across combinatorics, topology, and applied mathematics.

### REFERENTS

Salah Bouzina
Maikel Vazquez

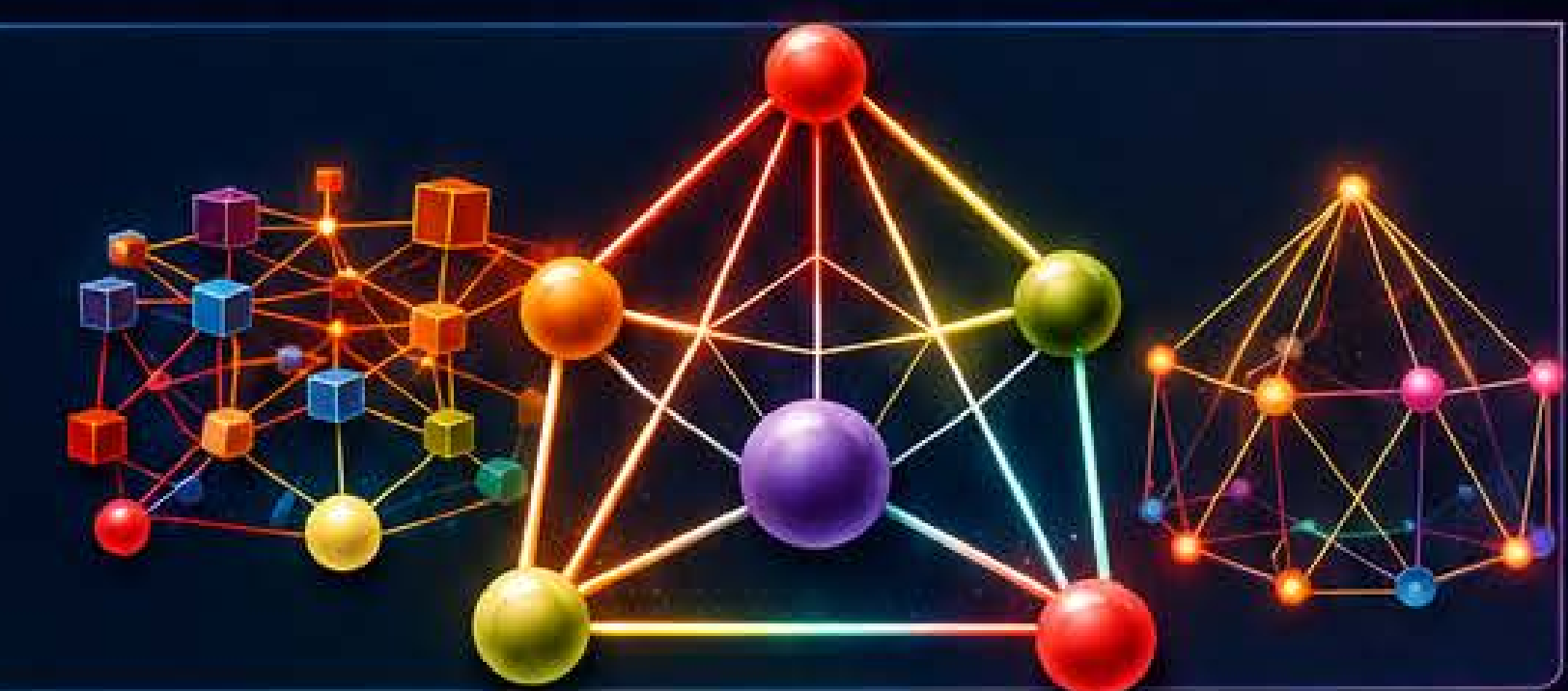